\documentclass[a4paper,11pt, twoside]{article}

\usepackage{setspace}

\newenvironment{changemargin}[2]{%
\begin{list}{}{%
\setstretch{1.35}
\setlength{\voffset}{-0.75in}
\setlength{\leftmargin}{#1}%
\setlength{\rightmargin}{#2}%
}%
\item[]}{\end{list}}

\pdfoutput=1

\usepackage[bf]{caption}
\usepackage{enumitem}
\usepackage[]{subfigure}
\usepackage[latin1]{inputenc}
\usepackage[english]{babel}
\usepackage{amssymb}
\usepackage{amsmath}
\usepackage[]{graphicx}
\usepackage[]{subfigure}
\usepackage[titletoc]{appendix}
\usepackage{tensor}
\usepackage{color}
\usepackage{cancel}
\usepackage{setspace}
\usepackage{fancyhdr}
\usepackage{etaremune}
\numberwithin{equation}{section}
\usepackage{epigraph}
\usepackage{changes}
\definechangesauthor[name=Alice, color=blue]{ADT}
\usepackage{subfiles}
\usepackage{amsthm}
\usepackage[hypertexnames=false]{hyperref}
\hypersetup{linktocpage, colorlinks=true, plainpages = false, citecolor = blue,  linkcolor=blue, urlcolor = blue, filecolor = blue}

\begin{document}
\allowdisplaybreaks

\pagenumbering{roman}

{\large
	
	\begin{center}\parindent0pt
		
		\vspace*{0cm}
		
		{\Large \bf    Path integrals, saddle points 
		and the beginning of the universe}\medskip\\[25pt]
		
			by\\[15pt]
	Alice Di Tucci \\[45 pt]

	\end{center}

\noindent Dissertation\\	
zur Erlangung des akademischen Grades\\
doctor rerum naturalium (Dr. rer. nat.)\\
im Fach Physik\\
Spezialisierung: Theoretische Physik\\
eingereicht an der\\
Mathematisch-Naturwissenschaftlichen Fakult\"at \\
der Humboldt-Universit\"at zu Berlin\\
Pr\"asidentin der Humboldt-Universit\"at zu Berlin:
Prof. Dr.-Ing. Dr. Sabine Kunst\\
Dekan der Mathematisch-Naturwissenschaftlichen Fakult\"at:
Prof. Dr. Elmar Kulke\\[25 pt]
\\[45 pt]

\begin{center}

\includegraphics[scale=.35]{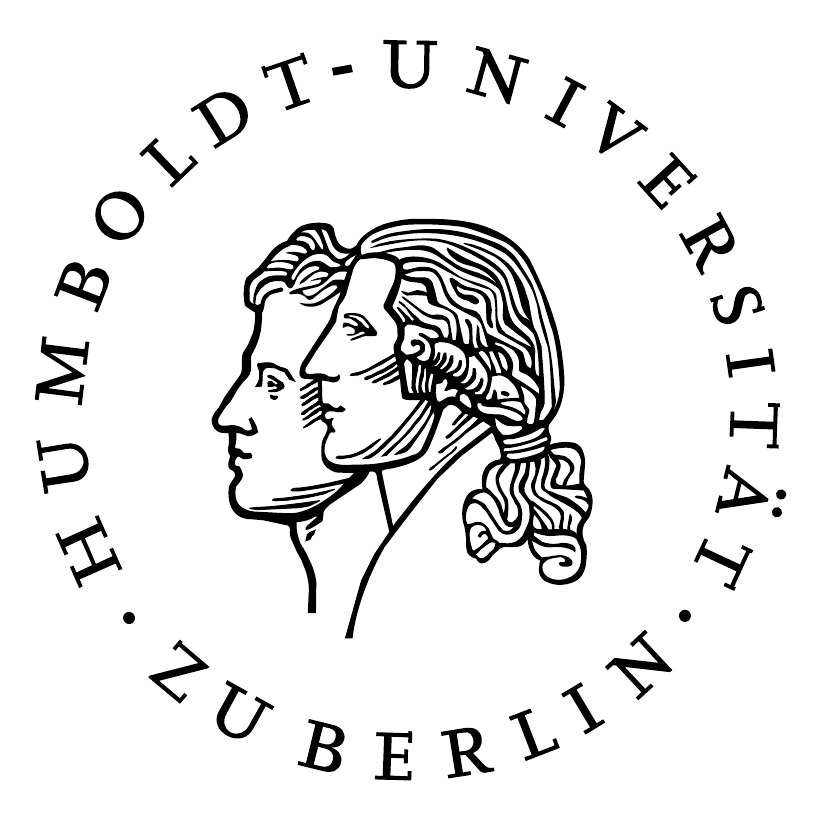}

\end{center}

\thispagestyle{empty}

\clearpage

\noindent Tag der Disputation am 10.08.2021\\
Betreuer:
Prof. Dr. Hermann Nicolai, Dr. Jean-Luc Lehners \\
Gutachter:
Prof. Dr. Hermann Nicolai, Prof. Dr. Claus Kiefer, Dr. Olaf Ohom 	\\	
}

\clearpage

\shipout\null

\vfill

\epigraph{If you wish to make an apple pie from scratch, you must first invent the universe.}{\textit{Carl Sagan}}

\clearpage

\shipout\null

%%%%%%%%%%%%%%%%%%%%%%%%%%%%%%%%%%%%%%%%%%%%%%%%%%%%%

\section*{Abstract}

%1 page abstract in English ad German

%the early universe cosmology qft in curved space time. this deicate limit not well understood. we calculate explicitly the saddle point approximation of integrals in the minisuperspace approximation. directly for cosmology, no boundary proposal. Somewhat indirectly studying negative lambda to connect with holography. Our main finding is that neumann works for everyone and is a regularity condition.

The very early universe is successfully described by quantum field theory in curved spacetime where the classical background spacetime is typically an FLRW cosmology and the quantum fields which propagate on it include gravitational waves and energy density fluctuations.
This regime is however little understood from a theoretical point of view because part of the gravitational degrees of freedom, and only part of them, are quantized. In this work we study this limit by assigning quantum properties both to the background universe and the fluctuations and then focusing on the limit where the background universe behaves nearly classically. The quantization is realized in the framework of quantum general relativity through Feymann's path integrals.
We study the saddle point approximation of gravitational path integrals in the cases of a positive and a negative cosmological constant making use of the minisuperspace approximation.
Our first findings are two important negative results concerning path integrals with Dirichlet boundary conditions: inflation does not allow for Bunch-Davies initial conditions if no pre-inflationary phase is admitted and the no boundary proposal is ill-defined as a sum of regular geometries which start at zero size. This motivates us to study the impact on the path integral of other classes of boundary conditions such as those of Neumann and Robin types. 
We find that Robin types of boundary conditions can be used to reconciliate inflation with the Bunch-Davies initial condition and are also useful to describe large homogeneous scalar field fluctuations in the eternal inflation regime.
Our main finding is that, for both the no boundary proposal and black holes in Euclidean anti-de Sitter space, the path integral needs to be defined with Neumann initial conditions. The Neumann condition is in fact necessary to recover sensible black holes thermodynamics and to stabilize the no boundary proposal. At the same time, it can be seen, in both cases, as a regularity requirement on the geometries entering the sum. The need for Neumann conditions implies that the interpretation of the no boundary wavefunction is very different from Hartle and Hawking's original intuition, since the initial expansion rate of the universe is fixed rather than its size. Our results for black holes stands in support of this implementation of the no boundary proposal, where regularity is the primary requirement, and allows for a well-defined QFT in curved spacetime limit. Moreover, in the case of black holes, we find that when the asymptotic AdS spacetime is cut off at a finite radius additional saddle points contribute to the path integral. The possibility of testing this result in the dual picture gives an element of falsifiability to the minisuperspace approximation, crucial for the reliability of the entire paradigm.

\clearpage

\begin{changemargin}{0cm}{-0.48cm}
\section*{Zusammenfassung}
\noindent Das sehr fr\"{u}he Universum wird erfolgreich durch die Quantenfeldtheorie in gekr\"{u}mmter Raumzeit beschrieben, wobei die klassische Hintergrundraumzeit typischerweise eine FLRW-Kosmologie ist und die Quantenfelder, die sich darauf ausbreiten, beinhalten Gravitationswellen und Energiedichtfluktuationen. 
Dieses Regime ist jedoch aus theoretischer Sicht wenig verst\"{a}ndlich, da nur ein Teil der Gravitationsfreiheitsgrade, und nur ein Teil davon, quantisiert werden. 
In dieser Arbeit untersuchen wir diese Begrenzung durch das Zuweisen von Quanteneigenschaften, sowohl zum Hintergrunduniversum als auch zu den Fluktuationen und konzentrieren uns dann auf dem Limes, an der sich das Hintergrunduniversum fast klassisch verh\"{a}lt. 
Die Quantisierung wird im Rahmen der allgemeinen Quantenrelativit\"{a}t durch Feymanns Pfadintegralen beschrieben. 
Wir untersuchen die Sattelpunktsn\"{a}herung von Gravitationspfadintegralen in den F\"{a}llen einer positiven und einer negativen kosmologischen Konstante, unter Benutzung der Minisuperspace Ann\"{a}herung. 
Unsere ersten Ergebnisse sind zwei wichtige negative Ergebnisse in Bezug auf Pfadintegrale mit Dirichlet-Randbedingungen: Die Inflation erlaubt keine Bunch-Davies Anfangsbedingungen, wenn keine pr\"{a}inflation\"{a}re Epoche zugelassen ist und der no-boundary Vorschlag 
%(der Vorschlag ``ohne Grenzen") 
als Summe regul\"{a}rer Geometrien, die bei einer Gr\"{o}{\ss}e von Null beginnen, schlecht definiert ist.
Dies motiviert uns, die Auswirkungen auf das Pfadintegral anderer Klassen von Randbedingungen, wie Neumann- und Robin-Arten zu untersuchen.
Wir finden heraus, dass Robin Randbedingungen verwendet werden k\"{o}nnen, um die Inflation mit der Bunch-Davies Anfangsbedingung in Einklang zu bringen und auch n\"{u}tzlich sind, um gro{\ss}e homogene Skalarfeldschwankungen im Ewigen Inflationsregime zu beschreiben.
Unsere wichtigste Erkenntnis ist, dass sowohl f\"{u}r den no-boundary Vorschlag als auch f\"{u}r die Schwarzen L\"{o}cher im Euklidischen anti-De-Sitter Raum das Pfadintegral mit Neumann-Randbedingungen ben\"{o}tigt wird.
Die Neumann-Bedingung ist in der Tat 
notwendig, um eine vern\"{u}nftige Thermodynamik f\"{u}r Schwarze L\"{o}cher wiederherzustellen und um den no-boundary Vorschlag zu stabilisieren.
Gleichzeitig kann man dies in beiden F\"{a}llen als Regelm\"{a}{\ss}igkeitsanforderung an die Geometrien zusammenfassen.
Die Notwendigkeit von Neumann-Bedingungen impliziert, dass sich die Interpretation der no-boundary Wellenfunktion stark von Hartles und Hawkings urspr\"{u}nglicher Idee unterscheidet, da die anf\"{a}ngliche Expansionsrate des Universums eher als seine Gr\"{o}{\ss}e bestimmt wird.
Unsere Ergebnisse f\"{u}r Schwarze L\"{o}cher unterst\"{u}tzen diese Implementierung des no-boundary Vorschlags, bei dem Regelm\"{a}{\ss}igkeit die Hauptanforderung ist, und erm\"{o}glichen einen genauen definierten Limes der QFT im gekr\"{u}mmten Raumzeit.
Dar\"{u}ber hinaus stellen wir im Fall von Schwarzen L\"{o}chern fest, dass zus\"{a}tzliche Sattelpunkte zum Pfadintegral beitragen, wenn die asymptotische AdS Raumzeit mit einem endlichen Radius abgeschnitten wird.
Die M\"{o}glichkeit, dieses Ergebnis im dualen Bild zu testen, verleiht der Minisuperspace Ann\"{a}herung ein Element der Falsifizierbarkeit, das f\"{u}r die Zuverl\"{a}ssigkeit des gesamten Paradigmas entscheidend ist.

\end{changemargin}

\clearpage

%%%%%%%%%%%%%%%%%%%%%%%%%%%%%%%%%%%%%%%%%%

\section*{Declaration of Authorship}

I hereby confirm that I have authored this PhD
thesis independently and without use of others than the indicated
sources. All passages which are literally or in general matter
taken out of publications or other sources are marked as such. I declare that I have completed the thesis independently using only the aids and tools specified. I have not applied for a doctor's degree in the doctoral subject elsewhere and do not hold a corresponding doctor's degree. I have taken due note of the Faculty of Mathematics and Natural Sciences PhD Regulations, published in the Official Gazette of Humboldt-Universit\"{a}t zu Berlin no. 42/2018 on 11/07/2018.
\vspace{1cm}

\noindent Berlin, March 16, 2021 \vspace{0.5cm}

\noindent Alice Di Tucci

\clearpage

\shipout\null

%%%%%%%%%%%%%%%%%%%%%%%%%%%%%%%%%%%%%%%%%%%%%%

\tableofcontents 

 \clearpage

\shipout\null

%%%%%%%%%%%%%%%%%%%%%%%%%%%%%%%%%%%%%%%%%%%%%%%%%%%

\clearpage

%%%%%%%%%%%%%%%%%%%%%%%%%%%%%%%%%%%%%%%%%%%%%%%%%%%%%%%%%%%%%%%%%%

\addcontentsline{toc}{section}{List of publications}

\section*{Pubblications}

\noindent This thesis is based on the following publications:\\

\begin{etaremune}

\item \cite{DiTucci:2018fdg} A. Di Tucci, J. L. Lehners, \textit{Unstable no-boundary fluctuations from sums over regular metrics}, Phys.Rev.D 98 (2018) 10, 103506, \href{https://arxiv.org/abs/1806.07134}{arXiv: 1806.07134 [gr-qc]} \label{p6}\\

\item \cite{DiTucci:2019dji} A. Di Tucci, J.L. Lehners, \textit{No-Boundary proposal as a Path Integral with Robin Boundary Conditions}, Phys.Rev.Lett. 122 (2019) no.20, 201302, \href{https://arxiv.org/abs/1903.06757}{arXiv: 1903.06757[hep-th]} \label{p2} \\

\item \cite{DiTucci:2019xcr} A. Di Tucci, J. Feldbrugge, J.L. Lehners, N. Turok, \textit{Quantum incompleteness of Inflation}, Phys.Rev.D 100 (2019) no.6, 063517, \href{https://arxiv.org/abs/1906.09007}{arXiv: 1906.09007 [hep-th]} \label{p3} \\

\item \cite{Bramberger:2019zks} S.F. Bramberger, A. Di Tucci, J.L. Lehners, \textit{Homogeneous transitions during Inflation: a Description in Quantum Cosmology}, Phys.Rev.D 101 (2020) no.6, 063501,
\href{https://arxiv.org/abs/1907.05782}{ arXiv: 1907.05782 [gr-qc]} \label{p4}\\
 
  \item \cite{DiTucci:2019bui} A. Di Tucci, J.L. Lehners, L. Sberna, \textit{No-boundary prescriptions in Lorentzian quantum cosmology}, Phys.Rev.D 100 (2012) no.12, 123543, \href{https://arxiv.org/abs/1911.06701}{arXiv: 1911.06701 [hep-th]} \label{p5}\\
  
    \item \cite{DiTucci:2020weq} A. Di Tucci, M.P. Heller, J.L. Lehners, \textit{Lessons for quantum cosmology from anti-de Sitter black holes}, Phys.Rev.D 102 (2020) 8, 086011, \href{https://arxiv.org/abs/2007.04872}{arXiv: 2007.04872[hep-th]} \label{p6}\\
  \vspace{-.5\baselineskip}  
  
\end{etaremune}

\clearpage 

\shipout\null

%%%%%%%%%%%%%%%%%%%%%%%%%%%%%%%%%%%%%%%%%%%%%%%%%%%%%%%%%%%%%

%\setcounter{page}{1}  
\pagenumbering{arabic} 

\section{Introduction}

\begin{figure}
    \centering
    \includegraphics[width = 6.9 cm]{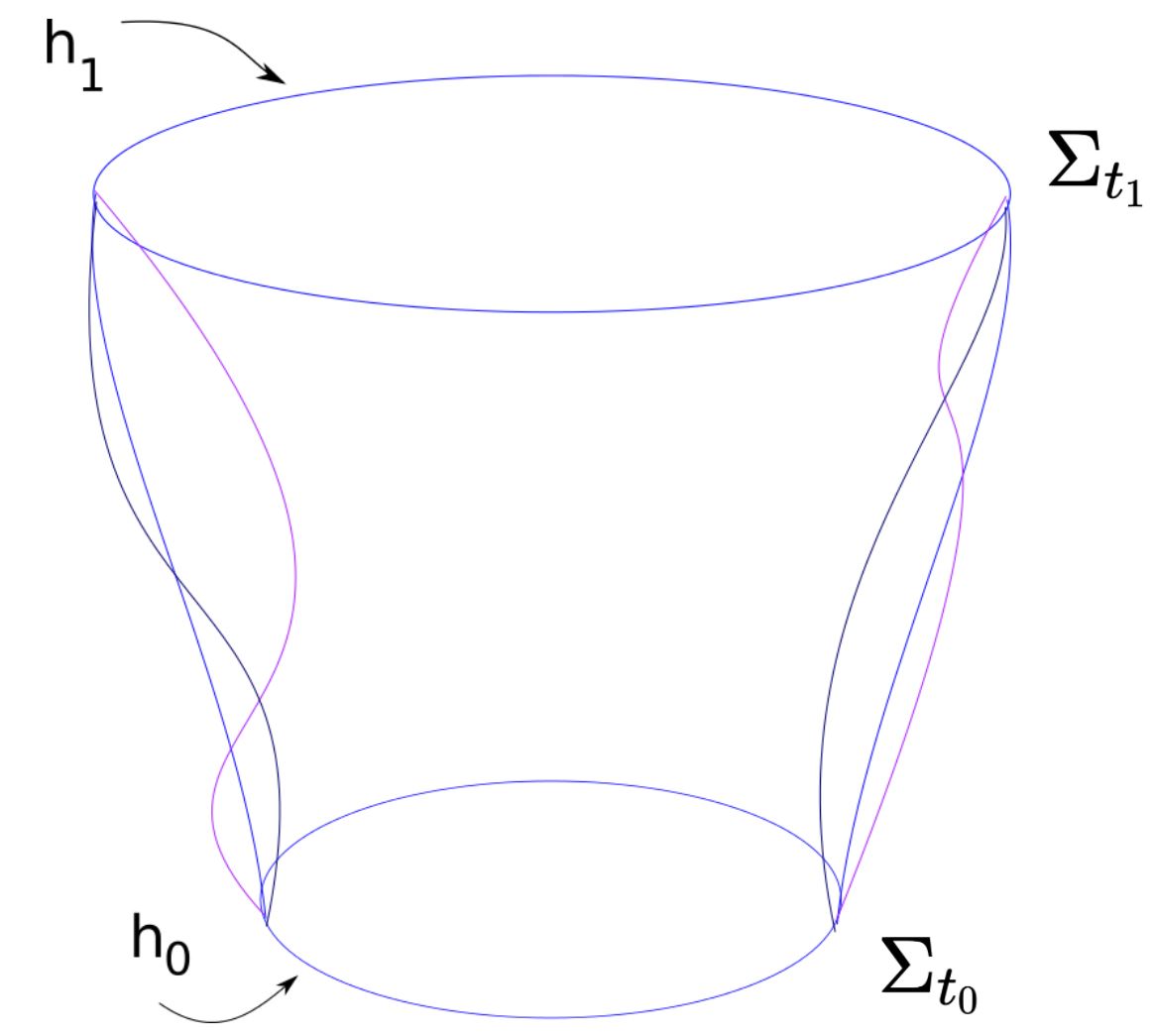}
    \hspace{1.1em}
        \includegraphics[width = 7.5 cm]{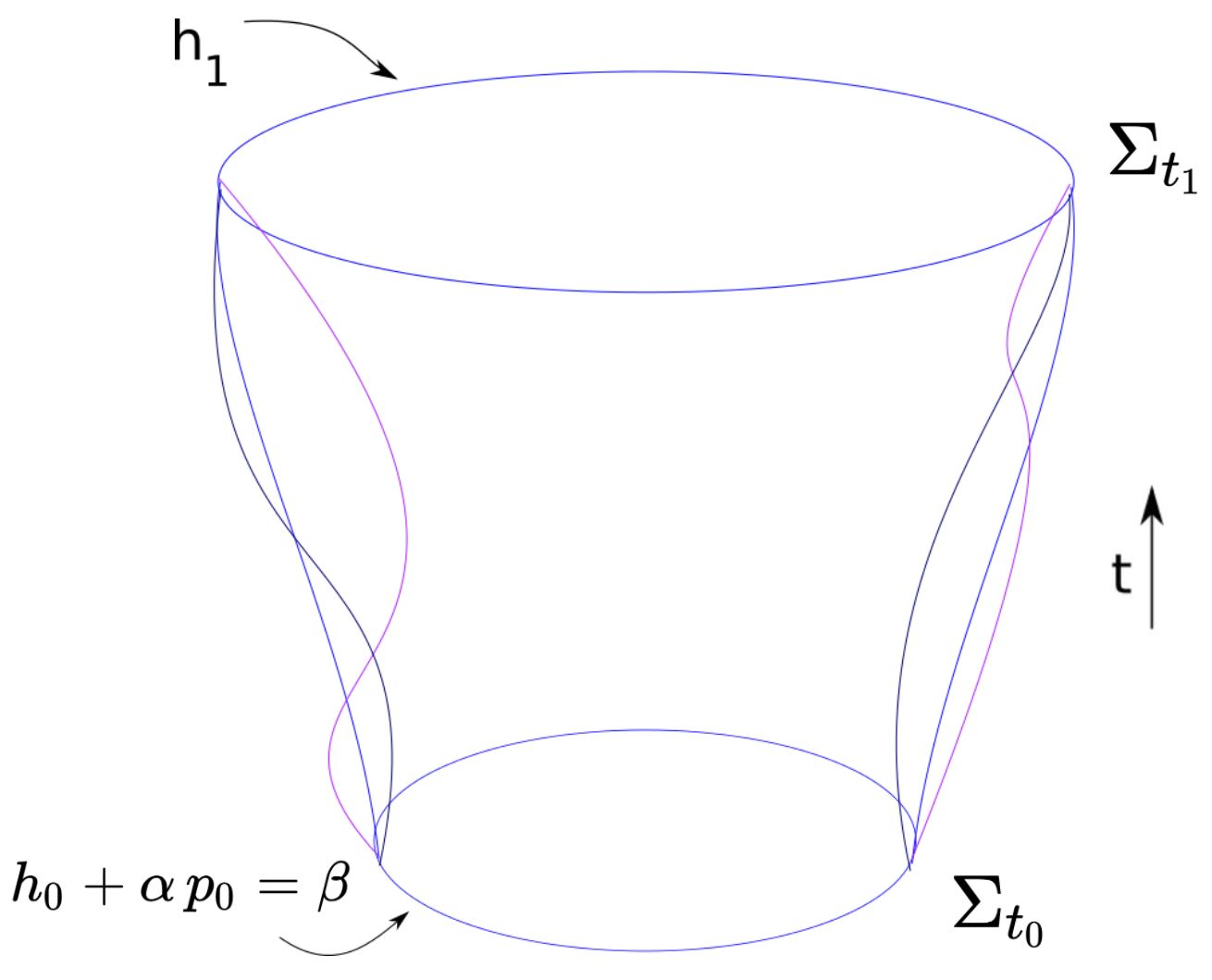}\\
    \vspace{1cm}
          \includegraphics[width = 7 cm]{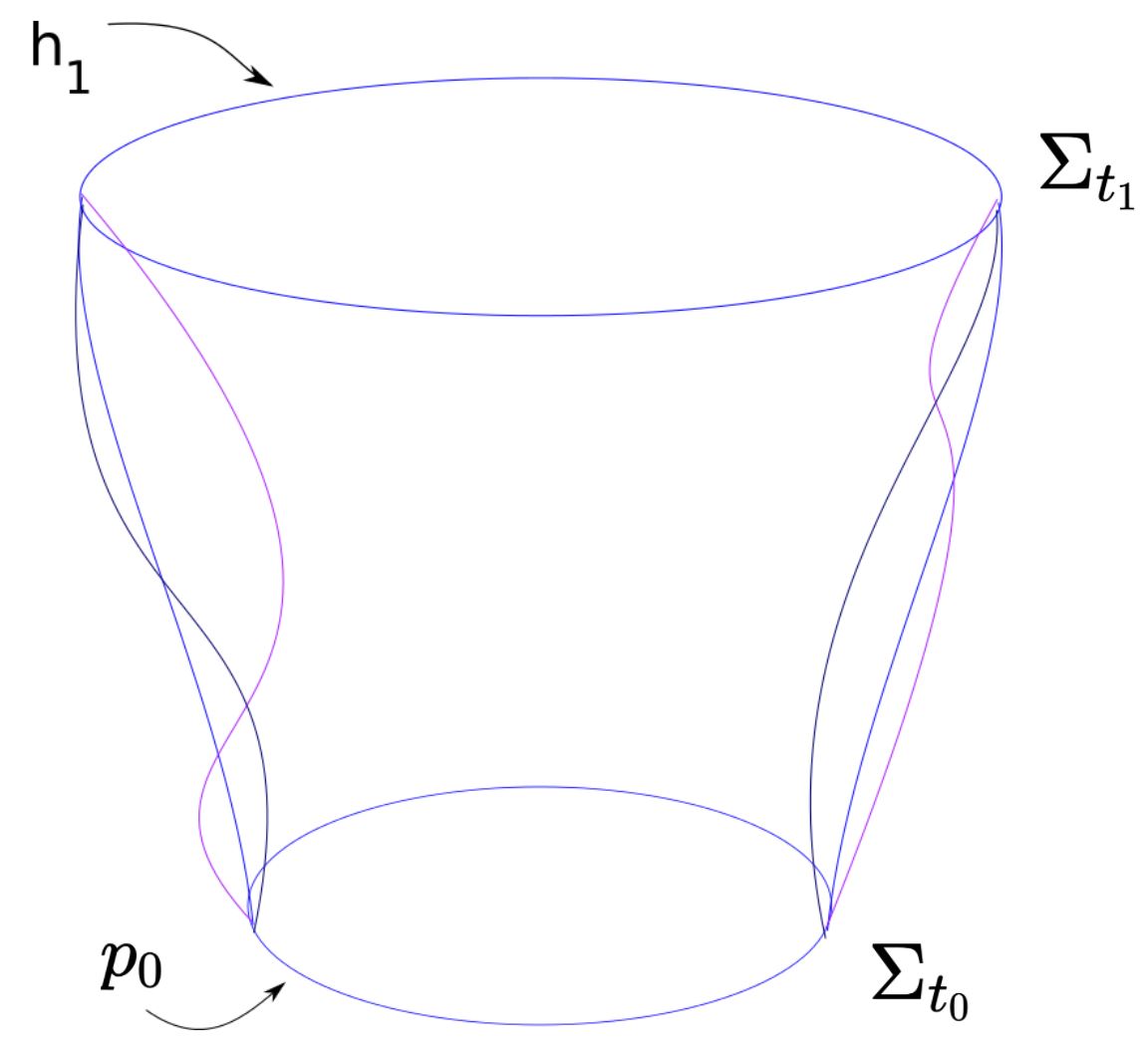}
           \hspace{1em}
           \includegraphics[width = 6.7 cm]{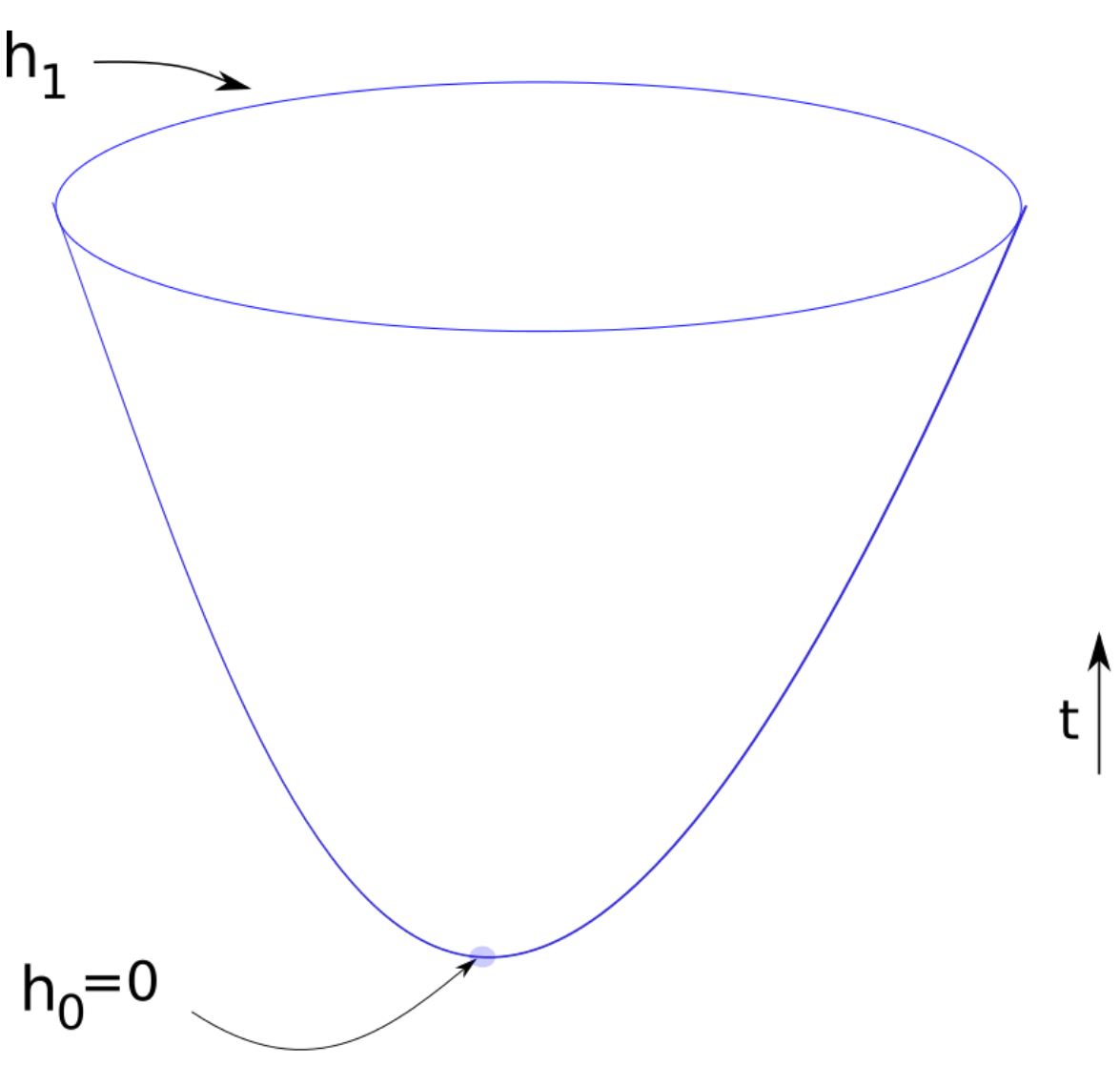}
    \caption{The wavefunction of the unvierse $\Psi[h_1]$ and partition function $Z[h_1]$ are functionals of the three-geometry $h_1$. The path integral is given by the sum over four-geometries which interpolate between $h_1$ and the quantity fixed on $\Sigma_{t_0}$ which can be taken to be the induced metric $h_1$ (top left panel), the conjugate momentum $p_0$ (bottom left panel) or a linear combination of the two (top right panel where $\alpha$ and $\beta$ are constants.). The pictures are intended to clarify our setting but do not capture some of the features of the path integral: the sum over four-geometries includes also a sum over proper time separations between the two surfaces and in the cases of Neumann and Robin conditions contains geometries of different initial sizes. The saddle point geometry shall have no boundary other that $\Sigma_{t_1}$ (bottom right panel).}.
\label{ideafig}
\end{figure}

\setcounter{page}{1}
In a classical world, there is only one real trajectory which links two events: the one that minimizes the action. In the quantum world, the probability for the transition is associated with the sum over all possible paths linking the two boundary configurations. 
%Each such path is weighted with a factor of $e^{i S/\hbar}$ which depends on the action $S$ evaluated along the path and the probability of the transition is given by the modulus of the path integral.
In the semi-classical limit the full sum is well approximated by a single trajectory, a solution to the classical equations of motion. If the boundary conditions are classically forbidden, the solution will be complex giving a quantum weighting to the transition.  What happens if we apply these concepts to the entire universe? Very large progress in understanding the semi-classical properties of gravity has been done by looking at the contribution of the dominant saddle point of gravitational path integrals. We talk of semi-classical gravity as the regime where part of the system is well described by the saddle point approximation of the path integral while the rest is not. The classical part of the system provides a notion of classical spacetime where the quantum part of the system lives and quantum field theory (QFT) in curved spacetime applies. 
%In particular we talk of a semi-classical limit when part of the system behaves classically while the rest doesn't. In this case the rest of the world could be use to make the background classical? cite Kiefer?. This is a very important limit?, because it is the one where .. and the quantumness of the system is there because the solution to the classical equations of motion can be complex. ?\todo{levare sta frase?}
This is the case for example of the no boundary \cite{Hawking:1981gb, Hartle:1983ai, Hawking:1983hj, Hartle:2008ng} and tunneling \cite{Vilenkin:1982de, Vilenkin:1983xq, Vilenkin:1984wp, Vilenkin:1986cy} proposals in cosmology and the Hawking-Page phase transition with the associated thermodynamic properties for what concerns black holes \cite{Hawking:1982dh, Witten:1998zw, Hawking:1974rv, Hawking:1974sw, Bardeen:1973gs, Hartle:1976tp, Bousso:1996au}. Another notable framework where these concepts are truly foundational is that of the AdS/CFT correspondence~\cite{Maldacena:1997re,Gubser:1998bc,Witten:1998qj}, where the saddle point approximation of the gravity path integral gives the partition function of the dual quantum field theory, in the appropriate limit. The holographic principle is by now widely applied in cosmology too \cite{Hertog:2004rz, Hertog:2005hu, Harlow:2010az, Maldacena:2010un, McFadden:2009fg, Casalderrey-Solana:2020vls, Hertog:2011ky}.  In all these cases there is an open question how the full integral is defined and under what conditions the sum over all paths is well approximated by a specific saddle point contribution. In this thesis we investigate such questions with particular focus on fundamental issues which find their setting in cosmology.
This type of questions are in fact crucial in cosmology since our current understanding of the very early universe is based on the treatment of background and perturbative gravitational degrees of freedom on a different footing within the semi-classical framework of QFT in curved space time~\cite{Mukhanov:1981xt,Starobinsky:1979ty,Starobinsky:1982ee, Guth:1982ec,Hawking:1982cz,Bardeen:1983qw, Lehners:2007ac, Khoury:2001wf, Khoury:2001zk, Gratton:2003pe, Boyle:2004gv}. The idea of our work is to test the validity of this assumption allowing for quantum properties of the background universe in the most conservative manner. We include in the path integral a sum over background geometries with a weight which depends on the action of general relativity only. We are after a systematic study of gravity path integrals within the minisuperspace approximation where we can handle the calculations in full detail~\cite{Halliwell:1988wc, Halliwell:1988ik, Halliwell:1989vu, Halliwell:1990tu, Kiefer:1990ms, Feldbrugge:2017kzv}. We focus on characterizing well-defined convergent integrals and look for their saddle point approximation to verify that in the semi-classical limit we indeed recover the standard description. As we will see, the outcome is in many cases somewhat unexpected. 
We will focus in particular on the impact of various boundary conditions on the path integral. In the case of AdS/CFT it is well known what the condition on the boundary of AdS means: with a Dirichlet condition we fix the geometry on which the dual QFT lives, while with different types of boundary conditions one can allow for this geometry to be dynamical \cite{Papadimitriou:2005ii, Compere:2008us}. To solve second order equations of motion two conditions are needed. In the case of AdS/CFT it is however obvious what the other condition should be: AdS spacetime has only one boundary and the second requirement in holographic calculations is that fields behave regularly in the interior. The situation is rather different in cosmology. Here, we have a future space-like boundary where the wavefunction of the universe lives, which could be for example the surface where inflation ends. Then any predictions of the model under study will depend on what condition one imposes on the past space-like boundary. We can think of cutting the spacetime in the past at some finite radius and fine tune the desired initial conditions there. Then the question arises of how these conditions where generated. One would need to consider another space-like surface where to fine tune the right conditions in order to get the desired initial conditions and the repeat the procedure for the new surface. If we think of the case of de Sitter space, it truly has two disjoint (past and future) boundaries and there is no way out of this argument. The idea of Hartle and Hawking is to cut out the problem of initial conditions all together by closing off de Sitter space in such a way that it has only one boundary, the future boundary where the wavefunction of the universe takes values (see the bottom right panel of Fig. \ref{ideafig}). This is the background saddle point geometry which shall approximate the wavefunction of the universe in the semi-classical limit. Then a regularity condition is naturally imposed on the fields living on this geometry in resemblance to the case of AdS. The question we ask in this work is how to construct path integrals which admit such saddle point approximation.  We will make use of an ADM splitting of spacetime~\cite{Arnowitt:1959ah} which allows us to identify two surfaces and evaluate the sum over histories which interpolated between the induced quantities fixed on the two surfaces. The idea is sketched in Fig. \ref{ideafig} where the parameter $t$ used to slice the spacetime is a time-like variable in the case of a positive cosmological constant and can be thought of as a radial variable if the cosmological constant is negative. One surface will be the future boundary for the wavefunction of the universe or the ``outer'' boundary for a partition function in asymptotically AdS spacetime. We called this surface $\Sigma_{t_1}$ in Fig. \ref{ideafig} and the wavefunction or partition function are functionals of the three-geometry $h_1$ induced on this surface. We will study boundary conditions on other surface, $\Sigma_{t_0}$ in the figure, where in both cases we need to enforce the requirement that the saddle point geometry does not have any boundary other that $\Sigma_{t_1}$. We will discuss in particular Dirichlet, Robin and Neumann type of boundary conditions fixed on this past or ``inner'' boundary surface, for a positive and negative cosmological constant respectively, corresponding to fixing the three-metric $h_0$ induced on $\Sigma_{t_0}$, its conjugate momentum $p_0$ or a linear combination of the two. The question is what type of condition gives the desired saddle point approximation of the path integral. As we will see this will not correspond to the requirement that all of the geometries in the sum have no boundary. 

We will start, in chapter \ref{introchapter}, by reviewing well-understood aspects of cosmology which will serve as a motivation for our work. We discuss the standard model of cosmology and highlight some of the open questions related to it. Then we introduce inflation as a possible answer to such questions. In chapter \ref{sec:quantum} we review the methods of canonical and path integral quantization of the gravitational field which we will use throughout the thesis. In chapter \ref{quantuminitial} we see our method at work for the first time. We show how inflation is in fact not robust against quantum corrections since two background solutions contribute in general to the path integral leading to a breakdown to the QFT in curved spacetime picture and an instability of fluctuations. One way out of this problem is to provide a semi-classical explanation for the initial conditions of inflation which allows inflation to start and develop already well within the realm of QFT in curved spacetime, with perturbations in their Bunch-Davis vacuum. This explanation is 
possibly provided by the no boundary proposal, which we introduce in chapter \ref{nbDirichlet}. In chapters \ref{nbDirichlet}, \ref{robincosmology} and \ref{chapterneumann} we study minisuperspace path integrals with Dirichlet, Robin and Neumann boundary conditions, respectively. In chapter \ref{nbDirichlet}, we show that the Hartle-Hawking wavefunction cannot be represented as a path integral with Dirichlet boundary conditions. We provide possible path integral representations of the Hartle-Hawking wavefunction using Robin boundary conditions in chapter \ref{robincosmology}. In this chapter we show also how Robin boundary conditions can be used to describe large quantum fluctuations of the inflaton in the eternal inflation regime. In chapter \ref{chapterneumann} we propose what we believe is the best implementation of the no boundary proposal using Neumann boundary conditions for FLRW and the Bianchi IX model. In this case, the initial expansion rate of the universe is fixed rather than its size. Given that momentum and position are conjugate variables in quantum mechanics, fixing the momentum corresponds effectively to a sum over all possible initial sizes. We thus learn that a radical change in our understanding of the no boundary proposal is needed when the gravitational path integral is carefully analyzed. This motivates us to study similar Neumann path integrals for the case of a negative cosmological constant in the second part of chapter \ref{chapterneumann}, where we focus in particular on the Hawking-Page phase transition. As we will see, an analogous Neumann condition is needed to recover the usual thermodynamic description of black holes in anti-de Sitter space. We thus find a nice agreement on the requirements for a well-defined path integral for AdS/CFT and the no boundary proposal. We summarize our findings and discuss possible future research directions and extensions of our studies in chapter \ref{chapconclusion}.

Throughout this work we make use of the minisuperspace approximation to evaluate the saddle point approximation of gravitational path integrals.
Our main message is that the the saddle point approximation of a path integral is not merely obtained by taking one's favorite solution to Einstein's equation and evaluating $e^{i S/\hbar}$ using the action $S$ along the trajectory, as the approximation to some not better defined  
or definable sum over geometries. The point is that a single solution to the classical equations of motion can be a saddle point of different types of integrals and the quantum state of a system is really defined only by the full path integral. Focusing on the saddle point geometry only can in fact be misleading: in chapter \ref{chapterneumann} we will be interested in saddle points which correspond to black holes in Euclidean anti-de Sitter space. One might be tempted to think that the sum shall be given by a sum over Euclidean geometries but we will see this is not the case and the Euclidean integral is explicitly divergent. Similarly, the no boundary saddle point instanton has no boundary in the past but the integral cannot be thought of as a sum over geometries with no boundary. Can then one still think of the Hartle-Hawking quantum state as describing the nucleation of the universe out of nothing, as the saddle point geometry suggests?\\
We describe a number of minisuperspace path integrals with positive and negative cosmological constant specifying boundary conditions and integration contours and in each case we discuss the meaning of the sum and the shortcomings of the implementation. By the end of the thesis, the reader shall be able to associate, for the cases studied, a specific saddle point to an explicitly characterized integral. In this sense, we achieve what we think is a systematization of minisuperspace path integrals.

\clearpage

%%%%%%%%%%%%%%%%%%%%%%%%%%%%%%%%%%%%%%%%%%%%%%%%%%%%%%%%%%%%%%%%%%%%%%%%%%%%%%%%%%%%%%%%%%%

\section{Classical cosmology}\label{introchapter}

\begin{figure}
  \centering
      \includegraphics[width = 6 cm]{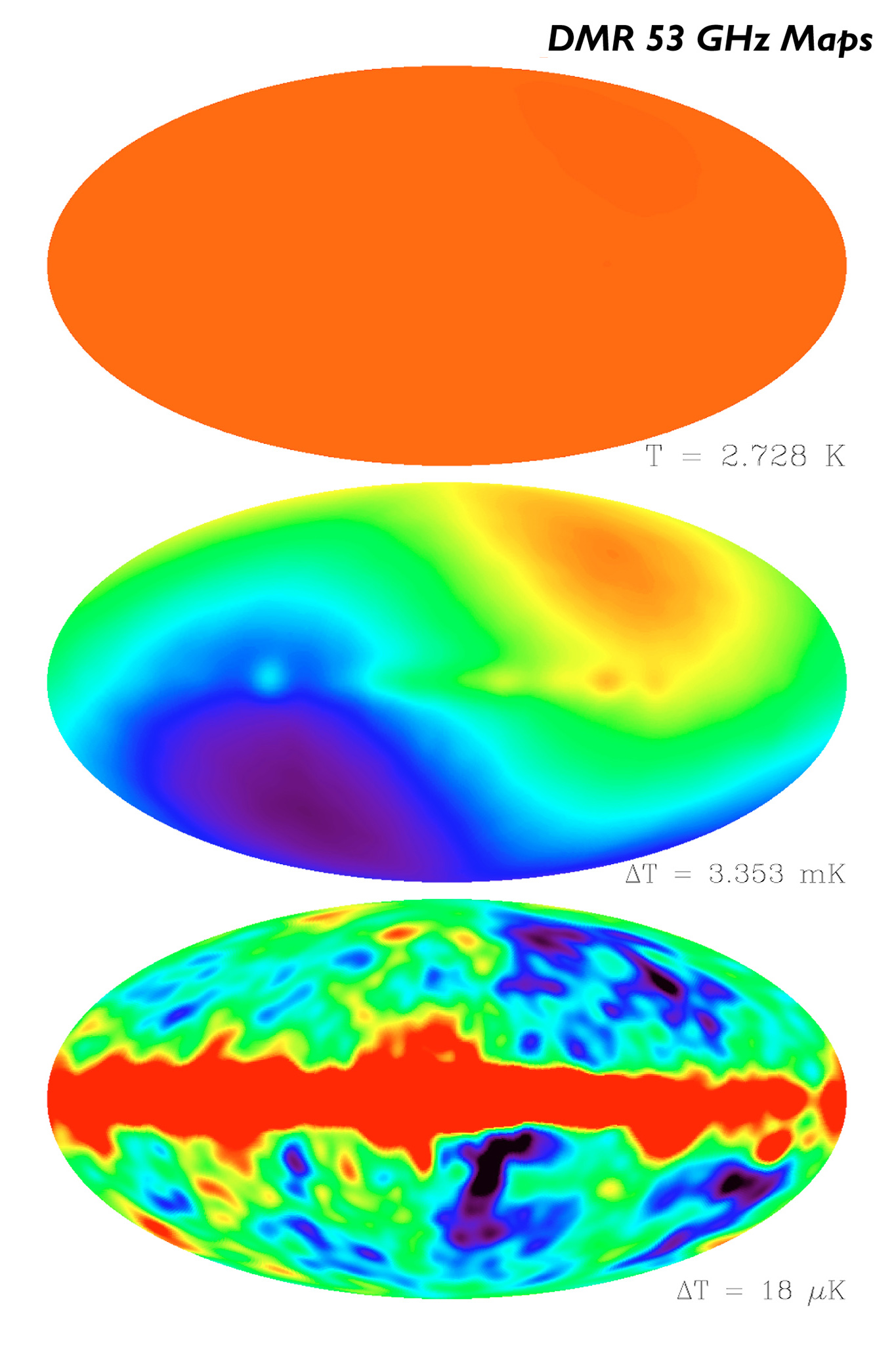}
    \includegraphics[width = 8 cm]{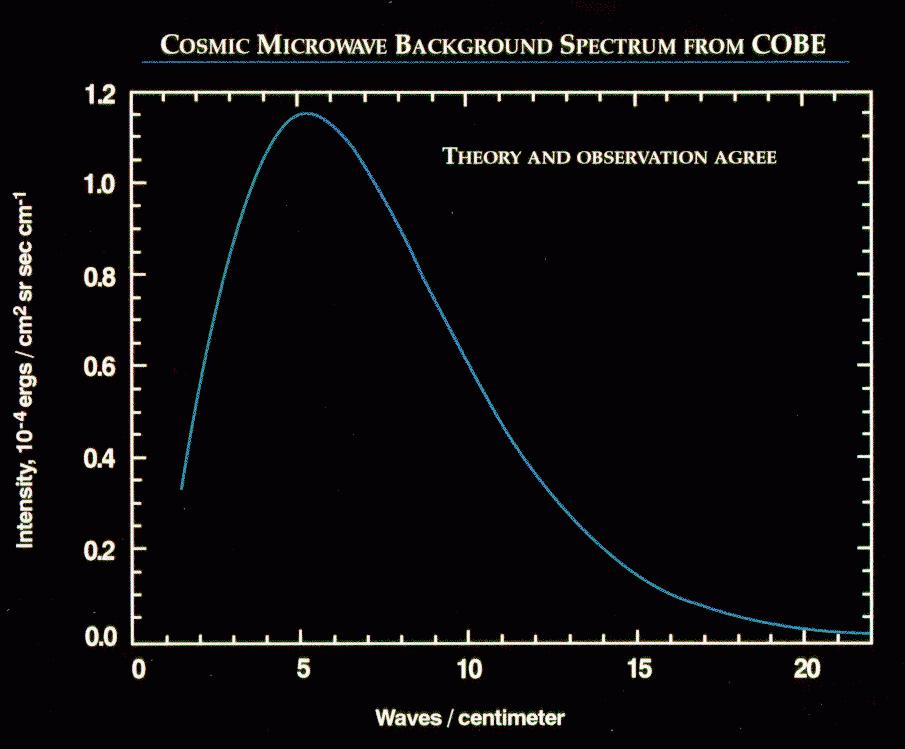}
    \caption{Left panel: The CMB temperature map as observed by COBE. Right panel: The CMB black body spectrum. Error bars are too small to be seen in this picture. The figures are taken from the NASA FIRAS/COBE website.}
    \label{Cobe}
\end{figure}

\begin{figure}
    \centering
     \includegraphics[width = 6 cm]{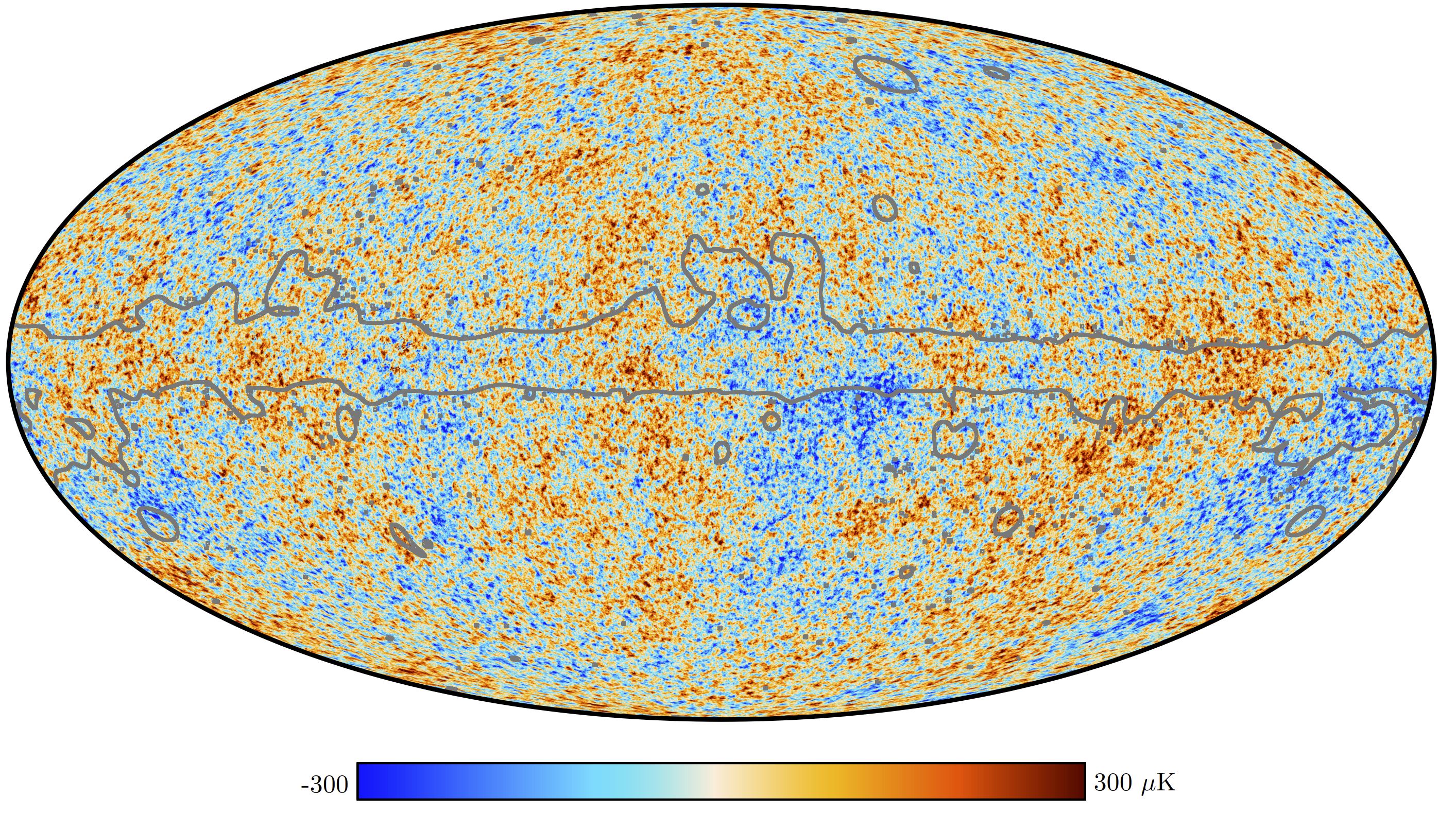}
      \includegraphics[width = 8 cm]{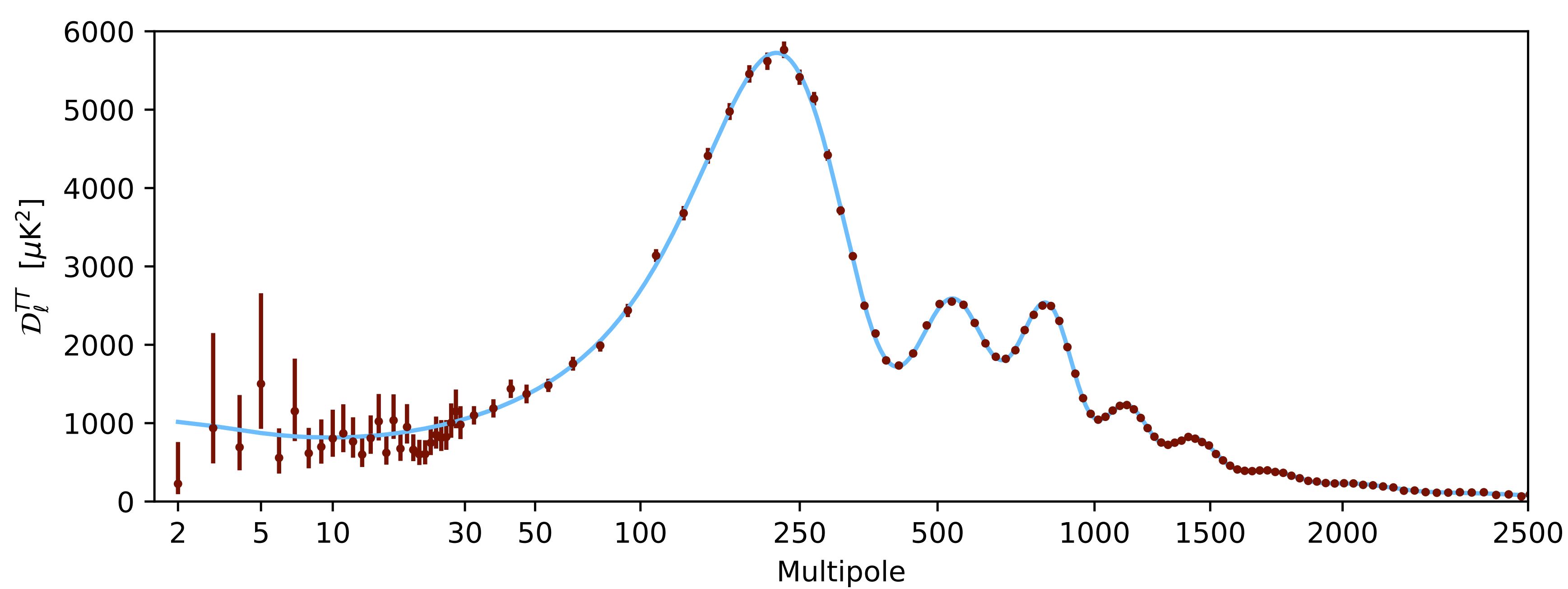}
    \caption{
   Left panel: The CMB temperature map as observed by PLANCK. The region in between the gray lines represents our galaxy. Right panel: The CMB temperature power spectrum.
    The figures are taken from \cite{Akrami:2018vks}.}
    \label{Planck}
\end{figure}

Cosmology studies the characteristics and the history of our universe as a whole. The starting point are the observed features of the universe as provided by the Cosmic Microwave Background radiation (CMB) and detected by the COBE, WMAP and PLANCK satellites \cite{Fixsen:1996nj, Bennett:2012zja, Akrami:2018vks}. The first relevant data for us comes from the COBE's observations and is showed in Fig. \ref{Cobe}. The left panel of the figure shows the temperature of the CMB on the full sky projected onto a oval. We learn that the temperature of the CMB is in first approximation the same in every direction and thus our universe is homogeneous and isotropic. This confirms the intuition given by the cosmological principle: we occupy no special place in the universe, in fact any point is just like any other. The right panel shows that the CMB spectrum matches very precisely that of a black body at T$= (2.72548 \pm 0.00057 )$K \cite{Mather:1993ij, Fixsen11} and represents a striking piece of evidence in support of the Big Bang theory for cosmology: the early universe was a very hot black body in thermal equilibrium where the low temperature detected today is due to the fact the universe cools down as it expands. 
The central figure in the left panel shows the first degree of anisotropy observed in the CMB and is a Doppler shift due to the motion of the Solar System with respect to the CMB. The second relevant piece of data for us is showed in the bottom left oval of Fig. \ref{Cobe} and, in a refined version, in Fig. \ref{Planck}. When resolving to a level of 1 part in $10^5$, the CMB shows temperature anisotropies with a spectrum represented in the right panel of Fig. \ref{Planck}. From this spectrum we learn that the universe is filled with 5\% ordinary baryonic matter, 27\% dark matter, 68\% dark energy and is approximately flat \cite{Aghanim:2018eyx}. The inhomogeneous perturbations can be traced back via well established physics to primordial nearly scale invariant perturbations. The goal of theoretical cosmology is to provide a consistent history which explains these observed features. We will begin, in the rest of the chapter, by reviewing the well understood part of this story, namely the one corresponding to the $\Lambda \mbox{CDM}$ model (the standard Big Bang cosmology). We will then continue our journey discussing the theory inflation, which aims at describing a phase in the history of the universe prior to the $\Lambda \mbox{CDM}$ cosmology. The bulk of this thesis will deal with the problem of initial conditions of inflation and the universe, going all the way back to very early times when semi-classical aspects of gravity might have played a crucial role.

\subsection{The standard model of cosmology} 
The standard model of cosmology results from the application of general relativity to the entire universe and describes very successfully most of its history. We are going to review the main features in what follows with the goal of clarifying the notation and at the same time setting the scene for the core of our work by highlighting the aspects and open questions which will be relevant for us.\\  

Denoting with $M$ a manifold equipped with a metric $g_{\mu \nu}$ with $(-, +, + , +)$ signature, the action functional for general relativity reads\footnote{In the following the greek letters refer to the spacetime components $\mu = 0,1,2,3$. The Latin letters denote the spatial components and run from 1 to 3. Also, we use units such that $c=1$.}
\begin{equation}
    S = \frac{1}{16 \pi G } \int_M d^4 x \sqrt{-g} [ R - 2 \Lambda + \mathcal{L}_m]  +  \frac{1}{8 \pi G} \int_{\partial M} d^3y \, \epsilon \sqrt{|h|}  K   \label{einsteinhilbert}
\end{equation}
where $g= det(g_{\mu \nu})$ is the determinant of the metric,  $\Lambda$ is the cosmological constant and $\mathcal{L}_m$ is the matter Lagrangian. The Ricci scalar $R= g^{\mu \nu } R_{\mu \nu}$ is the trace of the Ricci tensor 
\begin{equation}
    R_{\mu \nu} = \Gamma^{\rho}_{\mu \nu , \rho} - \Gamma^{\rho}_{\mu \rho, \nu}+ \Gamma_{\mu \nu}^{\sigma}\Gamma_{\rho \sigma}^{\rho} - \Gamma_{\rho \mu}^{\sigma} \Gamma_{\sigma \nu}^{\rho}
\end{equation}
where the Christoffel symbols are given by
\begin{equation}
    \Gamma_{\mu \nu}^{\sigma} = \frac{1}{2} g^{\sigma \rho} (g_{\nu \rho, \mu} + g_{\rho \mu , \nu} - g_{\mu \nu , \rho})
\end{equation}
The second term in eq. \eqref{einsteinhilbert} is the Gibbons-Hawking-York boundary term \cite{York:1972sj, Gibbons:1976ue} and is necessary to properly implement the Dirichlet variational principle on manifolds with a boundary and will be discussed in more details in the next chapter. $K= h^{ij}K_{ij}$ is the trace of the extrinsic curvature 
\begin{equation}
   K_{\alpha \beta} =  h_{\alpha}^{\gamma} h^{\delta}_{\beta} \nabla_\gamma N_{\delta}
\end{equation}
where $N^{\alpha}$ is the normalized vector normal to the boundary $\partial M$ of $M$,  $h_{\alpha  \beta} = g_{\alpha  \beta} - \epsilon N_{\alpha} N_{\beta}$ is the induced metric on the boundary with determinant $h$. The constant $\epsilon= N^{\alpha}N_{\alpha}$ equals $+1$ or $-1$ depending on if the boundary is space-like or time-like.\\
The field equations of general relativity are Einstein's equations, a set of ten partial non-linear differential equations for the ten independent component of the metric tensor $g_{\mu \nu}$, which extremize the Einstein-Hilbert action \ref{einsteinhilbert}:
\begin{equation}
R_{\mu \nu} - \frac{1}{2} R \, g_{\mu \nu}+ \Lambda g_{\mu \nu} = 8 \pi G \, T_{\mu \nu}  \label{einsteinequations}
\end{equation}
where the matter stress-energy tensor $T_{\mu \nu}$ is given by 
\begin{equation}
 T_{\mu \nu} := - 2 \frac{\delta \mathcal{L}_m}{\delta g^{\mu \nu }} + g_{\mu \nu } \mathcal{L}_m
\end{equation}
 %$R_{\mu \nu} = g^{\rho \sigma} R_{\rho \mu \sigma \nu}$ and $R_{\alpha \beta \mu \nu}$ is the Riemann tensor.\\
According to the observations of the CMB radiation, our universe is on large scales mainly homogeneous and isotropic. As a consequence, the four-dimensional line element that describes the geometry of our universe can be taken to be of the FLRW (Friedmann-Lema\^{i}tre-Robertson-Walker) form
\begin{equation}
ds^2 = g_{\mu \nu} dx^\mu dx^\nu = - dt^2 + a(t)^2 \Bigl[ \frac{dr^2}{1 - k r^2} + r^2 d\theta^2 + \sin^2(\theta) d\phi^2 \Bigr] \label{frw} =  - dt^2 + a(t)^2 \Bigl[ h^{RW}_{ij} dx^i dx^j  \Bigr]
\end{equation}
with a suitable choice of coordinates $\{ t, r, \theta , \phi \}$ .\\
The constant $k$ represents the curvature of the spatial surfaces which, for an appropriate choice of units for $r$, takes the values $0, +1, -1$ corresponding to flat, positively curved and negatively curved universes, respectively. The scale factor $a(t)$ encodes how proper distances change with time and thus parameterizes the expansion of the universe.
Under this ansatz,
Einstein's equations determine the dynamics of the universe fixing the scale factor $a(t)$, for a given value of $k$. 
The Ricci tensor non-vanishing components and trace are given by
\begin{align}
    R_{00} &= - 3 \frac{\ddot{a}}{a}\\
    R_{ij} &=  \Bigl( \frac{\ddot{a}}{a} + \frac{2 \dot{a}^2}{a^2} + \frac{2 k }{a^2}\Bigr)\, h_{ij}^{RW}\\  
    R &=  6 \Bigl(\frac{\ddot{a}}{a}+ \frac{\dot{a}^2}{a^2} + \frac{k}{a^2} \Bigr) \label{Ricciscalar}
\end{align}
The macroscopic behaviour of the universe thermal bath is well-described by a sum of homogeneous and isotropic perfect fluids whose stress-energy tensor is given by
\begin{equation}
T_{\mu \nu} = (\rho + p ) u_\mu u_\nu + p g_{\mu \nu} \label{tmunu}
\end{equation}
$u_\mu$ being the fluid velocity, $\rho$ and $p$ its energy density and pressure, respectively.\\
Einstein's equations reduce in this case to the Friedmann's equations
\begin{align}
\Bigl (\frac{\dot{a}}{a}  \Bigr)^2 &= \frac{8 \pi G}{3} \rho - \frac{k}{a^2} \label{einstein1} \\
\frac{\ddot{a}}{a} &= - \frac{4 \pi G}{3} (\rho + 3 p ) \label{einstein2}
\end{align}
Using the first equation, the second can be re-written as the continuity equation 
\begin{equation}
    \dot{\rho} + 3 \frac{\dot{a}}{a} (\rho + p) = 0 \label{einstein3}
\end{equation}
The cosmological fluid can be described by an equation of state $p = \omega \rho$ where $\omega=0$ corresponds to non-relativistic matter, $w=\frac{1}{3}$ to radiation or relativistic matter, $\omega = -1 $ to dark energy.
Equation (\ref{einstein3}) gives the energy density of the fluid as function of the scale factor 
\begin{equation}
    \rho(a) \propto a^{-3(\omega + 1)} \label{energysclae}
\end{equation}
To determine the dynamics of the universe, it is useful to introduce the dimensionless density parameters
\begin{equation}
    \Omega_r = \frac{\rho_{r, 0}}{\rho_c}, \; \; \;   \Omega_m = \frac{\rho_{m, 0}}{\rho_c},  \; \; \;  \Omega_{\Lambda} = \frac{\rho_{\Lambda}}{\rho_c} , \; \; \;  \Omega_k = - \frac{k}{H_0^2}
\end{equation}
where $H := \frac{\dot{a}}{a}$ is the Hubble parameter, $\rho_c = \frac{3 H_0^2}{8 \pi G}$ is the energy density of a flat universe and quantities with subscript ``0" are measured at the current age of the universe $t_0$ with the convention that $a(t_0) = 1$.\\
Equation (\ref{einstein1}) can then be written as 
\begin{equation}
\Bigl(\frac{H}{H_0} \Bigr)^2 = \frac{\Omega_r}{a^4 } + \frac{\Omega_m}{a^3} + \frac{\Omega_k}{a^2} + \Omega_{\Lambda}
\end{equation}
from which we see that the various components dominate the expansion of the universe in different epochs since their energy density scales with different powers of the size of universe.\\
As measured by the PLANCK satellite \cite{Aghanim:2018eyx}, the universe we live in is characterized (with $68\%$ confidence region) by
\begin{equation}
\begin{split}
    \Omega_k &= - 0.001 \pm 0.002\\
    \Omega_\Lambda &= 0.680 \pm 0.013 \\
    \Omega_m &= 0.321 \pm 0.013\\
    \Omega_r &= (9.14 \pm 0.34)10^{-5}\\
    H_0 &= (66.88 \pm 0.92) \mbox{ Km} \, \mbox{s}^{-1} \mbox{Mpc}^{-1}
    \end{split}
\end{equation}
The picture of the $\Lambda\mbox{CDM} $\footnote{ $\Lambda\mbox{CDM} $ stands for a universe filled with dark energy ($\Lambda$) and cold dark matter (CDM).} for cosmology is then that of a flat universe which initially expands dominated by radiation; as the expansions proceeds the matter component comes to dominate and at late times dark energy takes over. \\
Analytic solutions to Einstein's equations for a flat universe in the phases when only one of the components is relevant are given by 
\begin{align}
    a(t) &= t^{\frac{2}{3 (1 + \omega)}} \; \; \; \; \mbox{ for  } \;  \omega \neq  - 1   \label{timescale} \\
    a(t) &=  e^{H t} \; \; \; \;  \; \; \; \, \,  \mbox{  for  }\;  \omega = - 1 
\end{align}
where for $ \omega = - 1 $, the Hubble rate $H := \frac{\dot{a}}{a} = \sqrt{\frac{8 \pi G}{3} \rho_{\Lambda}} = \sqrt{\frac{\Lambda}{3}}$ is constant. Notice that the PLANCK measurements point at a flat universe with experimental error bars which allow for closed and negatively curved cosmologies. For this reason we will discuss in the following FLRW models of all three types.\\
The radiation dominated universe hits the Big Bang singularity at $t=0$ where $a=0$ and the scalar curvature \eqref{Ricciscalar} and energy density of the fluid \eqref{energysclae} diverge.\\
As a final remark let us introduce the concept of comoving particle horizon. For this, it will be useful to rewrite the flat FLRW line element in conformal time
\begin{equation}
    ds^2 = a(\eta)^2 [ - d\eta^2 + dr^2 + r^2 (d \theta^2 + \sin^2 \theta d \phi^2)]
\end{equation}
A null geodesic corresponds to $ds^2 = 0$ and thus radial photon trajectories are given by
\begin{equation}
 dr = \pm d \eta
\end{equation}
Thus the maximum comoving distance light can travel between time $0$ and time $t$ is
\begin{equation}
    \eta = \int_0^t \frac{dt'}{a(t')} = \int_0^{a} d \ln a' \Bigl(\frac{1}{a' H}\Bigr) \label{horizonc}
\end{equation}
This is a causal horizon in the sense that two regions separated by a distance larger than $\eta$ could never have communicated with each other. Both the horizon $\eta$ and the comoving Hubble radius $(a H)^{-1}$ grow with the expansion of the universe in the $\Lambda \mbox{CDM}$ model. This means that the large scales on the CMB, which only recently entered the horizon, could have not been in causal contact with each other when the CMB was emitted or, in other words, that, when we look at the CMB sky, we look at a large number of causally disconnected regions. If we assume, for the sake of the argument and with a tremendous simplification\footnote{The number of causally disconnected regions one gets from the fully accurate calculation is of the same order of magnitude.}, that the universe was matter dominated all the way back to the Big Bang than $\eta \propto a^{1/2}$ and the number of disconnected regions, which goes as the volume, is \begin{equation}
    \Bigl( \frac{r_0}{r_{CMB}} \Bigr)^3 =  \Bigl( \frac{a_0}{a_{CMB}} \Bigr)^{3/2} \approx 35000
    \end{equation} 
where $a_{CMB} \approx 9.08 \, 10^{-4}\, a_0$ according to the PLANCK data.
How come then that these 35000 causally disconnected regions share the same temperature up to a part in $10^5$? \cite{Dodelson:2003ft, Montani:2011zz, Baumann:2009ds} One possible answer to this question is that the universe simply started out in a very special state, homogeneous and isotropic to a very large degree with only tiny primordial deviations from that. The level of fine tuning of these initial conditions required for this explanation to work is however extremely high. That is the reason why cosmologists are after a dynamical explanation, beyond the  
$\Lambda \mbox{CDM}$  model, to explain the observed features of the CMB.

\subsection{Inflation and the cosmological perturbations}

If the comoving Hubble radius $(a H)^{-1}$ was shrinking in the very early universe, the large-angle isotropy of the CMB could possibly be explained without requiring fine-tuned initial conditions: the comoving horizon \eqref{horizonc} would have received its largest contribution at early times so that regions which cannot communicate today (because they are outside each other's Hubble sphere) could have been in causal contact in the past (being inside each other's horizon).\\
It follows from the Friedmann's acceleration equation \eqref{einstein2} that the comoving Hubble radius shrinks in an accelerated universe or, equivalently, in a universe dominated by matter with negative pressure:
\begin{equation}
    \frac{d}{d t} \Bigl( \frac{1}{a H}\Bigr)<0, \; \; \; \ddot{a}>0, \; \; \; \rho + 3 p <0
\end{equation}
According to the theory of inflation \cite{Guth:1980zm,Linde:1981mu,Albrecht:1982wi,Mukhanov:1981xt,Starobinsky:1979ty,Starobinsky:1982ee, Guth:1982ec,Hawking:1982cz,Bardeen:1983qw}, our universe underwent a phase of such accelerated expansion prior to the standard $\Lambda\mbox{CDM}$ story described in the previous section. If this phase lasted long enough, it could prepare our universe to be already extremely flat, homogeneous and isotropic at the beginning of the radiation-dominated phase, potentially answering some of the questions left open by the $\Lambda\mbox{CDM}$ model. Crucially, inflation provides also a framework where the classical gaussian temperature fluctuations of the CMB are generated as primordial quantum fluctuations, as we will show in the next section. It however is important to keep in mind that while inflation can provide the suitable initial conditions for the standard Big Bang cosmology, the question arises of how peculiar the initial conditions of the universe must be for inflation to happen in the first place. If such initial conditions must be highly fine-tuned as well, the question of how our universe found itself in such a special state is simply moved from the beginning of the $\Lambda\mbox{CDM}$ cosmology to the beginning of inflation. This thesis will deal with our understanding of the initial conditions for inflation especially focusing on the semi-classical aspects of the problem. It is worth mentioning also that while accepted by most of the scientific community and in agreement with all current observations, the framework of inflation still lives in the realm of the theory and will be treated as such in this work (see for example \cite{Gibbons:2006pa,Ijjas:2013vea, East:2015ggf,Clough:2016ymm,Marsh:2018fsu} for details and discussions).\\

\subsubsection{De Sitter space}
The simplest model of inflationary universe is provided by de Sitter space which is the solution to vacuum Einstein's equation with a positive cosmological constant.\\
De Sitter space is an Einstein manifold i.e. a pseudo-Riemannian differentiable manifold whose Ricci tensor is proportional to the metric with
\begin{align}
R_{\mu \nu} &= \Lambda g_{\mu \nu}\\
R &= 4 \Lambda
\end{align}
The constant $\textit{l} := \frac{1}{H} =  \sqrt{\frac{3}{\Lambda}}$ has the units of length and is called the ``De Sitter radius''.\\
Introducing the flat five dimensional space $\mathbb{R}^{1,4}$ with metric 
\begin{equation}
ds^2 = - dx_0^2 + dx_1^2 + dx_2^2 + dx_3^2 + dx_4^2  
\end{equation}
de Sitter space can be thought of as the hyperboloid
\begin{equation}
- x_0^2 + x_1^2 + x_2^2 + x_3^2 + x_4^2 = \frac{3}{\Lambda}
\end{equation}
The de Sitter metric is then the metric induced on the hypersurface by the Lorentzian geometry of the five-dimensional Minkowski space.
The four-dimensional de Sitter metric can take the form of all three possible FLRW cosmologies with suitable choices of coordinates.
\begin{figure}[htbp]
\begin{center}
\includegraphics[width = 14 cm]{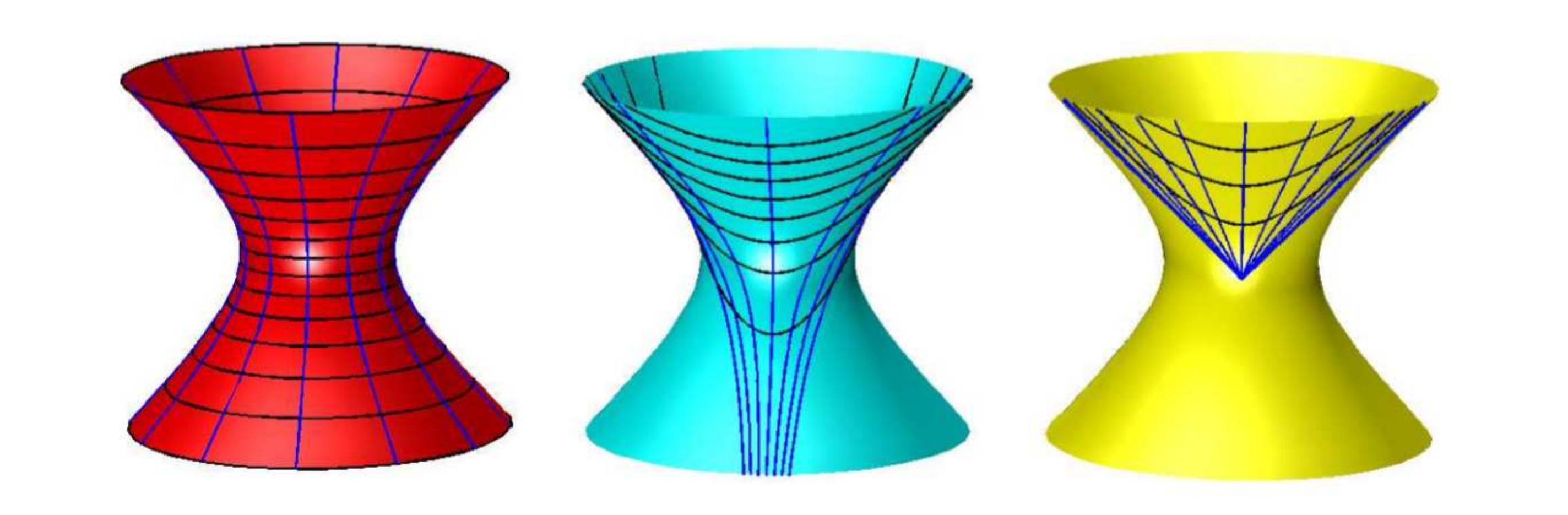}
\caption{The de Sitter hyperboloid with different choices of coordinates. The blue lines represent time-like geodesics. The black lines are hypersurfaces of constant cosmic time. The red, blue and yellow manifolds show the embedding of closed, flat and hyperbolic cosmological models respectively. The figure is taken from \cite{Moschella:2006pkh}.}
\label{fig:desitter}
\end{center}
\end{figure}
\\
Consider the following set of coordinates
\begin{align}
x_0 &= \sqrt{\frac{3}{\Lambda}} \sinh(\sqrt{\frac{\Lambda}{3}} t), \\
x_1 &= \sqrt{\frac{3}{\Lambda}} \cosh(\sqrt{\frac{\Lambda }{3}} t)\, \cos \chi,  \\
x_2 &= \sqrt{\frac{3}{\Lambda}} \cosh(\sqrt{\frac{\Lambda }{3}} t)\, \sin \chi \cos \theta,\\  \;
x_3 &= \sqrt{\frac{3}{\Lambda}} \cosh(\sqrt{\frac{\Lambda }{3}} t)\, \sin \chi \sin \theta \cos \phi, \\
x_4 &= \sqrt{\frac{3}{\Lambda}} \cosh(\sqrt{\frac{\Lambda }{3}} t)\, \sin \chi \sin \theta \sin \phi 
\end{align}
with $  - \infty < t < \infty $, $0 \leq \chi \leq \pi $, $0 \leq \theta \leq \pi$, $0 \leq \phi \leq 2 \pi$, the de Sitter line element describes a closed FLRW model
\begin{equation}
ds^2 = - dt^2 + \frac{3}{\Lambda} \cosh^2 (\sqrt{\frac{\Lambda}{3}} t) [ d \chi^2 + \sin^2 \chi (d \theta^2 + \sin^2 \theta d \phi^2)] 
\end{equation}
where the hypersurfaces of constant time are spheres $\mathbb{S}^3$ and $d \Omega_3 =  d \chi^2 + \sin^2 \chi (d \theta^2 + \sin^2 \theta d \phi^2) $ is the line element of the unit three-sphere. 
\begin{figure}[htbp]
\begin{center}
\includegraphics[width = 7 cm]{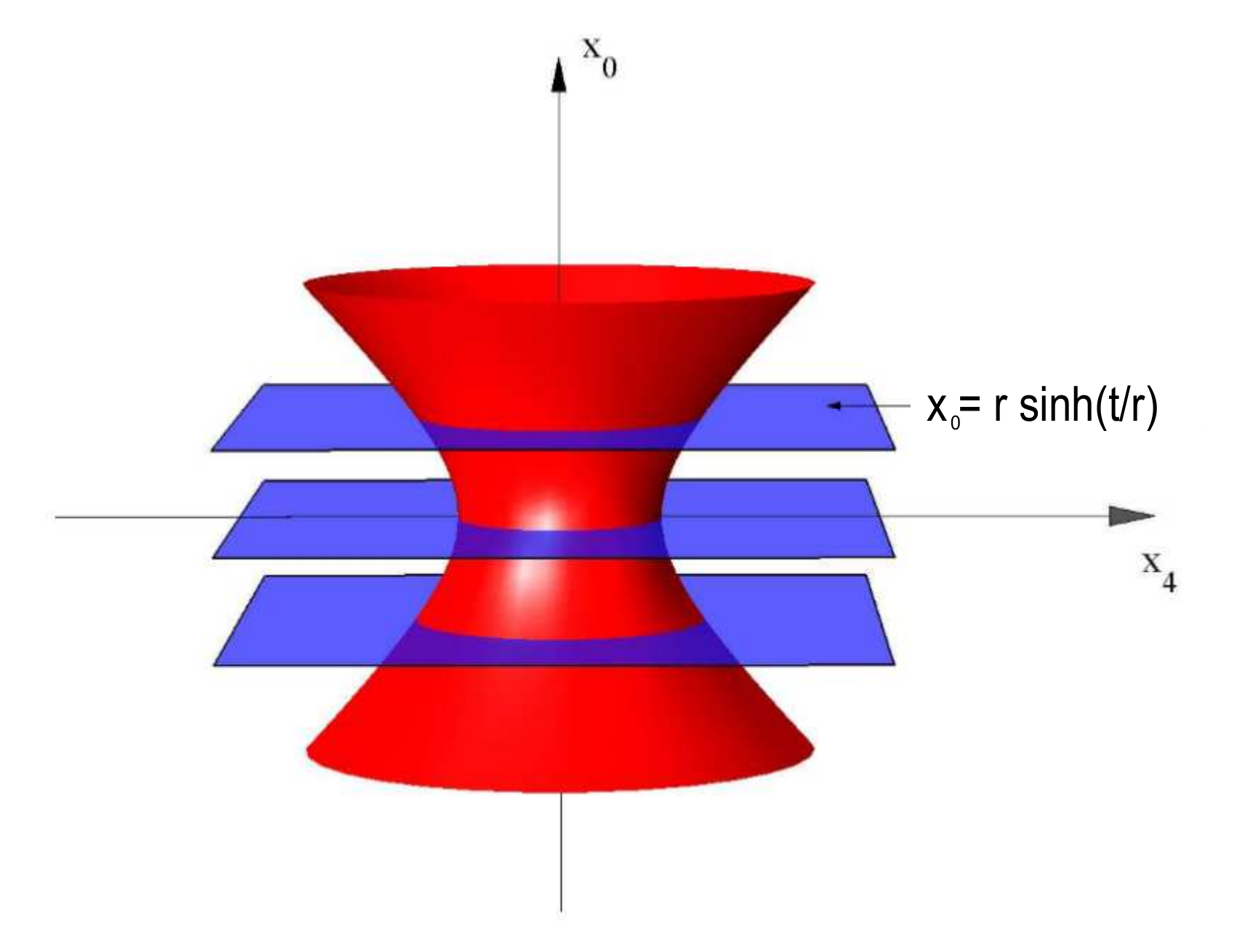}
\includegraphics[width = 7 cm]{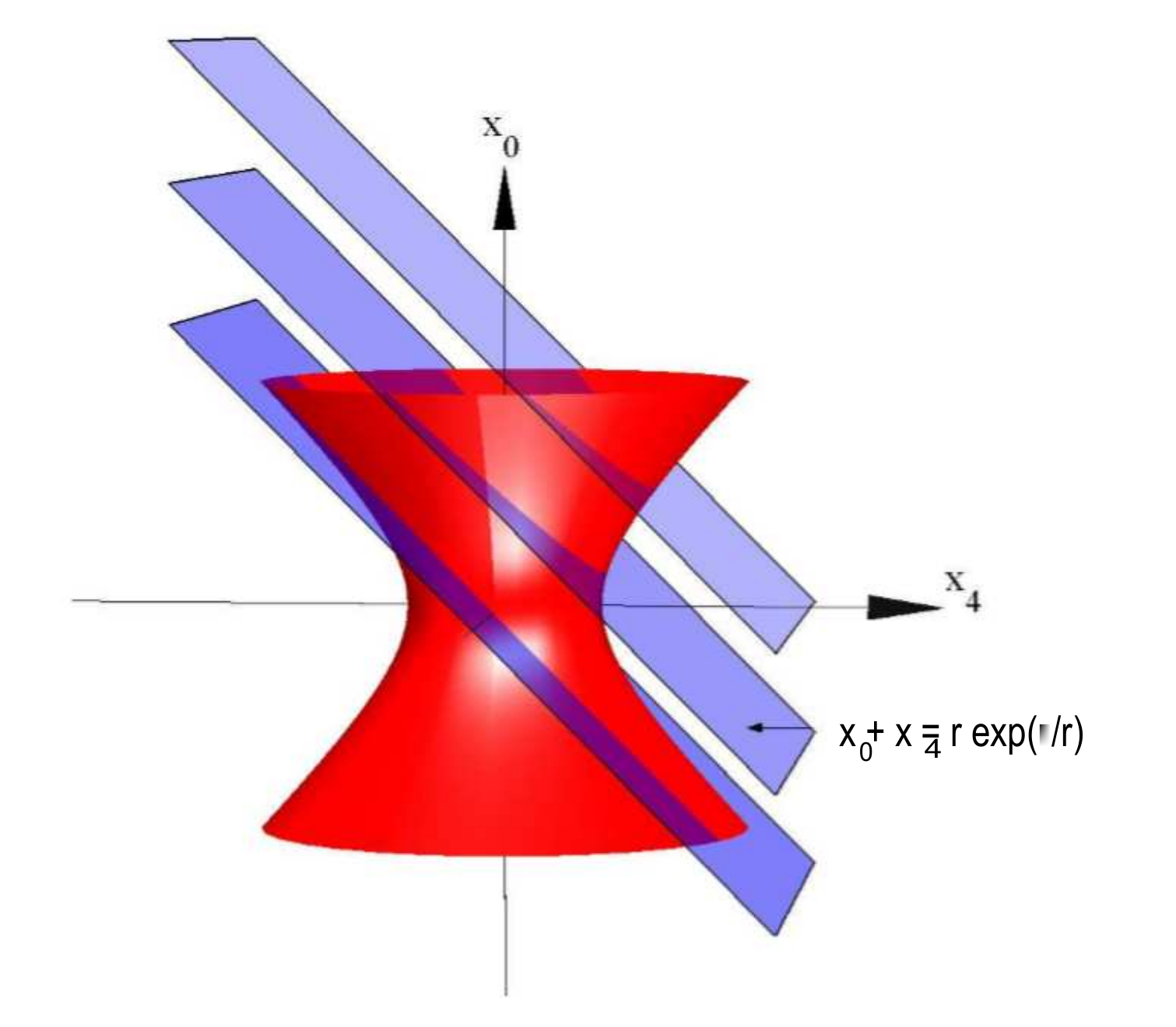}
\caption{Left panel: The de Sitter space as a closed FLRW model. The hypersurfaces of constant time are spheres given by the intersection of the de Sitter hyperboloid with the hyperplanes $x_0 = const$. Right panel: The de Sitter space as a flat FLRW model. The hypersurfaces of constant time are intersections of the de Sitter hyperboloid with the hyperplanes $x_0 + x_4 = const$. The figures are taken from \cite{Moschella:2006pkh} and $r$ represents the de Sitter radius $l$.}
\label{fig:desitter2}
\end{center}
\end{figure}
\\ One can also introduce the coordinates
\begin{equation}
t = \sqrt{\frac{3}{\Lambda}} \log \Bigl( \frac{ x_0 + x_4}{\sqrt{3/\Lambda}}\Bigr), \; x = \sqrt{\frac{3}{\Lambda}} \frac{x_1}{x_0 + x_4}, \; y = \sqrt{\frac{3}{\Lambda}} \frac{x_2}{x_0 + x_4}, \; z = \sqrt{\frac{3}{\Lambda}} \frac{x_3}{x_0 + x_4}
\end{equation}
In these coordinates the constant time surfaces are copies of $\mathbb{R}^3$ and the De Sitter line element is the analogous of a flat FLRW model
\begin{equation}
ds^2 = - d t^2 + \exp \Bigl(2 \sqrt{\frac{\Lambda}{3} } t \Bigr)(dx^2 + dy^2 + dz^2)
\end{equation}
with $- \infty < t < \infty$.
Note that these coordinates cover only half of the hyperboloid since $t$ is defined only for $x_0 + x_4 = l \, e^{t/l} > 0$. \\
Most often in the literature, de Sitter space in flat slicing is written in terms of conformal time $d \tau = \frac{dt}{a} $ with $\tau \in (- \infty , 0^{-})$. 
\begin{equation}
    ds^2 = a(\tau)^2 [ - d \tau^2  + dx^2 + dy^2 + dz^2], \; \; \; \; \; a(\tau) = - \frac{1}{H \tau}
\end{equation}

In the following we will consider de Sitter space both in flat and closed slicing as toy-models for inflation. It is important to keep in mind that this realization is evidently not realistic because it does not allow inflation to come to an end and it does not generate scalar perturbations, in contradiction with the CMB observations. One can however study tensor perturbations around this background, which are exactly gaussian distributed in this case, to gain a qualitative understanding of many aspects of slow-roll inflation. The de Sitter approximation will greatly simplify the calculations without altering the qualitative analysis and provide a favorable framework for the purposes of our work.    \\

\subsubsection{Scalar field inflation}
Realistic models of inflation are realized by minimally coupling gravity with a scalar field $\phi$, the inflaton, for a suitable choice of the potential $V(\phi)$:
\begin{equation}
 S =  \int d^4 x \sqrt{-g} [ \frac{1}{16  \pi G} R - \frac{1}{2}g^{\mu \nu} \partial_\mu \phi \partial_\nu \phi - V(\phi) ]  + S_B   
\end{equation}
For a homogeneous scalar field $\phi = \phi(t)$ in a flat FLRW universe (eq. (\ref{frw})) the stress-energy tensor is that of a perfect fluid with 
 \begin{align}
     \rho_\phi &= \frac{1}{2}  \dot{\phi}^2 + V(\phi)  \\ 
     p_\phi &= \frac{1}{2}  \dot{\phi}^2 - V(\phi) 
 \end{align}
If the potential energy $V(\phi)$ dominates over the kinetic energy $\frac{1}{2}\dot{\phi}^2$ the scalar field drives an accelerated expansion ($\omega_{\phi} < - \frac{1}{3}$). 
The equations of motion for this system are 
\begin{align}
    &\ddot{\phi } + 3 H \dot{\phi} + V_{, \phi}  = 0 \label{inflation} \\
   & \frac{3}{8 \pi G} H^2 = \frac{\dot{\phi}^2}{2} + V(\phi)
\end{align}
from which it follows that
\begin{equation}
    \frac{\ddot{a}}{a} = H^2 (1 - \epsilon) \; \; \mbox{  with  } \epsilon := \frac{3}{2} (1 + \omega_{\phi}) = 8 \pi G  \frac{\dot{\phi}^2}{2 H^2}
\end{equation}
With a suitable choice of the potential it is then possible to build a dynamical system in which the expansion of the universe is accelerated as long as the slow-roll parameter $\epsilon< 1$ and where inflation comes to an end when this condition fails to be satisfied ($\epsilon(\phi_{end})=1$). \\
Note that de Sitter spacetime represents the ``no roll" limit $\epsilon= const = 0$.\\ 
For inflation to last long enough it is also required that
\begin{equation}
|\ddot{\phi}| \ll |3 H \dot{\phi}|, |V_{, \phi}|
\end{equation}
and thus that
\begin{equation}
    |\eta| := | - \frac{\ddot{\phi}}{H \dot{\phi}}| < 1
\end{equation}
The slow-roll regime is realized when 
\begin{equation}
    \epsilon \ll 1 \; \; \; \; |\eta| \ll 1
\end{equation}
in which case the slow-roll parameter is approximated by
\begin{equation}
    \epsilon \approx \frac{1}{16 \pi G} \Bigl( \frac{V_{, \phi}}{V} \Bigr)^2
\end{equation}
We then learn that slow-roll inflation can be realized if the inflaton potential is sufficiently flat.

%%%%%%%%%%%%%%%%%%%%%%%%%%%%%%%%%%%%%%%%%%%%%%%%%%

\subsubsection{Cosmological perturbation theory}\label{sec:QFT}

The Universe we observe today is homogeneous and isotropic on scales larger than about 300 Mly \cite{Scrimgeour:2012wt}. As a first approximation, many phenomena of physical interest can be described by FLRW type of metrics (\ref{frw}). However, the CMB tells us that this is not the whole story. At the time of the decoupling our universe was nearly homogeneous and isotropic with small inhomogeneities.
Indeed the testability of cosmological models relies mostly on the prediction of the correct amount of these inhomogeneities. The biggest success of the theory inflation is to provide a mechanism through which they can be generated. In the following we will describe the primordial fluctuations during inflation and their quantization within the framework of QFT in curved spacetime using the cosmological perturbations theory \cite{Mukhanov:1990me}. \\
Let us consider small generic perturbations $ \delta g_{\mu \nu}(t , \mathbf{x})$ around a homogeneous and isotropic background $g_{\mu \nu}^0(t)$
\begin{equation}
g_{\mu \nu} (t, \mathbf{x}) = g_{\mu \nu}^0 (t) + \delta g_{\mu \nu}(t , \mathbf{x})
\end{equation}
associated with the line element 
\begin{equation}
ds^2 = g_{\mu \nu} dx^\mu dx^\nu = - N^2(1 + 2A) dt^2 + 2 a B_i dt dx^i + a^2 [(1 - \psi) h_{ij} + E_{ij}]dx^i dx^j \label{metri}
\end{equation}
In a spacetime filled with other homogeneous fields $\phi(t)$ they get perturbed too $\phi(t, \mathbf{x}) = \phi(t) + \delta \phi(t, \mathbf{x}) $ as a consequence to Einstein's equations. \\
%We will not investigate any further these aspects of theory since we are studying pure (empty) De Sitter space. \\
We will deal with first order perturbations whose dynamics is given by Einstein's equation linearized around the background where all the terms of second order in perturbations are neglected. Their action is given by an expansion to the second order of the Einstein-Hilbert action for gravity. In this case, the treatment of perturbations is rather simplified by the properties of symmetry of the homogeneous and isotropic background. The translation invariance of the linear equations of motion for the perturbations allows one to work in Fourier space rather than in real space (see \cite{Baumann:2009ds} for further details). The Fourier modes of the perturbations do not interact and can be treated independently. Moreover, the metric and stress-energy perturbations can be the decomposed in their scalar, vector and tensor components (SVT decomposition), according to their helicity. Given a wave vector $\mathbf{k}$, a rotation by a angle $\theta $ around it effects a perturbation of helicity \textit{m} by a dilatation of his amplitude by a factor $e^{i m \theta}$. 
Helicity scalars are then defined to have $m=0$, vectors correspond to $m = \pm 1$, tensors to $m = \pm 2$. The rotational invariance of the FLRW background implies that the three different sectors evolve independently and thus we can treat them separately. \\
The SVT decomposition applies to the elements $E_{ij}$ and $B_i$ of the metric (\ref{metri}) and allows us write them as follows
\begin{equation}
B_i \equiv \partial_i B  - S_i \, \, \mbox{where} \, \, \partial^i S_i = 0
\end{equation}
\begin{equation}
E_{ij} \equiv 2 \partial_{ij} E + 2 \partial_{(i} F_{j)} + \eta_{ij} \, \, \mbox{where} \, \, \partial^i F_{i} =0 ,  \,\, \eta^i_i= \partial^i \eta_{ij} = 0  \label{svt}
\end{equation}
The scalar sector of the perturbations is described by the four scalars A, $\psi$, B, E and is related to the temperature fluctuations of the CMB. The tensor components ($\eta_{ij}$) can be observed as gravitational waves. The vector fluctuations, given by $S_i$ and $F_i$, always decay during the expansion of the universe and are usually ignored in cosmology.\\
While tensor perturbations are gauge invariant, the four functions A,$\phi$, B,E are affected by changes of coordinates.\\
Under the transformation
\begin{align}
t &\rightarrow t + \alpha \\
x^i &\rightarrow x^i + \delta^{ij} \beta_{, j } 
\end{align}
the metric scalar perturbations transform as
\begin{align}
A &\rightarrow A - \dot{\alpha}\\
B &\rightarrow B + a^{-1} \alpha - a \dot{\beta} \\
E &\rightarrow E - \beta \\
\psi &\rightarrow \psi + \frac{\dot{a}}{a} \alpha
\end{align}
and the scalar field perturbation transforms as $\delta\phi \to \delta\phi - \dot\phi \alpha.$
The two functions $\alpha$ and $\beta$ can be chosen in such a way that two scalar degrees of freedom vanish, say $E$ and $\psi$. At linear order the constraints, which can be thought of as the $00$ and $0i$ Einstein's equations, are given by (see e.g. \cite{Koehn:2015vvy})
\begin{align}
A=& \frac{\dot\phi}{2 H}\,\delta\phi  = \sqrt{\frac{\epsilon}{2}} \delta \phi \label{eq:alpha}\\
\partial^i \partial_i B =& -\frac{1}{2 H}(V_{,\phi}+\frac{\dot\phi}{H}V)\delta\phi-\frac{\dot\phi}{2 H} \dot{\delta\phi} = -\epsilon \frac{\mathrm{d}}{\mathrm{d}t}\left( \frac{\delta\phi}{\sqrt{2\epsilon}}\right),
\end{align} 
where in the constraint for $B$ we have already used \eqref{eq:alpha} to replace $A.$  In pure de Sitter space, where $\epsilon=0$, scalar perturbations are thus forced to vanish. Said differently, there are no scalar degrees of freedom in an empty de Sitter spacetime and thus no scalar perturbations can be generated. Gravitational waves can instead always be produced and will be the subject of our study.

%%%%%%%%%%%%%%%%%%%%%%%%%%%%%%%%

\subsubsection{Tensor perturbations} 
For tensor perturbations in a flat de Sitter universe, it is conventional to define the following Fourier expansion
\begin{equation}
\eta_{ij} (\eta, \mathbf{x})=\int \frac{d^3 k}{(2 \pi)^3} \sum_{s =+, \times} \epsilon_{ij}^s (\mathbf{k}) \phi_{\mathbf{k}}^s(\eta) e^{i \mathbf{k \cdot x}} 
\end{equation}
where the polarization tensor satisfies $\epsilon_{ii} = k^i \epsilon_{ij} = 0$ and $\epsilon^s_{ij}(\mathbf{k}) \epsilon^{s'}_{ij}(\mathbf{k}) = 2 \, \delta_{s s'}$.\\
Given that different Fourier modes do not interact with each other at linear order, the study of tensor perturbations reduces to the study of a single scalar $\phi_{\mathbf{k}}^s(\eta)$ of wave number $k$ and polarization $s$, where the overall sum over all modes will be kept implicit\footnote{For simplicity of notation, we will mostly use the notation $\phi(\eta)$ rather than $\phi_k^s(\eta)$}. It will also be useful to introduce the canonically normalized field 
\begin{equation}
v_{\mathbf{k}}^s = \frac{a}{2 \sqrt{8 \pi G}} \; \phi_{\mathbf{k}}^s \label{defcanvar}
\end{equation}
With this construction the second variation of the Einstein-Hilbert action (\ref{einsteinhilbert}) can be written in conformal time as follows
\begin{equation}
\delta_2 S =\frac{1}{2} \sum_s \int d \eta \, d^3 k \, \Bigl[(v_{\mathbf{k}}^{'s} )^2 - \Bigl(k^2 - \frac{a''}{a}\Bigr) (v_{\mathbf{k}}^s)^2 \Bigr]
\end{equation}
Therefore the action for a single mode of wave number $k$ and a given polarization is
\begin{equation}
S^{(2)} = \frac{1}{2}  \int d \eta \, \Bigl[(v_k^{'s} )^2 - \Bigl(k^2 - \frac{a''}{a}\Bigr) (v_k^s)^2 \Bigr] \label{actionperturbations}
\end{equation}
where $\frac{a''}{a} = \frac{2}{\eta^2}$ in de Sitter space.\\
When dealing with the closed FLRW model, we consider the harmonic expansion in terms of the normalised eigenfunction $Q_{nlm }$ of the Laplacian operator
\begin{align}
\Delta Q_{n l m } &= - n(n +2) Q_{nlm}\\
\int d^3x \sqrt{\eta} \, Q_{nlm}(x) Q_{n'l'm'} (x) &= \delta_{n n'} \, \delta_{l l'} \, \delta_{m m'}
\end{align}
where $n$ is an integer $n\geq 2$.\\
The expansion in modes reads
\begin{equation}
\phi(\eta, \mathbf{x}) = \sum_{nlm} \phi_{nlm}(\eta) Q_{nlm}(\mathbf{x})  
\end{equation}
In this case the action for a single canonically normalized mode takes the form
\begin{equation}
S^{(2)} =\frac{1}{2} \int d \eta \, \Bigl[(\phi_k^{'} )^2 - \Bigl( n(n+2) - \frac{a''}{a}\Bigr) \phi_n^2 \Bigr] \label{pertk1}
\end{equation}		
where we dropped the $l$ and $m$ subscripts.\\
We can recognize in eq. \eqref{actionperturbations} and \eqref{pertk1} the action of a harmonic oscillator with a time-dependent frequency. The quantization of cosmological perturbations is thus simply the quantization of a series of independent harmonic oscillators.

%%%%%%%%%%%%%%%%%%%%%%%%%%%%%%%%%%%%%%%%%%%%%%%%%%%%%%%%%%%%%%%

\subsubsection{Quantization and the Bunch-Davies vacuum}\label{sec:QFT}
Perturbations during inflation are assumed to start out in their vacuum state. The question is: which vacuum? 
In quantum field theory in Minkowski spacetime there is a unique vacuum state defined as state over which the expectation value of Hamiltonian is minimized. If the spacetime evolves with time, this definition gives a vacuum state which depends on the time at which 
the expectation value is calculated. Hence the ground state at time $t_0$ might not be the state of lowest energy at the time $t_1$. Said differently, for a quantum field theory in de Sitter space there exists an entire class of quantum states which are invariant under de Sitter isometries known as $\alpha$-vacua. This leads to an ambiguity in the choice of the vacuum state which is usually solved by identifying the Bunch-Davies vacuum \cite{Bunch:1978yq} as the preferred one.
In what follows we are going to review the quantization of cosmological perturbations in flat space and introduce the Bunch-Davies vacuum in the Schr\"{o}dinger picture \cite{Martin:2012pea, Brizuela:2015tzl, Lehners:2020pem}. \\
We start by promoting the perturbative modes $v_{\mathbf{k}}$ and their conjugate momenta $\pi_{\mathbf{k}} = v_{\mathbf{k}}^{'} $ to quantum operators  $\hat{v}_{\mathbf{k}}$, $ \hat{\pi}_{\mathbf{k}} $ with 
\begin{align}
    [\hat{v}_{\mathbf{k}}, \hat{\pi}_{\mathbf{p}}] &= i \, \delta^{(3)} (\mathbf{k} - \mathbf{p}) \\
    \hat{v}_{\mathbf{k}} \Psi = v_{\mathbf{k}} \Psi,&  \; \; \; \; \;   \hat{\pi}_{\mathbf{p}} \Psi = - i \frac{\partial \Psi}{\partial v_{\mathbf{k}}}
\end{align}
The wavefunction $\Psi_{\mathbf{k}} (\eta, v_{\mathbf{k}}) $ satisfies the Schr\"{o}dinger equation 
\begin{equation}
    i \frac{\partial \Psi_{\mathbf{k}}}{\partial \eta} = \mbox{H}_{\mathbf{k}} \Psi_{\mathbf{k}} \label{schroe}
\end{equation}
where the quantum Hamiltonian which follows from the action (\ref{actionperturbations}) is
\begin{equation}
    \mbox{H}_{\mathbf{k}} = - \frac{1}{2} \frac{\partial^2}{\partial v_{\mathbf{k}}^2} + \frac{1}{2} \left(k^2 - \frac{2}{\eta^2} \right)  v_{\mathbf{k}}^2
\end{equation}
for each mode.\\
Since we are after the ground state of a harmonic oscillator, we can solve the Schr\"{o}dinger equation using a gaussian ansatz
\begin{equation}
    \Psi_{\mathbf{k}} = N_{\mathbf{k}} (\eta) \, e^{- \frac{1}{2} \Omega_{\mathbf{k}} (\eta) v_{\mathbf{k}}^2 } \label{answave}
\end{equation}
Plugging this ansatz into eq. \eqref{schroe} we obtain the two equations
\begin{align}
    i\, \dot{N}_{\mathbf{k}} &= \frac{1}{2} N_{\mathbf{k}} \Omega_{\mathbf{k}} \\
    i \, \dot{\Omega}_{\mathbf{k}} &= \Omega_{\mathbf{k}}^2 -  \left(k^2 - \frac{2}{\eta^2} \right) \label{eqpert2}
\end{align}
where the first simply fixes the normalization factor. To solve the second equation we change variable to 
\begin{equation}
    \Omega_{\mathbf{k}} = - i \frac{f^{*'}_{\mathbf{k}}}{f_{\mathbf{k}}^{*}} \label{ansatzpert}
\end{equation}
The $f_{\mathbf{k}}^{*}$ are simply the complex conjugate of the mode functions one encounters in the more standard quantization of cosmological quantization in the Heisenberg picture. In fact, with this ansatz, eq. \eqref{eqpert2}
becomes the familiar Mukhanov-Sasaki equation
\begin{equation}
f_k^{\prime\prime} + \left(k^2 - \frac{2}{\eta^2} \right) f_k = 0\,,
\end{equation}
which admits two linearly independent solutions, one of positive and one of negative frequency
\begin{equation}
f_k = c_1\, e^{-ik\eta}\left(1 - \frac{i}{k\eta} \right) + c_2\, e^{+ik\eta}\left(1 + \frac{i}{k\eta} \right)\,, \label{eq:BD} 
\end{equation}
The ambiguity in the definition of the vacuum in de Sitter lies in the freedom in the choice of the coefficients $c_{1}$ and $c_2$. The standard choice is to select the Bunch-Davies vacuum noticing that in the far past, \textit{i.e.} in the limit $\eta \rightarrow - \infty,$ the equation of motion becomes that of a fluctuation in Minkowski spacetime. Since the stable positive frequency solution to the wave equation in Minkowski spacetime is of the form $v_k \propto e^{-ik\eta}$, it is usually argued that the mode functions should satisfy
\begin{equation}
\lim_{\eta \to -\infty} f_k(\eta) = \sqrt{\frac{1}{2k}} e^{-ik\eta}
\end{equation}
which leads one to set $c_2=0$ so that one obtains the Bunch-Davies vacuum \cite{Bunch:1978yq}\footnote{A number of cosmologists have pointed out the dangers of this assumption in the past, see in particular the description of the trans-Planckian problem in \cite{Martin:2000xs}.}
\begin{equation}
f_k = \sqrt{\frac{1}{2k}} e^{-ik\eta}\left(1 - \frac{i}{k\eta} \right) \,.
\end{equation}
Physically this is related to the fact that in the far past all the modes of astrophysical interest today had a physical wavelength smaller than the Hubble radius. This allows one to impose the initial condition when
$\eta \rightarrow - \infty$ or $|k\eta| \gg 1$ or $k \gg a \mbox{H}$ and the curvature of the spacetime is not felt and thus it becomes a physical requirement that the mode functions should limit to the Minkowski vacuum solutions. \\
With this choice we obtain that $\Omega_{\mathbf{k}}$ is given by
\begin{equation}
    \Omega_{\mathbf{k}} =+ \frac{k^3 \eta^2}{1 + k^2 \eta^2} + \frac{i}{\eta (1 + k^2 \eta^2)}
\end{equation}
At early times, $ |k \eta| \gg 1$, the wavefunction \eqref{answave} resembles the gaussian ground state of an ordinary of oscillator in Minskowsi spacetime with $\Omega_{\mathbf{k}} \approx k$. Had we chosen the complex conjugate mode we would have obtained a minus sign in front of the real part of  $\Omega_{\mathbf{k}}$ resulting in a nonsensical inverse gaussian distribution for the wavefunction. While at early times the wavefunction is real, at late times, when $ |k \eta| \ll 1$, it becomes increasingly oscillatory with the real part of $\Omega_{\mathbf{k}}$ shrinks to zero. The transition happens at the horizon exit, when $|k \eta| = 1$. We will see in section \ref{classicality} that a system can be said to behave classically in the WKB sense when the phase of its wave function varies rapidly as compared to its amplitude. We thus learn that primordial perturbations become classical at the horizon exit.\\    
The power spectrum of the $v_{\mathbf{k}}$ is then given by \cite{Martin:2012pea} 
\begin{equation}
    P(k) := \frac{k^3}{2 \pi^2} \frac{1}{2 \mathfrak{Re}(\Omega_{\mathbf{k}})}
\end{equation}
with $\mathfrak{Re}(\Omega_{\mathbf{k}})$ evaluated at late times $|k \eta| \ll 1$ . Plugging in the definition of the canonical variable $v_{\mathbf{k}}$ \eqref{defcanvar}, we obtain the standard result for the power spectrum of tensor perturbations \cite{Peter:2013avv}
\begin{equation}
    P_T := 2 \Bigl( \frac{32 \pi G}{a^2}\Bigr)  \frac{k^3}{2 \pi^2} \frac{1}{2 k^3 \eta^2} = \frac{16 G H^2}{\pi}
\end{equation}
where the factor of 2 in front comes from the fact that each tensor mode has two polarizations and we used $a \, \eta = - H^{-1}$.\\
The Bunch-Davies vacuum quantum fluctuations are amplified into a Gaussian distribution of late-time fluctuations $v_k/a$ that reach a constant value  on super-horizon scales and exhibit a scale-invariant spectrum ($|v_k/a|^2 \propto \hbar \,H^2/k^3$). Hence, according to this argument, primordial perturbations with the correct features are naturally produced in inflation and thus a universe which starts out in an initial (quasi-) de Sitter inflationary phase nicely matches current observations. It is important to note that within this framework of QFT in curved spacetime, the early universe is treated as a set of quantum harmonic oscillators on a classical spacetime. We will challenge this framework in chapter \ref{quantuminitial} allowing for quantum properties of the background spacetime and we will see that the Bunch-Davies vacuum is not recovered unless extra ingredients are introduced because the background quantum effects force the choice of the bad behaved mode, complex conjugate to the Bunch-Davies.

%%%%%%%%%%%%%%%%%%%%%%%%%%%%%%%%%%%%%%%%%%%%%%%%%%%%%%%%%%%%%%%%%%%%%%%%%%%%%%%%%%%%

\subsubsection{Eternal inflation}\label{eternalintro}

The usual description of inflationary fluctuations uses the framework of quantum field theory (QFT) in curved spacetime described above, in which quantum fluctuations are superimposed on a classical background spacetime. Even for large fluctuations, such as those envisioned during a regime of eternal inflation~\cite{Vilenkin:1983xq, Guth:2000ka, Linde:1994gy}\footnote{Our discussion of eternal inflation here and in section \ref{papersebastian} will be largely based on \cite{Bramberger:2019zks}. To keep the same notation of the reference in this section we use units such that $8 \pi G =1$.}, this framework is frequently used \cite{Steinhardt:1982kg,Vilenkin:1983xq}. 
Within this regime the quantum fluctuations of the inflaton become comparable to the field displacement due to the classical evolution. In this case, in certain patches of the universe the inflaton might be effectively jumping up its potential, instead of classically rolling down, giving new fuel to the inflationary evolution. If this regime was reached in the early universe, inflation might have ended in the part of the universe we live in but would be still happening somewhere else becoming, at the global level, eternal.\\
 The Einstein's constraint equations (\ref{eq:alpha}) show that when the slow-roll parameter is very small, $\epsilon \ll 1,$ the metric perturbations are negligible compared to the scalar field fluctuations $\delta\phi$ since they are suppressed by factors of $\sqrt{\epsilon}.$ This is the basis for the standard intuition that in slow-roll inflation one may think of the background spacetime as being constant, with only the scalar field fluctuating.

This picture is reinforced by the fact that at cubic order in interactions, up to a numerical factor of order one the leading contribution in the Lagrangian is a term of the form $\sqrt{\epsilon}(\dot{\delta\phi})^2\delta\phi,$ which is also small in the slow-roll limit. Hence, in the presence of a very flat potential, the system is perturbative. In other words, to a first approximation the system is described by free scalar field fluctuations in a fixed geometry. 

In flat gauge the comoving curvature perturbation is given by $\mathcal{R} = \psi - \frac{H}{\dot \phi}\delta  \phi = - \frac{H}{\dot \phi}\delta  \phi \approx -\frac{1}{\sqrt{2\epsilon}} \delta\phi.$ An analogous calculation to the one presented in the previous section shows that inflation amplifies scalar quantum fluctuations and induces a variance of the curvature perturbation which on super-Hubble scales and in the slow-roll limit is given by \cite{Mukhanov:1981xt,Guth:1982ec,Hawking:1982cz,Bardeen:1983qw}
\begin{align}
\Delta_{\mathcal{R}}^2 = \frac{H^2}{8\pi^2 \epsilon}\,.
\end{align} 
The relation between the curvature perturbation and the scalar field perturbation then implies that the variance of the scalar field is given by 
\begin{equation}
\Delta\phi_{qu} \equiv \langle (\delta \phi)^2 \rangle^{1/2} = \frac{H}{2\pi}\,.
\end{equation}
This is the typical quantum induced change in the scalar field value during one Hubble time. By comparison, the classical rolling of the scalar field during the same time interval induces a change
\begin{equation}
\Delta\phi_{cl} \equiv \frac{|\dot\phi|}{H} 
\end{equation}
Note that the quantum change dominates over the classical rolling when 
\begin{equation}
\Delta\phi_{qu} > \Delta\phi_{cl} \quad\leftrightarrow \quad\frac{H^2}{2\pi |\dot\phi|} \approx \frac{H}{\sqrt{8\pi^2 \epsilon}}>1 \quad\leftrightarrow\quad \Delta_{\mathcal{R}}^2 > 1\,,
\end{equation}
i.e. precisely when the variance of the curvature perturbation is larger than one, and when perturbation theory becomes questionable. In this regime inflation is thought to be eternal, and this leads to severe paradoxes in its interpretation \cite{Ijjas:2014nta, Guth:2000ka, Linde:1994gy}. The entire framework of inflation might in fact get in trouble because the eternal possibility: if an inflating region of spacetime enters the eternal regime, pocket universes, which expand extremely fast, are created as the scalar field jumps up the potential. Within each such pocket universe, new pocket universes are created in a process which goes on forever. We thus end up with a picture of infinitely many disconnected universes, each with possibly different physics and predictions for cosmology. Since the eternal regime is not yet well understood, inflation is in danger of loosing its predictive power. One motivation of this work is to verify the intuitions from QFT in curved spacetime: does quantum cosmology, where the scale factor of the universe is also quantized, support the view that the scalar field fluctuations evolve in a fixed background spacetime? Does this picture become better or worse as the potential becomes flatter? Is there a qualitative difference between the eternal and non-eternal regimes?

%%%%%%%%%%%%%%%%%%%%%%%%%%%%%%%%%%%%%%%%%%%%%%%%%%%%%%%%%%%%%%%%%%%%%%%%%%%%%%%%%%%%%%%%%%%%%%%%%%%%%%%%%%%%%%%%%%%%%%%%%%%%%%%%%%%%

\clearpage

%%%%%%%%%%%%%%%%%%%%%%%%%%%%%%%%%%%%%%%%%%%%%%%%%%%%%%%%%%%%%%%%%%%%%%%%%%%%%%%

\section{Quantization}\label{sec:quantum}

We take here a break from cosmology to introduce all the necessary mathematical and conceptual tools which will be used in the next chapters in our analysis of the semi-classical properties of the early universe. Using the ADM formalism, we will introduce the canonical quantization of the gravitational field \textit{\`{a}} la Wheeler-de Witt. We will also describe the path integral approach underlining the differences between the Euclidean and the Lorentzian formulation, focusing on the notion of causality and the impact of various boundary conditions. In the rest of the work we mainly will make use of path integral techniques and only occasionally refer to the analogue results that can be derived in the canonical framework.

Let us consider a manifold $M$ equipped with the metric $g$. The couple $(M,g)$ describes a globally hyperbolic spacetime if it admits a Cauchy hypersurface i.e. a hypersurface such that the future and the past evolution of the metric is uniquely determined given the initial conditions defined on it.
If this is the case, it is possible to define a slicing of the full spacetime introducing a time-like four-vector field $\overrightarrow{\eta}$. The vectors orthogonal to $\overrightarrow{\eta}$ define a sub-manifold and $M$ is diffeomorphic to a manifold $\Sigma_t \otimes \mathbb{R} $ where $\Sigma_t$ denotes hypersurfaces of equal time \cite{Geroch:1970uw}.    \\
We introduce the four-vector $N^\mu = (N, N^i)$ which describes the deformation that connects the surface of time $t$ with that of time $t + \delta t$. The deformation vector joining the points $(t, x^i)$ and $(t + \delta t, x^i)$ has non vanishing projections on the normal vector $\overrightarrow{\eta}$, given by lapse function $N$, and on the tangent basis vectors corresponding to the components of the shift vector $N^i$. The lapse function measures the proper-time separation of surfaces of constant $t$ while the shift vector measures the deviation of the lines of constant $x^i$ from the normal to the surface $\Sigma$.
\begin{figure}[htbp]
\begin{center}
\includegraphics[width = 7 cm]{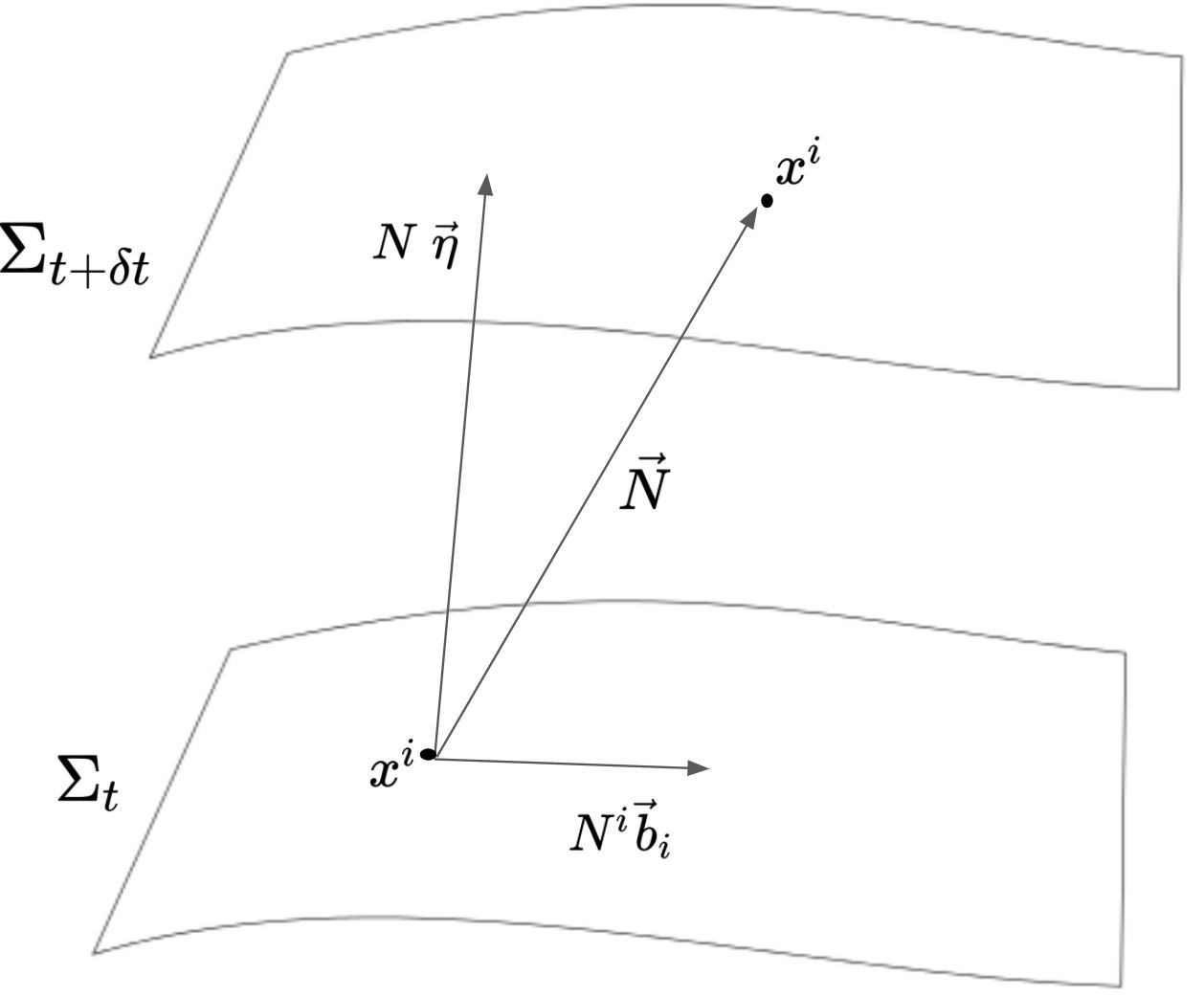}
\caption{ADM slicing of a four-dimensional manifold. The deformation vector, $\stackrel{\rightarrow}{N}$ in the picture, which connects two points with same coordinates $x^i$ has non vanishing projection on the normal vector $\stackrel{\rightarrow}{\eta}$, the lapse function $N$, and on the tangent basis vectors $\stackrel{\rightarrow}{b}_i$, the components of the shift vector $N^i$.}
\label{fig:integrando}
\end{center}
\end{figure}
This defines the so-called Arnowitt-Deser-Misner (ADM) variables which will be useful derive the Hamiltonian formulation of general relativity~\cite{Arnowitt:1959ah}:
\begin{equation}
ds^2 = - (N^2 - N_i N^i) dt^2 + 2N_i dx^i dt + h_{ij} dx^i dx^j \label{aDm}
\end{equation}
The Einstein-Hilbert Lagrangian density in ADM variables \eqref{einsteinhilbert} reads
\begin{equation}
\mathcal{L}_{ADM} = \frac{c^3}{16 \pi G}  N \sqrt{h} (R^{(3)} + K_{ij}K^{ij} - K^2) \label{ladm}
\end{equation}
where $R^{(3)}$ is the Ricci scalar on the three-dimensional constant time hypersurfaces and the extrinsic curvature $K_{ij}$ reads\footnote{$D$ denotes the covariant derivative on the hypersurface $D_i A^k = \partial_i A^k + ^3\Gamma_{i r}^k A^r$ where  $^3\Gamma_{i r}^k$ are the three-dimensional Christoffel symbols.}
\begin{equation}
K_{ij}= \frac{1}{2N} \Bigl( \partial_t h_{ij} - D_i N_j - D_j N_i \Bigr)\label{curv}
\end{equation}
Let us now switch to the Hamiltonian formalism defining the conjugate momenta associated to the 10 variables $h_{ij}, N^i,N$.
The momenta conjugated to $h_{ij}$ read
\begin{equation}
\Pi^{ij} := \frac{\partial \mathcal{L}_{ADM}}{\partial( \partial_t h_{ij})}= \frac{c^3}{16 \pi G} \sqrt{h} \, \Bigl(h^{ri } h^{sj} - h^{ij} h^{rs}\Bigr) \, K_{rs}
\end{equation}
It follows from eq.(\ref{ladm}) and eq.(\ref{curv}) that the momenta conjugated to $N$ and $N^i$ vanish. Thus the dynamics satisfies the four primary first-class constraints
\begin{align}
\Pi &= \frac{\partial \mathcal{L}_{ADM}}{\partial( \partial_t N)} \approx 0  \label{cont1}\\
\Pi_{i} &= \frac{\partial \mathcal{L}_{ADM}}{\partial (\partial_t N^i)} \approx 0  \label{cont2}
\end{align} 
Here and in the following the symbol $\approx $  indicates a weak equality that is, $f$ is weakly equal to $g$ ($f \approx g$ ) if they differ for an arbitrary linear combination of the constraints.\\ 
The total Hamiltonian density can be written introducing four Lagrangian multipliers $ \lambda, \lambda_i$ 
\begin{equation}
\begin{split}
H_{ADM} &= \Pi^{ij} \partial_t h_{ij} + \lambda \Pi + \lambda^i \Pi_i - \mathcal{L}_{ADM} \\
&= N \mathcal{H} + N^i \mathcal{H}_i  +  \lambda \Pi + \lambda^i \Pi_i - 2 D_i (\Pi^{ij} N^k h_{kj})
\end{split}
\end{equation}
where the last term vanishes with suitable boundary conditions.\\
The Hamiltonian of the system, which coincides with the total Hamiltonian on the primary constraints surface, reads
\begin{equation}
H = N \mathcal{H} + N^i \mathcal{H}_i \label{hamil}
\end{equation}
where the quantities $\mathcal{H}$ and  $\mathcal{H}_i$ are known as the ``super-Hamiltonian''  and the '`super-momentum'' respectively
\begin{align}
\mathcal{H} = &\frac{16 \pi G}{2}  G_{i j k l}  \Pi^{ij} \Pi^{kl} - \frac{\sqrt{h}}{16 \pi G} \, (^{(3)}R)  \label{HH} \\
\mathcal{H}_i = &- 2  D_j \Pi_i^j \label{hi} 
\end{align}
\\
The ``supermetric'' $  G_{i j k l} $ is given by
\begin{equation}
    G_{i j k l} = \frac{1}{\sqrt{h} } (h_{ik}h_{jl} + h_{il} h_{jk} - h_{ij} h_{kl})
\end{equation}
As the primary constraints are conserved along the time evolution we get the four secondary first-class constraints 
\begin{align}
\partial_t \Pi &= \{ \Pi, H_{ADM}\} = - \mathcal{H} \approx 0 \label{conny}\\
\partial_t \Pi_i &= \{ \Pi_i, H_{ADM}\} = - \mathcal{H}_i \approx 0 \label{conny1}
\end{align}
As a consequence, the total Hamiltonian of General Relativity is a linear combination of constraints and weakly vanishes too
\begin{equation}
H_{ADM} \approx 0
\end{equation}
The evolution of $N$ and $N^i$ is completely arbitrary as it follows from the equations of motion
\begin{align}
\partial_t N &= \{N, H_{ADM}\} = \lambda \\
\partial_t N^i &= \{N^i, H_{ADM}\} = \lambda^i
\end{align}
All the eight constraints (\ref{cont1}, \ref{cont2}, \ref{conny}, \ref{conny1}) are first-class, since the Poisson brackets of anyone of them with any other weakly vanishes, and thus are generators of gauge transformations.

%%%%%%%%%%%%%%%%%%%%%%%%%%%%%%%%%%%%%%%%%%%%%%%%%%%

\subsection{Canonical Quantization} \label{canonical}

The canonical quantization \textit{\`{a}} la Wheeler-de Witt applies Dirac's quantization procedure \cite{Diracqm} to the gravitational field \cite{Dirac:1958sc, DeWitt:1967yk, Wheeler:1988zr}.
The configurations variables and momenta are promoted to quantum operators acting on some Hilbert space. Observables quantities are represented by Hermitian operators acting on this space. The commutator of two operators is an operator corresponding to the Poisson brackets of the two observables.\\
Ehrenfest's theorem states that quantum expectation values behave almost as classical space phase functions\footnote{We use Dirac's notation: $F= \langle \Psi | \hat{F} | \Psi \rangle$ is the expectation value of the quantum operator $\hat{F}$ on the state $|\Psi \rangle$ }
\begin{equation}
 \langle \dot{F} \rangle = \langle \Psi | \{ \widehat{F, H} \} | \Psi \rangle = \frac{1}{i \hbar} \langle \Psi | [\widehat{F}, \widehat{H}] | \Psi \rangle + O(\hbar) = \frac{\partial }{\partial t} \langle \widehat{F}\rangle + O(\hbar)
\end{equation}
This should in particular holds for the constraints.\\
Denoting with $\phi_a$ ($a=1,..,8$) all the constraints of the theory the following relation must hold for every value of $a$ and any element $|\Psi \rangle$ of the Hilbert space
\begin{equation}
\langle \Psi | \widehat{\phi_a}|\Psi \rangle = 0
\end{equation}
We interpret this quantum constraint as a restriction to be imposed on the state $|\Psi \rangle$. That is, we define the ``physical state space'' as the linear subspace of the representation space of states that are annihilated by the constraints. Since first class constraints are generators of gauge transformations, physical states are gauge invariant quantities. Note that the physical state space is not a Hilbert space yet. Once the constraints are solved at the quantum level one has to define a proper scalar product on the space of the solutions to make sense of the quantum theory. This is however a not well understood issue in the canonical quantization program for general relativity.\\ 
Let us apply now Dirac's program to Einstein theory of gravity.\\
The space of states is that of proper functionals of configuration variables, the ``wave functionals"
 \begin{equation}
\Psi = \Psi [N,N^i, h_{ij}]
\end{equation}
In the standard representation, the configuration variables and the conjugate momenta act as multiplicative and derivative operators respectively:
\begin{align}
\widehat{h}_{ij}(x) \Psi &= h_{ij}(x) \Psi \; \; \; \; \widehat{\Pi}^{ij}(x) \Psi= - i \hbar \frac{\delta \Psi}{\delta h_{ij}(x)}\\
\widehat{N}(x) \Psi &= N(x) \Psi \; \;  \; \; \;\widehat{\Pi}(x) \Psi=- i \hbar \frac{\delta \Psi}{\delta N(x)} \\
\widehat{N^i}(x) \Psi &= N^i(x) \Psi \; \;  \; \; \;\widehat{\Pi}_i(x) \Psi=- i \hbar \frac{\delta \Psi}{\delta N^i(x)} 
\end{align} 
The implementation of the constraints (\ref{cont1},\ref{cont2}, \ref{conny}, \ref{conny1}) on a quantum level leads to the following set of equations
\begin{align}
- i\hbar \frac{\delta}{\delta N} \Psi[N,N^i, h_{ij}] &= 0 \label{c1} \\
- i\hbar \frac{\delta}{\delta N^i} \Psi[N,N^i, h_{ij}] &= 0 \label{c2}\\
\widehat{\mathcal{H}_i} \Psi[N,N^i, h_{ij}] &=  0  \label{c3}\\
\widehat{\mathcal{H}} \Psi[N,N^i, h_{ij}]  &= 0 \label{c4}  
\end{align}
Eq.(\ref{c1}) and eq.(\ref{c2}) imply that the physical states do not depend on $N$ and $N^i$ i.e. they do not depend on the slicing of the spacetime. Thus the wave functional $\Psi$, which describes the quantum state of the universe, is a function on the infinite-dimensional manifold $W$ of all three-metrics $h_{ij}$. \\
The super-momentum constraint is satisfied by wave functionals which depend on three-geometries $ \{ h_{ij} \}$ rather than any specific representation in a given coordinate system.
This can be seen for example considering the variation of $\Psi$ under an infinitesimal spatial translation $x^i \rightarrow x^i + \xi^i(x^l)$ is
\begin{equation}
\delta \Psi = -2 \int d^3x  \, \frac{\delta \Psi}{\delta h_{ij}(x)} D_j \xi_i = 2 \int d^3x \, D_j \Bigl[ \frac{\delta \Psi}{\delta h_{ij}(x)} \Bigr]\xi_i
\end{equation}
Thus requiring that $\delta \Psi =0$ implies that
\begin{equation}
\widehat{\mathcal{H}_i} \Psi[N,N^i, h_{ij}] = - 2 i \hbar D_j \Bigl[ \frac{\delta \Psi}{\delta h_{ij} (x)}\Bigr] = 0 \label{hii}
\end{equation}
where we used the expression definition \eqref{hi} of $ \widehat{\mathcal{H}_i}$\footnote{Note that the last equality holds for compact manifolds. For manifold with a boundary the same result is achieved introducing suitable boundary terms.}. \\
The physical space is thus made up of equivalence classes of metrics connected by a spatial coordinate transformation i.e. geometries $\{ h_{ij} \}$. This space is called ``superspace''. Note that the super-momentum constraint is trivially satisfied in cosmology where the universe is described with a homogeneous and isotropic model of spacetime.
Eq.(\ref{c4}) is called Wheeler-deWitt (WdW) equation and is a functional differential equation. It follows from expression (\ref{HH}) that $\widehat{H}$ contains products of operators evaluated at the same point and it is general not tractable or ill-defined. In this work we will only consider highly symmetric classes of metrics $h_{ij}$ ( in most cases homogeneous and isotropic metrics $h_{ij} = h_{ij}(t)$). This means that the wave functional will not take values in the entire infinite dimensional superspace but in the smaller space known as ``minisuperspace'', with a finite number of degrees of freedom. Notice that this restriction corresponds to a symmetry reduction performed at the classical level, before the canonical quantization, which will allow us to deal with solvable models. This procedure is strictly speaking in tension with the Heisenberg's uncertainty principle: for example the quantization of a geometry of FLRW type corresponds to setting to zero at the quantum level both all of the in-homogeneous quantum degrees of freedom and their conjugate momentum \cite{Kuchar:1989tj}. We will come back to this issue in chapter \ref{chapterneumann} where we will discuss a possible path to verify the validity of this approximation, or possibly disprove it, making use of holography.\\

%WdW equation can be derived from a path integral using Einstein-Hilbert action for gravity \cite{kiefer}, \cite{hal}. In the following we will focus on the Path-Integral formulation of Quantum Cosmology and relate our results with those of the canonical approach showing under which assumptions the wave function constructed via path integral satisfies the super-Hamiltonian constraint (see Section \ref{wheldewitt}). \\
%As mentioned before, the Hamiltonian Eq.(\ref{hamil}) is a linear combination of constraints. Therefore it also annihilates physical states
%\begin{equation}
%\widehat{H} \Psi = 0 \label{wdwequation}
%\end{equation}  
%This equation expresses what is known as the problem of time in quantum gravity \cite{rovelli}, \cite{thiemann}. It can be seen as a Schroedinger equation for a quantum state which does not evolve in time giving rise to what is called the \textit{frozen formalism}.
%The wave function $\Psi$ does not depend explicitly on the time $t$ because $t$ is just a coordinate which can be given arbitrary values by different choices of the Lagrangian multipliers $N$ and $N^i$. \\

\subsection{Path Integral} \label{sectionpathintegral}
In the path integral formulation of quantum mechanics one defines the quantum amplitude for a transition of a particle from a position x at time t to a position x' at time t' as a sum over all possible paths linking the initial and the final points
\begin{equation}
\langle x', t' | x, t \rangle = \int \delta x(t) \, e^{i S[x(t)]/\hbar} \label{pathi}
\end{equation}
where all the path in the summation are continuous but might be non differentiable.\\
By construction, this formalism is useful in defining the classical limit of the theory. As $\hbar \rightarrow 0$, two close histories associated with two close values of the action $S$ correspond to huge changes in the phase $e^{iS/\hbar}$. As a consequence most of the paths interfere destructively. The path integral will then be peaked on the classical histories which by definition render the action stationary $\delta S = 0$.\\ 
The quantum mechanics of the gravitational field can also be formulated through path integral methods and deals with the transition amplitude from one three-dimensional geometry to another.
These three-dimensional geometries are defined on spatial hypersurfaces labelled by the time coordinate and separated by a local proper time interval, as given by the ADM slicing (\ref{aDm}). One must consider the class of all the four-geometries in which these two spacelike surfaces occur but which are in general different off the surfaces. The path integral is defined as a sum over all such four-geometries joining the two boundary geometries. Notice that the local proper time separation between the surfaces is not specified as this quantity is be different for every four-geometry in the sum and thus the path integral includes also a sum over all the possible proper time separation between the boundary hypersurfaces.\\
The quantum gravitational path integral for a transition from a three-geometry $g_0$ to another $g_1$ \cite{Misner:1957wq} is then defined to be
\begin{equation}
G[g_1;g_0]=\int \delta g_{\mu \nu}(t,x)\, e^{i S [g_{\mu \nu}(t,x)/ \hbar} \label{path}
\end{equation}
where $S$ is the action functional for gravity with the Gibbons-Hawking-York boundary term \eqref{einsteinhilbert}. As we will see in section \ref{sec:boundaryterms}, the GHY boundary term is required to fix Dirichlet type of boundary conditions i.e. if we wish to fix the two boundary geometries $g_1$ and $g_0$. When other quantities than the induced three-geometry are fixed at the boundaries, different boundary terms must be introduced accordingly. \\
The action \eqref{einsteinhilbert} is invariant under diffeomorphisms i.e. the gauge transformations generated by the first class constraint $\mathcal{H},\mathcal{H}_i$ . When computing the path integral one should sum only over physically different histories  and hence keep only one representative of the class of equivalence of geometries related to each other by these transformations. This means that the domain of integration is a gauge fixing hypersuperface in the configuration space and not the whole space. Hence we should fix the gauge in the summation and introduce a Faddeev-Popov type of ghost contribution according to the usual procedure for systems with gauge freedom. The whole treatment of Batalin-Fradkin-Vilkovisky (BFV) ghost is showed in  \cite{Teitelboim:1983fk, Teitelboim:1981ua, Teitelboim:1983fh, Teitelboim:1983fi, Halliwell:1988wc} and here we will only make some comments on the  key points useful for our purposes.\\
The path integral with the ghost contribution reads
\begin{equation}
G[g_1;g_0]=\int \delta h_{ij} \, \delta \Pi^{ij}  \delta N \delta N^i \, \delta c \, \delta p \, f(N,N^i)e^{i (S + S_{ghost})/ \hbar} \label{pathintegral}
\end{equation}
where $f(N,N^i)$ is gauge fixing condition and $c$ and $p$ are the ghosts and their conjugate momenta.\\
As explained in \cite{Teitelboim:1983fk, Teitelboim:1981ua}, the simplest gauge condition we can impose is the so called ``proper time gauge'' condition
\begin{align}
 \dot{N}(x,t) = 0 \\
 N^i(x,t) = 0
\end{align}
Importantly, with this choice of gauge fixing, all of the ghosts decouple in minisupersace \cite{Teitelboim:1981ua}.\\
The first condition implies that pointwise the proper time separation between two neighbour surfaces does not depend on the position of the surfaces. From a practical point of view this gauge choice implies that the functional integral over the lapse reduces to a ordinary one.\\
The latter condition means that the vector joining a point on the spatial hypersurface at $t$ to the point with the same spatial coordinates at $t + dt$ is normal to $\Sigma_t$. \\
If $\langle g_1, T | g_0 , 0\rangle$ is the usual propagator in quantum mechanics, the path integral is an object of the form
\begin{equation}
\int d T \langle g_1, T | g_0, 0 \rangle = G (g_1;g_0 | E)|_{E = 0}
\end{equation}
where the energy Green function is defined to be
\begin{equation}
G (g_1,g_0| E) = \int d T \, e^{i E T} \langle g_1, T | g_0, 0 \rangle
\end{equation}
The integration over the whole four-metric, including the lapse, implies then that the quantum-gravitation path integral resembles more a energy Green function than the propagator of ordinary quantum mechanics.

We discussed in the previous section how the constraints \eqref{c1}-\eqref{c4} have a functional nature and are therefore difficult if not impossible to solve. The same problem arises trying to define quantum transitions for gravity via path integral. The full integral is in general not manageable, when not ill-defined.  
It is useful in this sense to formulate the problem performing a symmetry reduction. This allows us to deal with systems with a  finite number of degrees of freedom or more tractable problems with still infinitely many of them. The former case includes homogeneous models for which the configuration space is finite-dimensional and we reduce the superspace to a minisuperspace. The term midisuperspace refers to the latter case, which includes for example systems with spherical symmetry, which we will discuss in the last chapter. In this work we are going to consider symmetry reduced path integrals with FLRW, Kantowski-Sachs and Bianchi IX ans\"{a}tze. We discussed in section \ref{canonical} how the minisuperspace approximation is in tension with the uncertainty principle in the canonical quantization. The symmetry reduction of the path integral is clearly equally problematic. The questions we ask here are: how much of the physical information is lost by excluding all of the in-homogeneous histories from the path integral? Is the final result heavily dependent on this restriction? Does the minisuperspace path integral already encode the key elements to describe the system? We are going to keep these questions in mind throughout the work. We will see that, for example, the no boundary proposal implemented with Dirichlet boundary conditions is ill-defined because it leads to unstable inhomogeneous perturbations. This instance is clearly signaling that the minisuperspace approximation is not enough in this case since the inhomogeneous degrees of freedom are not suppressed and one runs into inconsistencies when trying to ignore them.
\\

\subsubsection{Comments on causality}\label{sec:causality}
In quantum field theory Feynman's path integral defines a causal propagator given that there is causal structure in the histories summed over.
Here, once the gravitational field is quantized, there is no more notion of time and apparently we cannot save any clear concept of causality. As pointed out in \cite{ Teitelboim:1983fh}, we can impose a causal structure to the histories that form the path integral partially breaking the gauge invariance of the classical theory. 
In order to describe a transition from $g_0$ to $g_1$, one calculates the quantum amplitude for having a three-geometry $g_0$ on a given spacelike hypersurface and the three-geometry $g_1$ on another one when the proper time separation between two points with spatial coordinate $x$ on the two hypersurfaces is $N(x)(t_1 - t_0)$. Then one integrates over all possible values of the lapse. To recover a causal structure of the propagator, we can choose to allow a summation over only those histories for which $g_1$ lies in the future of $g_0$. This is implemented by the request that $N(x)>0 $ $\forall x$.
Recall that the timelike diffeomorphisms, through the Lie derivative, push the initial hypersurface backward and forward. Thus, with such restriction in the integration over the diffeomorphism group elements, one averages the amplitude only over half of the space of the possible normal deformations of the initial surface. As a consequence the path integral is not invariant under the action of the generator of the time translations i.e.
\begin{equation}
\hat{H}(x)\, G_{+}[g_1,g_0] = - i \delta(g_1 - g_0) \neq 0
\end{equation}
Thus the path integrals gives a Green function of the WdW operator i.e. a propagator.\\
If one allows $N$ to run over the whole real axis one recovers indeed that
\begin{equation}
\hat{H}(x)\, G[g_1,g_0]  = 0
\end{equation}
but the amplitude is now a-causal. \\
Thus if one considers ($ - \infty < N < + \infty$) the path integral is a real solution to Wheeler-DeWitt equation that is, a wave function, with no reference to any underling causal structure.\\
In this work we will consider both propagators and wave functions and will always specify when unclear what type of object the path integrals represents.

\subsubsection{Euclidean vs Lorentzian formulation}\label{sec:conffactor}
In quantum field theory it is often appropriate to perform a rotation from the original formulation in spacetime to the four-dimensional Euclidean speace. This procedure is known as ``Wick rotation" and is implemented by just sending the time coordinate $t \rightarrow - i \tau$. Among other advantages, the Euclidean formulation of the theory improves the convergence properties of the path integral as the integrand changes as follows
\begin{equation}
e^{i S/ \hbar} \rightarrow e^{- S_E}
\end{equation}
and the Euclidean action $S_E$ is bounded from below.\\ 
In analogy with the Wick-rotated quantum field theory, the gravitational path integral is also often formally defined over positive definite Euclidean metrics. Euclidean path integrals are typically advocated in quantum cosmology and the no boundary proposal, on which we will focus later, was originally formulated in this fashion (see \cite{Hartle:1983ai}). 
However in a diffeomorphism invariant theory the analytic continuation of time has not clear interpretation as the 'time' coordinate is not uniquely determined.
A general Lorentzian metric will not have a sector in the complexified spacetime manifold where it is real and positive definite. That is, it is in general not possible to change a metric with a Lorentzian signature (-,+,+,+) to an euclidean signature (+,+,+,+) just sending $t \rightarrow -i \tau$. 
Moreover in the case of gravity, unlike the cases of scalar or Yang-Mills fields, if one takes the Euclidean nature to be fundamental ab initio, the action is unbounded from below because of the so called ``conformal factor problem'' \cite{Gibbons:1978ac}.
The Euclidean action can indeed be written as
\begin{equation}
S_{E} = - \frac{c^4}{16 \pi G} \int d^4 x \sqrt{g} R - \frac{c^4}{8 \pi G} \int d^3 y \, \sqrt{h} K 
\end{equation}
for a positive definite metric $g_{\mu \nu}$.
Considering a conformal decomposition
\begin{equation}
g_{\mu \nu} = e^{2 \phi(x)}\overline{g}_{\mu \nu}
\end{equation}
the gravity action becomes
\begin{equation}
S_E = - \frac{1}{16 \pi G} \int d^4 x e^{2 \phi} \sqrt{\overline{g}} \Bigl[\overline{R} + 6 (\nabla \phi )^2 \Bigr] - \frac{1}{8 \pi G} \int d^3 y \, \sqrt{h} e^{2 \phi}K 
\end{equation}
Since the kinetic term of the conformal mode is positive definite the action can assume arbitrarily  large negative values for rapidly varying $\phi$. As a consequence, the path integral over metrics spanning a given fixed boundary is in general ill-defined and one has to specify a complex contour of integration to get a finite result. There is of course some arbitrariness in the choice of such a contour which affects the final results. In particular this might influence which saddle points are relevant to the path integral. \\
For these reasons we consider misleading the passage to the Euclidean theory and will be focusing in this work on Lorentzian path integrals for quantum cosmology. The Lorentzian formulation is in our opinion much more natural and of more direct physical interpretation. In the following, we will evaluate the corresponding oscillating integral using the mathematical tools of ``Picard Lefschetz theory'' (see section \ref{picard}).  We will show in particular that in the case of a positive cosmological constant, which is relevant for cosmology, minisuperspace Lorentzian path integrals indeed converge. This will let us derive precise and un-ambiguous results for the no boundary proposal. 
We will see in chapter \ref{chapterneumann} that the Lorentzian integral fails to converge if the cosmological constant is taken to be negative and we will be forced to introduce suitable complex integration contours. In this case the path integral gives the partition function for the dual
Euclidean quantum field theory and it is not clear a priori what is the natural set of geometries one should sum over. In this work we adopt the point of view that path integrals for cosmology, where the cosmological is positive, which shall give a semi-classical description of our world and where the lapse function has a clear interpretation in terms of proper time distance between event, should by Lorentzian. We remain agnostic regarding signature requirements for other cases including in particular asymptotically AdS spacetimes.

\subsubsection{Boundary conditions}\label{sec:boundaryterms}
The $d$ dimensional solutions to the Einstein's equations \eqref{einsteinequations} are critical points of the Einstein-Hilbert action
\begin{equation}
    S = \frac{1}{16 \pi G } \int_M d^d x \sqrt{-g} [ R - 2 \Lambda]  +  S_B
\end{equation}
 The boundary contributions is necessary to obtain a consistent variational problem $\delta S=0$ upon variation with respect to the metric.
Fixing different quantities requires different boundary terms. Dirichlet and Neumann are widely known and adequate conditions under most circumstances. While less common, Robin conditions have already proven useful for some gravitational problems, e.g. the formal definition of perturbation theory in Euclidean gravity~\cite{Witten:2018lgb}. The following are the usual options:
\begin{itemize}
\item \textit{Dirichlet} boundary conditions: the required boundary term is the well-known Gibbons-Hawking-York (GHY) term \cite{York:1972sj, Gibbons:1976ue}
\begin{equation}
     S_B = \frac{1}{8 \pi G} \int_{\partial M} d^{(d-1)}y \, \epsilon \sqrt{|h|}  K 
\end{equation}
with $K$ the trace of the intrinsic curvature and $h$ the determinant of the metric induced on the boundary.\\
The variation of the action yields
\begin{equation}
\begin{split}
    \delta S =   &+\frac{1}{16 \pi G } \int_M d^d x \sqrt{-g} \Bigl( R_{\mu \nu } - \frac{1}{2} g_{\mu \nu } R + \Lambda g_{\mu \nu} \Bigr) \, \delta g^{\mu \nu} + \\
    &-   \frac{1}{16 \pi G } \int_{\partial M} d^{(d-1)}y \, \epsilon \sqrt{h} \Bigl(K^{ij} - K h^{ij} \Bigr) \delta h_{ij} 
    \end{split}
\end{equation}
Recalling that the conjugate momentum is given by 
\begin{equation}
    \Pi^{ij}:= \frac{\delta \mathcal{L}}{\delta \dot{g}_{ij}}  = -   \frac{1}{16 \pi G }  \epsilon \sqrt{h} \Bigl(K^{ij} - K h^{ij} \Bigr)
\end{equation}
we can write that variational principle as
\begin{equation}
    \delta S =   
    \frac{1}{16 \pi G } \int_M d^d x \sqrt{-g} \Bigl( R_{\mu \nu } - \frac{1}{2} g_{\mu \nu } R + \Lambda g_{\mu \nu} \Bigr) \, \delta g^{\mu \nu} +\int_{\partial M} d^{(d-1)}y \, \Pi^{ij} \, \delta h_{ij} =0 \label{Dirichlet}
\end{equation}
Thus we find that the principle of least action is compatible with vanishing variation of the metric induced on the boundary $\delta h_{ij} = 0$ or, in other words, that the boundary metric  $h_{ij}$ can be fixed to an arbitrary value. 
\item \emph{Neumann} boundary conditions demand the boundary term~\cite{Krishnan:2016mcj}
\begin{equation}
    S_B= \frac{4 - d}{8 \pi G} \int_{\partial M} d^{(d-1)}y \, \epsilon \sqrt{|h|}  K  
    \end{equation}
The variation of the action is 
\begin{equation}
    \delta S =   
    \frac{1}{16 \pi G } \int_M d^d x \sqrt{-g} \Bigl( R_{\mu \nu } - \frac{1}{2} g_{\mu \nu } R + \Lambda g_{\mu \nu} \Bigr) \, \delta g^{\mu \nu} +\int_{\partial M} d^{(d-1)}y \, h_{ij} \, \delta \Pi^{ij} \label{Neumann}
\end{equation}

implying that we can set the momentum $\Pi^{ij}$ to any desired value at the end points, without fixing the boundary metric itself. Note that for $d=4$ the boundary term vanishes so that in this case Neumann boundary conditions can be imposed adding no boundary term to the Einstein-Hilbert action. 

\item For \emph{Robin} boundary conditions \cite{Krishnan:2017bte}

\begin{equation}
  S_B= \frac{4 - d}{8 \pi G} \int_{\partial M} d^{(d-1)}y \, \epsilon \sqrt{h} K     - \xi (d-3) \int_{\partial M} d^{(d-1)}y \, \epsilon \sqrt{|h|} 
\end{equation}
In this case the variation of the boundary action yields a condition on a combination of both the field value and the momentum, since the variation gives
\begin{equation}
     \delta S =   
    \frac{1}{16 \pi G } \int_M d^d x \sqrt{-g} \Bigl( R_{\mu \nu } - \frac{1}{2} g_{\mu \nu } R + \Lambda g_{\mu \nu} \Bigr) \, \delta g^{\mu \nu} - \int_{\partial M} d^{(d-1)}y \, \delta \Bigl( \Pi^{ij}+ \xi \sqrt{|h|}  h^{ij} \Bigr)
\end{equation}

so that the combination $\Pi^{ij}+ \xi \sqrt{|h|}  h^{ij}$ can be set to any desired value on the boundary. Note that a Robin boundary condition is in no way exotic: when we specify the Hubble rate on a given hypersurface, we are effectively imposing a Robin condition: say we specify $H = \frac{a_{,t_{p}}}{a} \equiv {\cal H},$ then we can rewrite this condition in Robin form as $a_{,t_{p}} - {\cal H} a = 0,$ where $t_{p}$ denotes the physical time. 

\item \emph{Mixed} boundary conditions are obtained introducing different types of boundary terms and thus fixing different quantities on disjoint parts of the boundary.

\item \emph{Special} boundary conditions arise when the prefactor of the variation is set to zero. For instance, we may obtain a Special Neumann condition if in \eqref{Dirichlet}, instead of setting $ \delta h_{ij}=0,$ we set $\Pi^{ij}  =0$ at the boundary. The variational problem is then also well-defined, but note that this only works for the special case where the momentum is set to zero. If one wants to set the momentum to a non-zero value, one must use the Neumann condition \eqref{Neumann} instead, and this requires a different boundary term. In a similar vein, one may set $h_{ij}=0$ on the boundary thereby converting \eqref{Neumann} into a Special Dirichlet condition. We will encounter slightly more general examples of such Special boundary conditions in section \ref{robincosmology}, where we will implement a Special Robin condition. \\

\end{itemize}

\subsubsection{Picard-Lefschetz theory} \label{picard}

Picard-Lefschetz theory can be seen as a systematic way of evaluating conditionally convergent integrals which we will use for computing the saddle point approximation of oscillatory integrals.
We will briefly review the salient features in this section where part of the text is based on \cite{Bramberger:2019zks}. \\
We are interested in integrals of the form
\begin{equation}
    \int_{\mathcal{C}} dz \, e^{i S(z)/\hbar}
\end{equation}
with $\mathcal{C}$ a real infinite domain. Evidently, this integral does not converge absolutely. The main idea of Picard-Lefschetz theory is to complexify the integral of interest and then deform the original contour of integration in such a way as to render the resulting integral manifestly convergent. It may be useful to consider a simple example, say $\tilde{I}=\int_{\mathbb{R}} dx e^{ix^2}.$ Along the defining contour, namely the real line, this is a highly oscillating integral. But now we can deform the contour by defining $x=e^{i\pi/4}y,$ such that $\tilde{I}=e^{i\pi/4}\int dy e^{-y^2}.$ Along the new contour, the integral has stopped oscillating, and in fact the magnitude of the integrand decreases as rapidly as possible. The integral is now manifestly convergent, and one may check that the arcs at infinity linking the original contour to the new one yield zero contribution. Note that along the steepest descent path, there is an overall constant phase factor (here $e^{i\pi/4}$) -- this is a general feature of such paths.\\
More formally, we can write the exponent $iS[z]/\hbar$ and its argument, taken to be $z$ here, in terms of their real and imaginary parts, $iS/\hbar=h+i H$ and $z=x + i y$ --  see Fig.~\ref{fig:thimble} for an illustration of the concepts. The complex function $iS[z]/\hbar$ is holomorphic if its Wirtinger derivative with respect to the complex conjugate of $z $ vanishes or, equivalently, if its real and imaginary part are solutions to the Cauchy-Riemann equations
\begin{equation}
\begin{split}
\frac{\partial \mathcal{S}}{\partial \overline{z}} = 0 \; \Longleftrightarrow \; &\frac{\partial h}{\partial x} = +  \frac{\partial H}{\partial y} \\
&\frac{\partial h}{\partial y} = - \frac{\partial H}{\partial x} 
\end{split} \label{cr}
\end{equation}
Then the critical points $z_\sigma$ of the action are also critical points of the real part of the exponent $h = \operatorname{Re}(\mathcal{S}) $, the so called the Morse function
\begin{equation}
\frac{i}{\hbar}\frac{\partial S }{\partial z}  = 2 \frac{\partial h}{\partial z}
= 0  \Longleftrightarrow \frac{\partial h}{\partial x} = 0, \; \frac{\partial h}{\partial y}=0
\end{equation}
The critical points $z_\sigma$  are in fact saddle points of $h$ since its Hessian matrix is indefinite
\begin{equation}
     {\mathcal H} = det  \left( \begin{array}{cc} \frac{\partial^2 h}{\partial x^2} & \frac{\partial^2 h}{\partial x \partial y} \\  
\frac{\partial^2 h}{\partial x \partial y} &  \frac{\partial^2 h}{\partial y^2}
\end{array} \right) \Bigr|_{z_{\sigma} }=0 
\end{equation}
Therefore, the solutions to the classical equation of motion, which are stationary points $z_{\sigma}$ of $S$, are saddle points for $h$, from which a steepest descents and a steepest ascent paths start. Downward flow of the magnitude of the integrand is then defined by 
\begin{equation}
\frac{\mathrm{d}u^i}{\mathrm{d}\lambda} = -g^{ij}\frac{\partial h}{\partial u^j}\,,
\label{eq:dw}
\end{equation}
with $u^i=(x,y)$ and $\lambda$ denoting a parameter (along the flow) and $g_{ij}$ denoting a metric on the complexified plane of the original variable $z$ (here we can take this metric to be the trivial one, $ds^2 = d|u|^2$). The Morse function decreases along the flow, since  $\frac{\mathrm{d}h}{\mathrm{d} \lambda} = \sum_i\frac{\partial h}{\partial u^i}\frac{\mathrm{d}u^i}{\mathrm{d}\lambda} = -\sum_i\left(\frac{\partial h}{\partial u^i}\right)^2<0.$ The downward flow Eq. (\ref{eq:dw}) can be rewritten as 
\begin{equation}
\frac{\mathrm{d}u}{\mathrm{d}\lambda} = - \frac{\partial {\bar{\cal I}}}{\partial \bar{u}}, \quad \frac{\mathrm{d}\bar{u}}{\mathrm{d}\lambda} = - \frac{\partial {{\cal I}}}{\partial {u}}\,,
\end{equation} 
and this form of the equations is useful in that it straightforwardly implies that the phase of the integrand,  $H = \text{Im}[iS/\hbar],$ is conserved along a flow,
\begin{equation}
\label{eq:imh}
\frac{\mathrm{d} H}{\mathrm{d}\lambda} = \frac{1}{2i}\frac{\mathrm{d}({\cal I} - \bar{\cal I})}{\mathrm{d}\lambda} = \frac{1}{2i}\left( \frac{\partial {\cal I}}{\partial u}\frac{\mathrm{d}u}{\mathrm{d}\lambda} - \frac{\partial \bar{\cal I}}{\partial \bar{u}}\frac{\mathrm{d}\bar{u}}{\mathrm{d}\lambda}\right) = 0\,.
\end{equation}
Thus, along a flow the integrand does not oscillate, rather its amplitude decreases as fast as possible. Such a downwards flow emanating from a saddle point $z_\sigma$ is denoted by $\mathcal{J}_\sigma$ and is often called a ``Lefschetz thimble''. They define convergent contours for the integral $\int dz \, e^{i S/\hbar}$ under very general assumptions.

\begin{figure}[h] 
		\includegraphics[width=0.85\textwidth]{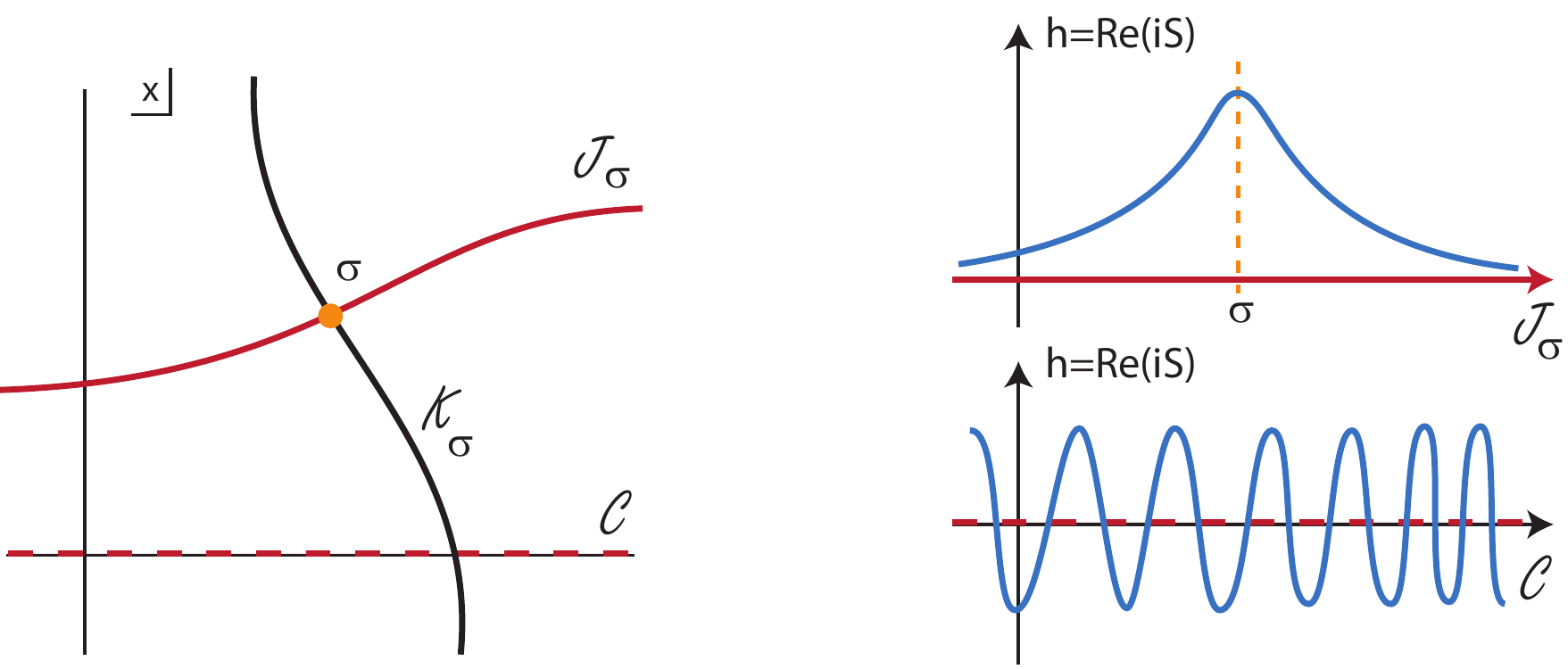}
	\caption{Picard-Lefschetz theory instructs us how to deform a contour of integration such that an oscillating integral along a contour $\mathcal{C}$ gets replaced by a steepest descent contour (or in general a sum thereof) along a Lefschetz thimble $\mathcal{J}_\sigma$ associated with a saddle point $\sigma$. Only those saddle points contribute for which the flow of steepest ascent $\mathcal{K}_\sigma$ intersects $\mathcal{C}.$. The figure is taken from \cite{Bramberger:2019zks}.}
	\protect
	\label{fig:thimble}
\end{figure} 

In much the same way one can define an upwards flow
\begin{equation}
\frac{\mathrm{d}u^i}{\mathrm{d}\lambda} = +g^{ij}\frac{\partial h}{\partial u^j}\,,
\label{eq:uw}
\end{equation}
with $H$ likewise being constant along such flows. Upwards flows are denoted by ${\cal K}_\sigma,$ and they intersect the thimbles at the saddle points. Thus we can write
\begin{equation}
{\rm Int}({\cal J}_\sigma, {\cal K}_{\sigma'})=\delta_{\sigma \sigma'}. \label{eq:intersection}
\end{equation}
Our goal then is to express the original integration contour $\mathcal{C}$ as a sum over Lefschetz thimbles, 
\begin{equation}
\label{eq:contourexp}
{\cal C} = \sum_\sigma n_\sigma {\cal J}_\sigma\,.
\end{equation}
Multiplying this equation on both sides by ${\cal K}_{\sigma}$ we obtain that $n_\sigma= {\rm Int}(\mathcal{C}, {\cal K}_{\sigma}).$ Thus a saddle point, and its associated thimble, are relevant if and only if one can reach the original integration contour via an upwards flow from the saddle point in question. Intuitively, this makes sense: we are replacing an oscillating integral, with many cancellations, by one which does not contain cancellations, and thus the amplitude along the non-oscillating path must be lower. Putting everything together, we can then re-express the conditionally convergent integral by a sum over convergent integrals, 
\begin{align}
\label{eq:contour}
\int_{\cal C} \mathrm{d} x \, e^{iS[x]/\hbar} & = \sum_\sigma n_\sigma \int_{{\cal J}_\sigma} \mathrm{d} x \, e^{iS[x]/\hbar} \\
& = \sum_\sigma n_\sigma \, e^{i \, H(x_\sigma)}\int_{{\cal J}_\sigma} e^h \mathrm{d}x \\
& \approx \sum_\sigma n_\sigma \, e^{i S(x_\sigma)/\hbar}\,.
\end{align}
The last line expresses the fact that the integral along each thimble may easily be approximated via the saddle point approximation, the leading term being the value at the saddle point itself.  If required, one can then evaluate sub-leading terms by expanding in $\hbar,$ but in the present work this will not be necessary. This concludes our mini-review of Picard-Lefschetz theory -- for a detailed discussion see \cite{Witten:2010cx} and \cite{Behtash:2015loa}, and for applications in a similar context than the present one see \cite{Feldbrugge:2017kzv,Feldbrugge:2017fcc,Feldbrugge:2018gin}.

\subsection{Classicality}\label{classicality}

The saddle point approximation of the wavefunction of the universe can be evaluated in a precise way thanks to Picard-Lefschetz theory. But under what conditions can we recover a classical universe from the wavefunction?
\cite{Vilenkin:1988yd, Kiefer:1990pt, Kiefer:1993fg, Halliwell:1984eu}\\
Let us consider a vacuum gravitational system described by a set of variables $q^a$. The wavefunction of the system satisfies the WdW equation \eqref{c4}

\begin{equation}
\Bigl[    - \frac{1}{2} G^{a b} \frac{\partial^2 }{\partial q^a \partial q^b} + V(q^a) \Bigr] \Psi = 0 
\end{equation}
where the potential $V$ is given by $V= \frac{\sqrt{h}}{2 (8 \pi G)^2} \, (^{(3)}R - 2 \Lambda) $.
We can identify the real and the oscillatory part of the wavefunction introducing the two real functions $A(q^a)$ and $P (q^a)$
\begin{equation}
    \Psi(q^a) = e^{ - A(q^a) + i P (q^a)}
\end{equation}
Separating the real and in imaginary part of the WdW equation we obtain 

\begin{align}
    \frac{1}{2} G^{ab} \Bigl[ \frac{\partial^2 A}{\partial q^{a} \partial q^b} - \frac{\partial A}{\partial q^a} \frac{\partial A}{\partial q^b}  + \frac{\partial P}{\partial q^a} \frac{\partial P}{\partial q^b}   \Bigr] + V &= 0\\
 \frac{1}{2} G^{ab} \Bigl[ \frac{\partial^2 P}{\partial q^{a} \partial q^b}  - 2\frac{\partial A}{\partial q^a} \frac{\partial P}{\partial q^b}  \Bigr]   &= 0  \label{wdwRI}
\end{align}
If the phase of the wavefunction varies much faster than its amplitude as a function of all variables $q^a$ 
\begin{equation}
    \frac{\partial A}{\partial q^a} \ll  \frac{\partial P}{\partial q^a}  \label{classicalitycondition}
\end{equation}
the first equation becomes the Hamilton-Jacobi equation for gravity 
\begin{equation}
      \frac{1}{2} G^{ab}\frac{\partial P}{\partial q^a} \frac{\partial P}{\partial q^b}   + V = 0 
\end{equation}
where $P$ can be identified as the classical action. With this identification one obtains that the canonical momentum is related to the action in the usual way
\begin{equation}
    p_a = \frac{\partial P}{\partial q^a} \label{momentumcl}
\end{equation}
Knowing $P$, one can intregrate this equation to find the classical dynamics of the system. Note that the Hamilton-Jacobi equation is equivalent to Einstein's equations and thus when the classicality condition \eqref{classicalitycondition} is matched the solution is a classical spacetime.  The WdW equation, which is equivalent to a Klein-Gordon equation, is associated with the conserved current
\begin{equation}
    J^a = - \frac{i}{2} G^{ab} \Bigl[\Psi^* \frac{\partial \Psi}{\partial q^b} - \Psi \frac{\partial \Psi^*}{\partial q^b} \Bigr]
\end{equation}
The second equation \eqref{wdwRI} enforces the conservation of this current and fixes the amplitude associated to the wavefunction.
With the identification \eqref{momentumcl} we obtain that the modulus squared of the wavefunction is roughly given by $1/p$: in ordinary quantum mechanics the amplitude gives the probability for a particle to be found in a certain space interval and it is expected that in the classical limit this probability is given by the inverse of the particle's velocity \cite{Landau:1991wop}.\\
This perturbative method to solve differential equations such as the Wheeler-deWitt and the Schr\"{o}dinger equations was developed by Gregor Wentzel, Hendrik Anthony Kramers and L\'{e}on Brillouin (WKB) and boils down to a familiar concept in quantum mechanics: the solution to the Schr\"{o}dinger equation for a particle in a potential is oscillatory in the classically allowed region and decays exponentially in the classically forbidden region. Throughout this work we will say that a system is becoming classical in a WKB sense when the phase of its wavefunction varies rapidly as compared to its amplitude.\\
In semi-classical gravity we are interested in the regime where part of the system behaves classically while the rest does not. For this case, we shall consider a set of genuinely quantum variables $v^{\alpha}$ together with the classical ones $q^a$. The WdW equation will then be

\begin{equation}
   \Bigl[    - \frac{1}{2} G^{a b} \frac{\partial^2 }{\partial q^a \partial q^b} + V(q^a) + H_v(q^a , v^{\alpha}) \Bigr] \Psi = 0 
\end{equation}
with $H_v(q^a , v^{\alpha})$ is the part of the Hamiltonian which involves the $v^{\alpha}$ variables. In the case of cosmology, we could think for example of the $q^a$ as associated with the FLRW background and the $v^\alpha$ as rescribing the quantum fluctuations.\\
The wavefunction can be written as
\begin{equation}
    \Psi(q^a, v^{\alpha}) = e^{- A(q^a) + i P(q^a)} \chi(q^a , v^{\alpha})
\end{equation}
where to introduce a separation of scales we can think of $\chi$ as given by
\begin{equation}
     \chi(q^a , v^{\alpha}) = e^{i \epsilon \, S(q^a , v^{\alpha})  }
\end{equation}
with $\epsilon$ a small parameter\footnote{This procedure is often called the Born-Oppenheimer approximation of the wavefunction of the universe and the expansion parameter is taken to be the inverse Planck mass \cite{Kiefer:1990pt}.}.\\
The WdW equation for the full system is given by the two equations \eqref{wdwRI}, which are of order $\epsilon^0$, plus the higher order equation
\begin{equation}
    \frac{1}{2}G^{ab} \Bigl[ - \frac{\partial^2 \chi}{\partial q^a \partial q^b} + 2 \frac{\partial \chi}{\partial q^a} \frac{\partial A}{\partial q^b} - 2 i \frac{\partial P}{\partial q^a} \frac{\partial \chi}{\partial q^b} \Bigr] + H_v \chi =0
\end{equation}
In the limit of $\epsilon \ll 1$ and  $ \frac{\partial A}{\partial q^a} \ll  \frac{\partial P}{\partial q^a} $, the first two terms are negligible and we obtain

\begin{equation}
   -G^{ab} i \frac{\partial P}{\partial q^a} \frac{\partial \chi}{\partial q^b}  + H_v \chi =0 \label{wkb2}
\end{equation}
Since the background part of the system, described by the $q^a$ only, is classical and $p_a = \frac{\partial P}{\partial q^a} $, we can define the WKB time as 
\begin{equation}
     \frac{\partial}{\partial \eta}=G^{ab}   \frac{\partial P}{\partial q^a} \frac{\partial }{\partial q^b}
\end{equation}
in which case the eq. \eqref{wkb2} becomes a Schr\"{o}dinger equation for the quantum part of the system
\begin{equation}
    i     \frac{\partial \chi}{\partial \eta} = H_v \chi
\end{equation}
When we quantized the primordial fluctuations in section \ref{sec:QFT} we solved precisely this Schr\"{o}dinger equation. We can in fact think of that Schr\"{o}dinger equation as emerging from solving the WdW 
equation for the full system, including background and perturbations, in the semi-classical limit, that is, when the background FLRW universe behaves classically in a WKB sense.\\
The WKB classicality is based on the saddle point approximation of the wavefunction since in this limit the wavefunction its given by $\Psi \propto e^{iP}$, with $P$ the action evaluated along a solution to the classical equations of motion. It could happen that there is more than one such saddle point, in which case the wavefunction is given by a sum of WKB type of contributions
\begin{equation}
\Psi = \sum_{k}   e^{- A_k(q^a) + i P_k(q^a)}\, ( \chi_k(q^a, v^{\alpha})) \label{stato}
\end{equation}
This superposition does not describe a classical world since interferences come into play and a notion of classical universe seems difficult to recover. Decoherence \cite{Giulini:1996nw} provides a dynamical mechanism to describe the emergence of a classical world considering a separation between relevant and irrelevant degrees of freedom in a measurement.
In cosmology, we can for example take as relevant degrees of freedom the scale factor and some homogeneous scalar fields. Irrelevant degrees of freedom will be 'all the rest of the world' such as gravitational waves and density fluctuations or gas molecules, photons, etc. \cite{Kiefer:1987ft}. 
The interaction of the inhomogeneous fields with the homogeneous ones can render the latter classical. In a sense, then, a classical space-time arises from a 'self-measurement' of the universe \cite{Giulini:1996nw}. 
In the high-dimensional configuration space which includes both homogeneous and in-homogeneous fields, states like (\ref{stato}) are highly correlated. However, most of the environmental degrees of freedom are inaccessible to a localised observer. This leads to a reduced density matrix which is obtained from (\ref{stato}) by integrating out the huge number of irrelevant degrees of freedom $\rho (q_k, q_k') =\int \, dv^\alpha \psi \, \psi^* $. 
After the integration over the irrelevant degrees of freedom, the different branches of the superposition will be correlated to orthogonal states of the 'rest of the world' and hence locally there would be no observed interference at all. The off-diagonal terms of $\rho (q_k, q_k')$ are vanishingly small due to the interaction with the environment and the system is said to decohere. If this happens, the density matrix can be considered to describe a mixture of non-interfering WKB branches.\\
Taking these notions into account, one can attempt to characterize a classical behaviour by using the following criteria \cite{Paz:1991nd}: a system can be regarded as behaving classically when the wave function predicts the existence of correlations between coordinates and momenta \cite{Halliwell:1987eu}. These correlations should be such that the classical equations of motion are satisfied. Moreover, in the classical limit the system decoheres and can be written as a statistical mixture of non-interfering branches.\\
In cosmology, the very concept of time makes sense only after decoherence has occurred. In fact, the suppression of interferences selects a unique background spacetime. Such background acts as a clock which defines the time with respect to which the other fields evolve. In fact, in this limit we recovered an approximate Schroedinger equation for quantum fields from the timeless Wheleer-deWitt equation. This is the limit where QFT in curved space time applies. The concept of decoherence is often advocated in the context of the boundary proposal (see chapter \ref{nbDirichlet}): the real Hartle-Hawking wavefunction is peaked around the two time-reversed versions of the Hartle-Hawking geometry. A single classical world with a definite time direction is said to emerge in this picture due to decoherence. In the next chapter we will see another example where the saddle point approximation of the gravitational path integral is given by two distinct background histories. There we will see that the two histories interfere in a catastrophic way and decoherence cannot fix the problem. The reason for that is that one of the two solutions is always associated with unsuppressed inhomogeneous fluctuations and this leads to a non normalizable wavefunction. In this case one cannot trace over the inhomogeneous degrees of freedom to obtain a separation of WKB branches and the limit of QFT in curved spacetime is ill-defined.

\clearpage

%%%%%%%%%%%%%%%%%%%%%%%%%%%%%%%%%%%%%%%%%%%%%%%%%%%%%%%%%%%%%%%%%%%%%%%%%%%%%%%%%%%%%%%%%%%%%%%%%%%%%%%%%%%%%%%%%%%%%%%%%%%%%%%%%%%

\section{The need for quantum initial conditions}\label{quantuminitial}

One often thinks of gravity as describing the very large, and quantum mechanics the very small. However, one of the most stunning ideas to emerge from contemporary cosmology is that the largest visible structures in the universe originated through the amplification of primordial quantum fluctuations. It is frequently argued that inflation \cite{Guth:1980zm,Linde:1981mu,Albrecht:1982wi} provides a natural mechanism to achieve this amplification, in a rather generic manner. The calculations underpinning this assertion are performed in the framework of quantum field theory (QFT) in curved spacetime, where the background spacetime is treated classically and only the fluctuations are quantized \cite{Mukhanov:1981xt,Starobinsky:1979ty,Starobinsky:1982ee, Guth:1982ec,Hawking:1982cz,Bardeen:1983qw}. There is an implicit assumption that this classical/quantum split between background and perturbations is a good approximation to a more fundamental theory of quantum gravity, where the complete four-geometry is treated quantum mechanically. In this Chapter, which will be largely based on \cite{DiTucci:2019xcr}, we test this assumption by going beyond QFT in curved spacetime and studying inflationary quantum dynamics using the path integral for Einstein gravity within a semi-classical expansion. 

There is a fundamental difference in the behavior of quantum and the classical relativistic and diffeomorphism-invariant theories, which lies at the heart of our work. A classical relativistic particle, for example, cannot change the sign of the time component of its four-velocity. If it is travelling forwards in time at one moment, it will always travel forwards. However, a quantum relativistic particle cannot be so constrained: it can ``turn around'' in time. Indeed, including such amplitudes is essential to the final consistency of the theory (for a nice discussion see, for example, Feynman's Dirac Memorial lecture ~\cite{Feynman:1987gs}). 

The scale factor of the universe, being a time-like coordinate on superspace, is analogous to the time coordinate for a particle. The usual, classical picture of inflation is of a universe which always expands. However, when we study the quantum dynamics of the cosmological background, we have no right to exclude trajectories for which the scale factor turns around, {\it i.e}, they may start out contracting before expanding to attain their very large, final size. Even in classical general relativity, there is the example of de Sitter spacetime in the closed (global) slicing.  If one fixes the initial and the final radius to be both greater than the de Sitter radius, then there evidently exist {\it two} classical saddle point solutions to the gravitational path integral, one of which undergoes a ``bounce.'' It seems self-evident that {\it both} such saddle point solutions  must be included in the transition amplitude between the initial and final three-geometries. 

In the case of the flat slicing of de Sitter (where we take a toroidal spatial universe in order to keep the action finite), it turns out there are likewise two classical solutions, both of which are relevant saddle points for the gravitational path integral giving the amplitude for an initial small universe to become a large one. One of the solutions contracts to touch zero size before inflating to the final size. We shall show that this background comes along with perturbations which are unstable and out of control. Unfortunately, since we are doing quantum mechanics we cannot exclude this background solution. One might hope that by taking the size of the initial universe to zero, one could obtain a solution which only expanded ``from nothing.'' We shall carefully study this limit and show that the unstable perturbations are unavoidable, at least within semi-classical gravity. Likewise, we shall study the flat limit of a closed de Sitter universe. Again, we find no way to avoid these problematical perturbations. The only way out of the problem, as we explain in Section \ref{sec:stableperts},  appears to be to choose an ``off-shell,'' localized initial state for the universe, so that the universe is, at the start, sufficiently large and expanding that contributions from initially contracting universes are suppressed.  

%NT edit : next three paras
Our discussion extends recent studies of the no-boundary proposal~\cite{Hartle:1983ai}, based on the Lorentzian path integral for gravity~\cite{Feldbrugge:2017kzv,Feldbrugge:2017fcc,DiazDorronsoro:2017hti,Feldbrugge:2017mbc,DiazDorronsoro:2018wro,Feldbrugge:2018gin}, see also \cite{Kamenshchik:2018rpw}. In those works, we found an interesting interplay between the cosmological background and the perturbations, when both are treated quantum mechanically. In particular, we found that imposing a ``no boundary'' initial condition, {\it i.e.}, that the universe began on a three-geometry of zero size, results in unstable perturbations. In this work we generalize these calculations to scenarios for the beginning of inflation. We show that if an initial inflationary patch is assumed to start out much smaller than the Hubble volume, but still much larger than the Planck volume, there are generally {\it two} relevant semi-classical backgrounds, one of which ``bounces.'' We then show that the quantum perturbations about the bouncing background are badly behaved and out of control. This finding is closely related to our recent demonstration that the path integral formulations of the  ``no boundary" proposal of Hartle and Hawking, as well as the ``tunneling" proposal of Vilenkin, as originally proposed, lead to unacceptable quantum fluctuations~\cite{Feldbrugge:2017kzv,Feldbrugge:2017fcc,Feldbrugge:2017mbc,Feldbrugge:2018gin}. Here we make the statement more general: if one assumes a (quasi-) de Sitter phase all the way back to the beginning of the universe, and attempts to describe this phase using semi-classical Einstein gravity, then one picks out the {\it unstable} fluctuation mode rather than the stable, Bunch-Davies mode. We conclude that, when both the background and the fluctuations are quantized, inflation does {\it not} automatically pick out the Bunch-Davies vacuum, with associated nearly Gaussian-distributed fluctuations.  At face value, our finding invalidates the usual predictions of inflationary models unless some additional mechanism is invoked to explain the ``initial'' Bunch-Davies vacuum state. In other words, inflation on its own cannot explain the origin of the primordial perturbations. It is in this sense that we conclude that inflation is quantum incomplete.

One may then wonder under what circumstances the usual treatment of inflation, using QFT in a classical curved background spacetime, may be recovered. We shall show that a standard QFT description may be recovered provided that suitable initial conditions are imposed. Namely, we assume a localized initial quantum state describing an {\it expanding} universe of a prescribed size. We propagate this state forward using the Feynman path integral propagator. Provided the assumed initial size (three-volume) of the universe is sufficiently large,  then indeed only one, monotonically expanding background solution is relevant to the path integral for gravity. We can, in this way, recover a description of inflation in terms of QFT in curved spacetime. However, we emphasize that a prior, pre-inflationary phase of the right type must be assumed in order to complete inflation as a consistent framework for cosmology. 

The plan of this chapter is as follows: in section \ref{sec:path integral}, we introduce the path integral formalism for gravity, both for the background and the perturbations. Here we will show that explicit mode functions for the perturbations can be found for all values of the lapse function, a feature that enables us to carry out novel analytical calculations. In section \ref{sec:vanishing} we explore the consequences of taking a vanishingly small initial three-geometry, and establish our result that this limit inevitably leads to unstable fluctuations with an inverse Gaussian distribution (in linear cosmological perturbation theory -- see also \cite{Hofmann:2019dqu} for related work that supports this conclusion). We verify this result by relating it to the instability of the no-boundary proposal in section \ref{sec:nb}. Faced with this problematic result, we explore possible resolutions in section \ref{InitialConditions}. Introducing a pre-inflationary localized, expanding state, we recover the standard description in terms of a single classical background. We furthermore show, again by assuming an appropriate initial quantum state, that stable, Gaussian-distributed perturbations can also be recovered. We discuss our results, their implications and future directions in section \ref{sec:discussion}.

%%%%%%%%%%%%%%%%%%%%%%%%%%%%%%%%%%%%%%%%%%%%%%%%%%%%%%%%%%%%%%%%%%%%%%%%%%%%%%%%%%%%
\subsection{Semi-classical gravity} \label{sec:path integral}

We now wish to reconsider this calculation in semi-classical gravity, meaning that we should evaluate the Feynman path integral amplitude \eqref{path}
\begin{equation}
G[h_{ij}^1,\varphi_1;h_{ij}^0,\varphi_0] = \int \mathcal{D} g\, \mathcal{D}\varphi \ e^{\frac{i}{\hbar} S[g,\varphi]}\label{eq:Tprop}
\end{equation}
to propagate from an initial three-geometry with metric $h_{ij}^{0}$ and, potentially, matter fields $\varphi_0$ to a final three-geometry with metric $h_{ij}^{1}$ and matter fields $\varphi_1$. The action $S$ is taken to be the Einstein-Hilbert action $S_{EH}[g]$ along with a matter action $S_m[\varphi,g]$. The path integral is performed over all four-metrics $g$ and matter fields $\varphi$ consistent with the specified boundary conditions. 
For a homogeneous and isotropic background universe with only small fluctuations $\phi$ (which in the simplest case consist only of gravitational waves), the background spacetime can be described by two zero modes (moduli) namely the lapse $n(t)$ and the scale factor $a(t)$: the line element is given by
\begin{equation}
\mathrm{d}s^2 = -n (t)^2 \mathrm{d}t^2 + a(t)^2 \gamma_{i j} (x) dx^i dx^j,
\end{equation}
where $\gamma_{i j} (x)$ is the three-metric for a maximally symmetric space. In this chapter, for the most part we shall consider a spatially flat, FRW cosmology with $\gamma_{i j} (x)=\delta_{ij}$ and a toroidal topology, {\it i.e.}, periodic boundary conditions in comoving coordinates. In Section \ref{sec:nb} we generalize this to a spherical three-geometry for comparison. In these variables, using BFV quantization \cite{Batalin:1977pb} one may impose the proper-time gauge $\dot{n}=0$, and the Feynman propagator \eqref{eq:Tprop}  can be rigorously  expressed as
\begin{equation}
G[a_1,\phi_1;a_0,\phi_0] = \int_{0^+}^\infty \mathrm{d}n 
\int_{a_0}^{a_1} \mathcal{D} a\
\int_{\phi_0}^{\phi_1}
\mathcal{D} \phi\ e^{\frac{i}{\hbar}S[g,\phi;n]}
\end{equation} 
as derived by Teitelboim \cite{Teitelboim:1981ua, Teitelboim:1983fk} and Halliwell \cite{Halliwell:1988wc}. Note that in this gauge, the integral over the gauge-fixed lapse $n$ amounts to an integral over the proper time between the initial and the final three-geometry. In this section we first focus on the propagator for the background. In subsequent sections, we analyze the  fluctuations. 

\subsubsection{The background}\label{sec:background}

To set the scene, recall the description of the classical background -- de Sitter spacetime in the flat slicing. Taking the line element as $\mathrm{d}s^2 = -n(t)^2 \mathrm{d}t^2+a(t)^2 \mathrm{d}{\bf x}^2$, the action is
\begin{equation}
{S}^{(0)}= V_3 \int_0^1 dt \left(-3 M_{Pl}^2 n^{-1} a \dot{a}^2  -n a^3 \Lambda\right),
\label{frwact}
\end{equation}
where $M_{Pl}^2\equiv 1/(8 \pi G)$ is the reduced Planck mass and $\Lambda$ is the cosmological constant which we shall assume to be positive. To avoid clutter, we generally set $M_{Pl}=1$ in what follows, restoring it by dimensions when helpful. Without loss of generality we may choose the coordinate $t$ to run from $0$ to $1$ and the spatial volume  $V_3$ of the torus, in comoving coordinates, to be unity. The action (\ref{frwact}) is inconvenient because it is cubic in $a$ and $\dot{a}$. However, if we redefine $N(t)=a(t) n(t)$ and $q(t)=a(t)^2$, so that the line element is $\mathrm{d}s^2 = -N(t)^2 \mathrm{d}t^2/q(t)+q(t) \mathrm{d}{\bf x}^2$, the action becomes quadratic~\cite{Halliwell:1988ik} and hence easier to analyze\footnote{As we are considering semi-classical gravity, we will ignore factor ordering ambiguities and Jacobian factors in the measure, which will lead to corrections at subleading orders in $\hbar$.} :
\begin{equation}
S^{(0)}= \int_0^1 dt \left(-{3\over 4}  N^{-1} \dot{q}^2  - N q \Lambda\right),
\label{frwact2}
\end{equation}
showing that $q$ behaves as the coordinate of a particle moving in a linear potential. The canonical momentum conjugate to $q$ is $p=-{3\over 2} N^{-1} \dot{q} $.  In natural units, the  coordinates $t$ and ${\bf x}$ are dimensionless, $n$ and $a$ have dimensions of length, $N$ and $q$ have dimensions of length$^2$ and $p$ has dimensions of length$^{-2}$. These scalings are helpful in our later analysis in section \ref{sec:stableperts}. The Hamiltonian is $H=N\left(-{1\over 3} p^2 +q \Lambda\right)$. It vanishes, when the equations of motion are satisfied, as a consequence of time reparameterization invariance. 

\begin{figure}[h]
\centering
\begin{minipage}{0.45\textwidth}
\includegraphics[width=\linewidth]{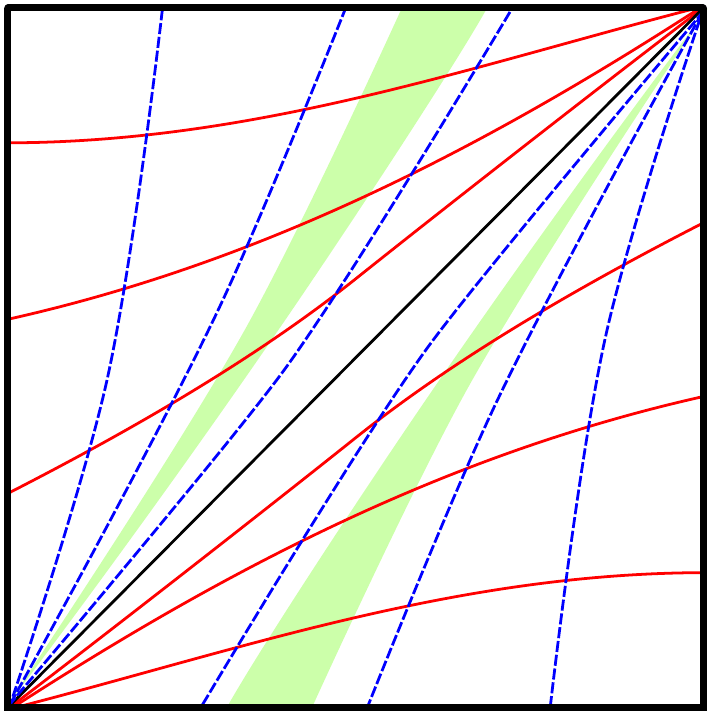}
\end{minipage}~
\begin{minipage}{0.45\textwidth}
\includegraphics[width=\linewidth]{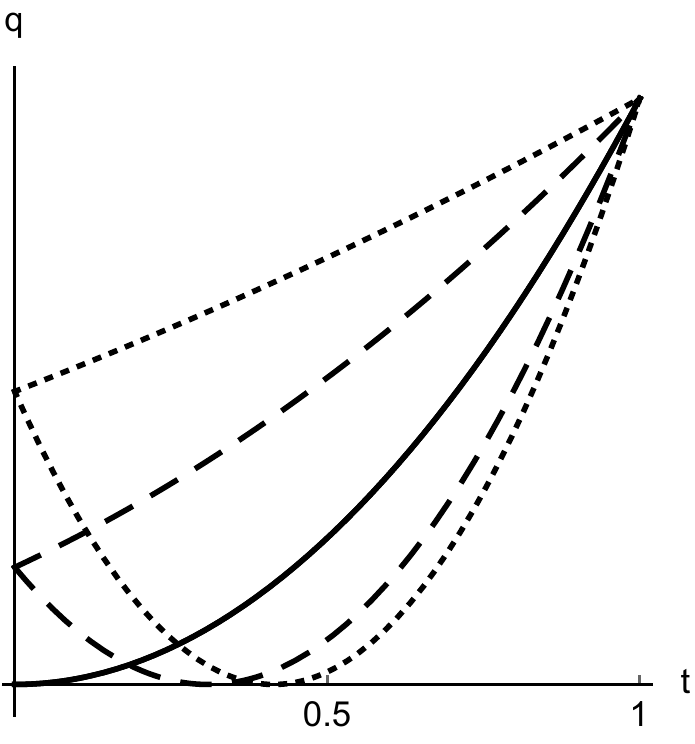}
\end{minipage}~
\caption{{\it Left panel:} The Penrose diagram of de Sitter space-time in the flat slicing. The red lines denote time slices and the blue lines denote space slices. The two background solutions relevant to the propagator consist of either pure expansion (a finite portion of the green patch on the left) or first contraction followed by expansion (a portion of the green patch on the right disappearing at the top right, reappearing bottom left and ending up at the same location as the purely expanding solution). {\it Right panel:} The two background solutions relevant to the propagator, shown as a function of $t$ for successively decreasing initial scale factor squared $q_0$ (dotted, then dashed, and finally solid at $q_0=0$). As the initial scale factor is decreased, the two solutions approach each other and merge when $q_0=0.$ This last solution would correspond to the limit where one starts at the bottom left corner (or equivalently the top right corner) in the Penrose diagram on the left.}\label{fig:bgd}
\end{figure}

Evidently $q$, the scale factor squared, behaves like the coordinate of a particle moving with zero energy in a linear potential. It will be convenient to pick a gauge in which the lapse is constant, $\dot{N}=0$. Classically, the value of $N$ encodes the total proper time between the initial and final three-geometry. Quantum mechanically, when performing the path integral we must integrate over all positive values of $N$. Clearly, for a linear potential $-q \Lambda$ there are two classical solutions which travel, at zero energy, from some initial $q_0$ to a final, larger $q_1$. Either $q$ increases all the way or it starts out decreasing, ``bounces" off the potential at $q=0$, and increases to $q_1$. These are the two  trajectories, alluded to in the introduction of this chapter, which we shall study semi-classically in some detail, along with their associated perturbations. 

The existence of the ``bouncing" trajectory, with a larger real value of $N$ and, correspondingly, a larger proper time,  is closely related to the fact that the flat slicing only covers half of de Sitter spacetime. In a bouncing trajectory, $q$ vanishes as $t^2$ near the bounce so the line element behaves (up to constants) as $-\mathrm{d}t^2/t^2+t^2 \mathrm{d}{\bf x}^2$. Defining the conformal time $\eta=-1/t$, the line element becomes $\eta^{-2}(-\mathrm{d}\eta^2 +\mathrm{d}{\bf x}^2)$, the familiar expression for de Sitter spacetime in the flat slicing. As $t$ runs from $-\infty$ to $+\infty$ through all real values, we have an infinite universe collapsing to zero size and rebounding to infinity. This includes both the contracting and expanding halves of de Sitter: see Fig.~\ref{fig:bgd}. We see this by analytically continuing in $t$ (or $\eta$) around the point $t=0$. The given range of $t$ corresponds to $\eta$ running from $0^-$ to $-\infty$, through (or around) the point at infinity and from  $+\infty$ to $0^+$. Standard treatments of inflation are able to ignore one half of de Sitter spacetime by treating the background as classical. However, as we shall show,  when the background is treated quantum mechanically, the bouncing solutions are in general relevant and must be included.

We shall show later that the ``bouncing" trajectory leads to a disastrous probability distribution for perturbations -- either gravitational waves or scalar density perturbations -- and hence must somehow be made irrelevant through a choice of initial conditions. Our main result, that there is a minimum initial size for an inflating patch, follows from this consideration. 
Consider a quantum mechanical particle with coordinate $q$ moving in a potential $-\Lambda q$,  with a positive initial velocity. We can describe it with a wavepacket, assumed Gaussian for convenience, with central value $q_i$, standard deviation $ \sigma\equiv \sqrt{\langle \Delta q^2\rangle}$ and central value of the momentum $p_i$. In order that this initial quantum wavefunction describes a real, classical inflating universe with high probability, we must have $q_i \gg \sigma$ so that the spacetime metric has the desired signature (recall $q\equiv a^2$), we must take $p_i$ to be the expanding solution of the Hamiltonian constraint (Friedmann equation), $p_i=-\sqrt{3 q_i \Lambda}$, and also impose that the uncertainty $\Delta p \sim \hbar/\sigma$ given by the Heisenberg uncertainty relation, is smaller than $|p_i|$ in order that we can be sure that the initial universe is really expanding. Writing $q_i =\nu \sigma$ and $|p_i|=\nu \Delta p$, with $\nu\gg1$ measuring the number of standard deviations by which we can be assured that the initial $q_i$ is positive and that the initial universe is expanding, neglecting numerical factors we find $\sqrt{q_i \Lambda}>\nu^2 \hbar /q_i$. Rearranging this, we find the initial comoving volume of our background torus has to satisfy
 \begin{equation}
V_3 a_i^3 > {\nu^2 \hbar \over M_{Pl} \sqrt{\Lambda}},
\label{estbd}
\end{equation}
where we restored the coordinate three volume and the Planck mass. We shall rederive this bound much more carefully and rigorously in section \ref{InitialConditions}, exhibiting a Stokes phenomenon whereby the ``bouncing" solution becomes irrelevant as the initial size of the universe is raised. These more detailed considerations confirm the scaling exhibited in (\ref{estbd}) and, furthermore, yield an accurate numerical coefficient. 

How large should we take $\nu$ to be? Recall that {\it any} admixture of the ``bouncing" background results in disastrous perturbations. Therefore, to be conservative, such a background should be excluded for {\it all} perturbation modes we consider, that is, all the modes which exited the Hubble radius during inflation, and are now encompassed by region we currently observe. The number of such modes is proportional to $e^{3 N_I}$ where $N_I$ is the number of efoldings of inflation our current Hubble volume underwent.
If we divide this volume up into $e^{3 N_I}$ identical cubes, we need to ensure that none of them underwent the ``bouncing" evolution. The probability for any one cubical region to ``bounce" is suppressed by  $e^{-\nu^2/2}$ for our assumed Gaussian distribution. Therefore, in order to get {\it no} bouncing region we require $\nu> \sqrt{6 N_I} \sim 20$ for $N_I=60$ efolds of inflation. For an inflationary scale $\Lambda^{1\over 4}$ of order the GUT scale $\sim 10^{-3} M_{Pl}$, Eq. (\ref{estbd}) requires an initial inflating volume of around $10^8$ Planck volumes.  

Finally, let us note that the estimate given above may well be generous to inflation. What we have done is treat the isotropic moduli,  {\it i.e.}, the scale factor and the lapse, non-perturbatively, but all other modes perturbatively. Our analysis could, with some effort, be extended to treat the other homogeneous modes -- the {\it anisotropy} moduli -- non-perturbatively as well. Since the inclusion of anisotropies tends to counteract inflation and strengthen the onset of singularities, this may make it harder to choose quantum initial conditions which avoid singular semi-classical trajectories of the type we have shown to lead to uncontrolled perturbations.

\subsubsection{Background path integral and saddles}\label{sec:backgroundpi}

The Feynman propagator for the background, in these variables, is
\begin{align}
G[q_1;q_0] = \int_{0^+}^\infty dN \int_{q_0}^{q_1} {\mathcal D}q\ e^{\frac{i}{\hbar}S^{(0)}[q;N]}\,,
\label{feynprop}
\end{align}
with $S^{(0)}[q;N] = \int_0^1 dt  \left( -\frac{3}{4N} \dot{q}^2 -Nq\Lambda \right)\,$. Since the action is quadratic in $q$, the path integral over the scale factor $q$ can be expressed in terms of the classical action. With the specified boundary conditions, the solutions to the equation of motion $\ddot{q}=\frac{2\Lambda}{3}N^2$ are given by
\begin{equation}
\bar{q}(t) = H^2 N^2 (t - \alpha)(t -\beta),
\end{equation}
with
\begin{equation} 
\alpha,\beta = \frac{1}{2 N^2} \left[N^2 - N_{s-}N_{s+} \pm \sqrt{( N^2 - N_{s-}^2)( N^2 - N_{s+}^2 )} \right] \,, \label{ab}
\end{equation}
where
\begin{equation}
N_{s\pm} = \frac{\sqrt{q_1} \pm \sqrt{q_0}}{H}\,.\label{eq:saddle}
\end{equation}
The corresponding classical action is given by
\begin{align}
\bar{S}^{(0)}[q_1;q_0;N] = V_3 \left[ \frac{\Lambda^2}{36}N^3 -\frac{\Lambda}{2}(q_0+q_1)N -\frac{3(q_1-q_0)^2}{4N}\right]\,, \label{Naction}
\end{align} 
where $V_3$ denotes the spatial three-volume at $q=1,$ which we assume to be finite. In the subsequent calculation we will choose spatial coordinates such that $V_3=1,$ though we will re-instate $V_3$ explicitly in section \ref{InitialConditions}. Using the classical action $\bar{S}^{(0)}$, the Feynman propagator reduces to an oscillatory integral over the lapse in the proper-time gauge
\begin{align}
G[q_1;q_0] = \sqrt{\frac{3 i}{4\pi \hbar}} \int_{0^+}^\infty \frac{\mathrm{d}N}{\sqrt{N}}  e^{\frac{i}{\hbar}\bar{S}^{(0)}[q_1;q_0;N]}\,. \label{Nintegral}
\end{align}

\begin{figure}
%\centering
\begin{minipage}{0.5\textwidth}
\includegraphics[width=\linewidth]{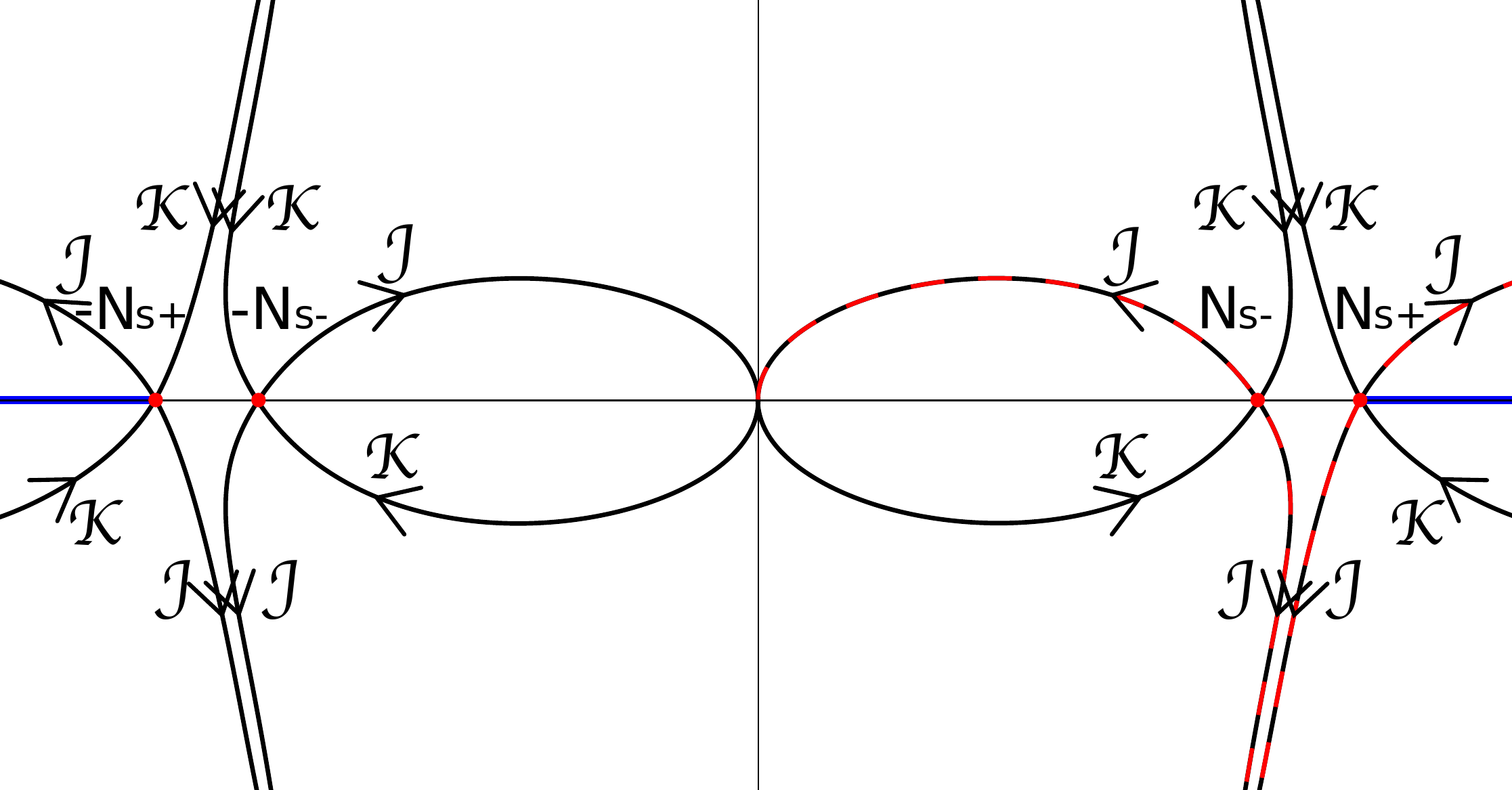}
\end{minipage}~
\begin{minipage}{0.5\textwidth}
\includegraphics[width=\linewidth]{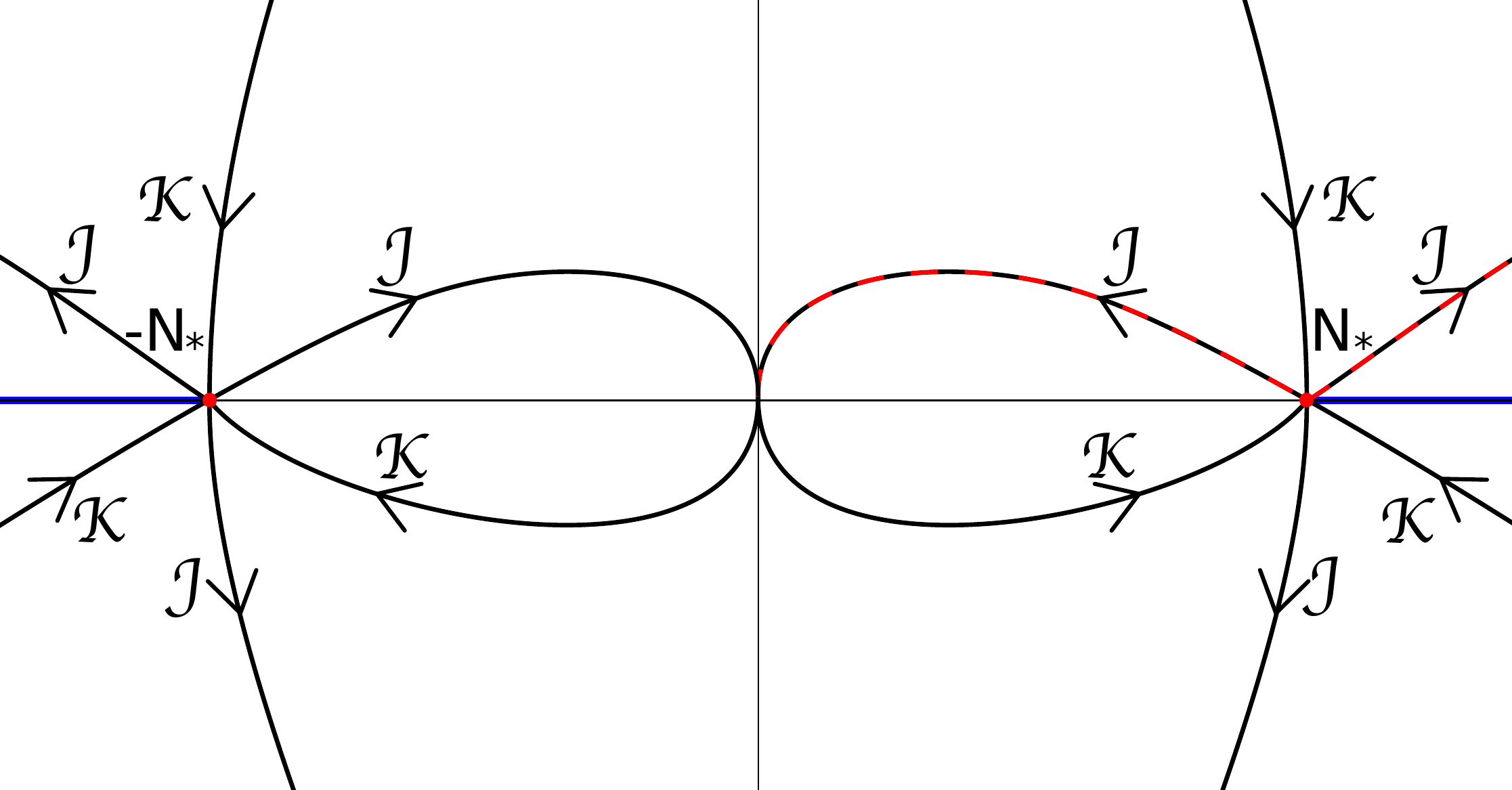}
\end{minipage}
\caption{The lines of steepest ascent $\mathcal{K}$ and descent $\mathcal{J}$ for the lapse integral, shown in the complexified plane of the lapse $N$. Left: for generic boundary conditions with $q_0,q_1 \neq 0$. Right: the limit of vanishing initial boundary $q_0=0$.}\label{fig:PLbackground}
\end{figure}

We approximate the lapse integral in the saddle point approximation using Picard-Lefschetz theory \cite{Feldbrugge:2017kzv}.
The exponent $i \bar{S}^{(0)}/\hbar$ has four saddle points in the lapse $N$, located at $\pm N_{s\pm}$. The lines of steepest ascent and descent emanating from the saddle points run through the complex plane to the essential singularities at the origin and complex infinity (see figure \ref{fig:PLbackground}). Since we are integrating the lapse over the positive real line, only the two saddle points with positive real part $+N_{s\pm}$ are relevant to the integral (assuming $q_0 \leq q_1$ as we will henceforth). For generic boundary conditions, with $q_0,q_1 \neq 0$, the saddle points are non-degenerate. The saddle point approximation of the Feynman propagator is given by
\begin{equation} 
G[q_1;q_0] \approx \sqrt{ \frac{3i}{4\Lambda \sqrt{q_0 q_1}}}
 \left[ \, e^{-i\frac{\pi}{4}} e^{i\bar{S}^{(0)}[N_{s-}]/\hbar} +   \, e^{i\frac{\pi}{4}}e^{i\bar{S}^{(0)}[N_{s+}]/\hbar} \right]\,,
\end{equation}
with the classical action at the saddle points
\begin{equation}
\bar{S}^{(0)}[N_{s\pm}]=   -2 \sqrt{\frac{\Lambda}{3}}  \left( q_1^{3/2} \pm q_0^{3/2} \right)\,.
\label{nbwf_classical}
\end{equation} 

The propagator consists of the interference of two classical solutions. At the relevant saddle points  $N_{s\pm}$, the background solutions simplify to 
\begin{align}
\bar{q}\mid_{N_{s\pm}} = H^2 N_{s\pm}^2 (t-\alpha)^2\,, \qquad \alpha = \beta = \pm \frac{\sqrt{q_0}}{HN_{s\pm}}\,.
\end{align} 
At $N_{s-}$, the parameter $\alpha$ is negative and the background continuously expands from $q_0$ to $q_1$. In physical time $t_p$ this expansion takes the usual exponential form $a(t_p)=\frac{1}{H}e^{+Ht_p}$ for a suitable range of $t_p$. At $N_{s+}$ the parameter $\alpha$ is positive and the universe first contracts from $q_0$ to zero size and then re-expands to $q_1$. In proper time $t_p$ this amounts to a contracting phase given by $a(t_p) = \frac{1}{H} e^{-H t_p}$ followed by an expanding phase described by $a(t_p) = \frac{1}{H} e^{+H t_p}$ (see figure \ref{fig:bgd}). In QFT on curved spacetime one only works with the solution corresponding to the inner saddle point $N_{s-}$, restricting the calculation to the upper triangle of the Penrose diagram of de Sitter spacetime in the flat slicing (see the upper left triangle of figure \ref{fig:bgd}). In quantum gravity we cannot restrict our analysis to a fixed evolution of the scale factor of de Sitter space time and need to consider the solutions corresponding to both $N_{s-}$ and $N_{s+}$ located in the entirety of de Sitter spacetime.

As we saw in section \ref{sec:QFT}, the Bunch-Davies vacuum is selected by considering the limit $\eta \to -\infty$. In semi-classical gravity, this corresponds to the limit where the initial boundary shrinks to a point $q_0 \rightarrow 0$,
%NT edit next phrase 
while $q_1$  is held fixed at a large positive value. This is an interesting limit for the background universe since the two classical solutions coincide and the two saddle points $N_{s\pm}$ merge, thus forming a degenerate saddle point of order $2$ located at
\begin{align}
N_\star \equiv \sqrt{\frac{3q_1}{\Lambda}} = \frac{\sqrt{q_1}}{H}\,\,.
\end{align}
See figures \ref{fig:bgd} and \ref{fig:PLbackground} for an illustration. The saddle point approximation to the Feynman propagator in this limit involves an integral over cubic fluctuations around the (degenerate) saddle point, as explained in~\cite{Feldbrugge:2017kzv}, and reads
\begin{equation} \label{nbwf_classical_limit}
G[q_1;q_0=0]  \approx \frac{e^{i\frac{\pi}{4}}3^{17/12}\Gamma(\frac{4}{3})}{2 \pi^{1/2}\hbar^{1/6}\Lambda^{5/12}q_1^{1/4}} \, e^{-i2\sqrt{\frac{\Lambda}{3}} q_1^{3/2}/\hbar} \,. 
\end{equation} 
We thus observe that generic boundary conditions lead to an interference of an expanding and a bouncing solution. In the limit of a vanishing initial boundary, however, the interference changes form. The propagator is now dominated by a single classical solution corresponding to a degenerate saddle point in the saddle point approximation. 

For completeness, let us note that the path integral \eqref{Nintegral} admits an exact expression in terms of products of Airy functions. It can be shown (see for instance \cite{Halliwell:1988ik,Feldbrugge:2017kzv}) that it satisfies the inhomogeneous WdW equation 
\begin{equation}
\hbar^2 \frac{\partial^2 G[q_1 ; q_0]}{\partial q_1^2} + 3 (\Lambda q_1 - 3 k ) G[q_1 ; q_0] = - 3 i \hbar \,  \delta(q_1 - q_0)\,.
\end{equation}
Defining $z \equiv \frac{(-3)^{1/3}}{(\hbar \Lambda)^{2/3}} (\Lambda q - 3 k),$ this equation is solved by 
\begin{equation}
G[q_1 ; q_0] = \frac{\pi (-3)^{2/3}}{(\hbar \Lambda)^{1/3}} \{ Ai(z_1) [Ai (z_0) - i Bi(z_0) ] \Theta(q_1 - q_0) + Ai(z_0) [Ai(z_1 )- i Bi(z_1)] \Theta(q_0 - q_1) \}
\end{equation}
The above expression then reduces to (\ref{nbwf_classical_limit}) for $k=0$, $q_0 = 0$ in the limit of large $q_1$.

It is important to note that the degeneracy of the saddle point in $N$ in the limit $q_0 \to 0$ stems from the fact that we do not a priori know whether the boundary conditions of the propagator lie in the upper or lower triangle of the Penrose diagram (see figure \ref{fig:bgd}). Since the analytically continued scale factor $a$ changes sign at the singularity, one might argue that the degeneracy can be lifted by specifying the relative sign of the scale factors $a_0$ and $a_1$. However it should be noted that the the metric -- over which we integrate in the path integral -- is insensitive to the sign of the scale factor. It is for this reason appropriate to work in terms of the squared scale factor $q=a^2$. Moreover, in section \ref{sec:nb} we derive the same result from the no-boundary proposal in the limit of vanishing curvature, where no such ambiguities exist as the scale factor is everywhere non-negative.

%%%%%%%%%%%%%%%%%%%%%%%%%%%%%%%%%%%%%%%%%%%%%%%%%%%%%%%%%%%%%%%%%%%%%%%%%%%%%%%%%%%%
\subsubsection{The fluctuations} \label{subsec:fluctuations}

In the coordinates introduced in the previous section, the action for the perturbations (\ref{actionperturbations}) is given by
\begin{align}
S^{(2)}[\phi,q;N] = \frac{1}{2} \int_0^1 dt \int d^3x \left( \frac{q^2}{N}\dot\phi_k^2 - N k^2 \phi_k^2 \right)\,, \label{actionperturbations2}
\end{align}
where we focus on a single mode with wavenumber $k$. A sum over Fourier modes is straightforward to implement, but will be kept implicit. One may think of the fluctuation as the component of a gravitational wave, or an additional massless scalar field. The equation of motion, ignoring the backreaction of the fluctuations on the metric, is given by
\begin{equation}
\ddot{\phi} + 2 \frac{\dot{\bar{q}}}{\bar{q}} \dot{\phi}+ \frac{k^2 N^2}{\bar{q}^2} \phi = 0 \label{212}\,.
\end{equation}
The solutions are of the form
\begin{equation}
\phi(t) = \frac{a f(t) + b g(t)}{\sqrt{\bar{q}(t)}} \label{linearcomb}
\end{equation}
with
\begin{equation}
f(t) = \Bigl[\frac{t - \beta}{t - \alpha}\Bigr]^{\mu/2}   \Bigl[(1 - \mu)(\alpha - \beta) + 2(t - \alpha) \Bigr]\,,
\end{equation}
\begin{equation}
g(t) = \Bigl[\frac{t - \alpha}{t - \beta }\Bigr]^{\mu/2}   \Bigl[(1 + \mu)(\alpha - \beta) + 2(t - \alpha)\Bigr]\,.
\end{equation}
Here the exponent $\mu$ is given by
\begin{equation}
\begin{split}
\mu^2 &= 1 -\Bigl( \frac{2k  }{ (\alpha - \beta)  H^2 N}\Bigr)^2 \\
& \equiv \frac{(N^2 - \overline{N}_{-}^2)(N^2 - \overline{N}_{+}^2)}{(N^2 - N_{s-}^2)(N^2 - N_{s+ }^2)}\,, \label{defmu}
\end{split}
\end{equation}
where the zeros of $\mu$ are specified by
\begin{equation}
\pm \overline{N}_{\pm} = \pm \frac{\sqrt{q_1^2 H^2 + k^2} \pm \sqrt{q_0^2 H^2 + k^2}}{H^2} 
\end{equation}
The integration constants $a$ and $b$ in the solution \eqref{linearcomb} can be determined by imposing the boundary conditions $\phi(t=0) = \phi_0$ and $\phi(t=1)=\phi_1,$ leading to
\begin{equation}
a =+  \frac{1}{D}\left[ \left[\frac{1 - \alpha}{1 - \beta }\right]^{\mu/2} (2 - (\alpha + \beta) + \mu (\alpha - \beta)) \sqrt{q_0} \phi_0 + \left[\frac{ \alpha}{\beta }\right]^{\mu/2} (\alpha + \beta - \mu (\alpha - \beta)) \sqrt{q_1} \phi_1 \right]
\end{equation}
\begin{equation}
b = - \frac{1}{D} \left[ \left[\frac{1 - \beta}{1 - \alpha}\right]^{\mu/2} (2 - (\alpha + \beta) - \mu(\alpha - \beta))\sqrt{q_0} \phi_0 + \left[\frac{\beta}{ \alpha}\right]^{\mu/2} (\alpha + \beta + \mu (\alpha  - \beta)) \sqrt{q_1} \phi_1 \right]
\end{equation}
\begin{equation}
\begin{split}
D =&+ \left[\frac{(1 - \beta)\alpha}{(1 - \alpha) \beta}\right]^{\mu/2} (2 - (\alpha + \beta) - \mu(\alpha - \beta))(\alpha + \beta - \mu(\alpha - \beta))  \\
&- \left[\frac{(1 - \alpha)\beta}{(1 - \beta) \alpha }\right]^{\mu/2}  (2 - (\alpha + \beta) + \mu(\alpha - \beta))(\alpha + \beta + \mu(\alpha - \beta))\,.
\end{split}
\end{equation}

\begin{figure}
\includegraphics[width=0.4 \textwidth]{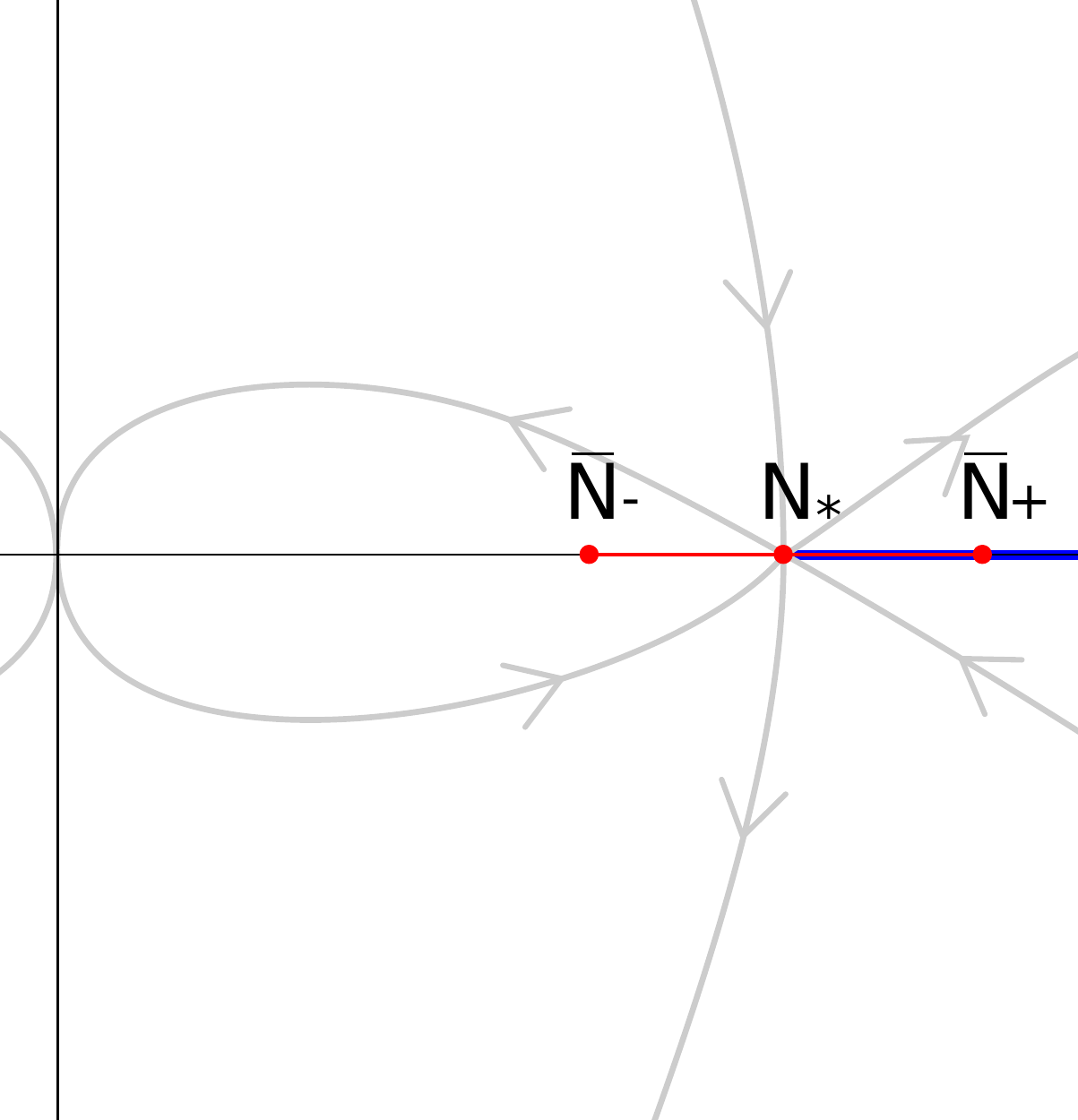}
\caption{The complex $N$ plane with the degenerate saddle point $N_*$ and the branch points $\bar{N}_{+}$ and $\bar{N}_{-}$, relevant to the case where the initial size of the universe is taken to zero. The red line represents the branch cut. The blue line represents the $N$ for which the background $\bar{q}$ passes through $0$ for a time $t \in (0,1]$.}\label{fig:plot}
\end{figure}

Since the perturbation action $S^{(2)}$ is quadratic in $\phi$, the classical action $\bar{S}^{(2)}$ is given by the boundary terms
\begin{equation}
\bar{S}^{(2)}[\phi_1,q_1;\phi_0,q_0] = \frac{\bar{q}^2(t) \phi(t) \dot{\phi}(t)}{2 N} \bigg|_{t=0}^1\,. \label{eq:boundary}
\end{equation}
Equation \eqref{eq:boundary} holds for all $N$ in the complex plane except for part of the real line $|N| \geq N_{\star}$, where $q$ passes through zero and additional singularities appear (see figure \ref{fig:plot} and a closely related discussion in \cite{Feldbrugge:2017mbc}). These parts of the real $N$ line must be excluded from the domain of integration in the integral, since the action becomes infinite there. Away from these line segments, the action is explicitly given by
\begin{equation}
\bar{S}^{(2)} = \frac{n}{d}
\end{equation}
with the numerator
\begin{align*}
n = \frac{ k^2}{H^2 N } \Bigl \{ &+ \left [\frac{(1 - \alpha) \beta}{(1 - \beta) \alpha} \right]^{\mu/2} \left[ (2 - (\alpha + \beta) + \mu (\alpha - \beta))q_0 \phi_0^2  + (\alpha + \beta + \mu (\alpha - \beta))q_1 \phi_1^2\right]   \\
& - \left [\frac{(1 - \beta) \alpha}{(1 - \alpha) \beta} \right]^{\mu/2} [(2 - (\alpha + \beta) - \mu (\alpha - \beta)) q_0 \phi_0^2  + (\alpha + \beta - \mu (\alpha - \beta)) q_1 \phi_1^2  ]  \\
& - 4 \mu (\alpha - \beta) \sqrt{q_1 q_0}\phi_1 \phi_0 \Bigr \}\,,
\end{align*}
and the denominator
\begin{align*}
d =& +   \left [\frac{(1 - \beta) \alpha}{(1 - \alpha) \beta} \right]^{\mu/2} ( 2 -( \alpha + \beta) - \mu (\alpha - \beta)  )( \alpha + \beta - \mu (\alpha - \beta )) \\
& -  \left [\frac{(1 - \alpha) \beta}{(1 - \beta) \alpha} \right]^{\mu/2} (+ 2 -( \alpha + \beta) + \mu (\alpha - \beta)  )( \alpha + \beta + \mu (\alpha - \beta ))\,.
\end{align*}

Despite the occurrence of roots in $\alpha, \beta$ and $\mu,$ the action does not contain branch points as long as $q_0, q_1 \neq 0.$ To see this, it is useful to express the root as an explicit function of $N$,
\begin{equation}
\left [\frac{(1 - \alpha) \beta}{(1 - \beta) \alpha} \right]^{\mu/2} =  \left [\frac{N_{s+}^2 + N_{s-}^2 - 2 N^2 + 2 \sqrt{(N^2 - N_{s+}^2)(N^2 - N_{s-}^2)}}{N_{s+}^2 + N_{s-}^2 - 2 N^2 - 2 \sqrt{(N^2 - N_{s+}^2)(N^2 - N_{s-}^2)}} \right]^{ +\frac{1}{2} \sqrt{\frac{(N^2 - \overline{N}_{+}^2)(N^2 - \overline{N}_{-}^2)}{(N^2 - N_{s+}^2) (N^2 - N_{s-}^2)}}} \label{root}
\end{equation}
As one considers a closed loop in $N$ which includes for example $N_{s-}$, the numerator and the denominator in this last expression exchange their roles. However, since the exponent changes its sign at the same time, the action remains unchanged, and the saddle points $\pm N_{s\pm}$ are not branch points. Similarly, the net effect of completing a loop around $ \pm \overline{N}_{\pm}$ (\textit{i.e.} changing the sign of $\mu$) is to send  $n \rightarrow - n$ and $d \rightarrow - d$ and once again the action remains unaffected.

%%%%%%%%%%%%
\subsection{The limit of a vanishing initial three-geometry} \label{sec:vanishing}

In section \ref{sec:QFT} we showed that the Bunch-Davies vacuum is selected by assuming that the (quasi-) de Sitter phase extends back to the ``beginning'' of the universe and considering the limit of early conformal time $\eta \to -\infty$. In the path integral formulation it is more natural to consider the limit where the initial scale factor goes to zero $q_0\to 0$. From the explicit mode functions \eqref{linearcomb} it follows that the combination $q_0 \phi_0^2$ appearing in the action tends to zero in this limit. The classical action for the fluctuations simplifies to
\begin{equation}
\bar{S}^{(2)}[\phi_1,q_1;\phi_0,q_0=0;N] = - \frac{k^2 N q_1 \phi_1^2}{H^2} \frac{1}{N^2 + N_\star^2 - \sqrt{(N^2 - \overline{N}_{+}^2)(N^2 - \overline{N}_{-}^2)}}\,.\label{actionzeroq0}
\end{equation}	
Two important consequences emerge:
\begin{itemize}
\item The action now contains branch points at $\pm \overline{N}_{\pm}$. It is convenient to place the associated branch cut on the real $N$ line between $\overline{N}_-$ and $\overline{N}_+$ and similarly on the negative $N$ axis (see figure \ref{fig:plot}). This cut appears only in the limit $q_0 \rightarrow 0$ since the terms that would be necessary to compensate for the sign changes when considering closed loops around $\pm \overline{N}_{\pm}$ are now absent. It turns out that we are interested in the action $\bar{S}^{(2)}$ evaluated at $N_\star$ which lies on the branch cut. In the analysis below we study the action at $N_\star$ in more detail. 
\item The classical action is independent of $\phi_0$. Hence we obtain a unique result for the final fluctuations, regardless of what the initial fluctuations are. In other words, in the limit where the de Sitter phase, or inflation, started the universe, we obtain a unique result for the quantum vacuum. Conversely, we may use this calculation to test whether inflation can be considered as a theory of the initial phase of the universe.
\end{itemize}

In the analysis of the background, we saw that the two saddle points $N_{s\pm}$ merge into a degenerate saddle point at $N_\star$. The fact that the action \eqref{actionzeroq0} contains a branch cut indicates that it may be delicate to evaluate it at $N_*$. It is indeed clearer to start with the case where $q_0 \neq 0$ and only then take the limit where the initial scale factor tends to zero. Moreover, this limiting procedure allows one to make contact with the calculation of inflationary fluctuations in the framework of QFT in curved spacetime. 

The evaluation of the action and the classical solutions at the saddle points must be taken with care, since at the saddle points we have $\alpha = \beta$ while the action contains negative powers of $(\alpha - \beta)$.  Using the limit $\lim_{x\rightarrow \infty}(1+\frac{1}{x})^x = e$, we obtain the following identities
\begin{align}
\mu(\alpha - \beta) & \rightarrow \,  \frac{2ik}{H^2 N_{s\pm}} \\
\left[ \frac{t-\beta}{t-\alpha} \right]^{\mu/2} = \left[1+ \frac{\alpha-\beta}{t-\alpha} \right]^{\mu/2} & \rightarrow \, e^{ \frac{ik}{H^2 N_{s\pm}(t-\alpha)}} \\
\left[ \frac{(1-\beta)\alpha}{(1-\alpha)\beta} \right]^{\mu/2} = \left[1+ \frac{\alpha-\beta}{(1-\alpha)\beta} \right]^{\mu/2} & \rightarrow \, e^{\frac{ik}{H}(\frac{1}{\sqrt{q_1}}\pm \frac{1}{\sqrt{q_0}})} \\
\left[ \frac{\alpha}{\beta} \right]^{\mu/2}= \left[1- \frac{\alpha-\beta}{\beta} \right]^{\mu/2} & \rightarrow \, e^{\pm \frac{ik}{H\sqrt{q_0}}}
\end{align}
at the saddle points $N_{s\pm}$ (where the upper/lower signs are always correlated). The mode functions may be re-expressed as
\begin{align}
\frac{HN_{s\pm}}{2}\frac{f(t)}{\sqrt{q}}\bigg|_{N_{s\pm}} & \rightarrow \,e^{\frac{ik}{H^2 N_{s\pm}(t-\alpha)}} \left(1 - \frac{ik}{H^2 N_{s\pm}(t-\alpha)} \right) \\
\frac{HN_{s\pm}}{2}\frac{g(t)}{\sqrt{q}}\bigg|_{N_{s\pm}} & \rightarrow \,e^{-\frac{ik}{H^2 N_{s\pm}(t-\alpha)}} \left(1 + \frac{ik}{H^2 N_{s\pm}(t-\alpha)} \right)\,. 
\end{align}
These are the familiar mode functions since identifying conformal time $\eta$ as 
\begin{equation}
\eta = -\frac{1}{H\sqrt{\bar{q}}} = - \frac{1}{H^2 N_{s-}(t-\alpha)}
\end{equation}
with $\eta_0 = -\frac{1}{H\sqrt{q_0}}$ and $\eta_1 = -\frac{1}{H\sqrt{q_1}}$\footnote{The boundary conditions in conformal time $\eta_0$ and $\eta_1$ correspond to the expanding solution and should be considered as a short hand for the condition in terms of $q_0$ and $q_1$.}, we recover the standard Bunch-Davies mode functions in conformal time,
\begin{equation}
\frac{HN_{s-}}{2}\frac{f(t)}{\sqrt{\bar{q}}}\bigg|_{N_{s-}}  \rightarrow \,e^{-ik\eta} \left(1 + ik\eta \right)\,,
\quad
\frac{HN_{s-}}{2}\frac{g(t)}{\sqrt{\bar{q}}}\bigg|_{N_{s-}}  \rightarrow \,e^{ik\eta} \left(1 - ik\eta \right)\,.
\end{equation}

The saddle point $N_{s+}$ resides precisely at the edge of the region where no finite action perturbative solutions exist. Since the integration contour for the background passes through this point, it makes sense to deform the contour ever so slightly away from the excluded half-line $N>N_{s+}, (N \in \mathbb{R}),$ and evaluate the action at
\begin{align}
N^2 = N_{s+}^2 + \delta^2\,,
\end{align}
where $\delta$ is a small complex number. Then the action can be evaluated by expanding in $\delta,$ \textit{e.g.}
\begin{align}
\left[ \frac{(1-\beta)\alpha}{(1-\alpha)\beta} \right]^{\mu/2}\bigg|_{\sqrt{N_{s+}^2+\delta^2}} & \rightarrow \, e^{-ik(\eta_0 + \eta_1)} \left( 1-\frac{i \delta k N_{s+}H^{3/2}}{(q_0 q_1)^{3/4}}\right)\,, 
\end{align}
and subsequently taking the limit $\delta \rightarrow 0$.
With this prescription, we evaluate the classical action $\bar{S}^{(2)}$ evaluated at both saddle points,
\begin{align}
\bar{S}^{(2)}|_{N_{s\pm}}&=\frac{k^2}{2H^2}\frac{num}{ 
e^{ik(\eta_0\pm\eta_1)}(i k \eta_0 -1)(i k \eta_1\mp 1)
-e^{-ik(\eta_0\pm \eta_1)}(ik\eta_0+1)(ik\eta_1 \pm 1)
}
\label{eq:actionPrecise}
\end{align}
with the numerator
\begin{align}
num&=-4 i k \phi_0 \phi_1\\
& \mp e^{i k(\eta_0 \pm \eta_1)} \left[ \frac{\phi_0^2}{\eta_0}(ik\eta_1\mp 1) + \frac{\phi_1^2}{\eta_1}(ik \eta_0 -1)\right]\\
& \mp e^{-ik(\eta_0 \pm \eta_1)} \left[ \frac{\phi_0^2}{\eta_0}(ik\eta_1 \pm 1) + \frac{\phi_1^2}{\eta_1}(i k \eta_0 +1)\right]\,.
\end{align} 
Note that equation \eqref{eq:actionPrecise} is a precise form of equation \eqref{actionzeroq0} evaluated in $N_{s\pm}$ when approached from above and below the branch-cut.

This finally enables us to take the limit of vanishing initial scale factor, used in selecting the Bunch-Davies vacuum. This is equivalent to the limit where $\eta_0$ approaches minus infinity since the scale factor and conformal time are related by $q=1/(H\eta)^2$. As is clear from the above expressions, the classical action does not converge in the limit of $\eta_0\to -\infty$ along the real axis. One can regularise the limit by adding a (vanishingly) small imaginary part to the initial scale factor. This however leads to two inequivalent results. In the limit $\eta_0\to - \infty (1-i \epsilon)$ for positive $\epsilon$ or equivalently $q_0 \rightarrow 0$ with $\text{Im } q_0 > 0$ -- normally considered for the Bunch-Davies calculation -- the classical action reduces to
\begin{align}
\bar{S}^{(2)}|_{N_{s\pm}}&\to 
\frac{k^2}{2H^2}
\frac{
\mp e^{-i k(\eta_0 \pm \eta_1)} \left[ \frac{\phi_0^2}{\eta_0}(ik\eta_1\pm 1) + \frac{\phi_1^2}{\eta_1}(ik \eta_0 +1)\right]
}
{
(-1)e^{-ik(\eta_0\pm\eta_1)}(i k \eta_0 + 1)(i k \eta_1\pm 1)
}\\
&=\frac{k^2\phi_1^2}{2H^2\eta_1 (\pm i k \eta_1 + 1)}\\
&=\frac{q_1k^2\phi_1^2}{2 (\pm i k - H\sqrt{q_1})}\\
& \approx -\frac{\sqrt{q_1}k^2}{2H}\phi_1^2 \mp i \frac{k^3}{2H^2} \phi_1^2 \,.
\end{align}
Since the semi-classical approximation amounts to exponentiating the classical action, \textit{i.e.} $e^{i \bar{S}^{(2)}/\hbar}$, we observe that the saddle point $N_{s-}$ corresponds to a Gaussian while $N_{s+}$ corresponds to a non-normalizable, inverse Gaussian, mode. Meanwhile, taking the limit with negative $\epsilon$ or equivalently with $\text{Im } q_0 < 0$ instead, we obtain the complex conjugate result. 
The saddle point $N_{s+}$ then corresponds to Gaussian and $N_{s-}$ to inverse Gaussian fluctuations. In either case we observe that the fluctuations are always stable at one saddle point, and unstable at the other.

\begin{figure}[h] 
\centering
\begin{minipage}{0.45\textwidth}
\includegraphics[width=\linewidth]{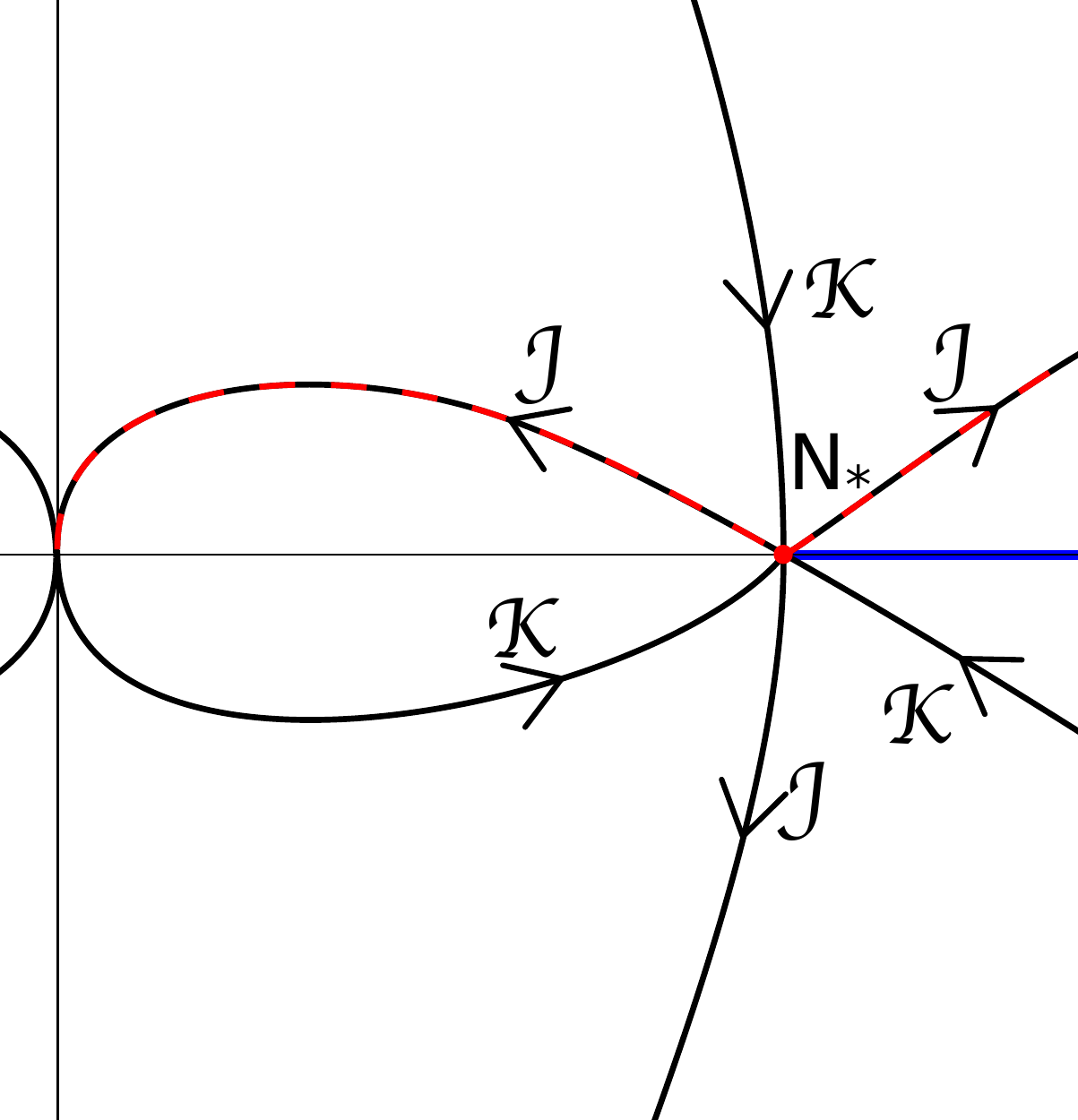}
\end{minipage}~
\begin{minipage}{0.45\textwidth}
\includegraphics[width=\linewidth]{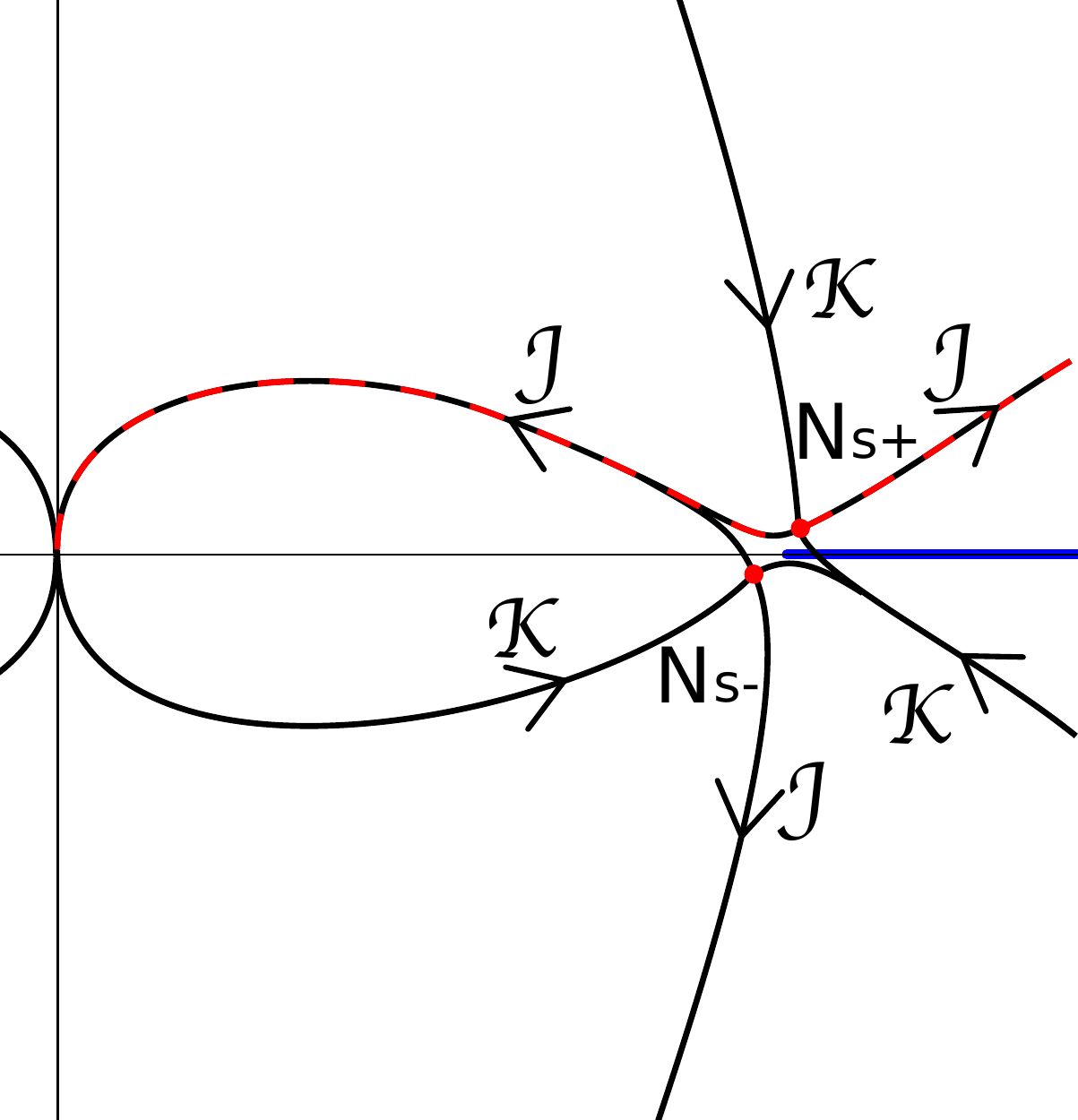}
\end{minipage}
\caption{Picard-Lefschetz theory for the lapse integral in the limit $q_0\to 0$. The dark lines show the lines of steepest ascent/descent indicated by $\mathcal{K}$ and $\mathcal{J}$. Using Picard-Lefschetz theory the real half-line $(0^+,\infty)$ is deformed to the red dashed curves. Left: the Lefschetz thimbles for degenerate saddle point $N_*$ when setting $q_0=0$. Right: the Lefschetz thimbles in the limit $q_0\to 0$ approached from the upper half plane $\text{Im } q_0 >0$. We observe that when approaching $q_0=0$ from above, the saddle point $N_{s+}$ is relevant while the saddle point $N_{s-}$ is irrelevant. When approaching $q_0$ from below the saddle points switch positions.}
\label{fig:Perturb}
\end{figure}

In order to determine the relevant saddle points in the saddle point approximation, we apply Picard-Lefschetz theory to the limits $q_0\to 0$ with $\text{Im } q_0 <0$ and $\text{Im } q_0 >0$ (see figure \ref{fig:Perturb}). We observe that the saddle points move in the complex plane. For the limit $\text{Im } q_0>0$ we observe that the lines of steepest ascent of the $N_{s+}$ saddle point intersect the positive half line while the lines of steepest ascent of $N_{s-}$ curve away from the real line. We thus conclude that only $N_{s+}$ is relevant in this limit. In the limit $\text{Im } q_0<0$ the saddle points switch making $N_{s-}$ relevant. We thus observe that the propagator always selects the \emph{unstable} saddle point that contributes to the Feynman propagator, while the stable saddle point is irrelevant to the propagator in the limit of zero initial scale factor. We conclude that the semi-classical description of an initial (quasi-) de Sitter phase does not lead to the Bunch-Davies vacuum often assumed in inflation when studied as a QFT on curved spacetime. Instead the Feynman propagator selects the unsuppressed (inverse Gaussian) fluctuations. Before discussing the implications of this result, it is useful to verify it by relating it to a similar calculation arising for the no-boundary proposal.

%%%%%%%%%%%%%%%%%%%%%%%%%%%%%%%%%%%%%%%%%%%%%%%%%%%%%%%%%%%%%%%%%%%%%%%%%%%%%%%%%%%%
\subsection{Flat space limit of the no-boundary proposal} \label{sec:nb}

An alternative manner in which to study the early time limit of de Sitter space in the flat slicing is to set $q_0=0$ from the start and let the spatial curvature parameter $K$ be positive at first, and then approach zero. This effectively corresponds to the flat space limit of the no-boundary proposal, in which one sums over compact metrics between an initial point $q_0=0$ and a final 3-surface specified by $q_1$. The relevant calculation has been performed in the papers \cite{Feldbrugge:2017kzv,Feldbrugge:2017fcc,Feldbrugge:2017mbc} and will be discussed in more details in the next chapter. To include positive spatial curvature, we generalise the metric to read $\mathrm{d}s^2=-\frac{N^2}{q}\mathrm{d}t^2 + q \mathrm{d} \Omega^2,$ where $\mathrm{d}\Omega^2$ is the metric on a 3-sphere of curvature $6K.$ Taking the limit of zero spatial curvature can then be thought of as enlarging the sphere to reduce the local curvature, and in the infinite size limit where flat space is reached we assume that the volume is suitably regularised (\textit{e.g.} by imagining that the flat 3-space has the topology of a torus). Alternatively, one may take the limit where at fixed $K$ the final scale factor $q_1$ tends to infinity, as in this limit the spatial curvature also becomes insignificant. For the purposes of illustration, we will take the point of view that $K$ is reduced with $q_1$ held fixed.

In the presence of spatial curvature, the saddle points are qualitatively different from the flat case in that they reside at complex values of the lapse function,
\begin{align}
N_s &= \pm \frac{1}{H^2} \left[ (H^2 q_1 - K)^{1/2} \pm i K^{1/2}\right]\,.
\end{align}
As the curvature $K$ is reduced, the saddle points approach the real $N$ line, and in the limit of zero curvature they reach it at $\pm N_\star = \pm \frac{\sqrt{q_1}}{H}.$ Note that $N_\star$ also happens to be the saddle point of the perturbative action \eqref{actionzeroq0}. Fig. \ref{fig:thimble} illustrates the positions of the saddle points, and the corresponding lines of steepest ascent and descent of the 
%JF
real part of the exponent $h=\text{Re}(iS)$\footnote{Note that our previous terminology, in which the $h$-function was called the Morse function, was unfortunately inconsistent with the common meaning of that term in the mathematical literature, since the critical points of the $h$-function can be degenerate.}.  In the limit of $K=0$ a degenerate saddle point of order 2 is obtained.

\begin{figure}
\begin{minipage}{150pt}
		\includegraphics[width=0.9\linewidth]{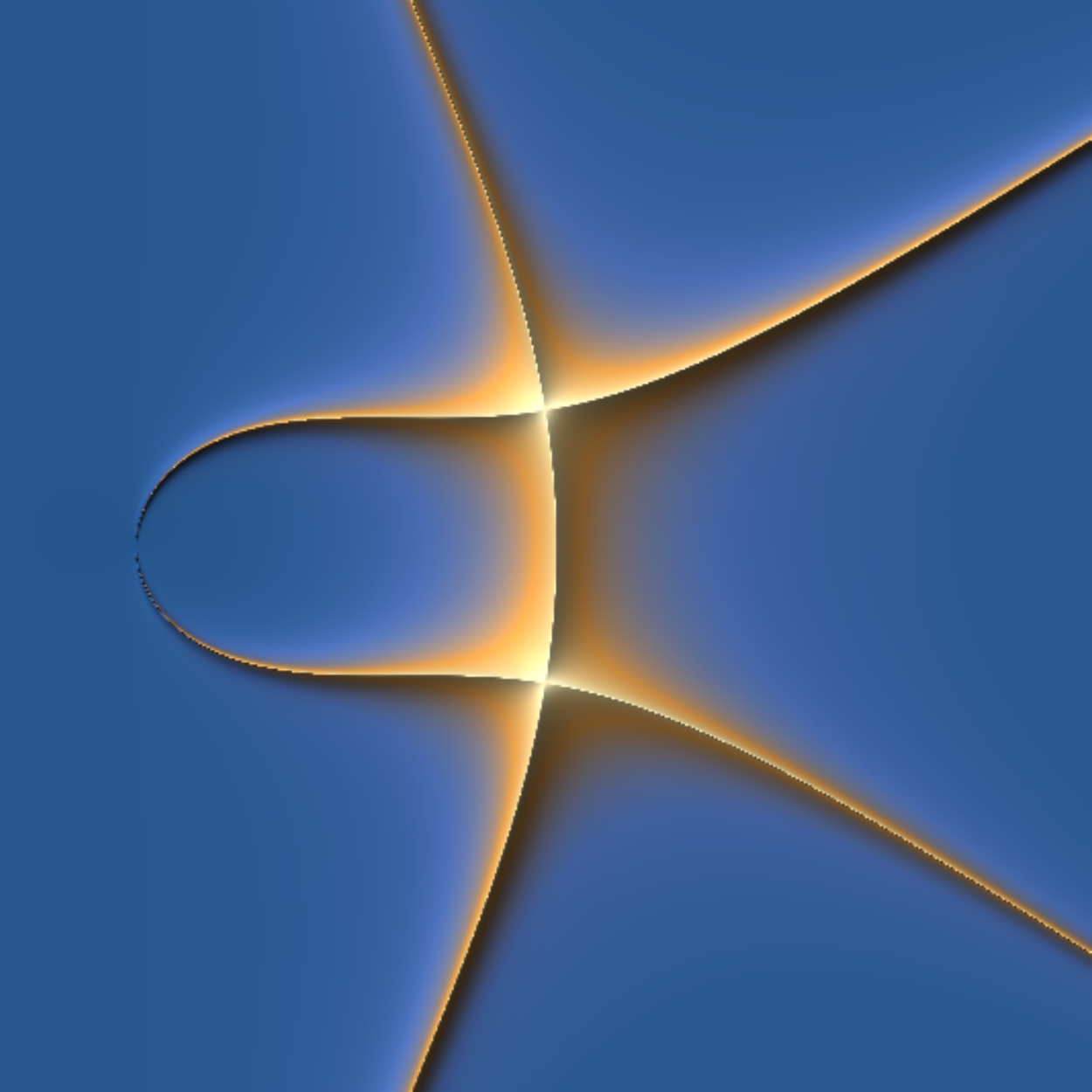}
	\end{minipage}%
	\begin{minipage}{150pt}
		\includegraphics[width=0.9\linewidth]{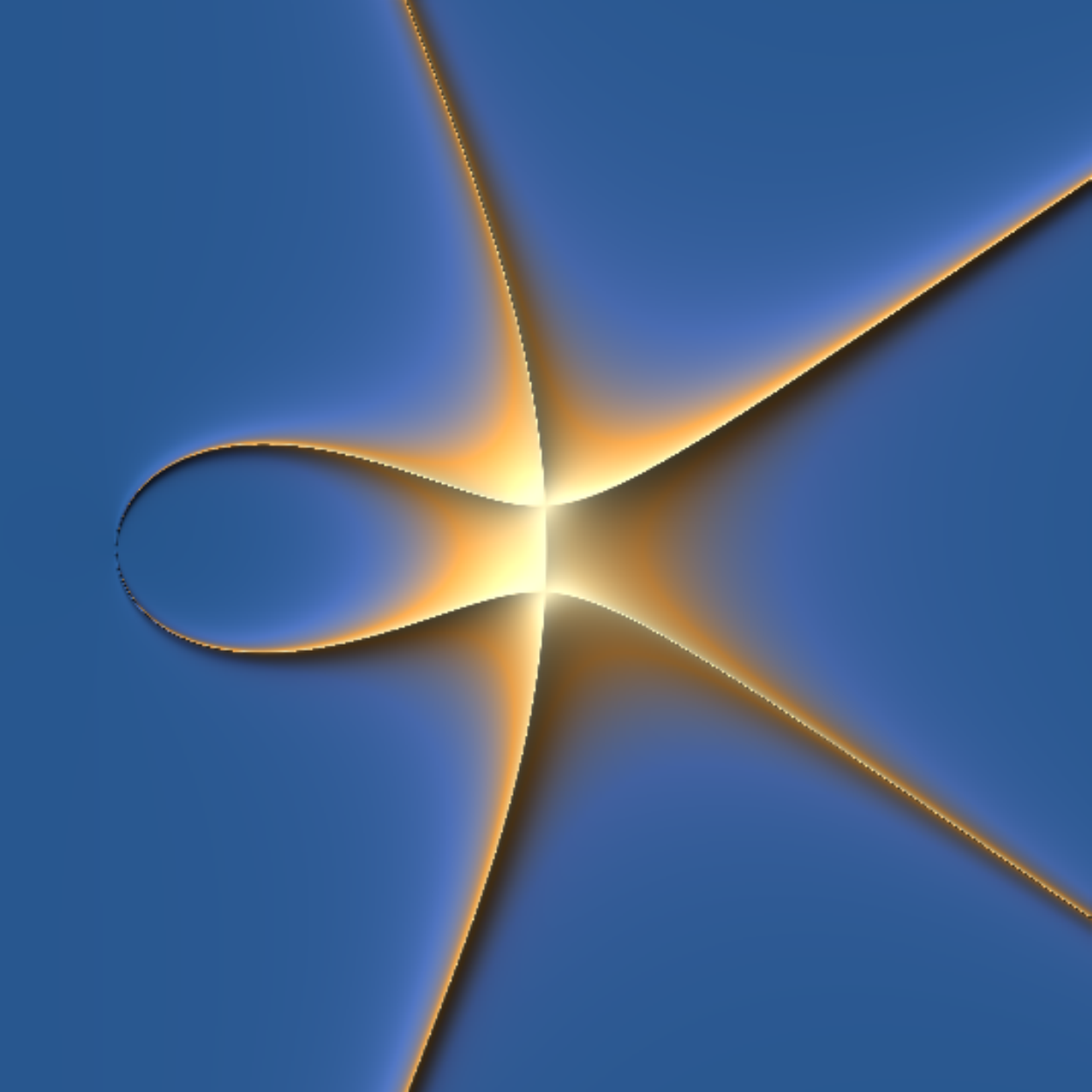}
	\end{minipage}%
	\begin{minipage}{150pt}
		\includegraphics[width=0.9\linewidth]{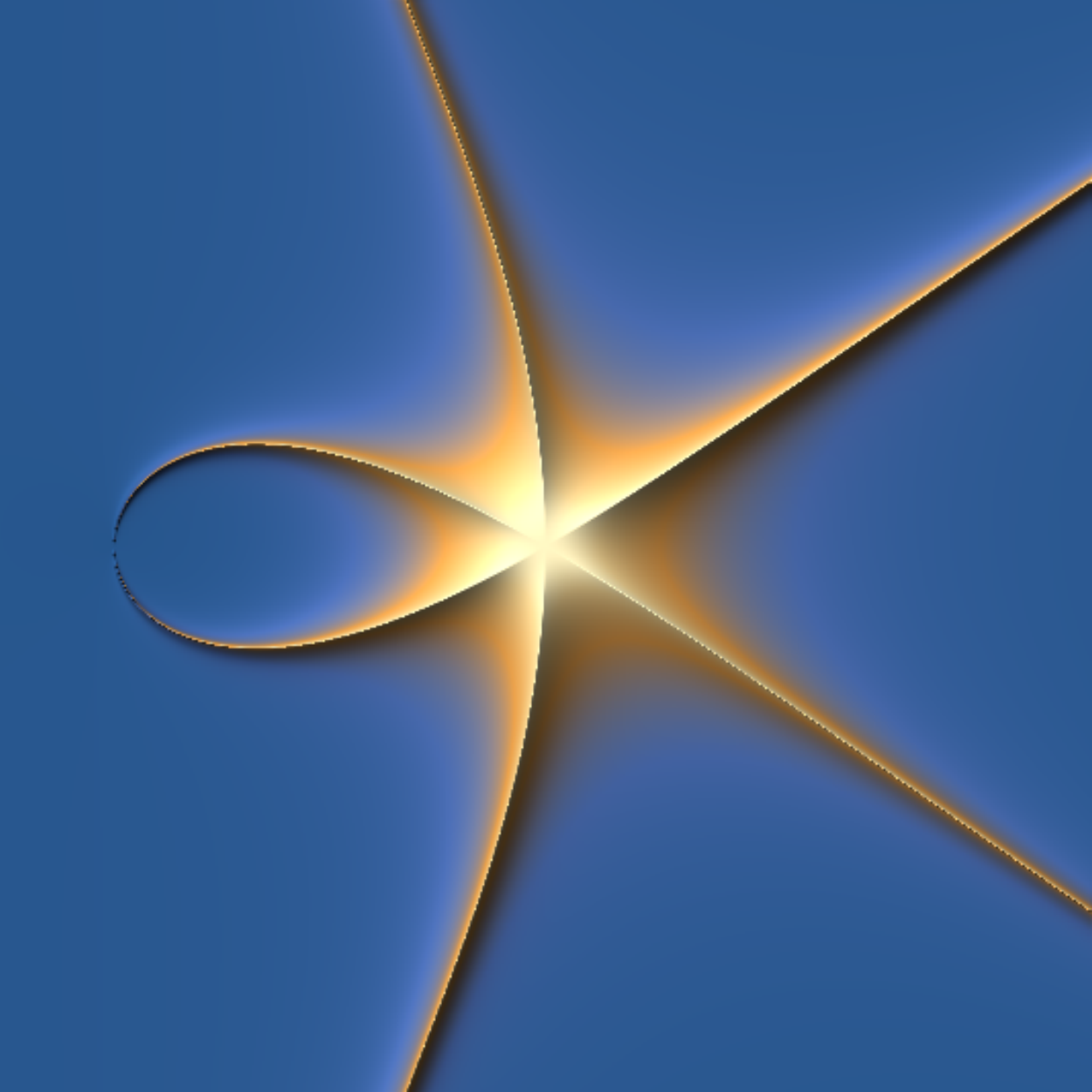}
	\end{minipage}%
	\caption{Picard Lefschetz theory in $N$ with $q_0=0,$ in the limit $K \to 0,$ with $\Lambda=3$ and $q_1=10$. The curves are the lines of steepest ascent and descent emanating from the saddle point at the crossings of the lines. The origin the point where the two lines meet at the left. Note that the figure is symmetric around the real line. {\it Left panel:} Spatial curvature $K=1$: this is the standard no-boundary graph. {\it Middle panel:} Spatial curvature $K=1/10$.  {\it Right panel:} Spatial curvature $K=0$.}
	\protect
	\label{fig:thimble}
\end{figure} 

The form of the action \eqref{Naction} makes it clear that the integrand $e^{iS}$ is convergent near the origin ($N=0$) and at large $N$ in the regions just above the real $N$ line. Thus the integration contour of steepest descent from $N_\star$ resides entirely in the upper half plane\footnote{The steepest descent line emanating from $N_\star$ and approaching $-i\infty$ is traversed in both directions, so that its contribution to the integral cancels out.}. Then in the Lorentzian integral we are forced to approach $N_\star$ from above, where in the large $k$ limit the fluctuation action \eqref{actionzeroq0} approaches
\begin{equation}
\bar{S}^{(2)}(N_\star^+)  \rightarrow - \frac{i}{2} k q_1 \phi_1^2\,,
\end{equation}
implying a weighting $e^{+k q_1 \phi_1^2/(2\hbar)},$ where $N_\star^+$ denotes $N_\star$ approached from above. Note that the action at $N_\star^+$ is independent of $\Lambda,$ and scales as $k \phi_1^2$ rather than $k^3 \phi_1^2/\Lambda$. The large $k$ limit corresponds to modes with short wavelengths, which are not yet frozen.  Another interesting limit is that of large final scale factor $\sqrt{q_1}$, where
\begin{equation}
\bar{S}^{(2)}(N_\star^+)  \rightarrow  - \frac{\sqrt{q_1}}{2H} k^2 \phi_1^2 -  \frac{i}{2H^2} k^3 \phi_1^2\,. \label{Sabovecut}
\end{equation}
The frozen modes acquire a scale-invariant \emph{inverse} Gaussian distribution, with weighting $e^{+k^3 \phi_1^2/(2H^2 \hbar)}$. This confirms the conclusion that a phase of de Sitter (or quasi-de Sitter) expansion considered all the way into our past does not lead to the Bunch-Davies vacuum, and hence cannot by itself explain the origin of structure in our universe.

%%%%%%%%%%%%%%%%%%%%%%%%%%%%%%%%%%%%%%%%%%%%%%%%%%%%%%%%%%%%%%%%%%%%%%%%%%%%%%%%%%%%

\subsection{Inflatable initial conditions} \label{InitialConditions}

The discussion above indicates that, in a consistent semi-classical treatment of both the background and the perturbations, sending the initial size of the universe to zero in a well-defined way results in an unacceptable, unbounded probability distribution for the perturbations.  One may, however, wonder whether there might be some other way to specify the initial conditions for inflation in a natural manner, which would avoid the interference of backgrounds described above and would be able to recover the limit of QFT in curved spacetime. 

In Ref.~\cite{Feldbrugge:2019sew}, the authors discuss the use of localized initial and final quantum states in relativistic, diffeomorphism invariant (and hence constrained) quantum mechanics. It is shown there that, in order to prescribe states which are localized, it is necessary to employ ``off-shell" wavefunctions, namely, wavefunctions which are {\it not} annihilated by the Hamiltonian constraint. These states represent the quantum amplitude resulting from a preparation process and {\it not} the dynamics of the system itself, hence the system Hamiltonian alone does not annihilate the state. Once such states are evolved with the Feynman propagator, outside of the preparation region they become a superpositon of ``on-shell'' physical states of the system, independent of the preparation device. In the present section we shall make use of this formalism to define an inflating initial state which avoids the problem of multiple contributing backgrounds. The initial state will be localized in superspace at small scale factor, and endowed with a positive expansion velocity. In the localized region of phase space in which it is initially prepared, the wavefunction does not satisfy the Hamiltonian constraint (or Wheeler-deWitt equation). However, once this initial wavefunction is propagated to large scale factor, it becomes that for a large, expanding universe and satisfies the Wheeler-deWitt equation.  

In section \ref{sec:background}, we obtained the Feynman propagator for the background as an ordinary integral over $N$, \textit{i.e.} 
\begin{align}
G[q_1;q_0] = \sqrt{\frac{3V_3 i}{4\pi \hbar}} \int_{0^+}^\infty \frac{\mathrm{d}N}{\sqrt{N}}  e^{\frac{i}{\hbar}\bar{S}^{(0)}[q_1;q_0;N]}\,, \label{Nintegral2}
\end{align}
in terms of the classical action
\begin{align}
\bar{S}^{(0)}[q_1;q_0;N] = V_3 \left[ \frac{\Lambda^2}{36}N^3 -\frac{\Lambda}{2}(q_0+q_1)N -\frac{3(q_1-q_0)^2}{4N}\right]\,. \label{classicalaction}
\end{align}
Implicitly, this calculation assumed that the universe was initially in a position eigenstate, with the scale factor squared $q$ being precisely equal to $q_0.$ As a direct consequence of the uncertainty principle, the initial momentum of the universe was thus maximally unknown -- one may interpret the resulting interference of an expanding and a bouncing solution as a reflection of this fact. We can now consider a more general initial state $\psi_0,$ which is evolved  by convolution with the propagator,
\begin{align}
G[q_1;\psi_0] = \int G[q_1;q_0] \psi_0(q_0) \mathrm{d}q_0 = \int_{0^+}^\infty \int G[q_1;q_0;N] \psi_0(q_0) \mathrm{d}q_0 \mathrm{d}N\,.
\end{align}
Here we will consider the choice 
\begin{align} \label{Initialstate}
\psi_0(q_0) = \frac{1}{\sqrt[4]{2\pi \sigma^2}}e^{\frac{i}{\hbar} p(q_i) q_0 - \frac{(q_0-q_i)^2}{4 \sigma^2}}\,,
\end{align}
which expresses the idea that the universe starts with a certain size $q_i$ (with uncertainty $\sigma$), and that we choose the momentum $p(q_i)= \pm V_3 \sqrt{3\Lambda q_i}$ so that the universe is initially contracting (positive sign in $p(q_i)$) or expanding (negative sign in $p(q_i)$). Below we will specialize to the expanding case. Note that this state simply represents a {\it choice}, not explained by inflation. Our strategy here is rather to imagine the physical situation in which we know with some confidence that the universe is expanding, and is likely to be of a certain size already, and then explore the consequences of this assumed initial state. The mathematical form of the initial state is that of a generalized coherent state, which distributes the uncertainty between ``position'' $q$ and momentum $p$ depending on the value of the spread $\sigma.$ (We have performed analogous calculations with different phase correlations in $\psi_0(q_0),$ in particular with the momentum factor $e^{ip(q_0)q_0/\hbar}$ instead of $e^{ip(q_i)q_0/\hbar},$ and have checked that qualitatively similar results are obtained. Similarly, one may consider states which have support only at positive $q_0$, again with analogous results. However, the present choice is technically the simplest.)

The convolution of the propagator with the state \eqref{Initialstate} is Gaussian and can thus be evaluated exactly, yielding
\begin{align}
\int G[q_1;q_0;N] \psi_0(q_0) \mathrm{d}q_0 = 
 \frac{1}{\sqrt[4]{2\pi}}
  \sqrt{\frac{i 3 V_3 \sigma N}{\hbar N + 3 i V_3 \sigma^2}}
e^{\frac{i}{\hbar} \bar{S}^{(0)}[q_1;\bar{q}_0;N]+ \frac{i}{\hbar} p(q_i) \bar{q}_0 - \frac{(\bar{q}_0-q_i)^2}{4 \sigma^2}}
\end{align}
with
\begin{align}
\bar{q}_0=
\frac{N q_i - i (N^2 \Lambda V_3 -2 N p(q_i) - 3 q_1 V_3) \sigma^2/\hbar}{N + 3 i \sigma^2 V_3 /\hbar}\,. \label{barq0}
\end{align}
Note that in the limit $\sigma \to 0$, we localize $\bar{q}_0 \to q_i$.

\begin{figure}[h] 
\centering
\begin{minipage}{0.49\textwidth}
\includegraphics[width=\linewidth]{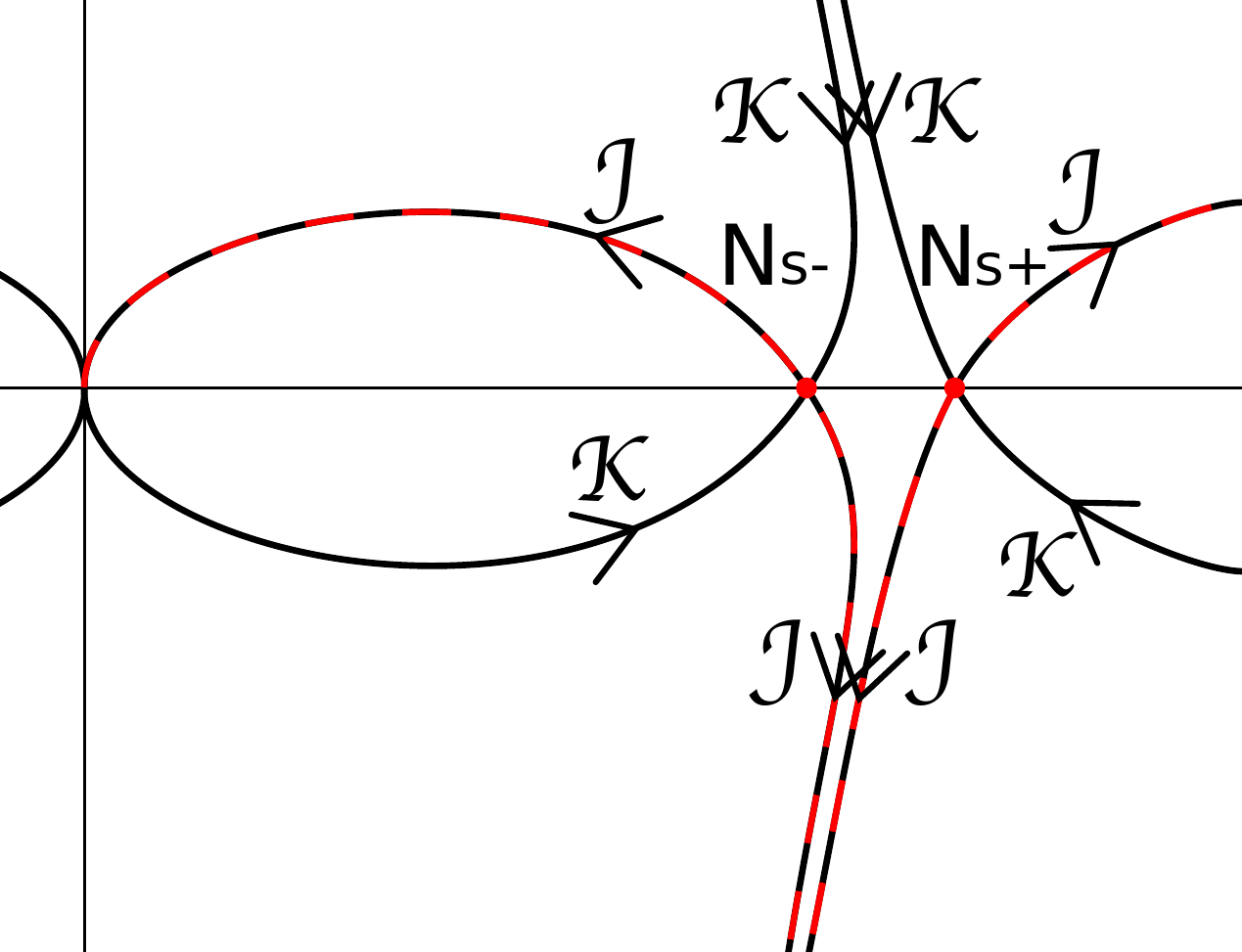}
\end{minipage}~
\begin{minipage}{0.49\textwidth}
\includegraphics[width=\linewidth]{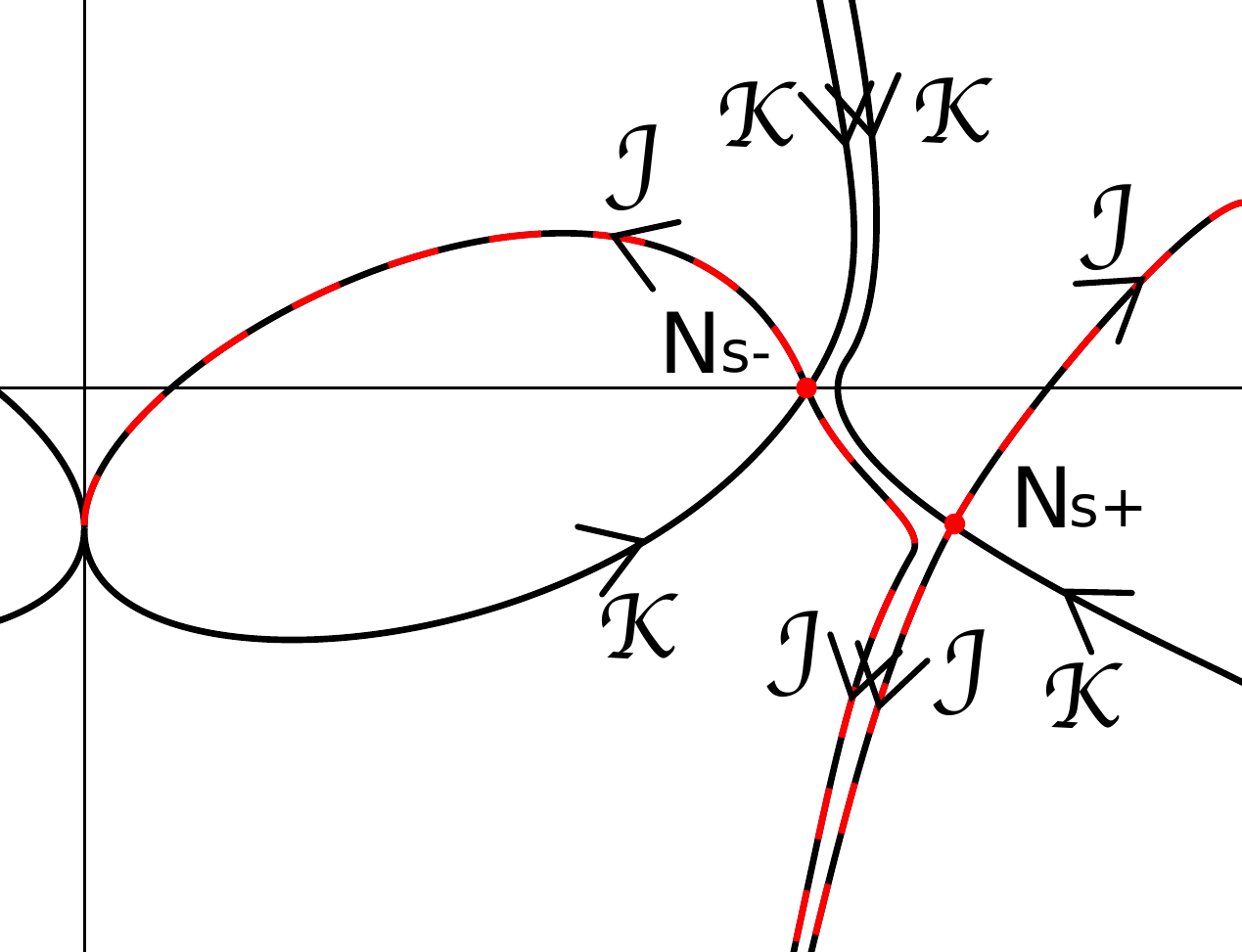}
\end{minipage}\\
\begin{minipage}{0.49\textwidth}
\includegraphics[width=\linewidth]{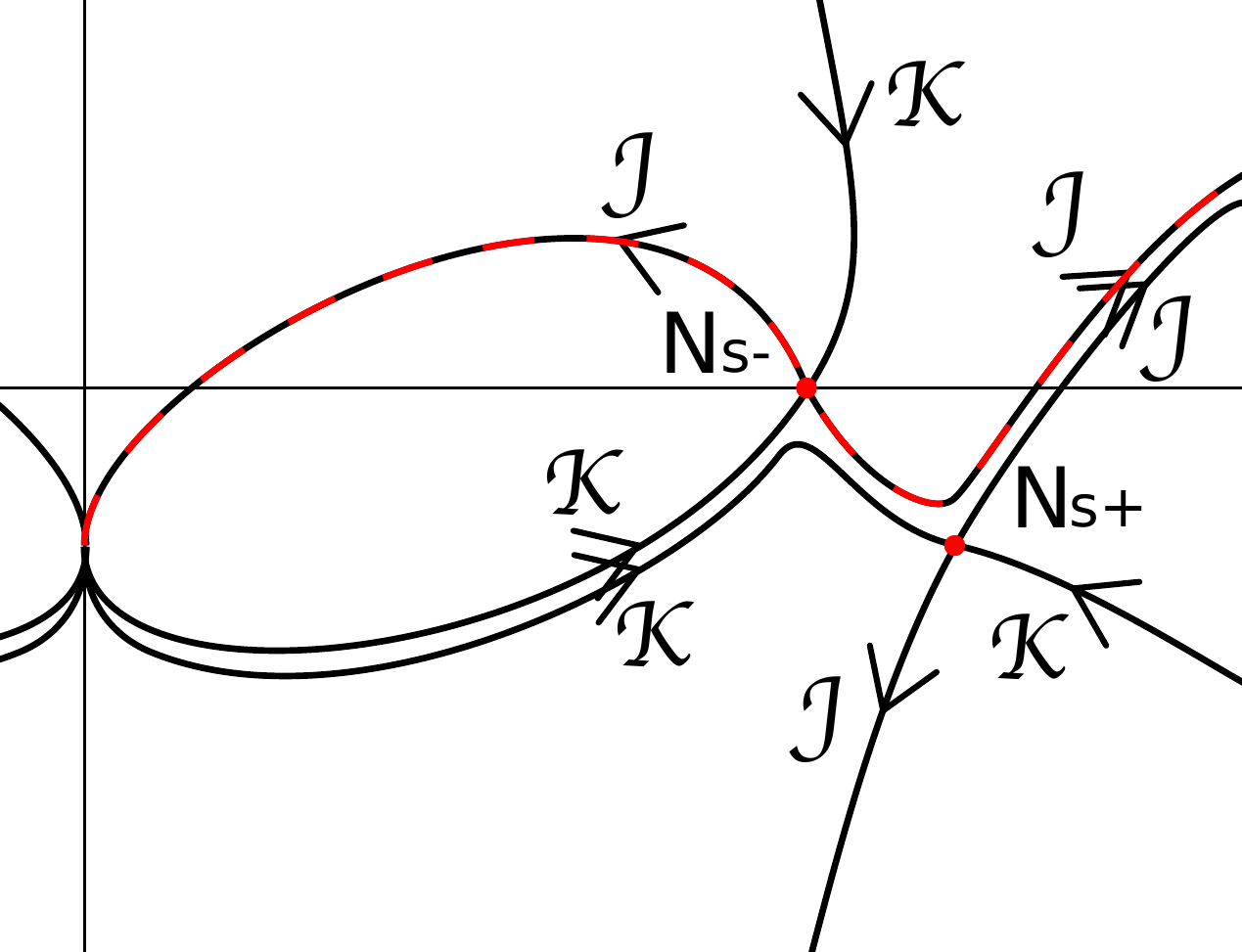}
\end{minipage}~
\begin{minipage}{0.49\textwidth}
\includegraphics[width=\linewidth]{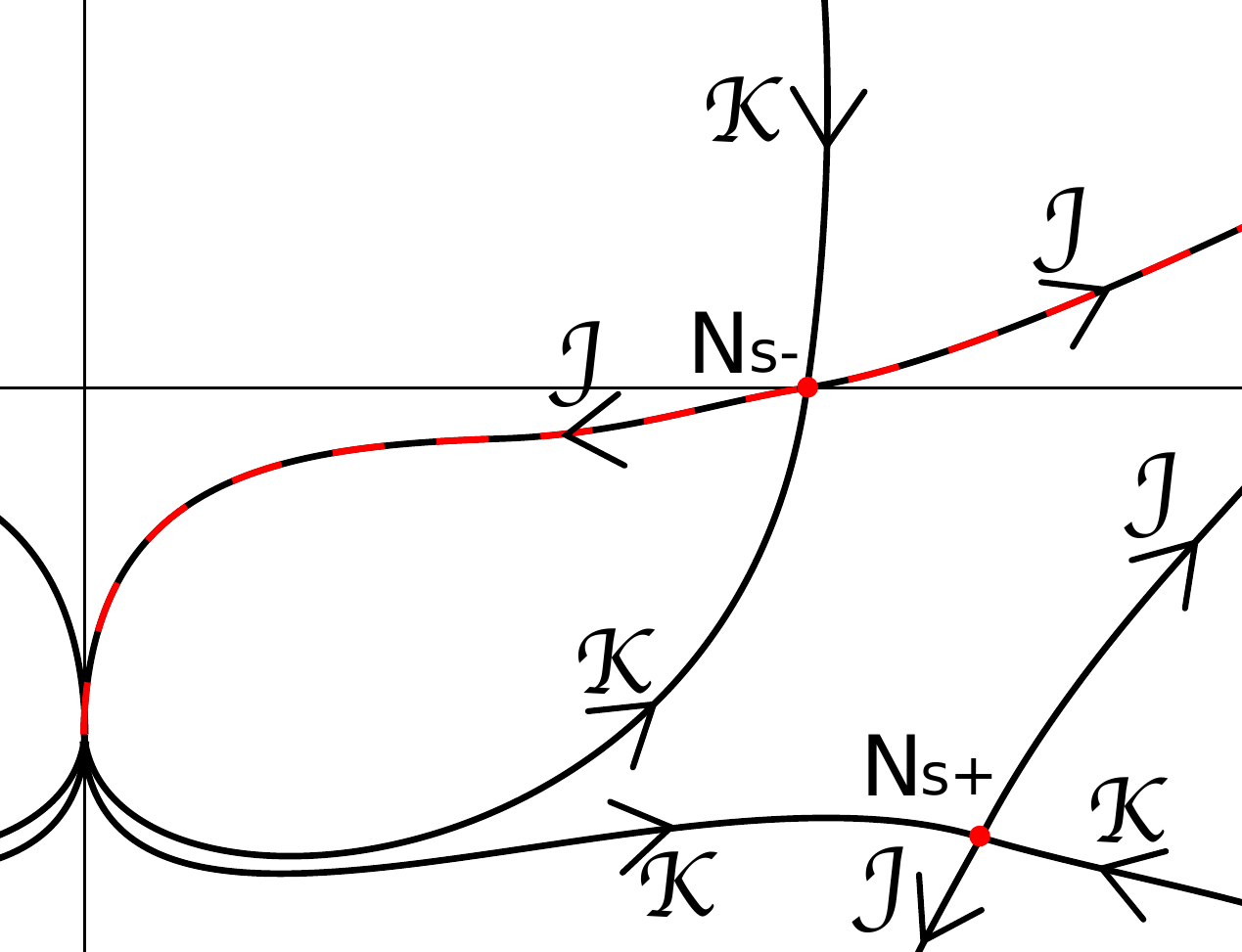}
\end{minipage}
\caption{The lines of steepest ascent and descent for the integral over $N$ as a function of an increasing initial localization $\sigma$ (the localization $\sigma$ increases from top left to top right, then lower left and finally lower right). When $\sigma$ becomes larger than the critical value $\sigma_c$ (which occurs in the transition from the top right to the lower left graph), a Stokes phenomenon happens: the line of steepest ascent $\mathcal{K}$ from the saddle point $N_{s+}=N^\sigma_{1,1}$ corresponding to the bouncing solution no longer intersects the original integration contour on the real $N$ line. Beyond this critical value, only one saddle point (namely the purely expanding saddle point $N_{s-}=N^\sigma_{1,-1}$) remains relevant to the semi-classical path integral and moreover this saddle point resides on the real $N$ line, so that it describes classical evolution. At this point standard quantum field theory in curved spacetime is recovered.}
\label{fig:PLsigma}
\end{figure}

The subsequent integral over $N$ can be evaluated using a saddle point approximation, with the help of Picard-Lefschetz theory. There are four saddle points of $\bar{S}[q_1;\bar{q}_0;N]$ in $N$ located at
\begin{align}
N^\sigma_{c_1,c_2}  &=  -\frac{3 i \sigma^2 V_3}{\hbar} +c_1 \sqrt{ \frac{3}{\Lambda}}\left(\sqrt{q_1} + c_2 \sqrt{q_i + \frac{2 i p(q_i) \sigma^2}{\hbar} - \frac{3 \Lambda \sigma^4 V_3^2}{\hbar^2}} \right) \\ &= \sqrt{ \frac{3}{\Lambda}} \left( c_1 \sqrt{q_1}+ c_2\sqrt{q_i}\right)  +(\mp c_2-1) \frac{3 i \sigma^2 V_3}{\hbar} \,, \label{Nssigma}
\end{align}
with $c_1,c_2=\pm 1$ and where we have used $p(q_i) = \mp V_3 \sqrt{3\Lambda q_i}$ for the initially expanding/contracting cases respectively. The corresponding Lefschetz thimbles for the boundary conditions $0<q_i \ll q_1$ are given in Fig. \ref{fig:PLsigma}, for the case of an expanding initial state. 

In the limit of zero spread $\sigma =0$ (the upper left panel of Fig. \ref{fig:PLsigma}), we saw that $\bar{q}_0 \to q_i$ and thus we recover our earlier results, namely that the propagator is dominated by both an expanding and a bouncing solution. The expanding solution is given by $(c_1,c_2)=(+1,-1)$, while the bouncing solution is given by $(c_1,c_2)=(+1,+1),$ and we will keep referring to these saddle points as $N_{s-}=N^\sigma_{1,-1}$ and $N_{s+}=N^\sigma_{1,1}$ respectively, by analogy with our discussion in earlier sections. As $\sigma$ is increased, the saddle point $N_{s-}$ corresponding to expansion stays put, and in fact for this saddle point $\bar{q}_0=q_i,$ that is to say the expanding solution starts from the central value of $q.$ Meanwhile, the saddle point $N_{s+}$ corresponding to a bounce starts to move off the real axis. For sufficiently small $\sigma$, both saddle points are relevant (top right panel). However, as $\sigma$ reaches a critical value $\sigma_c$ we observe a Stokes phenomenon after which solely the expanding solution is relevant for the Feynman propagator (see the lower left panel of Fig. \ref{fig:PLsigma} for an illustration). Thus, for a localization with $\sigma > \sigma_c,$ the propagator is dominated by a single saddle point, which moreover resides on the real $N$-axis, implying that the treatment of both the background and the fluctuations will be well approximated by quantum field theory on curved space-time. If we \emph{assume} the appropriate Bunch-Davies state for the perturbations of the (expanding) initial state, we will recover the predictions of inflation, as long as the perturbative action remains stable and well-defined along the entirety of the thimble. We will explore this issue in the next section. Note however that in this framework inflation does not explain the Bunch-Davies state by itself -- rather it must be put in by hand from the outset and eventually an additional theory will be required to explain it. 

The critical localization $\sigma_c$ can be found by solving for the condition that the two saddle points are linked by a line of steepest ascent/descent,
\begin{align}
\text{Im}\left[i\bar{S}^{(0)}+ip(q_i)\bar{q}_0-\frac{\hbar(\bar{q}_0-q_i)^2}{4\sigma^2}\right]_{N_{s-}}=\text{Im}\left[i\bar{S}^{(0)}+ip(q_i)\bar{q}_0-\frac{\hbar(\bar{q}_0-q_i)^2}{4\sigma^2}\right]_{N_{s+}}\,. \label{Stokes}
\end{align}
The action at ${N}_{s-}$ is very simple (and real), and is given by
\begin{equation}
\begin{split}
\left[\bar{S}^{(0)}+p(q_i)\bar{q}_0+i\frac{\hbar(\bar{q}_0-q_i)^2}{4\sigma^2}\right]_{N_{s-}}= - V_3 \sqrt{\frac{\Lambda}{3}} \left( 2 q_1^{3/2} + q_i^{3/2}\right)\,.
\end{split}
\label{actionnob}
\end{equation}
For the bouncing saddle point the action reads
\begin{equation}
\begin{split}
\left[\bar{S}^{(0)}+p(q_i)\bar{q}_0+i\frac{\hbar(\bar{q}_0-q_i)^2}{4\sigma^2}\right]_{N_{s+}} =  &- V_3 \sqrt{\frac{\Lambda}{3}} \left[2 q_1^{3/2} +  \sqrt{q_i} \Bigl(\frac{ 5 q_i \hbar^2 - 36 \Lambda \sigma^4 V_3^2 }{\hbar^2}  \Bigr) \right] + \\
& - i \left(\frac{ \Lambda \sigma^4  V_3^2}{\hbar^2} - q_i\right)12 \Lambda \sigma^2 \frac{V_3}{\hbar}
\end{split}
\end{equation}
Therefore, we find that the two saddles are linked by a Stokes line if
\begin{equation}
4 \frac{V_3}{\hbar^2} \sqrt{\frac{q_i \Lambda}{3}} (\hbar^2 q_i - 9 V_3^2 \Lambda \sigma^4) = 0\,,
\end{equation}
implying that the critical localization is given by 
\begin{align}
\sigma_c = \left( \frac{q_i \hbar^2}{9 \Lambda V_3^2} \right)^{1/4}\,. \label{sigapprox}
\end{align}
Note that the critical localization $\sigma_c$ is independent of the final state $q_1,$ but that it depends on the inflationary vacuum energy $\Lambda$.

\begin{figure}
\centering
\includegraphics[width=0.5\textwidth]{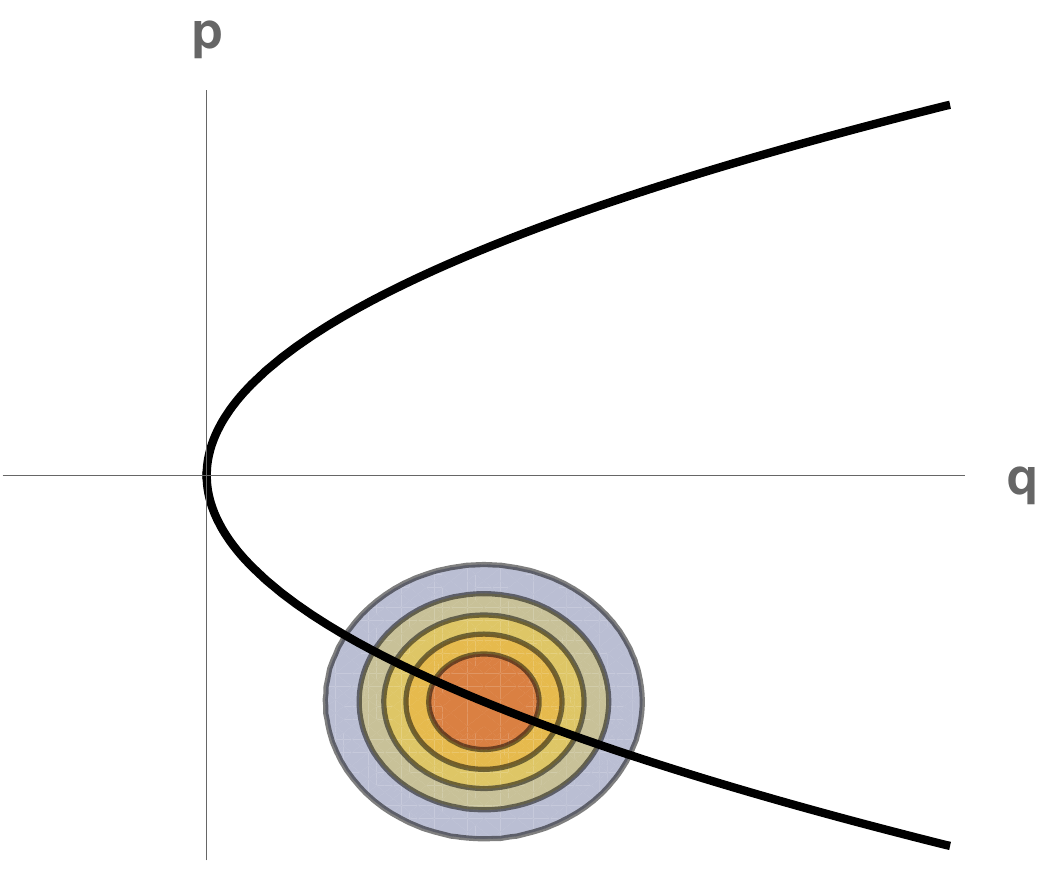}
\caption{The probability distribution of the initial wave function in phase-space $(q_i,p_i)$. The black line is the Hamiltonian constraint. The horizontal axis is $q_i$ and the vertical axis  $p_i$. In order to have only an expanding universe, the wave function needs to have support only in the lower right quadrant.}
\label{fig:wavefunction}
\end{figure}

We can set up initial conditions for an expanding universe whenever 
\begin{align}
q_i >  \sigma_c\,,
\end{align}
%\begin{align}
%q_i > \nu \sigma_c
%\end{align}
%for sufficiently large $\nu$, 
since this will allow us to specify the momentum $p(q_i)$ with sufficient accuracy, \textit{i.e.}
\begin{align}
|p(q_i)| > \frac{\hbar}{2 \sigma_c}\,.
\end{align}
This is confirmed by evaluating the Wigner function of the initial state 
\begin{align}
P(q,p) = \frac{1}{\pi \hbar} \int \psi_0^*(q+y)\psi_0(q-y) e^{2i p y/\hbar} \mathrm{d}y\,,
\end{align}
which is plotted in Fig. \ref{fig:wavefunction}. The size of the initial spatial slice $a_i$ depends on the value of the vacuum energy, as the condition $q_i > \sigma_c$ translates into
\begin{align}
a_i = \sqrt{q_i} > \frac{\hbar^{1/3}}{(9\Lambda)^{1/6}V_3^{1/3}} \qquad \quad \text{or} \qquad \quad a_i^3 V_3 > \frac{\hbar}{3 M_{Pl} \sqrt{\Lambda}},
\label{stokesbound}
\end{align}
%\begin{align}
%a_i = \sqrt{q_i} > \frac{\hbar^{1/3}\nu^{2/3}}{(9\Lambda)^{1/6}V_3^{1/3}} \qquad \quad \text{or} \qquad \quad a_i^3 V_3 > \frac{\hbar \nu^2}{3 M_{Pl} \sqrt{\Lambda}},
%\label{stokesbound}
%\end{align}
where in the last expression we have restored $M_{Pl}$ by dimensions. This agrees with the bound we derived through a heuristic argument in Eq.~(\ref{estbd}) above.  

The implied bound means, for example, that for inflation at the grand unified scale $\Lambda \sim \left( 10^{-3}\right)^4 M_{Pl}^4,$ one can only describe the beginning of inflation in the context of QFT in curved spacetime when the initial size is larger than about a million Planck volumes, or about a hundred Planck lengths in linear size. Note that the current analysis is conservative, in that we have neglected any constraints that might arise when adding perturbations -- as we will see, the inclusion of perturbations leads to a small strengthening of the bound.

\subsubsection{Stable Perturbations} \label{sec:stableperts}

Having established that with a suitable initial state the background evolution can be consistently reduced to a configuration describing an expanding universe only, we would like to see if the perturbations are also well-behaved in this background. The effect of the convolution with the initial state is to substitute the initial scale factor $q_0$ with an ``effective'' initial size $\overline{q}_0$ defined as 
\begin{equation}
\overline{q}_0= \frac{N q_i - i ( 3 N^2 H^2 V_3 - 2 N p(q_i) - 3 q_1 V_3) \sigma^2/\hbar}{N + i 3 \sigma^2 V_3/\hbar}\,,
\end{equation}
with $p(q_i)= -  V_3 \sqrt{3 \Lambda q_i}$ and $H =\sqrt{ \frac{\Lambda}{3}}$. With this initial condition, the background solution becomes
\begin{equation}
\begin{split}
\bar{q}(t) = & H^2 N^2 t^2 + \frac{q_1 - \sqrt{q_i}(\sqrt{q_i} - i 6 H \sigma^2  V_3/\hbar) - H^2 N^2}{N + 3 i \sigma^2 V_3/\hbar} N t \\
& + \frac{N q_i + 3 i (q_1 - 2 H N \sqrt{q_i} - H^2 N^2)\sigma^2  V_3/\hbar}{N + 3 i \sigma^2 V_3/\hbar}\,.
\end{split}
\end{equation}
In analogy with the background solution that we obtained in the absence of an initial state, we may again write this as
\begin{equation}
\bar{q}(t) = H^2 N^2 (t - \alpha)(t - \beta)\,,
\label{qsol2}
\end{equation}
but this time with the $\sigma$-dependent coefficients
\begin{align}
\alpha &= \frac{ \left[ N^2 - (N^\sigma_{1,-1}N^\sigma_{1,1} + N^\sigma_{-1,1} N^\sigma_{-1,-1}) /2 + \sqrt{(N - N^\sigma_{1,-1})(N - N^\sigma_{1,1})(N - N^\sigma_{-1,1})(N - N^\sigma_{-1,-1})} \right] }{2 N (N + 3 i \sigma^2 V_3/\hbar)}\,, 
\label{eq:alphaGen}\\
\beta &= \frac{ \left[ N^2 - (N^\sigma_{1,-1}N^\sigma_{1,1} + N^\sigma_{-1,1} N^\sigma_{-1,-1}) /2- \sqrt{(N - N^\sigma_{1,-1})(N - N^\sigma_{1,1})(N - N^\sigma_{-1,1})(N - N^\sigma_{-1,-1})} \right] }{2 N (N + 3 i \sigma^2 V_3/\hbar)}\,.
\label{eq:betaGen}
\end{align}
Here the lapse values $N^\sigma_{c_1,c_2}$ with $c_1,c_2 = \pm 1$ correspond to the saddle points of the background action, determined earlier in Eq. \eqref{Nssigma}. Note that the definitions of $\alpha$ and $\beta$ are a simple generalization of the definitions in Eq. \eqref{ab}, the latter being recovered in the limit of $\sigma \to 0.$ It also remains the case that $\alpha = \beta$ at the saddle points.  

The equation of motion for the perturbations also remains identical in form, cf. Eq. \eqref{212}, but now with the $\sigma-$dependent coefficients $\alpha$ and $\beta$. The two linearly independent solutions are then once again $f(t)/\sqrt{q(t)}$ and $g(t)/\sqrt{q(t)}$ with
\begin{align}
f(t) &= \left[ \frac{t - \beta}{t  - \alpha } \right]^{\mu/2} \left[ (1 - \mu)(\alpha - \beta) + 2 (t - \alpha)\right]\,, \label{pertmode1}\\
g(t) &= \left[ \frac{t - \alpha}{t  - \beta } \right]^{\mu/2} \left[ (1 + \mu)(\alpha - \beta) + 2 (t - \alpha)\right]\,, \label{pertmode2}
\end{align}
where
\begin{equation}
\mu^2 = 1 - \left( \frac{2 k}{(\alpha - \beta) H^2 N} \right)^2\,.
\end{equation}
The two solutions correspond to the two square roots $\pm \mu$. At the (expanding) saddle point $N_{s-}=N^\sigma_{1,-1}$, which stays put on the real $N$ line as the localization $\sigma$ is varied, the solution given by $g(t)$ retains the form appropriate to the stable Bunch-Davies state. This is the solution that any initial state must single out in order to recover the standard description of inflationary fluctuations. 

\begin{figure}
\includegraphics[width=0.45\textwidth]{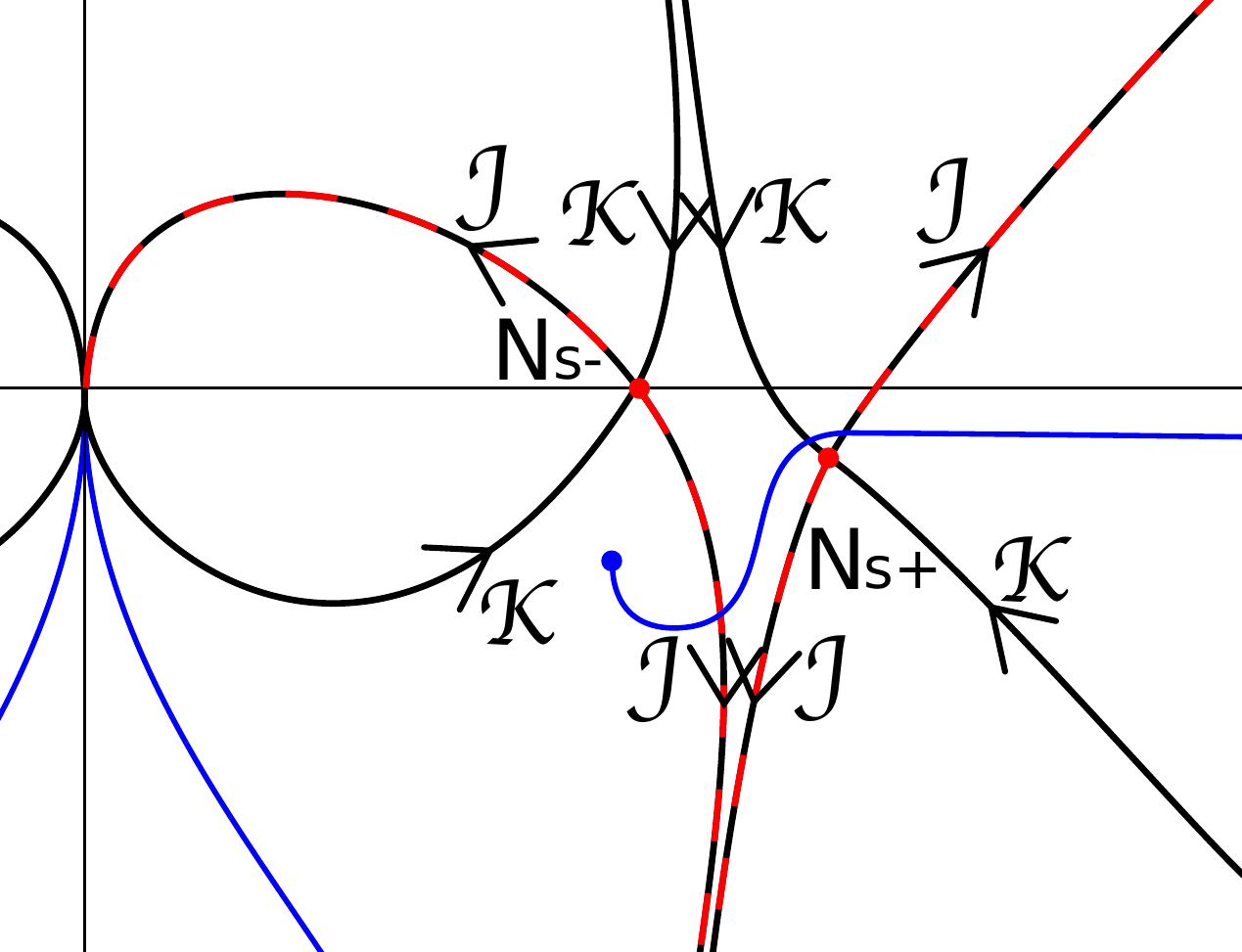} \hspace{.5cm}
\includegraphics[width=0.45\textwidth]{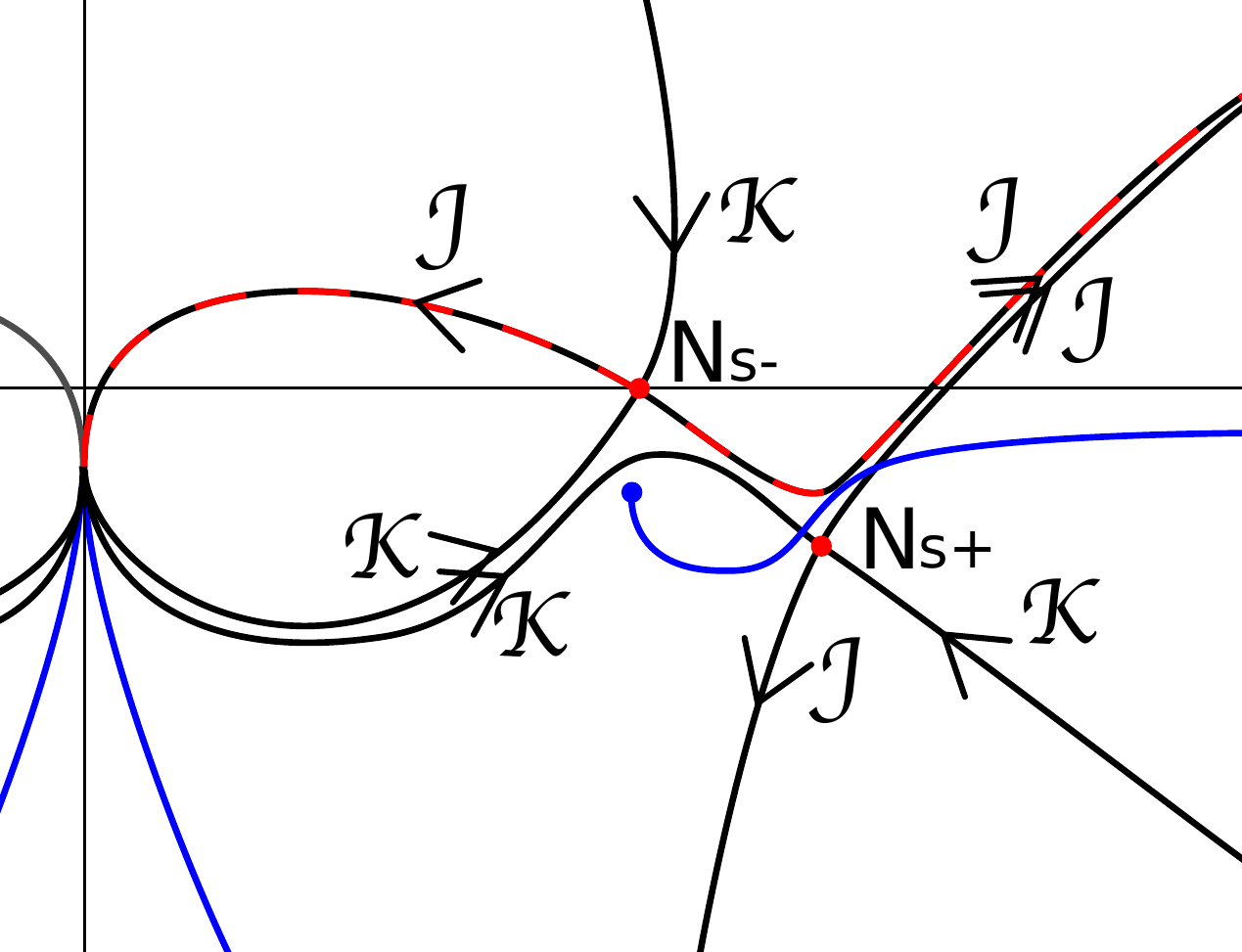}
\caption{{\it Left panel:} When the initial spread $\sigma$ is small, the relevant thimbles are forced to pass through the ``singular curve'' where the perturbative action blows up due to regions of vanishing scale factor in the geometries that are summed over.   {\it Right panel:} For a sufficiently large localization $\sigma \gtrsim \sigma_c,$ the relevant thimble stays above the singular curve, the path integral for the perturbations is well-defined and, with a judicious choice of initial state, stable fluctuations may be imposed.} \label{fig:SingularLines}
\end{figure}

However, selecting the appropriate saddle point solution is not enough. An important issue for the consistency of the calculation is that the perturbative action ought to be well-defined along the entire Lefschetz thimble that one integrates over in order to obtain the Feynman transition amplitude. An obstruction occurs whenever an (off-shell) background spacetime that is being integrated over contains a region where the scale factor of the universe reaches zero or passes through zero. At such locations the perturbations blow up, and the perturbative action $S^{(2)}$ becomes infinite, rendering the path integral ill-defined. Note that in such a case the surface form of the action \eqref{eq:boundary} would be augmented by one or more (infinite) surface terms at the intermediate times where $\bar{q}=0.$ Thus, in order for the path integral for background and fluctuations to be well-defined, we must ensure that the whole Lefschetz thimble does not include any backgrounds containing regions where $\bar{q}(t)=0,$ for $0 < t < 1.$ The occurrence of singular regions depends on the value of the localization $\sigma,$ according to Eq. \eqref{qsol2}. As discussed in subsection \ref{subsec:fluctuations}, when $\sigma=0$ the locus of singular backgrounds consists of the parts of the real $N$ line with $|N| \geq N_\star.$ Thus, with vanishing localization we not only obtain interference between two background solutions, but the thimbles also cross a region where the perturbative action is ill-defined since $N_{s+}>N_\star$. For small non-zero localization $\sigma,$ the singular curve moves away from the real $N$ line into the lower right quadrant in the complex $N$ plane, see the left panel of Fig. \ref{fig:SingularLines}. At this point, the thimbles still cross the singular curve. As the localization is further increased, we find that there exists a critical value which, as we will shortly see, is closely related to (though slightly larger than) the critical value $\sigma_c$ for the Stokes phenomenon. At this value the single remaining thimble just touches the singular curve of infinite $S^{(2)},$ and beyond this value of the localization the thimble is everywhere well-defined. This configuration is illustrated in the right panel in Fig. \ref{fig:SingularLines}.

In order to determine the critical value of the localization quantitatively, we will consider the physically relevant limit in which the final scale factor $a_1 = \sqrt{q_1}$ becomes large. In this limit we are able to obtain an analytic description of the thimble. Then we shall work out the locus of the singular curve consisting of all points in the complex $N$-plane for which $q(t)$ has a zero for some $0<t<1.$ This will allow us to find the intersections of the thimble with the singular curve, as a function of the localization $\sigma$. Since the algebra is a little involved, it is helpful to scale out dimensions. We therefore set $\Lambda=3 H^2 M_{Pl}^2$, $N=\tilde{N} H^{-2}$, $q=\tilde{q} H^{-2}$, $a=\tilde{a} H^{-1}$, $p=-3 \tilde{a}_i V_3$ (the classical value corresponding to an expanding universe), $\tilde{\hbar}=\hbar H^2/(M_{Pl}^2 V_3)$,  and $\sigma=H^{-2} \tilde{a}_i^{1\over 2} s \sqrt{\tilde{\hbar}}$. It is also helpful to define $\tilde{a}_1 = \gamma \tilde{a}_i$ and $\tilde{N}=\tilde{a_i} n$. We now calculate the total exponent minus that for the saddle point $N_{s -}$, in the limit of large $\gamma$, obtaining 
\begin{align}
& i\bar{S}^{(0)}+ip(q_i)\bar{q}_0-\frac{\hbar(\bar{q}_0-q_i)^2}{4\sigma^2} - \left[i\bar{S}^{(0)}+ip(q_i)\bar{q}_0-\frac{\hbar(\bar{q}_0-q_i)^2}{4\sigma^2}\right]_{N_{s-}} \nonumber \\ \approx \, &  i a_i^3(\delta n^2(-3 +9i s^2)+\delta n^3 )/\tilde{\hbar} +O(1/\gamma),
\label{acteq}
\end{align}
where we have defined $\delta n\equiv n-(\gamma-1)$ as the deviation of $n$ from its saddle point value. Requiring that the imaginary part of (\ref{acteq}) is zero is the defining condition for the thimble, and provides a relation between the real and imaginary parts of $\delta n$, which in turn determines the locus of the thimble in the complex $n-$plane. 

It follows from (\ref{qsol2}) that the condition for $q(t)$ to possess a zero for $0<t<1$ is that $\alpha$ or $\beta$, given in (\ref{eq:betaGen}), are real and lie between zero and one, at the given value of $N$.  But $\alpha$ and $\beta$ are just the two roots $g$ of a quadratic equation,
\begin{equation}
g^2+{\gamma^2-1-n^2+6 i s^2 \over n(n+3 i s^2) }g +{n+3 i s^2(\gamma^2-2 n -n^2)\over n^2(n+3 i s^2)}=0.
\label{gameq}
\end{equation}
Since we are interested in values of $n$ for which $g$ is real and lies between zero and one, we can set the imaginary part of (\ref{gameq}) to zero and obtain $g$ in terms of $n$. Substituting this expression for $g$ into the real part of (\ref{gameq}) gives a relation between the real and imaginary parts of $n$ which determines the locus of the points in the complex $n-$plane for which $q(t)$ vanishes for some $0<t<1$. (One must also check that indeed $g$ lies between zero and one).  Setting $n=\gamma-1+\delta n$ as above, and again taking the limit of large $\gamma$, we find a relation between the real and imaginary parts of $\delta n = x+i y$: 
\begin{equation}
6 s^2 y^3+y^2+6 s^2 y(x^2-x)+9 x^2s^4=0.
\label{gameq1}
\end{equation}
In terms of the same variables, it follows from setting zero the imaginary part of the exponent in eq. \eqref{acteq} to its value at the saddle that the locus of the thimble is given by
\begin{equation}
-3 x^2 + x^3-18 s^2 x y+3 y^2-3 x y^2=0.
\label{thimblelocus}
\end{equation}
The smaller root $y=(9 s^2 x-\sqrt{3}\sqrt{3 x^2+27 s^4 x^2-4 x^3+x^4})/(3(1-x))$ of (\ref{thimblelocus})  corresponds to the right hand portion of the thimble. Substituting this back into the left hand side of (\ref{gameq1}), we plot the result against $x$ for various values of $s$. At low values of $s$ there are two nonzero roots which move towards each other as $s$ is increased. There is a critical value of $s$ which corresponds to the thimble just touching the line of zeroes of $q(t)$. Above this value there is no real, nonzero root. We find the critical value  $s_c \approx 0.46937$. Restoring the units in which our bound (\ref{stokesbound}) is expressed, we are able to conclude that the Lefschetz thimble relevant to the background solution encounters no zero in $q(t)$, and hence the stable mode persists across the entire relevant Lefschetz thimble, provided that
\begin{align}
a_i^3 V_3 \gtrsim C \frac{\hbar}{M_{Pl} \sqrt{\Lambda}},
\label{moreaccbound}
\end{align}
%\begin{align}
%a_i^3 V_3 > \frac{\hbar \nu^2}{C M_{Pl} \sqrt{\Lambda}},
%\label{moreaccbound}
%\end{align}
with the constant $C\approx 0.382$. This bound represents a modest strengthening of the Stokes bound given in (\ref{stokesbound}) and characterizes the parameter range over which the path integral for the perturbations is well-defined, once the background has been integrated out.

\subsection{Discussion} \label{sec:discussion}

When inflation was discovered, there were hopes that it would explain the initial conditions of the universe. In fact, it was thought that inflation, being an attractor, would explain the starting point of the hot big bang model irrespective of what came before it. According to this view,  if inflation had taken place, we would never have to, nor would we ever be able to, understand what occurred at earlier times. In recent years it has become increasingly clear that, at the classical level, the situation is more nuanced: inflation requires special initial conditions to get underway, and it is still actively debated whether such initial conditions are likely or unlikely (see, {\it e.g.}, \cite{Gibbons:2006pa,Ijjas:2013vea}, and the recent numerical works \cite{East:2015ggf,Clough:2016ymm,Marsh:2018fsu} which, however, neglect the crucial effect of the initial scalar field velocity). 

Our work adds an additional, purely quantum consideration: if inflation were truly  independent of what came before it, then we ought to be able to consider inflation all the way back to the singularity when the size of the universe approached zero.  And we ought to be able to assume that there was nothing prior to inflation. As we have shown, the de Sitter propagator from an initial vanishingly small three-geometry to a large final three-geometry does indeed become independent of the initial fluctuations, and provides an answer depending solely on the final fluctuations. At first sight, this would seem to reinforce the hopes described above, and might even suggest that at a quantum level inflation really can stand on its own and explain the state of the universe. However, our calculations reveal that, semi-classically at least, there is a tension between the background and  the fluctuations, and the wrong-sign Bunch-Davies mode functions are selected for the fluctuations. This result immediately implies a breakdown of the model, {\it i.e.}, that one cannot describe the origin of an inflationary universe of vanishingly small initial size. Our calculation demonstrates that quantum gravity effects cannot be ignored at the beginning of inflation -- put differently, the beginning of inflation is highly sensitive to UV effects, not just in the sense of its potential being sensitive to curvature corrections etc., but also in terms of its quantum vacuum. Since all predictions of inflation depend sensitively on the quantum vacuum, this is not a small issue.

We believe that these results change the status of inflation: instead of ``creating'' a flat universe, and ``creating'' the primordial fluctuations, inflation may better be thought of as a mechanism reinforcing flatness, and processing fluctuations out of a pre-existing (but assumed) quantum vacuum into classical density fluctuations. Inflation can exist as a phase of cosmological evolution, sandwiched between other phases, but it does not by itself explain the initial state. On the one hand, this significantly weakens our ability to test inflation observationally. After all, other types of cosmological evolution, such as ekpyrosis \cite{Lehners:2008vx}, can perform analogous processing tasks. On the other hand, we reach the highly welcome prospect that early universe cosmology offers us a window onto full quantum gravity, and not just QFT in curved spacetime.

Can one think of ways to avoid the negative result just described? An obvious possibility is to consider a pre-inflationary phase. In fact, this was a popular idea in the early days of inflation. A pre-inflationary phase would have to set up the initial conditions for inflation, both classically (in terms of preparing a region of the universe that is sufficiently flat and at a sufficiently high and homogeneous energy density) and quantum mechanically (in terms of preparing the Bunch-Davies vacuum). We have provided the first steps in this direction by showing that quantum field theory in curved spacetime may be recovered as long as the spatial volume at the onset of inflation is large enough, more specifically as long as it is larger than 
\begin{align} 
a_i^3 V_3 \gtrsim \frac{\hbar \nu^2}{\sqrt{\Lambda}}\,,
\end{align}
assuming an expanding initial state. This result shows that the background may be treated classically from this size onward, but the quantum state of the fluctuations remains unexplained by these arguments. These results set the scene for the introduction of the no-boundary proposal which will be discussed in the next chapter. The main goal of the proposal is in fact precisely to explain how such a large classical inflationary universe, with Bunch-Davies quantum fluctuations, can be generated out of nothing due to a quantum effect.

\clearpage
%%%%%%%%%%%%%%%%%%%%%%%%%%%%%%

\section{The no-boundary proposal: not all is well with Dirichlet path integrals}\label{nbDirichlet}

\begin{figure}
	\centering
	\includegraphics[width=0.4\textwidth]{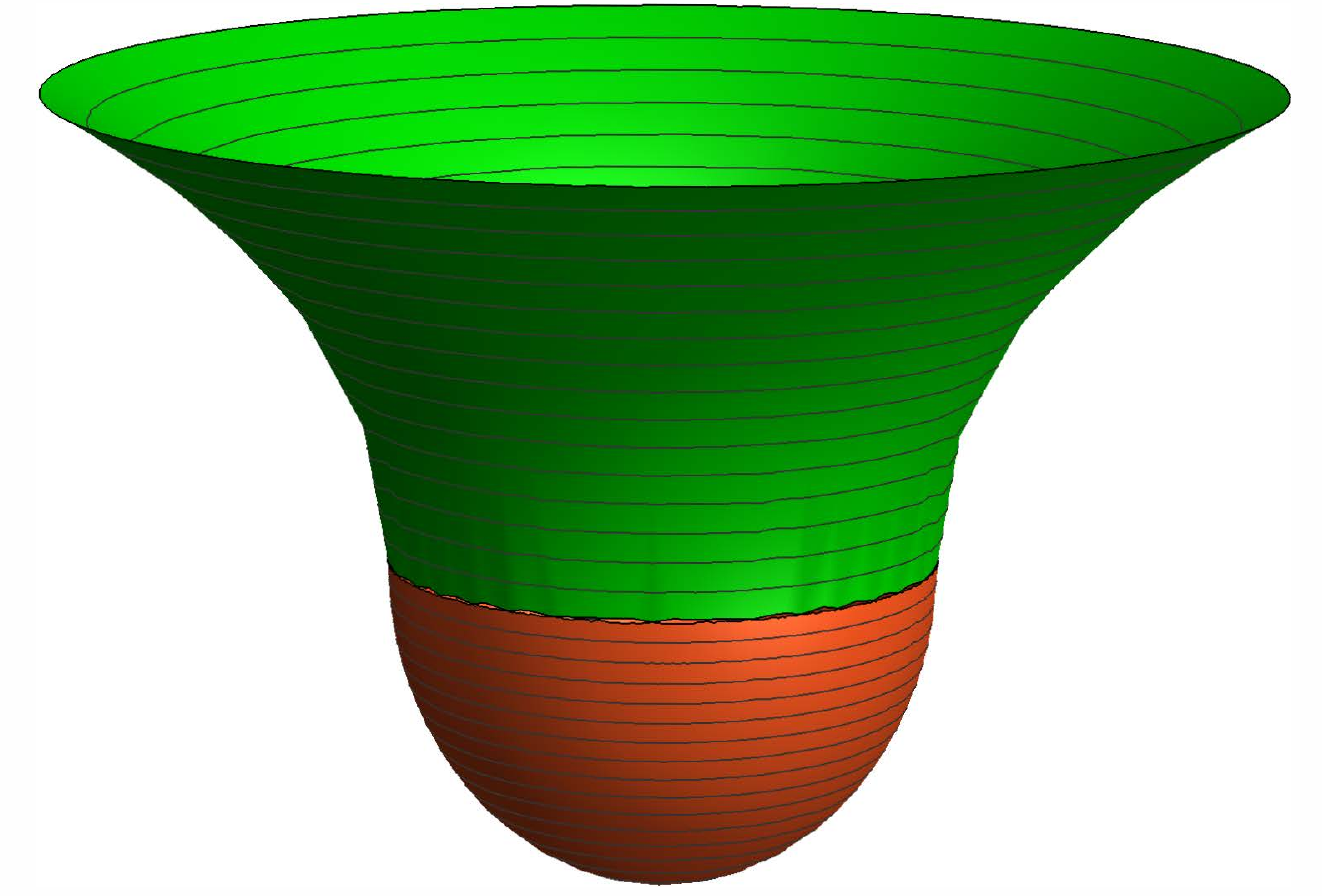}
	\includegraphics[width=0.4\textwidth]{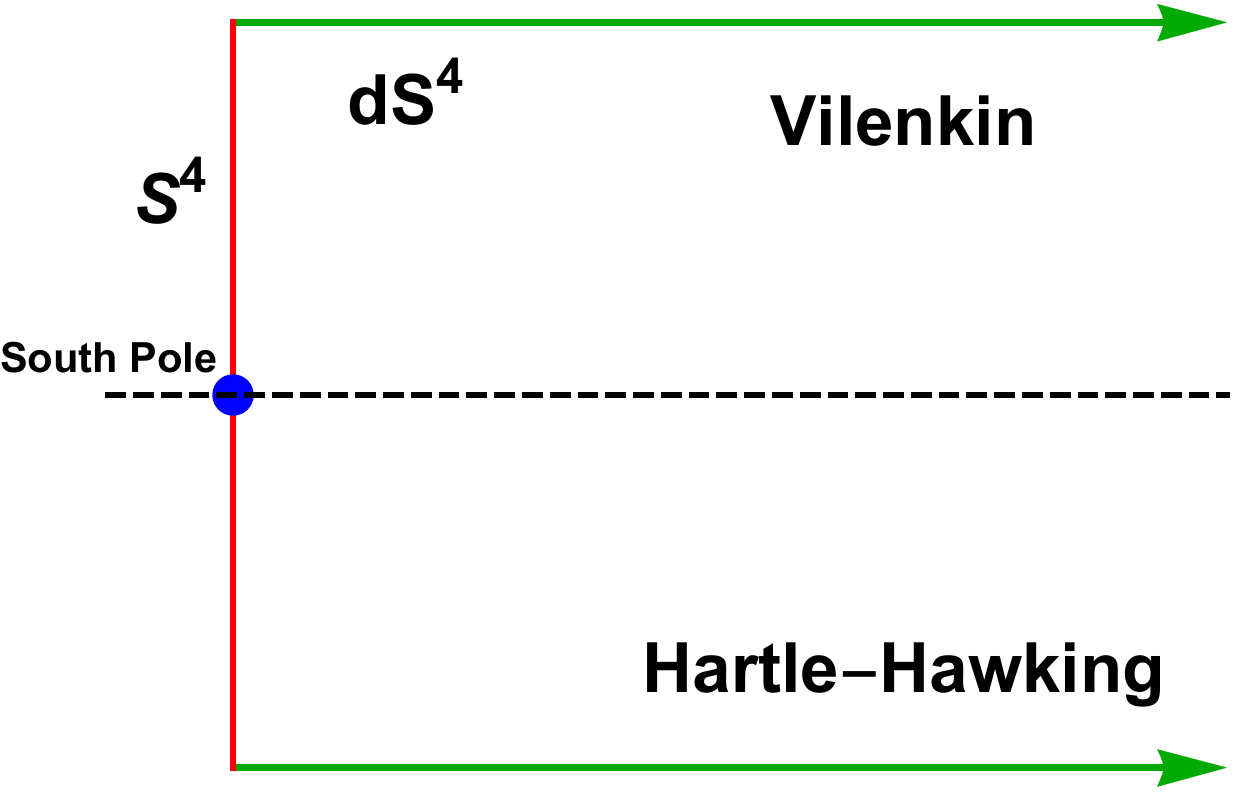} 
	\caption{{\it Left panel:} The Hartle-Hawking/Vilenkin geometry, typically represented as a Lorentzian geometry (in green) glued onto a Euclidean geometry (in red). {\it Right panel:}  Green lines with arrowheads indicate directions of increasing real/Lorentzian time, red lines depict imaginary/Euclidean time directions. The Euclidean extensions of the geometries imply specific Wick rotations \cite{DiTucci:2019bui}.
	}
	\label{fig:Wick}
\end{figure}

 The universe is a system different from any other in the fact that it is a one-time experiment for which the initial conditions where set once and for all, with no observer being there to measure them. Thus in this case the definition of the initial conditions becomes part of the research program itself. In this sense, quantum cosmology stands as a theory of initial conditions for the universe which aims at giving a definite answer stemming first principles. The question we ask then is if our universe started out in a very special state and if so whether this was a necessary condition or a fortunate accident.
The no-boundary proposal \cite{Hartle:1983ai}, also known as Hartle-Hawking proposal (HH), provides a possible answer to this question based on the idea that among the infinitely many possible initial states of the universe there is one which plays a distinct role and as such should be identified as the correct one. It selects a specific wavefunction of the universe which shall provide suitable initial conditions for our universe as described by inflationary and the standard cosmology. Such wave function describes a (quasi-)de Sitter space being nucleated out of a geometry of zero size or, in other words, 'nothing'.\\
%In the particle analogy, there is a classically forbidden region $0<a<a_0$. A zero-energy particle in $a=0$ will be stuck there if only classical paths are allowed. However, due to quantum mechanics there is non-vanishing probability that the particle tunnels through the barrier and emerges at the classical turning point $a=a_0$, from which it evolves classically.
%The probability of the tunneling is given by $e^{- S_{E}}$ where $S_E$ is the Euclidean action associated with the corresponding instanton solution\cite{kolb} \todo{maybe $e^{- 2 Im[S]}$? o qualcosa del genere}.\todo{elaborate}\\
Let us describe in what follows the proposal in its simplest version where an inflationary universe is approximated by a portion of de Sitter space in closed slicing
\begin{align}
ds^2 = - dt_p^2 + \frac{1}{H^2}\cosh^2 (Ht_p) d\Omega_3^2\,. \label{eq.dS}
\end{align}
where $d\Omega_3^2$ is the line element of a unit three sphere $V_3=1$ and $\Lambda = 3\,H^2,$ is the cosmological constant.\\
In this case, the minimum classically allowed size of the universe corresponds to the waist of the hyporboloid de Sitter (where the scale factor $a$ takes the value $a(t_p=0) = H^{-2}$) and at the classical level it is impossible to start from 'nothing' ($a=0$) and end up in de Sitter space. We can however think about a quantum tunneling from zero size to the minimum classically allowed size that is, to the waist of the de Sitter hyperboloid. The corresponding instanton is then given by the Euclidean version of de Sitter space. Concretely, the Euclidean de Sitter space in four dimensions is a four-sphere $S^4$ and the instanton is given by half of the sphere, where the equator corresponds to the size $H^{-2} $ 
\begin{align}
ds^2 = d\eta^2 + \frac{1}{H^2} \sin^2 (H\eta) d\Omega_3^2\,, \quad t_p = - i \left(\eta - \frac{\pi}{2H}\right)\,, \quad \eta \in [0 , \frac{\pi}{2 H}] \label{eq.S4}
\end{align}
and the geometry smoothly closes of at the South Pole $\eta = 0$.\\
Importantly, this instanton solution\footnote{Here and in the following we will call, with a slight abuse of language, "instantons" the saddle points solutions around which the semi-classical wavefunction is peaked not only when these solutions are purely Euclidean but also when they are generically complex.} is not meant to describe the actual geometry of the early universe but represents solely a mathematical tool to describe a tunneling probability.\\ Classical solutions exist for values of the scale factor larger than the de Sitter waist. Thus, the nucleation from nothing of a universe of size larger than the waist $H^{-2}$ will be described by a complex geometry which can be thought of as half of the Euclidean de Sitter four-sphere, describing the classically forbidden region, glued onto a portion of the Lorentzian de Sitter hyperboloid, corresponding to the classically allowed region. This gives rise to the famous Hartle-Hawking geometry represented in Fig. \ref{fig:Wick}. The no-boundary proposal states that the wavefunction of the universe ought to be peaked around it. This wavefunction then provides,``the amplitude for the Universe to appear from nothing" \cite{Hartle:1983ai} as given by $e^{iS}$, where $S$ is the action evaluated along the HH geometry. Notice that, while the Lorentzian de Sitter space has two boundaries, one in the future and one in the past, with this construction, we obtained a geometry which has no boundary in the past, from which the name of the proposal. Importantly, for this to work it is essential that the space-like surfaces of the universe are closed which becomes a genuine prediction of the no-boundary proposal.\\ 
While the separation into a Euclidean and a Lorentzian section is useful for grasping the physical intuition, the full geometry is really a complex smooth geometry. In order to understand that, it is useful to switch time coordinate to $t \in [0,1]$ with \cite{Halliwell:1988ik}
\begin{equation}
\begin{split}
&ds^2 = - \frac{N_{HH}^2}{q(t)} dt^2 + q(t) d \Omega_3^2, \quad q(t) = H^2 N_{HH}^2 + (R_3^2 -H^2 N_{HH}^2 )t \\ 
    &\sinh(H t_p) = H^2 N_{HH} t + i   , \quad N_{HH} = \frac{\sqrt{H^2 q_1 - 1}}{H^2} - \frac{i}{H^2} \label{eq.nbgeometry}
    \end{split}
\end{equation}
With this choice of coordinates the entire HH complex geometry is covered where $q_1$ is size of the final three-sphere $S^3$ and is the argument of the Hartle-Hawking wave function.\\
In order to understand the relation of the metric \eqref{eq.nbgeometry} to the Hartle-Hawking geometry given by the appropriate gluing of~\eqref{eq.dS} with~\eqref{eq.S4}, it is useful to measure the comoving ``distance''~$D$ traversed in the geometry~\eqref{eq.nbgeometry} as $t$ varies from~0 to~1\footnote{The next paragraph is taken from Ref. \cite{DiTucci:2020weq}.}
\begin{equation}
\label{eq.HHdistance}
D = \int_{0}^1 dr \, \sqrt{-\frac{N_{HH}^2}{q}}.
\end{equation}
If $q_1 \leq 1/H^2$, then $D$ is real and positive and one can check that the geometry~\eqref{eq.nbgeometry} is Euclidean and represents a portion of the four-sphere~\eqref{eq.S4}. In the case when $q_1 > 1/H^2$, one obtains complex~$D$. The real part of~$D$ is then always equal to $\frac{\pi}{2\,H}$, which corresponds to the Euclidean distance traversed in the half-sphere part of the Hartle-Hawking geometry~\eqref{eq.S4}. The imaginary part of~$D$, which in the absence of a real part of~$D$ would correspond to a time-like separation in the Lorentzian signature, turns out to be nothing else than the proper time elapsed in the de Sitter geometry~\eqref{eq.dS} between $t = 0$ and $t = H^{-1} \, \mathrm{arcosh}{(H\, \sqrt{q_1})}$, i.e. the time in which the three-sphere reaches proper size~$q_1$. This implies that our complex saddle-point geometry~\eqref{eq.nbgeometry} for $q_1 > 1/H^2$ interpolates ``diagonally'' in the complex metric plane between the locus where~$q = q_1 > H^{-2}$ and the locus where~$q = 0$ with $q$ in between these points being a complex function of $t$. The Hartle-Hawking geometry of~\eqref{eq.S4} and~\eqref{eq.dS} achieves the same end point for $q$ by first moving in the real direction of~$D$ (the four-sphere part) and then in the imaginary direction of~$D$ (the de Sitter part), in such a way that $q$ takes real values everywhere (see right panel of Fig. \ref{fig:Wick}). This geometry, which may be seen as a gluing of geometries with two different lapse values, is related to our saddle point geometry by a complex diffeomorphism and should be regarded as equivalent in our formalism\footnote{It is only in the case of a cosmological constant or adiabatic matter that the Hartle-Hawking geometry has a representation in which the scale factor is everywhere real \cite{Feldbrugge:2017fcc}. When more general matter is added, such as a scalar field in a non-constant potential, it is necessarily complex \cite{Lyons:1992ua}.}. 

Hartle and Hawking's idea is extremely powerful in describing the initial conditions of our universe. The classical (de Sitter) inflating universe is in fact closed off in the past in a smooth way by means of a Euclidean region: What was the Big Bang singularity in the classical picture is now replaced by the smooth sphere with the South Pole playing no special role. \\
Quoting Hawking's lecture at the Pontifical Academy in 2016
\begin{quote}
    Asking what came before the Big Bang is meaningless, according to the no-boundary proposal, because there is no notion of time available to refer to. It would be like asking what lies south of the South Pole.
    \end{quote}
At the same time, the Hartle-Hawking wavefunction of the universe $\Psi \approx e^{iS(q_1)}$ describes the quantum birth of a classical universe. The Einstein-Hilbert action \ref{einsteinhilbert} evaluated along the HH geometry is
\begin{align}
S(q_1) = \frac{2 V_3}{H^2} [- i + (H^2 q_1 -1 )^{3/2}]\,. \label{hhaction}
\end{align}
The imaginary part determines the weighting and thus the relative probability of nucleation of a universe with this value of the cosmological constant while the real part of the action is associated with the classical growth of the universe up to a radius $q_1.$ This classical growth is seen as a phase in the wavefunction. Thus the larger the final size the more oscillating is the wavefunction. The fact that this phase grows fast as the universe expands, while the weighting remains constant, is then an indication that the universe has become classical in a WKB sense. When the system is classical enough, the saddle point solution can be trusted to represent the actual trajectory of the system that is, the classical story of cosmology starts with a large universe which is inflating and the HH wavefunction effectively describes its quantum nucleation.\\
Since the weighting gets larger for smaller values of $H$, one would be tempted to conclude that the no boundary proposal favours the quantum birth of a large de Sitter universe. An analogous calculation to the one presented above \cite{Halliwell:1990uy} shows that when the inflationary phase is driven by a scalar field $\phi$ rolling down a potential $V(\phi)$, the HH saddle point gives the action \eqref{hhaction} where in this case $3 H^2(\phi)= V(\phi)$. As a consequence, a universe where the scalar field finds itself on a lower location on the potential seems to be quantum mechanically favoured, raising the question of whether the no boundary proposal is compatible with an inflationary phase which is long enough to match observations. In this sense it is fair to say that, despite the nice properties discussed above, the question of how to extract precise physical predictions from the no boundary proposal and whether it predicts the universe we live in still lacks a definite answer (see for example \cite{Hartle:2008ng, Vilenkin:1987kf, Hartle:2007gi, Lehners:2015sia, Matsui:2020tyd} for ideas and discussions). \\ 
So far we only discussed the no-boundary wavefunction of a homogeneous and isotropic universe. When perturbations are included in the picture they are forced by regularity to be in the Bunch-Davies vacuum. In the HH geometry, the Euclidean section is reached by means of a Wick-rotation in a specific direction that is, we chose the sign $t_p = - i \left(\eta - \frac{\pi}{2H}\right)$ over $t_p = + i \left(\eta - \frac{\pi}{2H}\right)$ in \eqref{eq.S4}. Upon such Wick-rotation, one of the perturbative modes in eq.\eqref{eq:BD} diverges and while the other vanishes as the universe reaches zero size, which one of the two depends on the direction of the rotation. The HH geometry is constructed in such a way that the Bunch-Davies mode is the regular one and thus perturbations are forced by regularity to start out in this state. The geometry given by the opposite Wick-rotation corresponds to the so called tunneling/Vilenkin proposal \cite{Vilenkin:1982de, Linde:1983mx} (on which we will comment further momentarily) and in our opinion does not represent a suitable choice precisely because it leads to unsuppressed perturbations.
%The direction of the Wick-rotation also effects the resulting tunneling probability giving rise to different interpretation of the semiclassical properties of the two proposals.\todo{vedi se elaborare}. \\
The Hartle-Hawking wavefunction thus predicts the nucleation out of 'nothing' of a classical inflating closed universe with gaussian distributed inhomogeneous perturbations providing the suitable initial conditions for inflation we were after~\cite{Halliwell:1984eu}.\\
Why should we pick this initial state of the universe? 
In Hawking words \cite{Hawking:1981gb}:
\begin{quote}
    There ought to be something very special about the boundary conditions of the universe, and what can be more special than the condition that there is no boundary?
\end{quote}
or put differently \cite{Hawking:1988qt}
\begin{quote}
     One could say: ``The boundary condition of
the universe is that it has no boundary." The universe would be completely self-contained and not affected by
anything outside itself. It would neither be created nor destroyed, It would just BE."
\end{quote}
This is ultimately the fundamental principle which selects the Hartle-Hawking wavefunction as the initial state of the universe and comes from an intuition provided by the HH saddle point geometry. \\
Classical inflationary spacetimes are not past-complete \cite{Borde:2001nh} and thus appropriate boundary conditions for every field in the universe, including the primordial perturbations, need to be put in by hand or explained by physics other than inflation. In chapter \ref{quantuminitial} we provided a semi-classical argument which points in the same direction: an external mechanism needs to be advocated to put gravity wave perturbations in the Bunch-Davies vacuum showing the 'quantum' incompleteness of inflation. The no-boundary proposal solves the problem of boundary conditions all together since the HH geometry has no boundary in the past.
So far, we only discussed the saddle point approximation of the Hartle-Hawking wavefunction. The full wavefunction  of the universe is formulated as a path integral according to Feymann's description of quantum mechanics, where here the sum is a sum over geometries:
\begin{equation}
\Psi[h_{ij}, \phi] =\sum_M\nu(M) \int_D \delta g \, \delta \phi \, e^{i S [g_{\mu \nu},\phi]} \label{hawk}
\end{equation}
where $\phi$ represents a generic matter degree of freedom and $S$ is the action for the matter fields and gravity including the cosmological constant. The integral is a sum over four-geometries with a compact boundary $\Sigma$ whose induced three-metric is $h_{ij}$ and all field configurations which match $\phi$ on the boundary. The path integral is often defined to include also a sum over all four-manifolds $M$ with measure $\nu(M)$. The sum \eqref{hawk} provides a path integral of a solution to the WdW equation \eqref{c4}. A specific wavefunction is single out when the class of geometries in sum $D$ is specified. Hartle-Hawking proposal is to include in the summation only compact geometries i.e. whose only boundary is $\Sigma$ with the fields regular on it. This idea extends the intuition suggested by the HH geometry that the universe has no boundary in space and time to the entire sum and shall give a wavefunction representing the amplitude for a three-geometry to arise from 'nothing' i.e. a three-geometry of zero size. In their '83 paper \cite{Hartle:1983ai}, Hartle and Hawking thought about the path integral \eqref{hawk} as a sum over compact Euclidean geometries. The reason behind that is that the oscillatory Lorentzian integral was thought to be ill-defined. The idea, very popular at the time, was that, just like in quantum field theory, a Wick rotation to Euclidean time would improve the convergence properties of the integral, as opposed to the sum over Lorentzian metrics, and thus that Euclidean quantum gravity would be a better defined theory than a hypothetical Lorentzian quantum gravity. While in general not each Lorentzian metric can be recasted into a Euclidean form by means of a Wick rotation, the idea was that the sum over all Euclidean metric should be equivalent to the sum over all Lorentzian metrics \cite{Hawking:1981gb}. However, the Euclidean action suffers of the conformal factor problem discussed in section \ref{sec:conffactor} which makes the path integral \eqref{hawk} diverge. The way out suggested by Hartle and Hawking was then to ensure the convergence of the integral with a suitable choice of complex contour of integration. However, different convergent complex contours give different wavefunctions. At the semi-classical level, each such contour can be peaked around a different saddle point geometry providing inequivalent semi-classical descriptions of our universe. Thus the prescription that the wavefunction of the universe should be given by a sum over compact geometries is incomplete as it gives different wave functions for different choices of integration contour. Various contours of integration for the no-boundary proposal have been analysed over the years in a large number of papers (see for example \cite{Louko:1988bk, Halliwell:1988ik, Halliwell:1989dy}). A breakthrough in the field happened thanks to Ref.  \cite{Feldbrugge:2017kzv} were the authors where able to demonstrate that the Lorentzian no-boundary path integral indeed converges and, surprisingly enough, does not give the desired answer. In what follows we will present the results of \cite{Feldbrugge:2017kzv} as reviewed in \cite{DiTucci:2019dji}.\\

The full no-boundary path integral \eqref{hawk} cannot be evaluated explicitly in full generality and it is normally studied within the minisuperspace approximation where the geometries considered are of the FLRW form already introduced in eq. \eqref{eq.nbgeometry}
\begin{equation}
ds^2 = - \frac{N^2 }{q(t)} dt^2 + q(t) d \Omega_3^2 \label{FLRW}
\end{equation}  
  with $ d \Omega_3^2$ representing the line element of a three sphere of volume $V_3$ and $t \in [0,1]$.  The function $q(t)$ is the squared scale factor which determines physical lengths in the universe, while $N(t)$ is the lapse function. As mentioned already in section \ref{sec:causality} the range of the lapse integration determines whether the path integral represents a propagator or a wavefunction. In \cite{Feldbrugge:2017kzv}, the authors study the propagator, corresponding to an integration over the positive real $N$ line, and so we will do here but it is important to stress that, qualitatively, the negative result we are going to present applies to both cases.
  The propagator for the no boundary proposal $G[q_1 , 0]$ describes a transition from a spatial 3-geometry of zero size, $q_0 = 0$, to another one which lies in the future of the first and where the scale factor takes the value $q_1 \in \mathcal{R}^+$. The propagator can be evaluated as a path integral, i.e. as a sum over 4-geometries which interpolate the two 3-geometries mentioned above. Each such 4-geometry enters in the sum with a weighting $e^{iS}$ given by Einstein-Hilbert action with a cosmological constant $\Lambda=3H^2$. The various steps needed to define the gravitational path integral can be found in \cite{Teitelboim:1981ua} and \cite{Halliwell:1988wc}. We report here only the final result which in constant lapse gauge $\dot{N} = 0 $ reads 
  \begin{equation}
  G[q_1 , 0 ] = \int_{0^+}^\infty dN \, \int_{q(t = 0)= 0}^{q(t = 1) = q_1} \delta q \, e^{i S / \hbar}  \label{prop}
  \end{equation}
where the action takes form (with $8 \pi G =1$)   
  \begin{equation}
  S = \int d^4 x \sqrt{-g} \left( \frac{R}{2} - \Lambda \right) = V_3 \int_0^1 dt \, [ - \frac{3 \dot{q}^2}{4 N} + 3 N( 1 - H^2 q)] \label{S}
  \end{equation}
The functional integral is evaluated over functions $q(t)$ which take the value $q_1$ at $t=1$ and $q_0 = 0$ at $t= 0 $, for a fixed value of $N$. The integral over the lapse $N$ is an ordinary one due to the gauge fixing condition $\dot{N} = 0 $.
 The lapse is in fact a real variable which takes values along the positive real line $N \in (0^+ , \infty)$. This ensures that the geometries in the sum have a Lorentzian signature. The considered domain of integration makes the path integral a propagator in the sense that it solves the inhomogenous Wheleer-deWitt equation $\hat{\mathcal{H}} G[q_1  , 0 ] = - i \delta(q_1)$ where $\hat{\mathcal{H}} $ is the quantum Hamiltonian.
 
It has been shown in \cite{Feldbrugge:2017kzv} that the result of the various integrations to leading order in the saddle point approximation is 
 \begin{equation}
 G[q_1 , 0] = e^{- \frac{2 V_3}{H^2 \hbar} - i \frac{2 V_3 H}{\hbar} (q_1 - \frac{1}{H^2})^{3/2}} \label{GGG}
\end{equation}  
where it is assumed that $q_1 > \frac{1}{H^2}$. The obtained negative weighting $ e^{- \frac{2 V_3}{H^2 \hbar} } $ is characteristic of the tunnelling proposal \cite{Vilenkin:1982de}, whereas the conjectured no-boundary result (at the same level of approximation) is \cite{Hartle:1983ai}
\begin{equation}
\Psi (q_1) =  e^{+\frac{2 V_3}{H^2 \hbar}} \cos[  \frac{2 V_3 H}{\hbar} (q_1 - \frac{1}{H^2})^{3/2}]\,. \label{hh}
\end{equation}  

The meaning of this result together with the difference between the two proposals may be elucidated by considering the saddle points of the full path integral \eqref{prop}. In fact, there are four saddle points, with geometries given by
\begin{equation}
q(t)= H^2 N^2 t (t - 1) + q_1 t \label{q}
\end{equation}
with the lapse taking the values
\begin{equation}
N_{c_1,c_2} = c_1 \frac{\sqrt{H^2 q_1  - 1} + c_2 i}{H^2} \label{saddle}
\end{equation}
for $c_1 , c_2 \in \{ -1 , 1 \}$. \\
We recognize that the HH geometry \eqref{eq.nbgeometry} corresponds to $c_1 = 1, c_2 = -1$. The geometry corresponding to $c_1 = - 1, c_2= 1$ can be interpreted as the time reversed of the HH geometry. The no-boundary wave function is often required to be peaked around both the HH geometry and its time reversed so that it becomes the real function of the three-sphere of radius $q_1$ given in eq. \eqref{hh}. The complex conjugate geometries, associated with the complex conjugate saddle points $c_1 = 1, c_2 = - 1$ and $c_1 =- 1, c_2 = 1$, give the Vilenkin geometries. Coming back to the representation of these geometries as the gluing of a Lorentzian~\eqref{eq.dS} with a Euclidean~\eqref{eq.S4} section, the Vilenkin geometry amounts to reaching Euclidean time with a opposite Wick-rotation  $t_p = + i \left(\eta - \frac{\pi}{2H}\right)$. Indeed the Hartle-Hawking ($N_{HH}$) and Vilenkin geometries ($N_V$) correspond to opposite signs of initial Euclidean momentum $p := \frac{\partial \mathcal{L}}{\partial \dot{q}}$
\begin{align}
p_{HH}^0=\frac{3 V_3 \, \dot{q}}{2N_{HH}}\mid_{t=0} = + 3 V_3 \, i \,,\qquad p_{V}^0=\frac{3 V_3 \, \dot{q}}{2N_{V}}\mid_{t=0} = - 3 V_3 \, i\,, \label{initialmomentum}
\end{align}

The action evaluated on those solutions is
\begin{equation}
S(N_{c_1 , c_2}) = c_1 \frac{2 V_3}{H^2} [ c_2 i - (H^2 q_1 - 1)^{3/2}]
\end{equation}
From this expression we see that the propagator \eqref{GGG} is given by the contribution of the saddle point with $c_1 =1$ and $c_2 = 1,$ in the upper right quadrant of the complex N plane. It is in fact possible to show, applying Picard-Lefschetz theory, that the integral along the positive real $N$ line is equivalent to the integral along the complex steepest descent path (``thimble'') running through this saddle point alone. As mentioned earlier, this saddle point  is however unstable against inhomogeneous perturbations around the FLRW background (see \cite{Feldbrugge:2017fcc} and \cite{Halliwell:1989dy}). Thus the result \eqref{G}, although mathematically correct, cannot describe our universe on physical grounds. 

The no-boundary result \eqref{hh} instead would have been obtained by considering the contribution of the two saddle points in the lower half of the complex $N$ plane corresponding to $c_1 = 1$, $c_2= -1$ and $c_1 = -1$, $c_2= 1$. There is however no convergent contour which can be deformed into a steepest descent path running through solely these two saddle points \cite{Feldbrugge:2017mbc}. In this sense, the wavefunction \eqref{hh} is not the saddle point approximation of the no boundary wavefunction for any Lorentzian path integral with these boundary conditions. With different boundary conditions, the situation may change, as we will discuss later.

To conclude this section, we would like to highlight that these results imply that the two definitions that "the no-boundary wave function is peaked around the Hartle-Hawking geometry" and "the no-boundary wavefunction is given by a sum over compact geometries" are in fact not compatible, the latter giving a wavefunction peaked around the Vilenkin geometry and thus representing the Vilenkin tunneling proposal. This wavefunction corresponds however to bad behaved perturbations. Notice that the issue with this definition is twofold. On the one hand, it predicts that larger and larger perturbations are quantum mechanically favored, in obvious disagreement with the CMB observations. On the other hand, at a more technical level, this conclusion undermines the assumptions of the computation itself. The full gravitational path integral \eqref{hawk} shall be given by the sum over all geometries which satisfy the imposed boundary conditions. Since the integral cannot be evaluated explicitly in full generality, we restricted the sum to the subset of highly symmetric geometries of FLRW type. The idea is that since our universe is indeed in first approximation homogeneous and isotropic, not much fundamental information is lost with this reduction. If however inhomogeneous perturbations turn out to be unsuppressed, the validity of the approximation itself gets into trouble. These facts lead to the claims of \cite{Feldbrugge:2017fcc, Feldbrugge:2017mbc, Feldbrugge:2018gin} that the no-boundary proposal is ill-defined and not tenable as a theory of initial conditions of the universe. In this work we will take however a different viewpoint of this issue. We will take as the definition of the Hartle-Hawking wavefunction that it must be peaked around the Hartle-Hawking geometry, with all the associated desired properties, and ask the questions: there exists a suitable path integral representation of the Hartle-Hawking wavefunction within the minisuperspace approximation? If so, what is the interpretation associated with this representation? As we will see, it is indeed possible to define the Hartle-Hawking wavefunction within the minisuperspace approximation imposing suitable (non Dirichlet) boundary conditions. These findings will however change drastically the status of the no-boundary proposal.

\clearpage

%%%%%%%%%%%%%%%%%%%%%%%%%%%%%%%%%%%%%%%%%%%%%%%%%%%%%%%%%%%%%%%%%%%%%%%%%%%%%%%%%%%%

\subsection{The no-boudary proposal as a sum over regular metrics}\label{sec:regmet}

The results shown in the previous section show that no-boundary proposal thought as a sum over compact geometries leads to unstable fluctuations, in that small fluctuations obey an inverse Gaussian distribution \cite{Feldbrugge:2017fcc}. This seemed to imply that the no-boundary proposal leads to unphysical results, and must be abandoned as a possible initial condition for the universe. We will show in chapter \ref{robincosmology} e \ref{chapterneumann} that a physically meaningful path integral definition of the Hartle-Hawking wavefunction can be given implementing different types of boundary conditions which however hint at a substantially different interpretation of the wavefunction itself.

For the moment will pursue a different avenue with the goal of strengthening the results presented above: in the minisuperspace calculations just mentioned, the no-boundary condition is imposed as the condition that the universe should start out at zero size. Then it was found that the saddle points of the integral are all complex, and can be represented as half of a Euclidean 4-sphere glued onto half of a Lorentzian de Sitter hyperboloid. Thus the saddle point geometries are indeed regular, with the locus of zero scale factor simply being a regular point on a sphere. However, the minisuperspace integral itself contains many off-shell geometries that are singular where the universe has zero size. Could it be that the instability of the fluctuations is due to these singular off-shell geometries? Should one sum only over purely regular geometries?  On the one hand, it seems unlikely that the instability will disappear, since the relevant saddle point of the Lorentzian path integral is regular, yet also unstable. On the other hand, it seems worthwhile to investigate a restricted sum over purely regular geometries, since this seems to be more closely in the spirit of the original formulation of the no-boundary proposal \cite{Hawking:1981gb,Hartle:1983ai}. 

\begin{figure}
\begin{center}
\includegraphics[scale=0.4]{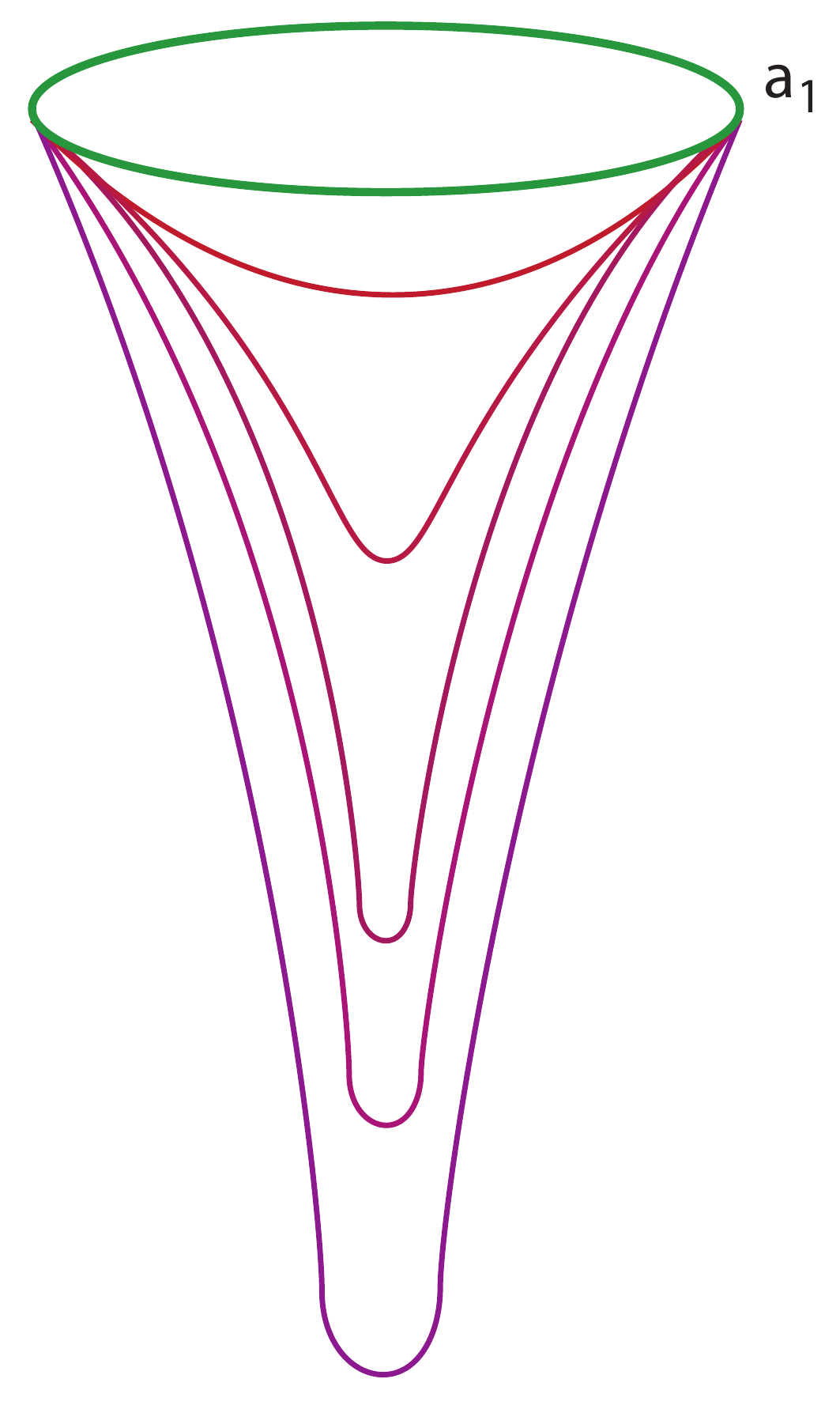}
\caption{We will consider a path integral over purely regular geometries, having as their only boundary a large late time universe with scale factor $a_1.$ The integral can be pictured as a sum over complexified 4-spheres, bearing in mind that a complexified 4-sphere contains a Lorentzian de Sitter section. One may argue that such a restricted sum is closest in spirit to the original idea of the no-boundary proposal \cite{Hawking:1981gb,Hartle:1983ai}.}
\label{fig:regular}
\end{center}
\end{figure}

Thus in the rest of this chapter, based on \cite{DL}, we will consider a sum over regular geometries, as pictured in Fig. \ref{fig:regular}. Note that this is a highly restricted class of 4-geometries, as gravity has a tendency of causing collapse to a singularity. We will implement this sum over regular geometries by taking as our starting point a paper by Halliwell and Louko \cite{Halliwell:1989vu}, where they considered an integral over complexified 4-spheres of various radii. We will extend their work both for the background in section \ref{sec:bgd} (it turns out that they only considered half of all such 4-sphere solutions)  and then by adding perturbations in section \ref{sec:perts}. In the minisuperspace calculations of \cite{Feldbrugge:2017kzv}, the path integral contained separate integrals over the scale factor of the universe $r$ and over the lapse function $N$. The present setting is so restrictive that these two integrals get combined into a single ordinary integral over a variable $z$ which depends on both $r$ and $N$. One consequence of this is that it becomes impossible to define a purely Lorentzian integral, but we will discuss how one can define an integration contour that corresponds to the Lorentzian path integral as closely as possible, and we will also discuss other choices of contour. 

Going beyond a treatment of the background, we find that the action for the perturbations contains poles in the complex $z$ plane (where $z$ is the variable combining scale factor and lapse), with the location of the poles being wavenumber-dependent. This imposes additional restrictions on possible integration contours for the $z$ integral, as the contour must remain sensible and well defined for all possible choices of fluctuations. Here we will show that the analogue of the Lorentzian integral remains well defined after adding perturbations, but that a closed contour for instance would be adversely affected.

Our final result of the perturbative analysis is that the saddle points that were unstable in the minisuperspace setting remain unstable here, though now in a context where every summed over geometry is regular. Moreover, it is straightforward to verify in the present context under what conditions the backreaction of the perturbations remains small. Our conclusion will be that the instability of the no-boundary proposal is robust and in full agreement with the minisuperspace results.

\subsection{Background} \label{sec:bgd}

We are interested in evaluating the (suitably gauge fixed \cite{Teitelboim:1981ua,Teitelboim:1983fk}) path integral
\begin{align}
G[a_1;0]=\int^{a_1} {\cal D} g \, e^{iS/\hbar}
\end{align}
where $S=\frac{1}{2}\int d^4x \sqrt{-g}\left( R - 2 \Lambda \right)+   \frac{c^3}{8 \pi G} \int_{\delta M} d^3y \, \sqrt{h} K$  is the action for gravity plus a positive cosmological constant $\Lambda$ equipped with the GHY boundary term. The integral should be over metrics which have as their only boundary a final 3-surface with scale factor $a_1.$ A spatially homogeneous, isotropic and closed metric can be written as 
\begin{equation}
ds^2 = -N_L^2 dt^2 + a(t)^2 d \Omega_3^2
\end{equation}
where $d \Omega_3^2 $  is the unit 3-sphere with volume $2\pi^2$ and we will scale the time coordinate such that $0 \leq t \leq 1$. In light of the results of \cite{Feldbrugge:2017kzv,Feldbrugge:2017mbc}, we are interested in a sum over four metrics with Lorentzian signature ($N_L$ real). However, in order to sum over regular (though complexified) four spheres we will use coordinates with Euclidean signature, as this choice greatly facilitates the imposition of the no-boundary condition. Note that this is simply a coordinate choice (and since we will consider complexified metrics anyway this choice is truly arbitrary) -- the requirement that the path integral should be the Lorentzian one will then be implemented as a restriction on the possible integration contours. With the metric above, the action reads
\begin{equation}
S = 2 \pi^2 \int_0^1 N_L dt \left[ - 3 \frac{a \dot{a}^2}{N_L^2} - \Lambda a^3 + 3 a \right]\,,
\end{equation}
or, in terms of the Euclidean lapse $N_L = - i N_E$, 
\begin{equation}
S =2 \pi^2 i \int_0^1 N_E dt \left[ - 3 \frac{a \dot{a}^2}{N_E^2} + \Lambda a^3 - 3 a \right]\,.
\end{equation}
The  associated constraint then reads
\begin{equation}
3 \frac{\dot{a}^2}{a^2} - \frac{3 N_E^2}{a^2} + \Lambda N_E^2 = 0 
\end{equation}
and, with the boundary conditions $a(t=0)=0$ and $a(t=1)=a_1$ its solution is of the form 
\begin{equation}
a(t)= \pm \sqrt{\frac{3}{\Lambda}} \sin\left(\sqrt{\frac{\Lambda}{3}} N_E t\right) \label{asaddle}
\end{equation}
provided we choose $N_E$ such that $a(t = 1) = \pm \sqrt{\frac{3}{\Lambda}} \sin \left(\sqrt{\frac{\Lambda}{3}} N_E \right) = a_1 $. For both choices of sign, this metric then represents a 4-sphere of radius $\sqrt{3/\Lambda},$ which in general will be complexified since $N_E$ will take complex values whenever $a_1 > \sqrt{3/\Lambda}$. 

In the path integral we are not directly interested in the solutions to the equations of motion or the constraints, rather we first want to state the class of metrics that are to be summed over. Following \cite{Halliwell:1989vu} we will consider the simplest and most symmetric possibility, namely we will sum over 4-spheres with given  boundaries $a_0=0$ and $a_1 >0$ and arbitrary radius, 
\begin{equation}
a(t)= \pm r \sin \left(\frac{N_E \, t}{r}\right)  \label{eq:summedmetrics}
\end{equation}
with $a_1 = \pm r \sin \left( \frac{N_E}{r}\right)$ and $\dot{a}_1 = \pm N_E \cos \left( \frac{N_E}{r}\right)$. Accordingly, one should think of $N_E$ as being fixed by the boundary conditions and the sum to be over $r$. Given that $N_E$ will in general be a complex number, we should also expect $r$ to be complex, and that the integral will be over a contour in the complex $r$ plane. We note that in \cite{Halliwell:1989vu} only one choice of sign in \eqref{eq:summedmetrics} was considered. As we will see below, it is important to keep both signs at first, and then we will see how the sign may be fixed according to the integration contour chosen.

The action for the positive/negative choice of sign in \eqref{eq:summedmetrics} reads respectively
\begin{equation}
S = \pm \frac{i}{3 N_E } \left( 3 r^2 \left[ - 4 + 3 \cos \left(\frac{N_E}{r}\right)+ \cos^3 \left(\frac{N_E}{r}\right)\right] +  \Lambda r^4 \left[ 2 - 3 \cos \left(\frac{N_E}{r}\right)+ \cos^3 \left(\frac{N_E}{r}\right)\right]  \right) \label{eq2}
\end{equation}
It is possible to simplify the analysis by defining a new variable $z$ such that 
\begin{equation}
z = 1 + \cos \left( \frac{N_E}{r} \right) = 1 \pm \frac{\dot{a}_1}{N_E}\,, \label{z}
\end{equation}
implying the useful relation
\begin{align}
r^2 = \frac{a_1^2}{z(2 -z)}\,.
\end{align}
The action given by Eq. (\ref{eq2}) then reads in this variable
\begin{equation}
S = \pm 2 \pi^2 i a_1^2 \left[-z+ 1  + \left(\frac{a_1^2 \Lambda}{3}  - 4 \right)\frac{1}{z} + \frac{\Lambda}{3} \frac{a_1^2}{z^2} \right]\,,
\end{equation}
which diverges for $z \rightarrow 0 $ and $z \rightarrow \infty$.

Depending on the argument of $z $ the action diverges to $+ i \infty$ or $- i \infty$. This divides the complex $z$ plane into regions of convergence and divergence of the path integral $\int dz \, e^{ i S}$ (we will take the simplest measure in $z$, since we will only be interested in the leading terms in $\hbar$).  For $|z| \rightarrow \infty$, $ S \approx \mp i z \equiv \mp i R e^{i \theta}$, therefore the integrand $e^{ i S}$ diverges or vanishes in this limit depending on $Arg(z)$. For the choice $a = - r \sin ( \frac{N t}{r}),$ for instance, we have the limits
\begin{align}
\lim_{R \rightarrow \infty } e^{ R \cos \theta } e^{ i R \sin \theta} = \infty \; \; \; \; \; \; & \mbox{ for     }  \; \;  \frac{\pi }{2} < \theta < \frac{3 \pi }{2} \\
\lim_{R \rightarrow \infty } e^{ R \cos \theta } e^{ i R \sin \theta} = 0 \; \; \; \; & \mbox{ for } \; \;\;   - \frac{\pi }{2} < \theta < \frac{ \pi }{2}
\end{align}
Therefore, as $z$ goes to infinity along a direction exactly parallel to the imaginary line, the integrand $e^{iS}$ is purely oscillating. As soon as it slightly deviates from that direction, the path integral is either convergent or divergent. For $|z| \rightarrow 0,$ the action can be approximated by $S \approx \pm \frac{i}{z^2} = \pm \frac{i}{R^2} e^{ - 2 i\theta}.$ Thus the convergence regions in the small $z$ limit are the wedges $- \frac{\pi}{4} < \theta < \frac{\pi}{4}$ and $- \frac{3 \pi}{4} < \theta < \frac{5 \pi}{4}$ for $ a = + r \sin \left( \frac{N t}{r} \right)$ and respectively $\frac{\pi}{4}<  \theta < \frac{ 3 \pi}{4}$ and $\frac{5 \pi}{4}<  \theta < \frac{ 7 \pi}{4}$ for $a = - r \sin \left( \frac{N t}{r} \right)$.

The three saddle points for each sign of the action are located at 
\begin{align}
z_2 &= - 2 \\
z_{1,3} &= 1 \pm i \sqrt{\frac{a_1^2 \Lambda}{3} - 1} \,,
\end{align}
while the action at the saddle points is given by (where the signs are correlated with \eqref{eq:summedmetrics})
\begin{align}
S(z_2) &= \pm 2 \pi^2 i \, a_1^2 \left[ 5 - \frac{a_1^2 \Lambda}{12}\right]\,, \label{s1} \\
S(z_1) & = \mp \frac{ 12 \pi^2}{\Lambda} \left[i - \left(\frac{a_1^2 \Lambda}{3} -1 \right)^{3/2}\right]\,, \label{s2} \\
S(z_3) & = \mp \frac{ 12 \pi^2}{\Lambda} \left[i + \left(\frac{a_1^2 \Lambda}{3} - 1 \right)^{3/2}\right]\,. \label{s3}
\end{align}

\begin{figure}
\begin{center}
\includegraphics[width=0.4\linewidth]{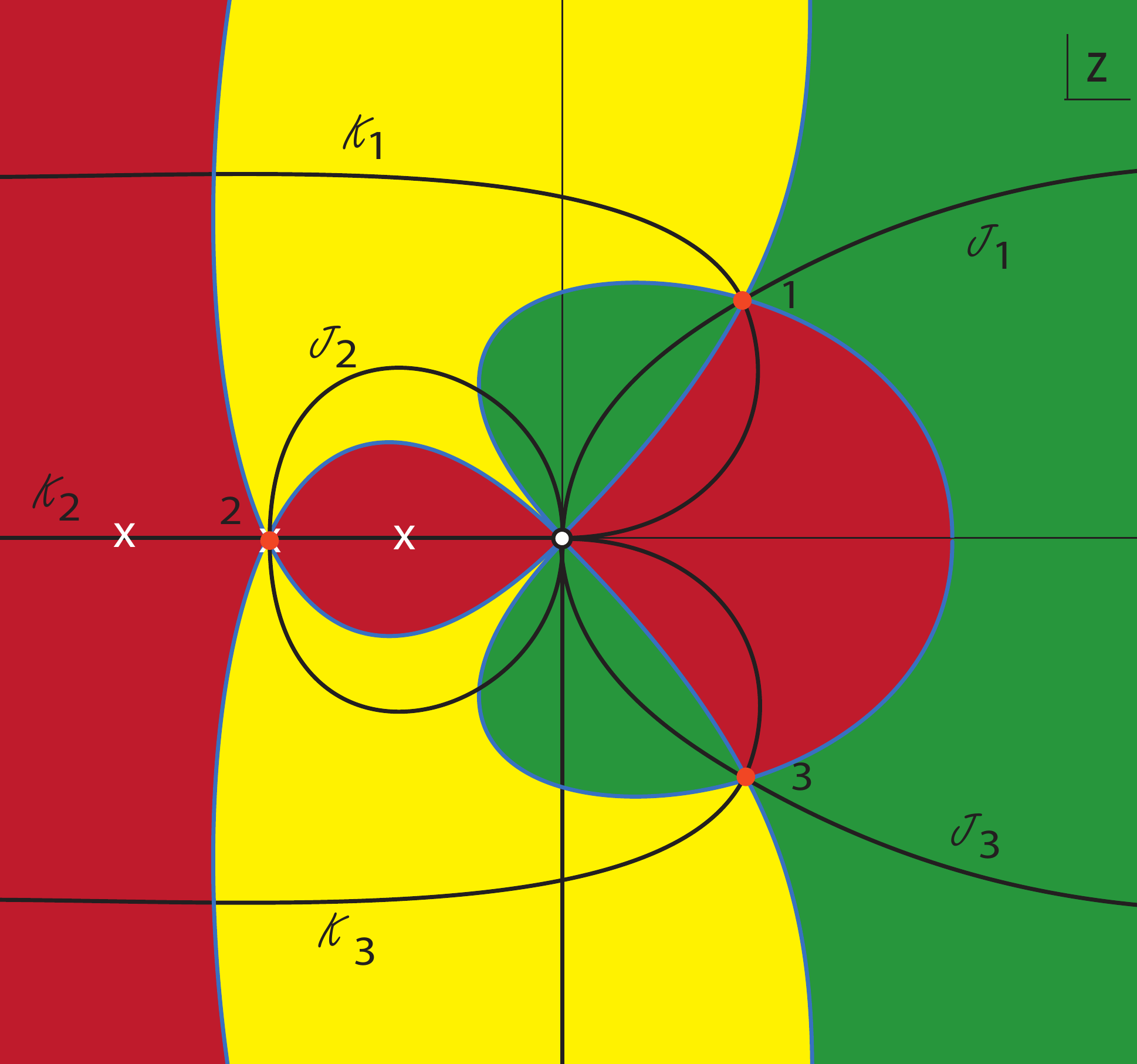}
\hspace{1cm}
\includegraphics[width=0.4\linewidth]{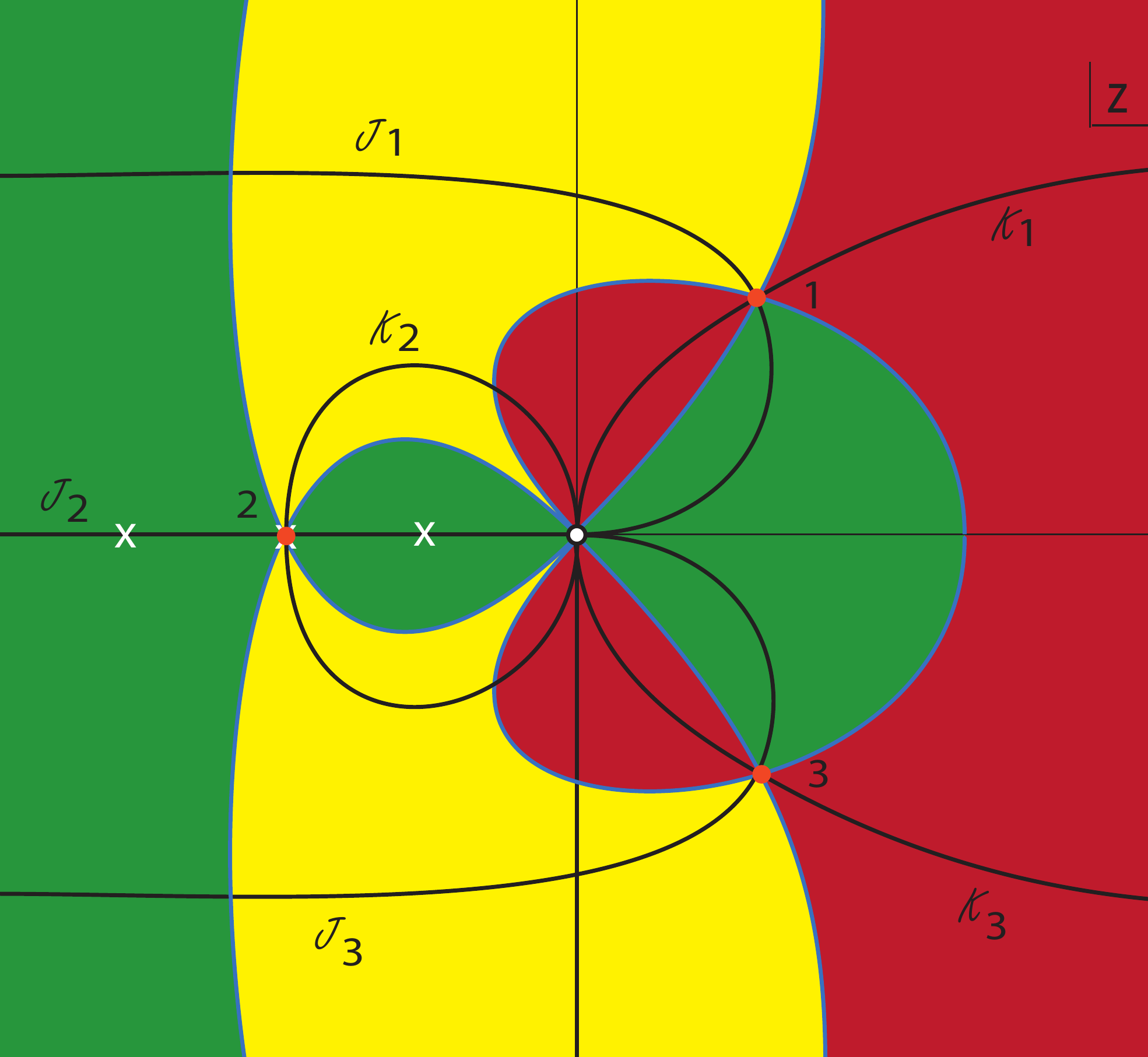}
\caption{The figure shows the qualitative structure of the Morse function in the complex $z$ plane, for boundary conditions  $1 < \frac{a_1^2 \Lambda}{3} \lessapprox 4.51.$ The left panel corresponds to $a(t) = - r \sin \left(\frac{N t}{r} \right)$, the right panel to the opposite choice of sign.  The flow lines are shown for the background. We will see that, when perturbations are added, poles  arise in the action (marked by white crosses here). A full description is provided in the main text.}
\label{fig:flows}
\end{center}
\end{figure}

\begin{figure}
\begin{center}
\includegraphics[width=0.4\linewidth]{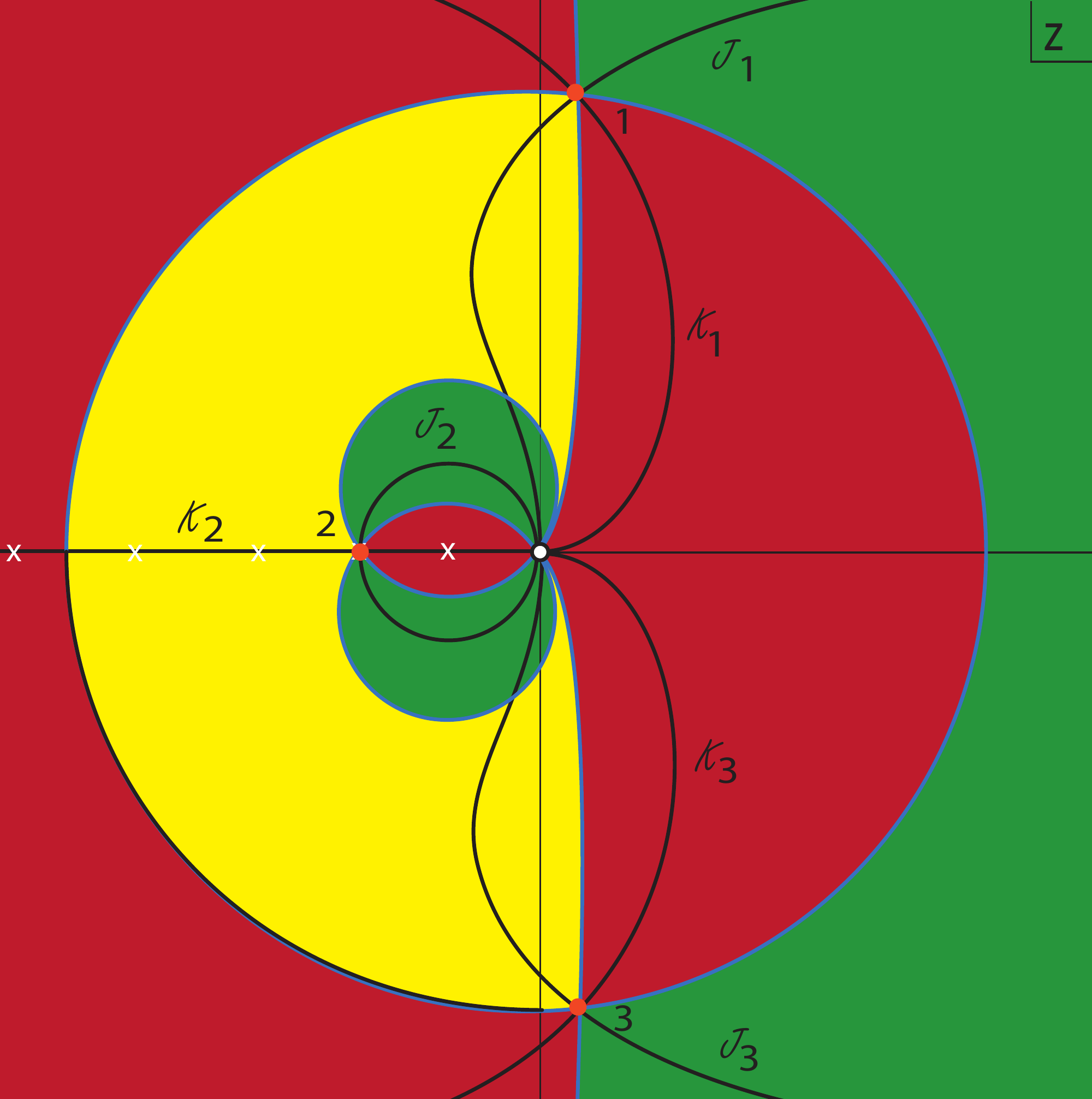}
\hspace{1cm}
\includegraphics[width=0.4\linewidth]{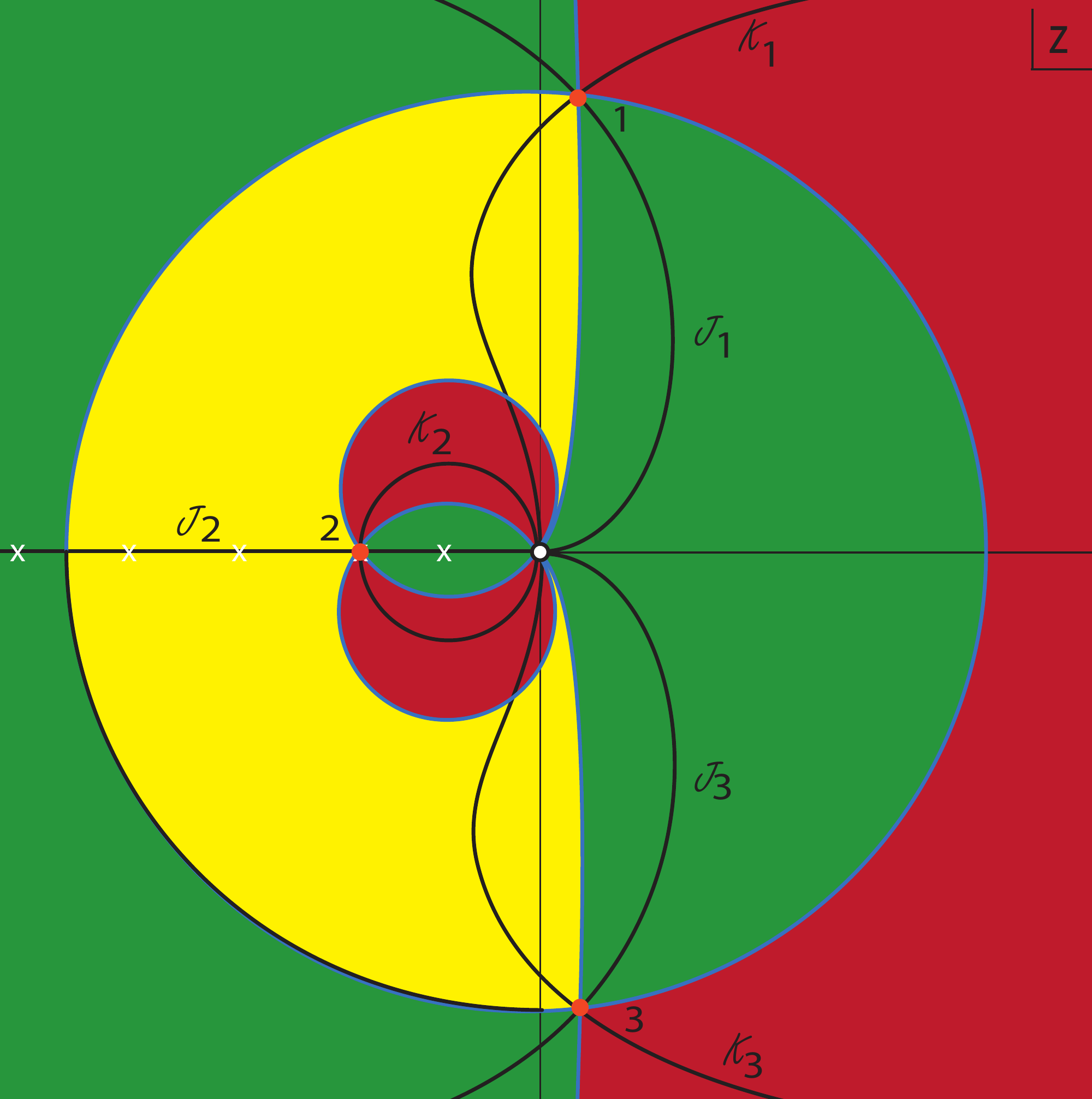}
\caption{The figure shows the qualitative structure of the Morse function in the complex $z$ plane, this time for boundary conditions  $\frac{a_1^2 \Lambda}{3} \gtrapprox 4.51.$  The left panel corresponds to $a(t) = - r \sin \left(\frac{N t}{r} \right)$, the right panel to the opposite choice of sign.  The flow lines are shown for the background and the locations of the poles, which arise when perturbations are added, are marked by white crosses here.}
\label{fig:flows2}
\end{center}
\end{figure}

The four saddle points at $z_{1,3}$ are those also seen in the minisuperspace calculation \cite{Feldbrugge:2017kzv}. These saddle points can be pictured as half of a 4-sphere glued onto half of the de Sitter hyperboloid, with a radius $r$ determined by the cosmological constant, $r^2 = \frac{3}{\Lambda},$ thus they are all four complex solutions to the Einstein equations. They differ in the way that the implied Wick rotation from the de Sitter geometry to the sphere is implemented, while two of the saddle points are the time reverses of the other two. In the present context these four saddle points arise for two possible sign choices for the complexified scale factor, whereas the minisuperspace calculation already includes a sum over both possible choices. By contrast, the saddle points at $z=-2$ are of a different character. As already discussed by Halliwell and Louko \cite{Halliwell:1989vu}, these are spurious solutions which do not satisfy the Einstein equations. Moreover, they do not lead to classical evolution, as they do not yield a phase in the exponent $e^{iS/\hbar}.$ 

Note that the position of the saddle points is the same for the two choices of sign for $a(t)$. However the value of the action is the opposite. As a consequence, the flow lines are the same, with exchanged roles of the steepest descent and ascent paths. This has important consequences for Picard-Lefschetz theory and the choice of integration contour. The locations of the saddle points in the complex $z$ plane, along with the paths of steepest ascent and descent emanating from them, are shown in Figs. \ref{fig:flows} (for small values of the final scale factor $1 < \frac{a_1^2 \Lambda}{3} \lessapprox 4.51$) and and \ref{fig:flows2} (for larger values of $a_1$).

We will now describe these figures -- for more details about the general procedure see \cite{Feldbrugge:2017kzv,Feldbrugge:2017mbc}. Figs. \ref{fig:flows} and  \ref{fig:flows2} show the qualitative behaviour of the Morse function, defined as the magnitude of the integrand. More specifically, one rewrites the integrand, now seen as a holomorphic function of the fields, as $e^{iS/\hbar} \equiv e^{h+iH},$ where $h, H$ are real functions. The Morse function $h$ then determines the amplitude of the integrand, while $H$ describes the phase. Critical points (in fact saddle points) of $h$ are also critical points of the total action, and the lines which have the same phase as that of a saddle point are the lines of steepest ascent/descent from that saddle point. Along these lines the Morse function changes most rapidly, and monotonically, away from the saddle points. The saddle points are marked by orange dots, the steepest ascent (${\cal K}$) and descent (${\cal J}$) lines are black, while the blue lines have the same value of the Morse function as the saddle points which they link up to. The green regions have a lower value of the Morse function than the adjacent saddle point, while the red regions have a higher Morse function. Yellow regions have a value of the Morse function in between the values at the two adjacent saddle points. The action has an essential singularity at $z=0.$ Approaching this within a green region thus implies a converging integral, while approaching it in a red region leads to divergence. Picard-Lefschetz theory aims to replace a highly oscillating integral by an equivalent absolutely convergent one along lines of steepest descent, if possible. The oscillating integral involves many cancellations due to the oscillations, while along a steepest descent line no such cancellations occur. Thus the Morse function along the steepest descent path must be lower than along the original integration contour. In other words, a steepest descent path (also called Lefschetz thimble) is relevant to the integral only of it can be reached by flowing the original integration contour down towards it. Equipped with these tools, we can discuss possible integration contours.

\begin{figure}
\begin{center}
\includegraphics[width=0.3\linewidth]{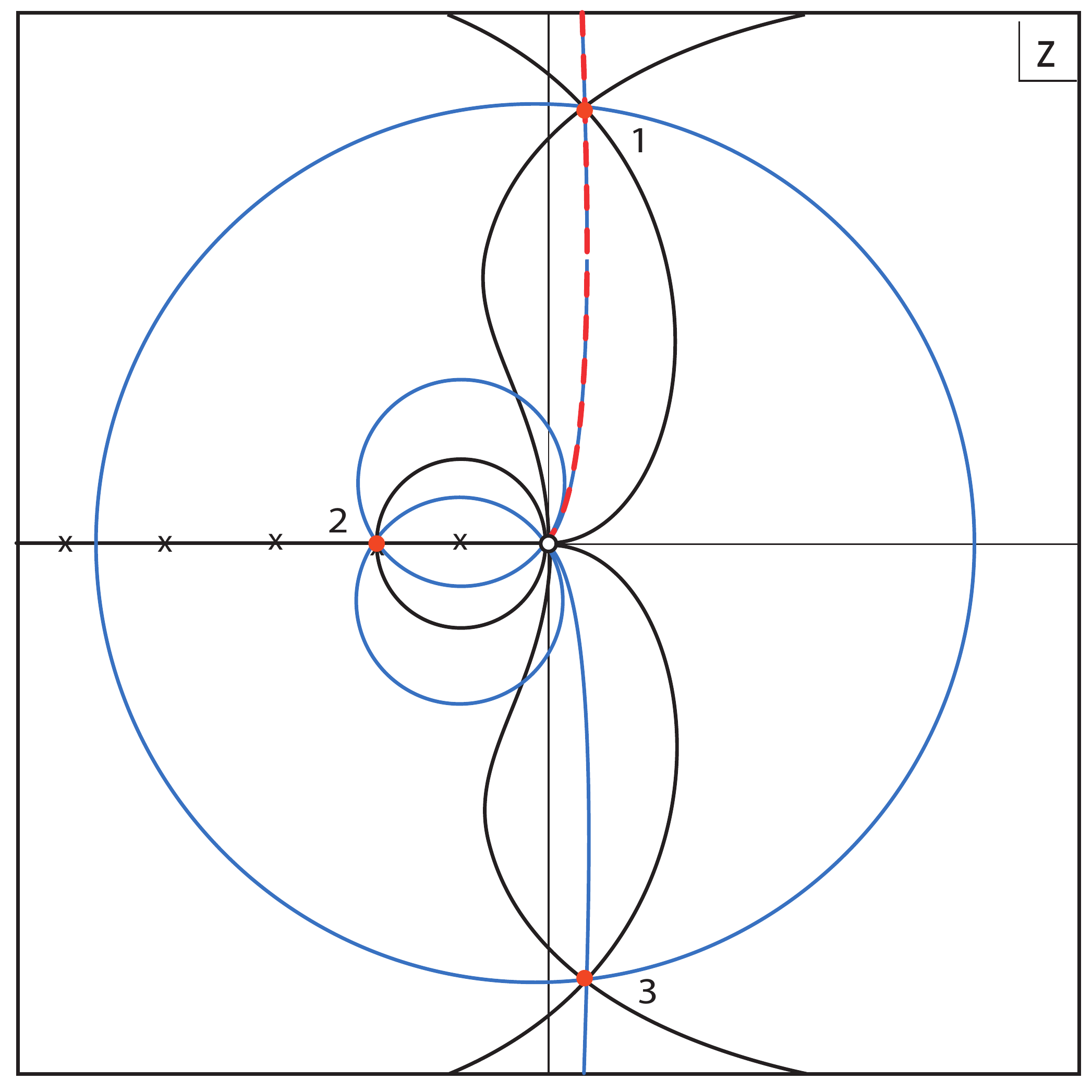}
\includegraphics[width=0.3\linewidth]{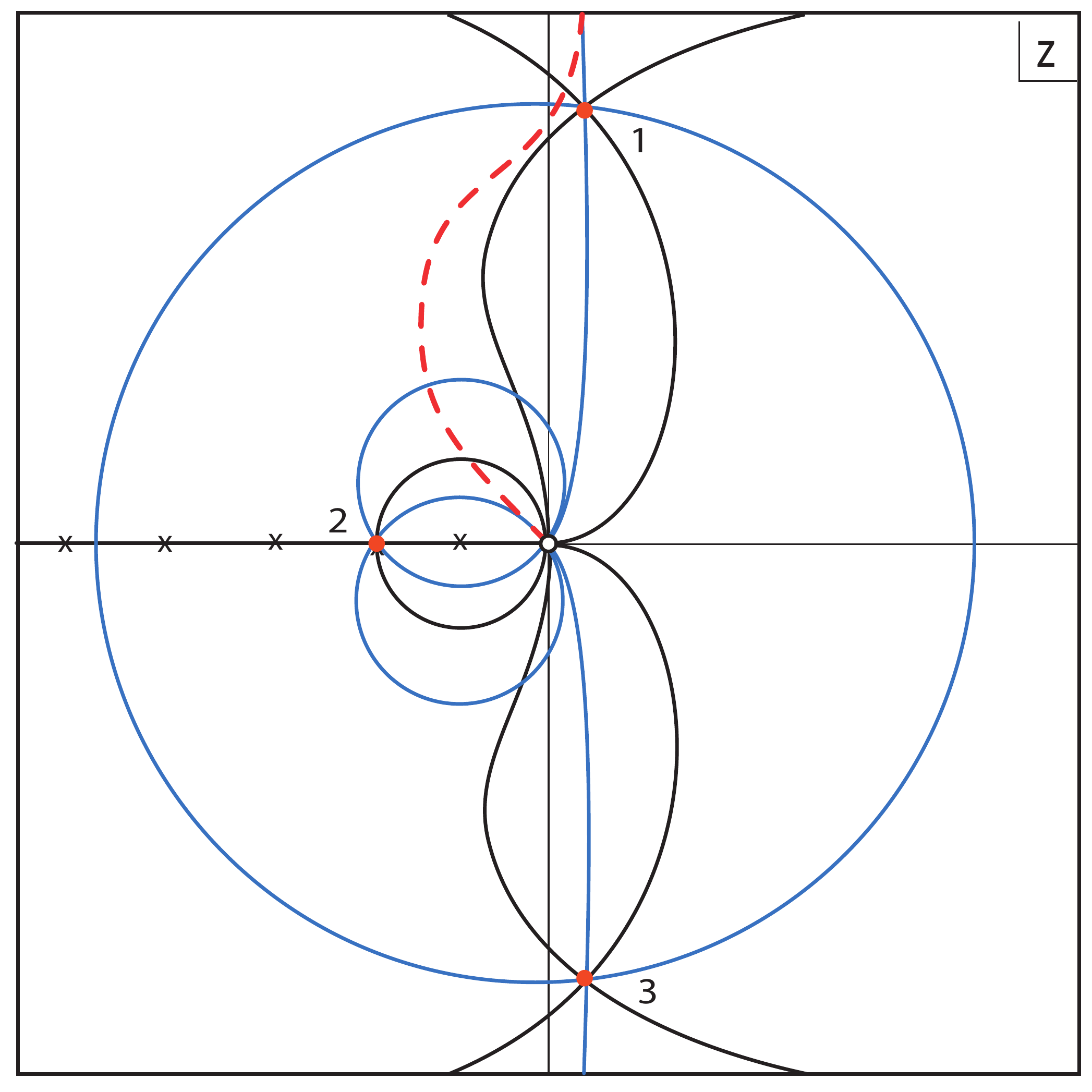}
\includegraphics[width=0.3\linewidth]{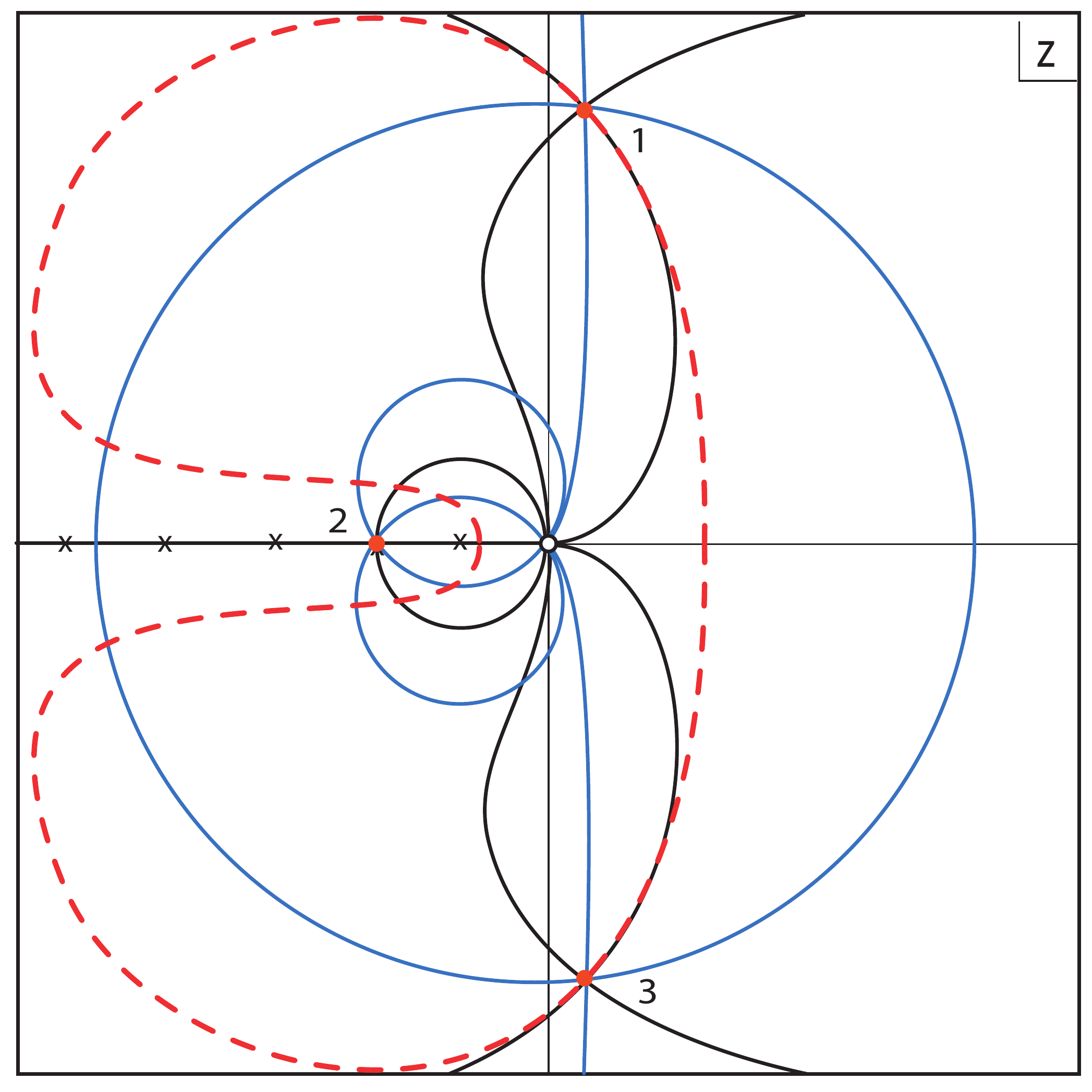}
\caption{Possible contours of integration are shown by the dashed orange lines. The figures here correspond to the physically most relevant boundary conditions where $\frac{a_1^2 \Lambda}{3} \gtrapprox 4.51.$ Analogous contours also exist for smaller values of $a_1.$}
\label{fig:contours}
\end{center}
\end{figure}

But before doing so, we should point out an important difference with minisuperspace calculations: the restricted version of the path integral considered here is in fact not straightforwardly related to the actual gravitational path integral. There, one defines the Lorentzian path integral so as to include a sum over all possible (real and positive) values of the lapse function $N_L$. This results in a Green function of the Wheeler-deWitt (WdW) operator which possesses a well defined underlying causal structure \cite{Feldbrugge:2017kzv,Feldbrugge:2017mbc}. By contrast, the object constructed as a sum over regular metrics is not explicitly related to the WdW equation, as there is no explicit integration over the lapse function. As a consequence it is not obvious in this case what the appropriate integration contour should be.  In what follows we will define an appropriate integration contour by taking guidance from the gravitational path integral in order to identify desirable properties. First, the integration contour should run between singularities of the action. This implements the idea of summing over all possible configurations, since there is no reason to end the contour at any particular configuration along a path. Also, the convergence of the path integral with a finite boundary configuration highly depends on the measure. Since the measure is not uniquely defined here, we will rather consider endpoints where the integrand is exponentially suppressed, $iS \rightarrow - \infty$. Secondly, we will try to define an analogue of a Lorentzian contour. Indeed there is no contour in $z$ for which the 4-geometries have a purely Lorentzian signature. However, it is possible to require that the metrics become Lorentzian at late times, near the final boundary. This is equivalent to imposing that the velocity on the final boundary must be purely imaginary.  One can see this by rewriting the metric near the final boundary as
\begin{equation}
ds^2 = N_E^2 dt^2 + a(t)^2 d \Omega_3^2 = N_E^2 \left( dt^2 + \frac{a(t)^2}{N_E^2 \Delta t^2} \Delta t^2 d \Omega_3^2 \right)
\end{equation}
Indeed this line element has a Lorentzian signature near the final boundary if $\frac{\dot{a}_1}{N_E}$ is purely imaginary. According to the definition of $z,$ eq. (\ref{z}), this requirement singles out the line $z = 1 + i s $ with $s$ real. Since opposite values of $s$ correspond to opposite final velocities, it is physically meaningful to restrict ourselves to the upper half complex $z$ plane, i.e. $s >0$. In fact, $s = 0 $ is a branch point for the map $r \rightarrow z$ (given by $z-1= \pm \sqrt{1-\frac{a_1^2}{r^2}}$) which must not be crossed for a proper definition of the $z$ variable. The line $1+i s$ can also be seen as the region where $r < a_1$. The appropriate integration contour should indeed approach this line for $ |z|\gg1$ (or equivalently when $r \ll a_1$). In this limit $N_E \rightarrow 0$ along the imaginary axis. Therefore the singularity of the action at infinity can be interpreted as the necessity of an ever bigger kinetic energy for a transition of the scale factor from zero to a finite value in ever smaller amounts of time. But where should the other end point of integration lie?  The only other singularity is at $z=0,$ hence it must lie there. By definition (see Eq.(\ref{z})), the limit $z \rightarrow 0 $ is equivalent to $ \cos \left( \frac{N_E}{r}\right) \rightarrow - 1$. Thus $ N_E \approx \pm \pi r$ there. Since  $z$ vanishes when $r \rightarrow \infty$, the lapse blows up in this limit, and thus the singularity at $z=0$ corresponds to the limit of infinite $r$ and $N_E.$ As an aside, note that $r$ also diverges near $z = 2$. However in this case there is no condition on the lapse, and the action remains perfectly finite there (in fact, for $z = 2$, $a(t) = \pm N_E t $ and $a_1 = \pm N_E$). 

In the present case, the end points alone are not enough to completely fix the contour of integration yet, as there are inequivalent directions of approach to the essential singularity at $z=0$. By analogy with the Lorentzian path integral, we will require our defining contours to lie in a region where the integral is conditionally convergent, i.e. a region that asymptotically borders the regions of manifest divergence and convergence. In Figs. \ref{fig:flows} and \ref{fig:flows2} the lines which asymptotically demarcate regions of convergence and divergence are the blue lines, and hence when approaching $z=0$ from above we have two choices: either approaching at an angle of $\pi/4$ or at an angle of $3\pi/4.$ The first possibility is shown in the left panel of Fig. \ref{fig:contours}, and consists of the dashed line passing through saddle point $1$. This line can be deformed into the thimble ${\cal J}_1$ for both choices of sign of the scale factor $a(t).$ If one were to sum over both choices of sign, then the resulting amplitude would, in the saddle point approximation, consist of a sum of the two saddle point contributions at $z_1,$ i.e. a sum of a suppressed saddle point with weighting $e^{-12\pi^2/(\hbar\Lambda)}$ and an enhanced saddle point with weighting $e^{+12\pi^2/(\hbar\Lambda)}.$ We will comment on this contour again at a later stage when discussing perturbations\footnote{We note that this contour would be related to the infinite complex contour proposed by Diaz Dorronsoro et al. in \cite{DiazDorronsoro:2017hti} -- in fact it would give ``half'' of that result. However, in minisuperspace a contour which would yield a similar result would not be Lorentzian at all, as it would have to run from $-i\infty$ to $+\infty$ in the complex plane of the lapse function, see e.g. Fig. $5$ in \cite{Feldbrugge:2017kzv}.}. For now, let us just re-iterate a comment already made in \cite{Feldbrugge:2017mbc}, which is that the enhanced saddle point does not obey the correspondence principle: in the limit that $\hbar \rightarrow 0,$ its weighting becomes larger and larger, so that this quantum effect becomes \emph{more} dominant in the classical limit, rather than less. This strongly suggests that the upper sign in eq. \eqref{eq:summedmetrics} should in fact be discarded.

The second possibility of interest is shown in the middle panel of Fig. \ref{fig:contours}. Here the contour (orange dashed line) leaves the essential singularity along a blue line at an angle of $3\pi/4$ and asymptotically joins the ``Lorentzian'' $1+is$ line. This contour turns out to be the closest analogue of the Lorentzian contour in minisuperspace \cite{Feldbrugge:2017kzv}. For the choice $a(t)=-r\sin{N_E t/r},$ it is equivalent to the previous contour (as they are separated by a green region of convergence near $z=0,$ so that an arc linking the two contours at $z=0$ yields a vanishing contribution to the integral), see also the left panel of Fig. \ref{fig:flows2}. Again, it can be deformed into the thimble ${\cal J}_1$ and it will yield a propagator that can be approximated by the contribution of the saddle point at  $z_1,$ giving an amplitude $\propto e^{-12\pi^2/(\hbar\Lambda)}.$ This coincides with the result of the minisuperspace analysis. We will take this contour to be our preferred contour. For the opposite choice of sign for $a(t),$ this contour is inequivalent to the one in the left panel of Fig. \ref{fig:contours}, as can be seen very clearly in the right panel of Fig. \ref{fig:flows}. Now the two contours are separated by a red region of divergence near $z=0.$ The ``preferred'' contour now only crosses the ${\cal K}_2$ steepest ascent line, and consequently only the spurious (and in this case highly enhanced) saddle point $z_2$ contributes to the integral. (Moreover, as we will see below, the action for the perturbations develops a pole at $z=-2$ for perturbation modes with wavenumber $k=3$.) This unphysical result can be avoided by considering only the lower choice of sign in the sum over metrics \eqref{eq:summedmetrics}, in agreement with the comment made at the end of the last paragraph. With that restriction on the sum over metrics, both contours described above yield identical results.

\subsection{Perturbations at leading order} \label{sec:perts}

We are now in a position to add perturbations, i.e. we would like to evaluate the propagator
\begin{align}
G[a_1,\phi_1;0,0] = \int {\cal D}z\int {\cal D}\phi \, e^{iS^{tot}/\hbar}
\end{align}
where the total Lorentzian action, including a gravitational wave of wavenumber $k,$ amplitude $\phi$ and fixed polarisation, reads
\begin{equation}
S^{tot} = 2 \pi^2 \int_0^1 N_L dt \left[ - 3 \frac{a \dot{a}^2}{N_L^2} - \Lambda a^3 + 3 a  + \frac{a^3 \dot{\phi}^2}{2 N_L^2} - \frac{a}{2} (k^2 -1) \phi^2 \right]
\end{equation}
It would be straightforward to include sums over wavenumbers and polarizations, but in order to avoid clutter we omit this extension.

\begin{figure}
\includegraphics[width=0.45\linewidth]{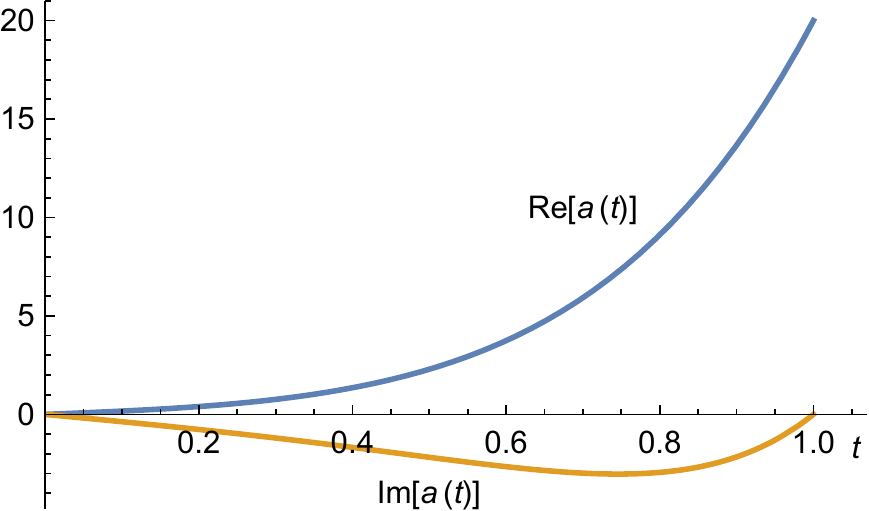}
\hspace{.5cm}
\includegraphics[width=0.45\linewidth]{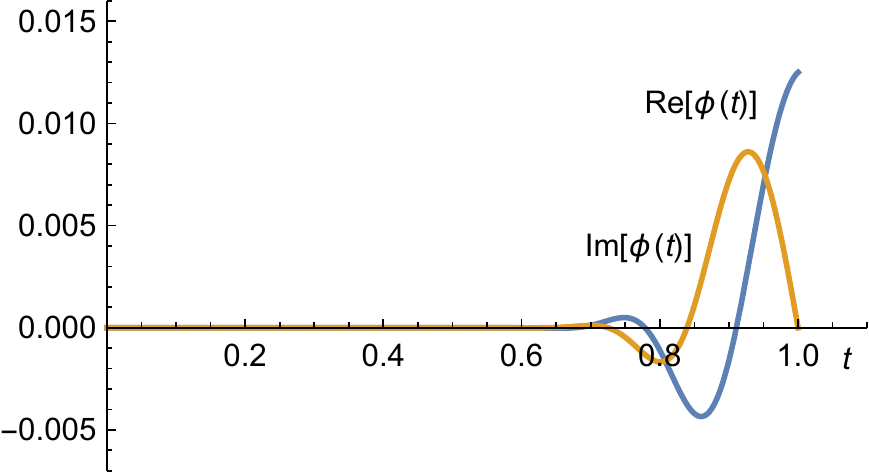}
\caption{The field evolution of the scale factor and of a gravitational wave perturbation at the saddle point $z_1 = 0.9997 + 19.97 i $ for the parameter values $\Lambda =3 $, $a_1 = 20 $, $\phi_1 = 1/80$, $k = 80$.}
\label{fig:fields}
\end{figure}

With a Euclidean lapse, the equation of motion for $\phi$ is given by
\begin{equation}
\frac{\ddot{\phi}}{N_E^2} + 3 \frac{\dot{a}}{a} \frac{\dot{\phi}}{N_E^2} - \frac{(k^2 - 1)}{a^2} \phi = 0   \label{phi}
\end{equation}
Note that this equation does not depend on the sign of $a(t)$. The general solution for both choices is given by \cite{Feldbrugge:2017fcc}
\begin{equation}
\begin{split}
F(t) = &+c_1 \frac{\left(1 - \cos \left( \frac{N_E t}{r}\right) \right)^{\frac{k-1}{2}} \left(\cos \left( \frac{N_E t}{r}\right)  + k\right)}{\left( 1 + \cos \left( \frac{N_E t}{r}\right)\right)^{\frac{k + 1}{2}}} +\\
  &+c_2  \frac{\left(1 + \cos \left( \frac{N_E t}{r}\right) \right)^{\frac{k-1}{2}} \left(\cos \left( \frac{N_E t}{r}\right)  - k\right)}{\left( 1 - \cos \left( \frac{N_E t}{r}\right)\right)^{\frac{k + 1}{2}}}\,,
\end{split}
\end{equation}
where $c_{1,2}$ are integration constants. The solution which is regular at $t =0 $ corresponds to setting $c_2 = 0,$ and thus the field perturbation is in fact zero at $a=0$. The boundary condition $\phi(t = 1) = \phi_1$ implies $c_1 = \frac{\phi_1}{F(1)}$ and $\phi(t) = c_1 F(t).$ An example of the field evolution is shown in Fig. \ref{fig:fields}.

The perturbative action reads
\begin{equation}
\begin{split}
S_{k}^{(2)} &=i \int_0^1 dt \left( \frac{a^3 \dot{\phi}^2}{2 N_E} + \frac{N_E}{2} a (k^2 - 1) \phi^2 \right)= \\
& = i\frac{d}{dt}\int_0^1 dt \, \frac{a^3 \phi \dot{\phi}}{N_E} - i\int_0^1 dt \frac{\phi(\ddot{\phi} + 3  H \dot{\phi})}{2 N_E} + i\int_0^1 dt \, N_E \frac{a}{2} (k^2 - 1) \phi^2
\end{split}
\end{equation}
Since the last two terms vanish along the solution to the e.o.m. for $\phi,$ the action takes the remarkably simple form
\begin{equation}
\begin{split}
S_{k}^{(2)} &= i\frac{a^3 \dot{\phi } \phi}{2 N_E} \mid_{t =  1} =i\frac{a_1^3 \dot{\phi_1 } \phi_1}{2 N_E}   = \pm i\frac{a_1^2 \phi_1^2}{2 } \frac{(k^2  -1) }{ \left[ \cos \left( \frac{N_E }{r}\right)+ k \right] }\\ & = \pm i\frac{a_1^2}{2} \frac{(k^2 - 1)}{(z + k - 1)} \phi_1^2 \label{eq:actionpert}
\end{split}
\end{equation}
The perturbative action evaluated at the saddle points reads respectively 
\begin{align}
S_{k}^{(2)}(z_2) &= \pm i\frac{(k^2 -1 ) \phi_1^2}{2 (k -3)} \\
S_{k}^{(2)} (z_1) & = \pm i\frac{a_1^2 (k^2 - 1)}{2 (k + i \sqrt{\frac{a_1^2 \Lambda}{3} - 1})} \phi_1^2 \, \approx\,  \pm \frac{\sqrt{3} (k^2 - 1) a_1 \phi_1^2}{2 \sqrt{\Lambda}} \pm i\frac{3k(k^2-1)\phi_1^2}{2\Lambda} \label{actionz1}\\
S_{k}^{(2)} (z_3) & = \pm i\frac{a_1^2 (k^2 - 1)}{2 (k - i \sqrt{\frac{a_1^2 \Lambda}{3} - 1})} \phi_1^2 \, \approx \, \mp \frac{\sqrt{3} (k^2 - 1) a_1 \phi_1^2}{2 \sqrt{\Lambda}} \pm i\frac{3k(k^2-1)\phi_1^2}{2\Lambda}\,,
\end{align}
where the approximate expressions correspond to the limit of a large final scale factor value $a_1.$ 

Two properties stand out immediately: first, the implied weighting of the perturbations is Gaussian for the upper sign (i.e. also for the upper sign in eq. \eqref{eq:summedmetrics}), and inverse Gaussian ($\sim e^{+k^3\phi_1^2/{\hbar \Lambda}}$) for the lower sign. Since the Lorentzian saddle point corresponds to the lower sign, we find that the ``preferred''  contour, i.e. the one that coincides most closely with the Lorentzian minisuperspace contour, yields a propagator of the form
\begin{align}
G^{Lorentz}[a_1,\phi_1;0,0] \approx e^{- \frac{ 12 \pi^2 }{\hbar\Lambda} \left[1 + i \left(\frac{a_1^2 \Lambda}{3} -1 \right)^{3/2}\right]  + \frac{3k(k^2-1)\phi_1^2}{2\hbar\Lambda} - \frac{i \sqrt{3} (k^2 - 1) a_1 \phi_1^2}{2 \hbar\sqrt{\Lambda}}} 
\end{align}
The present calculation confirms that the no-boundary transition amplitude from nothing to a large final universe is unstable, even when only regular geometries are summed over. This is our main result. It demonstrates that the instability of the no-boundary proposal is not due to off-shell singularities, but is an intrinsic feature of the Lorentzian no-boundary/tunneling saddle points. 

\begin{figure}
\begin{center}
\includegraphics[width=0.45\linewidth]{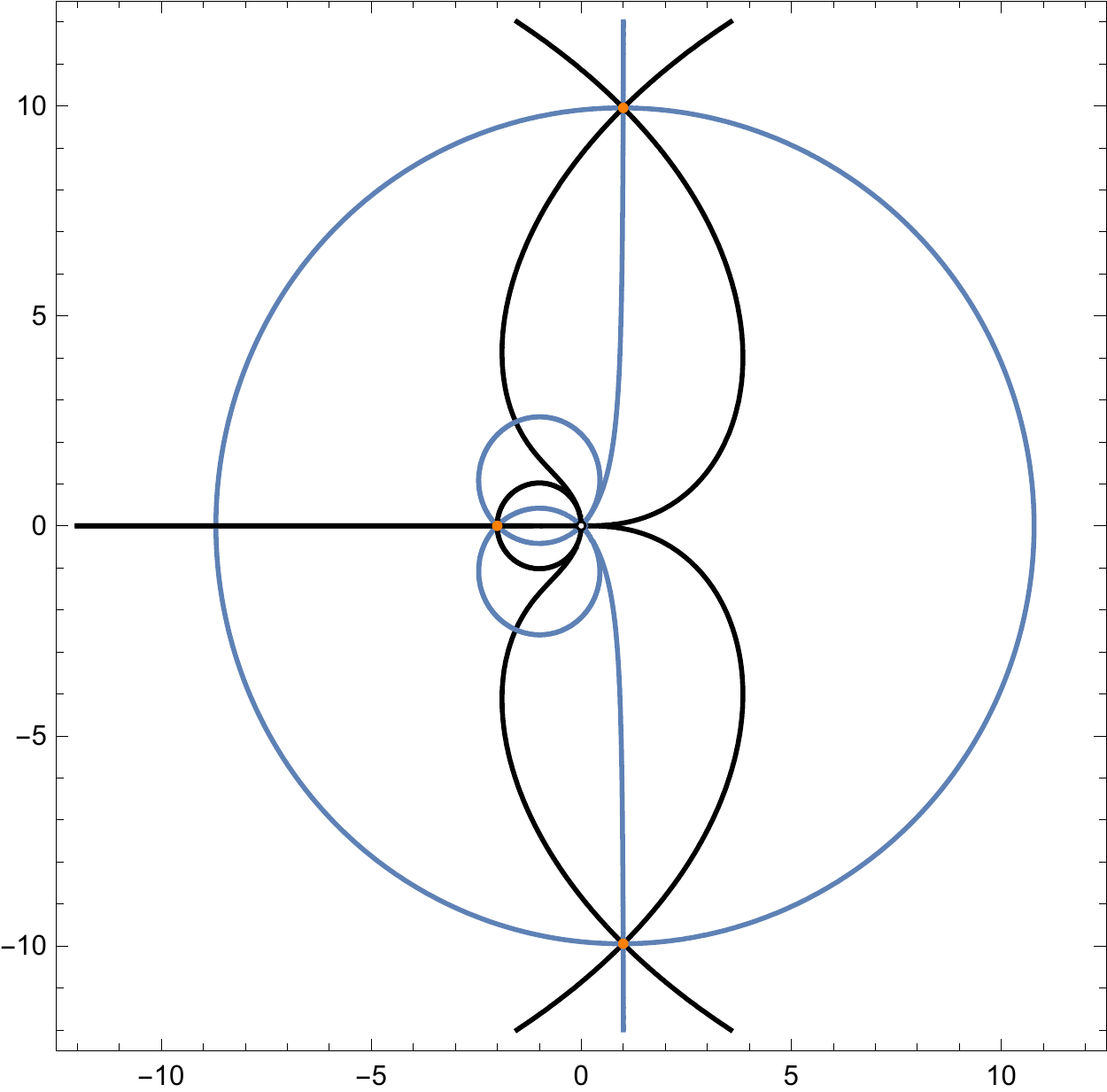}
\hspace{.5cm}
\includegraphics[width=0.45\linewidth]{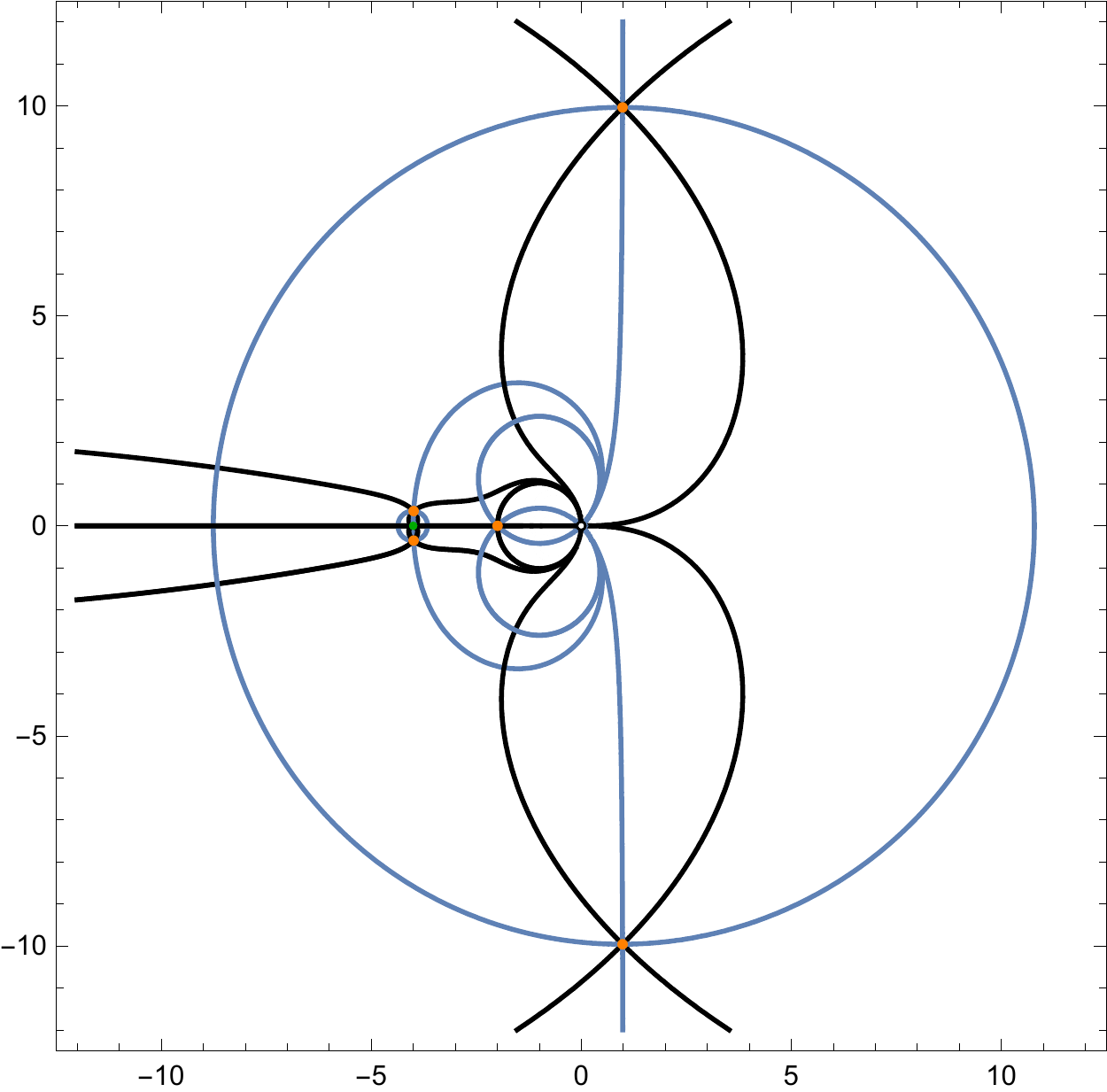}
\caption{Left panel:  flow lines for large values of the final scale factor, cf. Fig. \ref{fig:flows2}. In the figure the parameters are $\Lambda=3, a_1 = 10.$ Saddle points are marked in orange, black lines are steepest ascent/descent lines (the directions of descent, say, can be inferred straightforwardly by comparison with the two panels in Fig. \ref{fig:flows2}), and blue lines have the same value of the Morse function as the saddle points that they cross. Right panel: adding a gravitational wave perturbation of wavenumber $k$ results in two additional saddle points near the new pole of the action at $z=-k+1$. Here the parameters that were used are $\Lambda=3, a_1=10, k=5, \phi_1=1/5,$ with the pole (in green) located at $z=-4.$ The figure shows that the new flow lines, associated with the two additional saddle points, are irrelevant for our preferred contour of integration.}
\label{fig:contourslargea1}
\end{center}
\end{figure}

The second feature is that the perturbative action contains a wavenumber-dependent pole at $z=-k+1,$ as is evident form eq. \eqref{eq:actionpert}.  One consequence is that two new saddle points appear near this pole with $Re(z_{saddle}) < 0$. The flow lines associated to those saddle points end at the new pole $ z = 1 - k$ and otherwise remain close to the negative $z$ axis -- see Fig. \ref{fig:contourslargea1} for an illustration. As $k$ increases, the pole moves away from the origin along the negative real line. The contours discussed so far receive no contribution from these new saddles, while the positions of their relevant saddle points are merely shifted by negligible amounts. However, any contour that crosses the negative real $z$ line is liable to non-trivial corrections. In particular, a circular contour such as that proposed in \cite{DiazDorronsoro:2018wro} becomes essentially untenable: for it to be well defined, it must encircle the origin at a radius smaller than the closest pole. Given that small $|z|$ corresponds to large $r,$ this would imply that a circular contour can only sum over very large $r$ geometries (which do not contain a Lorentzian region near the final boundary), i.e. certainly not over a representative set of regular geometries. In addition such a contour could not be deformed into the two thimbles ${\cal J}_{(1,2)},$ as envisioned in \cite{DiazDorronsoro:2018wro}, but would receive additional (wavenumber-dependent) contributions from the perturbative poles, as illustrated in the right panel of Fig. \ref{fig:contours}. This type of contour is thus of no physical interest, as argued before in \cite{Feldbrugge:2018gin}.

A further consequence concerns the contour presented in the left panel of Fig. \ref{fig:contours}. With this choice of contour, both signs of the scale factor are allowed, and the propagator becomes a sum over both contributions. This propagator then consists of a stable and an unstable saddle point, and for large $a_1$ it has a weighting of the form
\begin{align}
|G[a_1,\phi_1;0,0]| \approx e^{-\frac{ 12 \pi^2 }{\hbar\Lambda} + \frac{3k(k^2-1)\phi_1^2}{2\hbar\Lambda}} + e^{+ \frac{ 12 \pi^2 }{\hbar\Lambda} - \frac{3k(k^2-1)\phi_1^2}{2\hbar\Lambda}} 
\end{align}
For large enough wavenumbers $k$ and final amplitudes $\phi_1$, the unstable part dominates and the overall probability distribution becomes unstable. Subject to verifying that this result resides within the limits of linear perturbation theory, this confirms the results of \cite{Feldbrugge:2017mbc} obtained in minisuperspace.

This last observation brings us to the subject of backreaction, as it is important to know the range of validity of linear perturbation theory. In order to discuss backreaction, it is useful to re-iterate the calculational strategy employed in the present work: our aim is to sum over regular geometries. These are simply off-shell geometries, chosen with the unique criterion that they should be regular. For the background, we look at a particularly simple sub-class, namely 4-spheres with varying radii (or even just a subset of those, given the choice of sign of the scale factor). A priori these geometries do not satisfy any equations of motion, they are just a particular subset of geometries that are summed over in the path integral. Then we add perturbations, subject to two criteria: they should not destroy the regularity, and they should satisfy the linear equation of motion around the off-shell background geometries. Hence one should think of them as saddle points of the $\phi$ integral. Only after both integrals over $\phi$ and $z$ have been performed do we expect the final saddle point to be a solution to the full Einstein equations. Thus it only makes sense to check backreaction at the final saddle point.  This is different for the minisuperspace case studied in \cite{Feldbrugge:2017mbc}, where the lapse integral was over geometries that were already saddle points of the integral over the scale factor $q$. Hence each such configuration was already a solution to the $q$ equation of motion, and it made sense to check whether there was a large backreaction or not on those solutions, before the lapse integral was performed. But in the present context, we are only interested in whether the final overall saddle point is trustworthy.

The action for the background and perturbations leads to a system of two coupled differential equations, where the equation of motion for the scale factor $a(t)$ is modified, compared to the background, by $\phi$-dependent terms 
\begin{equation}
- 2 \frac{\ddot{a}}{a N_E^2} - \frac{\dot{a}}{a^2 N_E^2 } + \frac{1}{a^2} = \Lambda + \frac{1}{2 }\frac{\dot{\phi}^2}{N_E^2} + \frac{(k^2 - 1)}{6 a^2} \phi^2 \label{back1}
\end{equation}
Absence of backreaction corresponds to neglecting the kinetic and the gradient terms for $\phi$ in this equation. Thus a conservative view is to demand that these additional terms remain small at every value of $t$, i.e. that
\begin{equation}
|\frac{1}{2 }\frac{\dot{\phi}^2}{N_E^2}| , \, |\frac{(k^2 - 1)}{6 a^2} \phi^2| \, \ll \Lambda \qquad  \forall t \in [ 0 , 1] \label{back2}
\end{equation}

\begin{figure}
\begin{center}
\includegraphics[width=0.45\linewidth]{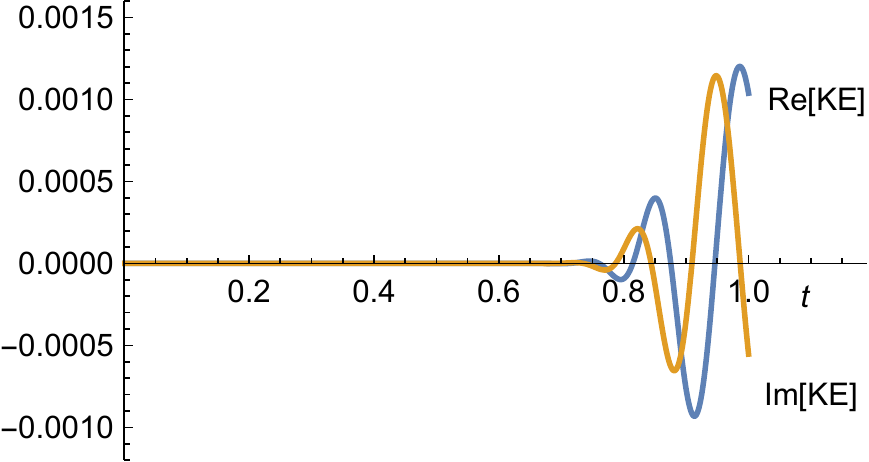}
\hspace{.5cm}
\includegraphics[width=0.45\linewidth]{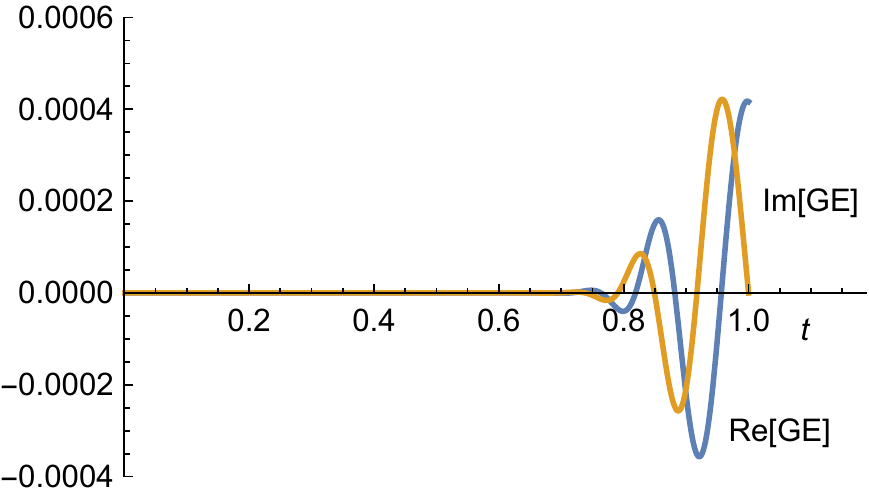}
\caption{The figure shows the behaviour of the backreaction terms involving the kinetic energy (KE)  $ \frac{1}{2 }\frac{\dot{\phi}^2}{N_E^2} $  (left panel) and the gradient energy (GE) $ \frac{(k^2 - 1)}{6 a^2} \phi^2$  (right panel) at the relevant saddle point  for $ \Lambda = 3, a_1 = 20, k=80, \phi_1 = 1/80$ (the field evolutions were shown in Fig. \ref{fig:fields}). These terms are everywhere small compared to $\Lambda$. }
\label{fig:backreaction}
\end{center}
\end{figure}

For large $k$ sub-Hubble modes, i.e. for modes that obey $k > a_1 \sqrt{\Lambda/3},$ the perturbation grows fastest right at the end, near $t=1.$ Thus the backreaction is also largest there, see for instance the numerical example in Fig. \ref{fig:backreaction}. Still, the backreaction remains negligibly small, staying well below a tenth of a percent at all times for this example where the parameters that are used satisfy $k \phi_1 = 1,$ so that the contribution of the perturbations to the total weighting is in fact large (it is larger than that of the background). We may in fact find analytic expressions for the backreaction at $t=1,$ using the results of the calculation of the perturbative action in eq. \eqref{eq:actionpert},
\begin{align}
\frac{\dot\phi^2}{2N_E^2}(t=1) = \frac{(k^2-1)^2}{2 (z_1+k-1)^2} \left( \frac{\phi_1}{a_1}\right)^2\,, \qquad \frac{(k^2 - 1)}{6 a^2} \phi^2 (t=1) = \frac{(k^2 - 1)}{6} \left( \frac{\phi_1}{a_1}\right)^2\,.
\end{align}
The backreaction at $t=1$ scales as $\left( \frac{k \phi_1}{a_1}\right)^2,$ and it will be small compared to $\Lambda$ as long as 
\begin{align}
\phi_1 \ll \frac{a_1}{k \sqrt{\Lambda}}\,,\, \qquad \left( k > a_1 \sqrt{\frac{\Lambda}{3}}\right) \,.
\end{align}
Note that this bound does not preclude a large contribution from the perturbative action \eqref{actionz1} to the total weighting.

\begin{figure}
\begin{center}
\includegraphics[width=0.32\linewidth]{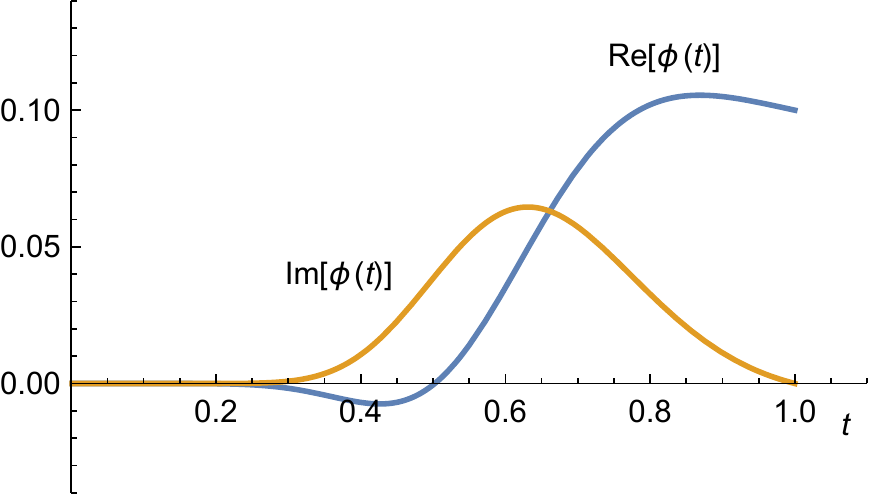}
\includegraphics[width=0.32\linewidth]{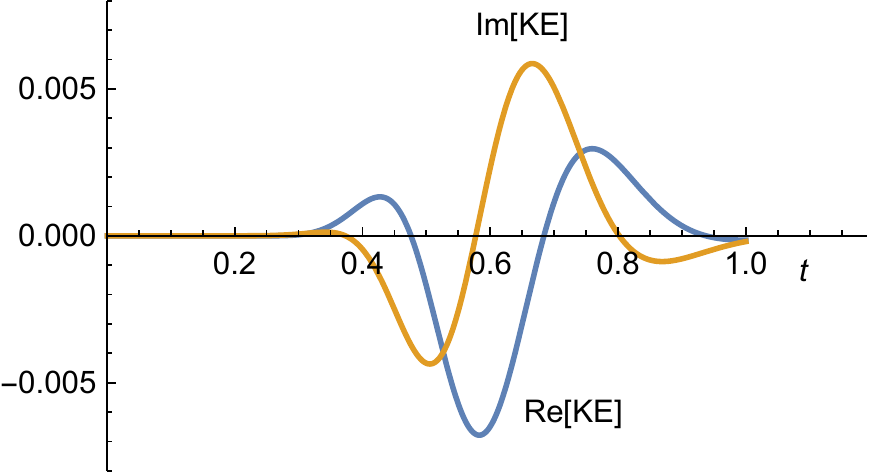}
\includegraphics[width=0.32\linewidth]{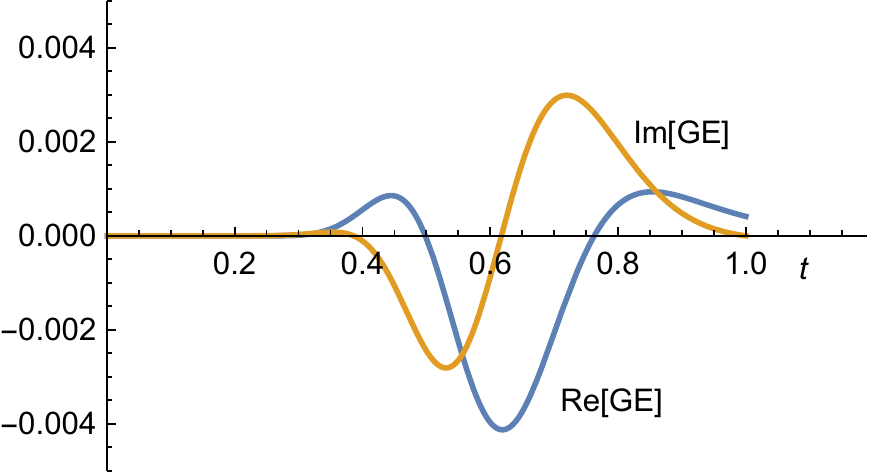}
\caption{The field evolution and backreaction terms for a perturbation with $ \Lambda = 3, a_1 = 20, k=10, \phi_1 = 1/10.$}
\label{fig:backreaction10}
\end{center}
\end{figure}

For super-Hubble modes, with $k \lesssim a_1 \sqrt{\Lambda/3},$ the mode functions grow earlier and then freeze out. Thus their main contribution to backreaction occurs significantly before the final hypersurface is reached. A numerical example is shown in Fig. \ref{fig:backreaction10},  where the same background was used as above, but now with wavenumber $k=10$ and final amplitude $k=1/10.$ Again, one can see that the backreaction terms stay below two tenths of a percent, as compared to the cosmological term $\Lambda=3.$ For smaller $k$ we find that the maximally allowed amplitude to stay within the regime of validity of linear perturbation theory does not decrease, such that overall linear perturbation theory is a very good approximation for a large parameter range.

\subsection{Discussion}

The quest to find a theory of initial conditions for the universe is intimately related to the quest of learning how to apply quantum theory to the universe. An appealing framework in this regard is the path integral approach to gravity. While this formulation is in all likelihood intrinsically limited to the semi-classical domain, it is highly useful since it provides a direct link to non-gravitational formulations of quantum theory, and since it allows one to use geometrical intuitions and methods.  Out of these geometrical considerations arose the no-boundary proposal of Hartle and Hawking, which may be seen as a proposal for replacing the big bang by sums over compact, regular geometries. By construction, the curvature singularity at the big bang is then avoided, and one may hope to find finite, well-defined results, ideally in agreement with observations.

A difficulty has been to define the gravitational path integral precisely, in particular to properly define the class of geometries that should be summed over. Mostly, interest has focussed on minisuperspace models, which still contain many singularities off-shell, but where the saddle points correspond to the smooth geometries that Hartle and Hawking had in mind. Unfortunately, a mathematically precise implementation of the no-boundary idea in terms of a Lorentzian path integral resulted in the conclusion that perturbations around these geometries are unstable, and that the proposal is untenable as a theory of the initial conditions of the universe, at least in the context of a universe dominated by a positive cosmological vacuum energy. 
In follow-up work there have been several attempts to modify the definition of the no-boundary proposal in order to avoid this negative conclusion. One attempt by Diaz-Dorronsoro et al. was to consider intrinsically complex contours of integration for the lapse function \cite{DiazDorronsoro:2017hti} (also in conjunction with a specification of the initial momentum \cite{DiazDorronsoro:2018wro}). However, these modifications either still included unstable \cite{Feldbrugge:2017mbc} or led to inconsistencies \cite{Feldbrugge:2018gin}. Another approach by Vilenkin and Yamada (in the context of the tunnelling proposal) was to modify the boundary conditions for the perturbations to be of Robin type \cite{VY1,VY2}, but in our view this leads to uncontrolled backreaction on the background geometry \cite{Feldbrugge:2018gin}. A final proposition by Halliwell, Hartle and Hertog was to simply abondon the path integral, and concentrate on solutions of the Wheeler-deWitt equation with desirable properties \cite{Halliwell:2018ejl}. However, the path integral neatly captures quantum interference, and focussing on solutions that cannot be described via a path integral may thus not correctly reproduce central quantum effects. 
A question which arose from this discussion was whether the instability was in fact caused by the off-shell singularities. For this reason, we investigated the restriction to summing over only manifestly regular geometries, in the simplest possible case of summing only over complexified 4-spheres of various radii. This approach was pioneered by Halliwell and Louko, and we extended their calculations by finding a complementary set of 4-spheres to be summed over, and by adding perturbations. Our calculations confirm that the no-boundary condition leads to an instability, well within the limits of applicability of linear perturbation theory.  

Thus we can conclude that the reason for the failure of the no-boundary proposal was not the breakdown of perturbation theory, nor the inclusion of off-shell geometries that included singular configurations and led to a non-analytic structure of the perturbative action (which incidentally, might be avoided by performing the minisuperspace path integral using Picard-Lefschetz theory extended to the infinite-dimensional case). We will see in the next chapter that imposing suitable boundary conditions, different from the above studies Dirichlet ones, one can indeed formulate that Hartle-Hawking wavefunction as a path integral in semi-classical gravity.

\clearpage

%%%%%%%%%%%%%%%%%%%%%%%%%%%%%%%%%%%%%%%%%%%%%%%%%%%%%%%%%%%%%%%%%%%%%%%%%%%%%%%%%%%%

\section{Robin path integrals for cosmology}\label{robincosmology}

If quantum theory is universal, and there currently is no reason to think otherwise, then the universe should be describable by a quantum state just like any other system. While its quantum properties might be hidden today they may well have played a crucial role in an early phase of its evolution. In this chapter, based on \cite{DiTucci:2019dji}, \cite{DiTucci:2019bui} and \cite{Bramberger:2019zks}, we continue our study of such semi-classical effects in the early universe defining path integrals with Robin type of boundary conditions (see section \ref{sec:boundaryterms}). We will study two different physical situations where these boundary conditions are useful: first, following \cite{DiTucci:2019dji} and \cite{DiTucci:2019bui}, we will return to the question of how to define a no-boundary path integral, which we have seen in the last chapter fails to be well-defined with Dirichlet type of boundary conditions; then, in section \ref{papersebastian}, we will study homogeneous quantum transitions during inflation \cite{Bramberger:2019zks}. There, we will see that a Robin boundary condition is needed to recover the correct semi-classical limit, since both the expansion rate and the size of the universe need to be specified to a certain degree. Part of the discussion will be reminiscent of section \ref{InitialConditions} of \ref{quantuminitial} where a similar type of Robin boundary conditions was implemented to the define suitable initial conditions for inflation.

%%%%%%%%%%%%%%%%%%%%%%%%%%%%%%%%%%%%%%%%%%

\subsection{The no boundary proposal with canonical Robin boundary conditions}

Let us consider the usual FLRW ansatz \eqref{FLRW} augmenting the Dirichlet Einstein-Hilbert action \eqref{S} with a (Special) Robin boundary term at the initial surface specified by $q_0,$
\begin{equation}
S_{tot}= S + \alpha q_0 + \frac{q_0^2}{2 \beta} + \gamma\,. \label{eq:Robin}
\end{equation}
The constant $\gamma$ plays no role in the boundary value problem and we add it to the action just to keep the discussion general. Note also that it changes the value of the action by the same factor for all the geometries so that it does not affect relative probabilities. The variation of the full action is 
\begin{equation}
\delta S_{tot} = V_3 \int_0^1 dt \Bigl[ \frac{3 \ddot{q}}{2 N} - 3 N H^2 \Bigr] \, \delta q  - \frac{3 V_3}{2 N} \dot{q}_1 \, \delta q_1 + \Bigl[  \frac{3 V_3}{2 N} \dot{q}_0  + \alpha + \frac{q_0}{\beta}  \Bigl]  \,\delta q_0\,.
\end{equation}
Thus we can see that the variational principle is well defined if we impose 
\begin{align}
 \delta q_1 &= 0 \\
 \frac{3 V_3}{2 N} \dot{q}_0 + \alpha + \frac{q_0}{\beta}  &= 0 
\end{align}
In other words we fix the value of the field at $t=1$ to be $q(t= 1) = q_1$ corresponding to a Dirichlet boundary condition, while at $t= 0$ we impose a condition on the linear combination of $q_0$ and $\dot{q}_0$. This is called a Special Robin boundary condition. (Special because in general one can set that linear combination to any constant value, here it is set to zero.) This way of implementing Robin boundary conditions as given by the boundary term in \eqref{eq:Robin} has very important implications for quantum cosmology and within the minisuperspace approximation which will be discussed below. It is however important to note that  this term is not covariant. Despite this drawback in what concerns diffeomorphisms, canonical Robin conditions have nice properties from a quantum mechanical point of view. The boundary terms can in fact be interpreted as initial and final states of (in general complexified) Gaussian form.

With these boundary conditions, the solution to the equation of motion reads
\begin{equation}
q(t)= H^2 N^2 t^2 - \frac{(\alpha \beta + q_1 - H^2 N^2)}{3 \beta V_3 - 2 N} 2 N t + \frac{(q_1 - H^2 N^2) 3 V_3 + 2 N \alpha}{3 \beta V_3 - 2 N} \beta \label{qsol}
\end{equation}
Plugging this solutions back into the action we find that the saddle points are
\begin{equation}
N_s = \frac{3 V_3 \beta}{2} + c_1 \frac{\sqrt{H^2 q_1 -1}}{H^2} + c_1 c_2 \frac{ \sqrt{9 \beta H^4 V_3^2 - 4 (1 + \alpha \beta H^2  )}}{2 H^2} \label{sadab}
\end{equation}
A crucial requirement in order to obtain an implementation of the no-boundary idea is now that at the saddle point, the geometry should be of Hawking type, and in particular it should start at zero size. From \eqref{qsol} one can see that the initial size $\bar{q}_0$ vanishes at one (or more) of the saddle points if $\alpha = \pm 3 i V_3$ or $\beta = 0 $. These are indeed the values for the initial momentum of the Hartle-Hawking and Vilenkin geometries found in eq. \ref{initialmomentum}.
For $\alpha = + 3 i V_3$, the initial geometries vanishes at the ``tunnelling'' saddle points
\begin{equation}
N_{1,2} = \frac{i}{H^2} \pm \frac{\sqrt{H^2 q_1 -1}}{H^2}
\end{equation}
These geometries are unstable; thus in the following we will consider $\alpha = - 3 i V_3$ (we will discuss the meaning of $\beta$ momentarily). This gives that $\overline{q}_0=0$ for
\begin{equation}
N_{3,4} = \frac{- i }{H^2} \pm \frac{ \sqrt{H^2 q_1 - 1}}{H^2}\,,
\end{equation}
which are precisely the Hartle-Hawking saddle points. The other two saddle points are now located at
\begin{equation}
N_{1,2} =  \frac{i }{H^2} \pm \frac{ \sqrt{H^2 q_1 - 1}}{H^2} + 3 \beta V_3 
\end{equation}
and their initial size is $\overline{q}_0 = 3 \beta V_3 (2 i + 3 \beta V_3 H^2)$. 

Therefore requiring that at least one of the saddle point geometries starts out at zero size corresponds to a specific value of $\alpha$ but leaves $\beta$ free.   The value of $\beta$ in fact determines which saddle point(s) are relevant to the path integral. 

\begin{figure}
\centering
\includegraphics[scale=0.5]{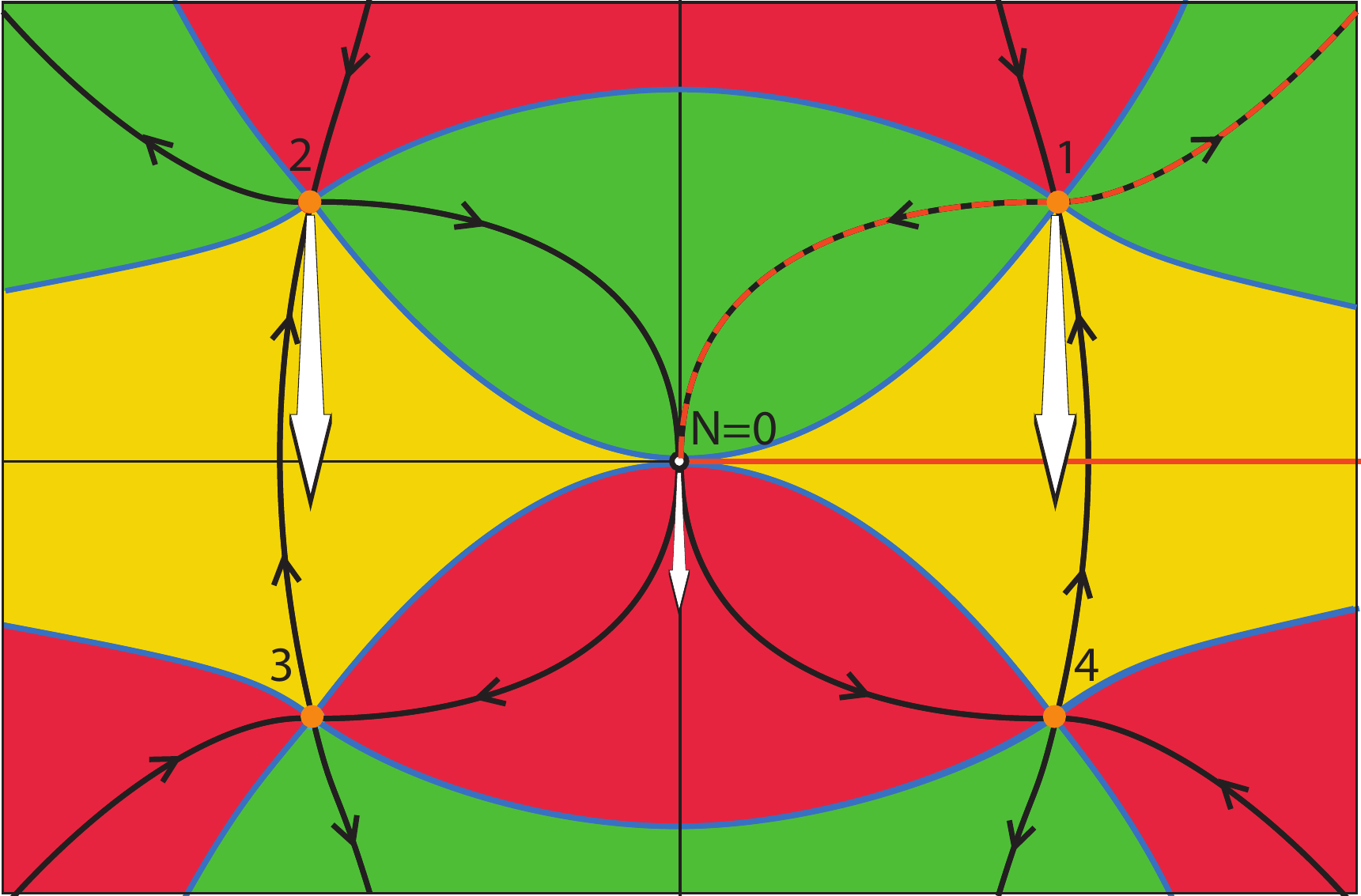}
\caption{The minisuperspace path integral contains $4$ saddle points (orange dots) in the plane of the complex lapse $N$. Steepest descent/ascent lines of the magnitude of the action are drawn as black lines, with the arrows indicating directions of descent. Asymptotic regions of convergence are shown in green, while divergent ones are red. For the path integral with Dirichlet boundary conditions, the defining contour of positive real lapse (orange line) can be deformed to the (orange dashed) thimble flowing through saddle point $1$ only. This saddle point however admits unstable perturbations. With Robin boundary conditions, saddle points $1,2$ and the singularity of the action at $N=0$ move. Shown here is the direction of motion for negative imaginary $\beta.$} \label{fig:Robin1}
\end{figure}

For $\beta = 0 $, the relevant saddle point is the Lorentzian one $N_1 = \frac{i + \sqrt{H^2 q_1 - 1}}{H^2}$. In this limit the Robin boundary condition reduces to the Dirichlet boundary value $q_0 = 0 $. For non-zero $\beta$ the saddle points in the upper half plane move, and the singularity in the action, which originally resides at $N^\star=0,$ is also shifted to $N^\star= \frac{3 \beta V_3}{2}$. Importantly, the lower saddle points (numbers $3$ and $4$ in Fig. \ref{fig:Robin1}), which correspond to the desired Hartle-Hawking geometries, stay put. 

Our strategy will be to define the path integral on a thimble, for the simple reason that it is then manifestly convergent. This has the important consequence that the partial integrations over the scale factor, the lapse and any other fields that might be present, can be performed in any desired order without changing the end result (i.e. in the case of absolute convergence, Fubini's theorem applies). For completeness, we show in Appendix \ref{WdWproof} that the
path integral satisfies the Wheeler-deWitt equation. At vanishing $\beta$ our defining integration contour is simply the Lorentzian one, i.e. along real positive values of $N.$ This can then be deformed, using Picard-Lefschetz theory, to the thimble passing through the unstable saddle point $1,$ as shown in Fig. \ref{fig:Robin1}. As $\beta$ is turned on, the singularity of the integral at $N=0$ shifts to $N^\star,$ and thus, in order to maintain an invariant definition, we will consider as our integration contour the thimble(s) emanating from $N^\star.$ 

\begin{figure}
\centering
\includegraphics[scale=0.5]{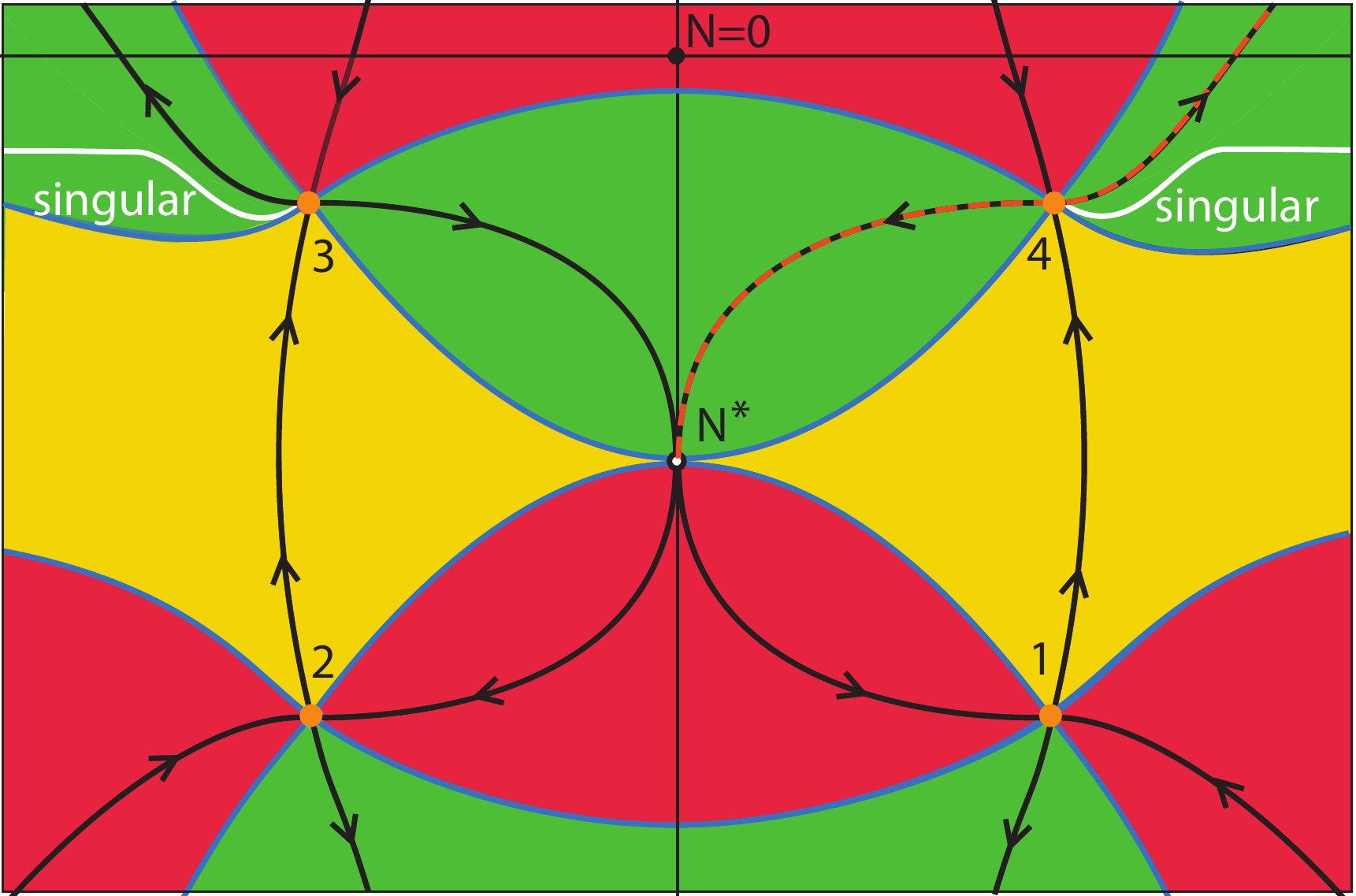}
\caption{For sufficiently large negative imaginary $\beta,$ the unstable saddle point $1$ moves below the Hartle-Hawking saddle point $4$. The thimble emanating from the singularity of the action at $N^\star$ now only passes through this stable saddle point, and the problem with instabilities is avoided. In white we indicate the locus of geometries that contain a singularity, and which we require the thimble to avoid.} \label{fig:Robin2}
\end{figure}

But which value should $\beta$ take? Roughly speaking, the unstable saddle points remain relevant until $\beta$ is large enough in magnitude so that they have moved ``out of the way''. For real negative $\beta$ this approximately means that they have to move further to the left than $N^\star,$ which marks the origin of the integration contour. But since the real part of the saddle points depends on $q_1,$ this is non-sensical from a physical point of view, as the initial conditions would have to keep being readjusted as the universe keeps expanding! On the other hand, in the imaginary $N$ direction, the unstable point (with an original location at $Im(N_{1,1})=+i/H^2$) only has to move beyond the stable one at $Im(N_{1,-1})=-i/H^2.$ This condition is independent of the final size of the universe, and thus we will only consider imaginary $\beta$. The minimal magnitude of $\beta$ is determined precisely by the condition that the unstable saddle point move below the stable one,
\begin{align}  
|\beta| > \beta_{min} = \frac{2}{3 V_3 H^2}\,.
\end{align}
For $\beta$ being negative imaginary and of magnitude larger than this minimal value, the Hartle-Hawking saddle point becomes the only relevant one -- see Fig. \ref{fig:Robin2} for an illustration. Thus we have successfully isolated the Hartle-Hawking saddle point, leading to the propagator
 \begin{equation}
 G[q_1 , 0]_{Robin} = e^{+ \frac{2 V_3}{H^2 \hbar} - i \frac{2 V_3 H}{\hbar} (q_1 - \frac{1}{H^2})^{3/2}} \label{G}
\end{equation} 
In fact, we also have the possibility of including both thimbles emanating from $N^\star,$ i.e. the ones passing through saddles numbered $3$ and $4$ in Fig. \ref{fig:Robin2}. In this case we obtain a linear combination of the Hartle-Hawking saddle point and its time reverse, and if this linear combination is taken with equal coefficients, as would for instance be the case if we defined the original integration contour to run over all real values of the lapse function, from minus infinity to plus infinity, the result will the the real no-boundary wavefunction \eqref{hh} proposed by Hartle and Hawking. As we will discuss next, there exists however one remaining criterion in order for the path integral to be well defined.

%%%%%%%%%%%%%%%%%%%%%%%%%%%%%%%%%%%%%%%%%%

\subsubsection{Avoiding singular geometries}

\begin{figure}
\centering
\includegraphics[scale=0.5]{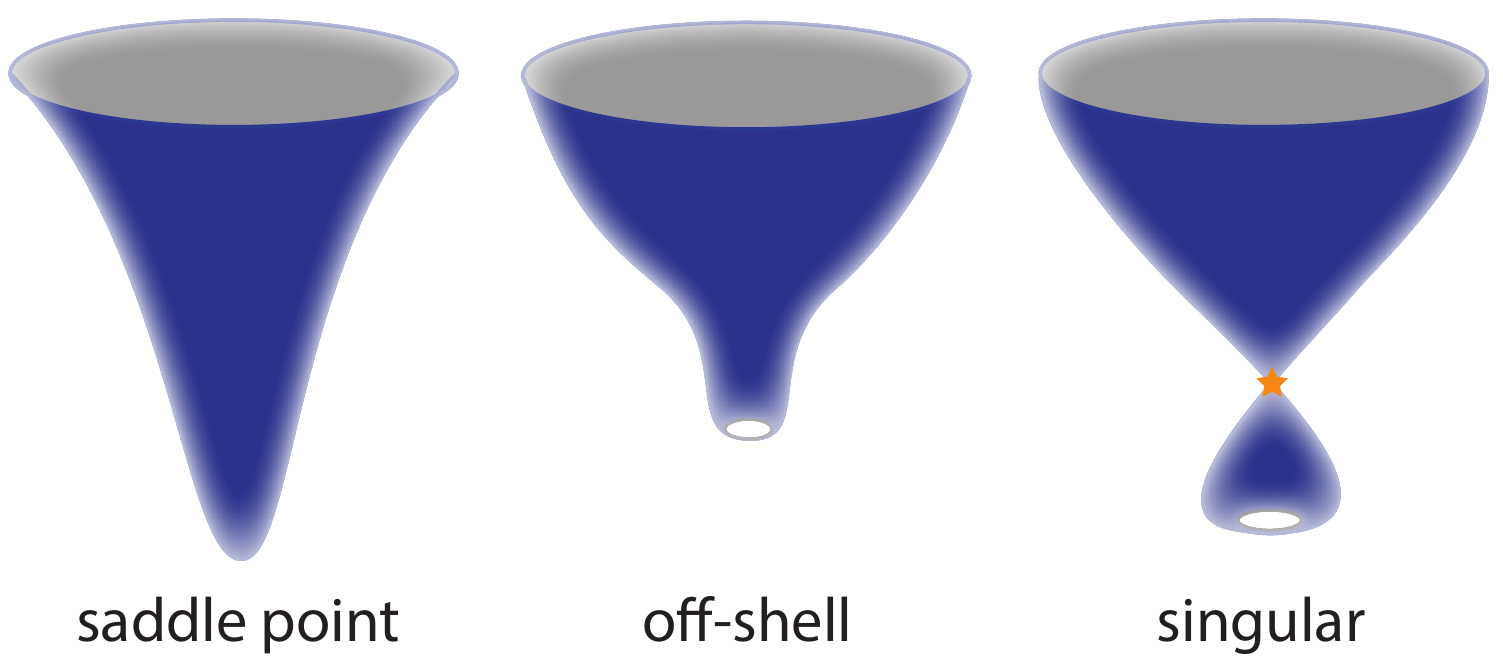}
\caption{On the left we have the smooth saddle point geometry of Hartle-Hawking type. By contrast, with Robin boundary conditions a typical off-shell geometry will not start at zero size (middle). Some off-shell geometries contain a recollapse to zero size, and it is these geometries that we would like to avoid summing over (right).} \label{fig:regsing}
\end{figure}

The Robin boundary condition implies a relationship between the initial size and the initial momentum of the geometries summed over in the path integral. It does however not necessarily avoid the appearance of singularities, in the sense that it may be possible that some of the off-shell geometries contain a region where the size of the universe passes through zero, see Fig. \ref{fig:regsing}. We would like to avoid summing over such geometries, as they are (infinitely) sensitive to the addition of higher curvature terms, thus rendering the result obtained so far potentially unreliable. We will then explore the possibility of defining a path integral that avoids such singularities along the thimble used to define the path integral. This will imply an upper bound on the magnitude of $\beta.$ Clearly, this upper bound will not be as strong as the lower bound found in the previous section: $|\beta|> \beta_{min}$ is essential for the no-boundary well defined, here we are investigating and discussing a potential problem which might arise with the inclusion of higher order curvature corrections.

Thus we must analyse the locus of geometries containing a region of zero volume. We start by noting that the imaginary part of the scale factor vanishes, $Im(q(\tau))=0,$ at
\begin{equation}
\tau = - \frac{3 i \beta V_3}{ H^2 m } \frac{[-q_1 + 3 i \beta V_3 + H^2(n^2 + m^2 + 3 i \beta m V_3 )]}{[4 n^2 + (2 m + 3i \beta V_3)^2 ]}\,, \label{tau}
\end{equation}
where we have split the lapse into real and imaginary parts as $N = n + i \, m $.  We must then determine if the real part of the scale factor \eqref{q} may simultaneously vanish, with $0 \leq \tau \leq 1$.    There is no concise analytic expression for this function. However, we are only interested in finding the slope of such a curve of singular geometries at the relevant saddle point. It is therefore sufficient to expand to first order around $N_4$ and then solve for the curve $m(n)$. The angle $\phi$ of the ``singular curve'' at the saddle point $N_4$ is 
\begin{equation}
\tan(\phi) = \frac{(- 4 + 3 i \beta H^2 V_3) \sqrt{H^2 q_1 - 1}}{(2 q_1 + 3 i \beta  V_3)H^2 - 4}\,.
\end{equation} 
Meanwhile, at large magnitudes of the lapse, this singular curve is approximately horizontal, see Fig. \ref{fig:Robin2} for a sketch.

Defining $e^{i S_{tot}} \equiv e^{h + i s}$ with $h , s \in \mathcal{R}$, the steepest descent path at the saddle point runs in the direction of the eigenvector of the Hessian matrix for $h(n,m) = - Im (S_{tot}(n , m))$ associated with the negative eigenvalue. At the saddle point $N_4,$ the angle $\theta$ of the thimble is given by
\begin{equation}
\tan(\theta) = \frac{-2 + 3 i \beta H^2 V_3}{2 \sqrt{H^2 q_1  - 1} + H \sqrt{4 q_1 - 12 i \beta V_3 - 9 V_3^2 H^2 \beta^2}}\,, \label{thetathimble}
\end{equation}
where we have assumed that $|\beta|> \beta_{min}$. It is then important that the thimble emanates from the saddle point at an angle that is larger than that of the singular curve, as the thimble runs off to infinity in the upper right quadrant at an angle that asymptotically reaches $\pi/6$, see \cite{Feldbrugge:2017kzv} and also Fig. ref{fig:Robin2}. Only then can the thimble avoid crossing the singular line.  Thus we must ensure that the condition $\tan(\theta) > \tan(\phi)$ is satisfied. For $q_1 > \frac{2}{H^2},$ this happens for negative imaginary $\beta$ with
\begin{equation}
|\beta| < \beta_{max} = \frac{{2}}{{3 V_3 H^2}}\frac{{ 3 H^2 q_1 -4}}{{ H^2 q_1 -2}} \approx \frac{2}{V_3 H^2}\,.
\end{equation}
Thus for negative imaginary $\beta$ with magnitude between $\beta_{min}$ and $\beta_{max}$ (e.g. $\beta = -i/(V_3 H^2)$) the path integral is well defined and contains only the Hartle-Hawking saddle point(s).

%%%%%%%%%%%%%%%%%%%%%%%%%%%%%%%%%%%%%%%%%%%%%%%

\subsubsection{Perturbations}

Let us now explicitly verify that the perturbations are indeed suppressed, and that the propagator (or wavefunction) describes a stable universe. In our model we only have gravitational wave perturbations to deal with. For a single mode with fixed polarisation, the action at quadratic order is given by (see e.g. \cite{Feldbrugge:2017fcc,Feldbrugge:2017mbc})
\begin{equation}
S^{(2)} = \frac{V_3}{2} \int_0^1 dt \, N \Bigl[q^2 \frac{\dot{\phi}^2}{N^2} - l (l + 2) \phi^2 \Bigr] 
\end{equation} 
where $\phi$ denotes the magnitude of the perturbation, which has been expanded in spherical harmonics with $l\geq 2$ being the principal quantum number. The extension to a sum over all modes is straightforward. The equation of motion for $\phi$ is thus
\begin{equation}
\ddot{\phi} + 2 \frac{\dot{q}}{q} \dot{\phi} + \frac{N^2 }{q^2} l (l + 2) \phi = 0\,,  \label{eqpert}
\end{equation}
where we must also specify boundary conditions. On the final hypersurface we will choose a Dirichlet boundary condition, $\phi(t=1)=\phi_1.$ Then, at the saddle point, regularity alone will automatically pick out the stable Bunch-Davies mode \cite{Feldbrugge:2017fcc}. However, in general we must specify a boundary condition for perturbations around the off-shell geometries too. Let us first look at the general solutions to the equations of motion. In order to do so, it is convenient to rewrite the scale factor as
\begin{equation}
q(t)= H^2 N^2 (t - \gamma)(t - \delta)\,, \label{background}
\end{equation}
where $\gamma, \delta$ can be read off from \eqref{qsol}. Plugging \eqref{background} into eq. \eqref{eqpert} we find that two linearly independent solutions for $\phi(t)$ are $f(t)/\sqrt{q}$ and $g(t)/\sqrt{q}$ with
\begin{align}
f(t) &= [\frac{t - \delta}{t - \gamma}]^{\frac{\mu}{2}} [(1 - \mu)(\gamma - \delta)+ 2 (t - \gamma)] \\
g(t) &= [\frac{t - \gamma}{t - \delta}]^{\frac{\mu}{2}} [(1 + \mu)(\gamma - \delta)+ 2 (t - \gamma)] \label{modes}
\end{align}
and
\begin{equation}
\mu^2 = 1 - \frac{4l(l+2)}{(\gamma-\delta)^2 N^2H^4}\,.
\end{equation}
At the saddle point the parameters $\gamma , \delta $ and $\mu$ reduce to
\begin{align}
\mu (N_4) = (l + 1), \quad \gamma (N_4) = 0, \quad \delta (N_4) = \frac{- 2 i}{\sqrt{H^2 q_1 -1} -i}\,.
\end{align}
It is straightforward to check that this implies that the mode $f$ blows up at $t=0$ whereas $g/\sqrt{q} |_{t=0} = 0 $. Thus the no boundary geometry automatically selects the perturbative mode $g$ and $\phi_0 = 0 $. 
The regular solution at the saddle point reads
\begin{equation}
g(t) = \Bigl[ \frac{t(- i + \sqrt{H^2 q_1 -1})}{2 i + t(- i + \sqrt{H^2 q_1 -1})} \Bigr]^{\frac{l + 1}{2}} \Bigl( \frac{2 i l + 4 i + 2t(- i + \sqrt{H^2 q_1 - 1})}{\sqrt{H^2 q_1 - 1} - i} \Bigr) \label{stablemode}
\end{equation} 
with associated action
\begin{equation}
S^{(2)} = -\phi_1^2  \frac{ l(l + 2) q_1}{2i (l + 1) + 2 \sqrt{H^2 q_1 - 1}} = i \, \phi_1^2 \frac{l (l + 1) (l +2) }{2 H^2} - \phi_1^2 \frac{l (l + 2) \sqrt{q_1}}{2 H} + \mathcal{O} \Bigl( \frac{1}{\sqrt{q_1}} \Bigr)
\end{equation}
The resulting amplitude $e^{i S^{(2)}} $ describes Gaussian distributed perturbations with a scale-invariant spectrum. We conclude therefore that the model is stable against small deviations since large values of $\phi_1$ are exponentially suppressed.  

More generally, we can specify Robin boundary conditions for the perturbations.
The full action for the perturbations is then
\begin{equation}
S_{tot}^{(2)} = \frac{V_3}{2} \int dt [q^2 \frac{\dot{\phi}^2}{ N^2} - l (l+2) \phi^2] + \alpha_{\phi} \phi_0 + \frac{\phi_0^2}{2 \beta_{\phi}}
\end{equation}
which leads to 
\begin{align}
\delta \phi_1 &= 0 \\
\delta \phi_0 [- \frac{V_3 q_0^2 \dot{\phi}_0}{N^2} + \alpha_{\phi} +  \frac{\phi_0}{\beta_\phi} ] &= 0 
\end{align}
Any Robin boundary condition with $\alpha_\phi = 0$ and arbitrary $\beta_\phi$ will then retain the stable mode at the saddle point, while also specifying initial conditions for the perturbations off-shell. Since we are only summing over non-singular geometries off-shell, this will leave the path integral well-defined, and stable.

%%%%%%%%%%%%%%%%%%%%%%%%%%%%%%%%%%%%%%%%%%%

\subsubsection{Interpretation and discussion}

As already pointed out in the case of perturbations by Vilenkin and Yamada \cite{VY1,VY2}, the Robin boundary condition can also be implemented as an additional Gaussian integral 
\begin{align}
\int \, dN \, dq \, e^{iS} \int \, dq_0 \, e^{i\alpha q_0 - \frac{q_0^2}{2|\beta|}} \,,
\end{align}
where we must include an integration over the initial scale factor $q_0.$ Owing to the fact that $\beta$ has to be negative imaginary, this may then also be interpreted as an initial coherent state, albeit one with a Euclidean momentum $\alpha$. 
\begin{align}
\Psi &= \int dN \, {\cal D}q \, dq_0  \, e^{iS/\hbar} \, \Psi_0\,, \quad
\Psi_0 \propto e^{ i  \alpha q_0 -  \frac{q_0^2}{2 |\beta|}} \, . \label{robinstate}
\end{align}
The presence of a Euclidean momentum not only implements the idea of closing the geometry off in Euclidean time, but it also adds a positive weighting to the associated geometries. In this way we can obtain a final result with an enhanced weighting $e^{+2V_3/(\hbar H^2)}$, which in the implementation with Dirichlet boundary conditions was simply impossible \cite{Feldbrugge:2017mbc}. Note that in this context $\sqrt{|\beta|}$ takes on the role of the uncertainty in the initial size $q_0.$ And because we have a coherent state, the uncertainty in the initial momentum is then simply its inverse,
\begin{align}
\Delta q_0 = \sqrt{|\beta|} \sim \frac{1}{H}\,, \quad \Delta p_0 = \frac{\hbar}{\sqrt{|\beta|}} \sim \hbar H\,.
\end{align}
Thus we see that in order to have a well defined path integral, the uncertainty must be shared between the initial size and the initial momentum, with the uncertainty in the initial size being of the order of the Hubble length.\\
One can think of the original formulation of the no-boundary wavefunction as having as initial state a delta function centred at zero size $\delta(q_0)$, inevitably leading to unsuppressed fluctuations. We are dealing here instead with a Gaussian state peaked around zero, with a spread whose range is determined solely by the cosmological constant. This can be considered as a successful, minor modification of the original no-boundary wavefunction, where the HH geometry is actually dominant and singular geometries are avoided. Note however that the original state already encodes fluctuations of the spacetime geometry, i.e. the universe does not arise out of pure nothingness, but rather out of spacetime fluctuations. In some sense this is to be expected on grounds of the uncertainty principle, when applied to the spacetime geometry. 

The (canonical) Robin boundary term in the action can equally well be interpreted as arising from a different state than \eqref{robinstate}
\begin{equation}
\Phi_0 \propto e^{i  \delta q_0 -  \frac{(q_0 - q_i)^2}{2 |\beta|}} \, ,
\end{equation}
if $\delta = - \frac{q_i}{\beta} + \alpha  = - \frac{q_i}{\beta}  -  i  $.  This is possible because $\alpha$ and $ \beta $ are imaginary in this implementation of the no-boundary wavefunction, and the imaginary piece $- \frac{q_i}{\beta}$ can be absorbed into a redefinition of the momentum. 
In particular, if $ \bar{q}_i=- \delta \, \beta $ one can rewrite the state as 
\begin{equation}\label{key}
\Phi_0 \propto e^{ -  \frac{(q_0 - \bar{q}_i)^2}{2 |\beta|}} \, ,
\end{equation}
where now the mean momentum is zero, and the mean size is $\bar{q_i} =i\beta \approx \frac{2}{H^2} \neq 0.$ In this case the central values of the scale factor and momentum are real, yet we still obtain complex saddle points of the path integral because the real values chosen are classically impossible (classically, the momentum is only zero at the waist of the de Sitter hyperboloid, where $q=1/H^2,$ while here we would demand the momentum to be zero at a larger scale factor value). This rewriting reinforces the point that we can no longer interpret this result as tunneling from nothing, even if the dominant saddle points are the complex Hartle-Hawking saddles.

%%%%%%%%%%%%%%%%%%%%%%%%%%%%%%%%%%%%%%%%%%%%%%%%%%%%%%%%%%%%%%%%%%%%%%%%%%%%%%%%%%%%%%%%%%%%%%%%%%%%%%%%%%%%%%%%%%%%%%%%%%%%%%%%%%%

\subsection{The final Hubble rate as a boundary condition}\label{sec:hubblerate}

We have seen in the previous section how, by fixing an initial Robin condition for the path integral, it is possible to define a path integral peaked around the HH saddle point(s). Interestingly, the Robin boundary term can be interpreted as a coherent state. However, we also mentioned that this boundary term is not covariant: it was shown in \cite{Krishnan:2017bte} that the Robin problem for gravity in 4 dimensions is obtained from the boundary term
\begin{equation}
S_B = \frac{1}{\xi} \int_{\partial M} d^3 y \, \sqrt{h} \, ,\label{actionRob}
\end{equation}
where $\xi $ is a constant. With the ansatz~\eqref{FLRW} the Robin boundary term becomes $S_B = \frac{V_3}{\xi} q_1^{3/2}$ which clearly differs from \eqref{eq:Robin}. As a result, it is not clear how this formulation could be extended beyond the minisuperspace approximation. \\
In this section we will provide a formulation of the quantum cosmology integral which avoids this problem making use of an initial Neumann condition and a final covariant Robin condition. One of the reasons why we pick an initial Neumann and a final Robin condition is that other ways to implement covariant Robin conditions (such as initial covariant Robin or Robin conditions at both sides) lead to nonsensical results or unmanageable expressions. This however is not simply of mathematical interest, but the main advantage is in fact that such conditions are physically sensible, as they allow one to specify the Hubble rate on the final hypersurface. One might argue that this is in any case more realistic, since the flatness of the presently observed universe does not allow us to measure the size of the universe whereas we can obtain the Hubble rate rather directly from redshift measurements. The price to pay will be a redefinition of the integral over the lapse $ N $: the Euclidean momentum imposed on the initial hypersurface requires the contour of integration of the lapse function to be shifted away from the Lorentzian contour by a constant imaginary offset. The other drawback of this approach is that, while the canonical Robin boundary term could be interpreted as an initial coherent state of the universe, the covariant Robin term has less clear interpretation from a quantum mechanical point of view. \\
Equipping the Einstein-Hilbert action with a Robin boundary term evaluated on the final spatial surface $ S=S_{EH}+S_B $, the variational principle requires the boundary conditions 
\begin{align}
\frac{\dot{q}_0}{2 N} &= \pi_0\,, \\
\frac{\dot{q}_1}{N } + \frac{2 \sqrt{q}_1}{\xi} & = \alpha  \label{rob} \, ,
\end{align}
with generic $\alpha$ and $\pi_0$. The solution to the equation of motion satisfying these boundary conditions is 
\begin{equation}
\overline{q}(t) = H^2 N^2 t^2 + 2 N \pi_0 t + \frac{\xi^2}{4} (\alpha - 2 (H^2 N + \pi_0))^2 - N (H^2 N + 2 \pi_0) \, ,
\end{equation} 
and the total classical action is 
\begin{align}
\frac{S}{V_3} = &N^3 H^4 (1 - H^2 \xi^2) + N^2 (\frac{3}{2} H^4 \xi^2 (\alpha - 2 \pi_0) + 3 H^2 \pi_0)  \nonumber \\
& + N (- \frac{3}{4} H^2 \xi^2(\alpha - 2 \pi_0 )^2 + 3  \pi_0^2  + 3) + \frac{\xi^2}{8} (\alpha - 2 \pi_0)^3  \, .
\end{align}
In our coordinates \eqref{FLRW}, the Hubble rate is given by $\frac{\dot{q}}{2 N \sqrt{q}},$ and thus we can see from \eqref{rob} that if we set $\alpha=0,$ we may interpret $H_1=-\frac{1}{\xi}$ as the Hubble rate on the final hypersurface. Note that due to the closed spatial slicing of de Sitter space we should require that $H_1 \le H$ or, equivalently, $\xi^2 H^2\ge 1$. With vanishing $\alpha,$ the action can be usefully rewritten as
\begin{equation}
\frac{S}{V_3} = H^2 (1 - H^2 \xi^2) (N + \frac{\pi_0}{H^2})^3 + 3 (N + \frac{\pi_0}{H^2})  - \pi_0 \frac{3  + \pi_0^2}{H^2}\,.\label{actionoffset}
\end{equation}
It is clear from this expression that for real boundary conditions $\pi_0 \in \mathbb{R}$, the integrand $e^{i S}$ oscillates along the real $N$ line and is conditionally convergent. Note that, since $(1 - H^2 \xi^2) \le 0 $, the asymptotic regions of convergence for the lapse integral lie in the wedges between the angles $(\frac{\pi}{3},\frac{2\pi}{3})$, $(\pi,\frac{4\pi}{3})$ and $(\frac{5\pi}{3},2\pi)$. Thus the Lorentzian integral can be defined and calculated, using Picard-Lefschetz theory to deform the real $N$ line to the appropriate steepest descent contour \cite{Feldbrugge:2017kzv}. For example, with the boundary condition $\pi_0=0$ the saddle points are located on the real axis at $N_{\pm} = \pm \frac{H_1}{H^2 \sqrt{H^2-H_1^2}}$ and they describe the expansion of the universe from the waist of the de Sitter hyperboloid where $q(t=0)=1/H^2$ to a final hypersurface with Hubble rate $H_1=-1/\xi,$ according to $q(t)=H^2 N_\pm^2 (t^2-1+\frac{H^2}{H_1^2}).$

\begin{figure}
	\centering
	\includegraphics[width=0.45\textwidth]{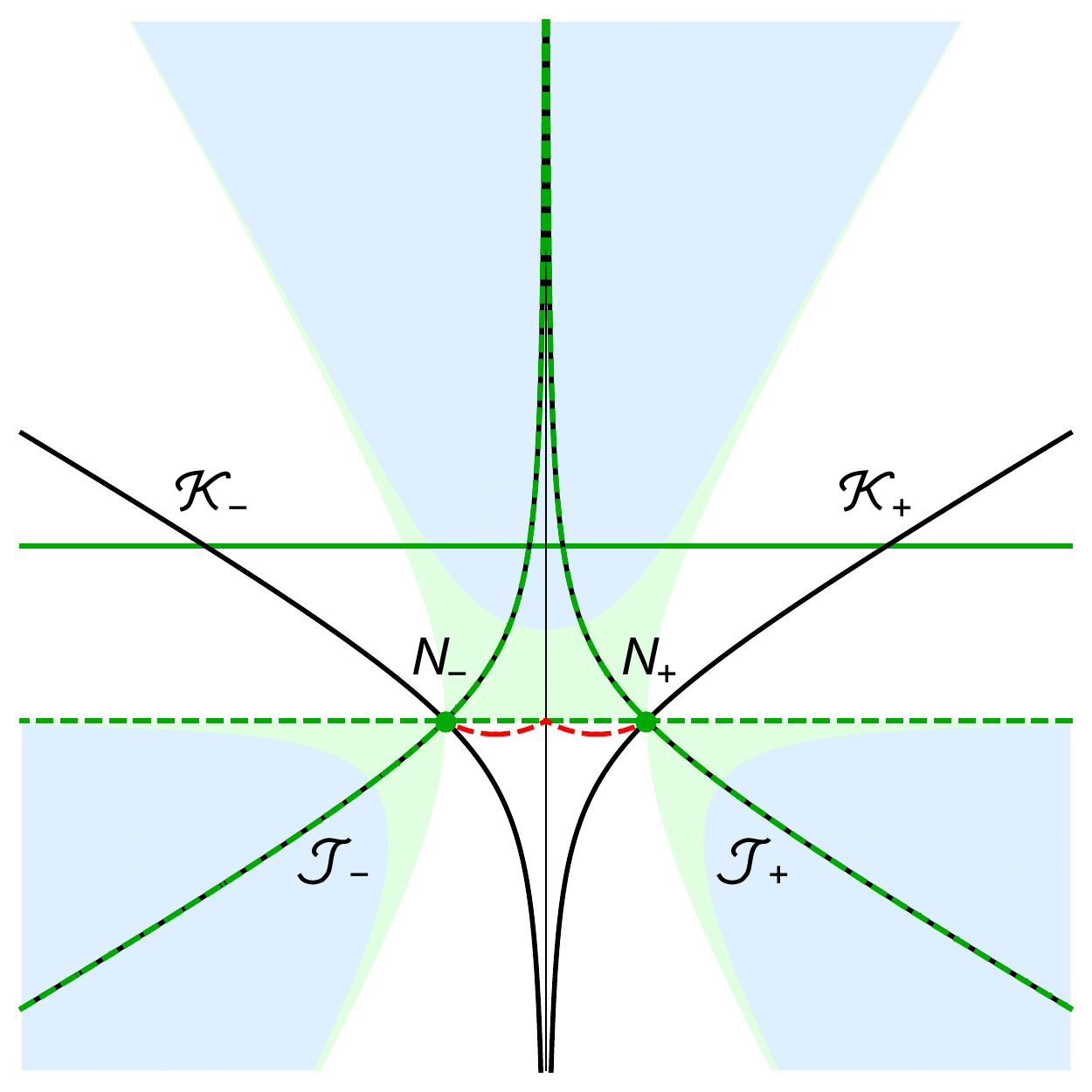} 
\includegraphics[width=0.45\textwidth]{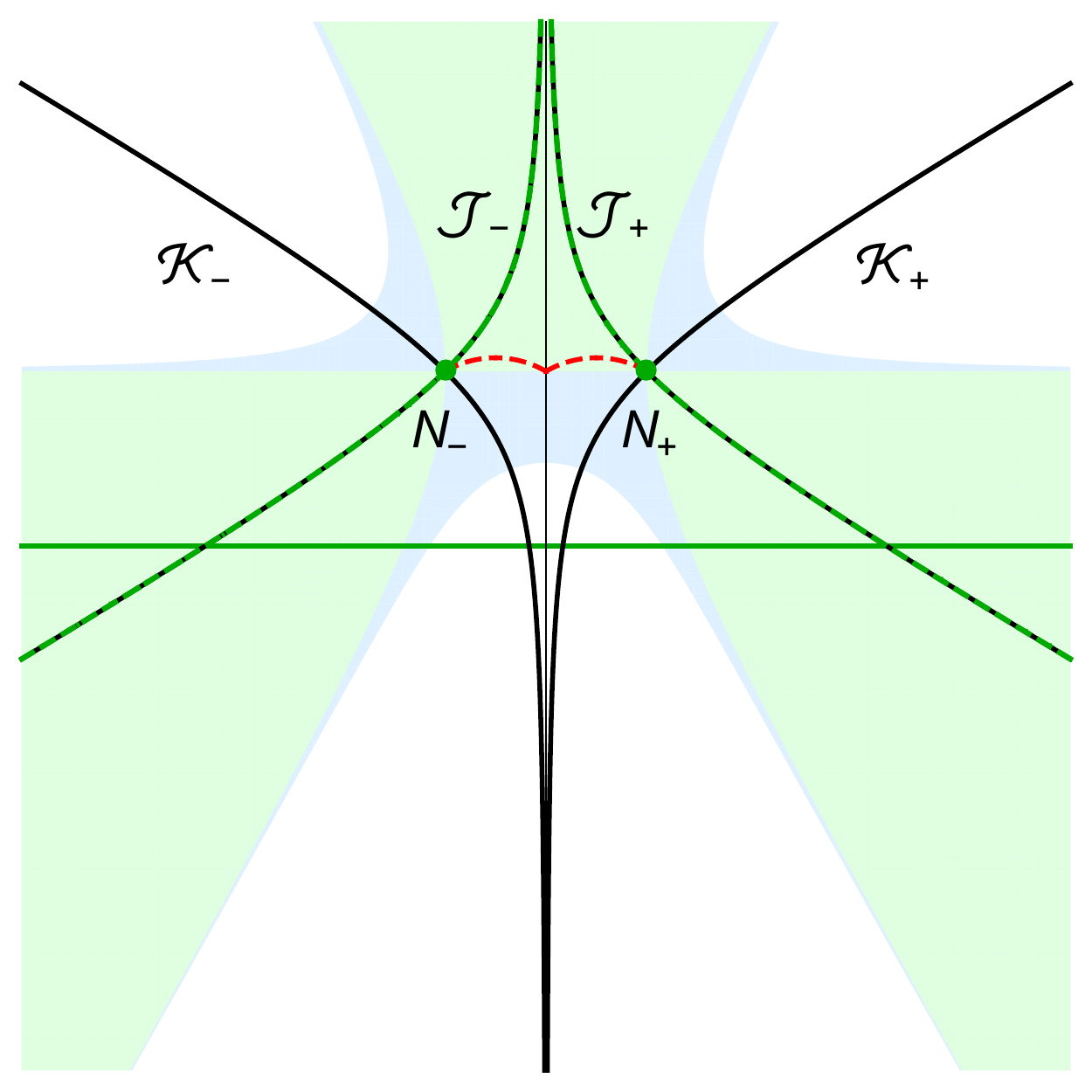}
	\caption{Flow lines and saddle points in the complex lapse plane, for a Neumann condition on $\Sigma_0$ and a covariant Robin condition on $\Sigma_1.$ {\it Left panel:} no-boundary wavefunction, $ \pi_0=i $.
	{\it Right panel:} Tunneling wavefunction, $ \pi_0=-i $. In both cases we set $H=1, H_1=1/2.$ In solid green, the real $N$ line; in dashed green, the shifted defining contour. We show in light green the regions of descent from the saddles and in light blue the regions where $ \text{Re}(iS)<0 $. The deformed contour runs along the steepest descent lines $ J_\pm $ and is shown in dashed green superimposed on the black steepest descent contours. The lines of zeroes (in red, dashed) are always avoided by the thimbles.
	}
	\label{fig:RobinFlow}
\end{figure}

For the no-boundary wavefunction however we need an imaginary initial momentum (see eq. \ref{initialmomentum}). When $\pi_0 \in i \mathbb{R}$ the integrand is oscillatory not along the real $N$ line, but rather (asymptotically) along the line parallel to the real axis with an offset given by $-\frac{\pi_0}{H^2},$ as can be seen by inspection from the action \eqref{actionoffset}. In fact, if $\pi_0 \in i \mathbb{R}^-$, the integral along the real line is convergent whereas for $\pi_0 \in i \mathbb{R}^+$ it explicitly diverges for large real $N$. Consequently, if we were to impose the Vilenkin momentum $\pi_0 = - i$ or any classically allowed boundary condition, the Lorentzian path integral would be mathematically well defined -- see also Fig. \ref{fig:RobinFlow}. However in order to implement the no-boundary wavefunction we need $\pi_0 = + i $, for which the integral along the real line diverges. 

A meaningful integral can be obtained by shifting the defining contour to the line $N=-\frac{i}{H^2} + x$ with $x \in \mathbb{R}$ (or potentially shifting the defining contour even below this line). This represents a departure from the exact Lorentzian integral, which is forced upon us by requiring the integral to be well defined. In some sense the departure is quite minimal, as the integration direction is still in the Lorentzian time direction. However, it is a clear departure as the defining sum is now over complex geometries. The extent to which this might constitute a problem may be debated. If we assume this new contour of integration, then the path integral will be equivalent to a sum over the two associated Lefschetz thimbles, as shown in Fig. \ref{fig:RobinFlow}. The thimbles are peaked on the HH saddle points
\begin{align}
N_{\pm} = \frac{1}{H^2}\left(\pm \frac{H_1}{\sqrt{H^2-H_1^2}} -i\right)\,,
\end{align} 
and consequently we again recover the Hartle-Hawking wavefunction
\begin{align}
\Psi \simeq e^{iS(N_-)/\hbar} + e^{iS(N_+)/\hbar}  = e^{+\frac{2V_3}{\hbar H^2}} \cos[  \frac{2 V_3 H}{\hbar} (q_1 - \frac{1}{H^2})^{3/2}]\,.
\end{align}

In analogy with the discussion of the previous section, we would now like to know whether the thimbles intersect any geometry that contains a singularity in the form of $q(t)=0$ for some real $t$ with $0<t<1.$  The equation $q(t) =0 $ is solved for 
\begin{equation}
t_{1,2} = - i \frac{H_1 \pm i \sqrt{H^2 - H^2_1} (1 - i H^2 N)}{H^2 N H_1}
\end{equation}
The imaginary part of $t_{1,2}$ vanishes respectively on the two circles $m^2 + n^2 + \frac{m}{H^2} \pm \frac{n H_1}{H^2 \sqrt{H^2 - H_1^2}} = 0$, where $ N=n + i m $. 
Moreover, the real part of $t_{1,2}$ should vary between $0$ and $1$, which imposes the condition $0< -\frac{H_1 m+\sqrt{H^2-H_1^2} \, n}{H^2 H_1 \left(m^2+n^2\right)} <1 $. This condition selects the arcs of the circumferences that link the saddle points (where $t _{1,2}= 0 $) to the point $ (n,m) = (0 , - \frac{1}{H^2})$ (where $t_{1,2} = 1$). 
Combining  the two conditions above, it is easy to see that the line of zeroes corresponds to the lower arcs of the two circles emanating from the saddles points, see Fig.~\ref{fig:RobinFlow}. In particular, one can verify that close to the saddle points the conditions impose $ m\leq \text{Im}(N_{\pm})=-1/H^2 $. Near the saddle point the curve of zeroes is approximated by the straight line
\begin{align}
q(\delta t)\mid_{N_{\pm} + \delta N} = 0 \rightarrow \delta N = \delta t  \, \frac{H_1^2}{(H^2 - H_1^2)} ( - \frac{i}{H^2} \mp \frac{\sqrt{H^2 - H_1^2}}{H^2 H_1})\,.
\end{align}
Thus at the two saddle points the line of zeroes forms an angle of $\tan(\theta) =\pm \frac{H_1}{\sqrt{H^2 - H_1^2}}$ with the horizontal, respectively. In other worlds, $ \theta \in (\pi, 3 \pi /2) $ for $ N_+ $ and $ \theta \in (0, -\pi/2) $ for $ N_- $.

As for the thimble, it is given by the equation
\begin{align}
\text{Re}\left( \mp 3\frac{H^2}{H_1^2} \sqrt{H^2-H_1^2}(\delta N)^2\right)=0 \, .
\end{align}
Since $H_1\le H$, the flow lines again point in the directions $e^{i\pi/4},e^{i3\pi/4}, e^{i5\pi/4}, e^{i7\pi/4}.$ In this case however the steepest descent path points at $3\pi/4$ radians away from the saddle point $N_+$ (and $\pi/4$ from $N_-$). 
Thus the thimble and line of zeroes avoid each other, and numerical calculations confirm this beyond leading order for the entire trajectory traced by these lines (see Fig.~\ref{fig:RobinFlow}).

In summary, the no-boundary wavefunction can be defined with an initial Neumann condition and a final covariant Robin condition. The latter has the physical interpretation of fixing the  Hubble rate on the final slice. The thimbles associated with the HH saddles also avoid singular geometries everywhere, so that the curvature is everywhere bounded and general relativity can be trusted at every step. The price to pay is a redefinition of the lapse integral. With the standard choice, a contour coinciding with the real line, the integral would have been divergent. The integral is convergent if the contour is shifted by an imaginary offset of at least $-i/H^2$. The important feature is that the initial Neumann condition eliminates the saddle points with unstable fluctuations, leaving only the stable HH saddles. Meanwhile, the physically attractive final condition of imposing the current Hubble rate rather than the (currently unobservable) size of the universe eliminates any potential interference of singular geometries.

%%%%%%%%%%%%%%%%%%%%%%%%%%%

\subsection{Homogeneous Transitions during Inflation}\label{papersebastian}

Robin types of boundary conditions proved to be useful for defining the no boundary proposal in the previous sections, and the initial conditions of inflation in section \ref{InitialConditions}. This fact is perhaps not so surprising. The canonical Robin condition can in fact be seen as an initial coherent state.
Given that coherent states saturate the uncertainty principle and are the most classical of quantum states, it is no wander that they play an important role in defining the semi-classical limit of quantum cosmology. In the rest of this chapter, which is based on \cite{Bramberger:2019zks}, we are going to use Robin types of boundary conditions to recover the limit of QFT in curved spacetime for cosmological models where large scalar field fluctuations are allowed.
 In the context of inflation, the QFT in curved spacetime approach is used not only for small fluctuations, but also for large fluctuations deep in the tails of the distribution. This is especially relevant for eternal inflation, where it is assumed that the quantum fluctuations of the inflaton can be larger than its changes due to classical evolution \cite{Steinhardt:1982kg,Vilenkin:1983xq}. Although such large fluctuations are rare, they may play an important role in the cosmological context as they can alter the global structure of spacetime: in a region where the inflaton jumps up the potential, the expansion rate of the universe will be larger than before, and this will cause that region to grow significantly more than the classical evolution would have suggested. It is notoriously hard to make predictions for observables under these circumstances (see e.g. \cite{Aguirre:2007gy,Johnson:2011aa} and references therein), and this provides further motivation for trying to understand such large quantum fluctuations in more detail.

In this work we will undertake a first step in the direction of understanding inflationary fluctuations in semi-classical gravity, where the background is quantized alongside the fluctuations. We achieve this by working with exactly solvable minisuperspace models in which gravity is coupled to a scalar field with a specific inflationary potential \cite{Garay:1990re}. The fact that we are working in minisuperspace, and that we consequently only consider homogeneous fluctuations of the fields, is a restriction that we hope to improve on in future work. However, on super-Hubble scales such an approximation should be rather accurate by simple virtue of causality (cf. also the stochastic picture of super-Hubble fluctuations \cite{Starobinsky:1986fx}). 

Our goal then is to describe homogeneous inflationary transitions, both small and large, in a fully quantum manner. This will be particularly useful in the context of eternal inflation where the universe's history is not well approximated by classical evolution. The framework that we employ allows us to see how the fields evolve ``during'' a quantum transition, and we will see how the transition amplitude depends not only on the change in the scalar field, but also (though to a lesser extent) on the change in the scale factor. The key feature of our calculation is the use of (canonical) Robin boundary conditions. This allows us to follow the semi-classical evolution of a universe which has a large enough initial size and is initially inflating. In order for these requirements to be compatible with Heisenberg's uncertainty principle, the initial size and velocity are specified only with some uncertainty. This is implemented by the canonical Robin condition which is in fact equivalent to an initial coherent state. A general feature that we observe is that the transition amplitude is governed by contributions from two saddle points when the uncertainty in the initial value of the scalar field is small, but with large uncertainty in the inflaton velocity. In this case a description in terms of QFT in curved spacetime in fact breaks down, as two separate backgrounds contribute significantly. However, as soon as the uncertainty in the field value is increased to the expected level ($H/(2\pi)$) while the uncertainty in the field momentum is correspondingly reduced, we generically see that a so-called Stokes phenomenon happens: this is a topological change in the (steepest descent) flow lines, beyond which only a single saddle point remains relevant to the path integral, and where consequently the approximation in terms of QFT in curved spacetime is vindicated. However, in the flattest region of the potential even this is not quite enough, and some additional uncertainty in the size of the universe is required in order to obtain consistent results.

The rest of this chapter will be based on \cite{Bramberger:2019zks} and is organized as follows: in section \ref{model} we present our formalism and the specific minisuperspace model we will focus on. In order to test this formalism, we will apply it in section \ref{test} to boundary conditions that correspond to a scalar field classically rolling down an inflationary potential. This example turns out to be non-trivial already, in that it demonstrates the need for, and the use of, an appropriate initial state. Equipped with these realisations we can then explore transitions during which the scalar field evolves up the potential, in section \ref{jump}. A further constraint on the validity of our calculations is analysed in section \ref{zeros}. We conclude with a discussion of our results in section \ref{discussion}. 

%%%%%%%%%%%%%%%%%%%%%%%%%%%%%%%%%%%%%%

\subsubsection{Exactly Soluble Scalar Field Minisuperspace Models} \label{model}

For gravity minimally coupled to a scalar field with a potential, the Feynman propagator in minisuperspace is given by
\begin{align}
G[a_1,\phi_1;a_0,\phi_0] = \int_{0^+}^{\infty} dN \int_{a_0}^{a_1}  \int_{\phi_0}^{\phi_1} Da D\phi e^{iS(a,\phi,N)/\hbar}\,.
\end{align}
This propagator describes the amplitude to go from an initial 3-surface with scale factor $a_0$ and scalar field $\phi_0$ to a final 3-surface specified by $a_1$ and $\phi_1$. The action here is given by the Einstein-Hilbert functional with a minimally coupled scalar field and the Gibbons-Hawking-York boundary term. Note that the last term is crucial to make the variational principle compatible with the mentioned Dirichlet boundary conditions. The full action reads
\begin{align}
\label{action-phys-coord}
S = 6 \pi^2 \int dt_p N \left( -\frac{a \dot{a}^2}{N^2} + a + \frac{a^3}{3} \left(\frac{1}{2}\frac{\dot{\phi}^2}{N^2} - V \right) \right) 
\end{align}
where we used the usual metric of a closed FLRW universe with lapse $N$
\begin{align}
ds^2 = -N^2 dt_p^2 + a(t_p)^2d\Omega_3^2\,.
\end{align}
We take the range of integration of the lapse function to be over strictly positive and real values only. (A detailed discussion of the attractive properties of the Lorentzian path integral was provided in section \ref{sectionpathintegral} and discussed in a number of papers \cite{Feldbrugge:2017kzv,Feldbrugge:2017mbc}.) While the path integral is a very intuitive tool in computing amplitudes for the evolution of the universe, it is not used very much because in most situations it is difficult or impossible to compute it explicitly. In particular it is impossible to solve the above analytically for generic potentials of the scalar field $V(\phi)$. For certain specific forms of $V(\phi)$, however, exact solutions may be obtained. One class has been studied in \cite{Garay:1990re} and we shall review their approach here. Our goal is to transform the action \eqref{action-phys-coord} into a form that is quadratic in its variables such that we can solve the resulting path integral exactly. To do this, first consider a rescaling of the time coordinate,
\begin{align}
ds^2 = -\frac{N^2}{a(t)^2}dt^2 +a(t)^2 d\Omega_3^2,
\end{align}
followed by a redefinition of the fields \cite{Garay:1990re},
\begin{align}
x(t) \equiv a^2(t) \cosh \left( \sqrt{\frac{2}{3}} \phi(t) \right)\,, \label{redef1}\\
y(t) \equiv a^2(t) \sinh \left( \sqrt{\frac{2}{3}}  \phi(t) \right)\,. \label{redef2}
\end{align}
The inverse transformations are given by 
\begin{align}
a(t)=\left(x^2(t)-y^2(t) \right)^{1/4}\,, \qquad \phi(t) = \sqrt{\frac{3}{2}} \tanh^{-1} \left( \frac{y(t)}{x(t)}\right)\label{invtrfm}\,.
\end{align}
Then, for a potential of the form
\begin{align}
V(\phi) = \alpha \cosh \sqrt{\frac{2}{3}}  \phi \,, \label{Pot}
\end{align}
the action reduces to the remarkably compact form \cite{Garay:1990re}
\begin{align}
S = V_3 \int_0^1 dt N\left[ \frac{3}{4 N^2}  \left(y'(t)^2 - x'(t)^2 \right) +3 - \alpha x(t)  \right]\,, \label{eq:action}
\end{align}
where a prime refers to derivation with respect to the coordinate time $t,$ and we are choosing the range of the time coordinate between the initial and final hypersurface to be $0 \leq t \leq 1.$ Here we wrote the coordinate volume of the three-dimensional spatial slice as $V_3$ -- for the standard three-sphere we have $V_3=2\pi^2$ but here, for notational simplicity, we will use re-scaled coordinates such that $V_3=1$ (since we will be interested in situations where the scale factor is large, our calculations also apply with good accuracy to FLRW metrics with flat spatial slices, as long as the spatial volume is regulated to a finite value). The resulting equations of motion are 
\begin{align}
x''(t) = \frac{2\alpha}{3}N^2\,,   \indent
y''(t) = 0\,.
\end{align}
Imposing Dirichlet boundary conditions $x(0) = x_0$, $x(1) = x_1$, and $y(0) = y_0$, $y(1) = y_1$ (where these boundary values are related to the original boundary conditions $a_{0,1}, \phi_{0,1}$ via the definitions \eqref{redef1} and \eqref{redef2}), the resulting solutions are given by
\begin{align}
\bar{x}(t) &= \frac{\alpha}{3} N^2 t^2 + (x_1 - x_0 - \frac{\alpha}{3} N^2)t + x_0\,, \label{sol1}\\
\bar{y}(t) &= (y_1 - y_0)t + y_0\,. \label{sol2}
\end{align}
A general path that is summed over in the path integral can now be written as $x(t) = \bar{x}(t) + X(t)$ and similarly for $y(t).$ The path integral over $x$ can then be performed by shifting variables to $X,$ where the integral over $X$ is a simple Gaussian that can be evaluated exactly. After solving the $x$ and $y$ integrals in this manner we are left with an ordinary one-dimensional integral over the lapse only,
\begin{align}
G[x_1,y_1;x_0,y_0] = \int_{0^+}^{\infty} dN  P(N) e^{iS_0(x_0,x_1,y_0,y_1,N)/\hbar} \label{lapseintegral}
\end{align}
where $P(N)$ is a non-exponential prefactor (scaling as $1/N$), and the action $S_0$ is obtained by substitution of the solutions \eqref{sol1} and \eqref{sol2}, yielding
\begin{align}
S_0 = \frac{\alpha^2}{36} N^3  + N \left( 3 - \frac{1}{2}\alpha (x_0 + x_1) \right) + \frac{3}{4N}  \left( (y_1 -y_0)^2 - (x_1 - x_0)^2 \right)\,. 
\label{action-integr}
\end{align}
In order to evaluate the above integral, which is a conditionally convergent integral, we will make use of Picard-Lefschetz theory, which has seen review in the section \ref{picard}. The first step in evaluating the propagator \eqref{lapseintegral} is to identify the saddle points of the integrand. Since we will be interested in the leading semi-classical approximation, we can neglect the prefactor $P(N)$ from this point onward, as it will not affect the saddle points of the integrand at leading order in $\hbar.$ The saddle points obey the condition
\begin{align}
\frac{\partial S_0}{\partial N} = \frac{\alpha^2}{12} N^2 + \left( 3 - \frac{1}{2}\alpha (x_0 + x_1) \right) - \frac{3}{4N^2}  \left( (y_1 -y_0)^2 - (x_1 - x_0)^2 \right) = 0\,,
\end{align}
which has four solutions
\begin{align}
N_{c_1,c_2} = c_1  \sqrt{\frac{3}{\alpha^2 }} \sqrt{-6 + \alpha(x_0 + x_1) - c_2 \sqrt{I} }\,, \label{saddles}
\end{align}
where 
\begin{align}
I = \alpha^2 \left((y_1 -y_0)^2 - (x_1 - x_0)^2 \right) + \left(6 -\alpha (x_0 + x_1) \right)^2
\end{align}
and $c_1,c_2 \in \{-1,1\}$. As we will see below, for the cases of interest to us, these saddle points will either be all real, or two real and two pure imaginary. The subsequent analysis depends on the boundary conditions that are chosen. 

We will be interested in inflationary evolution, in two distinct cases: first, to set up our calculation and to check its validity, we will investigate the description of purely rolling down the potential. Afterwards, we will consider the case where the universe inflates, and then we will demand that the scalar field jump up the potential. 

\begin{figure}[h]
	\centering
	\includegraphics[width=0.4\textwidth]{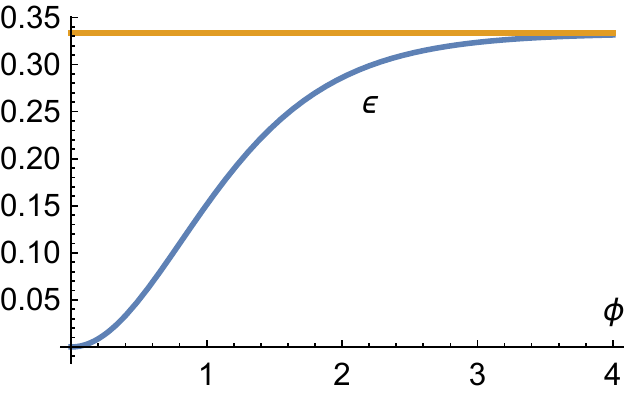} \hspace{1.5cm}
	\includegraphics[width=0.4\textwidth]{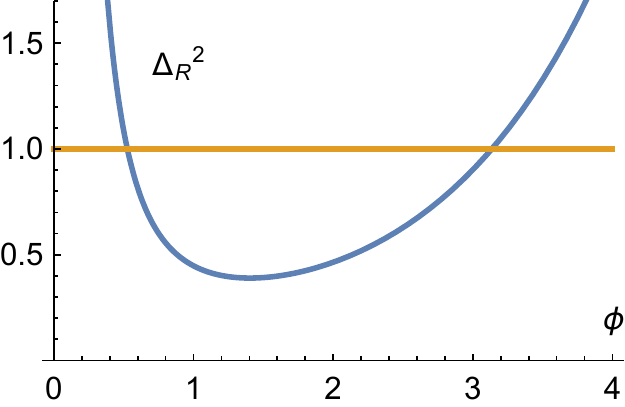}
		\caption{These plots show the flatness $V_{,\phi}^2/(2V^2)$ (left panel) and the variance of the curvature perturbation (right panel) for our potential \eqref{Pot} with $\alpha=1/10.$ Slow-roll is achieved only for small values of $\phi,$ and for small $\phi$ we are also in the conjectured regime of eternal inflation. There is a second regime of large variance at larger values of $\phi \gtrapprox 3,$ but here the potential quickly exceeds the Planck energy density, so that we will ignore this region in the present work. The yellow line on the left indicates the asymptotic value of $\epsilon$ for large $\phi$. On the right the yellow line separates the regimes where eternal inflation is expected from those where it is not.} \label{fig:eternal-infl-regime}
\end{figure}

Before continuing, we should add a note about the potential we are using, namely $V(\phi) = \alpha \cosh \left(\sqrt{\frac{2}{3}}\phi \right).$ In Fig. \ref{fig:eternal-infl-regime} we have plotted the flatness of the potential (more specifically, we have plotted $V_{,\phi}^2/(2V^2)$ which in the slow-roll limit coincides with $\epsilon$) as well as the variance of the curvature perturbation, for $\alpha=1/10$. Here we can see that inflationary solutions can be achieved throughout, but slow roll is only applicable for very small $\phi \lessapprox 0.2.$ Meanwhile the variance becomes large both for small field values $\phi \lessapprox 0.5$ and for very large values $\phi \gtrapprox 3,$ although these specific numbers will change for other choices of $\alpha.$

%%%%%%%%%%%%%%%%%%%%%%%%%%%%%%%%%%%%

\subsubsection{Inflation - Rolling Down the Potential} \label{test}

\begin{figure}
\begin{center}
\includegraphics[width=0.5\textwidth]{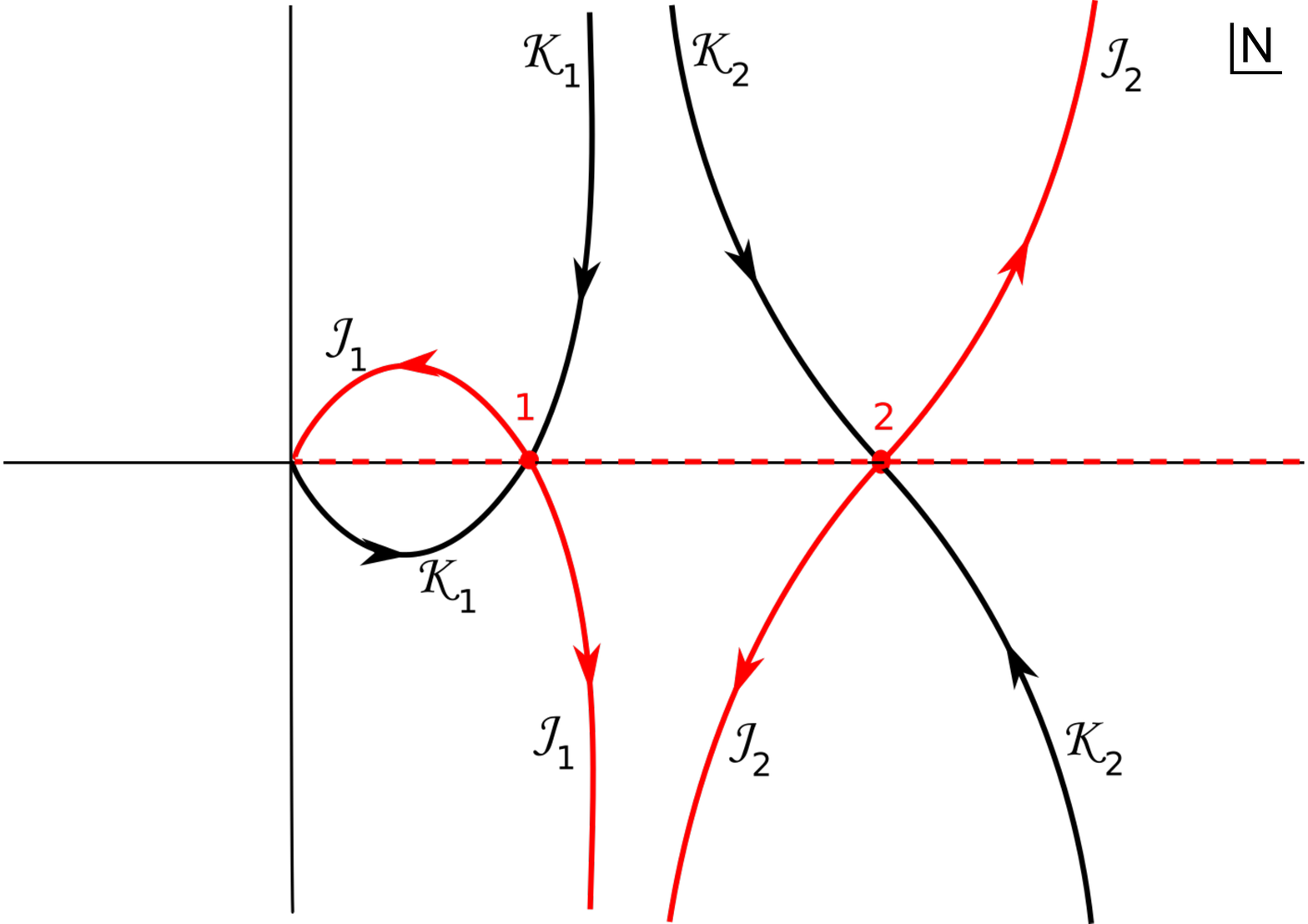}
\caption{The figure shows a typical example of the saddle points and the flow lines in the complex $N$ plane. The $\mathcal{J}_s$/$\mathcal{K}_s$ lines are the steepest descent/ascent paths associated with the saddle point $s,$ where arrows indicate downwards flow. The integral along the positive real $N$ line (dashed line) is equivalent to the integral along the path $\mathcal{J}_1 + \mathcal{J}_2$ (full red line). Both saddle points are relevant to the path integral.}
\label{fig:nfl-pl}
\end{center}
\end{figure}

Now that we have set up our model, we can evaluate transition amplitudes with various boundary conditions. In fact, in the present paper we will only look at homogeneous configurations. This is because on the one hand, this restriction brings about a significant technical simplification, and on the other hand it is suggested as a reasonable approximation (in a suitably sized patch of the universe) by the calculations of stochastic inflation, as discussed in the introduction. In order to test our formalism, we will start with a situation in which the universe is expanding while the scalar field is rolling down the potential, i.e. we start with a situation in which we expect there to exist a classical inflationary solution. Thus at first we will pick Dirichlet boundary conditions with
\begin{align}
a_1 > a_0 \,,\qquad  \qquad \phi_1 < \phi_0 \,,
\end{align}
where we will stick to the $\phi \geq 0$ side of the potential, and we will assume that the scale factors are larger than the de Sitter radius implied by the potential, $a_{0,1} > \sqrt{3/V(\phi_{0,1})}$. For boundary conditions such as these, the action \eqref{action-integr} admits four real saddle points, two at positive values of the lapse function, and two at negative values, as given by eq. \eqref{saddles}. The two saddle points at positive $N$ are trivially relevant to our path integral, since they lie on the original integration contour -- see Fig. \ref{fig:nfl-pl} for an illustration. The figure also shows the associated paths of steepest descent, and the original integration contour along $\mathbb{R}^+$ can indeed be deformed into the sum of these two steepest descent contours.  Superficially, it may be surprising that there are {\it two} relevant saddle points because we expect only the inflationary solution, but upon analysing the saddle point geometries it becomes clear what is happening.

\begin{figure}[h]
	\centering
	\includegraphics[width=0.45\textwidth]{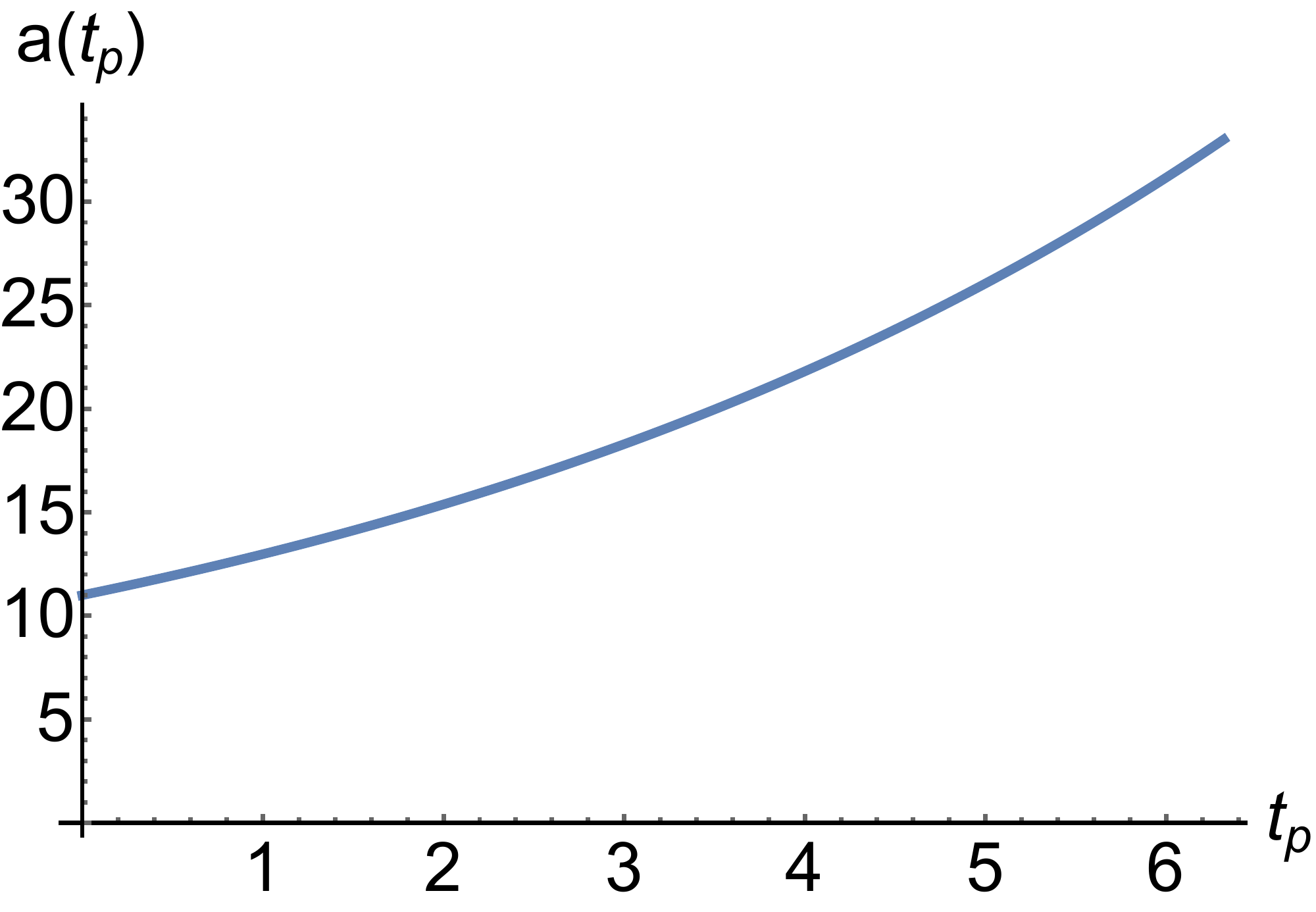}
	\includegraphics[width=0.45\textwidth]{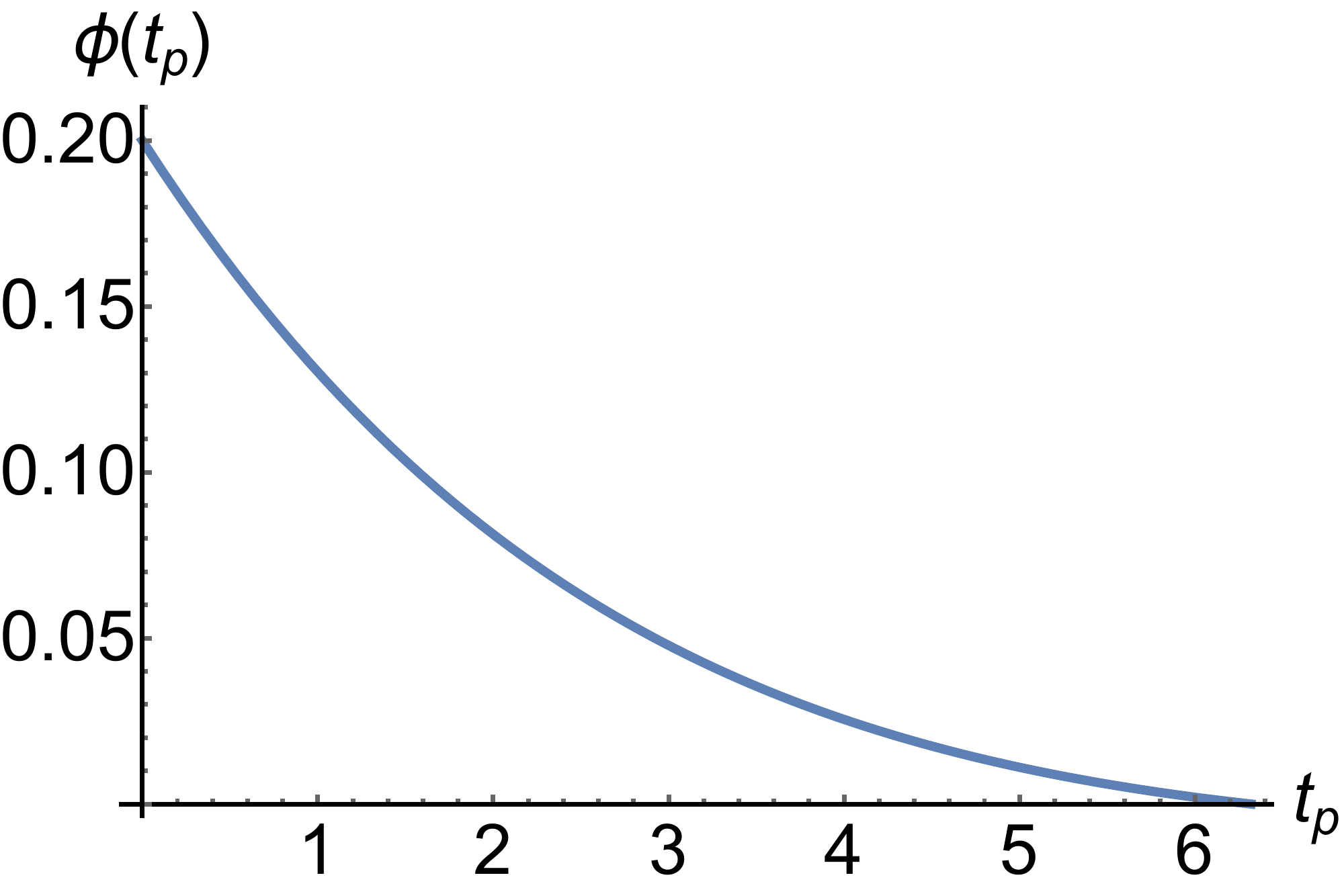}\\
	\caption{A typical example of the geometry at the saddle point $N_1$. In particular, here we have $\phi_0 = 2/10$, $\phi_1 = 0$, $a_0 = 11$, and $a_1 = 33$ corresponding to 1 e-fold of inflation and, as expected, we find inflationary behaviour of the scale factor and scalar field.} \label{fig:infl-sp1-geom}
\end{figure}

\begin{figure}[h]
	\centering
	\includegraphics[width=0.45\textwidth]{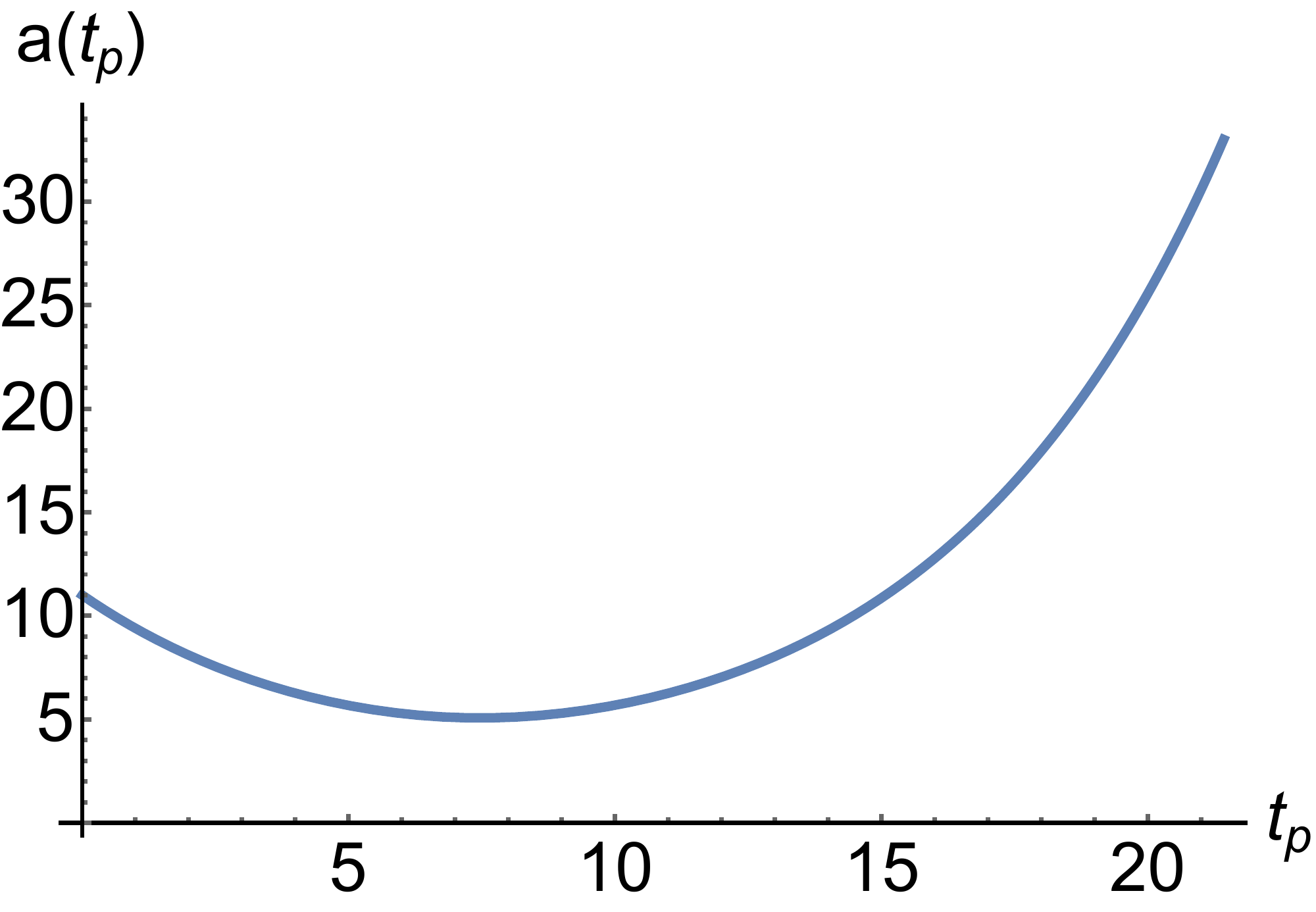}
	\includegraphics[width=0.45\textwidth]{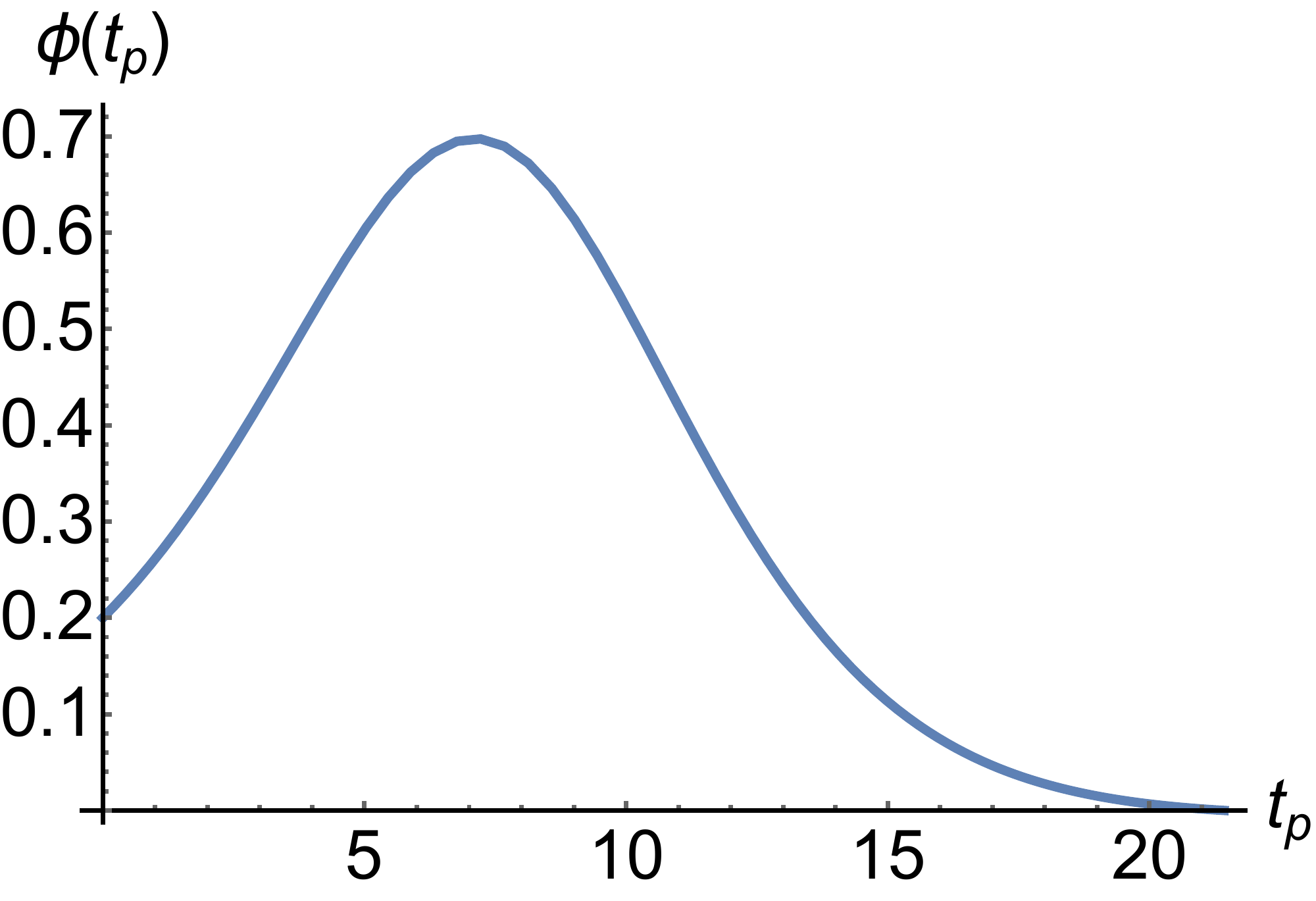}\\
	\caption{A typical example of the geometry of the saddle point $N_2$. In particular, here we have $\phi_0 = 2/10$, $\phi_1 = 1/10$, $a_0 = 11$, and $a_1 = 33$ corresponding to bouncing behaviour of the scale factor and scalar field.} \label{fig:infl-sp2-geom}
\end{figure}

The first solution, for smaller $N$, corresponds to an inflationary universe (an example of which is given in Fig. \ref{fig:infl-sp1-geom}). The second solution, the one for larger $N$, corresponds to a bouncing universe (see Fig. \ref{fig:infl-sp2-geom}). Note that due to the blue-shifting that occurs during contraction, the scalar field can initially roll up the potential, and then roll down again during the expanding phase. From these geometrical properties it also becomes clear why there are two solutions: the path integral simply finds all solutions corresponding to the given boundary conditions. It does not know about the prior evolution of the universe and hence picks out solutions consistent both with initial expansion and contraction. Note that a classical bouncing solution exists because we took the spatial sections of the metric to be closed, and hence the solution can be thought of as being a deformation of the de Sitter hyperboloid with the waist sitting in between the initial and final hypersurfaces. We should emphasise that in this situation, where two (real) saddle points contribute, an approximation in terms to QFT in curved spacetime does \emph{not} hold, since we are in the presence of \emph{two} relevant background spacetimes (cf. the analogous discussion regarding pure de Sitter space in chapter \ref{quantuminitial}, \cite{DiTucci:2019xcr}).

If we would like to single out the purely expanding inflationary solution we have to impose that the universe was already expanding with the scalar field rolling down the potential before we consider the transition computed through the path integral. In other words we need to include information not only about the initial values of the fields but also about their initial velocities. So far we have calculated the propagator with Dirichlet boundary conditions
\begin{align}
G[x_1,y_1;x_0,y_0] = \int_{0^+}^{\infty} dN  e^{iS(x_0,x_1,y_0,y_1,N)/\hbar}\,.
\end{align}
In this description we have complete certainty of the initial and final values of $x$ and $y$ or correspondingly of $a$ and $\phi$. On the other hand, the uncertainty principle implies that we have no knowledge of the initial and final velocities. 

We would now like to spread the uncertainty between positions and momenta imposing initial conditions where neither the value of the fields nor their conjugate momenta are specified but rather a linear combination of the two: 
 \begin{align}
c_1 x(0) + c_2 P_x(0) = c_3\,, \\
c_4 y(0) + c_5 P_y(0) = c_6\,.
\label{eq:robin-bcs-general}
 \end{align}
These are initial conditions of Robin type which require boundary terms in the action different from the Gibbons-Hawking-York one. To this effect, we will augment the action by additional boundary terms \cite{DiTucci:2019xcr,DiTucci:2019dji}, 
\begin{align}
S_{R} = S + p_x x_0 + p_y y_0 + \frac{i \hbar}{4 \sigma_x^2} (x_0 - x_i)^2 + \frac{i \hbar}{4 \sigma_y^2} (y_0 - y_i)^2\,,
\end{align}  
where $p_x ,p_y , \sigma_x$ and $\sigma_y$ are constants. The variation of the action now reads
\begin{align}
\delta S_{R} &= \int_0^1 N \left[ \frac{3}{2 N^2} \left( x''(t)\delta x - y''(t) \delta y \right) - \alpha \delta x \right] dt  \notag \\ &  - \frac{3}{2N}x'(t)  \delta x \Bigl |_0^1 +  \frac{3}{2N}y'(t)   \delta y \Bigl |_0^1+  \left(p_x + \frac{i \hbar}{2 \sigma_x^2} \left( x_0 - x_i\right) \right)  \delta x_0 +  \left(p_y + \frac{i \hbar}{2 \sigma_y^2} \left( y_0 - y_i\right) \right)  \delta y_0\,.
\end{align} 
Substituting the definitions of the momenta $P_x = - \frac{3}{2N}x'(t)$ and $P_y =  \frac{3}{2N}y'(t) $, the variational principle is satisfied if
\begin{align}
x_0 - \frac{2 \sigma_x^2}{i \hbar} P_x(0) = x_i - \frac{2 \sigma_x^2}{i \hbar} p_x \,,  \\
y_0 - \frac{2 \sigma_y^2}{i \hbar} P_y(0) = y_i - \frac{2 \sigma_y^2}{i \hbar} p_y 
 \label{eq:robin-bcs-special}
\end{align} 
at the initial boundary and if $x(1) = x_1, y(1) = y_1$ at the final boundary. Hence, comparing to the conditions (\ref{eq:robin-bcs-general}),
the action $S_R$ defines a mixed boundary value problem with a Dirichlet condition at $t=1$ and a Robin one at $t=0$. The Robin condition interpolates between Dirichlet (where the positions are known exactly) and Neumann (where the momenta are known exactly) as the parameters $\sigma_x$ and $\sigma_y$ are changed. 
For $\sigma_x, \sigma_y \to 0$ the boundary condition reduces to Dirichlet while for $\sigma_x, \sigma_y \to \infty$ it reduces to Neumann. 

In the following we will evaluate the path integral 
 \begin{equation}
 \int dN \int \,  \delta x  \, \int \, \delta y\, e^{i S_R/ \hbar} \label{robiN} 
   \end{equation}
with the mixed boundary conditions defined by $S_R$ for various values of $\sigma_x$ and $\sigma_y$ and explore the consequences in terms of the structure of the flow lines. Notice that the propagator (\ref{robiN}) can be interpreted as a convolution with an initial state 
\begin{align}
G[x_1,y_1;\psi_0] = \int \int G[x_1,y_1;x_0,y_0] \psi_0(x_0,y_0) dx_0 dy_0
\label{eq:convolved-propagator}
\end{align}
where $G[x_1,y_1;x_0,y_0]$ is the propagator evaluated with Dirichlet boundary conditions and the initial wave function reads
\begin{align}
\psi_0(x_0,y_0) = e^{\frac{i}{\hbar} \left(p_x x_0 + p_y y_0 \right) - \frac{(x_0 - x_i)^2}{4\sigma_x^2}- \frac{(y_0 - y_i)^2}{4\sigma_y^2}}\,.
\label{eq:wavefunction}
\end{align}
The functional form of this initial state is that of a coherent, Gaussian state, which allows us to express our knowledge of the initial uncertainty in the field values and their momenta. By construction the initial positions are peaked around the values $x_i$,$y_i$, with a Gaussian spread around them. In the limit where $\sigma_x = \sigma_y = 0$ the initial positions simply become $x_i$ and $y_i$ by construction. We are then back to the position representation which we were (implicitly) using up to now. Performing the Gaussian integrals over $x_0$ and $y_0$ gives us the saddle point solutions
\begin{align}
\bar{x}_0 &= \frac{\hbar N x_i - \alpha i N^2 \sigma_x^2 +2 i N p_x \sigma_x^2 + 3 x_1 i \sigma_x^2}{\hbar N+3 i\sigma_x^2 } \label{eq:x0-sp-approx} \,,\\
\bar{y}_0 &= \frac{\hbar N y_i +2 i N p_y \sigma_y^2 - 3 y_1 i \sigma_y^2}{\hbar N-3 i\sigma_y^2 }\,.
\end{align}
For small spreads $\sigma,$ we have $\bar{x}_0 \approx x_i, \bar{y}_0 \approx y_i,$ while for very large $\sigma$ we obtain 
\begin{align}
\bar{x}_0 \approx x_1 - \frac{\alpha}{3}N^2+ \frac{2N}{3} p_x\,, \qquad \bar{y}_0 \approx y_1 - \frac{2N}{3}p_y \qquad (\sigma_{x,y} \gg 1)\,. \label{largesigma}
\end{align}
Thus at large spreads $x_i, y_i$ disappear from the formula, which is an indication that the position is less well known. In fact at large $\sigma$ the momentum is determined with increasing precision.  To show this in more detail we focus on one of the momenta and variables ($p_x$ and $x$ respectively) but the result holds for both. Hamilton's equations give 
\begin{align}
P_x(t) = - \frac{3}{2 N}x'(t)= - N \alpha t - \frac{3}{2N} \left(x_1 - x_0 - \frac{1}{3} N^2 \alpha \right)
\end{align}
where the last line was obtained by plugging in the solution of the equations of motion for $x$. Thus, at the saddle points the initial momentum simply reduces to
\begin{align}
P_x(0) = - \frac{3}{2N} \left(x_1 - x_0 - \frac{1}{3} N^2 \alpha \right)\,,
\end{align}
which agrees with eq. \eqref{largesigma}. We may also find the sub-leading terms by making use of eq. \eqref{eq:x0-sp-approx}, plugging it into the general expression for the momentum and expanding for large $\sigma_x,$ to obtain
\begin{align}
P_x(0) = p_x + \frac{i \hbar}{\sigma_x^2}\frac{1}{6} \left( 3x_1 - 3x_i + 2Np_x - \alpha N^2 \right) + O\left(\frac{1}{\sigma_x^4} \right)\,,
\end{align}
which confirms that in the large $\sigma_x$ limit we reach the pure momentum representation.

\begin{figure}
\includegraphics[width=1\textwidth]{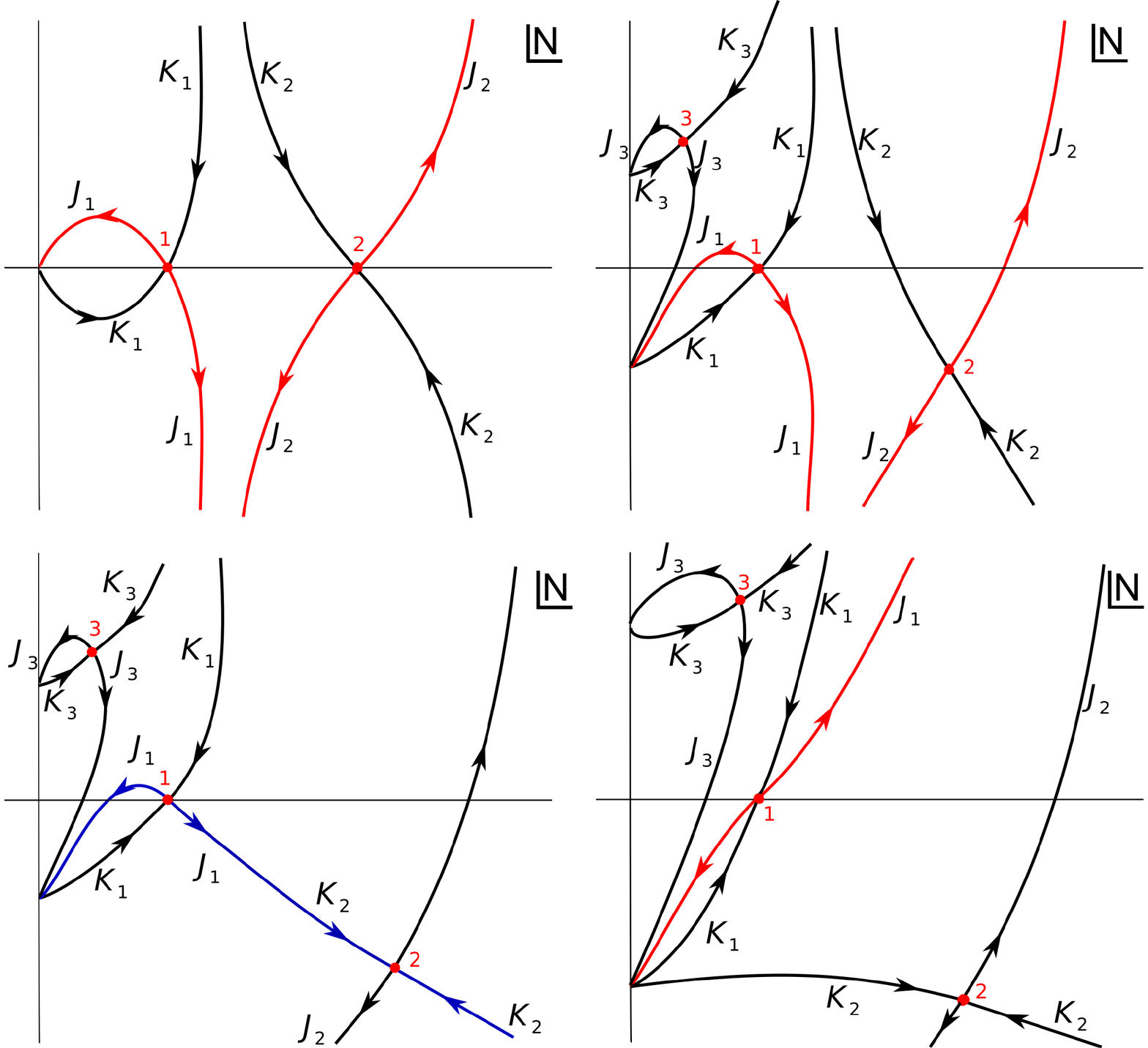}
\caption{The structure of the flow lines is shown as a function of the uncertainty $\sigma_\phi$ for inflationary boundary conditions, with $\sigma_{x,y}$ determined via eqs. \eqref{sigxphi}, \eqref{sigyphi}. In order to draw these graphs we have used the boundary conditions $a_0=100, \phi_0=1/10, a_1 = 200, \phi_1=1/100$ and $\alpha = 1/10$, with the corresponding momenta being given by $(\dot{a}(t_p),\dot{\phi}(t_p)) = (1.7953, -0.0820864)$. The numerically determined flow lines would have been difficult to put on a legible graph due to the large distances between saddle points, hence we have re-drawn these graphs to show the qualitative behaviour of the flow lines. Only the saddle points with $Re(N_i) >0$ are considered, being the only ones relevant for the flow analysis. Top left panel: For $\sigma_\phi=0$  both the expanding and the bouncing solutions are relevant to the path integral (corresponding to the saddle points $N_1$ and $N_2$). Top right panel: For non-zero $\sigma$ a new saddle point appears, the saddle point $N_2$ moves off the real line while $N_1$ maintains its original position (here $\sigma_\phi=0.0100$). For small enough $\sigma$ the original integration contour is deformed to the Lefschetz thimble $\mathcal{J}_1 + \mathcal{J}_2$. Bottom left panel: For a critical value of $\sigma = \sigma_c$ a Stokes phenomenon happens (here $\sigma_c \approx 0.0154$). The steepest descent path associated to $N_1$ ($\mathcal{J}_1$) coincides now with the steepest ascent through $N_2$ ($\mathcal{J}_2$). This is the Stokes line, the blue line in the figure. Bottom right panel: For $\sigma> \sigma_c$ the bouncing solution ($N_2$) no longer contributes to the path integral and only the inflating one ($N_1$) survives (here $\sigma_\phi=0.0200$).   }
 \label{fig:infl-pl-conv}
\end{figure}

Let us now return to our inflationary example. We choose initial momenta $p_x$ and $p_y$ such that $a$ is expanding and $\phi$ is rolling down the potential. The values of $p_x$ and $p_y$ are fixed such that they correspond to the classical inflationary solution that links our initial and final boundary conditions.   After performing the integrals over $x_0$ and $y_0$, we are again left with an integral over the lapse function $N,$
\begin{align}
G[x_1,y_1;\psi_0] \approx \int_{0^+}^{\infty} dN  e^{i\tilde{S}(x_i,x_1,y_i,y_1,p_x,p_y,\sigma_x,\sigma_y,N)/\hbar}
\end{align}
This new action results from having replaced $x_0$ and $y_0$ with their saddle point values $\bar{x}_0, \bar{y}_0$ in eq. \eqref{action-integr} and including the contributions from the initial state. More explicitly, we have 
\begin{align}
\frac{i}{\hbar}\tilde{S}& =\frac{i}{\hbar} S_0 + \frac{i}{\hbar}  (p_x x_0 + p_y y_0) - \frac{1}{4 \sigma_x^2} (x_0 - x_i)^2 - \frac{1}{4 \sigma_y^2} (y_0 - y_i)^2 \nonumber \\ &= 
\frac{i}{\hbar}\left[\frac{\alpha^2}{36} N^3  + N \left( 3 -  \frac{\hbar \alpha N (x_i+x_1) - \alpha^2 i N^2 \sigma_x^2 +2 i \alpha N p_x \sigma_x^2 + 6 \alpha x_1 i \sigma_x^2}{2(\hbar N+3 i\sigma_x^2) } \right) \right. \nonumber \\ & \left. \qquad+ \frac{3N}{4}  \left(\frac{\hbar  (y_i-y_1) +2 i  p_y \sigma_y^2}{\hbar N-3 i\sigma_y^2 }\right)^2 - \frac{3N}{4} \left(\frac{\hbar  (x_i-x_0) - \alpha i N \sigma_x^2 +2 i  p_x \sigma_x^2}{\hbar N+3 i\sigma_x^2 }\right)^2 \right] \nonumber \\
& + \frac{i}{\hbar}(p_x x_i + p_y y_i)  + \frac{ \alpha  N^2 \sigma_x^2 p_x -2  N p_x^2 \sigma_x^2 - 3 x_1 p_x \sigma_x^2}{\hbar(\hbar N+3 i\sigma_x^2) } +  \frac{-2  N p_y^2 \sigma_y^2 + 3 y_1  \sigma_y^2 p_y}{\hbar(\hbar N-3 i\sigma_y^2) } \nonumber \\
& + \left(\frac{- \alpha  N^2 +2  N p_x  + 3 (x_1-x_i)  }{4(\hbar N+3 i\sigma_x^2) }\right)^2 + \left( \frac{2  N p_y - 3 (y_1-y_i) }{4(\hbar N-3 i\sigma_y^2) }\right)^2 \label{actionfull}
\end{align}
The replacements of $x_0$ and $y_0$ have as a consequence that the dependence of the action on the lapse function $N$ has become more complicated. But once again we can solve this integral using Picard-Lefschetz theory. Let us start from small values of $\sigma_x$ and $ \sigma_y$ and investigate what happens as the spreads $\sigma_{x,y}$ are increased, see Fig. \ref{fig:infl-pl-conv}. At zero spread, we are in the pure position representation, with two relevant saddle points (upper left panel in the figure). But as soon as the spreads are turned on, the situation changes:  we now have six complex saddle points (three with positive real part, and three with negative real part) out of which two are relevant to the Lorentzian path integral, see the upper right panel in Fig. \ref{fig:infl-pl-conv}.  As we increase our certainty about the values of the initial momenta, the saddle points and flow lines change their location in the complex plane. Eventually a drastic transition occurs where the topology of the flow lines changes. This, so-called Stokes phenomenon, happens when a flow line connects two saddle points, for example in this case when
\begin{align}
Im(\tilde{S}(N_1)) = Im(\tilde{S}(N_2))
\end{align}
for two distinct saddle points $N_1$ and $N_2$. After this transition only one saddle point ($N_1$) remains relevant to the path integral, while the second one ($N_2$) has become irrelevant. The saddle point $N_1$, the only relevant critical point after the Stokes phenomenon, does not move at all as a function of $\sigma_{x}$ and $\sigma_{y}$. Furthermore the behaviour of the scale factor and scalar field at this location is inflationary as desired (see Fig. \ref{fig:infl-sp1-geom}), while the bouncing solution (Fig. \ref{fig:infl-sp2-geom}) has become irrelevant. This is entirely consistent with our interpretation of the initial state: as we increase our knowledge of the initial momentum (chosen to represent an expanding universe), only the expanding solution survives. Thus we see that the path integral gives sensible results for transitions in which the scale factor expands and the scalar field rolls down the potential. At the same time, we can appreciate the importance of the Robin initial condition in determining the outcome of future evolution.

%%%%%%%%%%%%%%%%%%%%%%%%%%%%%%%%%%%%%%%%%
\subsubsection{Jumping Up the Potential} \label{jump}

Inflation may be able to sustain itself indefinitely if the scalar field can jump up the potential, thus inducing a phase of enhanced accelerated expansion. In order to understand the true consequences of eternal inflation, it seems likely that a more fully quantum understanding of such transitions, and the associated issues of measures, must be developed. Here we take a step in that direction, by investigating the semi-classical geometries of such up-jumps. Thus we will now consider boundary conditions of the form
\begin{align}
a_1 > a_0 > \sqrt{\frac{3}{V(\phi_0)}}\,, \qquad  \qquad \phi_1 > \phi_0\,.
\end{align}
Again we must find the relevant saddle points, so that we can look at their geometries. Just as in the previous case in the Dirichlet limit $\sigma_{x,y} = 0$ we have four saddle points out of which two will be relevant for the path integral (the other two being the time reverses of the relevant two). 

\begin{figure}[ht!]
	\centering
	\includegraphics[width=0.9\textwidth]{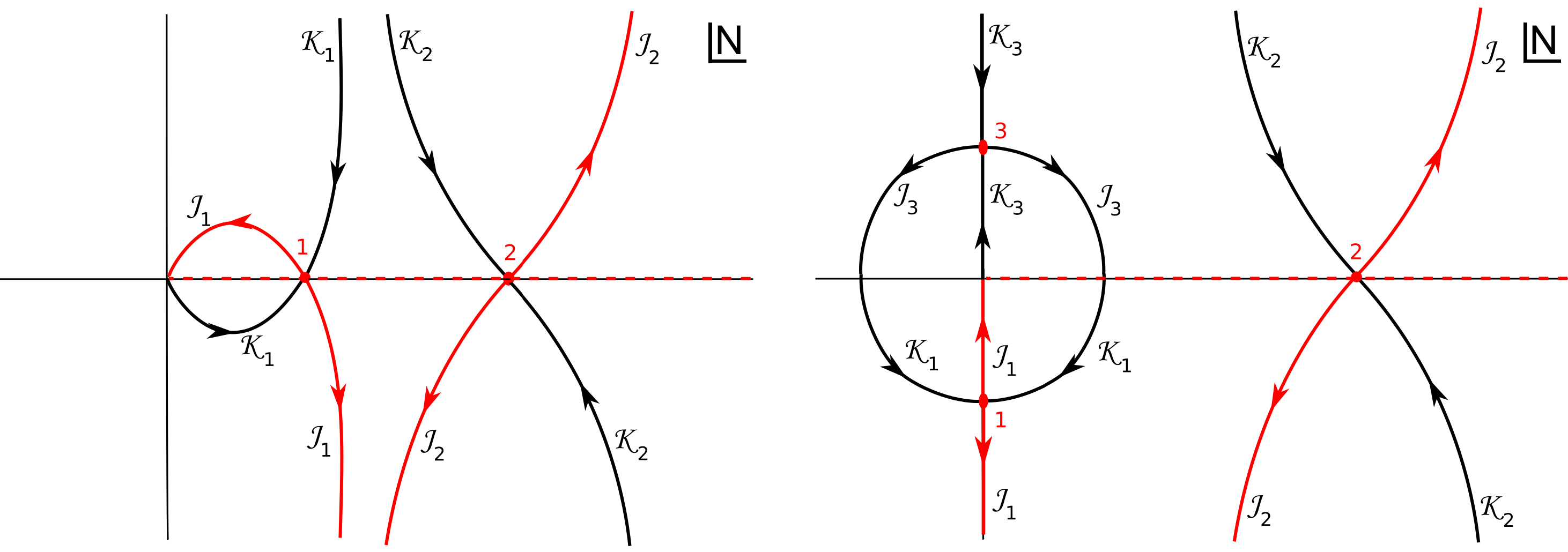}
	\caption{The figures show the structure of the steepest ascent and descent path ($\mathcal{J}_s$ and $\mathcal{K}_s$) for $\sigma_{x,y} = 0 $ and $ \phi_1 > \phi_0$. The left (case A) and the right (case B) panels show the only two inequivalent qualitative structures allowed for by up-jumping boundary conditions. In both cases the action has 4 critical points but only those with $Re(N_s) \geq 0$ are plotted. The original integration contour (dashed red line) can be deformed smoothly to the Lefschetz thimble $\mathcal{J}_1 + \mathcal{J}_2$ so that two saddle points contribute to the path integral. The geometries of these saddle points are plotted in Figs. \ref{fig:conv-sp-no-sigma} and \ref{fig:conv-sp2-no-sigma} for case B.}  
\label{fig:rollingup}
\end{figure} 

\begin{figure}[ht!]
	\centering
	\includegraphics[width=0.4\textwidth]{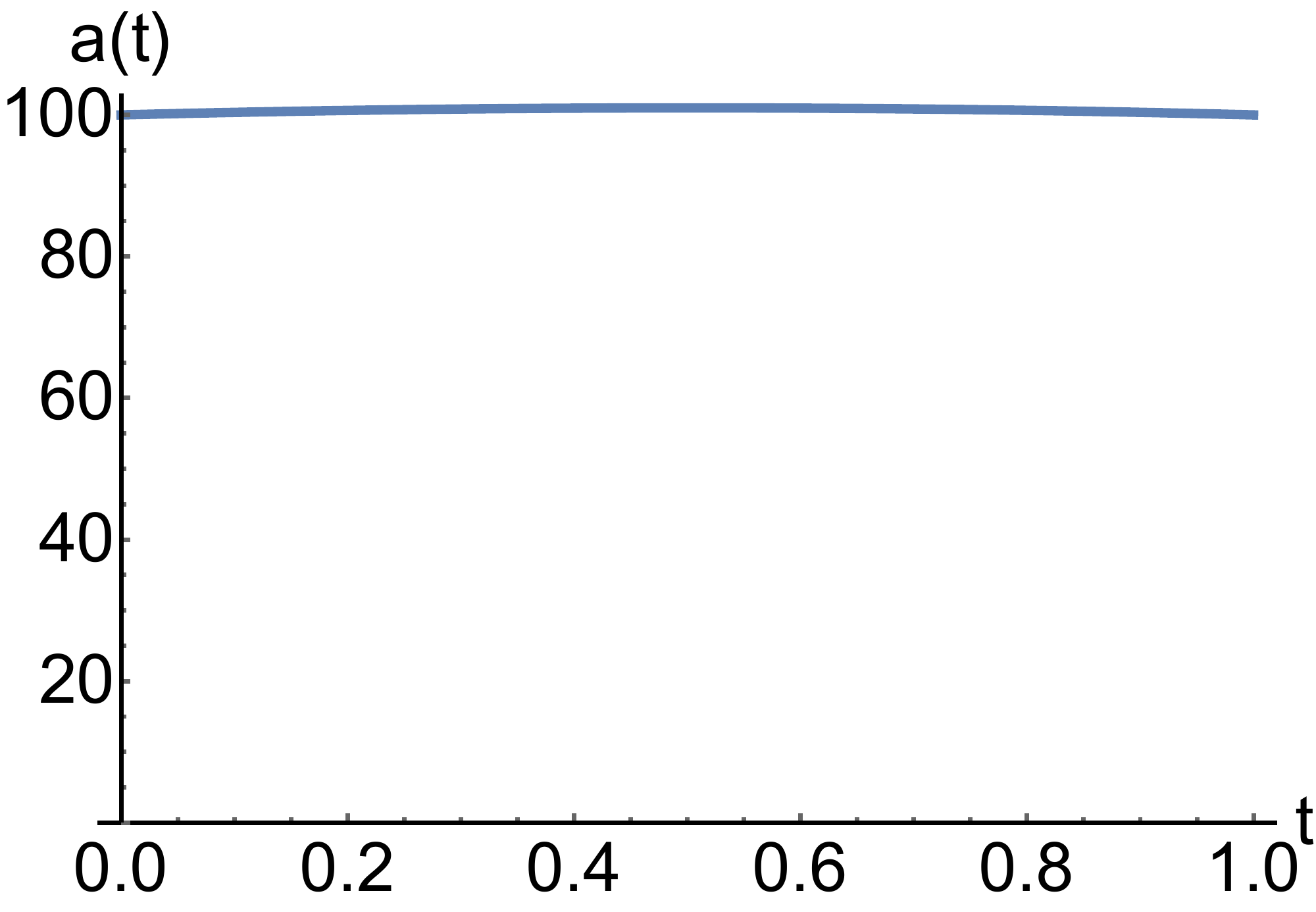}
	\includegraphics[width=0.4\textwidth]{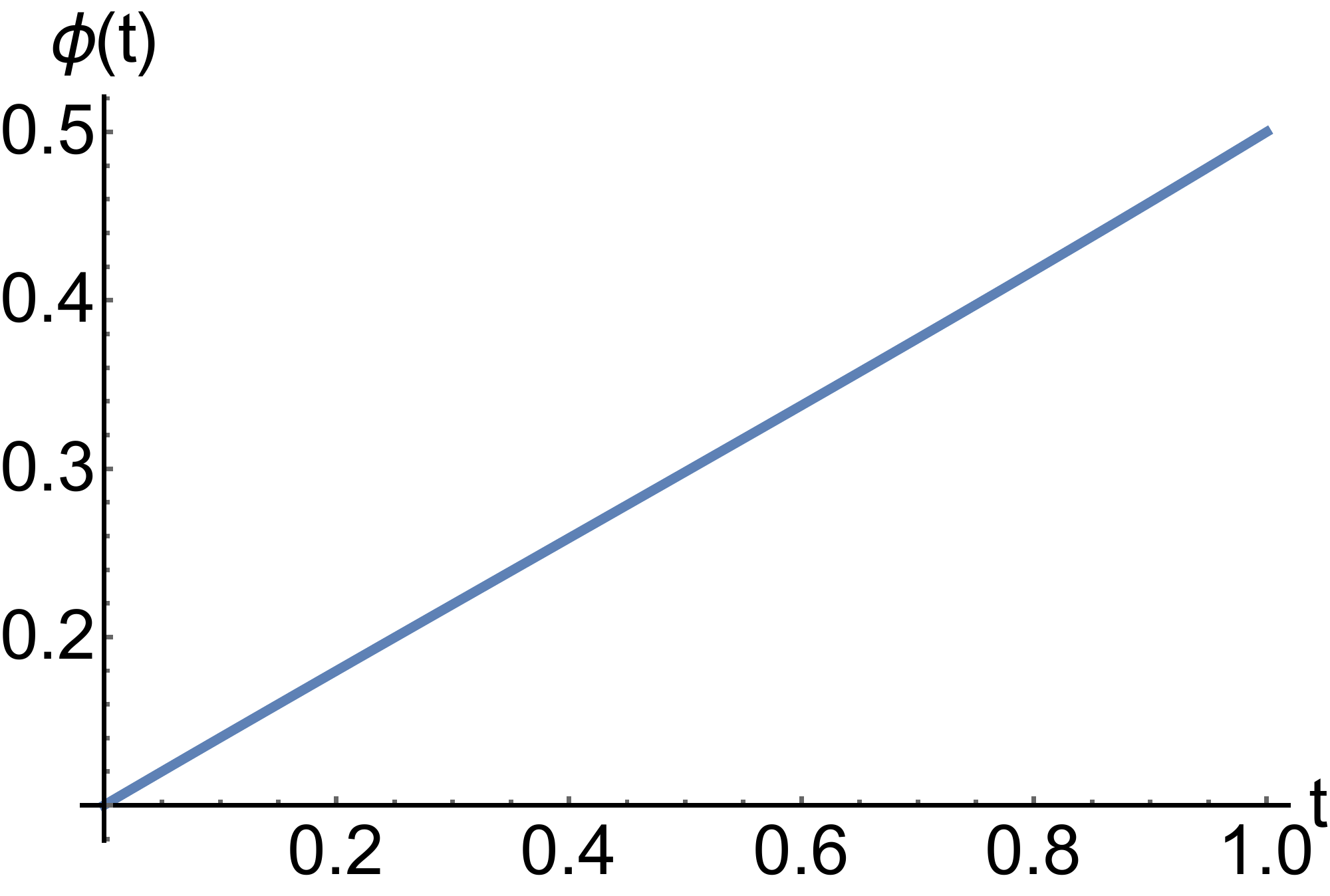}
	\caption{Geometry of saddle point number $1$ in the right panel of Fig. \ref{fig:rollingup}. Plotted here are the scale factor and scalar field with respect to coordinate time where we have chosen $\alpha = 1/10$, $\phi_i = 1/10$, $\phi_1 = 1/2$, $a_0 = 100$, $a_1 = 100$ and $\sigma_{\phi} = 0$. The saddle point is purely imaginary, and consequently the scale factor is Euclidean here.} \label{fig:conv-sp-no-sigma}
\end{figure}

\begin{figure}[ht!]
	\centering
	\includegraphics[width=0.4\textwidth]{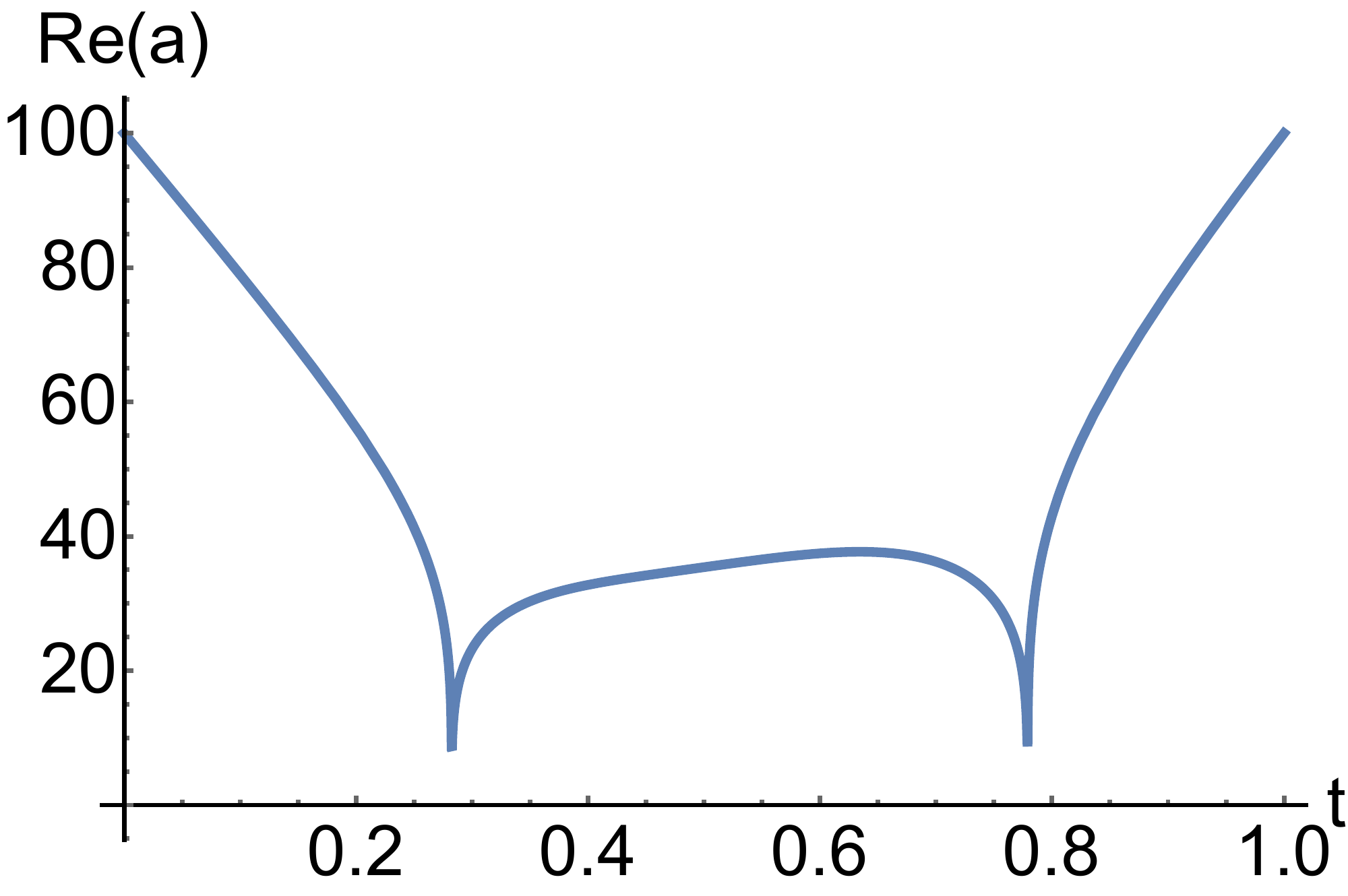}
	\includegraphics[width=0.4\textwidth]{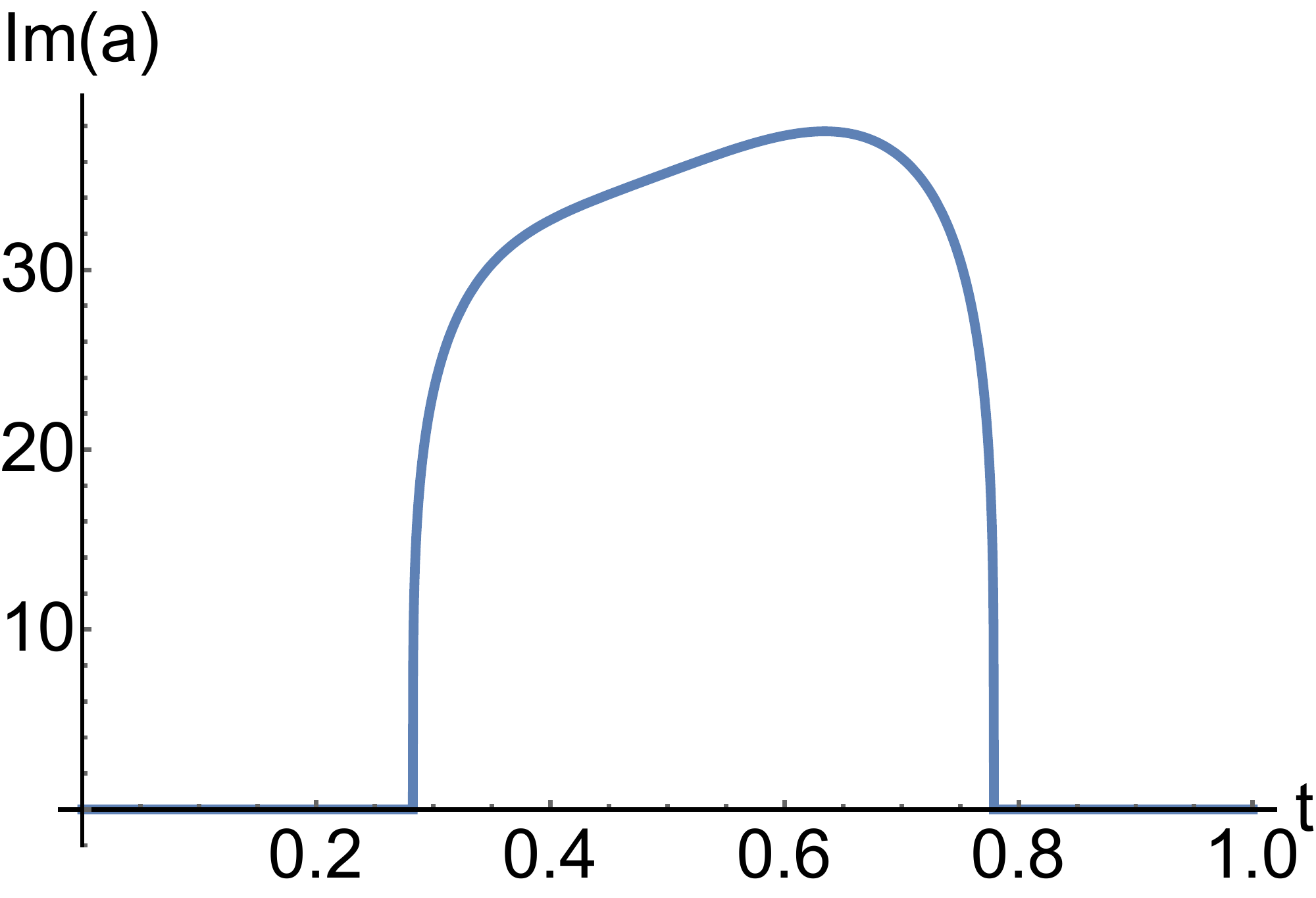}\\
	\includegraphics[width=0.4\textwidth]{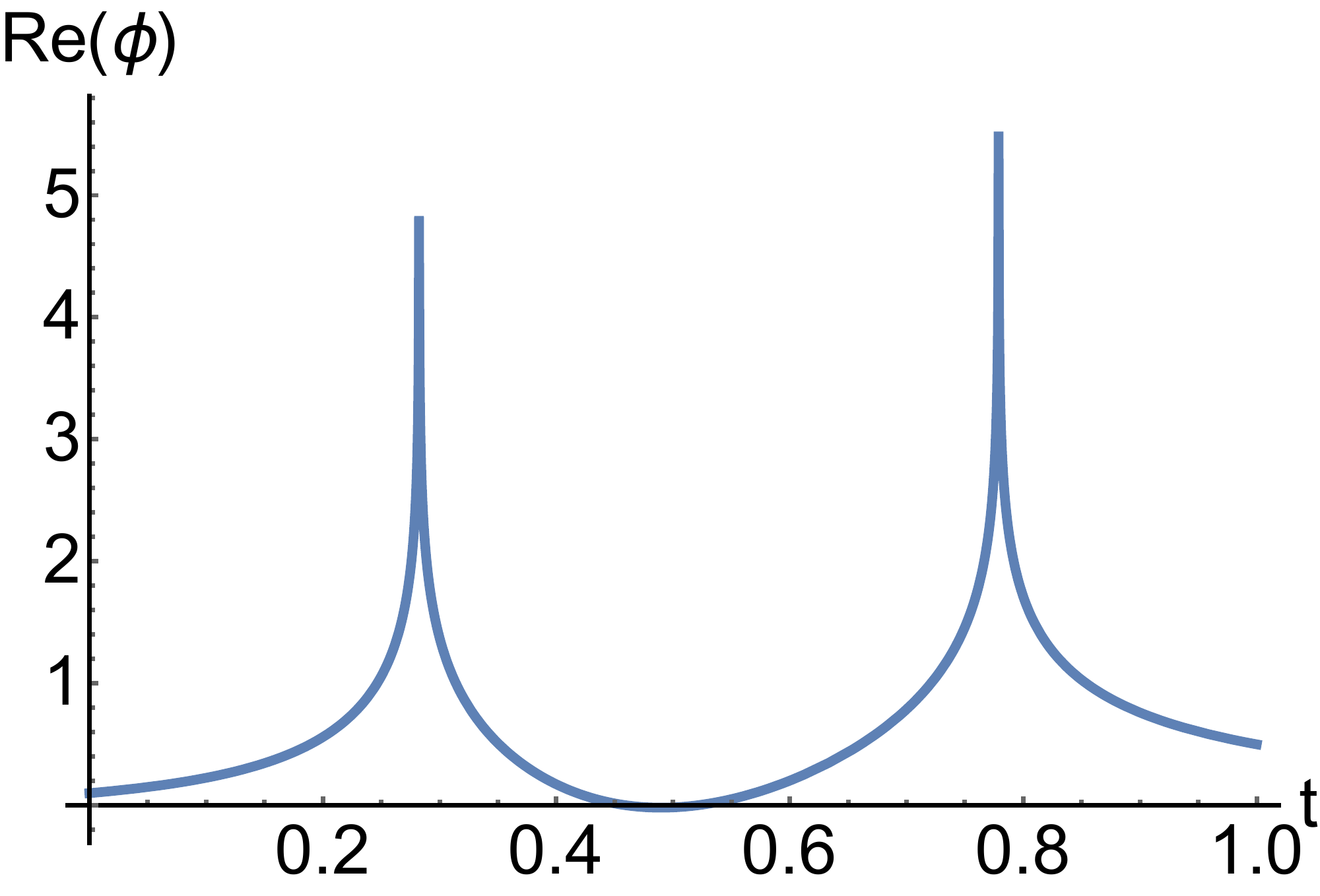}
	\includegraphics[width=0.4\textwidth]{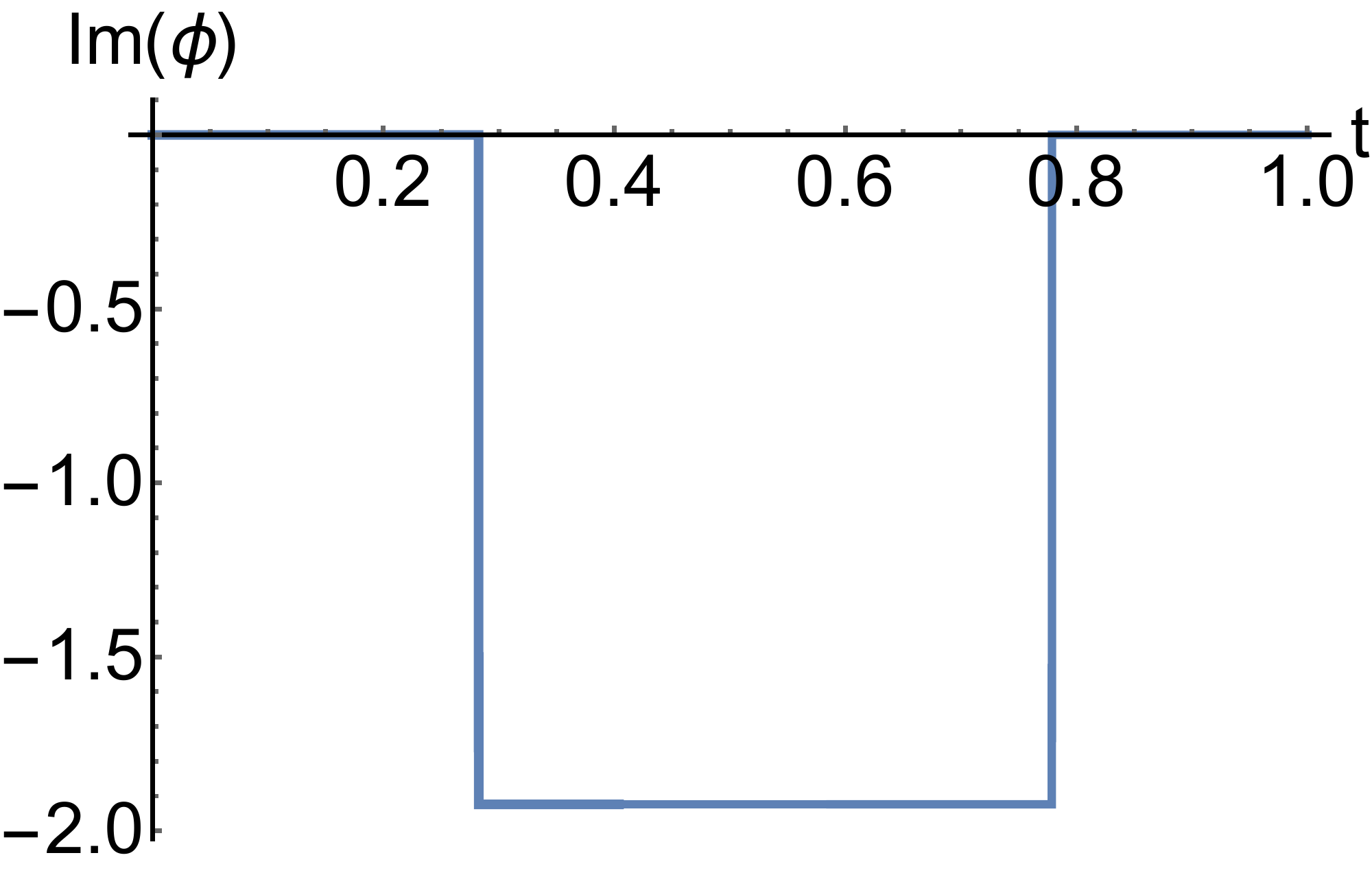}
	\caption{Geometry of saddle point number $2$ in the right panel of Fig. \ref{fig:rollingup}. The scalar field passes through zero twice, and at these singularities the scalar field blows up. Moreover, the scalar field starts out rolling up the hill, so that this geometry could only be relevant for a physical situation in which the scalar field would already have a large initial velocity up the hill, while we are interested in a prior state with the scalar slowly rolling down the potential.} \label{fig:conv-sp2-no-sigma}
\end{figure}

In fact, for different values of the initial conditions the saddle point $N_1$ can be either purely real or purely imaginary: we call these two possibilities case A and case B respectively. Case A is obtained for $a_1 \geq a_0 e^{\sqrt{6}(\phi_1-\phi_0)},$ otherwise we have case B. The second relevant saddle point $(N_2)$ always turns out to be real. Fig. \ref{fig:rollingup} shows the flow lines for the two possible inequivalent cases, while Figs. \ref{fig:conv-sp-no-sigma} and \ref{fig:conv-sp2-no-sigma} show the associated geometries for case B.

Note that the field values for all saddle points are strictly real, although for case B the saddle point $N_1$ is at a purely imaginary value, implying that the geometry is in fact Euclidean. This also means that the action at this saddle point is imaginary, and consequently the contribution to the path integral will be significantly suppressed compared to the saddle point $N_2.$ This second saddle point also has some peculiarities: it contains two singularities where the scale factor $a(t)$ passes through zero and where the scalar field blows up. On top of this difficulty, the scalar field starts out by rolling up the potential. Thus such a geometry may not be smoothly linked to a prior phase of inflation where the field is rolling down the potential. We can in fact show that no saddle point exists for which the scalar field is initially rolling down, but where it ends up higher in the potential. To see this consider the physical time derivatives of the scalar field and scale factor,
\begin{align}
\dot{\phi}(t_p) &= \frac{\phi'(t)a(t)}{N} = \sqrt{\frac{3}{2}}\frac{x(t)y'(t)-y(t)x'(t) }{N \left(x(t)^2 - y(t)^2 \right)^{3/4}} \,,\\
 \dot{a}(t_p) &= \frac{a'(t)a(t)}{N} = \frac{x(t)x'(t)-y(t)y'(t) }{2 N \left(x(t)^2 - y(t)^2 \right)^{1/2}}\,,
\end{align}
which at $t = 0$ reduce to
\begin{equation}
\begin{split}
\dot{\phi}_0 &= \frac{\alpha N^2 y_0 + 3x_0y_1 - 3x_1y_0}{\sqrt{6} N \left(x_0^2 - y_0^2 \right)^{3/4}}  \\
&= \frac{1}{\sqrt{6} N a_0}\left( N^2 \alpha \sinh\left( \sqrt{\frac{2}{3}} \phi_0\right) + 3 a_1^2\sinh\left( \sqrt{\frac{2}{3}} (\phi_1 - \phi_0) \right)  \right) \,,\label{phidot}
\end{split}
\end{equation}
\begin{equation}
\begin{split}
\dot{a}_0 &= \frac{x_0 (3 (x_1 - x_0) - \alpha N^2)- 3 y_0 (y_1 -y_0)}{6 N((x_0 - y_0)(x_0 + y_0))^{1/2}}  \\
&= \frac{1}{6 N } \left( 3 a_1^2 \cosh \left( \sqrt{\frac{2}{3} }(\phi_1 - \phi_0)  \right) - 3 a_0^2 -  \alpha N^2 \cosh \left( \sqrt{\frac{2}{3}} \phi_0   \right)  \right)\,.
\end{split}
\end{equation}
Since we assume a transition up the potential, $\phi_1 - \phi_0 > 0$ and this makes the second term in \eqref{phidot} positive. Thus $\dot{\phi}_0$ can never be real and positive for the considered boundary conditions. The reason for this stumbling block is simply that we are working in the pure position representation here, where we have not included any information about the momenta of the fields. But we are actually interested in the situation in which we have a prior inflationary state, with the scale factor growing and the inflaton rolling down the potential. Once we include this information, we will see that much more sensible results are obtained.

Thus we must repeat the same procedure as in the last section, i.e. we introduce Robin boundary conditions or, equivalently, convolve the propagator with an initial wavefunction as in eq. (\ref{eq:wavefunction}), yielding the effective action \eqref{actionfull}, where the momenta are chosen to correspond to an inflating universe. Let us be more specific about which form of the spreads $\sigma_{x,y}$ we will consider. From the definitions of the variables $x,y$ we have to leading order
\begin{align}
\sigma_x = 2 a_0 \cosh \left( \sqrt{\frac{2}{3}} \phi_0 \right)  \sigma_a + \sqrt{\frac{2}{3}} a_0^2 \sinh \left( \sqrt{\frac{2}{3}} \phi_0 \right) \sigma_\phi\,, \label{sigtrf1}\\
\sigma_y = 2 a_0 \sinh \left( \sqrt{\frac{2}{3}} \phi_0 \right)  \sigma_a + \sqrt{\frac{2}{3}} a_0^2 \cosh \left( \sqrt{\frac{2}{3}} \phi_0 \right) \sigma_\phi\,. \label{sigtrf2}
\end{align}
While these relations are only accurate for small spreads, we will simply use them as definitions, even when the spread is large. Our discussion in section \ref{eternalintro} indicated that we can expect that for flat potentials the metric changes little, and most of the perturbation is expressed as a change in the scalar field value. This would suggest the choice $\sigma_a=0$ with the entire spread relegated to $\phi.$ In this case
\begin{align}
\sigma_x = \sqrt{\frac{2}{3}} a_0^2 \sinh \left( \sqrt{\frac{2}{3}} \phi_0 \right) \sigma_\phi\,, \label{sigxphi}\\
\sigma_y = \sqrt{\frac{2}{3}} a_0^2 \cosh \left( \sqrt{\frac{2}{3}} \phi_0 \right) \sigma_\phi\,.\label{sigyphi}
\end{align}
Note that this case corresponds to a specific choice of initial state, a choice that is motivated by the calculations of section \ref{eternalintro}. If not specified otherwise, this will be our default choice of initial state. Thus when we quote results in terms of $\sigma_\phi$ alone, this should be understood as shorthand for $\sigma_{x,y}$ given by eqs. \eqref{sigxphi} and \eqref{sigyphi}. However, we will also be led to consider other choices, with both $\sigma_a$ and $\sigma_\phi$ turned on.

\begin{figure}
\includegraphics[width=1\textwidth]{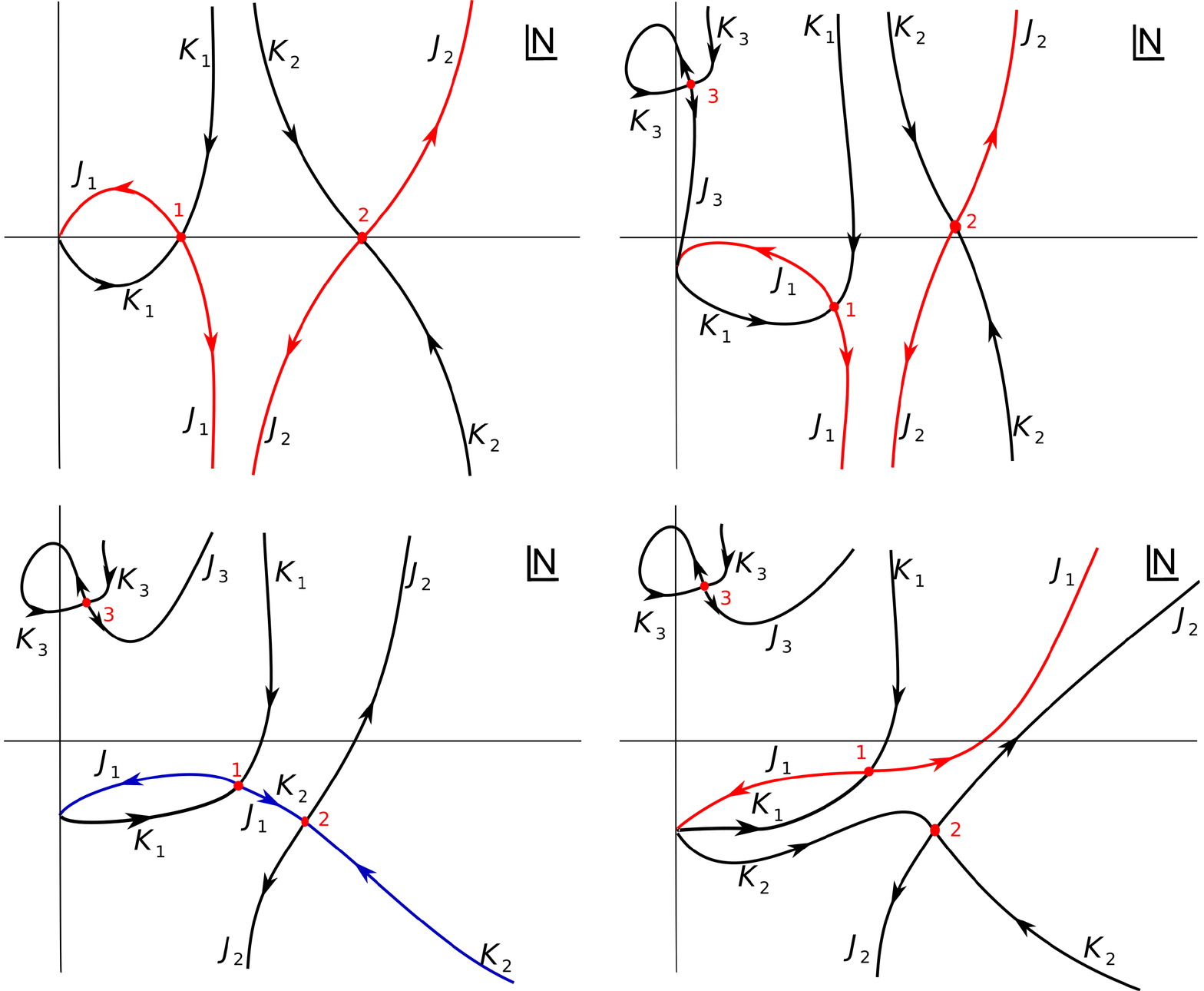}
\caption{Evolution of the saddle points and their associated flow lines in the complex $N$ plane as $\sigma_\phi$ is increased. The original integration contour (positive real line) can be smoothly deformed to the complex Lefschetz thimble (the red line) leaving the value of the path integral unchanged. The Lefschetz thimble runs through one or more saddle points of the action. One of the initially relevant saddle points ($N_1$) is relevant for all values of $\sigma_\phi$ but its position and therefore the geometry associated to it changes. The other saddle point becomes irrelevant after the Stokes phenomenon (the Stokes line is the blue line in the bottom left panel). The third saddle point never contributes to the path integral. In order to draw these graphs, we have used the boundary conditions $a_0=100, \phi_0=1/10, a_1=200, \phi_1=1/2,$ while the values of the spread for these four plots are respectively $\sigma_\phi=0, 0.0100, \sigma_c \approx 0.0154, 0.0700.$}
 \label{fig:conv-sp-evolution-srei}
\end{figure} 

\begin{figure}[ht!]
	\centering
	\includegraphics[width=0.4\textwidth]{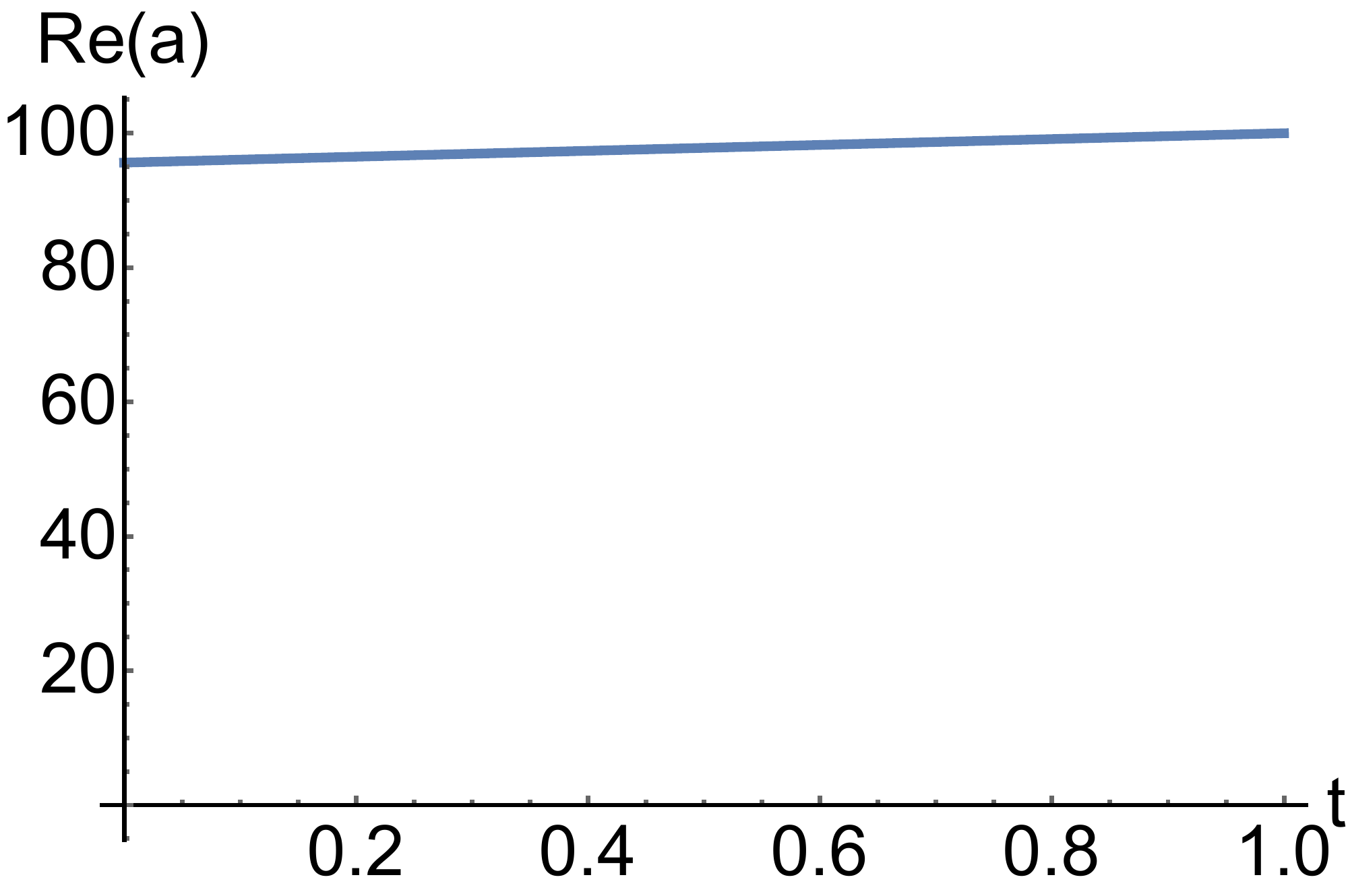}
	\includegraphics[width=0.4\textwidth]{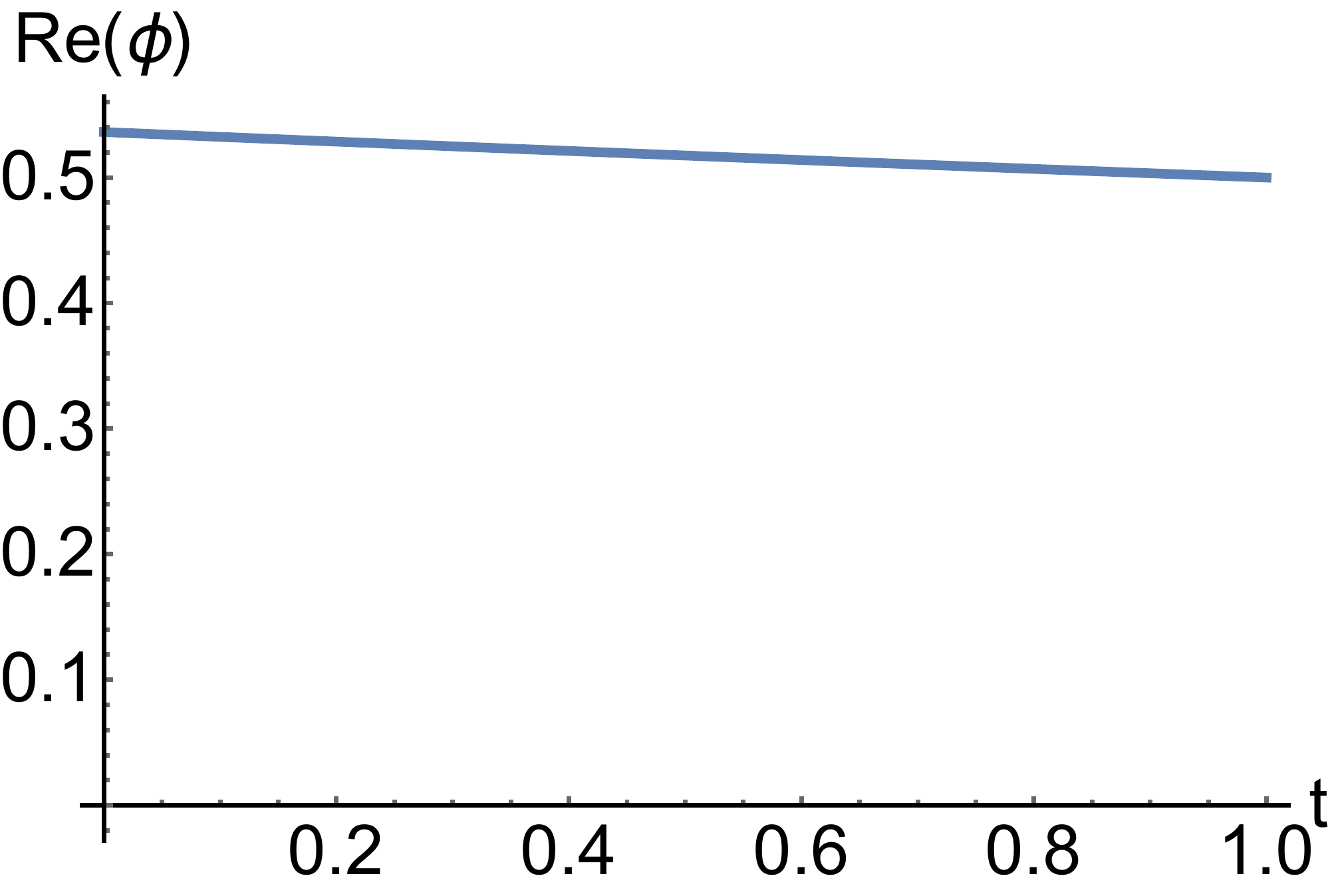}\\
	\includegraphics[width=0.42\textwidth]{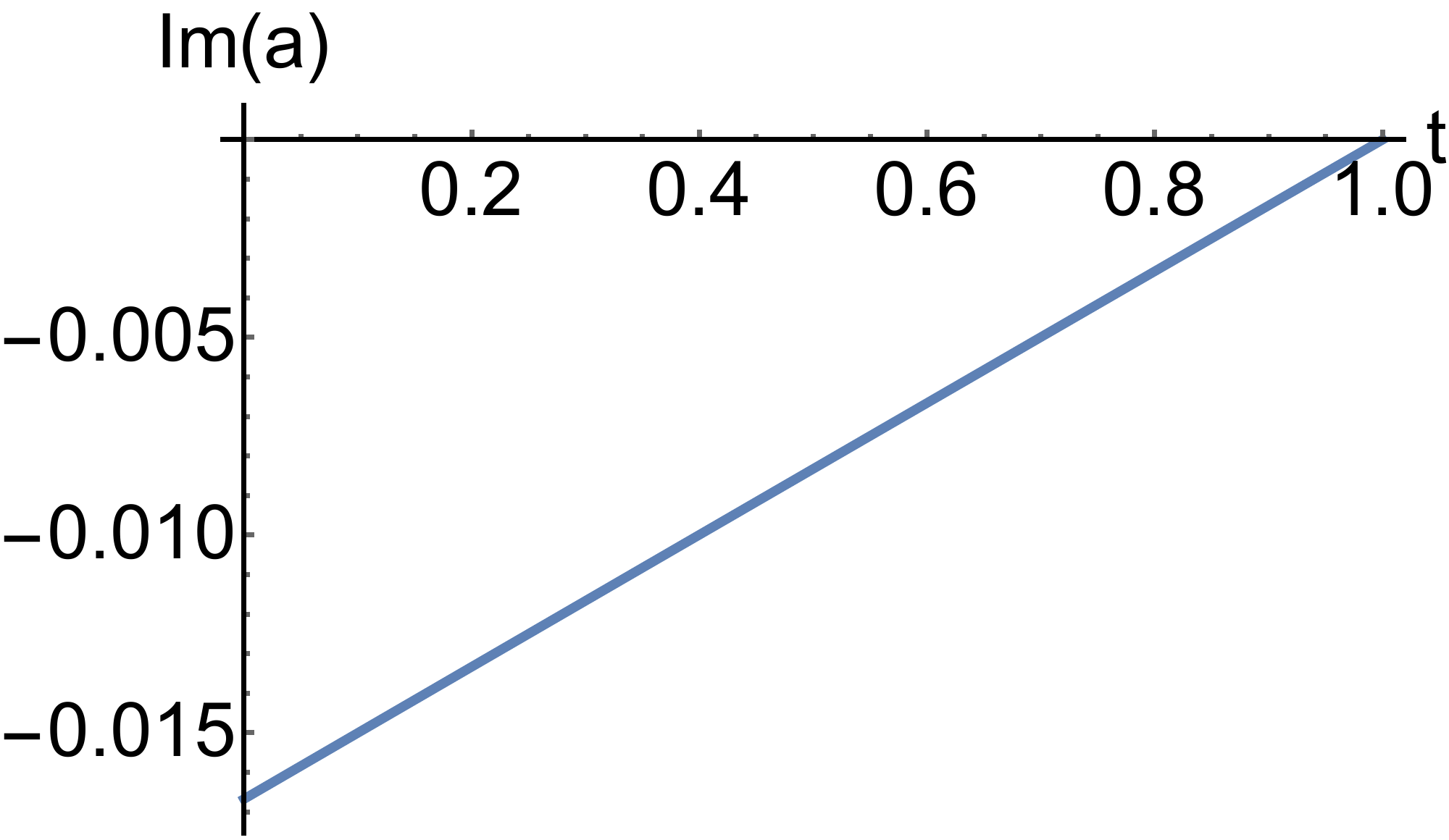}
	\includegraphics[width=0.42\textwidth]{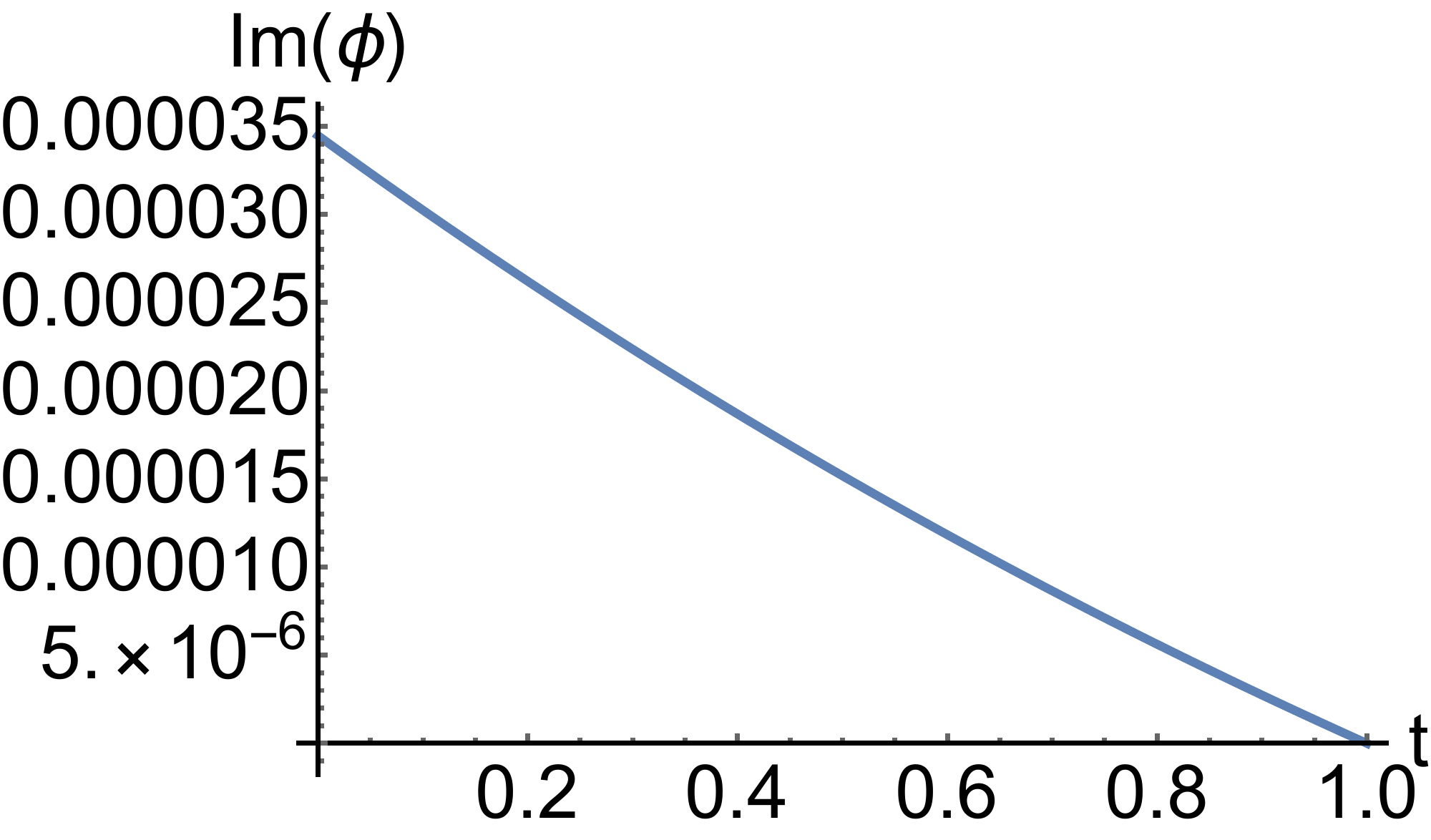}
	\caption{Geometry of the relevant saddle point for $\sigma > \sigma_c$. Plotted here are the real and imaginary parts of the scale factor and scalar field with respect to coordinate time where we have chosen $\alpha = 1/10$, $\phi_0 = 1/10$, $\phi_1 = 1/2$, $a_0 = 100$, $a_1 = 100$ and $\sigma_{\phi} = 2/100$. The final relevant solution is seen to be a slightly complexified version of an ordinary inflationary solution, with the scale factor expanding and the scalar field rolling \emph{down} the potential, even though our boundary conditions are such that we consider an up-jump from the central value of the inflaton.} \label{fig:conv-sp}
\end{figure}

The evolution of the saddle point locations and their associated flow lines as a function of $\sigma_\phi$ is illustrated in Fig. \ref{fig:conv-sp-evolution-srei}. For $\sigma_\phi = 0$, two of the four critical points of the action are relevant to the Lorentzian propagator. As we turn on $\sigma_\phi$, the four saddle points smoothly change their location in the complex $N$ plane and two extra saddle points appear, which however turn out to never give a dominant contribution to the path integral. For a critical value of the uncertainty $\sigma_\phi = \sigma_c,$ a Stokes phenomenon happens which changes the topology of the flow lines. The process is shown in Fig. \ref{fig:conv-sp-evolution-srei} for case A. The final result is entirely analogous for case B, the only difference lying in the fact that in that case the saddle point $N_1$ travels from the imaginary line to the real line as $\sigma_\phi$ increases. After the Stokes phenomenon ($\sigma > \sigma_c$), the only relevant saddle point is $N_1$ and this saddle point becomes more and more real as $\sigma_\phi$ is further increased. The geometry of the relevant saddle point $N_1$ after the Stokes phenomenon has occurred, i.e. for $\sigma_\phi > \sigma_c$, is shown in Fig. \ref{fig:conv-sp}. An interesting aspect is that the initial position of the scalar field $\bar{x}_0$ is no longer close to the original initial position $x_0,$ but is significantly larger -- in fact it has become larger than the final value $x_1$ (and $\phi$ also contains a small imaginary part, which is a reflection of the transition being a quantum transition).

What does this mean? In the Dirichlet formulation of the Feynman propagator we calculate a transition between two fixed geometries and matter content. In that setting it is not possible to continuously link an inflationary evolution with an evolution where the scalar field tunnels up the potential. However, by introducing Robin boundary conditions, thus allowing for a spread in field values and momenta, we do find solutions. Analysing them in more detail, we find that the scalar field \textit{already} starts higher up the potential and then simply rolls down according to an inflationary solution.  Thus, instead of choosing a solution that rolls up the potential, the system has picked out a (comparatively unlikely) configuration contained within the initial state in which the inflaton is already higher up in the potential than required, so as to allow a slow-roll solution to the final configuration. In complete analogy, the scale factor starts out at a smaller value and then grows as the scalar field rolls down. 

\begin{figure}[ht!]
	\centering
	\includegraphics[width=0.48\textwidth]{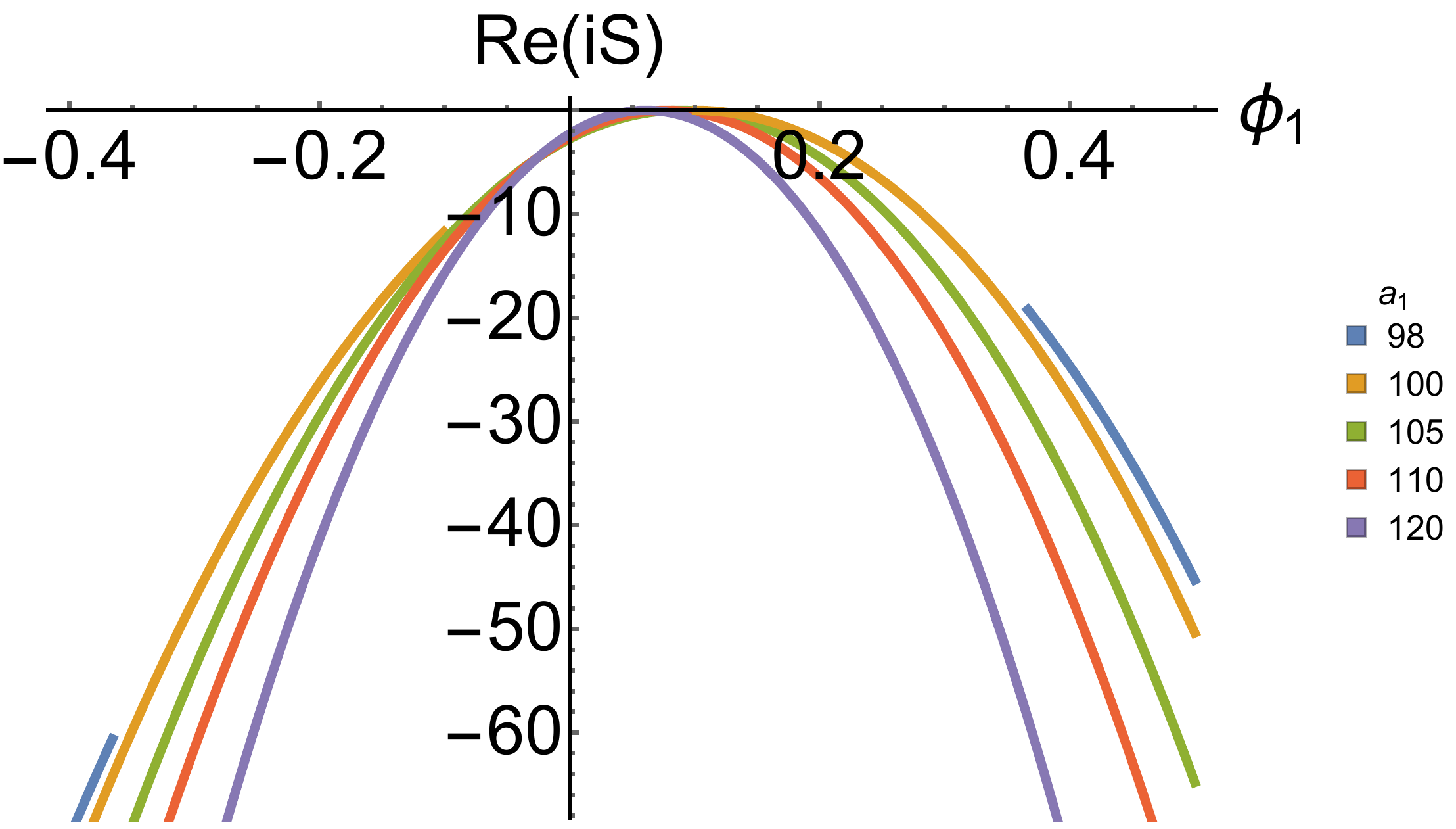} 
	\includegraphics[width=0.44\textwidth]{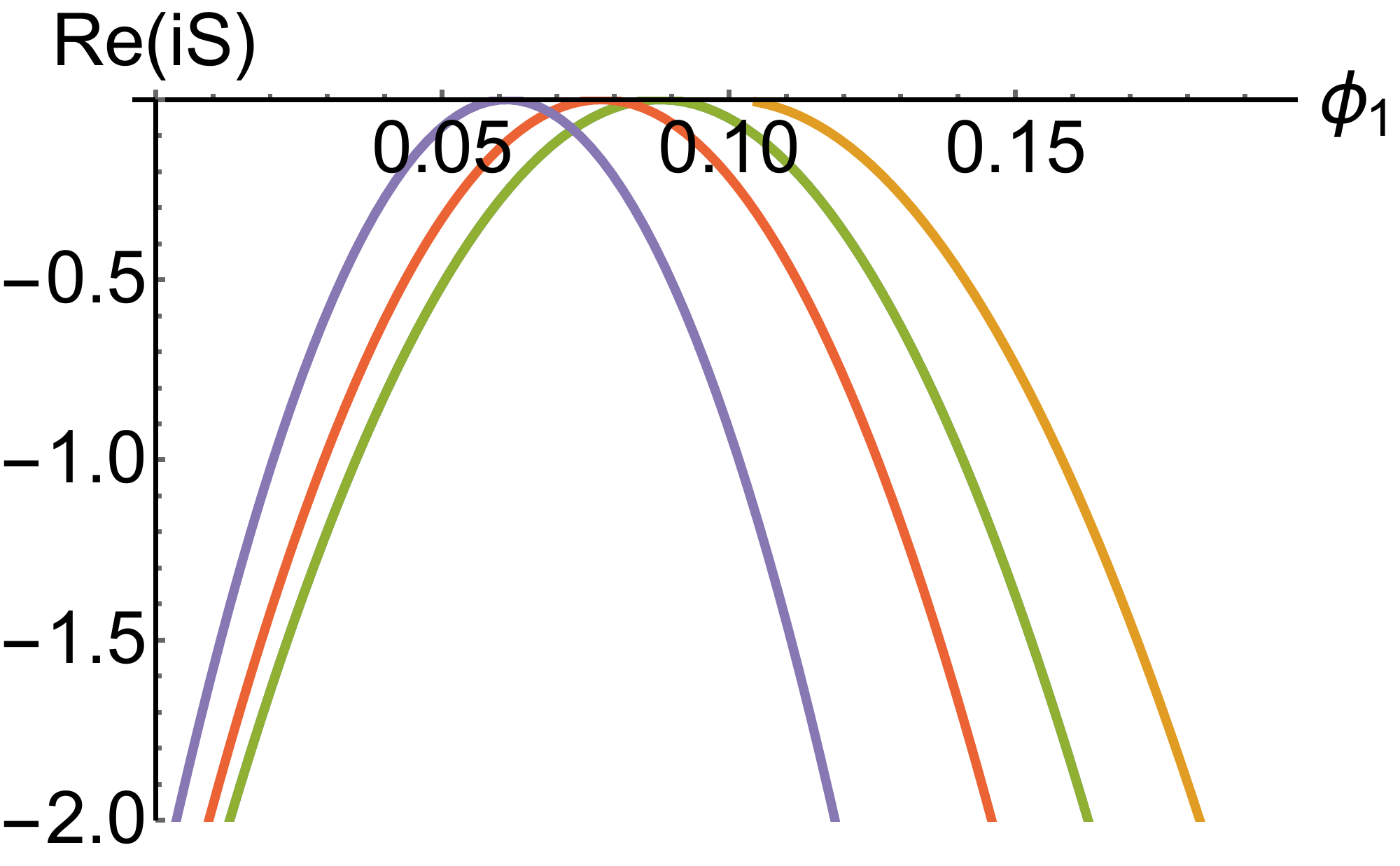}
	\caption{The logarithm of the transition amplitude going from $(a_0, \phi_0) = (100,1/10)$ to various values of $a_1$ and $\phi_1$. Here the x-axis represents $\phi_1$ while the different colours refer to different values of $a_1$, ranging between 98 and 120. The action was evaluated for a spread of $\sigma_\phi = \frac{H}{2\pi}$. Interestingly, for $a_1 \leq 100$ there are areas where no transition is possible and hence there are gaps in the parabola. This is because the usually relevant saddle point has negative real $N$ for these transitions, and no other saddle point is relevant. In such cases, the transition would be more than exponentially suppressed. The picture on the right is the same as on the left except zoomed in onto the top of the curves. As the final scale factor value $a_1$ is increased, the peak of the distribution shifts and the spread in $\phi$ narrows.} \label{fig-action-series-a-phi1}
\end{figure}

\begin{figure}[ht!]
	\centering
	\includegraphics[width=0.32\textwidth]{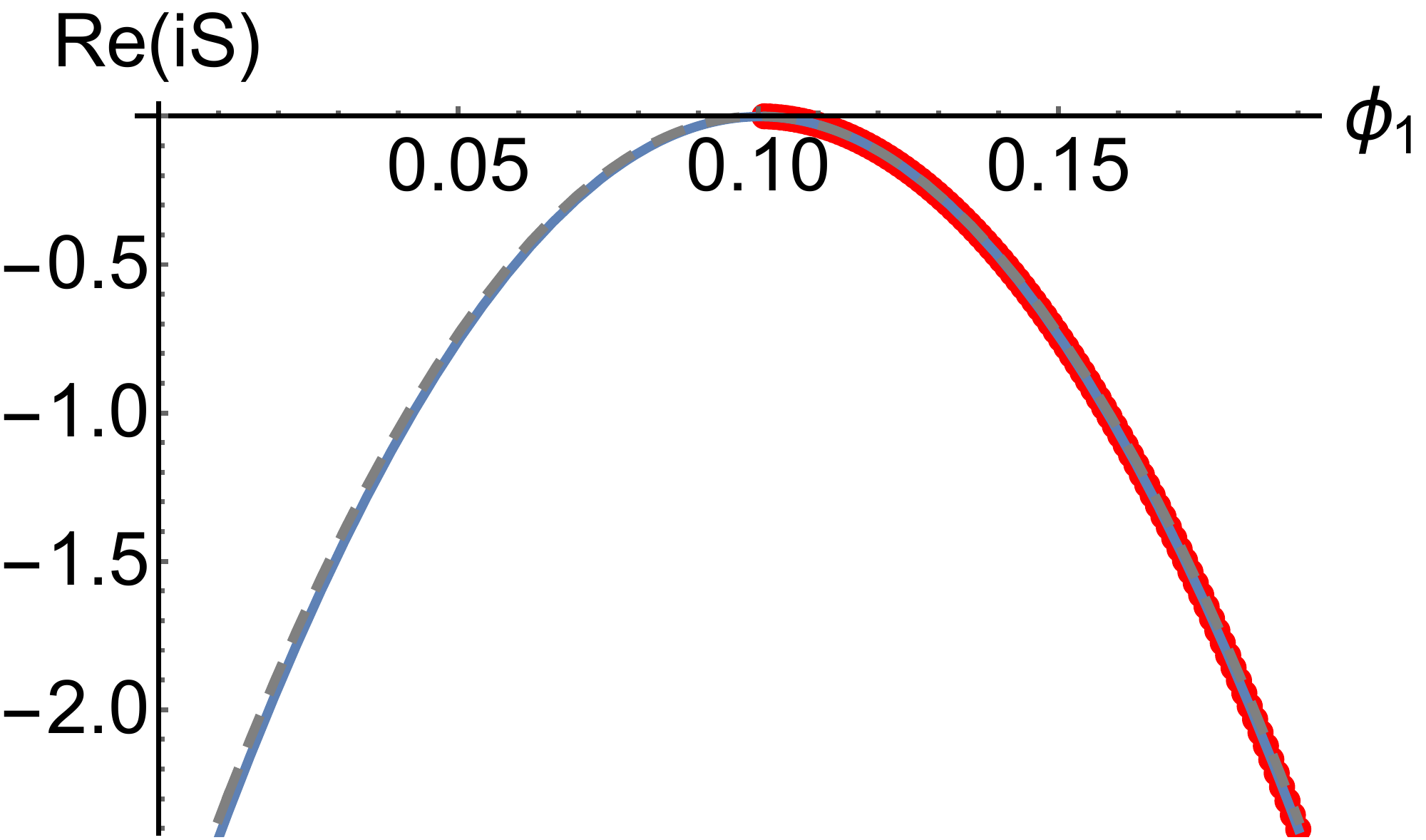}
		\includegraphics[width=0.32\textwidth]{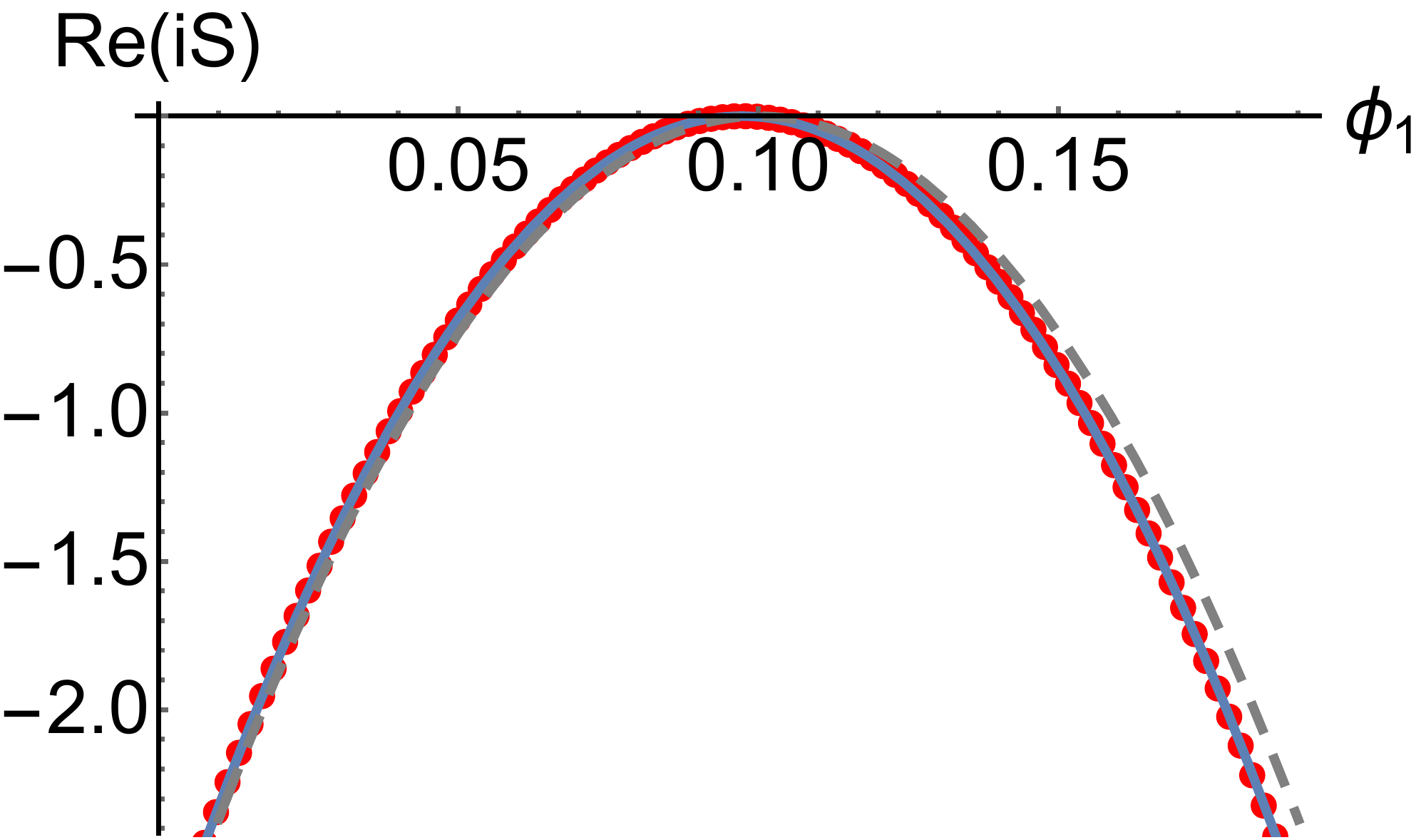}
		\includegraphics[width=0.32\textwidth]{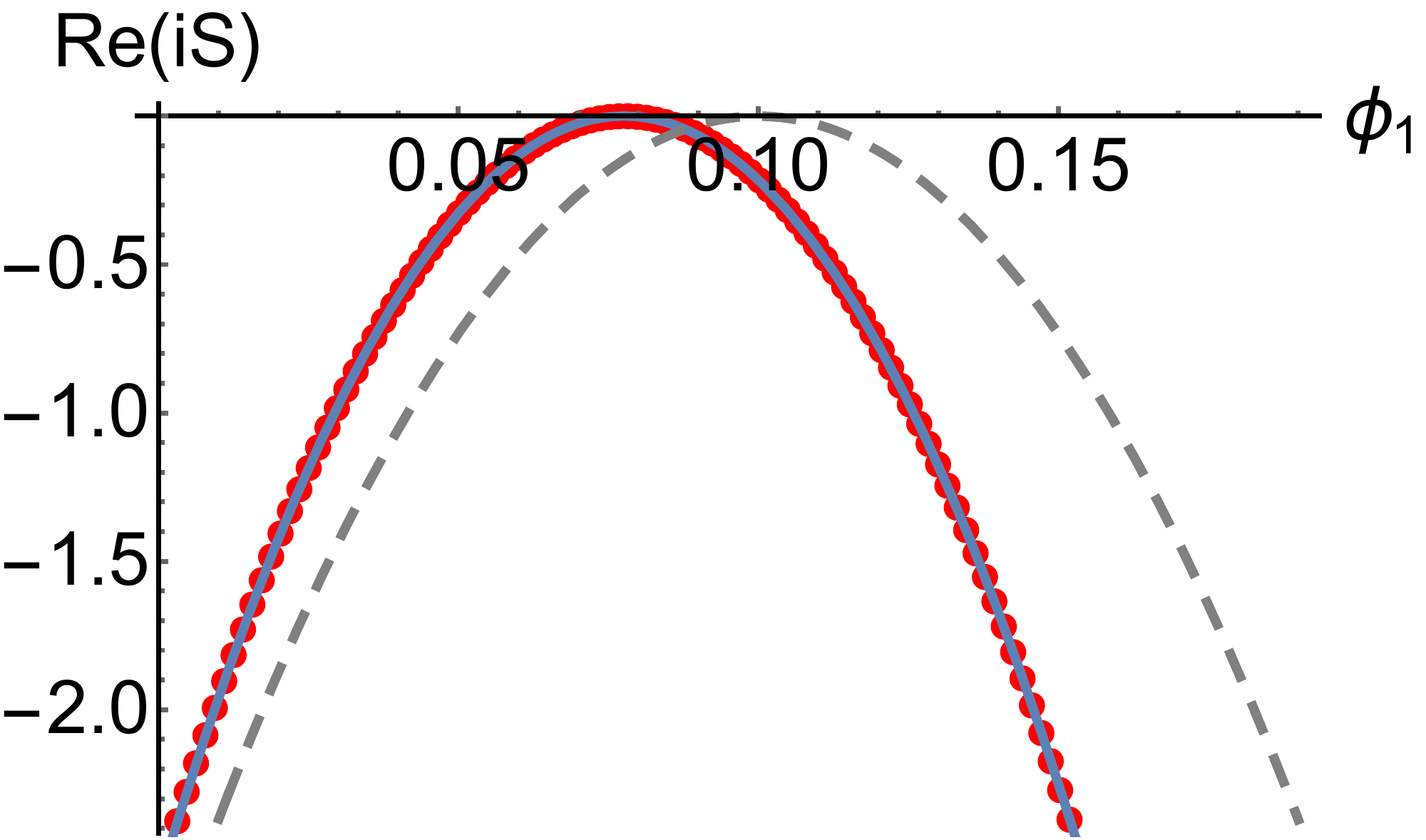}\\
	\caption{An example of the distribution of fluctuations after the quantum transition in red with a fitted parabola in blue. In all three graphs $\alpha = 1/10$, $\phi_0 = 1/10$, $a_0 = 100$. However $a_1 = 100, 101, 110$ in the left, centre and right graphs respectively. In dashed grey we have plotted the initial spread in $\sigma_\phi = H/(2 \pi)$ centred around the classical value $\phi_{top}$. From the picture it is clear that the peak of the distribution shifts and the spread in $\phi$ narrows as we increase the final scale factor -- see also Fig. \ref{fig-spread-af} for more details.} \label{fig-fluctuation-distribution-parabola}
\end{figure}

\begin{figure}[ht!]
	\centering
	\includegraphics[width=0.45\textwidth]{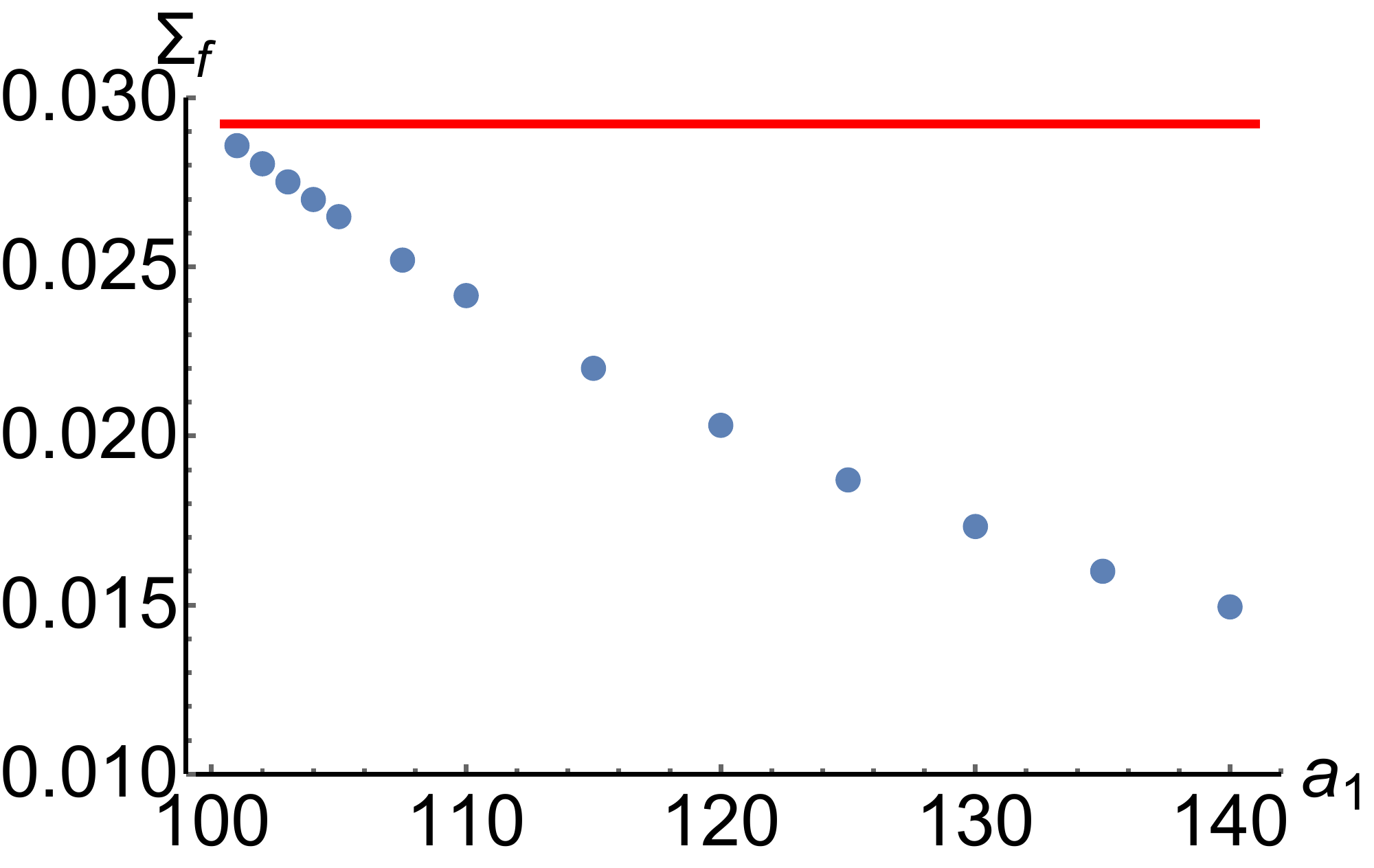}
	\caption{For transitions in which the final scale factor is only slightly larger than the initial one, the weighting for different final configurations is essentially equal to the weighting implied by the initial state. But as the final scale factor value $a_1$ increases the spread (of the weighting) is reduced as a result of the inflationary attractor. The numerical example shown here is the same one as in Fig.\ref{fig-action-series-a-phi1}. The red line is the spread in the inflaton value imposed before the transition occurs.} \label{fig-spread-af}
\end{figure}

This result can be further quantified by analysing probabilities of the geometry and scalar field undergoing transitions to various values of $\phi_1$ and $a_1$ as depicted in Fig. \ref{fig-action-series-a-phi1}. It is obvious that for larger values of $a_1$ and $\phi_1$, transitions become less and less likely. In fact, the most likely transitions occur for a tiny increase in the scale factor, in our example from $a_0=100$ to $a_1 \approx 101.$ This confirms the expectation from QFT in curved spacetime that the geometry ultimately changes very little when the scalar field jumps up the potential. Here we should note that when we impose a final scale factor value that is equal to or smaller than the initial one, then transitions to certain values of the scalar field are impossible (semi-classically). This is reflected in some of the curves in Fig. \ref{fig-action-series-a-phi1} having gaps in them. What happens in these cases is that the relevant saddle point moves to the region where $Re(N)<0,$ i.e. these solutions then actually correspond to time-reversed solutions. This is consistent with the fact that the system prefers to choose inflationary, expanding solutions and requiring the final scale factor to be small then clashes with this preference. In line with this observation is the fact that if we look at increasing values of the final scale factor, then the spread is actually reduced due to the inflationary attractor. In fact, the final weighting remains Gaussian to a good approximation, with only the peak value having shifted and the spread shrinking. We can see this more quantitatively in Fig. \ref{fig-fluctuation-distribution-parabola}, where we plot the final weighting alongside a fitted parabola with final width $\Sigma_f,$ defined via
\begin{align}
Re(i\tilde{S}/\hbar) = h(\phi_1) = h(\phi_{top}) - \frac{(\phi_1-\phi_{top})}{4\Sigma_f^2} + \cdots\,,
\end{align}
where $\phi_{top}$ denotes that value of $\phi$ at which the weighting (Morse function $h$) is maximal for a given final scale factor value $a_1.$ From the figure we can see that the parabola provides an excellent fit. The decrease in the width as the universe expands is plotted in Fig.\ref{fig-spread-af}.

\begin{figure}[ht!]
	\centering
	\includegraphics[width=0.75\textwidth]{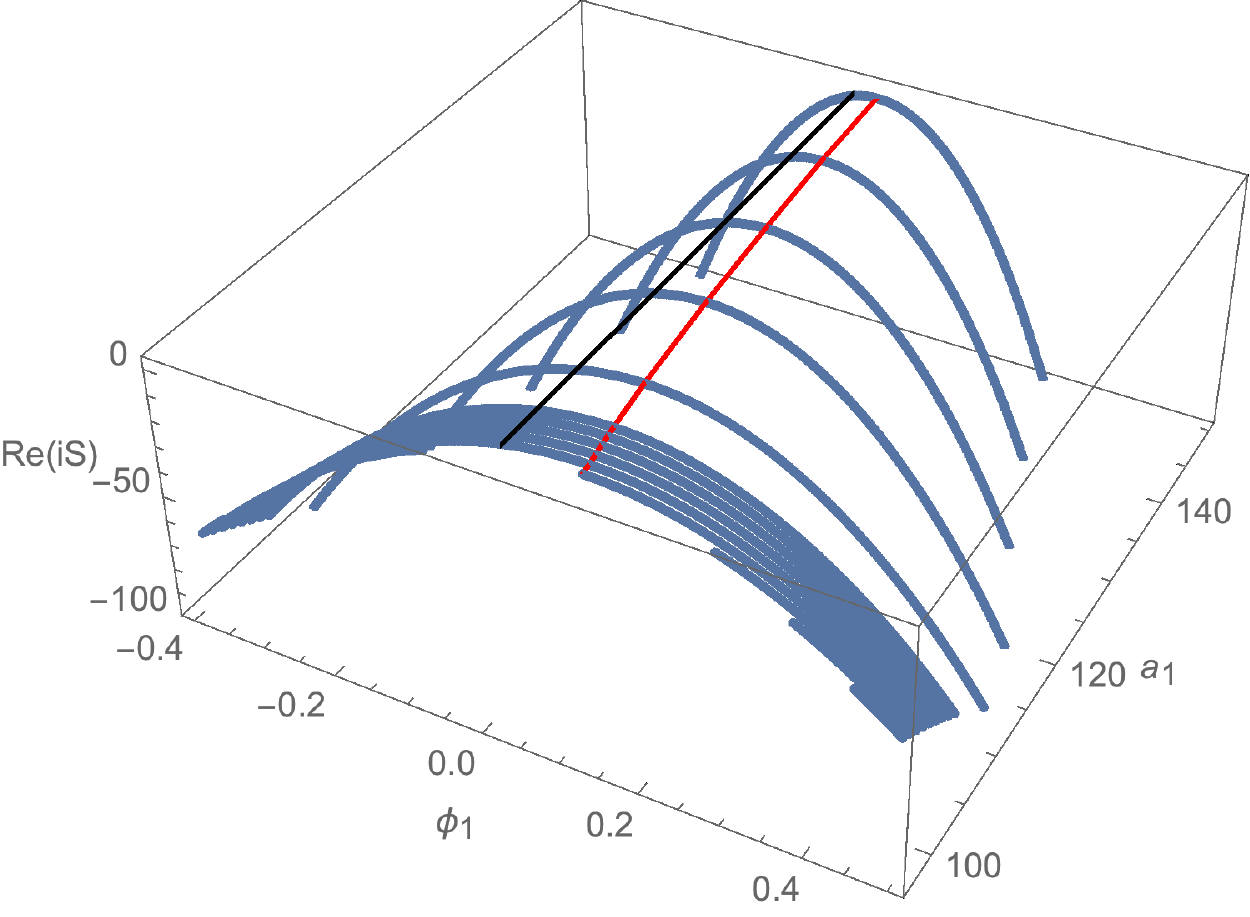}
	\caption{A 3-dimensional version of Fig. \ref{fig-action-series-a-phi1} illustrating how the peak of the weighting (red line) follows a slow-roll solution down towards the minimum of the potential at $\phi=0$ (black line), while large excursions of the inflaton away from the classical solution become less and less likely as the universe expands.} \label{fig-action-series-a-phi1-3d}
\end{figure}

All this is nicely visible in a 3-dimensional version of these plots in Fig. \ref{fig-action-series-a-phi1-3d}. Accompanied with this shrinking of the width is a displacement of the peak of the weighting. The 3-dimensional picture shows that the peak slowly approaches $\phi=0$ as the universe expands -- in other words, the peak of the weighting follows the classical slow-roll trajectory associated with the initial central values of the fields and their momenta that we imposed via the initial state. As the universe expands, the wavefunction narrows around this classical solution, and we attribute this feature to the inflationary attractor. Thus, starting from a fixed initial state and as the universe grows larger, inflaton excursions away from the classical solution become less likely.

\begin{figure}[h]
	\centering
	\includegraphics[width=0.45\textwidth]{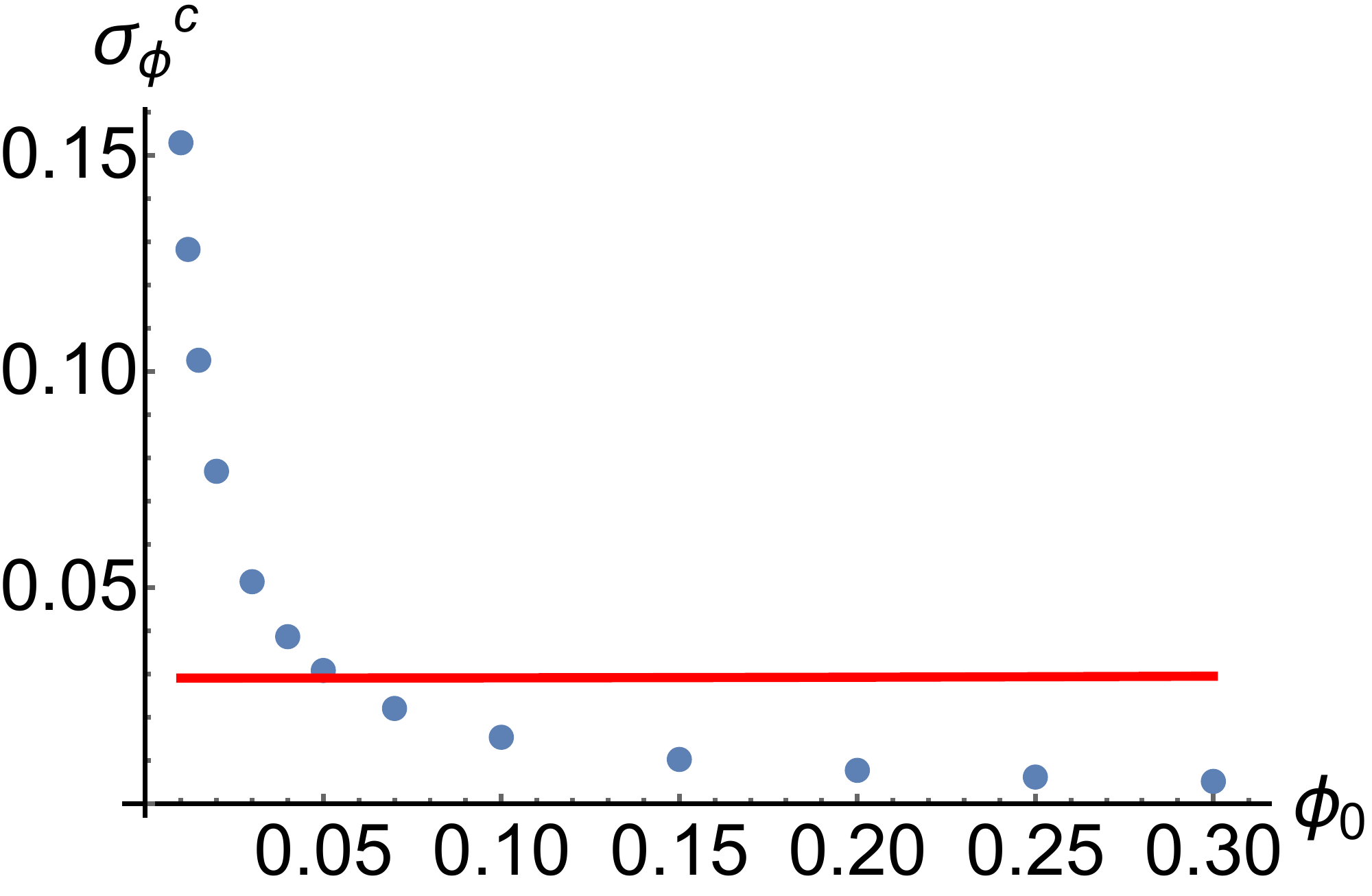}
	\includegraphics[width=0.45\textwidth]{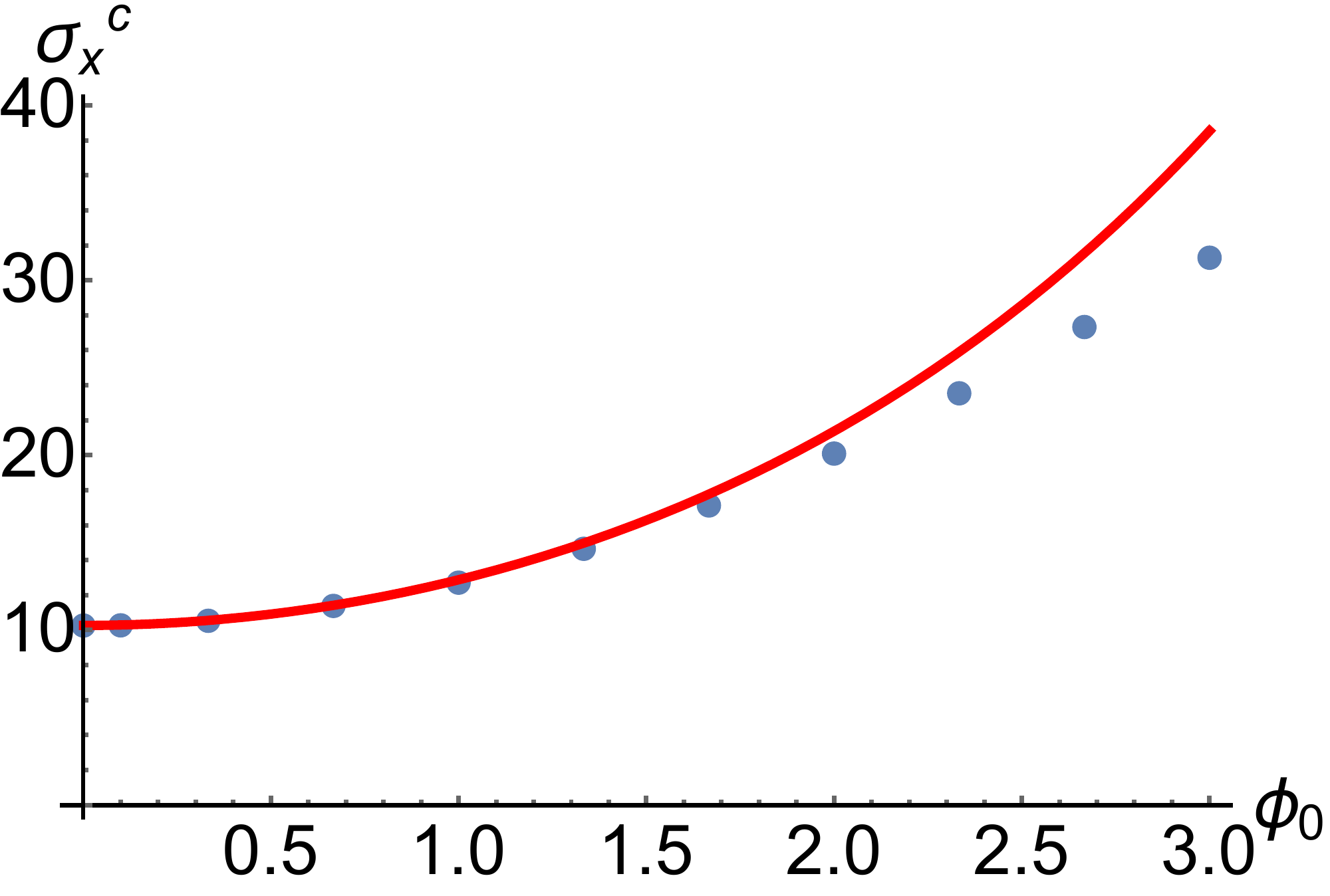}
	\caption{{\it Left panel:} The blue dots indicate the value of $\sigma_\phi$ at which the Stokes phenomenon appears while the value of $H/(2\pi)$ is given by the red line. Here $\alpha = 1/10$ and $a_0 = 100$. The critical value of $\sigma$ depends only on the initial value $\phi_0$ and not on the final values $a_1$ and $\phi_1.$ This graph shows that for a sufficiently large initial scalar field value, only one saddle point is relevant at $\sigma=H/(2\pi).$ {\it Right panel:} The critical spread expressed in terms of the canonical variables. This figure shows the critical value $\sigma_x^c$ as a function of the initial inflaton value $\phi_0.$ Near $\phi_0=0$ we recover the exact value for de Sitter space given in eq. \eqref{eq:CritdS}. The red curve shows the expected value if one were to assume only an uncertainty in the initial scale factor, and not in the inflaton value.  For larger $\phi_0$ we can see that the critical value lies below this curve, implying that for sufficiently large $\phi_0,$ where the potential is less flat, the uncertainty in the inflaton value can induce an earlier Stokes phenomenon. To calculate the points, we have fixed $a_0 = a_1 = 100$ with varying $\phi_0$ and a $\phi_1$ that corresponds to the field jumping up the potential. The initial state's momenta were calculated by keeping $\phi_0'$ fixed, finding the corresponding $a_0'$ via the Friedmann equation and then converting to the momenta in $x$ and $y$.} \label{fig-spread-af2}
\end{figure}

An important effect that we saw earlier was that beyond some critical value of the spread a Stokes phenomenon happens and only a single saddle point remains relevant. When this occurs, we automatically obtain a situation in which quantum field theory in curved spacetime is a reasonable approximation, as only a single background geometry is relevant to the path integral. We can now make this discussion more quantitative -- see Fig. \ref{fig-spread-af2}. An important aspect of this discussion concerns the relationship between the original variables $a, \phi$ and the canonical variables $x, y,$ expressed via the transformations eqs. \eqref{redef1} - \eqref{redef2} and the relations between the spreads \eqref{sigtrf1} and \eqref{sigtrf2}. We argued in section \ref{eternalintro} that the standard calculation in a fixed background suggests that the inflaton should have a significant spread, of order $H/(2\pi),$ with the scale factor being kept essentially fixed. This would amount to setting $\sigma_a = 0.$ The left panel in Fig. \ref{fig-spread-af2} shows the critical value of $\sigma_\phi$ that is required under those circumstances in order to obtain the Stokes phenomenon, as a function of the initial inflaton value $\phi_0.$ (An important point is that the critical spread does not depend on the final inflaton value $\phi_1$.) What the figure shows is that for large enough $\phi_0$ the Stokes phenomenon always occurs before the spread is increased to $H/(2\pi).$ Thus, in regions where the potential is not too flat (but including regions where the density perturbations that are generated may be large), the standard intuition is vindicated. 

\begin{figure}[ht!]
	\centering
	\includegraphics[width=0.4\textwidth]{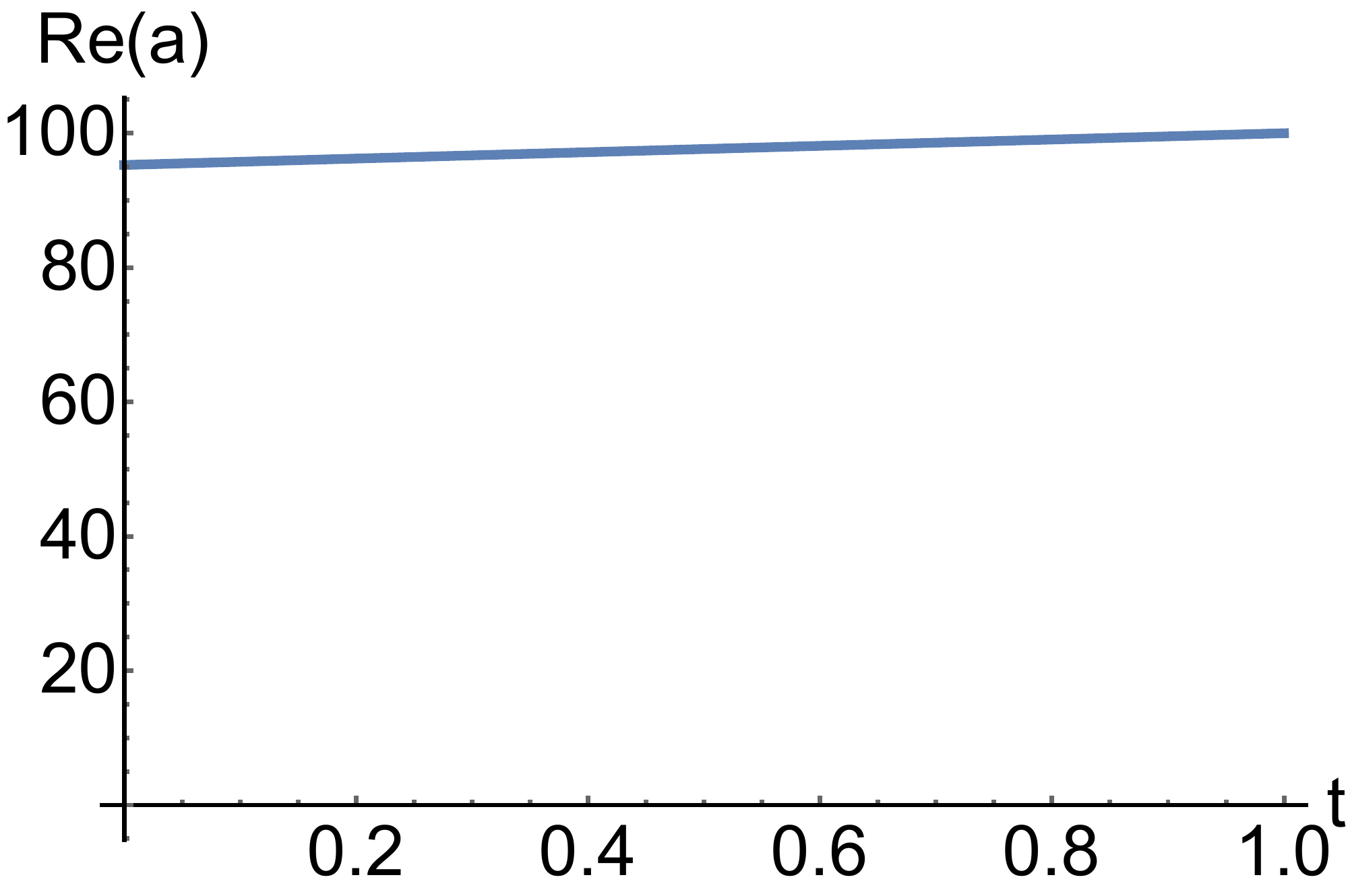} \includegraphics[width=0.4\textwidth]{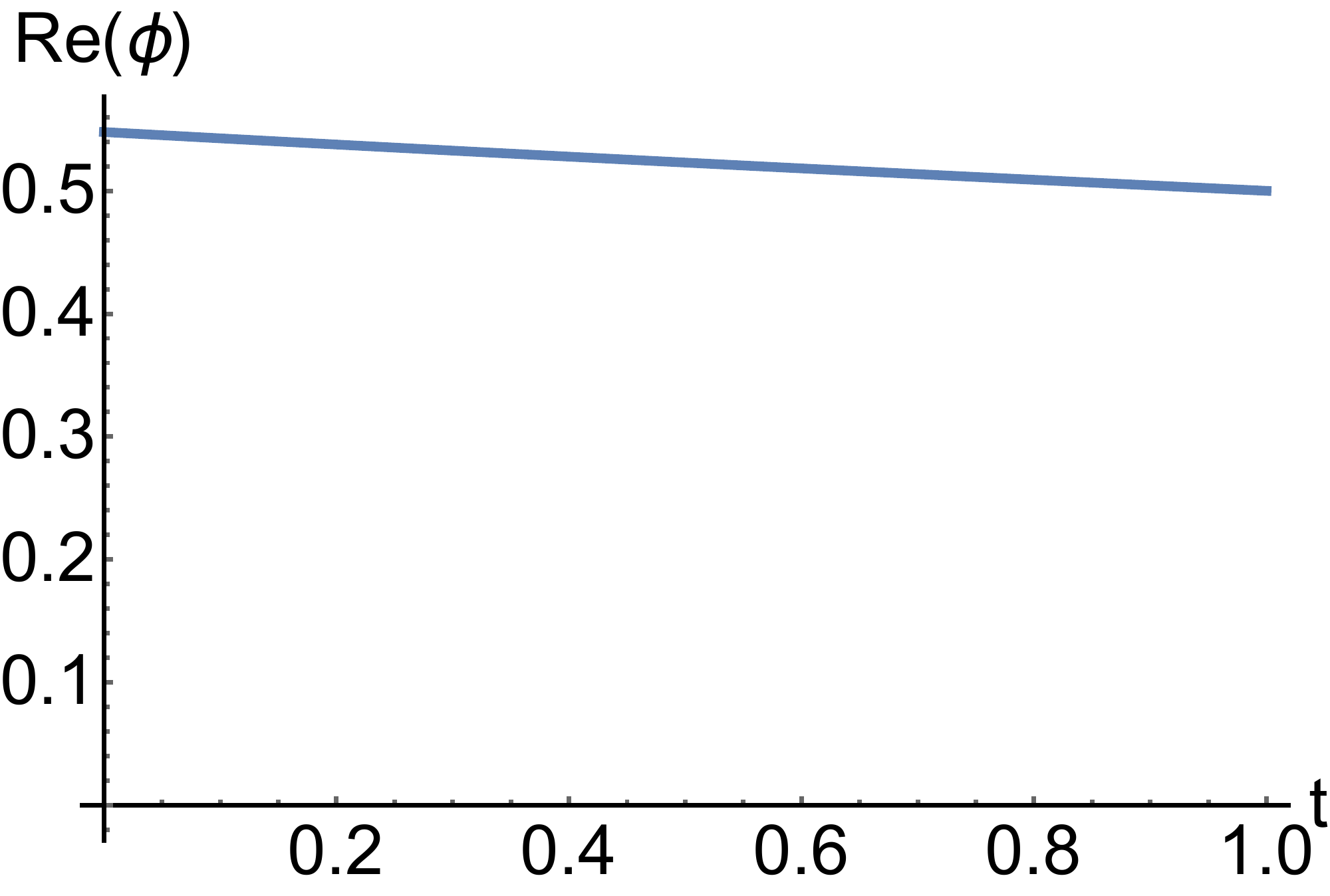}\\
	\includegraphics[width=0.4\textwidth]{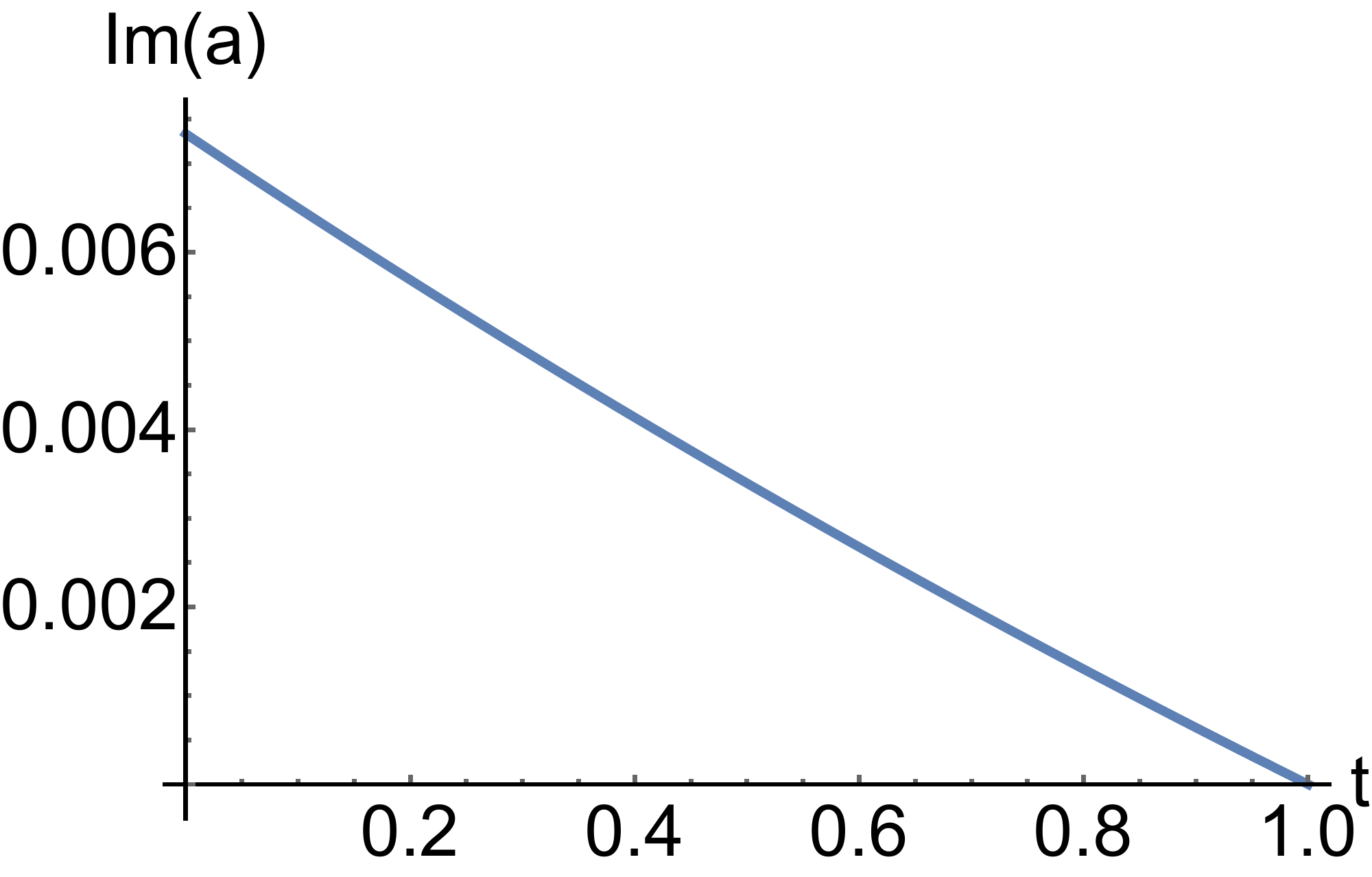} \includegraphics[width=0.4\textwidth]{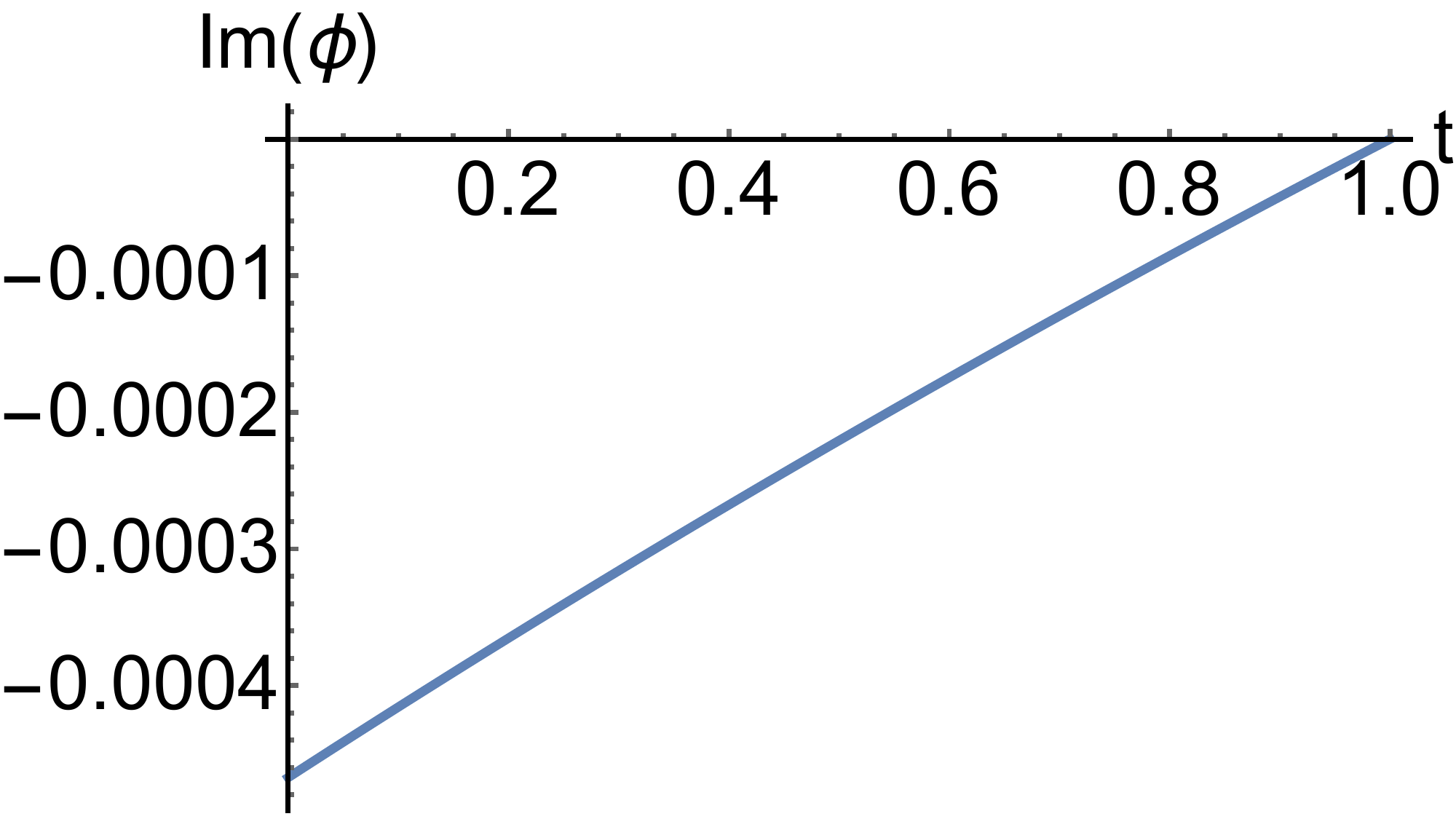}
	\caption{Geometry of the relevant saddle point where the scale factor is kept constant $a_0 = a_1 = 100$ and the scalar field transitions from $\phi_0 = 1/1000$ to $\phi_0 = 1/2$. The initial state's momenta were chosen to be $p_x \approx -54.7$ and $p_y \approx -0.0134$ with uncertainties $\sigma_x = 11$ and $\sigma_y = 100$ which implies that the Stokes phenomenon has already happened. Notice that this geometry closely resembles the one of Fig. \ref{fig:conv-sp}, where $\phi_0$ is larger.} \label{fig:largesigy}
\end{figure}

\begin{figure}[ht!]
	\centering
	\includegraphics[width=0.4\textwidth]{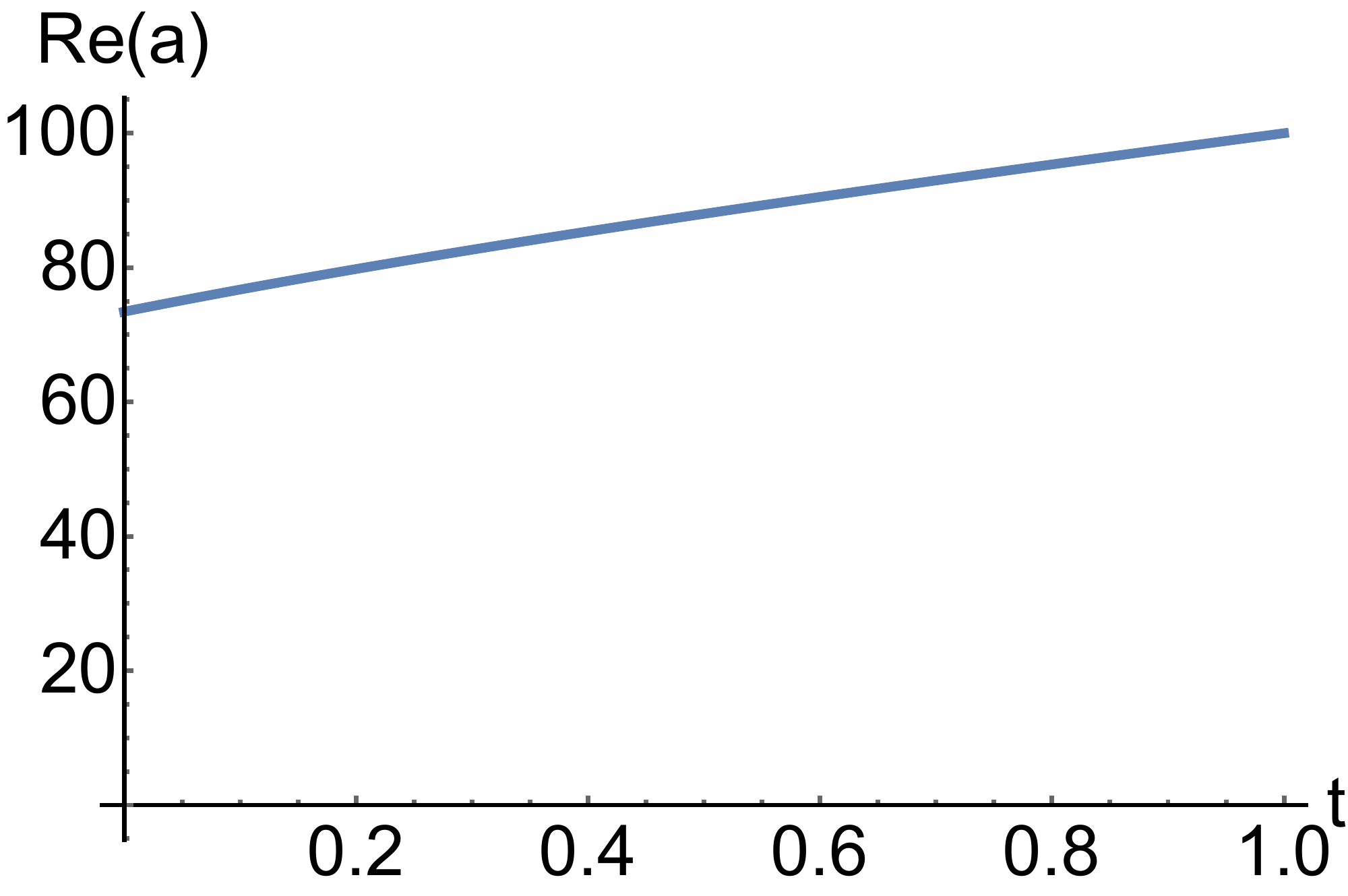} \includegraphics[width=0.4\textwidth]{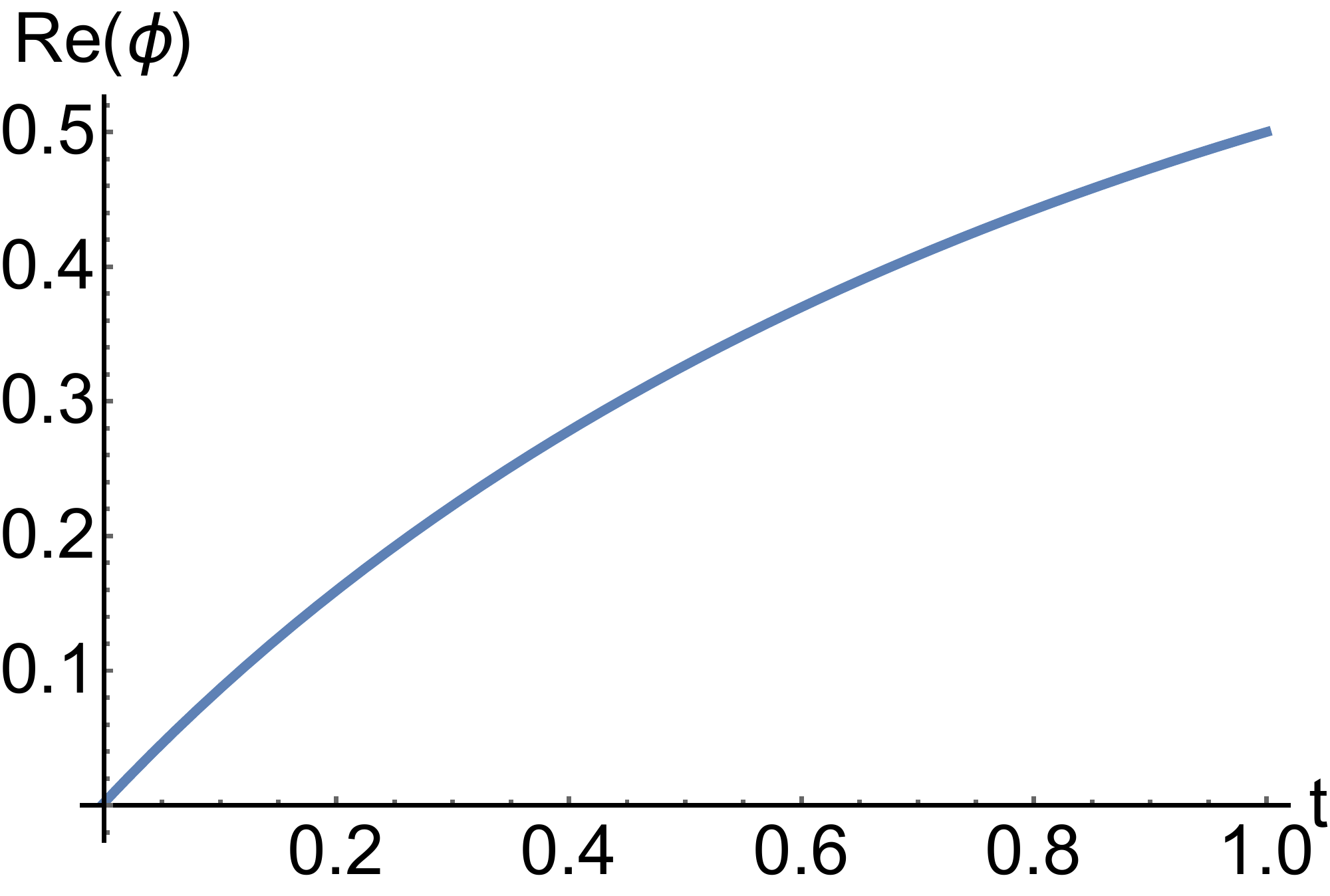}\\
	\includegraphics[width=0.4\textwidth]{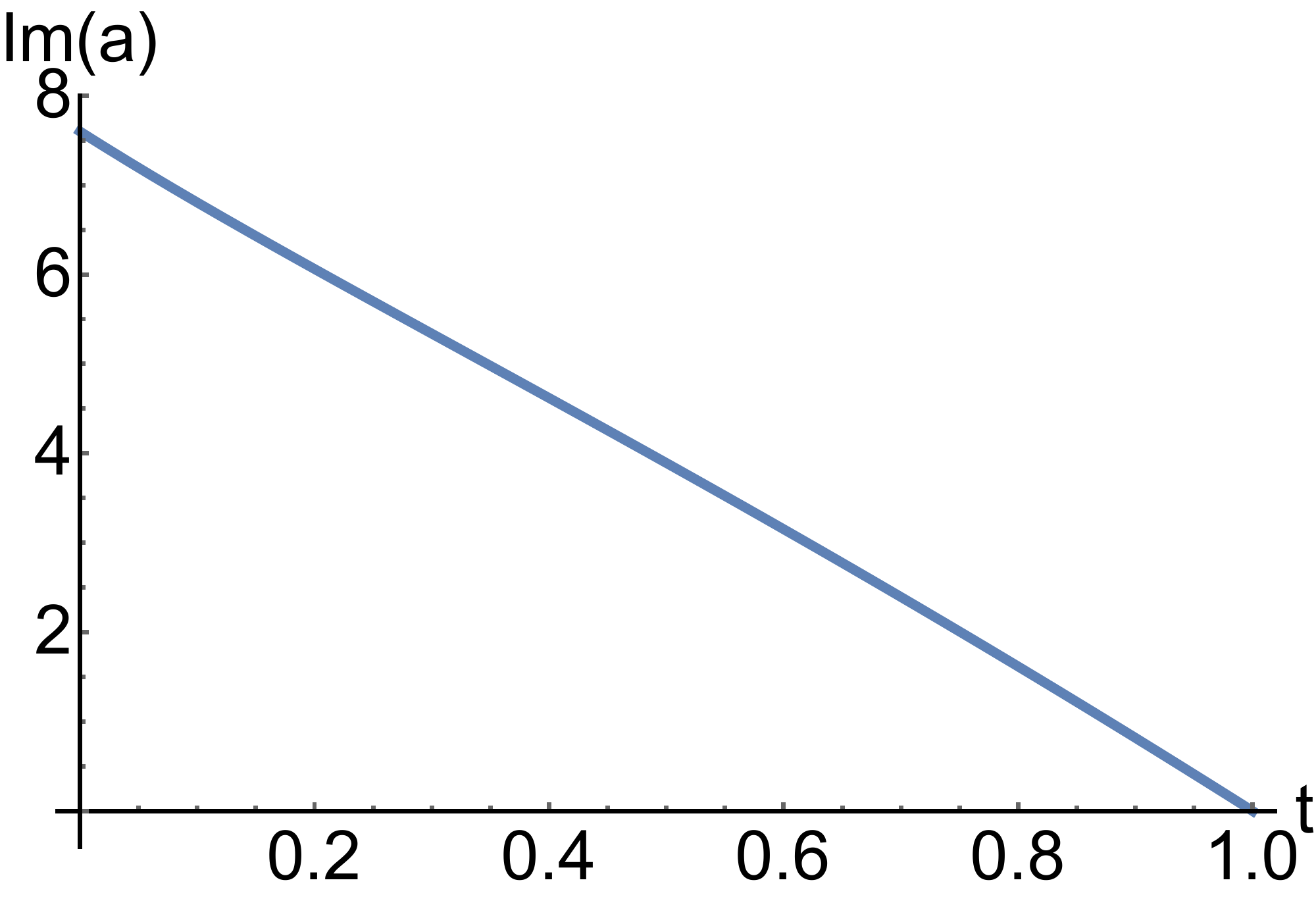} \includegraphics[width=0.4\textwidth]{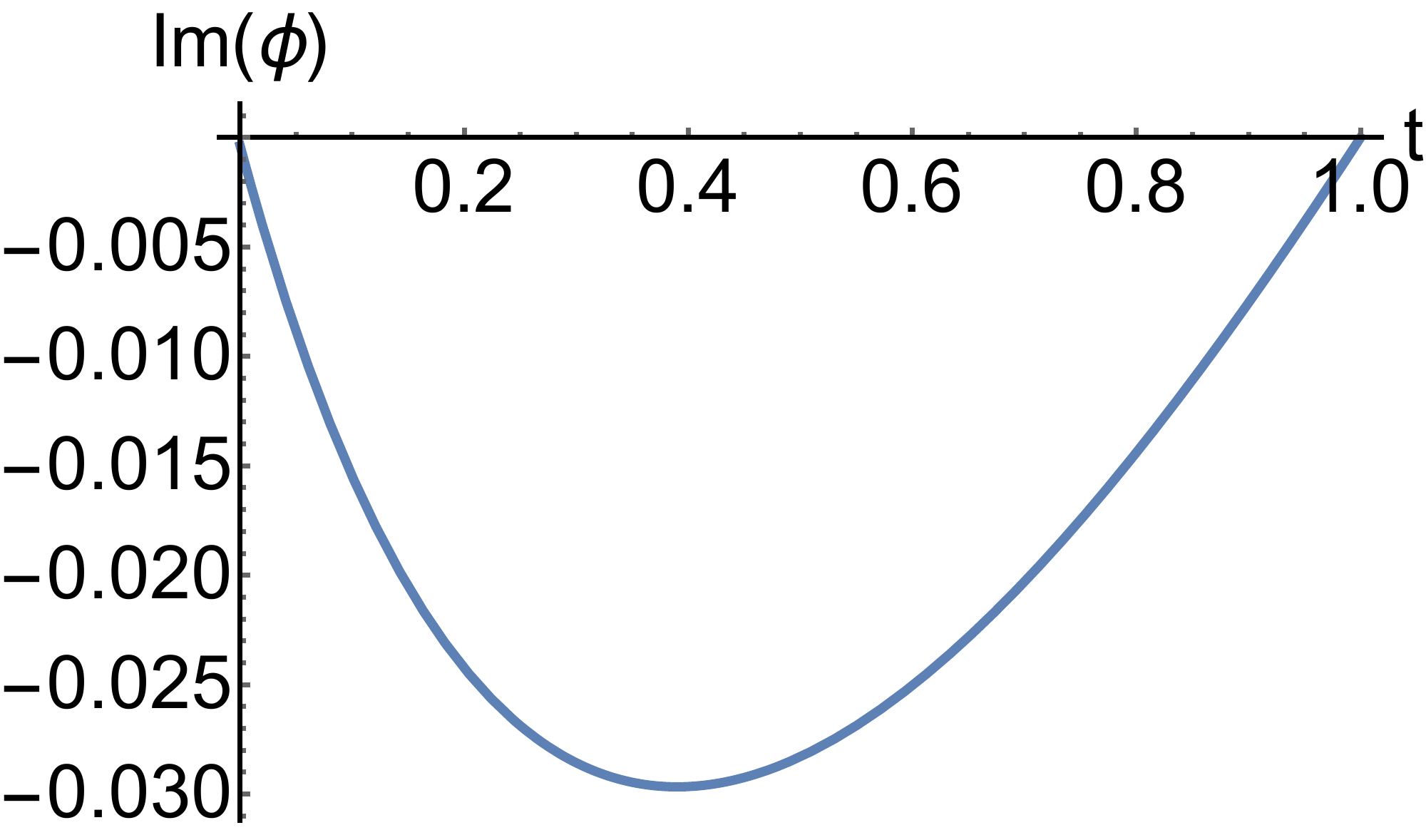}
	\caption{Geometry of the relevant saddle point where the numerical values are identical to the ones of  Fig. \ref{fig:largesigy} except that now $\sigma_y = 0$.} \label{fig:smallsigy}
\end{figure}

For small values of $\phi_0$ however we see a departure from this behaviour, in that the minimum value of $\sigma_\phi$ that would be required to obtain a Stokes phenomenon becomes larger and larger. At this point it is advantageous to switch to a description in terms of the canonical variables $x, y.$ Note that when $\sigma_a=0,$ we have that $\sigma_x \propto \sinh\left(\sqrt{2/3} \phi_0\right) \sigma_\phi$ and thus, for small $\phi_0$ a large inflaton uncertainty $\sigma_\phi$ may still correspond to a much smaller spread $\sigma_x.$ The right panel in Fig. \ref{fig-spread-af2} now shows the critical spread expressed in terms of $\sigma_x$ as a function of $\phi_0.$ Here we are departing from the assumption that $\sigma_a=0,$ and in fact in the plot we have chosen a constant value for $\sigma_y$.\footnote{It turns out that the precise value of $\sigma_y$ is not so important, except when $\sigma_y$ is very small (a case which we will discuss below). We believe that the relative insensitivity to $\sigma_y,$ and the importance of $\sigma_x,$ are simply a reflection of the fact that the potential depends solely on $x.$} What we see is that for small initial scalar field values the critical spread is reduced, rather than enhanced, compared to larger $\phi_0.$ Moreover, the limiting value at $\phi_0=0$ corresponds exactly to the critical value calculated for pure de Sitter space in \cite{DiTucci:2019xcr} and reported in section \ref{quantuminitial}, and where the scale factor of the universe was the only degree of freedom, 
\begin{align}
\sigma_x^c(\phi_0=0) = \left( \frac{a_0^2}{9 \alpha}\right)^{1/4}\,. \label{eq:CritdS}
\end{align}
Thus we see that in the region where the potential is flattest, we {\emph{require}} a minimum uncertainty in the size of the universe $\sigma_a \neq 0,$ and it appears not to be sufficient to only have a large enough uncertainty in the inflaton value. Based on the formula \eqref{eq:CritdS}, we might guess that the critical uncertainty should be given, as long as the slow-roll approximation holds, by replacing $\alpha$ by $V(\phi_0),$ and taking into account the transformation formula \eqref{sigtrf1}. Hence, if we assumed that now on the contrary $\sigma_\phi$ was set to zero, and we would consider only an initial spread in the scale factor, then we might expect the critical spread to be given by
\begin{align}
\sigma_x^c(\sigma_a \neq 0, \sigma_\phi=0) = \cosh\left(\sqrt{\frac{2}{3}\phi_0} \right) \, \left( \frac{a_0^2}{9 V(\phi_0)}\right)^{1/4} \,. \label{CritdS-corr}
\end{align}
The corresponding curve is plotted in red in the right panel of Fig. \ref{fig-spread-af2}. We can see that the true critical spread in fact lies somewhat below this curve. This can be understood in terms of the previous discussion where we showed that for large enough $\phi_0$ even a small $\sigma_\phi$ is already enough to cause the Stokes phenomenon. Thus, away from the very flat region of the potential near $\phi_0=0$ we find that an inflaton uncertainty $\sigma_\phi \lessapprox H/(2\pi)$ is sufficient to lead to a consistent description of quantum transitions, both down and up the potential.  However, in the flattest region of the potential  (where the slow-roll parameter is smaller than $\epsilon \lesssim 5 \times 10^{-4}$), which may be the region of most interest in terms of applications to eternal inflation, this is not enough, and the initial quantum state must contain a significant uncertainty in the scale factor too, of a magnitude indicated by the de Sitter result \eqref{eq:CritdS}. 

As discussed above and shown in Fig. \ref{fig:conv-sp}, the relevant saddle point geometry is typically similar to a standard slow-roll inflationary solution, albeit one with slightly complexified field values. This is certainly the case whenever the initial inflaton value $\phi_0$ is large enough, and $\sigma_\phi$ has been chosen to lie above the critical value $\sigma_c.$ However, as we just saw, in the flattest part of the potential a significant uncertainty in the size of the universe is also required in order to achieve a Stokes phenomenon. We may thus expect the relevant saddle point geometry to change character, and in closing this discussion we will briefly illustrate this effect. Near $\phi_0=0$ we still have the possibility of having a large uncertainty in the inflaton value too, i.e. we may still have a large $\sigma_\phi$ and thus, in combination with $\sigma_a,$ we may still have large values of both $\sigma_x$ and $\sigma_y.$ In this case we still have a roughly slow-roll saddle point geometry, where as before the inflaton starts with a comparatively unlikely value high up on the  potential and slowly rolls down - see Fig. \ref{fig:largesigy}. However, the uncertainty in the inflaton value could also be small, with a correspondingly well determined initial expansion rate, so that once again a Stokes phenomenon is achieved. This corresponds to having a very small (or vanishing) value for $\sigma_y.$ In this case the scalar field is forced to roll up the potential, since its initial and final values are specified with great certainty. But we showed in Eq. \eqref{phidot} that it is not possible for the inflaton to roll up, as long as the field values are real. The resolution is that in this case the saddle point becomes highly complex, and the field evolution also correspondingly complex - see Fig. \eqref{fig:smallsigy}. Moreover, at the end of the transition the scalar field is still rolling up the potential. These two cases thus nicely illustrate the importance of the initial Robin conditions, or equivalently the initial state, in determining the most likely subsequent evolutions. The most appropriate form of the initial state will of course depend on the physical situation under consideration, and  determining the appropriate form of the initial state will be the most important ingredient in applying our results to situations of interest, such as eternal inflation.

\subsubsection{Avoiding Off-Shell Singularities} \label{zeros}

For every value of $\sigma_\phi$ there are regions in the complex $N$ plane where the scale factor $a(t)$ vanishes for some $t \in [0,1]$ and the scalar field $\phi(t)$ correspondingly diverges \cite{DiTucci:2019xcr}. These configurations are irregular in terms of the physical variables $a$ and $\phi$ and as a consequence the action functional diverges. Note that this irregularity has no counterpart in terms of the canonical variables $x$, $y$ and the corresponding action is analytic. In fact, what becomes singular when the scale factor vanishes is the map which connects the two sets of variables. Thus, these singularities would  appear in the Jacobian factor the we have been ignoring in the saddle point approximation, because it usually plays a sub-leading role. However, in the special case where the map becomes singular, the Jacobian would render the path integral ill-defined. Therefore, in order to deal with a well defined path integral  we will require that such a curve of zeros (of the scale factor) in the complex $N$ plane does not lie on the defining integration contour, the Lefschetz thimble nor region in between the two. 

\begin{figure}
\includegraphics[width=0.32\textwidth]{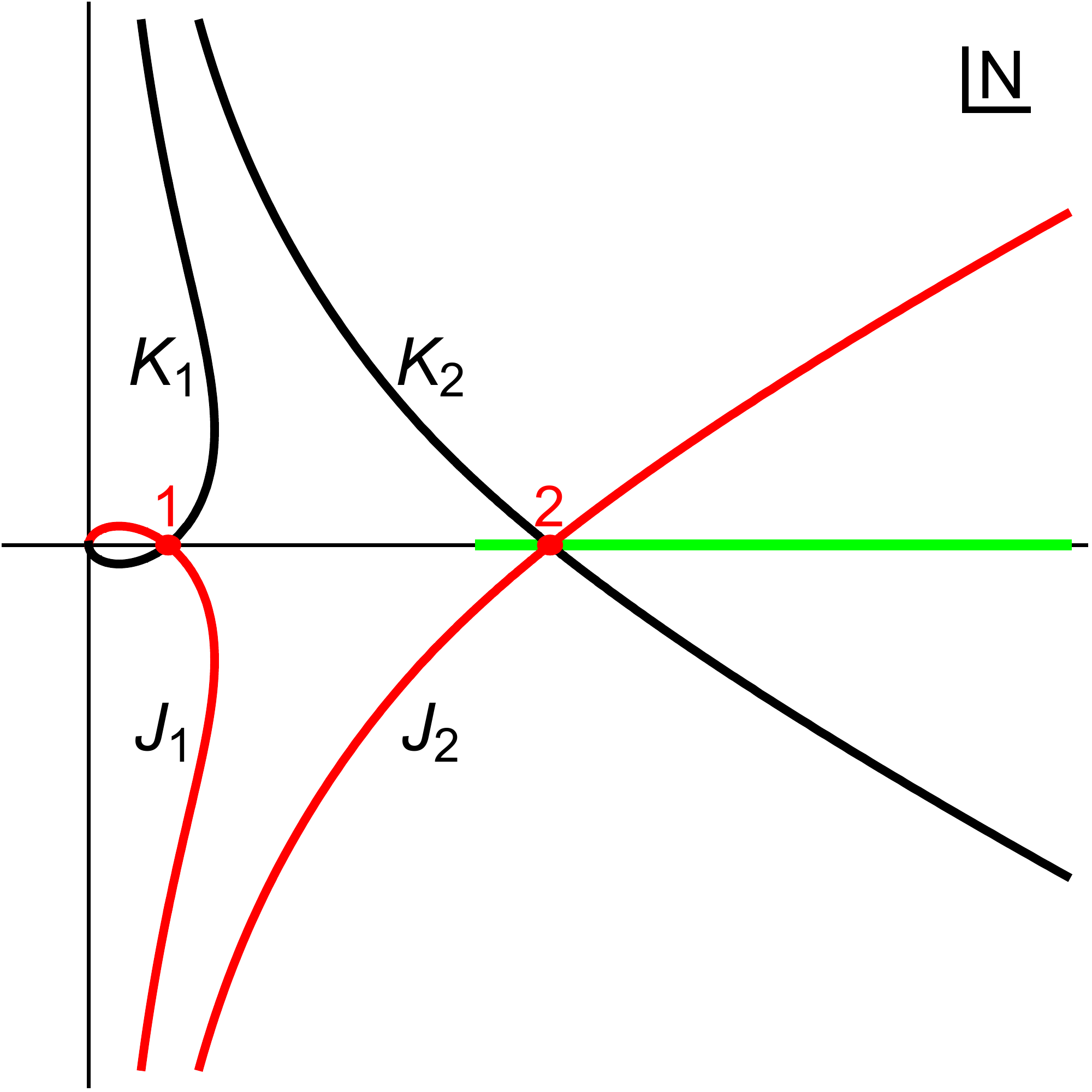}
\includegraphics[width=0.32\textwidth]{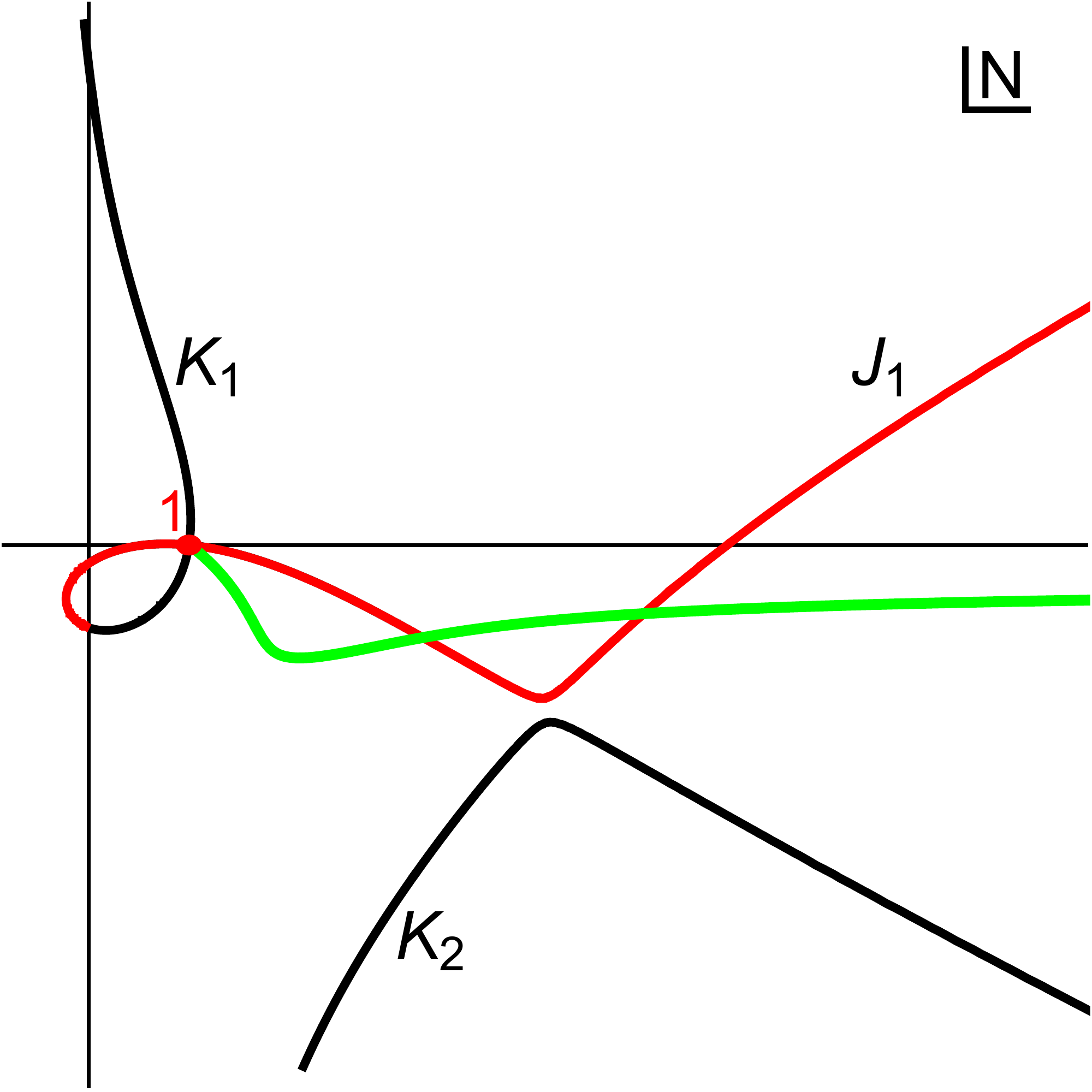}
\includegraphics[width=0.32\textwidth]{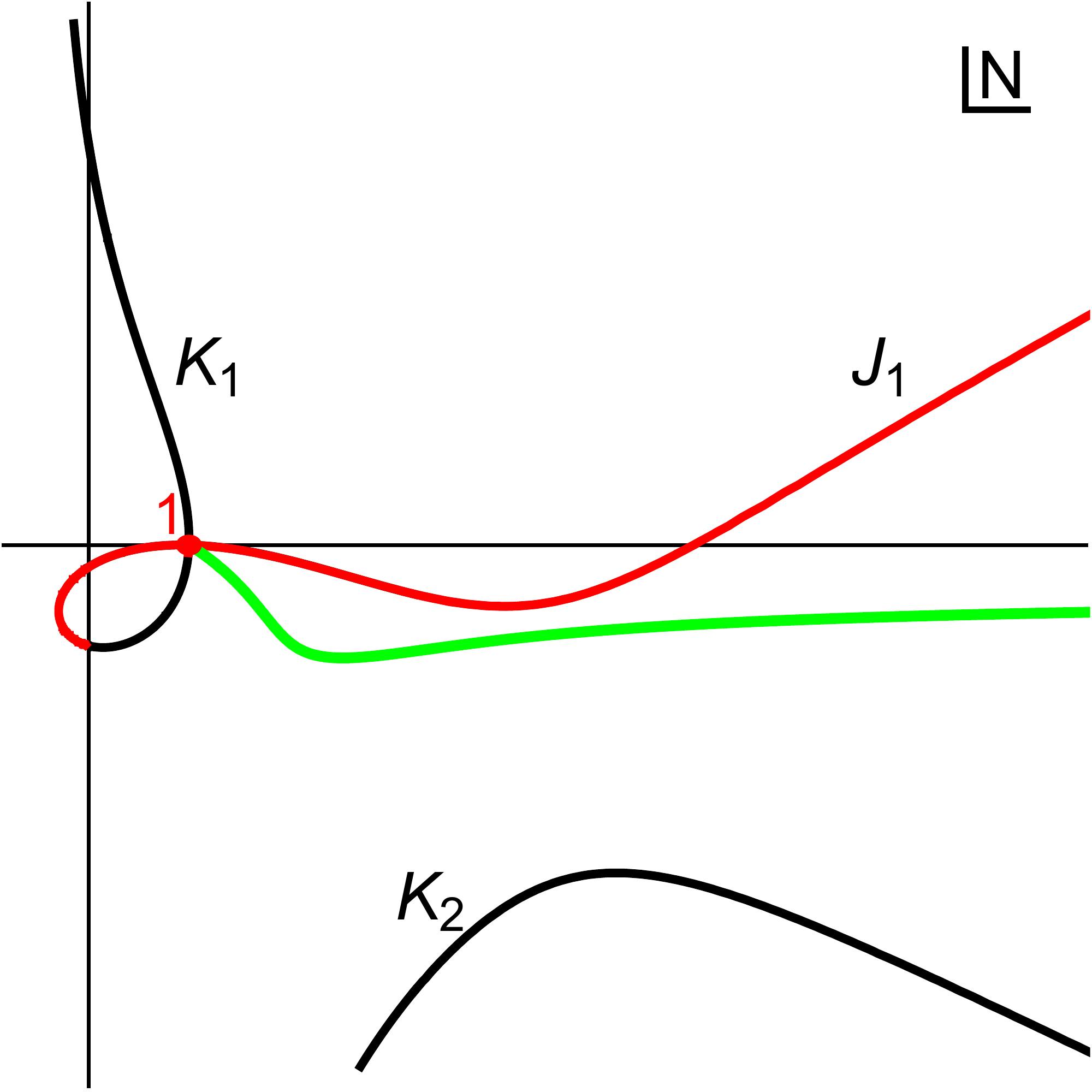}
\caption{The plots show the flow lines (in black) and the scale factor's curve of zeros (in green) for increasing values of $\sigma_{\phi}$. The left, middle and right panels correspond to $\sigma_\phi = 0$, $ \sigma_{\phi} = \sigma_c=0.0154$, and $\sigma_{\phi} = 0.0170 > \sigma_z$ respectively. The other parameters read $a_0 = 100$, $a_1 = 200$, $\phi_0 = 1/10$, $\phi_1=1/2$, $\alpha = 1/10$, $p_x = -54.79$, and $p_y= -1.34$ similar to the previous examples. The curves of zeros still crosses the Lefschetz thimble when the Stokes phenomenon happens, but after a further, modest increase in the spread to $\sigma_{\phi} \approx 0.0164$, the lines do not cross anymore, and the path integral is well defined.} \label{fig-zeros}
\end{figure}

The curve of zeros, just like the flow lines associated with the various saddle points, changes as a function of $\sigma_\phi$. The typical behaviour is shown in Fig. \ref{fig-zeros}. For $0 \leq \sigma_\phi \leq \sigma_c$, the curve of zeros crosses the Lefschetz thimble but there is no more crossing starting from $\sigma > \sigma_z > \sigma_c $. For larger values of the uncertainty, the path integral is well approximated by one saddle point and the variables $a$ and $\phi$ are well defined. From our numerical studies we found that $\sigma_z$ is only modestly larger than $\sigma_c,$ leading to a small increase of the spread required to recover QFT in curved space-time.

%%%%%%%%%%%%%%%%%%%%%%%%%%%%%%%%%%%%%%%%%%%%%%%
\subsubsection{Discussion} \label{discussion}

In this work we have taken the first steps in analysing inflationary quantum transitions in semi-classical gravity, more specifically in the path integral formulation of gravity. Such an analysis is of interest since inflationary fluctuations are regularly considered as having momentous implications: they may be the source of the primordial density fluctuations, and they are thought to be able to alter the global structure of spacetime. Since they are typically treated using the framework of QFT in curved spacetime, an important question is whether this approximate treatment is justified. We have analysed this question making use of a specific minisuperspace model containing a scalar field $\phi$ in a potential of the form $V(\phi) = \alpha \cosh\left( \sqrt{\frac{2}{3}} \phi\right),$ where the potential is chosen such that a transformation of variables is possible that enables the action to become quadratic. This potential has the interesting feature of interpolating between a very flat region near $\phi=0,$ where the potential is approximately constant, and a region with a larger slow-roll parameter $\epsilon \approx 1/3.$ 

Our results, which only deal with the simplified case of homogeneous transitions, in fact largely support the results of QFT in curved spacetime, under the assumption that an appropriate initial state of the universe is considered. The way in which the ``standard'' results are recovered is however rather surprising: for instance, we are led to think of a transition up the inflationary potential not so much as involving the inflaton rolling up the potential, but rather as the selection of an unlikely, but otherwise perfectly ordinary, inflationary solution that was already ``hidden'' in the initial state (our results share some conceptual similarities with the framework of Braden et al. in \cite{Braden:2018tky}). In other words, the semi-classical picture that is emerging is that an unlikely large value of the scalar field is picked out (typically containing a small imaginary part as well), such that the desired final value of the scalar field can be reached from it by ordinary slow-roll down the potential. 

In order to obtain consistent results, it is crucial however that an appropriate initial state is imposed. We have done this by using Robin initial conditions, which may equivalently be seen as the imposition of an initial coherent state for the canonical variables of the model. We find that in potential regions that are not too flat, the initial state must contain a sufficient uncertainty in the inflaton value in order for a single saddle point to be relevant to the transition amplitude, implying that an approximate description in terms of QFT in curved spacetime is justified. The critical minimal uncertainty in such potential regions is moreover below the expected scale $H/(2\pi),$ where $H$ denotes the Hubble rate at the start of the transition. More surprising is perhaps our finding that in very flat potential regions (in our model where the slow-roll parameter $\epsilon$ is smaller than about $5 \times 10^{-4}$), considering only an uncertainty in the inflaton is not sufficient: one must also allow for a sufficiently large uncertainty in the size of the universe. This may have consequences for models of eternal inflation, since it remains to be demonstrated that an appropriate state is generated prior to the up-jumping transitions that are usually considered in this framework. The generation of an appropriate initial state remains an interesting topic for future work.

There are in fact many other avenues for future work. An important extension of the present work will be to add inhomogeneous perturbations. Another aspect that will be worth studying will be the difference between transitions that occur while inflation is already underway, compared to transitions right at the beginning of inflation. This latter study will of course require the additional input from a theory of initial conditions, such as the no-boundary proposal \cite{Halliwell:2018ejl,DiTucci:2019dji}. In addition, it may be of interest to clarify what goes wrong when two saddle points remain relevant to a particular transition. Based on the earlier study in pure de Sitter space \cite{DiTucci:2019xcr} we expect fluctuations around the two saddle point geometries to be incompatible with each other and to lead to problematic interference effects or instabilities. Understanding such interference may help in clarifying what happens for transitions in very flat potential regions when the uncertainty in the size of the universe is insufficient. Finally, one can use the semi-classical techniques employed here to investigate other  physical setups, such as quantum transitions across the big bang \cite{Gielen:2015uaa,Bramberger:2017cgf}. We hope to report on progress along those lines in the future.

\clearpage

%%%%%%%%%%%%%%%%%%%%%%%%%%%%%%%%%%%%%%%%%%%%%%%%%%%%%%%%%%%%%%%%%%%%%%%%%%%%%%%%%%%%

\section{Neumann path integrals: cosmology and AdS black holes}\label{chapterneumann}

Quantum theory is based on calculating transition amplitudes. In particular, when a system is prepared in a certain state, we can ask: what is the probability for various outcomes? Quantum cosmology applies this framework to the universe. In that case, the question would be: if the universe is in a certain state at a certain time, what is the probability for it to evolve to different later configurations? Asking such questions presupposes that we know the state of the universe at a certain time. But as we extrapolate our knowledge of the universe back into the more distant past, we know less and less about the state of the universe. Yet, everything followed from these early conditions. 

When gravity is involved, transition amplitudes are calculated between 3-geometries, which in the cosmological context may usually be thought of as equal-time slices of the universe's evolution. Instead of having to specify conditions on ever earlier such initial slices, Hartle and Hawking had the beautiful idea that one could calculate transition amplitudes which have no boundary in the past, i.e. transitions amplitudes involving only a specification of the late time configuration and no initial 3-geometry \cite{Hartle:1983ai}. The idea was that this proposal could describe how the universe originated from nothing, i.e. how spacetime and matter arose from the absence thereof. Moreover, this proposal would implicitly fix the initial conditions of the universe \cite{Hartle:2008ng}.   

A question which has vexed quantum cosmologists since the appearance of this proposal has however been how to actually calculate such no-boundary amplitudes in practice. Transition amplitudes are naturally expressed as gravitational path integrals, in an extension of Feynman's path integral quantization programme to include gravity. Hartle and Hawking proposed that in this framework one should sum only over geometries that are compact and regular in the past, in order to implement their idea. Yet, it still remains difficult to perform such a calculation in practice. Even for simple examples in minisuperspace it has remained unclear how to precisely define the various required boundary conditions, integration ranges and integration contours (for an early investigation see \cite{Halliwell:1988ik}). A priori, it sounds like the absence of a boundary would eliminate the need for boundary conditions. But in fact one basic difficulty, which arises as a direct consequence of the $1+3$ split of spacetime, is that for each field that is considered one must impose \emph{some} conditions at the end points of integration, i.e. one must decide which boundary conditions best describe the absence of a boundary.

In the present chapter, based on \cite{DiTucci:2019bui} and \cite{DiTucci:2020weq}, we will study the consequences of imposing Neumann boundary conditions at the initial ``no-boundary'' hypersurface. In terms of choices of contour, we will stick to the most conservative choice, which is to integrate over real field configurations, and thus Lorentzian geometries\footnote{The integrals that we will consider typically do not contain a singularity at the origin $N=0$ of the complexified plane of the lapse function $N,$ and hence we will mainly be concerned with integrations over infinite ranges of the lapse. This means that our results will mostly pertain to wavefunctions, and not to Green's functions.}. 

Overall, we will take the point of view that the no-boundary wavefunction is a result in need of a definition. Thus, we will take the Hartle-Hawking saddle point geometries as the basic building blocks and we will seek a well-defined path integral that has these geometries (and ideally only these) as its relevant saddle points. We already saw in chapter \ref{robincosmology} that the specification of an (Euclidean) initial momentum is rather crucial to the success of the programme. There we showed how it is possible to define a meaningful Hartle-Harking path integral by partially fixing the initial momentum with a canonical Robin condition, the quantum uncertainty being distributed between the initial momentum and size of the universe. We also saw that one could fix the final Hubble rate of the universe by means of a covariant Robin condition if the initial momentum is fully specified with a Neumann condition. In this chapter we are going to elaborate more on the possibility of implementing an initial Neumann condition for the no-boundary path integral.
Early suggestions to use the Neumann condition can be found in \cite{Louko:1988bk, Halliwell:1988ik, Halliwell:1990tu}. The importance of specifying the initial momentum was also highlighted more recently in \cite{DiazDorronsoro:2018wro}. In the earlier works the Neumann condition was imposed at the level of the minisuperspace action. Here, as in \cite{DiTucci:2019bui}, we will be careful to consider the boundary conditions as arising from the full Einstein-Hilbert action, thus guaranteeing that our treatment is covariant.

Gravitational physics is arguably better understood in the presence of a negative cosmological constant than a positive one, yet there exist strong technical similarities between the two settings. These similarities can be exploited to enhance our understanding of the more speculative realm of quantum cosmology, building on robust results regarding anti-de Sitter black holes describing the thermodynamics of holographic quantum field theories. To this end, in section \ref{sec:blackholes} which will be based on \cite{DiTucci:2020weq}, we will study $4$-dimensional gravitational path integrals in the presence of a negative cosmological constant, and with minisuperspace metrics. We put a special emphasis on boundary conditions and integration contours. The Hawking-Page transition is recovered and we find that below the minimum temperature required for the existence of black holes the corresponding saddle points become complex. When the asymptotic anti-de Sitter space is cut off at a finite distance, additional saddle points contribute to the partition function, albeit in a very suppressed manner. These findings have direct consequences for the no-boundary proposal in cosmology, because the anti-de Sitter calculation can be brought into one-to-one correspondence with a path integral for de Sitter space with Neumann conditions imposed at the nucleation of the universe. Our results lend support to the implementation of the no-boundary proposal focusing on momentum conditions at the ``big bang''.

\subsection{The no-boundary term proposal}\label{sec:noboundaryterm}

Motivated by the instability associated with the ``wrong'' Wick rotation, we may try to sum only over geometries that contain the ``correct'' Wick rotation. We can do this by fixing the initial expansion rate both to be Euclidean and to possess the appropriate sign. Fixing the expansion rate requires a Neumann condition, which we have seen in section \ref{sec:boundaryterms} to arise automatically from the Einstein-Hilbert action, without having to introduce a boundary term. We will see below that, perhaps surprisingly, adding no boundary term will not lead to Neumann conditions for all metric deformations, but it does so for the scale factor. In fact, we will see that adding no boundary term at all leads to a viable implementation of the no-boundary idea. The idea for such a ``no boundary term'' proposal was already mentioned by Louko and Halliwell in early papers on the subject \cite{Louko:1988bk, Halliwell:1988ik, Halliwell:1990tu} and recently in \cite{DiazDorronsoro:2018wro}, but it has not been analysed in much detail so far. 

We will take the final condition to be given by a Dirichlet condition, fixing the size of the universe on the final hypersurface to $q(t=1)=q_1.$ Hence our boundary conditions are
\begin{align}
\frac{\dot{q}}{2N} \mid_{t=0} = + i\,, \qquad q \mid_{t=1} = q_1\,. \label{neumannconditions}
\end{align}
With these boundary conditions the equation of motion for $q$ is solved by
\begin{align}
q(t) &= H^2 N^2 t^2 + 2iNt + q_1 - 2iN - H^2 N^2\,, \label{NDq}
\end{align}
implying that after integrating over $q$ as described in the previous section, we are left with an action solely dependent on the lapse $N,$
\begin{equation}
\frac{S}{V_3} =  H^4 N^3 + 3 i H^2 N^2 - 3 H^2 q_1 N  - 3 i q_1\,. \label{NDaction}
\end{equation}
There are a few points to note about the form of this action: the first is that the action is explicitly complex. This is of course a direct result of imposing a Euclidean momentum at $t=0.$ This means however that even for real values of the lapse the weighting of the action (given by minus its imaginary part) will be non-zero. Picard-Lefschetz theory prescribes that a relevant saddle point of a path integral must be reached by flowing down from the original integration contour. For a purely real action this would imply that only saddle points with negative weighting could contribute, and this would preclude the Hartle-Hawking saddle point (which has a positive weighting $+4\pi^2/(\hbar H^2)$) from ever being relevant. The complexity of the action evades this obstruction. 

The second point to note about the action is that it does not contain a singularity at $N=0.$ Physically, this may be understood as follows: since we are fixing the initial momentum, and not the initial size of the geometries that are summed over, we are in effect summing over geometries of all possible initial sizes. This will include a geometry of size $q_1$ at $t=0,$ and the transition from this 3-geometry to the final hypersurface which also has $q=q_1$ can occur instantaneously, i.e. with $N=0.$ There is thus nothing singular occurring at $N=0$ \footnote{Note that, if the integration contour was taken to run from zero to infinity (to compute a propagator), the absence of a singularity would generate difficulties in the application of Picard-Lefschetz theory.}.
%In turn this implies that one may want to take the range of integration of the lapse $N$ to be the entire real line, rather than just the half-line starting at $N=0$.  

\begin{figure}
	\centering
	\includegraphics[width=0.6\textwidth]{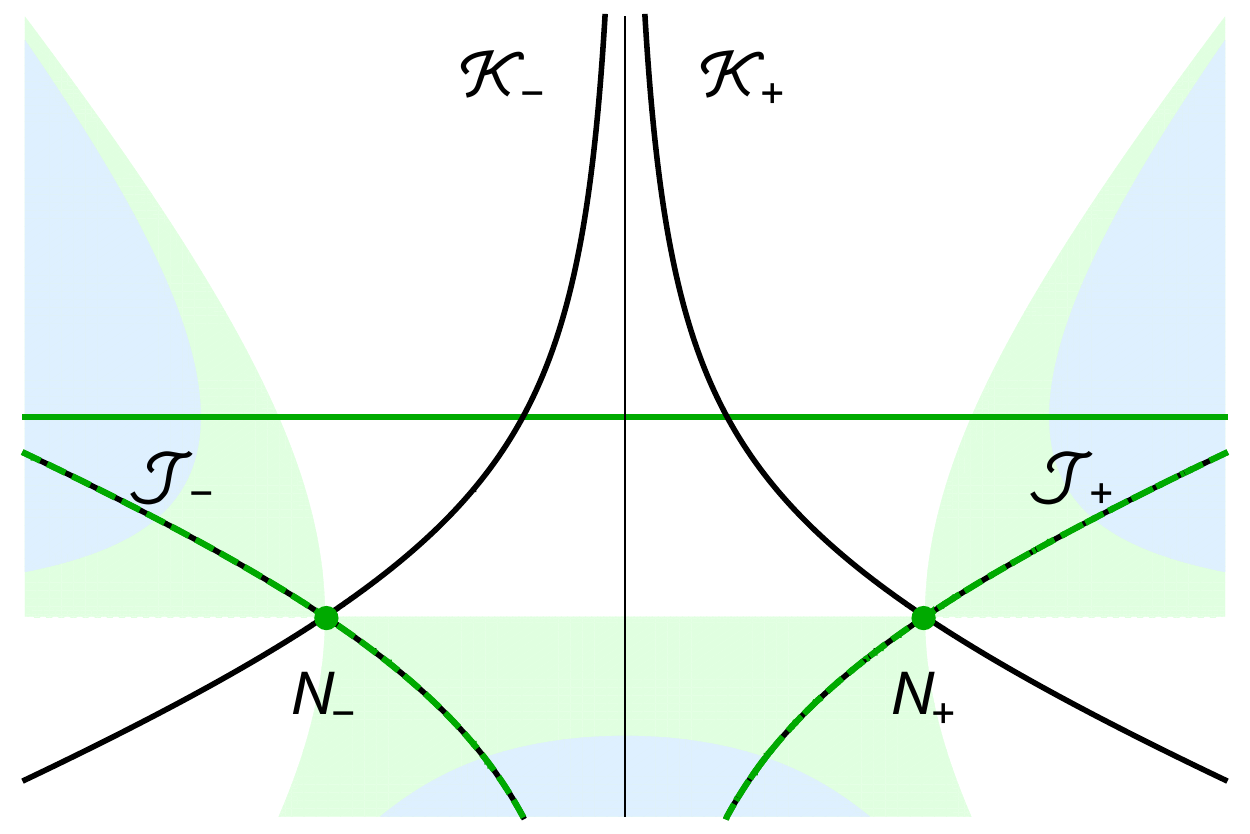} 
	\caption{Flow lines in the complex plane of the lapse function $N$ for the action \eqref{NDaction}, obtained by imposing boundary conditions with a Euclidean momentum at $t=0$ and a fixed final size at $t=1$ (the parameters used are $H=1$,  $\dot{q}_0/2N=i,$ $q_1=2$). The figure shows the paths of steepest ascent (black, ${\cal K}_\pm$) and steepest descent (black, ${\cal J}_\pm$) emanating from the saddle points (in green). Regions of asymptotic convergence are shown in blue, while regions of descent from the saddles are in light green. The dark green real line contour can be deformed into the dark green dashed thimbles ${\cal J}_- + {\cal J}_+.$}
	\label{fig:NeumannFlow}
\end{figure}

A third observation is that the action \eqref{NDaction} only contains two saddle points, located at
\begin{align}
N_\pm = \pm \frac{\sqrt{H^2 q_1 - 1}}{H^2} - \frac{i}{H^2}
\end{align}
These are precisely the Hartle-Hawking saddle points \eqref{eq.nbgeometry}. Compared to the calculation starting from zero size of chapter \ref{nbDirichlet}, which contained $4$ saddle points, the Neumann condition has eliminated the Vilenkin saddle points. Moreover, as Fig. \ref{fig:NeumannFlow} shows, the paths of steepest ascent/descent from these saddle points are such that they are both relevant to the path integral with Lorentzian contour of integration for the lapse. The real line contour for $N$ can in fact be deformed into the sum over both steepest descent paths ${\cal J}_+$,  ${\cal J}_-,$ while the arcs at infinity linking the thimbles to the real line yield zero additional contribution \cite{Feldbrugge:2017kzv}. Note that we could not have used a Euclidean integration contour, as this would have been divergent at large positive imaginary values of the lapse. For the Lorentzian contour, the saddle point approximation then yields the result
\begin{align}
\Psi &\simeq e^{iS(N_-)/\hbar} + e^{iS(N_+)/\hbar} \nonumber \\
& = e^{+\frac{2V_3}{\hbar H^2}} \cos[  \frac{2 V_3 H}{\hbar} (q_1 - \frac{1}{H^2})^{3/2}]\,.\label{PsiHH}
\end{align}
Thus we have recovered the Hartle-Hawking wavefunction \eqref{PsiHH} from a Lorenztian path integral. 

Before discussing some implications of this result, it is interesting to see what happens if we try to extend the ``no boundary term'' prescription to more general metrics. An obvious class of importance are anisotropic metrics. For instance, we can consider the Bianchi IX metric
\begin{align}
ds_{IX}^2 = - \frac{N^2(t)}{q} dt^2 + \sum_m \left( \frac{l_m(t)}{2} \right)^2 \sigma_m^2\,,
\end{align}
with the $\sigma_m$ being one-forms on the 3-sphere. The $l_m$ are direction-dependent scale factors. It is useful to rewrite them as 
\begin{align}
l_1(t) =\sqrt{q} e^{\frac{1}{2}\left(\beta_+(t) + \sqrt{3}\beta_-(t)\right)}\,, \quad l_2(t) = \sqrt{q} e^{\frac{1}{2}\left(\beta_+(t) - \sqrt{3}\beta_-(t)\right)}\,, \quad
l_3(t) = \sqrt{q} e^{-\beta_+(t)}\,,
\end{align}
such that $q$ denotes the average scale factor squared, while the $\beta_\pm$ functions parameterize the deformations/squashings of the spatial slices (the conventional subscript $ \pm $, indicating the polarisation of the gravitational wave, should not be confused with the label of the two relevant saddle points). In these variables the action is given by
\begin{align}
S = V_3 \int dt N \left[  \frac{3}{4N^2}\left( 2 q \ddot{q} + \dot{q}^2  + q^2(\dot{\beta}^2_+ + \dot{\beta}^2_-) \right)  - \left( q  \Lambda + U(\beta_+, \beta_-)\right)\right]\,, \label{actionB9}
\end{align}
with the anisotropy potential
\begin{align} \label{anisotropypotential}
U(\beta_+, \beta_-)  & = - 2 \left( e^{ 2 \beta_+ } + e^{-\beta_+ - \sqrt{3}\beta_-} + e^{-\beta_+ + \sqrt{3}\beta_-} \right) + \left( e^{ -4 \beta_+ } + e^{2\beta_+ - 2\sqrt{3}\beta_-} + e^{2\beta_+ + 2\sqrt{3}\beta_-} \right) \nonumber \\ & = -3 + 6\left(\beta_-^2 + \beta_+^2 \right) + {\cal{O}}(\beta_\pm^3)\,.
\end{align}
Inspection of the action \eqref{actionB9} shows that something interesting has happened: in contrast to $q,$ the anisotropy parameters $\beta_\pm$ only appear with at most single derivatives in the action. Variation of the action thus automatically leads to a Dirichlet condition $\delta \beta_\pm=0$ on the boundaries, without the need to add a Gibbons-Hawking-York term. In short, the ``no boundary term'' prescription leads to a Neumann condition for the scale factor, but to Dirichlet conditions for the anisotropies.  

The equations of motion and constraint following from varying the action are
\begin{align}
& \ddot{q} + q\left( \dot{\beta}^2_+ + \dot{\beta}^2_- \right)  -\frac{ 2N^2}{3}\Lambda  = 0\,, \label{eqq} \\
& \ddot{\beta}_\pm + 2\frac{\dot{q}}{q}\dot{\beta}_\pm  + \frac{2N^2}{3 q^2} U_{,\beta_\pm} = 0\,, \label{eqbeta} \\
& \frac{3}{4}\dot{q}^2 = \frac{3}{4} q^2 \left( \dot{\beta}^2_+ + \dot{\beta}^2_-  \right)  +  q N^2 \Lambda + N^2U(\beta_+, \beta_-)\,. \label{constraint}
\end{align}
By multiplying \eqref{eqbeta} through by $q^2$ and solving near $q=0$ one can immediately see that a regular solution requires $U_{,\beta_\pm}=0$ at $q=0,$ which, if we assume that the saddle point will still be the Hartle-Hawking one with $q(t=0)=0,$ translates into the requirement $\beta_\pm(0)=0.$ This is indeed a Dirichlet condition, which moreover ensures that the constraint \eqref{constraint} can also be satisfied at the South Pole. 

At linear order in the anisotropies, the equation of motion \eqref{eqbeta} reduces to the equation for a linear gravitational wave,
\begin{equation}
\ddot{\beta}_\pm + 2 \frac{\dot{q}}{q} \dot{\beta}_\pm + \frac{8N^2 }{q^2} \beta_\pm = 0\,.  \label{eqpert}
\end{equation}
Let us now focus, for simplicity, on the background geometry associated with the saddle point $ N_+ $.
With the Dirichlet conditions $\beta_\pm(0)=0,\,\beta_\pm(1)=\beta_{1\pm}$ and neglecting backreaction, the solution is given by
 $\beta_\pm(t)=\beta_{1\pm} \cdot g(t)/g(1)$ with
 \begin{equation}
 g(t)= \left(t^2 (\sqrt{H^2 q_1 -1} - i )+4it\right)\left(t (\sqrt{H^2 q_1 - 1}- i )+2i\right)^{-2}\,.
 \end{equation}
The quadratic action associated with this solution (at large final $q_1$) is 
\begin{align}
S^{(2)}_\pm &= \frac{V_3}{2} \int_0^1 dt \, N \Bigl[q^2 \frac{\dot{\beta}_\pm^2}{N^2} - 8 \beta^2_\pm \Bigr] \\
 & = -\beta_{1\pm}^2  \frac{ 4 V_3 q_1}{3i + \sqrt{H^2 q_1 - 1}} \nonumber \\ &  = i \, \beta_{1\pm}^2 \frac{12 V_3}{H^2} - \beta_{1\pm}^2 \frac{4 V_3 \sqrt{q_1}}{H} + \mathcal{O} \Bigl( \frac{1}{\sqrt{q_1}} \Bigr)
\end{align} 
thus explicitly verifying that the Hartle-Hawking saddle points lead to stable, Gaussian distributed, perturbations in a Bianchi IX spacetime.

%%%%%%%%%%%%%%%%

There remain two issues that require some discussion though: the first is one of interpretation. Based on the no-boundary geometry depicted in Fig. \ref{fig:Wick}, Hartle and Hawking proposed that the no-boundary wavefunction could be defined as a path integral where the sum over geometries is restricted to be over compact and regular metrics \cite{Hartle:1983ai}. A direct implementation of the sum over compact geometries can be achieved by using the Dirichlet condition that the scale factor vanishes at $t=0,$ i.e. $q(t=0)=0.$ This has been shown to ultimately fail \cite{Feldbrugge:2017fcc, Feldbrugge:2017mbc} in the sense that it leads to unstable perturbations. Here we have overcome this problem by fixing the initial momentum, rather than the initial size. Due to the uncertainty relation between scale factor and momentum this implies that the path integral sums over geometries with \emph{all} possible initial sizes, with some being larger than the current universe. The dominant geometry, of course, remains the HH geometry starting from ``nothing'', but off-shell all initial sizes are included. Hence, as defined here, the interpretation of the no-boundary wavefunction must change. In the present definition, imposing an appropriate Euclidean momentum takes precedence over the criterion of compactness. In the same spirit, it becomes questionable whether the no-boundary wavefunction truly describes tunneling out of nothing. Rather, as already discussed in \cite{DiTucci:2019dji}, it may describe a quantum transition from a prior state. If this is the case, then it may not be an ultimate theory of initial conditions. Despite this potential drawback, the no-boundary wavefunction retains highly appealing and non-trivial physical properties: in particular, it can describe how the universe becomes classical and provide an explanation for the Bunch-Davies state.

\begin{figure}
	\centering
	\includegraphics[width=0.6\textwidth]{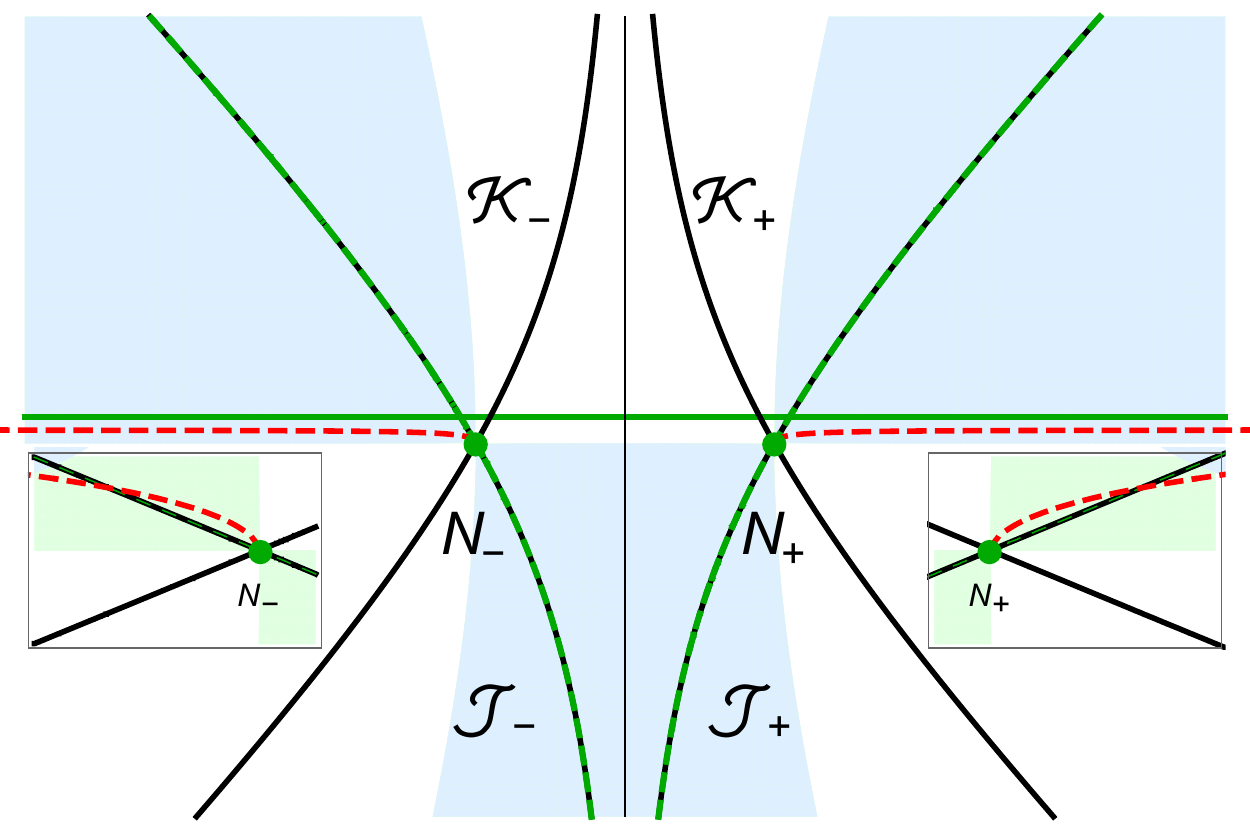}  
	\caption{The curves of zeroes (red, dashed), indicating the loci of geometries in which the scale factor vanishes at some time, emanate almost vertically from the saddle points (see inset) and then become horizontal, necessarily crossing the thimbles. The parameters used in making this figure are $H=1, \dot{q}_0 /2 N=i, q_1=100$.
	}
	\label{fig:Neumann}
\end{figure}

The second issue to be discussed is that of the potential singularities in the off-shell geometries that are being summed over. 
It is not entirely clear whether one should avoid off-shell singularities. On the one hand, they should be present since one is summing over all geometries, and already in quantum mechanics most paths that one sums over are ``singular'' as they are not differentiable. Moreover, if the singular geometries lead to infinite action they may automatically have zero weighting and thus provide no contribution to the path integral. On the other hand the calculation should be self-consistent and one should the integrations along Lefschetz thimbles to be mathematically well-defined and singularity free.
The no-boundary path integral with Dirichlet boundary conditions was plagued by the presence of many singular geometries in the sum: the saddle point geometries, representing portions of complexified de Sitter space, are indeed regular while all of the off-shell geometries, which also start at zero size, close off in a spiky way. Note that this fact does not cause any problem for the path integral at the level of the background because the singularity is integrable and the associated action is finite. However, as discussed in \cite{Feldbrugge:2017mbc}, many geometries along the real $N$ line start at zero size but also cross zero a second time for $t\in[0,1]$. In this case, the perturbative modes necessarily blow up at one of the two locus where $q=0$ leading to a divergence of the action and a break down of the minisuperspace approximation. The situation is significantly improved with a Neumann initial condition: in this case all of the off-shell geometries start with a non-vanishing (often complex) initial size, the saddle point geometries being the only ones which close off to zero. There are however geometries which vanish at different location for $t\in (0,1]$. Nevertheless, perturbations would correspondingly vanish too. Thus the presence of such singular geometries, whose associated Ricci scalar diverges, does not effect the self-consistency of the calculation within the framework of semi-classical general relativity. It is however important to keep in mind that this formulation is potentially sensitive to higher order quantum corrections, as terms involving higher powers of the Riemann curvature could significantly affect the integral.

A singularity will occur when the scale factor $q(t)$ vanishes somewhere along the geometry, i.e. for some real $t$ with $0 < t \leq 1.$ At the saddle point $N_+$ itself  (and similarly at $N_-$) the geometry starts out at $q(0)\mid_{N_{+}}=0$ and it is regular there by construction. But nearby we may expect the singularity to occur at a small value $\delta t$ (note that off-shell in $N$ the constraint, which ensures regularity, is not satisfied). Starting from \eqref{NDq}, a short calculation shows that
\begin{align}
q(\delta t)\mid_{N_+ + \delta N} = 0 \quad  \rightarrow \quad \delta N =  \frac{iN_+}{\sqrt{H^2 q_1 -1}} \delta t\,.
\end{align}
Thus the ``curve of zeroes'', representing the locus of geometries containing a singularity, emanates from the saddle point at an angle which is a rotation by $\pi/2$ compared to the angle subtended by the saddle point itself. At large $q_1,$ the saddle point is almost real and hence the curve of zeroes leaves the saddle point approximately vertically (in the positive imaginary lapse direction). This is confirmed by the numerical calculation shown in Fig. \ref{fig:Neumann}.  At large $N$ (for singularities occurring near $t=1$) the curve of zeroes runs off almost horizontally, just below the real $N$ line and with approximate imaginary part given by $-i/(2H^2).$

Meanwhile the thimbles are defined via the relation $Re(S(N))=Re(S(N_s)),$ since they correspond to stationary phase paths of the integrand. Again perturbing away from the saddle point, one finds that for small deviations $\delta N$ the thimble obeys the relation 
\begin{align}
\text{Re}\left( \frac{1}{2}S_{,NN}(N_+)(\delta N)^2\right)=\text{Re}\left( 3H^2 \sqrt{H^2q_1 -1}(\delta N)^2\right)=0\,.
\end{align}
This equation is quadratic in $\delta N$ because the first derivative of the action vanishes at the saddle point. Since $q_1 > 1/H^2$ we must have $(\delta N)^2$ purely imaginary, or in other words we need $\delta N$ to point in the directions $e^{i\pi/4},e^{i3\pi/4}, e^{i5\pi/4}, e^{i7\pi/4}.$ By inspection we can see that the thimble points at an angle $\pi/4$ away from the saddle point. Thus the thimble necessarily intersects the curve of zeroes, and will necessarily contain at least one singular geometry. Moreover, the deformation of the original integration contour along the real $N$ line will have to pass through the curve of zeroes to reach the thimble. As we said above, it is not entirely clear whether this is a serious obstruction or not as it does not effect the mathematical consistency of the present calculation. It is important, and we will leave it for future work, to verify that this formulation of the Hartle-Hawking wavefunction is indeed robust against higher order derivatives quantum gravity corrections (see for example \cite{Jonas:2020pos} for new results in this direction).

\subsection{Lessons for quantum cosmology from anti-de Sitter black holes}\label{sec:blackholes}

General relativity is a spectacularly successful theory of spacetime and gravity, but amongst its physically relevant solutions there are some that contain singularities and which thus predict the breakdown of the theory from which they originated. Most notable are the singularities inside black holes, and the big bang in cosmological solutions. A series of insights originating already in the 1970s and 1980s implied that black holes and the big bang, which may be seen as the most extreme manifestations of gravity in the universe, could be tamed when perceived from the point of view of Euclidean spacetime. In the case of black holes the Euclidean solution ends at the horizon, and the interior part containing the singularity is simply absent \cite{Hartle:1976tp}. For the big bang, the proposed resolution consists of rounding off the singularity by extending the spacetime to contain a non-singular Euclidean section near the big bang \cite{Hartle:1983ai}. 

The study of Euclidean black holes has been a fruitful way of deriving and elucidating thermodynamic properties of black holes, and thus also the link between gravity and quantum theory. In particular, Euclidean solutions offer the most pragmatic way of deriving the temperature of black holes. Another famous application for black holes in Anti-de Sitter (AdS) spacetime, on which we will focus here, is the Hawking-Page transition~\cite{Hawking:1982dh}. There one finds that, depending on the temperature, either empty AdS or black holes dominate the partition function while a minimum temperature is required for black hole solutions to exist at all. Via holography, gravitational physics in AdS acquires an alternative description in terms of more familiar quantum field theory (QFT) phenomena~\cite{Maldacena:1997re,Gubser:1998bc,Witten:1998qj}, which makes AdS the best understood instance of quantum gravity. In particular, the Hawking-Page transition in holography becomes a thermal phase transition between confined and deconfined phases in a dual QFT with the thermodynamic limit achieved by a very large number of underlying QFT degrees of freedom~\cite{Witten:1998zw}.

In the rest of this chapter, based on \cite{DiTucci:2020weq}, we will reproduce features of the Hawking-Page phase transition by calculating explicitly the gravitational path integral in the minisuperspace approach~\cite{Halliwell:1988wc}. Our studies for black holes in AdS space are motivated by recent developments in cosmology as there is a very close analogy between the calculations performed below and the novel studies of the no-boundary proposal discussed in the rest of this work. The idea is to make use of the firm results for the thermodynamics of black holes to learn lessons for path integrals in quantum cosmology. 

\begin{figure}[h]
\centering
\includegraphics[width=0.5\textwidth]{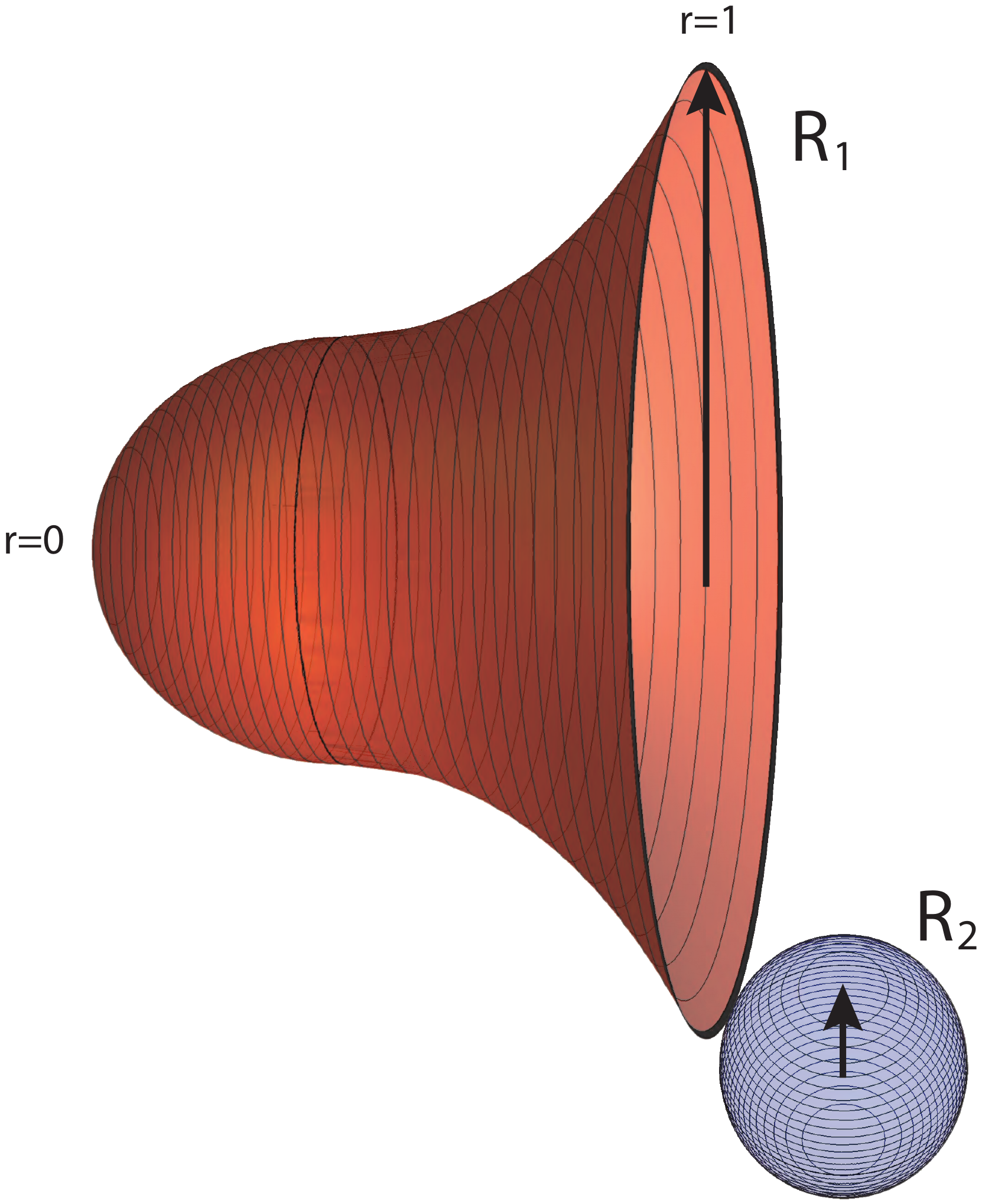}
\caption{We study partition functions that sum over all metrics with a fixed Euclidean boundary, where the boundary will either be a three-sphere or the product of a circle and a two-sphere. Illustrated here is the case with a fixed boundary consisting of the product of a circle of radius $R_1$ and a two-sphere of radius $R_2.$ The sphere is only shown at a single point on the circle. We will use coordinates in which the boundary resides at $r=1$ and we will sometimes refer to it as the ``outer'' boundary. We will assume that the geometry ends at $r=0$ and we will refer to the $r=0$ coordinate location as the ``inner'' boundary. Guided by holography, we will be interested in situations in which this location is not a true geometrical boundary but rather just the end point of a coordinate range.} \label{fig:partition}
\end{figure}

 To this end, we will consider path integrals over four-dimensional geometries within the minisuperspace class with weighting provided by the Einstein-Hilbert term + a negative cosmological constant + appropriate boundary terms. These geometries will be anchored on a Euclidean boundary and we will consider two separate cases of the latter: three-spheres and direct products of a circle and a two-sphere. According to holography, an appropriately understood gravitational path integral corresponds to evaluating the dual QFT partition function with the QFT living on the chosen boundary geometry. The latter we impose on the gravity side as a Dirichlet boundary condition for the four-dimensional metrics we path integrate over. Evaluating such path integrals in the minisuperspace approach requires us to introduce a coordinate~$r$ on which our metric will depend. Without loss of generality we can assume that this coordinate runs between $0$ and $1$ and we choose the Euclidean boundary to lie at $r = 1$. However, in order to make our calculation well-defined we will also need to provide information on how the metrics we integrate over behave at $r = 0$. To be more specific, for an ``outer'' boundary consisting of a circle of radius $R_1$ and a two-sphere of radius $R_2$ illustrated in Fig.~\ref{fig:partition} we will analyse the sum over geometries
\begin{align}
\label{eq.Zgravity}
Z(R_1,R_2) = \sum_{B} \int_{B}^{R_1,R_2} d[g_{\mu\nu}] e^{\frac{i}{\hbar}S}\,,
\end{align}
where $B$ encapsulates conditions imposed on the metric $g_{\mu \nu}$ at $r = 0$.

The first of our two main challenges will be to specify the boundary conditions~$B$ on the ``inner'' boundary, including possibly summing over a class of them. An important guiding principle for us in this quest will be the dual QFT interpretation. Following this thread, one thing that we do not want to do is to impose another Dirichlet boundary condition in which the locus $r = 0$ is a non-trivial three dimensional Euclidean space, since according to holography each such boundary corresponds to an independent copy of a QFT. As we will see later, interpreting the setup from Fig.~\ref{fig:partition} as calculating the thermodynamic partition function in a dual QFT implies that in our approach we must impose a Neumann condition on the metric at the inner boundary of the integration region at $r = 0$. Similarly, a Neumann boundary condition can also be utilised in the case of the boundary being a three-sphere, in which case the litmus test comes from a comparison with the exactly evaluated QFT partition function using supersymmetric localisation. Our AdS calculation in this case can be brought into one-to-one correspondence with the recent definition of the no-boundary proposal in \cite{DiTucci:2019bui} and shows that using momentum conditions at the big bang rather than summing over compact metrics is in fact quite natural from the holographic point of view.

The second main challenge in making sense of~\eqref{eq.Zgravity} will be evaluating the path integral itself. The appearance of meaningful Euclidean saddle point solutions, such as empty Euclidean AdS or a Euclidean AdS black hole, would naturally suggest that gravitational path integrals should be defined as sums over Euclidean geometries in the framework known as Euclidean Quantum Gravity~\cite{Gibbons:1976ue}. However, when using the minisuperspace path integral as a definition of~\eqref{eq.Zgravity}, with the exception of using a Neumann condition at $r = 0$ when the boundary is a three-sphere, we find that in general the sum over bulk Euclidean metrics is not given by a convergent integral and, therefore, is mathematically meaningless. Following~\cite{Feldbrugge:2017kzv}, we \emph{define} the path integral~\eqref{eq.Zgravity} as a sum over a class of sections of complex manifolds. This is necessary in order to turn the conditionally convergent integral \eqref{eq.Zgravity} into a sum of manifestly convergent integrals. These convergent integrals live on steepest descent contours (``Lefschetz thimbles'') and they fix both the meaning and the order of integration of the conditionally convergent sum over metrics. Our approach rests on the formalism of Picard-Lefschetz theory \cite{Witten:2010cx}, although we will only need the simplest, one-dimensional, version of the theory.

Let us already mention some consequences which we will explore. We find that in addition to well-known Euclidean saddle points describing empty AdS and AdS black holes, there always exist three other saddle points with Euclidean or complex bulk geometries. These play no role in the case where the outer boundary is sent off to infinity (i.e. $R_{1}$ and $R_{2}$ diverge with the ratio kept fixed) corresponding to dealing with a ultraviolet-complete holographic QFT [in our case, a holographic conformal field theory (CFT)]~\cite{Maldacena:1997re}. However, they do contribute, though in a very suppressed manner, to the path integral~\eqref{eq.Zgravity} when the boundary is brought to a finite radius (i.e. both $R_{1}$ and $R_{2}$ are finite) describing a particular class of effective holographic QFTs~\cite{McGough:2016lol,Hartman:2018tkw}. The latter arise as irrelevant deformations of CFT by an operator being a square of its energy-momentum tensor.

The plan of the rest of the chapter is as follows: in section~\ref{sec:metrics} we will first review different forms for the metrics of AdS space and AdS black holes, which we will require in our later calculations. Section~\ref{sec:ads} is devoted to the calculation of the partition function with a three-sphere boundary. In section~\ref{sec:bhs} we will then extend these results by changing the boundary topology to $S^1 \times S^2,$ which will allow us to include black holes to our discussion. The connections with cosmology are discussed in section \ref{sec:cosmology} and an outlook and some interesting open problems are provided in section~\ref{sec:discussionAds}. 

\subsubsection{Useful metrics for AdS and black holes} \label{sec:metrics}

The evaluation of gravitational path integrals is greatly simplified by choosing particularly well adapted metric ans\"{a}tze, which differ from the metric forms that are most often used in other contexts. In this section, for convenience we will present the AdS and asymptotically AdS black hole metrics both in a common form and in the form that we will employ later. 

We will consider general relativity in four spacetime dimensions in the presence of a negative cosmological constant $\Lambda,$ with action
\begin{align}
S = \frac{1}{16\pi G} \int d^4x \sqrt{-g} \left[ R - 2 \Lambda \right]\,,
\end{align}
where one may define 
\begin{equation}
\Lambda \equiv - \frac{3}{l^2}
\end{equation}
with $l$ denoting the radius of curvature of the maximally symmetric AdS solution. We specialise here and in the following to gravity in four dimensions because we want to draw lessons about cosmology in our universe. It would certainly be interesting to generalise the findings of our work to an arbitrary number of dimensions.

The Euclidean version of the empty AdS solution may be written as
\begin{align}
\label{eq.metricAdS}
ds^2 = d\rho^2 + l^2 \sinh^2 \left( \frac{\rho}{l}\right) d\Omega_3^2\,,
\end{align}
where $d\Omega_3^2$ is the metric on the unit three-sphere with volume~$V_3=2\pi^2$ and the (asymptotic) boundary resides at~$\rho \rightarrow \infty$. Via holography, the gravitational action evaluated on-shell on this solution and supplemented with appropriate counter terms to kill divergences incurred as $\rho \rightarrow \infty$ approximates the logarithm of a partition function (free energy) for a dual CFT living on the boundary. For the metric~\eqref{eq.metricAdS} the latter is a three-sphere. One reason why it is interesting to compute partition functions for three-dimensional QFTs on spheres stems from this quantity being a natural measure of the number of degrees of freedom in such QFTs~\cite{Jafferis:2011zi}.

We will find it useful to consider a metric of the form \cite{Halliwell:1988ik}
\begin{align}
\label{metric}
ds^2 = -\frac{N^2}{q(r)}dr^2 + q(r) d\Omega_3^2\,,
\end{align}
This ansatz is of course the same as \eqref{FLRW} but in what follows we will think of $r$ as a radial with a finite range, say $0 \leq r \leq 1.$ running from the interior to the boundary of the geometry. This choice of coordinate will to turn out to be useful as will consider situations in which there is a boundary at a fixed radius. Note that the minus sign in the above equations is not a standard convention in holography, but it will facilitate the comparison with cosmology. Also, in the end we will define Eq.~\eqref{eq.Zgravity} as a path integral over complex geometries, so this is just a choice of a convention. \\
A patch of the EAdS solution for $0 \leq \rho \leq \rho_{max} = l \, \mathrm{arcsinh}(\frac{R_{3}}{l})$ then corresponds to 
\begin{align}
N & = \pm i \, l\left( \sqrt{R_3^2+l^2} + l\right) \,, \\
q(r) &= \left(\sqrt{R_3^2 + l^2} + l \right)^2 r^2  - 2 l \left(\sqrt{R_3^2 +l^2} + l\right)r \,. \label{scalefactorads}
\end{align}
Note that the lapse function is imaginary, in accordance with the Euclidean nature of the solution. Here $R_3$ may be seen to fix the proper radius of the three-sphere at the outer boundary at coordinate location $r=1$. In particular, within this ansatz considering the solution all the way to the asymptotic boundary corresponds to blowing-up the radius $R_{3}$.

We will also consider metrics with $S^1 \times S^2$ topology on constant radial surfaces. A~corresponding metric for AdS space is
\begin{align}
ds^2 = \frac{d\rho^2}{\left( \frac{\rho^2}{l^2} + 1\right)} + \left( \frac{\rho^2}{l^2} + 1\right) d\tau^2 + \rho^2 d\Omega_2^2\,, \label{AdSmetric}
\end{align}
where 
\begin{equation}
d\Omega_2^2=d\theta^2 + \sin^2(\theta)d\phi^2
\end{equation}
is the metric on the unit two-sphere, and where we have chosen the ``time'' coordinate $\tau$ to be Euclidean. In these coordinates the empty Anti-de Sitter solution can be straightforwardly extended to include a (Euclidean) Schwarzschild black hole \cite{Carter:1973rla}, as
\begin{align}
ds^2 = \frac{d\rho^2}{\left( \frac{\rho^2}{l^2} + 1 - \frac{2M}{\rho}\right)} + \left( \frac{\rho^2}{l^2} + 1 - \frac{2M}{\rho}\right) d\tau^2 + \rho^2 d\Omega_2^2\,, \label{AdSbhmetric}
\end{align}
where $M$ denotes the mass of the black hole. The horizon radius $r_+$ of the black hole is given by the real root of 
\begin{align}
\frac{\rho^3}{l^2} + \rho -  2M=0\equiv \frac{1}{l^2}(\rho-r_+)(\rho-r_1)(\rho-r_2)\,, \label{cubic}
\end{align} 
while the other two roots $r_1,r_2$ form a complex conjugate pair, since the discriminant of this cubic equation is negative\footnote{The cubic roots have the properties that $r_++r_1+r_2=0$ and $r_1 r_2 = r_+^2 + l^2.$}. From this one immediately obtains an expression for the mass in terms of the horizon radius
\begin{align}
M=\frac{1}{2} r_+ \left( 1 + \frac{r_+^2}{l^2} \right)\,. \label{mass}
\end{align}
In order for the manifold to avoid a conical singularity at the horizon, one must impose that the $\tau$ coordinate is periodic (so that the near-horizon metric resembles that of the origin of flat space in polar coordinates) with period \cite{Hawking:1982dh}
\begin{align}
\beta = \frac{4\pi l^2 r_+}{3r_+^2+l^2}\,. \label{horizon}
\end{align}
The AdS/CFT correspondence maps the mass $M$ in~\eqref{mass} to the expectation value of the corresponding CFT Hamiltonian in a thermal state on a unit two-sphere at temperature equal to~$1/\beta$~\cite{Balasubramanian:1999re}.

Once again we would like to bring the metric \eqref{AdSbhmetric} into a form where the radial coordinate has finite range.  For this, we will first pick a radius $\rho=R_2,$ where $R_2$ denotes the radius of the two-sphere on the boundary. Our radial coordinate $r$ ranges from $0$ to $1,$ and should interpolate between $r_+$ and $R_2.$ Thus we will define
\begin{align}
\label{eq.defbBH}
\rho \equiv b(r) = r (R_2-r_+) + r_+.
\end{align}
Here one can see that sending $\rho$ to $\infty$ is equivalent to blowing up $R_{2}$, as we wrote earlier. Using the ansatz for the Kantowski-Sachs class of metrics~\cite{Halliwell:1990tu},
\begin{align}
\label{eq.bhmetric}
ds^2 = - \frac{b(r)}{c(r)} N^2  dr^2 + \frac{c(r)}{b(r)} d\tau^2 + b(r)^2 d\Omega_2^2\,,
\end{align}
the black hole geometry can now be rewritten as 
\begin{subequations}
\begin{align}
\label{eq.BHNc}
N & = \pm i (R_2 - r_+)\,, \\
c(r) & = \frac{1}{l^2}[b^3(r) + l^2 b(r) - r_+^3 - l^2 r_+]\,.
\end{align}
\end{subequations}
On the outer boundary at $r=1$ we have
\begin{align}
b(1) = R_2 \quad \mathrm{and} \quad c(1) = \frac{1}{l^2}(R_2^3 + l^2 R_2 - r_+^3 - l ^2 r_+ )\,. \label{b1c1}
\end{align}
If we denote the period of $\tau$ by $\beta$ then we can see that the size of the circle direction on the boundary is given by
\begin{align}
\sqrt{\frac{c(1)}{b(1)}}\, \beta \equiv R_1\,. \label{R1}
\end{align} 
Keeping $R_1$ and $R_2$ fixed specifies the size of the outer boundary. At the inner boundary at~$r=0,$ the metric \eqref{AdSbhmetric} implies
\begin{align}
b(0) = r_+\quad \mathrm{and} \quad c(0)=0\,.
\end{align}
The value $b(0)$ effectively specifies the mass of the black hole, according to \eqref{mass}. Again a conical singularity at $r=0$ is avoided provided the periodicity $\beta$ is given by
\begin{align}
\beta = \frac{4\pi b(0) |N|}{\dot{c}(0)} = \frac{4\pi l^2 r_+}{3 r_+^2 + l^2}\,, \label{noconical}
\end{align}
where a dot denotes a derivative w.r.t. $r.$ 

Pure Euclidean AdS space~\eqref{AdSmetric} is recovered in the limit $r_+ \to 0,$ and in that case the periodicity $\beta$ is arbitrary since the manifold is smooth in any case.

%%%%%%%%%%%%%%%%%%%%%%%%%%%%%%%%%
%%%%%%%%%%%%%%%%%%%%%%%%%%%%%%%%%

\subsection{$S^3$ boundary and Euclidean AdS$_4$ saddles}\label{sec:ads}

\subsubsection{Neumann condition at $r = 0$}\label{subsec:N}

We will first review how Euclidean AdS space is obtained as the saddle point of a gravitational path integral. This calculation was done previously by Caputa and Hirano in \cite{Caputa:2018asc}, and in three dimensions in \cite{Donnelly:2019pie}\footnote{See also Ref.~\cite{Hirano:2019szi} for another recent application of the minisuperspace approach in holography.}. Here we will perform the analogous calculation in a different style adopted from Ref.~\cite{Feldbrugge:2017kzv}, which has the advantage that it will allow us to extend the calculation to black holes in the next section. Also, motivated by the extension to black holes, we will impose here different boundary conditions at $r = 0$ than in \cite{Caputa:2018asc}. In subsection \ref{subsec:D} we will show how the results of~\cite{Caputa:2018asc} fit into our framework, and we will discuss some implications of our studies in \ref{subsec:lessons}.

The object we are interested in is the partition function
\begin{align}
\label{eq.defZ3}
Z(R_3) = \int^{R_3} d[g_{\mu\nu}] e^{\frac{i}{\hbar}S}\,,
\end{align} 
with a three sphere of radius $R_3$ at the fixed (outer) boundary. Note that the nature of the bulk metrics that we integrate over is going to be determined by the contour of integration in the lapse integral and will in general involve complex metrics. Also, the signature of the metric in which a dual QFT lives is fixed by the outer boundary condition and unaltered by this genuinely bulk phenomenon. To be more precise about our aim, we want to define Eq.~\eqref{eq.defZ3} within the minisuperspace approach so that it is mathematically meaningful and has features consistent with calculating a partition function in a dual QFT.

The action we will consider consists of the Einstein-Hilbert action with a negative cosmological constant, supplemented by the York-Gibbons-Hawking (YGH) surface term~\cite{York:1972sj,Gibbons:1976ue} at the (outer) boundary,
\begin{align}
\label{eq.SgravNeumann}
S = \frac{1}{16\pi G}\int d^4 x \sqrt{-g} \left[ R + \frac{6}{l^2} \right] + \frac{1}{8\pi G}\int_{outer} d^3 y \sqrt{h} K +S_{ct} \,,
\end{align}
where the counterterms $S_{ct}$ will be discussed below. Note that in this subsection we do not add any surface terms on the inner boundary, for reasons that will become clear. We work with minisuperspace metrics given the ansatz~\eqref{metric}. The coordinate $r$ interpolates between the inner boundary at $r=0$ and the outer boundary at $r=1,$ that is to say $0 \leq r \leq 1.$ Here $N(r)$ is the lapse function and $q(r)$ is the scale factor squared, which determines the size of the three-sphere. We will denote $q(r=0) \equiv q_0$ and $q(r=1)\equiv q_1=R_3^2.$ The reason for choosing this less familiar metric ansatz is that the action ends up being quadratic in $q,$ which will be very useful in evaluating the path integral over $q.$ In fact the action reduces to\footnote{One subtlety that we want to highlight is that in passing from the general action~\eqref{eq.SgravNeumann} to its form for the minisuperspace metrics~\eqref{metric} we made a choice of a branch in the expression~$\sqrt{-g}$. For purely Euclidean metric, this would be the standard choice one makes.} 
\begin{equation}
S = \frac{3\pi}{4 G} \int_0^1 dr \left[ - \frac{\dot{q}^2}{4 N} + N \left(1 + \frac{q}{l^2}\right) \right] - \frac{3 \pi q_0 \dot{q_0}}{8 GN} + S_{ct}\,, \label{NDaction}
\end{equation}
where a dot over a function denotes here and in the following a derivative w.r.t. $r$. Second derivatives acting on $q$ have been eliminated using integration by parts, and the resulting surface term at $r=1$ has eliminated the GHY surface term there while introducing a surface term~$- \frac{3 \pi q_0 \dot{q_0}}{8 GN}$ at $r=0.$ Variation of the action w.r.t. $q$ leads to 
\begin{align}
\delta S = \frac{3\pi}{4G} \int_0^1 dr \frac{\delta q}{2N} \left[  \ddot{q} + \frac{2 N^2}{l^2} \right] - \frac{3 \pi\dot{q}_1 \delta\left(q_1\right)}{8GN} - \frac{3 \pi q_0 \delta\left(\dot{q}_0\right)}{8GN} + \delta S_{ct}\,. 
\end{align}
Thus we obtain the equation of motion $ \ddot{q}  = - \frac{2 N^2}{l^2}\,, $ supplemented by the boundary conditions that we can hold $q$ fixed at the outer boundary, as desired (the variation $\delta S_{ct}$ will be consistent with this), while self-consistency upon not including the York-Gibbons-Hawking term on the inner boundary of the integration range forces us to fix $\dot{q}$ there. More properly we should say that it is the momentum conjugate to the scale factor, 
\begin{equation}
\Pi = \frac{\delta \mathcal{L}}{\delta \dot{q}} = -  \frac{3\pi}{8GN}\dot{q}\,,
\end{equation}
that will be fixed on the inner boundary.
This Neumann condition, understood as fixing the momentum to some value (not necessarily 0), is not strictly needed in the present calculation: as we will show in the next subsection, we could have used the Dirichlet condition here. However, a momentum condition will be necessary when including black holes, if we want to reproduce standard results of black hole thermodynamics in the canonical ensemble and we interpret the result of the bulk path integral as the thermodynamic partition function. It will be useful for us to parameterise the momentum at $r=0$ by a re-scaled parameter $\alpha,$ 
\begin{equation}
\Pi_0  = -  \frac{3\pi}{8GN}\dot{q}_0 \equiv -  \frac{3\pi}{4G}\alpha\,. \label{momentumcondition}
\end{equation}
To proceed, we must evaluate the path integral over the metric i.e. the integrals over the lapse and the scale factor, $\int d[g_{\mu\nu}] = \int d[N]d[q],$ where we will ignore Jacobian factors given that we will eventually evaluate the partition function in the saddle point approximation. Here and in the following we will make use of an old result of~\cite{Teitelboim:1981ua,Halliwell:1988wc}, namely that one can use the gauge freedom of general relativity to restrict the sum to run over manifolds in which the lapse $N$ does not depend on~$r$. This drastically simplifies the integral over $q$ and also transforms the functional integral over $N$ into an ordinary integral.  The procedure to evaluate the path integral over $q$ is to shift variables by writing $q = \bar{q} + Q$ \cite{Halliwell:1988ik}. Here $\bar{q}$ denotes a solution of the equation of motion for $q$ respecting the boundary conditions $\Pi(0)=\Pi_0$ and $q(1)=q_1=R_3^2.$ Explicitly, we have \begin{align}
\bar{q}(r) = - \frac{N^2}{l^2} (r^2-1)  +2N\alpha (r-1)  + R_3^2\,. \label{saddlescale}
\end{align}
Meanwhile $Q$ is an arbitrary perturbation (not necessarily small), which obeys the boundary conditions, i.e. vanishes at $r = 1$ and has a vanishing momentum at $r = 0$. Since the action is quadratic in $q,$ the path integral then turns into a factor given by the action evaluated along the $\bar{q}$ plus a Gaussian integral over $Q$. As we show in appendix \ref{sec:determinant}, this yields just a numerical prefactor. Thus, up to this numerical factor, we are left with
\begin{subequations}
\begin{align}
Z(R_3) & = \int dN e^{i(S_0+S_{ct})/\hbar}\,, \label{lapsepartition}\\
\frac{4G}{\pi} S_0(N)&=  \frac{N^3}{l^4}-3\alpha \frac{N^2}{l^2} +3N\left(\frac{R_3^2}{l^2}+1+\alpha^2 \right) -3\alpha R_3^2 \,, \label{sphereaction}
\end{align}
\end{subequations}
where it is important that the counter terms contain no dependence on the lapse. 
While the integral over $N$ from $- i\infty$ to $0$ converges in this particular case, below we take a route of Ref.~\cite{Feldbrugge:2017kzv} which will apply to all the cases considered in the present work\footnote{We will see below that $N=0$ is not a natural end point of integration, as the relevant steepest descent contour continues beyond this point.}.
 
\begin{figure}[h]
\centering
\includegraphics[width=0.5\textwidth]{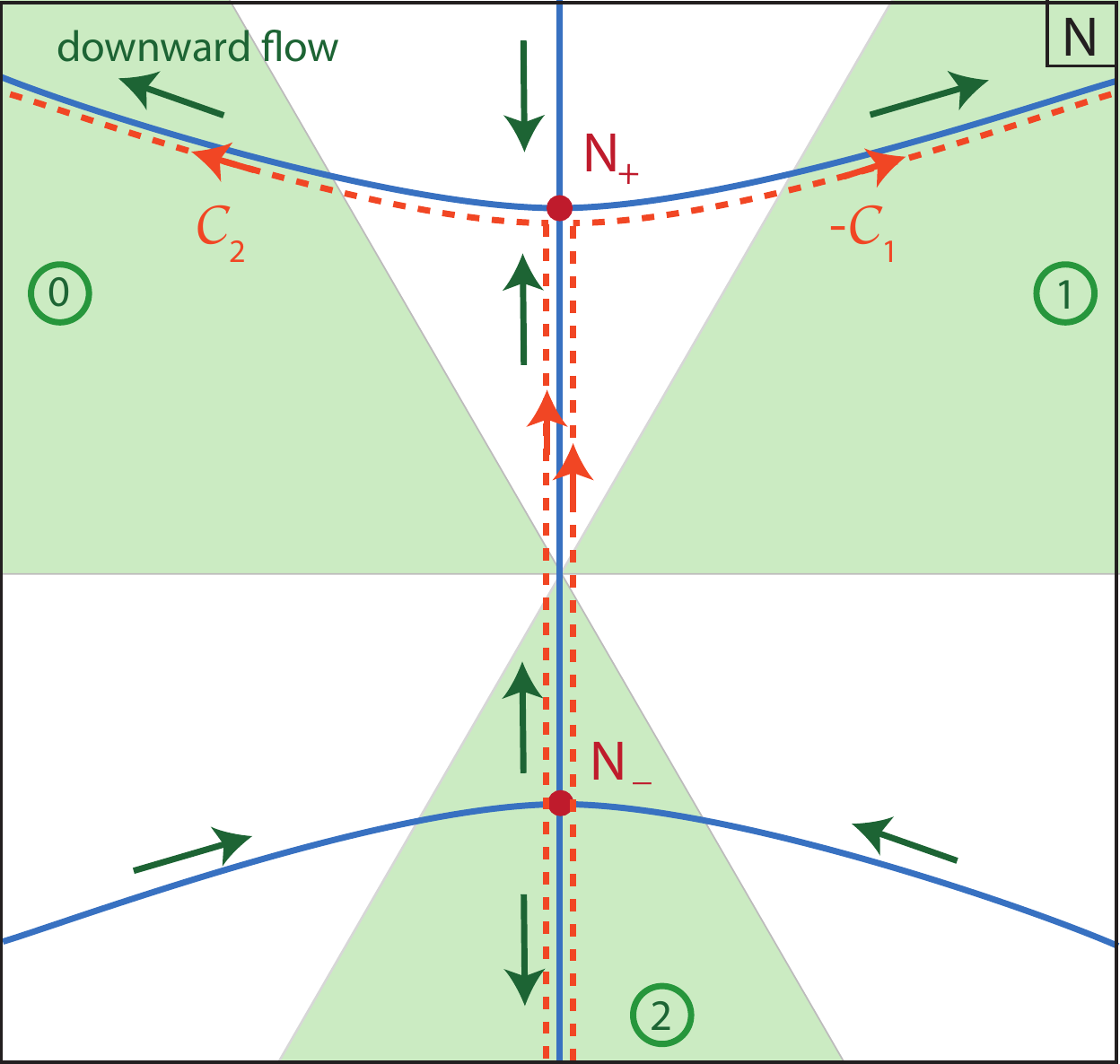}
\caption{The figure shows the structure of the flow lines with boundary conditions $\alpha=+i$, $q_1 = R_3^2$.  The saddle point $N_-$ represents the EAdS geometry, while $N_+$ represents a singular section of complexified AdS space. The asymptotic regions of convergence are shown in light green, and are labelled by the encircled numbers $0,1,2.$ We denote the contour of integration linking region $0$ to region $1$ by ${\mathcal C}_0 = 0 \to 1,$ and similarly ${\mathcal C}_1 = 1 \to 2$ and ${\mathcal C}_2 = 2 \to 0.$ Integration along ${\mathcal C}_0$ yields the Airy Ai function. Meanwhile, the combination ${\mathcal C}_2 - {\mathcal C}_1$ gives a result proportional to the Airy Bi function. This sum of contours is equivalent to summing the two contours shown by the orange dashed lines, which run via the saddle point $N_-$ to the saddle point $N_+$ and from there to opposite regions of convergence. Note that the required contours of integration are neither Lorentzian nor Euclidean and the combination ${\mathcal C}_2 - {\mathcal C}_1$ is the closest it gets to an effectively Euclidean path integral.} \label{fig:flowAdS}
\end{figure}

The lapse integral above can in fact be evaluated exactly. It is easiest to see this by shifting the integration variable to $\tilde{N} \equiv N-\alpha l^2,$ which leads to
\begin{align}
\label{eq.shiftedaction}
    \frac{4 G}{\pi} S = \frac{1}{l^4}\left[\tilde{N}^3 + 3 l^2 \left(R_3^2 + l^2 \right)\tilde{N} + \alpha (3+\alpha^2)l^6 \right]\,.
\end{align}
Then our path integral can be identified as an Airy integral (using $d\tilde{N}=dN$). The possible integration contours are defined in terms of the asymptotic regions of convergence labelled $0,1$ and $2$ in Fig. \ref{fig:flowAdS}. They are located at phase angles $\theta \equiv \arg(N)$: $0 \leq \theta \leq \frac{\pi}{3}$ (region $1$), $ \frac{2\pi}{3}\leq \theta \leq \pi$ (region $0$) and $ \frac{4\pi}{3}\leq \theta \leq \frac{5\pi}{3}$ (region $2$). We define the contours as ${\mathcal C}_0 = 0 \to 1$, ${\mathcal C}_1 = 1 \to 2$ and ${\mathcal C}_2 = 2 \to 0$, which define for us the two Airy functions as follows~\cite{Vallee},
\begin{equation}
\label{eq.defAiry}
Ai[z]=\frac{1}{2 \pi} \int_{\mathcal{C}_0}   dx \,  e^{ i \left(\frac{x^3}{3} + z x \right)} \quad \mathrm{and} \quad Bi[z]=i \frac{1}{2 \pi} \int_{\mathcal{C}_2 - \mathcal{C}_1 }   dx \,  e^{ i \left(\frac{x^3}{3} + z x \right)}\,.
\end{equation}

Since the CFT partition function is real, we should also expect the gravitational partition function to yield a real valued result. There are then two possible contours. The first is to integrate along ${\mathcal C}_0,$ and this yields a result proportional to the Airy function $Ai\left[ \left(\frac{3\pi l^2}{4 G \hbar} \right)^{\frac{2}{3}} \left( R_3^2+l^2\right)\right].$ At large $R_3$ this function is exponentially suppressed $\sim e^{-R_3^3}$ and thus cannot represent the desired answer. The other possibility is to integrate along ${\mathcal C}_2 - {\mathcal C}_1.$ This can be seen as follows: our integrand is odd in the lapse, which implies that under the transformation $N\to - N^*$ the integrand changes into its complex conjugate. An integration contour that is even under $N \to -N^*$ can then be split up into two pieces which are reflection symmetric across the imaginary $N$ axis, and, if they have the same orientation (e.g. left to right), they represent a sum of an integral and its complex conjugate -- thus, the result is real. This is the case for the contour ${\mathcal C}_0 = -({\mathcal C}_2 + {\mathcal C}_1),$ which gives the Airy Ai function. The other possible contour that is even under $N \to - N^*$ is the difference ${\mathcal C}_2 - {\mathcal C}_1.$ It corresponds to the difference of an integral and its complex conjugate and is pure imaginary; after multiplication with the imaginary unit $i$, it also yields a real result, namely the Airy Bi function.

Reinstating the spatial volume $V_3$ and using Eq.~\eqref{eq.defAiry}, the integral over ${\mathcal C}_2 - {\mathcal C}_1$ yields the following answer  
\begin{align}
    Z(R_3) &= e^{i \frac{V_3}{8\pi G \hbar}\alpha (3+\alpha^2)l^2} \, Bi\left[ \left(\frac{3V_3}{8\pi G \hbar l} \right)^{\frac{2}{3}} \left( R_3^2+l^2\right)\right] e^{\frac{i}{\hbar}S_{ct}}\,. \label{Z3result}
\end{align}
We will determine $\alpha$ momentarily, but first it is useful to look at the large $R_3$ limit of this expression. Naively our result diverges as the boundary is pushed to infinity, $R_3 \rightarrow \infty$. This is the usual infinite volume divergence found in the context of asymptotically AdS spacetimes. This divergence is cured by the introduction of counter terms~\cite{Balasubramanian:1999re,deHaro:2000vlm},
\begin{equation}
S_{ct} = \frac{i}{16\pi G} \int_{outer} d^3 y \sqrt{h} \left(\frac{4}{l} + l \,  R^{(3)}\right),
\end{equation}
where $h$ and $R^{(3)}$ are the determinant and the Ricci scalar of the three-metric on the outer boundary. For the metric ansatz \eqref{metric} they become $S_{ct} = +\frac{iV_3}{8 \pi G l} (2 R_3^3 + 3 R_3 l^2)\,,$ so that\footnote{Note that the counter terms depend only on the radius $R_3$ and not on its derivative, and hence the variation of the counter terms is consistent with our Dirichlet condition at $r=1.$} 
\begin{align}
e^{\frac{i}{\hbar}S_{ct}} = e^{-\frac{V_3}{8\pi G \hbar l} (2 R_3^3 + 3 R_3 l^2)}\,. \label{ct}
\end{align}
For large values of $R_{3}$ the gravitational path integral~\eqref{Z3result} takes the form
\begin{align}
    Z(R_3)  \approx e^{\frac{V_3}{8\pi G \hbar} \left[2 \left( R_3^2 + l^2\right)^{3/2} + i \alpha (3+\alpha^2)l^2 -2R_3^3 - 3 l^2 R_3\right]},
\end{align}
where we kept terms up to ${\cal O}(1/R_{3})$ in the exponent, and leads to
\begin{align}
\label{eq.ZS3Neumannexact}
Z = e^{i \frac{V_3}{8\pi G \hbar}\alpha (3+\alpha^2)l^2}
\end{align}
as $R_3 \to \infty$. We want to emphasise that, up to an ambiguity in the notion of the path integral measure, this is the exact result of the path integral within the minisuperspace ansatz.

On the QFT side of holography, the partition function on a three-sphere can be evaluated \emph{exactly} in the very special case of the ABJM theory~\cite{Aharony:2008ug} using localisation~\cite{Pestun:2007rz}. The result reads~\cite{Fuji:2011km,Marino:2011eh}
\begin{align}
\label{eq.ZABJM}
Z = Ai\left[ \left(\frac{3V_3l^2}{8\pi G \hbar}\right)^{2/3}\right],
\end{align}
where we utilised the holographic dictionary for the ABJM theory to reinstate $G$ instead of the number of underlying QFT degrees of freedom. In the limit when the gravity side is described in terms of classical gravity, i.e. when the argument in eq.~\eqref{eq.ZABJM} is very large, one gets
\begin{align}
Z \sim e^{-2V_3l^2/(8\pi G \hbar)}.
\end{align}
Thus we see that we recover the leading term of the ABJM result with the choice $\alpha=+i.$ At the current point in the calculation, this choice appears somewhat mysterious, but we will see shortly that it has a perfectly sensible physical origin. There are two immediate consequences however: the first is that this identification means that the momentum condition $\alpha=+i,$ corresponding to $\Pi_0 = - \frac{3\pi}{4G} i,$ must be fixed and should not be summed over in the partition function. The second is that with $\alpha=+i,$ the total partition function in eq. \eqref{Z3result} is real for any value of $R_3,$ as expected on general grounds.

\begin{figure}[h]
\centering
\includegraphics[width=0.5\textwidth]{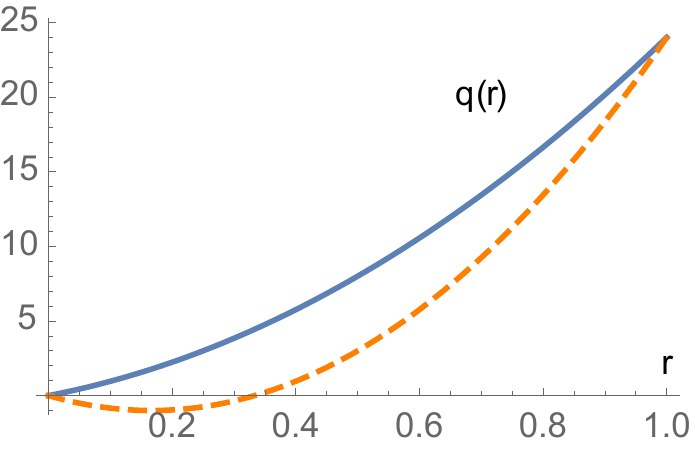}
\caption{This graph shows the profile of the scale factor squared $q(r)$ at the saddle points $N_- = -4i, N_+= 6i,$ obtained with the parameter values $l=1, R_3^2=24.$ The solid curve is given by $q(r)=8r+16r^2$ and represents EAdS space at $N_-$, while $N_+$ corresponds to a section $q(r)=12r+36r^2$ of (complexified) AdS space in which the scale factor turns imaginary in one region.} \label{fig:saddles1}
\end{figure} 

To gain more insight into this calculation, we will also perform the lapse integral \eqref{lapsepartition} in the saddle point approximation. For this it is useful to first study the nature of the saddle points. These are located at extrema of $S_0(N),$ i.e. at
\begin{subequations}
\begin{align} 
    N_\pm & = \alpha l^2 \pm i l \sqrt{R_3^2 + l^2}\,, \label{S3saddles} \\
    \frac{4 G}{\pi} S_0(N_\pm) &= \alpha (3+\alpha^2) l^2 \pm \frac{2i}{l}\left(R_3^2 + l^2 \right)^{3/2}\,, \label{saddleactionN}
\end{align}
\end{subequations}
where we also indicated the value of the action at the saddle points. We can obtain the saddle point geometry by inserting $N_\pm$ into Eq. \eqref{saddlescale},
\begin{align} \label{saddlemetric}
    \bar{q}(r)\mid_{N_\pm} = - \left( \alpha l \pm i \sqrt{R_3^2 + l^2}\right)^2 r^2 + 2\alpha \left( \alpha l^2 \pm i l \sqrt{R_3^2 + l^2}\right) r - l^2 (1+\alpha^2)\,.\end{align}
Here we can see that if we are to evade a physical boundary at $r=0,$ we need to restrict to the momentum conditions $\alpha=\pm i,$ so that $\bar{q}(0)=0.$ This consideration already reduces the possible values of $\alpha$ to just two. All saddle points consist of sections of complexified AdS spacetime. When $\alpha=+i,$ the saddle point $N_-$ corresponds to the usual Euclidean AdS space, which we expected to find. The other saddle point, $N_+,$ describes a section of Euclidean AdS glued onto a reversed-signature EAdS piece. For this last saddle point, the squared scale factor~$\bar{q}$ passes through zero and becomes imaginary, and thus we would expect perturbations to blow up there, cf. Fig. \ref{fig:saddles1}. Also note from Eq. \eqref{saddleactionN} that the EAdS solution has a higher weighting than the singular saddle point. By contrast, when $\alpha =-i$ the EAdS is at $N_+$ and the singular geometry at $N_-,$ and in this case the singular geometry dominates. In the limit of large $R_3$ the subdominant saddle points are suppressed, which indeed implies that we should choose $\alpha=+i.$ 

The saddle points, along with their steepest descent lines, are shown in Fig. \ref{fig:flowAdS}. The contour of integration ${\mathcal C}_2 - {\mathcal C}_1,$ which we chose above, can then be deformed into the sum of two contours that are symmetric w.r.t. the imaginary lapse axis, and which run from negative imaginary infinity either to the convergence region $0$ or $1.$ These contours follow the steepest descent path through the saddle point $N_-$ representing EAdS space, on to the saddle point $N_+$ and from there along either half of the steepest descent path associated with $N_+.$ At $N_+$ the two parts of the total integration contour run parallel to the real lapse axis, but in opposite directions, implying that the end result will not contain a contribution from the singular saddle point. In the saddle point approximation, including the counter terms, the partition function is then approximated as
\begin{align}
Z(R_3) & \approx e^{\frac{V_3}{8\pi G \hbar l}\left(-2l^3 + \frac{3l^4}{4R_3} + {\cal O}(R_3^{-3})\right)} \,,
\end{align}
in agreement with our earlier result~\eqref{eq.ZS3Neumannexact} for $\alpha = +i$.

We are thus able to define a partition function peaked around Euclidean AdS space, by using a Neumann condition at $r = 0$ and a Dirichlet condition at the boundary. Perhaps the most surprising aspect of this calculation is that the contour for the lapse integral can be neither Euclidean nor Lorentzian, but must be inherently complex, as shown in Fig. \ref{fig:flowAdS}. However, it is interesting to note that if one were to ``sum'' the contours together, then ${\mathcal C}_2 - {\mathcal C}_1$ is in fact equal to the Euclidean lapse axis\footnote{Note that the result of this calculation differs from the naive summation over the negative imaginary axis that one can do in this special case.}. This may be the closest one is able to come to a realisation of Euclidean quantum gravity with the caveat that we discussed before.

We will see later that many of these aspects persist when we extend our calculation to include black holes. For now, we will first compare our calculation with one using Dirichlet boundary conditions on both ends.

%%%%%%%%%%%%%

\subsubsection{Dirichlet condition at $r = 0$}\label{subsec:D}

To compare with Ref.~\cite{Caputa:2018asc} we have to compare our results with the calculation performed with Dirichlet boundary conditions
\begin{align}
q(r=0) = 0\,, \qquad q(r=1)=R_3^2\,.
\end{align}
Note that in this case the condition of starting at zero size is put in from the outset\footnote{As we have already mentioned, starting with a non-zero size would superficially imply including another holographic QFT. This is inconsistent with consideration of a partition function, hence our prescription.}. It will thus hold everywhere, i.e. also off-shell, and not just at the saddle points. However, this condition does not guarantee that at $r=0$ the geometry will be regular -- in fact it will only be so at the saddle points. For the Dirichlet calculation, we must use the Einstein-Hilbert action supplemented with the GHY terms at both $r=0,1,$ which reduces to the minisuperspace action\footnote{The Dirichlet condition $q_0=0$ is special in the sense that the surface term vanishes for this particular value, cf. eq.~\eqref{NDaction}, hence one does not necessarily need the GHY term at $r=0.$ However, if one thinks of this calculation as integrating from smaller and smaller initial sizes, then it makes sense to add the GHY term in order to ensure a smooth limit when $q_0 \to 0.$}
\begin{equation}
\frac{8\pi G}{V_3}S = 3  \int_0^1 dr \left[ - \frac{\dot{q}^2}{4 N} + N (1 + \frac{q}{l^2}) \right]\,.
\end{equation}
The second term in the action arises from the positive curvature of the three-sphere, and the last term from the cosmological constant. The GHY boundary term has eliminated all second derivatives, so that the variational problem will be well posed when imposing Dirichlet boundary conditions on $q.$ The trick to evaluate the path integral over $q$ is once again to shift variables by writing $q = \bar{q} + Q.$ Here $\bar{q}$ denotes a solution of the equation of motion for $q$ respecting the boundary conditions,
\begin{align}
\bar{q}(r)= - \frac{N^2}{l^2} r^2 + \left(\frac{N^2}{l^2} + R_3^2\right) r\,.
\end{align}
Meanwhile $Q$ is an arbitrary perturbation (not necessarily small) with vanishing value at the end points $Q(0)=Q(1)=0.$ Since the action is quadratic in $q,$ the path integral then turns into an integral over $\bar{q}$ which is just a given function of $r$ and can be integrated directly, plus a Gaussian integral over $Q$ which just changes the prefactor by a factor $1/\sqrt{N}$~\cite{Halliwell:1988ik}. Thus, up to an overall numerical factor that we were persistently neglecting throughout the text, we are left with
\begin{subequations}
\begin{align}
\Psi & = \frac{e^{i \frac{\pi}{4}}}{\sqrt{\pi \hbar}} \int \frac{dN}{\sqrt{N}} e^{iS_0/\hbar}\,, \label{lapseintegralDirichlet} \\
\frac{8\pi G}{3V_3} S_0 &=  \frac{N^3}{12 l^4} + \frac{N}{2 l^2}(R_3^2 +  2 l^2)  - \frac{R_3^4}{4 N} \,.
\end{align}
\end{subequations}
Here we have denoted the path integral by the new letter $\Psi,$ since the relation to the partition function of the previous section is a priori not clear. The asymptotic convergence regions at infinity are unchanged from the Neumann case, but in addition the action now contains a pole at $N=0,$ so that there are additional choices for the lapse integration contour. Intuitively the appearance of a singularity at $N = 0$ should not be all that surprising. In fact here we are summing over four-geometries which interpolate between two three-spheres of radii $q(r=0)=0$ and $q(r=1)=R_3^2>0$. When $N = 0$, then the proper distance between them vanishes and the singularity is signaling that the corresponding geometry is not smooth. Note also that this pole invalidates any attempts to perform the integral along the negative imaginary axis within the Euclidean quantum gravity approach. This singularity was not present in the case where we fixed the momentum at $r=0$ as this Neumann condition can be thought of as a sum over all possible sizes $q_0$. Indeed from eq. (\ref{saddlescale}) one can see that the size of the sphere located at $r=0$ changes with the lapse and the geometry with $N=0$ is regular and has $q(r=0)=q(r=1)$.% \\

It turns out that also in the present case the lapse integral in eq.~\eqref{lapseintegralDirichlet} can be evaluated exactly~\cite{Vallee}. The trick is to rewrite the measure factor as a Gaussian integral $e^{-i \,\frac{(q_1 - q_0)^2}{4 N}} e^{i\,\pi/4}\sqrt{\frac{\pi}{N}} = \int d\xi \, e^{i \,N\,\xi^2 +i \,(q_1 - q_0) \,\xi},$ and then, after a change of variables to $N \pm 2 \xi$, the integral \eqref{lapseintegralDirichlet} can be identified as a product of two Airy integrals. Thus the solution is given by the product of two Airy functions. The choice of integration contour for the lapse determines the type of Airy functions, where care is needed to ensure that all integrals converge. We once again require the resulting quantity to be real in order to interpret it as a partition function, and moreover, this time it must be symmetric w.r.t. the inner and outer boundaries since we imposed Dirichlet conditions on both ends. One possibility is to consider the real contour for the lapse running above the origin; correspondingly $\xi$ runs along the real axis and the path integral is given by $\Psi \propto Ai\left[\left(\frac{3V_3l^2}{8\pi G \hbar}\right)^{2/3}\right]Ai\left[\left(\frac{3V_3}{8\pi G \hbar l}\right)^{2/3}\left(R_3^2 + l^2\right)\right]e^{\frac{i}{\hbar}S_{ct}} $. This choice would however give a vanishing result in the limit where the boundary is pushed to infinity $R_3 \rightarrow \infty$. The solution with the right asymptotic behaviour is then given by
\begin{align}
\Psi \propto & \,\,\, Ai\left[\left(\frac{3V_3l^2}{8\pi G \hbar}\right)^{2/3}\right]Bi\left[\left(\frac{3V_3}{8\pi G \hbar l}\right)^{2/3}\left(R_3^2 + l^2\right)\right]e^{\frac{i}{\hbar}S_{ct}} \nonumber \\ & + Bi\left[\left(\frac{3V_3 l^2}{8\pi G \hbar }\right)^{2/3}\right]Ai\left[\left(\frac{3V_3}{8\pi G \hbar l}\right)^{2/3}\left(R_3^2 + l^2\right)\right] e^{\frac{i}{\hbar}S_{ct}}, \label{Dresult}
\end{align}
where we have also added the counter terms. Obtaining this solution requires an integration contour for the lapse which runs along the real $N$ line, but passes below the singularity at $N=0,$ cf. Fig.~\ref{fig:flows}. This will become very clear when considering the saddle point approximation below. 

In the result above, the second line is suppressed compared to the first. In fact in the infinite $R_3$ limit, the second line disappears completely, and in the first line the counter term compensates for the $Bi$ function, leaving the result
\begin{align}
    \Psi \to Ai\left[ \left(\frac{3V_3 l^2}{8\pi G \hbar}\right)^{2/3}\right]\qquad \left(\text{as }\,\,\, R_3 \to \infty \right).
\end{align}
Thus the Dirichlet calculation reproduces the exact ABJM result obtained in a superconformal field theory -- see the discussion around eq.~\eqref{eq.ZABJM}. This is rather surprising and is likely a coincidence, as our AdS calculation included only pure gravity, and was restricted to minisuperspace metrics. By comparison, the Neumann calculation in the same setting reproduced only the leading semi-classical term, which is truly all one could have hoped to recover in any case. As we will see below, the interpretation of the Dirichlet calculation is not entirely straightforward, as it does not reproduce the canonical ensemble once black holes are included. Still, it is interesting that it reproduces the associated CFT sphere partition function so precisely. This was already noticed in the work by Caputa and Hirano~\cite{Caputa:2018asc}.

As in the Neumann case, we can gain a little more insight into the Dirichlet calculation by evaluating the lapse integral in the saddle point approximation. There are now four saddle points, residing at the values 
\begin{equation}
N_{c_1 , c_2} = i \,  l \, c_1  [\sqrt{R_3^2 + l^2} + c_2 l]\,, \qquad c_1,c_2 =\pm 1\,.
\end{equation}
Thus all four saddle points reside on the imaginary axis. Note that we started with a Lorentzian metric ansatz in \eqref{metric}, but the lapse function at the saddle points nevertheless ends up being imaginary. Thus the saddle point geometries are Euclidean. In fact these are the four saddle points that we obtained in the calculation with Neumann conditions $\alpha=\pm i.$ Here they all appear together, because all four saddle points respect $q(0)=0.$ The two saddle points with $c_2 = 1$ are the singular bouncing solutions with the scale factor passing through zero. The two with $c_2 = -1$ are the EAdS geometries. The action at the saddle points once again reads
\begin{equation}
\frac{8\pi G}{V_3}S(N_{c_1 \, c_2}) = + c_1 \frac{2 i }{l} [(R_3^2 + l^2)^{3/2} + c_2 l^3 ]\,. \label{saddleactionD}
\end{equation}
It is purely imaginary. The saddle points with $c_1=+1$ will correspond to a suppressed weighting, while those with $c_1=-1$ will have an enhanced weighting, compared to a classical solution which would have a real action. 

\begin{figure}[h]
\centering
\includegraphics[width=0.8\textwidth]{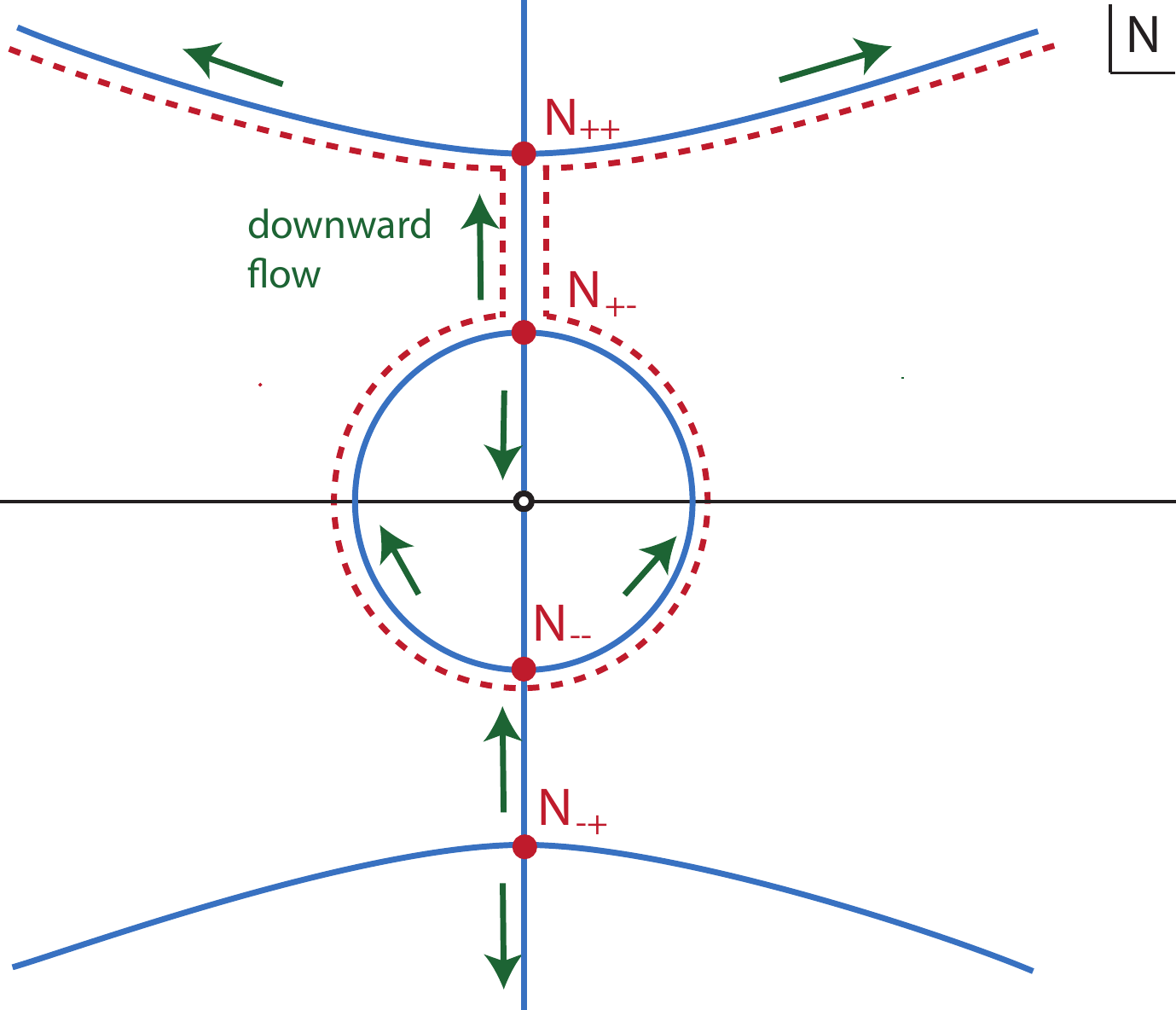}
\caption{Flow lines in the complexified plane of the lapse function, for Dirichlet boundary conditions. The saddle points closer to the origin have $c_2=-1,$ while those two that are further away have $c_2=+1.$ The dashed line in the figure indicates the preferred contour of integration. Note that the action contains a singularity at $N=0.$} \label{fig:flows}
\end{figure}

To see which saddle points are relevant to the path integral, we must analyse the upwards/downwards flow lines, i.e. the steepest ascent/descent lines of the weighting, and moreover we still have to specify the contour of integration for the lapse function. The flow lines are shown in Fig. \ref{fig:flows}. Even though we have obtained the same saddle points as in the Neumann calculation, the flow lines are different, not least because there is now a singularity of the action at $N=0,$ which acts as an essential singularity from the point of view of path integration.

All four saddle points are linked by steepest ascent/descent lines. There are several options for the contour of integration, though, as we have already stressed, one cannot define a Euclidean path integral over the negative imaginary axis, which would diverge due to the singularity at $N = 0$. One can however define integrals along the Lorenztian line of real $N$ values, but one must choose whether to pass above or below the singularity at the origin. Another option is to consider contours that run from the region of convergence at negative imaginary infinity out to the region of convergence between $0$ and $\pi/3$ radians, or between $2\pi/3$ and $\pi$. Then both saddle points in the lower half plane would be relevant. Let us try to figure out which contour is the most sensible by comparing again to the expected QFT result~\eqref{eq.ZABJM} in the semi-classical limit,  which reduces to $e^{-\frac{2V_3l^2}{8\pi G \hbar}}\,.$ Now recall that from Eq.~\eqref{saddleactionD}, the saddle point approximation to the path integral will be given by a sum over terms of the form
\begin{align}
e^{-\frac{2V_3 c_1}{8\pi G \hbar }\left[(R_3^2+l^2)^{3/2} +c_2 l^3\right]} \approx e^{-\frac{V_3 c_1}{l}\left[2 c_2 l^3 + 2 R_3^3 + 3 R_3 l^2 + {\cal O}(\frac{1}{R_3})\right]}
\end{align}
where we have set $q_0=0.$ The counterterm is $e^{\frac{i}{\hbar}S_{ct}} = e^{-\frac{V_3}{8\pi G \hbar l} (2 R_3^3 + 3 R_3 l^2)}\,.$ The divergence in the two saddle points in the lower half plane is then cancelled by the counterterm. To match the expected Airy function result (providing us with the correct leading order results), we must also have $c_2=-1.$ Thus we have to pick the third saddle from the top, i.e. the upper one in the lower half plane, which we called $N_{--}$. Reproducing this result requires using the contour over complexified metrics passing below the origin. In Fig.~\ref{fig:flows} we denoted a sample contour with a red dashed curve. Note that the contours that originate at negative imaginary infinity cannot be used, as they would also pick up the singular saddle point $N_{-+},$ which moreover would lead to a mismatch with the expected QFT result.

\subsubsection{Comments} \label{subsec:lessons}

Before we move on to considering black hole spacetimes in the next section, let us pause here and summarise the most salient features encountered so far in our exploration. To start with, if one were to trust the minisuperspace path integral as an exact statement, then one is either naturally (the Neumann case in section~\ref{subsec:N}) or necessarily (the Dirichlet case in section~\ref{subsec:D}) led to integrate over complex metrics, in the case of AdS quantum gravity. Such calculations require then an additional input regarding which contours in the space of complexified metrics to choose. These choices lead to different semi-classical limits and only some reproduce dual QFT expectations, such as the reality of the Euclidean path integral in dual QFT situations of interest or exact QFT results dictated by symmetries. It would clearly be desirable to have an entirely gravitational consistency criterion for a definition of the gravitational path integral, but for the moment we do not have one. Notice that this was not the case in cosmology where Lorentzian path integrals are well defined and there is no reason to depart from that.

Regarding more detailed findings in the two cases we consider, the results eventually gave rise to the same relevant saddle point upon adjusting the integration contours appropriately, but otherwise the two calculations differ. For example, the path integration measure is different in both cases, cf. eq.~\eqref{lapsepartition} vs.~Eq.~\eqref{lapseintegralDirichlet}. While the Dirichlet calculation can be made to match the exact ABJM result~\cite{Fuji:2011km,Marino:2011eh}, see Eq.~\eqref{eq.ZABJM}, as noted earlier in Ref.~\cite{Caputa:2018asc}, we believe this agreement is accidental. In particular, ambiguities in the integration measure, stemming from our uncertainty about the fundamental definition of an integration measure over metrics (which are the analogue of ordering ambiguities in the associated Wheeler-deWitt equations), alter the answer beyond the leading semi-classical exponent. Still, there exists a rather close link between the Neumann exponential \eqref{eq.ZS3Neumannexact} and the Dirichlet Airy function \eqref{Dresult}, which stems from the fact that the Fourier transform of the Airy function is indeed an exponential with a cubic exponent. More precisely, we have that \cite{Vallee}
\begin{align}
    \int_{- \infty}^{\infty} dq_0\, e^{i \, q_0 \, \Pi_0/\hbar} \, Ai\left[\Bigl(\frac{3\, V_3 }{8\, \pi \,G\, \hbar\, l } \Bigr )^{2/3} (q_0 + l^2)\right] = \Bigl(\frac{8 \pi \, G \hbar \, l}{3 \, V_3}\Bigr)^{2/3} e^{- \frac{i \,\Pi_0 \, l^2}{3 \,\hbar} \Bigl((\frac{8\, \pi \, G\,l}{3 \,V_3})^2\,\Pi_0^2 + 3 \Bigr)} \,,% \nonumber \\ &=  \Bigl(\frac{8 \pi G \hbar l}{3 V_3} \Bigr)^{2/3} e^{i \frac{V_3}{8 \pi G \hbar} \alpha (3 + \alpha^2)l^2}\,,
\end{align}
where the integral must be performed over all real $q_{0}$, i.e. all the sizes of three-sphere at $r = 0$, including also possible changes in signature\footnote{We want to remind the reader that the path integrals we consider involve in general complexified metrics.}. Upon using the relation~\eqref{momentumcondition} between $\Pi_{0}$ and $\alpha$ and up to an overall normalisation that we were persistently ignoring throughout the text we recognize in the outcome the partition function for the Neumann condition at $r = 0$ given by Eq.~\eqref{eq.ZS3Neumannexact}. Note that this relation applies only in the limit $R_{3} \rightarrow \infty$ and at finite $R_3$ it can only be approximate. \\
Thus, implicitly extending the Dirichlet result to $q_0 \neq 0,$ this relation provides a link between \eqref{eq.ZS3Neumannexact} and \eqref{Dresult} in the limit where $R_3 \to \infty,$ as in that limit the second line in \eqref{Dresult} disappears.
In other words, in this limit the Neumann calculation represents the momentum space wavefunction compared to the position space Dirichlet case as indeed the path integrals \eqref{eq.ZS3Neumannexact} and \eqref{Dresult} satisfy the Wheeler-deWitt equations in momentum and position space respectively.  \\
These considerations show that the close agreement between the Neumann and Dirichlet results is truly accidental, as conceptually these two calculations are very different. As we will argue below, the associated thermodynamic interpretations must therefore also differ. It will be interesting to understand the holographic interpretation of these two conditions. 

The calculations that we have presented so far have direct analogues in early universe cosmology. Before exploring the implications of this correspondence in section \ref{sec:cosmology}, we will however first deepen our results by considering the addition of black holes with AdS asymptotics as saddles.

%%%%%%%%%%%%%%%%%%%%%%%%%%%%%%%%%%%%%%%%%%
%%%%%%%%%%%%%%%%%%%%%%%%%%%%%%%%%%%%%%%%%%

\subsection{$S^2 \times S^1$ boundary and black holes as saddles} \label{sec:bhs}

\subsubsection{Preliminaries}

In the previous section we saw how to obtain Euclidean AdS$_{4}$ space from a path integral with a fixed three-sphere boundary. In order to include black holes in our discussion, and to see how classic results such as the Hawking-Page phase transition appear in our framework, we must change the topology of the boundary to a direct product of a two-sphere and a circle, as sketched in Fig. \ref{fig:partition}. We will proceed in much the same way as in the previous section, but the added complications of the metric ansatz make us focus on the saddle point approximation and the choice of contour in the underlying gravitational path integral. We will not include counterterms since we will use the empty AdS solution as our reference solution in the partition function, as is often the case in the holographic literature. Also, the question of which conditions should be used on the inner boundary of the integration range is rather subtle, and we will discuss it in detail.

The metric ansatz that we will use in the following is given in Eq.~\eqref{eq.bhmetric} and in the context of minisuperspace approaches appeared earlier starting from the work by Halliwell and Louko~\cite{Halliwell:1990tu}. We have adopted a convention for the lapse function $N$ such that for real $N$ and $b/c > 0$ the coordinate $r$ is time-like. However, in light of the black hole solutions presented in section \ref{sec:metrics}, we should expect the saddle point values of the lapse to turn out imaginary, thus rendering the metric Euclidean. There are two scale factors, $b(r),$ which determines the size of the $S^2,$ and $c(r)$ which determines the size of the Euclidean time direction $\tau.$ Moreover, we will take the $\tau$ direction to be periodically identified with period $\Delta \tau$, such that it will have the topology of a circle. We will once again assume a finite range of the $r$ coordinate, $0 \leq r \leq 1$ with the inner boundary at $r=0$ and the outer boundary at~$r=1.$

Let us immediately discuss the required boundary conditions. On the outer boundary at $r=1,$ we will impose Dirichlet boundary conditions, keeping the proper size of the outer boundary fixed. If we denote the size of the boundary circle by $R_1$ and that of the boundary two-sphere by $R_2,$ then that means that we will impose
\begin{align}
b(r=1) \equiv R_2 \quad \mathrm{and} \quad \sqrt{\frac{c(r=1)}{b(r=1)}}\Delta \tau \equiv R_1\,. \label{Dr1}
\end{align}
Note that both the form of the metric~\eqref{eq.bhmetric} and the above boundary condition are preserved under a residual diffeomorphism and redefinition of functions defining metric components:
\begin{equation}
\tau \rightarrow \gamma \, \tau, \quad c \rightarrow \gamma^{-2} \, c \quad \mathrm{and} \quad N \rightarrow \gamma^{-1} \, N. \label{residual}
\end{equation}
The easiest way to fix this gauge freedom is to fix the periodicity of the $\tau$ coordinate to a convenient value, as we will do below.

In order to obtain a variational problem consistent with Dirichlet boundary conditions as given by Eq.~\eqref{Dr1}, we will have to add the usual GHY term at the outer boundary.
As we will mention in a little more detail below, imposing Dirichlet boundary conditions on the inner boundary leads to results that are inconsistent with the interpretation of the gravitational path integral as the partition function. The Hawking-Page calculation of black hole thermodynamics in asymptotically AdS space in fact assumed that there was no surface term on the inner boundary (which coincides with the horizon location of the black holes) and indeed on shell the geometry smoothly caps off at $r = 0$. This suggests that off-shell we should impose Neumann conditions at $r = 0$, i.e. that we should fix the momenta rather than the field values as we did before in section~\ref{subsec:N}. In fact, we will view not including the GHY term at $r = 0$ and getting a well defined path integral in the minisuperspace approach as a covariant definition of imposing there the Neumann condition\footnote{In dimensions other than four, a surface term is required to obtain a Neumann boundary condition \cite{Krishnan:2016mcj}.}.

To get started, let us evaluate the extrinsic curvature that enters the GHY term for the metric~\eqref{eq.bhmetric}. To this end, at a fixed radius $r$ our ansatz describes a $S^1\times S^2$ manifold with a diagonal metric
\begin{align}
h_{ij} = \mathrm{diag}_{ij} \left(\frac{c}{b},\, b^2,\, b^2 \sin^2 \theta \right)\,.
\end{align}
The conjugate momenta are defined in terms of the extrinsic curvature $K_{ij}$ via
\begin{align}
\Pi^{ij} \equiv -\frac{\sqrt{h}}{16\pi G}\left(K^{ij} - h^{ij}K \right)\,,
\end{align}
which leads to
\begin{align}
\label{eq.PiBH}
\Pi^{ij} = -\frac{1}{16\pi G} \, \mathrm{diag}^{i j} \left[ 2 \, \frac{b\,\dot{b}}{N}, \frac{1}{2}\left(\frac{\dot{c}}{N\,b}+\frac{c\,\dot{b}}{N\,b^2}\right), \frac{1}{2 \sin^2\theta}\left(\frac{\dot{c}}{N\,b}+\frac{c\,\dot{b}}{N\,b^2}\right)\right]\,.
\end{align}
The total GHY surface term is given by the sum of the products of momenta and fields,
\small
\begin{align}
\Pi^{ij}h_{ij} = \frac{1}{8\,\pi \,G}\sqrt{h}\, K = - \frac{1}{16\,\pi\, G}\left( \frac{b\,\dot{c}}{N} + 3\frac{c\,\dot{b}}{N}\right)\,.
\end{align}
\normalsize
Based on these considerations we write the action as 
\begin{align}
S = \frac{1}{16\pi G}\int d^4 x \sqrt{-g}\left[ R + \frac{6}{l^2} \right] + \frac{1}{8\pi G}\int_{outer} d^3 y \sqrt{h} K \,.
\end{align}
With these mixed Neumann conditions at $r=0$ and Dirichlet conditions at $r=1$, the minisuperspace action reduces to
\begin{align}
\label{eq.SND}
S_{ND} = \frac{\Delta \tau}{2G} \int  dr \, \left[ -\frac{\dot{b}\,\dot{c}}{N} +N\left(1 + \frac{3 \, b^2}{l^2} \right)  \right] \, - \, \frac{\Delta\tau}{4 G} \left( \frac{b\,\dot{c}}{N} + 3\frac{c\,\dot{b}}{N}\right)\Bigg|_{r=0}\,.
\end{align}
Varying the action with respect to $b$ and $c$ gives
\begin{align}
\delta S_{ND} & = \frac{\Delta \tau }{2 G} \int dr \left[ \left(\frac{\ddot{c}}{N} + \frac{6 N b }{l^2}\right)\delta b + \left(\frac{\ddot{b}}{N}\right) \delta c \right] \nonumber \\ & - \frac{\Delta \tau }{2 G} \left( \frac{\dot{c_1}}{N} \delta b_1 +  \frac{\dot{b}_1}{N} \delta c_1\right) -\frac{\Delta \tau }{4 G} \left( b_0^2 \delta\left(\frac{\dot{c}_0}{N b_0}\right) + \dot{b}_0 \delta c_0  + 3  c_0 \delta \dot{b}_0  \right)\,.
\end{align}
Let us now explore possible boundary conditions that render the variational problem well-posed. As we anticipated, at $r = 1$ we will impose Dirichlet boundary conditions, which ensure~\eqref{Dr1} and make the respective boundary terms disappear. The situation at $r = 0$ is more subtle. Cancelling the last two terms at $r = 0$ in Eq.~\eqref{eq.SND} can be achieved by setting
\begin{equation}
\label{eq.c0condBH}
c(0) = c_{0} = 0.
\end{equation}
Note that despite the fact that it looks like a Dirichlet condition for $c$, we view it as a Neumann condition from the geometric point of view. Getting rid of the first term at $r = 0$ can be done either with
\begin{equation}
\label{eq.b0condBHve0}
b_{0} = 0 \quad \mathrm{or} \quad \frac{\dot{c}_0 \Delta \tau}{N \, b_0} = \mathrm{fixed}.
\end{equation}
In the following we will encapsulate both conditions in the form of the following single equation\footnote{There is a choice of sign on the right hand side, which is analogous to the choice of sign we encountered with the momentum condition \eqref{momentumcondition} in section \ref{sec:ads}. We choose this sign such that the black hole solutions are dominant over singular saddle points, rather than other way around (cf. the discussion below). This means that we will take $\omega$ to be a positive real number. \label{footomega}}
\begin{equation}
\label{eq.horcond}
\frac{\dot{c}_0 \, \Delta \tau}{4 \, \pi \,i \, N \, b_0} \equiv \omega = \mathrm{fixed}.
\end{equation}
This equation has a simple interpretation when dealing with Euclidean metrics for which $b_{0}, \, \dot{c}_0$ and $i \, N$ are positive, where $\omega$ is related to the deficit / excess angle spanned by the $S^{1}$ direction at $r = 0$. When $b_{0} = 0$, which corresponds to $\omega \rightarrow \infty$, the $S_{1}$ direction does not shrink to a zero size at $r = 0$. This is the case for the thermal AdS solution given by Eqs.~\eqref{eq.defbBH} and~\eqref{eq.BHNc} with $r_+ = 0$ and arbitrary periodicity in $\tau$. For $\omega = 1$ the geometry smoothly ends at $r = 0$ without a conical singularity, as is the case for the Euclidean AdS black hole. For any other value of $\omega$ we end up with a conical singularity at $r = 0$.

With the variational problem well-posed, we can proceed to perform the path integrals over the scale factors. Since the action is again quadratic in $b$ and $c$ we may evaluate the path integrals over these fields in analogy with the integration over $q$ in section \ref{sec:ads}, i.e. by shifting the variable of integration to the sum of a solution of the equations of motion plus a general fluctuation~\cite{Halliwell:1990tu}. The fluctuation integrals will be unimportant, since they will lead to an overall numerical prefactor in front of the partition function, see appendix \ref{sec:determinant}. The nontrivial physics lies in the solutions of the equations of motion for $b$ and $c,$ given by
\begin{align}
\label{eq.eomsthermal}
\ddot{b} = 0 \quad \mathrm{and} \quad \ddot{c} = -\frac{6}{l^2} N^2 b\,.
\end{align}
The solutions of Eqs.~\eqref{eq.eomsthermal} subject to the conditions~\eqref{Dr1}, \eqref{eq.c0condBH} and~\eqref{eq.horcond} take the form
\begin{subequations}
\begin{align}
b(r)  &= (R_{2}-b_0)\,r+b_0 \,,\label{bsol}\\
c(r)  &= \left( c_1 + \frac{(2 \, b_{0}+ R_{2})N^2}{l^2} \right) r - \frac{3 \, b_{0} N^2}{l^2} \, r^2 + \frac{(b_{0} - R_{2})N^2}{l^2} r^3\,, \label{csol}
\end{align}
\end{subequations}
where
\begin{equation}
b_{0} = \frac{\Delta\tau \left(l^2 \, c_{1} + N^2 \, R_2\right)}{4 \,\pi \, i \, \omega \, l^2 \, N - 2 \Delta\tau \, N^2}\,, \qquad  c_1 = \frac{R_1^2 R_2}{\Delta\tau^2}\,.
\end{equation}
With these solutions at hand, we may now perform the integrations over $b$ and $c,$ leaving us with an integral for the lapse function only. 

\subsubsection{Evaluation of the gravitational path integral}

The partition function once again reduces to an ordinary integral over the lapse function, with two additional features: first, we must include a suitable integral over the boundary conditions on the inner boundary, i.e. we must include a sum over $\omega$, and secondly, we will implement a background subtraction and use the AdS solution given by eqs.~\eqref{eq.defbBH} and~\eqref{eq.BHNc} with $r_+ = 0$ as a reference. Thus the partition function is given by
\begin{align}
Z(R_1,R_2) = \int d\omega\int dN e^{\frac{i}{\hbar}\left(S_{ND}(N) - S_{EAdS}\right)}
\end{align} 
with
\begin{align}
&\frac{8\,G \, l^2}{\Delta\tau}S_{ND} = \nonumber \\ &\frac{(3 R_2^2 + 4 l^2)\Delta\tau N^4 - 8 \pi i \omega l^2 (R_2^2 + l^2) N^3 - 6 l^2 R_2 c_1 \Delta \tau N^2 + 8 \pi i \omega l^4 R_2 c_1 N - l^4 c_1^2 \Delta \tau}{N^2 (\Delta \tau N -2\pi i \omega l^2)}\,. \label{lapseaction}
\end{align}
The partition function is a sum over interior geometries with the boundary at $r=1$ held fixed. When searching for saddle points, the only smooth geometries correspond to $\omega = 1$ (black holes) or the limit $\omega \rightarrow \infty$ (thermal AdS). We restrict ourselves to geometries that on-shell are smooth at $r = 0$, in which case the integral over $\omega$ thus reduces to a sum over just two values,
\begin{align}
  Z(R_1,R_2) =  \sum_{\omega = 1, \infty} \int dN e^{\frac{i}{\hbar}\left(S_{ND}(N) - S_{EAdS}\right)}\,.
\end{align}
It would be very interesting to include geometries with conical deficits in our framework and we leave it for future work.

We will analyse the full saddle point structure momentarily, but first we may check explicitly that the black hole and AdS solutions will arise from this action. It is not possible to solve for the saddle points of the action analytically, since the corresponding equation $\frac{d S_{ND}}{d N}=0$ is a quintic. There are generally five distinct saddle points. However one may verify by direct substitution that two of them are given by
\begin{subequations}
\begin{align}
\label{cbh}
N_s & = -i (R_2 - r_+)\,, \\ b(r) &= r (R_2-r_+) + r_+\,, \\ c(r)  &= \frac{1}{l^2}[b^3(r) + l^2 b(r) - r_+^3 - l^2 r_+],.
\end{align}
\end{subequations}
with $r_+$ taking the two possible values that solve \eqref{horizon} for a given $\beta,$ and where we have fixed the scaling ambiguity \eqref{residual} by choosing
\begin{equation}
\Delta \tau^{bh} = \beta.
\end{equation} 
These solutions satisfy the boundary condition $\omega=1.$ Evidently these are the sought after black hole solutions presented in section \ref{sec:metrics}. Their action is given by
\begin{align}
S^{bh} = -\frac{i}{4Gl}\sqrt{\frac{R_1^2 R_2}{(R_2 - r_+)(R_2^2 + l^2 + R_2 r_+ + r_+^2)}} \left(4 R_2 (R_2^2 + l^2) - 3 l^2 r_+ - r_+^3 \right)\,. \label{actionbh}
\end{align}
Note that the black hole solutions only arise on the negative imaginary lapse action. This is because our boundary condition \eqref{eq.horcond} has broken the invariance of the action under complex conjugation, cf. also footnote \ref{footomega}. 

We expect to recover the pure AdS solution as the limit where $\omega \to \infty.$ In this limit the action reduces to
\begin{align}
S_{ND,\omega \to \infty} = \frac{\Delta \tau}{2G l^2}\left( (R_2^2 + l^2) N - \frac{R_2 c_1 l^2}{N} \right)\,.
\end{align}
The saddle points are found to correspond to EAdS space, as expected, with 
\begin{align}
N_s =  \pm i R_2\,, \quad b(r) = R_2 r\,, \quad \frac{c(r)}{b(r)} = 1 + \frac{R_2^2 r^2}{l^2}\,.
\end{align}
The scaling ambiguity \eqref{residual} has been fixed in such a way that the AdS solution corresponds to the limit $r_+ \to 0$ of the black hole solution above. The periodicity in the $\tau$ direction must then be chosen such that the circle size on the outer boundary remains $R_1,$ namely
\begin{align}
\Delta \tau^{AdS} = \frac{R_1 l}{\sqrt{R_2^2 + l^2}}\,.
\end{align}
The action for these saddle points is
\begin{align}
S^{AdS} = \pm i \frac{R_1 R_2}{G l}\sqrt{R_2^2 + l^2}\,. \label{actionads}
\end{align}

\begin{figure}
\centering
\includegraphics[width=0.4\textwidth]{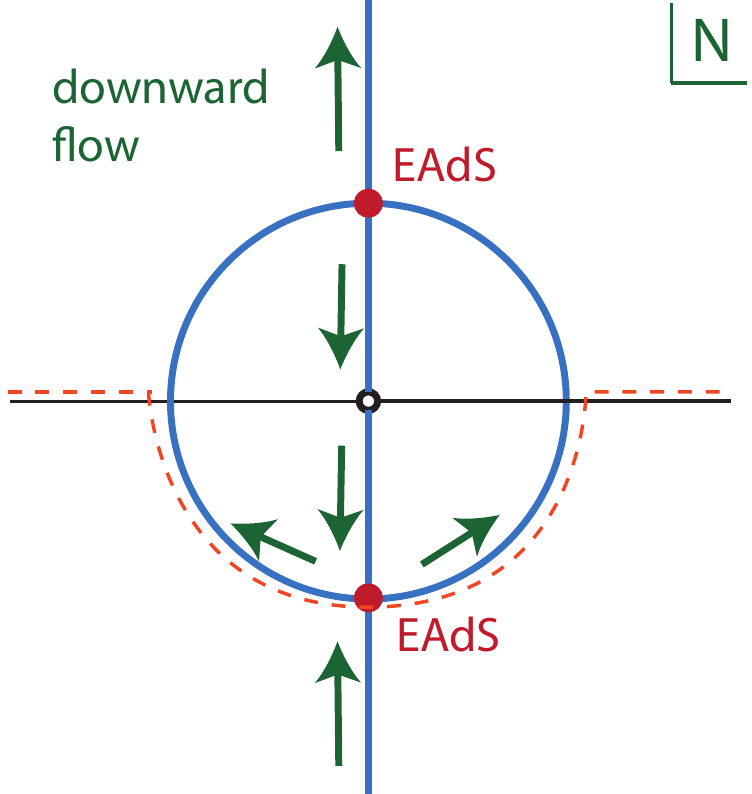}
\caption{Flows in the AdS case, which corresponds to the limit $\omega \to \infty.$ Arrows indicate the direction of steepest descent from the two saddle points. There is a singularity at $N=0.$ The dashed line indicates the required contour of integration which picks up a contribution from the enhanced EAdS saddle point in the lower half plane.} \label{fig:ads2}
\end{figure} 

The question now is which saddle points contribute? And what do the additional saddle points represent? We will look at the saddle point structure, and the associated paths of steepest descent, numerically. 

We start with the limiting case where $\omega \to \infty.$ As we have just derived, there are two saddle points in this case, which are complex conjugates of each other. They both describe EAdS space, but with different weightings. The corresponding flow lines are shown in Fig.~\ref{fig:ads2}, where the steepest descent paths emanating from the saddle points are drawn. The saddle point with the enhanced weighting is the one in the lower half plane. As in the case of the three-spheres partition function with Dirichlet boundary conditions, there is a singularity at $N=0.$ We may interpret the singularity in the action as a signal that the corresponding geometry does not exist. A suitable contour of integration is indicated by the dashed line in the figure. We must choose it such that at large $|N|$ it resides asymptotically in the upper half plane (for convergence) and such that it passes below the singularity so as to pick up a contribution from the enhanced EAdS saddle point. It would have been possible to define a Euclidean contour along the positive imaginary lapse axis, but such a contour would only have picked up the suppressed EAdS saddle point, which disappears in the limit of a large boundary size. Thus we are again forced to integrate over complex metrics.  

\begin{figure}
\centering
\includegraphics[width=0.5\textwidth]{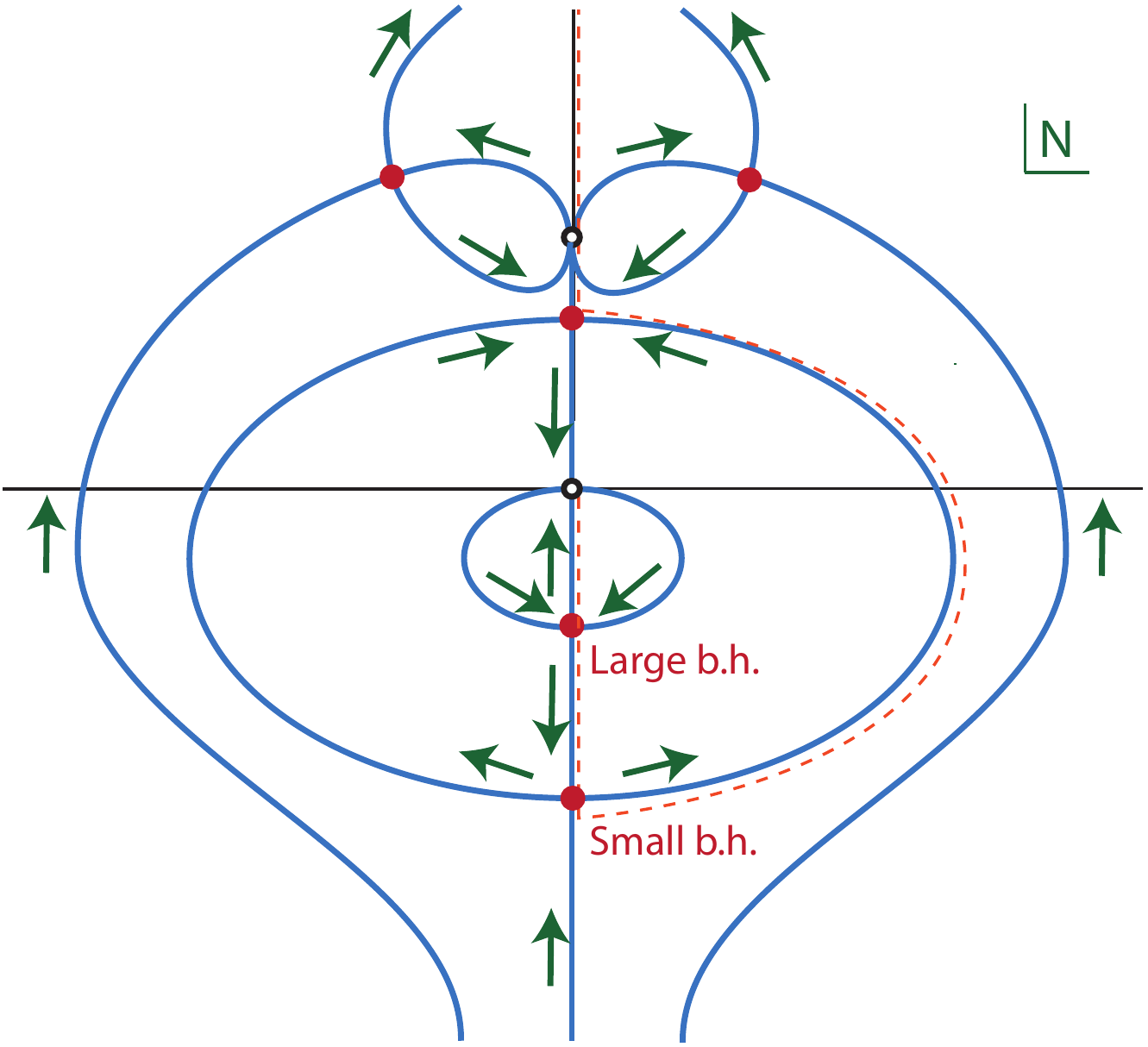}
\caption{Saddle points (in red) and steepest descent lines (in blue) for $R_1=5, R_2 = 10, l=1.$ Arrows indicate directions of steepest descent. There are two singularities, one at $N=0$ and one on the positive imaginary $N$ axis. In dashed orange is the required contour of integration.} \label{fig:bh1}
\end{figure} 

\begin{figure}
\centering
\includegraphics[width=0.3\textwidth]{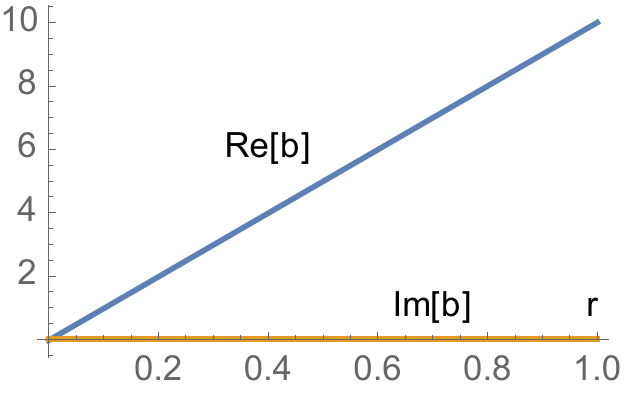}
\includegraphics[width=0.3\textwidth]{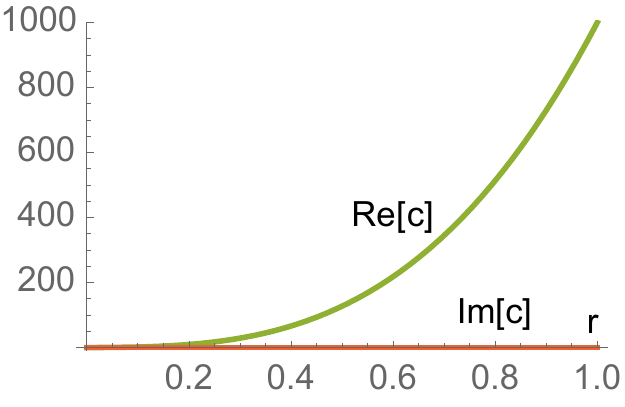}
\caption{The geometry of the Euclidean saddle point on the positive imaginary axis. Both $b(r)$ and $c(r)$ are real valued. At the origin $b$ has a small negative value and then passes through zero to reach the final value $b(1)=10.$ Here we used the same parameter values as in Fig. \ref{fig:bh1}.} \label{fig:bh1sad1}
\end{figure}

\begin{figure}
\centering
\includegraphics[width=0.3\textwidth]{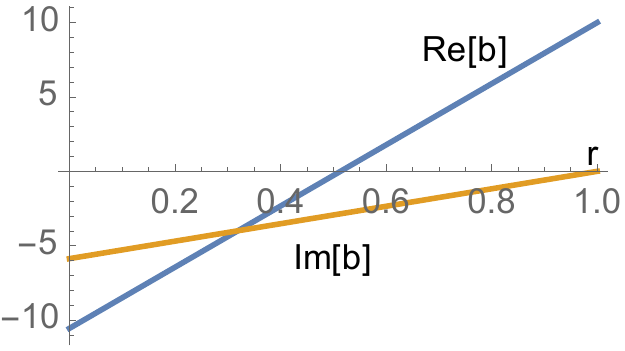}
\includegraphics[width=0.3\textwidth]{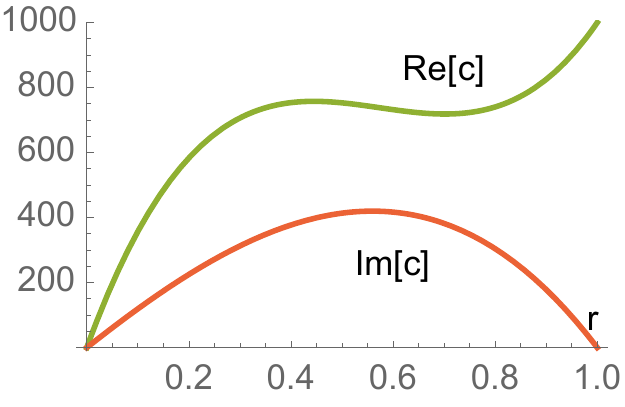}
\caption{The geometry of the complex saddle point in the first quadrant of Fig. \ref{fig:bh1}. Both $b(r)$ and $c(r)$ are complex valued. Here we used the same parameter values as in Fig. \ref{fig:bh1}.} \label{fig:bh1sad2}
\end{figure}

In the following figures, we will analyse the contributions from $\omega=1,$ at a fixed two-sphere radius $R_{2}$ and for increasing circle radii $R_1.$ The case of having a very small $R_1$ is shown in Fig. \ref{fig:bh1}. Three additional saddle points appear here: one near the origin, and two complex conjugate saddle points that move in from infinity (as $R_1$ is increased from zero) in the upper half plane. We will briefly describe the saddle points, starting with the two that already existed when $\omega \to \infty,$ i.e. what were the enhanced EAdS and the suppressed EAdS solutions. The enhanced EAdS solution now turns into the small Euclidean black hole, with horizon size $r_+$ growing from zero as $R_1$ is increased. Meanwhile, what was the suppressed EAdS solution turns into a geometry that starts out at a (small) negative value of $b,$ barely visible in Fig. \ref{fig:bh1sad1}. This means that the metric signature is $(-,-,+,+)$ near the origin, and then turns Euclidean after $b$ has crossed zero. Since $b$ crosses zero we may expect perturbations to blow up at that location. The most important saddle point is the one that appeared near the origin on the negative imaginary axis. This is the large black hole, with $r_+$ corresponding to the larger solution to \eqref{horizon}. This is the dominant saddle point, with the highest weighting. In fact the steepest descent line from this saddle point moves down towards the singularity, and on the other side of the saddle point down towards the small black hole and from there on to the Euclidean saddle point in the upper half plane. The two remaining saddle points in the upper half plane have a suppressed weighting, and their geometry is shown in Fig. \ref{fig:bh1sad2}. Their metric is complex throughout, with the exception of the final boundary at $r=1.$ A suitable contour of integration, capturing the large black hole, is indicated by the dashed line in Fig.~\ref{fig:bh1}. It has different characteristics than the one in the infinite $\omega$ case, namely it emanates from the origin in the negative imaginary direction, then winds around the singularity and ends up shooting off to infinity in the upper half plane. The contour has to start at the singularity in the direction of the lower half plane in order to capture the large black hole. This is possible because at finite $\omega$ the lapse integrand \eqref{lapseaction} near $N=0$ behaves as $e^{+\frac{1}{N^2}},$ implying that there exists a region of convergence in the wedge surrounding the imaginary axis at $\pm 45 ^o.$ A purely Euclidean contour is however not possible since the asymptotic region at large negative imaginary values of the lapse is a region of divergence, and hence the contour must wind around towards the upper half plane. Despite the fact that the contour differs from the infinite $\omega$ case, a similarity is that we are once again forced to integrate over complex metrics in order to obtain sensible results. 

\begin{figure}
\centering
\includegraphics[width=0.5\textwidth]{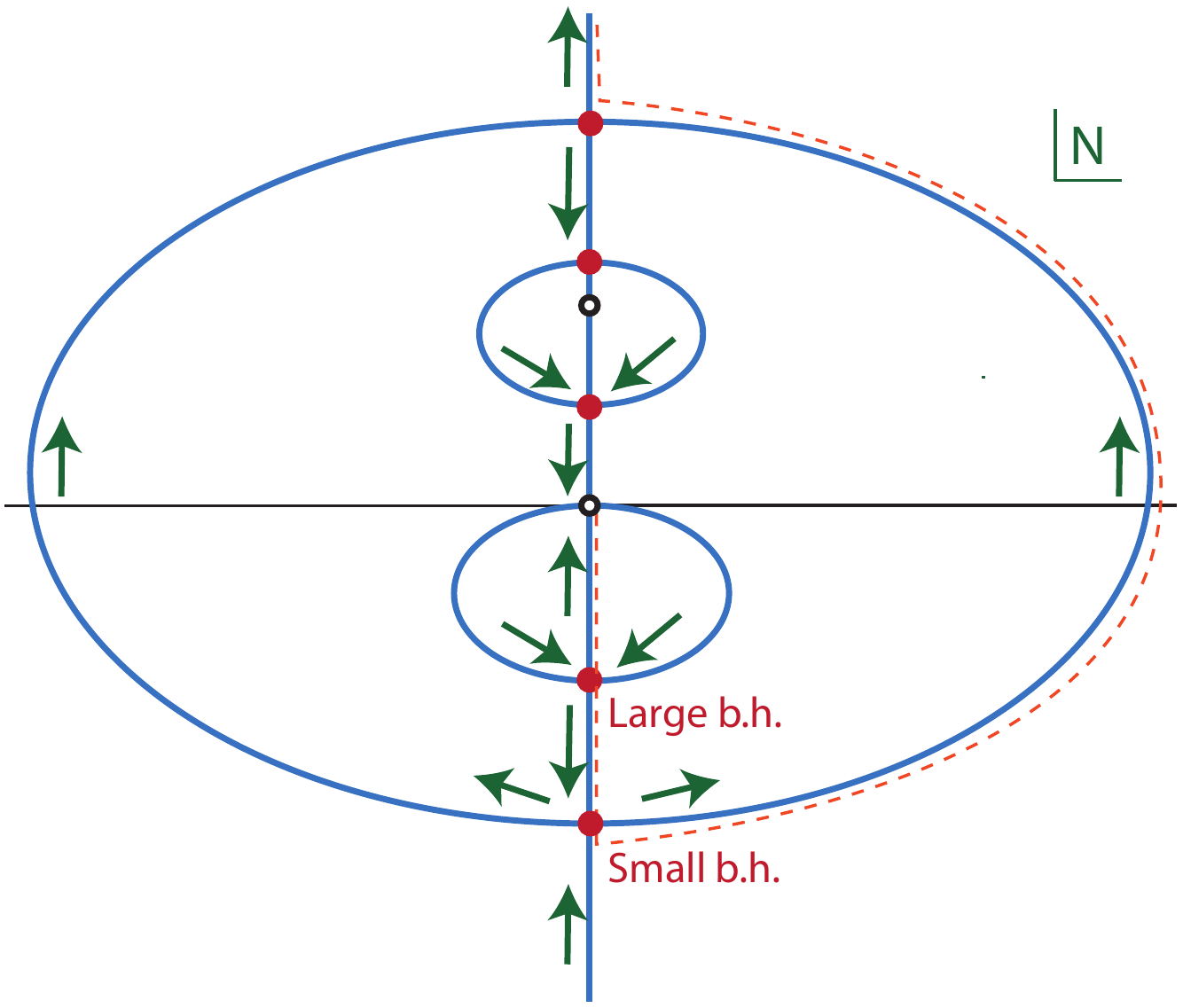}
\caption{Saddle points (in red) and steepest descent lines (in blue) for $R_1=12, R_2 = 10, l=1.$ Arrows indicate directions of steepest descent. There are two singularities, one at $N=0$ and one on the positive imaginary $N$ axis. In dashed orange is the required contour of integration.} \label{fig:bh2}
\end{figure}

\begin{figure}
\centering
\includegraphics[width=0.5\textwidth]{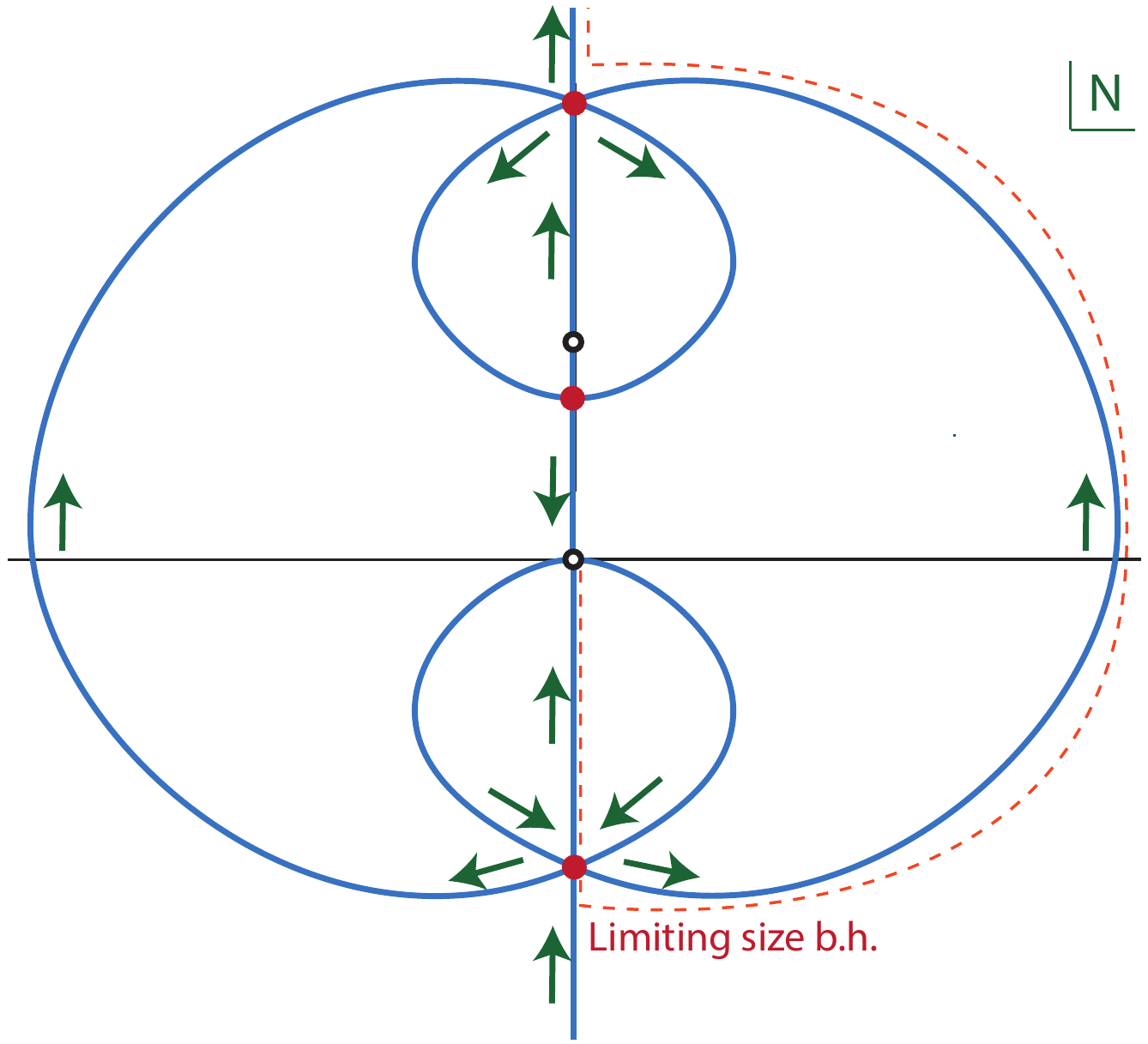}
\caption{Saddle points (in red) and steepest descent lines (in blue) for the limiting case $R_1= R_{1,limit}, R_2 = 10, l=1.$ Arrows indicate directions of steepest descent. In dashed orange is the required contour of integration.} \label{fig:bh3}
\end{figure}

\begin{figure}
\centering
\includegraphics[width=0.3\textwidth]{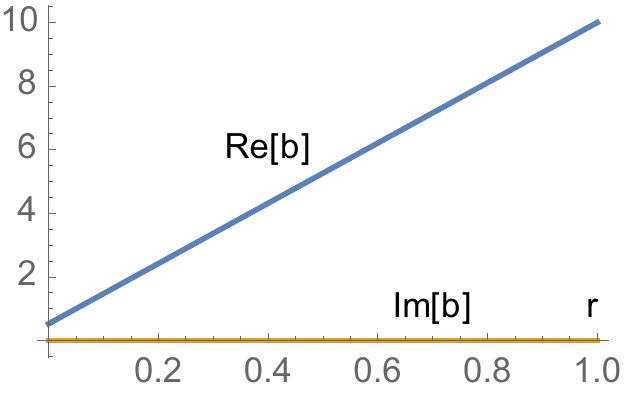}
\includegraphics[width=0.3\textwidth]{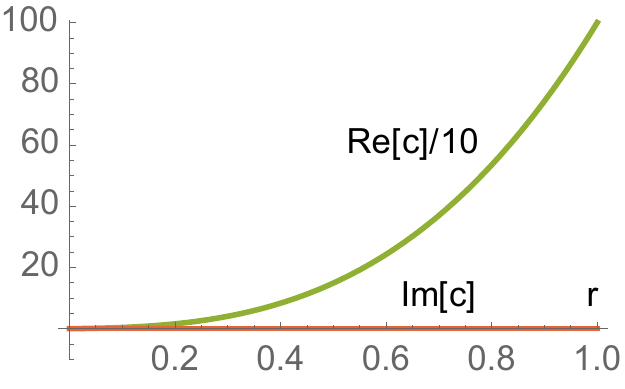}
\caption{The geometry of the limiting black hole, i.e. for the case where the small and large black holes have merged. Both $b(r)$ and $c(r)$ are real valued, and the saddle point geometry is Euclidean. Here we used the same parameter values as in Fig. \ref{fig:bh3}.} \label{fig:bh3sad}
\end{figure}

\begin{figure}
\centering
\includegraphics[width=0.5\textwidth]{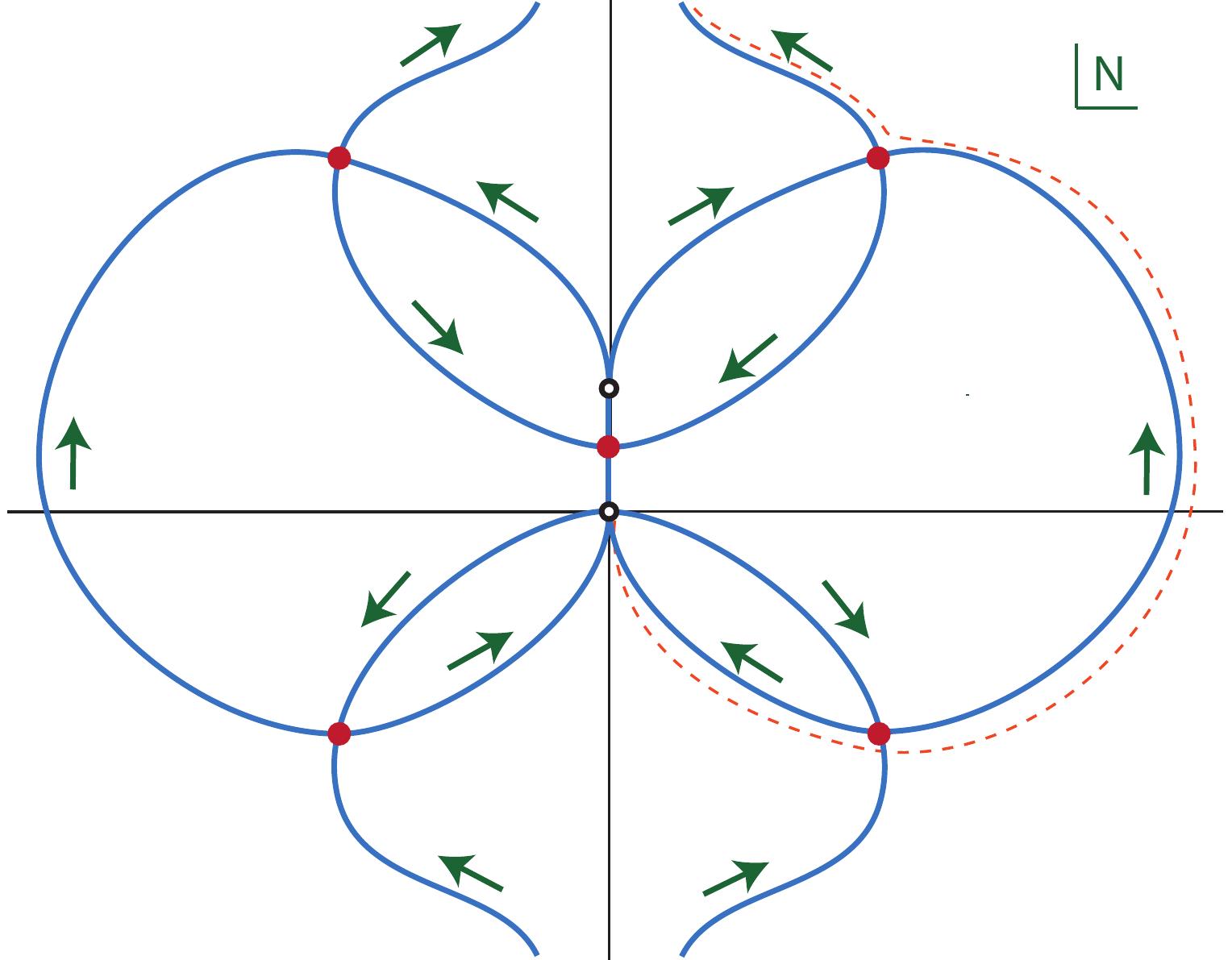}
\caption{Saddle points (in red) and steepest descent lines (in blue) for $R_1=50, R_2 = 10, l=1.$ Arrows indicate directions of steepest descent. There are two singularities, one at $N=0$ and one on the positive imaginary $N$ axis. In dashed orange is the required contour of integration.} \label{fig:bh4}
\end{figure}

\begin{figure}
\centering
\includegraphics[width=0.3\textwidth]{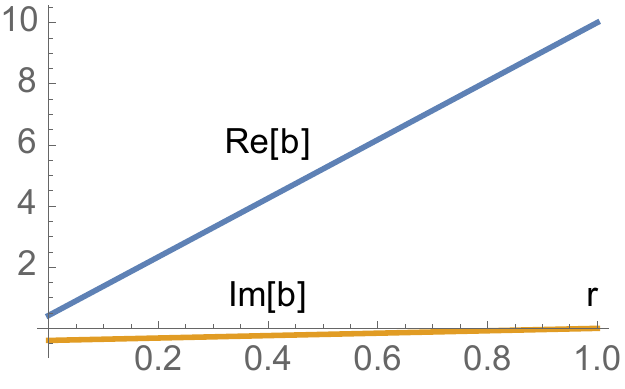}
\includegraphics[width=0.3\textwidth]{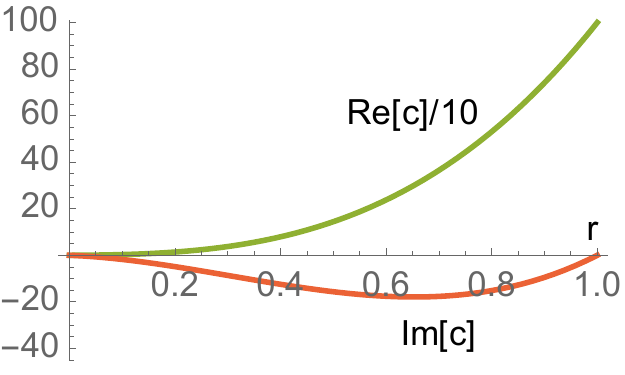}
\caption{At low temperature, when $R_1$ is large, the saddle points that used to correspond to black holes have moved into the complex plane. The associated geometry is no longer Euclidean, as imaginary parts of $b(r)$ and $c(r)$ develop. Here we used the values $R_1 = 50, R_2 = 10, l=1,$ just as in Fig. \ref{fig:bh4}.} \label{fig:bh4sad}
\end{figure}

As $R_1$ is increased, there are few relevant changes at first. The complex saddle points move towards the Euclidean axis, merge there and then separate again into two further Euclidean solutions similar to the one shown in Fig.~\ref{fig:bh1sad1}. The saddle points and their flow lines are shown in Fig.~\ref{fig:bh2}. The most important change occurs once $R_1$ reaches the limiting value $R_{1,limit}$ -- this case is shown in Fig.~\ref{fig:bh3}. At this radius, the two black hole solutions merge into a degenerate saddle point, representing the black hole at the minimum temperature (maximum radius) that is required for black holes to exist. This limiting black hole geometry is shown in Fig.~\ref{fig:bh3sad}. One may obtain an expression for the limiting radius by combining Eqs. \eqref{horizon}, \eqref{Dr1} and \eqref{cbh}, with $\Delta\tau=\beta$ and inserting the maximum value for the periodicity $\beta_{max} = \frac{2 \pi l}{\sqrt3}$, which leads to 
\begin{align}
\label{eq.R1limit}
R_{1,limit} = \frac{2\pi}{\sqrt{3}}R_2 \sqrt{1+ \frac{l^2}{R_2^2}- \frac{4l^3}{3\sqrt{3}R_2^3}}\,.
\end{align}
Once the circle radius $R_1$ is increased even further, the degenerate black hole saddle point, as well as two of the saddle points in the upper half plane, all move into the complex plane -- see Fig. \ref{fig:bh4} for a depiction of the saddle point locations and the associated steepest descent flows. At that stage there does not exist any Euclidean black hole solution anymore. The complex saddle points possess a fully complex geometry, which is shown in Fig. \ref{fig:bh4sad}. In our framework, this is the manifestation of the well-known fact that there exists a minimum temperature required for the existence of regular Euclidean black holes. The complex saddle points have a suppressed weighting, smaller in fact than the empty EAdS solution, as we will show below. 

In all cases one is forced to choose a contour of integration that starts at the singularity at $N=0,$ then follows the thimble associated with the large black hole saddle point, curves around and eventually flies off to infinity in the upper half plane, as required for convergence. Note that once again the integrand possesses the symmetry that for $N \rightarrow - N^*$ it changes to its complex conjugate. Thus in order to obtain a real partition function picked around the black hole saddle points one should consider the contour described above together with its reflection with respect to the imaginary lapse axis. As we commented on already for the case of the $S^3$ boundary, this symmetric contour is as close as it can get to the integral along the imaginary axis i.e. the Euclidean path integral, which in itself is divergent and thus ill-defined. The interesting point is that this contour neither corresponds to a sum over Euclidean metrics, nor a sum over Lorentzian metrics -- in order for the partition function to be mathematically meaningful as a minisuperspace statement, the sum must be defined over intrinsically complex metrics.

\begin{figure}
\centering
\includegraphics[width=0.5\textwidth]{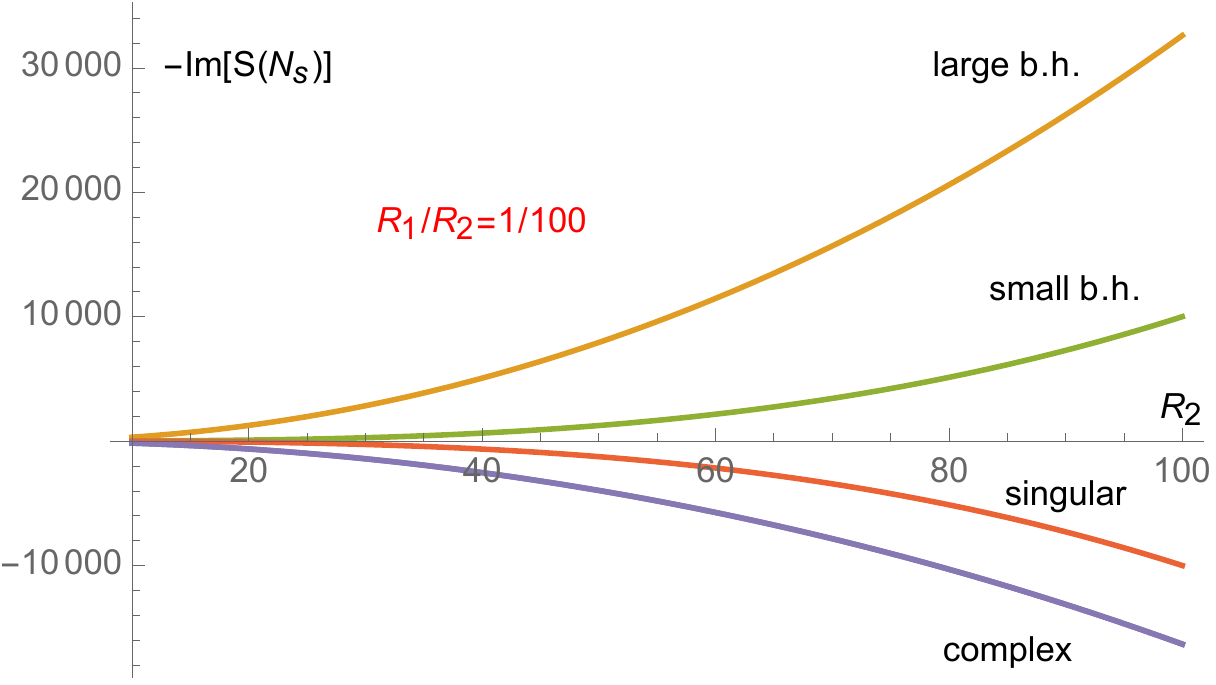}
\caption{Weighting of the saddle points $- Im[S(N_s)]$ with constant ratio $R_1/R_2=1/100,$ as a function of the boundary size and with $8 \, \pi \, G = 1$ and $l=1$. One can see that in the limit of infinite boundary size only the black hole saddle points will remain.} \label{fig:ratio}
\end{figure} 

\subsubsection{Thermodynamics from saddles}\label{thermodynamics}

Having discussed the saddle points and integration contours, we may now sketch how the usual  interpretation in terms of thermodynamics is recovered. In all cases we saw that the contribution to the partition function from $\omega = 1$ is dominated by the large black hole solution, provided $R_1$ is smaller than the limiting value~\eqref{eq.R1limit}. Thus, when approximating the partition function, the $\omega = 1$ contribution may be well approximated by the action of the large black hole solution. The difference in action between the black hole and AdS solution is given by the difference between Eq.~\eqref{actionbh} and Eq.~\eqref{actionads}, and as an expansion at large two-sphere radius $R_2$ is given by 
\begin{subequations}
\label{actiondiff}
\begin{align}
\Delta S & = -\frac{iR_1}{4Gl}\left[\frac{\sqrt{R_2}\left(4 R_2^3 +4 l^2 R_2  - 3 l^2 r_+ - r_+^3 \right)}{\sqrt{R_2^3 + l^2 R_2 - l^2 r_+- r_+^3}}   -4 R_2\sqrt{R_2^2 + l^2}\right]\\
&= \frac{i R_1}{4Gl R_2}(l^2 r_+ - r_+^3) -  \frac{i l R_1}{8G R_2^3}(l^2 r_+ - r_+^3)  + O(R_2^{-4})\,.
\end{align}
\end{subequations}
At leading order in a large $R_2$ expansion we may identify $R_1/R_2 \approx \beta/l,$ and with this substitution the leading order difference in actions at large $R_2$ recovers the classic Hawking-Page result~\cite{Hawking:1982dh}
\begin{align}
\Delta S_{HP} = -\frac{i \pi}{G}r_+^2\frac{r_+^2 - l^2}{3r_+^2 + l^2}  + O(R_2^{-1})\,.
\end{align}
The weighting of the AdS solution dominates when $-Im[\Delta S]<0.$ At large $R_2$ this is when $r_+< l,$ and there are corrections, implied by \eqref{actiondiff}, to this relation when $R_2$ is small. The phase transition thus occurs at the approximate radius $R_{1,HP} \approx \pi R_2.$ Thus the complex saddle points that replace the large black hole solutions at large $R_1>R_{1,limit} \approx  \frac{2\pi}{\sqrt{3}}R_2$ never play a dominant role, since the AdS solution has already become dominant by then.

The thermodynamic interpretation follows from an analysis of the partition function, approximated here by the difference in actions \eqref{actiondiff},
\begin{align}
\ln Z = i \frac{ \Delta S}{\hbar}\,.
\end{align}
It is important to keep in mind that we are considering the partition function as representing the canonical ensemble, i.e. we are considering a system that is kept at a fixed temperature~$T.$ At fixed boundary two-sphere with radius $R_2$ this temperature, which is redshifted as one moves away from the black hole horizon, is given by 
\begin{align}
R_1 =\beta \sqrt{1+ \frac{R_2^2}{l^2}-\frac{2M}{R_2}} = \Delta\tau^{AdS} \sqrt{1+ \frac{R_2^2}{l^2}} = \frac{1}{T}\,,
\end{align} 
where we denoted the Euclidean time periodicity of the $EAdS$ solution by $\Delta\tau^{AdS}.$ Thus, reintroducing the speed of light $c,$ we may usefully rewrite the partition function as 
\begin{align}
\ln Z = \frac{R_2}{T \emph{l}^2_P}\left(\sqrt{1+ \frac{R_2^2}{l^2}-\frac{2M}{R_2}} - \sqrt{1+ \frac{R_2^2}{l^2}} \right) + \frac{\pi r_+^2}{\emph{l}^2_P}\,.
\end{align}
where $\emph{l}_P = \sqrt{\frac{G \hbar}{c^3}}$ is the Planck length.\\
The expectation value of the energy is given by
\begin{subequations}
\label{energy}
\begin{align}
\langle E \rangle &=k_B T^2 \frac{\partial \ln Z}{\partial T} = \frac{k_B R_2}{\emph{l}^2_P} \left(\sqrt{1+ \frac{R_2^2}{l^2}} - \sqrt{1+ \frac{R_2^2}{l^2}-\frac{2M}{R_2}} \right) \\
&= \frac{ k_B }{ \emph{l}^2_P}\frac{ l M}{R_2} - \frac{k_B }{\emph{l}^2_P}\frac{ M l^3}{2 R_2^3}+  O(R_2^{-4}) 
\end{align}
\end{subequations}
and the entropy takes the form
\begin{align}
{\cal S} & =k_B \ln Z + \frac{\langle E \rangle}{T} =  \frac{k_B}{\emph{l}^2_P} \, \pi r_+^2 = \frac{k_B}{\emph{l}^2_P} \, \frac{Area}{4} \label{entropy}\,.
\end{align}
Note that this explicitly verifies the Quantum Statistical Relation \cite{Gibbons:1976ue}
\begin{align}
- k_B T \ln Z = \langle E \rangle - T{\cal S}\,.
\end{align}
Furthermore, the results derived above are in agreement with the Hamiltonian method employed by Brown et al. in Ref.~\cite{Brown:1994gs}. In deriving the energy \eqref{energy}, one may use the chain rule that $\partial \ln Z/\partial T = \partial \ln Z / \partial r_+ \left(\partial T/\partial r_+ \right)^{-1},$ with $M$ being thought of as a function of $r_+$ according to \eqref{mass}. The conserved mass differs from the energy by a factor of the lapse at $R_2$
\begin{align}
{\cal M} = \sqrt{1+ \frac{R_2^2}{l^2}-\frac{2M}{R_2}} \langle E \rangle = M + \frac{l^2}{2R_2^2}M - \frac{l^2}{R_2^3}M^2+ O(R_2^{-4})\,.
\end{align}
Note also that the entropy \eqref{entropy} is given precisely by a quarter of the horizon area, and that there are no corrections to this relation at finite $R_2.$ The specific heat at fixed boundary
\begin{align}
{\cal C} = \frac{\partial \langle E \rangle}{\partial T}
\end{align}
is negative for small black holes ($r_+ < \frac{l}{\sqrt{3}}$) and positive for large black holes ($r_+ > \frac{l}{\sqrt{3}}$), which implies that only large black holes are thermodynamically stable. This fits well with our flow diagrams which demonstrate that the large black hole always has a higher weighting than the small black hole, and is thus also more dominant in the canonical ensemble. 

One surprising aspect of our work is the appearance of additional saddle points. These have a weighting that is suppressed compared to the black hole saddle points, both large and small. Thus they do not play a large role. In fact, if the outer boundary is moved all the way to infinity, these additional saddle points disappear altogether -- see Fig. \ref{fig:ratio}. This implies that these extra saddles do not play any role in the original AdS/CFT correspondence and our study reproduces the results of Refs.~\cite{Hawking:1982dh,Witten:1998zw}. However, once the boundary is moved to a finite radius, they will provide a tiny additional contribution to the partition function. It would be very interesting to try to figure out if one can use the appropriate QFT description~\cite{McGough:2016lol,Hartman:2018tkw} to confirm their existence, or rule out the minisuperspace approach.

We would like to emphasise that our choice of mixed boundary conditions is crucial in obtaining an expression for the canonical ensemble. The Dirichlet condition on the outer boundary, which fixes the size of the Euclidean time circle, effectively fixes the temperature. However, this turns out not to be enough. Our premise that we wanted to sum over saddle point geometries that cap off smoothly in the interior led us to the Neumann boundary condition at the coordinate location $r=0.$ Had we used a Dirichlet condition on the inner boundary, i.e. at the black hole horizon, we would have obtained an additional boundary term of magnitude $\pi r_+^2$ contributing to the black hole action. Our path integral would then have been approximated by 
\begin{align}
    -k_B \ln Z \approx \frac{\langle E \rangle}{T} - {\cal S} + {\cal S} = \frac{\langle E \rangle}{T}\,,
\end{align}
which is reminiscent of the discussion in Ref.~\cite{deBoer:2015ija}. Thus, with Dirichlet conditions on both ends, the ``partition function'' would rather have looked like that of the microcanonical ensemble, where one sums over states of fixed internal energy. This is surprising, as in gravitational systems the energy is described by the asymptotic fall-off of the metric, and thus one would have expected the microcanonical ensemble to be given by a path integral with boundary conditions (at the outer boundary $r=1$) that are of Neumann form, where derivatives of the metric may be specified \cite{Brown:1992bq,Krishnan:2016mcj, Krishnan:2016tqj}. More precisely, one might have expected that the canonical and microcanonical ensembles would be related by a Legendre transform at the outer boundary, and not at the inner boundary. The extent to which this correspondence is accidental deserves further investigation.

%%%%%%%%%%%%%%%%%%%%%%%%%%%%%%%%%%%%%%%%%%
%%%%%%%%%%%%%%%%%%%%%%%%%%%%%%%%%%%%%%%%%%

\subsection{Implications for cosmology} \label{sec:cosmology}

The no-boundary proposal can be formulated as a path integral in a very similar fashion to the calculations above concerning asymptotically AdS spacetimes (this analogy should already be obvious by taking another look at Fig. \ref{fig:partition}, but rotating the figure by $90$ degrees counter-clockwise). In this context the path integral defines the wave function of the universe which sets the initial conditions for the universe.

The literature on no-boundary path integrals has a long history. 
Various works, both old and recent \cite{Louko:1988bk, Halliwell:1988ik, Halliwell:1989dy, DiazDorronsoro:2018wro, DiTucci:2019bui}, point to the necessity of specifying the initial momentum in the gravitational path integral in order to define a well-behaved Hartle-Hawking wave function. This implies that the path integral sums over geometries with all possible initial sizes while only the dominant Hartle-Hawking geometry has no boundary and starts from 'nothing'. As a consequence, this requirement implies a radical change in our interpretation of the Hartle-Hawking wave function as a theory for the initial conditions of the universe~\cite{DiTucci:2019bui}. \\
Another possibility is to consider an initial Robin condition with the quantum
uncertainty shared between initial size and momentum, as discussed in chapter \ref{robincosmology}. A covariant implementation of the initial Robin condition requires an extra boundary term \cite{Krishnan:2017bte} which was also discussed in chapter \ref{robincosmology}\footnote{Note that also in this case all of the off-shell geometries do have an initial boundary.}. \\
In this section we will review and elaborate on some of the well known results while highlighting the connections with the original calculations of the previous sections regarding the case of a negative cosmological constant. Our aim is to see what one may learn about the no-boundary proposal when viewed from the fresh perspective offered by calculations performed with AdS asymptotics.

The evaluation of the no-boundary path integral is in one-to-one correspondence with the calculations in section \ref{sec:ads} if one analytically continues the radius of curvature of AdS to an imaginary value $l = \frac{i}{H},$ where $H$ then denotes the Hubble rate of the corresponding de Sitter spacetime. Thus, in our coordinates, the correspondence is simply
\begin{align}
\Lambda = - \frac{3}{l^2} = + 3 H^2. \label{anacon}
\end{align}
It is now interesting to observe that in the case of a negative cosmological constant, evaluating the canonical ensemble in the black hole case required us to impose Neumann boundary conditions on the inner boundary. Not imposing a boundary term resonates well with the philosophy of the no-boundary proposal, as the name itself suggests. The exact same condition, given here by eq. \eqref{momentumcondition}, was also implemented in cosmology in eq. \eqref{neumannconditions}. While the saddle points were purely Euclidean with negative cosmological constant, as studied in sections~\ref{sec:ads} and~\ref{sec:bhs}, upon performing the analytic continuation in \eqref{anacon}, the saddle points \eqref{S3saddles} come to be in general complex.  The saddle points and their steepest descent flow lines are shown in Fig. \ref{fig:NeumannFlow}, which should be contrasted with Fig. \ref{fig:flowAdS}. The asymptotic regions of convergence for the lapse integral are unchanged, since they are determined by the leading term in $N,$ namely $e^{iN^3/l^4}= e^{iH^4 N^3},$ cf. eq. \eqref{NDaction}. Choosing a contour running from negative imaginary infinity to the first quadrant (which we called $-{\mathcal C}_1$) would pick up only a single saddle point $N_+.$ This would yield a perfectly acceptable wavefunction/partition function, $Z \approx e^{iS(N_+)/\hbar}.$ The contour running from negative imaginary infinity to the second quadrant (which we called ${\mathcal C}_2$) would yield (minus) the complex conjugate result. By combining these two contours we have the possibility of obtaining a real wavefunction, as originally advocated by Hartle and Hawking. Just as for the AdS case, there exist two options to do so. The first is to use the sum $-({\mathcal C}_1 + {\mathcal C}_2)={\mathcal C}_0,$ which is equivalent to the Lorentzian contour. Similar arguments to those presented in section \ref{subsec:N} imply that this yields the wavefunction
\begin{align}
    Z(R_3)\mid_{{\mathcal C}_0} = e^{\frac{V_3}{4\pi G \hbar H^2}} Ai\left[\left(\frac{3V_3}{8\pi G \hbar H^2} \right)^{2/3}\left(1-H^2 R_3^2 \right) \right]\,.
\end{align}
Meanwhile, summing ${\mathcal C}_{1,2}$ such that they both run towards the upper half plane yields the result
\begin{align}
    Z(R_3)\mid_{i({\mathcal C}_2-{\mathcal C}_1)} = e^{\frac{V_3}{4\pi G \hbar H^2}} Bi\left[\left(\frac{3V_3}{8\pi G \hbar H^2} \right)^{2/3}\left(1-H^2 R_3^2 \right) \right]\,.
\end{align}
Both of these results yield an acceptable no-boundary wavefunction, the only difference being a shift in the phase, given the asymptotic expansions for real $x$
\begin{align}
    Ai(-x) \sim \cos(\frac{2}{3}x^{3/2}-\frac{\pi}{4})\,, \qquad Bi(-x) \sim \cos(\frac{2}{3}x^{3/2}+\frac{\pi}{4})\,.
\end{align}
By contrast with the negative cosmological constant case, here both options are equally viable. They both contain a trigonometric factor that one can write as the sum of two phases $\sim \left(e^{iR_3^3} + e^{-iR_3^3}\right),$ where the phases arise due to the classical expansion at late times. One can then interpret these two phases as two time-reversed universes which would decohere quickly due to the cosmological expansion \cite{Hartle:2008ng}. It is noteworthy that with positive $\Lambda$ a genuine Lorentzian path integral is viable, while with negative $\Lambda$ it was not. This supports the concept that the cosmology of the 'real world' (where $\Lambda>0$) can indeed be described by Lorentzian path integrals. It will be interesting to see if this remains the case in more elaborate models, and in the presence of more general metrics\footnote{For $S^1 \times S^2$ boundary conditions and positive $\Lambda$, the saddle point geometries were already studied in \cite{Conti:2014uda}. It will be interesting to study the associated flow lines and integration contours.}.

The most important conclusion that we can draw from these observations is that the partition function in the presence of negative $\Lambda$ gives strong support to the implementation of the no-boundary proposal presented in section section \ref{sec:noboundaryterm}, which used an equivalent momentum condition to \eqref{momentumcondition}. In Ref.~\cite{DiTucci:2019bui} this was called the ``no boundary term'' proposal, as the Neumann condition is obtained by not adding any surface term to the Einstein-Hilbert action. Moreover, the choice of sign in specifying the initial Euclidean expansion rate is determined in the negative $\Lambda$ case by the requirement of obtaining a sensible thermodynamic interpretation once black holes are also included. On the cosmological side this sign choice translates into picking the HH no-boundary proposal rather than Vilenkin's tunneling proposal~\cite{Vilenkin:1982de}. Expressed as a one-line conclusion, one may say that black hole thermodynamics justifies the no-boundary proposal.

%%%%%%%%%%%%%%%%%%%%%%%%%%%%%%%%%%%%%%%%%%
%%%%%%%%%%%%%%%%%%%%%%%%%%%%%%%%%%%%%%%%%%

%%%%%%%%%%%%%%%%%%%%%%%%%%%%%%%%%%%%%%%%

\subsection{Discussion} \label{sec:discussionAds}

The no-boundary wavefunction remains a leading theory of the initial conditions of the universe, as it has the potential to describe the emergence of spacetime, the early approach to classicality, the Bunch-Davies vacuum of perturbations and to provide initial conditions for geometry and scalar field evolution \cite{Hartle:2008ng}. However, in some sense it still remains an answer waiting for the appropriate question. Our goal, which many people have pursued and to which we are adhering here, is to find a suitable path integral definition of the no-boundary idea. After all, it is in this context that the idea was originally formulated \cite{Hawking:1981gb}.

In this chapter, we have explored the case where one does not add any boundary term to the Einstein-Hilbert action. Despite the absence of a boundary term, boundary conditions are nevertheless implied for the fields. In particular we have reviewed how the absence of a boundary term implies a Neumann condition for the scale factor of the universe, allowing us to impose a suitable Euclidean value for the derivative of the scale factor at the initial ``time''. This Euclidean derivative encodes the idea that the geometry must contain a Euclidean (or approximately Euclidean) section, thus allowing the spacetime to be closed off smoothly.

We also provided a minisuperspace construction of gravitational partition functions in spacetimes with a negative cosmological constant, and with either $S^3$ or $S^1 \times S^2$ boundaries. Such partition functions are motivated both by classic results in black hole thermodynamics, and by the AdS/CFT correspondence. They also bear a close technical resemblance to studies in quantum cosmology, in particular in relation to the no-boundary proposal, which was another motivation for our study.

Our main findings are: 1. In the minisuperspace approach and with a negative cosmological constant, partition functions cannot be seen as sums over Euclidean metrics, but rather must  be defined as sums over certain complex classes of metrics. Despite this feature, the dominant saddle points representing AdS spacetime and AdS black holes always turn out to be Euclidean. In this way the semi-classical thermodynamic results are recovered, even though off-shell we are forced to sum over complex metrics. 2. Guided by black hole thermodynamics, we had to impose a Neumann boundary condition at the horizon of black holes in order to represent the canonical ensemble. A Dirichlet boundary condition fixing the size of the horizon would have led to a different interpretation of the partition function, more in line with calculations of the microcanonical ensemble (although this identification would require further justification). The Neumann condition, which is a condition on the expansion rate of the metric at the horizon, allows one to directly impose a regularity condition at the horizon. Even though our final result is expressed in terms of minisuperspace variables, we emphasise that our imposition of a Neumann condition is performed at the level of the full Einstein-Hilbert action, and is thus fully covariant. 3. When the outer boundary is sent to infinity, only the AdS and black hole saddle points remain of relevance. However, when the boundary resides at a finite radius, which is a scenario recently understood in terms of a dual QFT description~\cite{McGough:2016lol,Hartman:2018tkw}, three additional saddle points appear in the minisuperspace approach. Depending on parameters these subleading saddle points may be complex. As a result, providing an interpretation of these saddles\footnote{In holography, providing an interpretation of subleading saddle points in terms of dual QFT statements is challenging, yet not impossible, see Ref.~\cite{Balasubramanian:2016xho} for an example.} in the language of corresponding QFTs or ruling them out would add an element of falsifiability to the minisuperspace approach.

The fact that we had to use a Neumann boundary condition at the black hole horizons is noteworthy. In comparing with the case of a positive cosmological constant, this Neumann condition happens to be identical to the one used in the %recent 
'no-boundary term' implementation of the no-boundary proposal. There also, Dirichlet conditions proved unphysical, and a condition on the initial expansion rate of the universe is the key to obtaining a well-defined definition of the Hartle-Hawking wavefunction. As we wrote earlier, what we are finding here is that black hole thermodynamics justifies this choice, i.e. black holes thermodynamics supports the no-boundary proposal\footnote{Note that our setting is different from the ``holographic no-boundary measure'' of Hertog and Hartle \cite{Hertog:2011ky}. There they propose to use AdS/CFT on EAdS sections inside of the analytically continued saddle point geometries (which are the saddle points corresponding to a positive cosmological constant). By contrast, we start from AdS path integrals with negative cosmological constant and then simply let the cosmological constant evolve to positive values. Thus our expressions for the wave function of the universe involve a positive cosmological constant, while the ones of Hertog and Hartle involve the opposite value of the cosmological constant.}. One consequence of this is that one should no longer think of the no-boundary proposal as a sum over compact, regular metrics. Rather one should think of it as a sum over metrics with an initial Euclidean expansion rate. That the expansion rate must be Euclidean is then simply a manifestation that we are describing the quantum origin of the universe, which cannot be represented by a classical (real) solution.

Our work suggests many avenues for further study. One of them would be to understand if additional saddle points corresponding to complex geometries have any interpretation in the language of dual QFTs~\cite{McGough:2016lol}. If yes, this would provide an AdS/CFT indication about gravitational path integral including complex geometries, as we were forced to do in the minisuperspace approach. An important generalisation of our study would be to incorporate conical defects in our studies of partition functions in Section~\ref{sec:bhs}. Another very interesting direction to consider would be to generalise our minisuperspace studies away from four spacetime dimension, in which case the Neumann condition at $r = 0$ would have to be imposed differently~\cite{Krishnan:2016mcj}. More on this front, general relativity in three and especially in two spacetime dimensions does not contain dynamical gravitons and path integrals over asymptotically AdS geometries might then be well-defined. It would be very interesting to see what support such studies would give for including complexified metrics in the gravitational path integral beyond the minisuperspace approach.

\clearpage

%%%%%%%%%%%%%%%%%%%%%%%%%%%%%%%%%%%%%%%%%%%%%%%%%%%%%%%%%%%%%%%%%%%%%%%%%%%%%%%%%%%%

\section{Summary and outlook}\label{chapconclusion}

In this thesis we discussed various implications of including semi-classical effects in the description of the early universe. We focused on the path integral implementation of semi-classical gravity where the amplitude for a transition from a three-geometry to another, or similarly the wavefunction of the universe, is given by a sum over geometries weighted with the Einstein-Hilbert action. We saw in chapter \ref{quantuminitial} how this framework allows us to check the robustness of inflation against quantum effects. In the usual treatment of inflation, the expansion history of the universe is described by the classical solution to Einstein's equations and only primordial perturbations are given quantum properties, according to QFT in curved spacetime. We saw that this description breaks down if we assume no pre-inflationary phase and thus the beginning of inflation is pushed all the way back to a zero size universe. This work helps us answering the question of how much of a classical phase inflation actually is. A single classical inflating trajectory, over which quantum fields propagate according to QFT in curved spacetime, is recovered if inflation starts when the universe is already large and classical enough. We saw in section \ref{InitialConditions} how under these circumstances
primordial perturbations can then be put in the Bunch-Davies vacuum, a condition which comes about here as an external assumption with some associated level of fine-tuning. We concluded then that inflation, rather than generating primordial fluctuations with the correct features, is a mechanism to process such perturbations in the right way. With this motivation in mind we moved to an analysis of the no boundary proposal, a theory for the wavefunction of the universe, which aims at providing the initial condition for inflation describing the quantum birth of a large, classical, inflationary universe with gaussian distributed linear perturbations. We discussed how the no boundary wavefunction fails to be formulated as a path integral with Dirichlet initial conditions in chapter \ref{nbDirichlet}: if all geometries in the sum start out at zero size, primordial perturbations are necessarily unsuppressed. This means that the no boundary proposal is not given by a sum over compact geometries, questioning the idea that the Hartle-Hawking wavefunction represents the quantum state of a universe with no boundary in space and time. Indeed this conclusion applies within our framework that is, a minisuperspace approximation of the wavefunction of the universe as given by semi-classical general relativity. One cannot exclude that this prescription might give positive results within the yet-to-be-discovered theory of everything where quantum gravity is at work in its full power. It is however worth noting that no boundary solutions seem to be stable against higher order derivative quantum effects \cite{Jonas:2020pos}.
One of the points of strength of the no boundary proposal was indeed its simplicity and minimality: a full theory of quantum gravity might not be needed to answer many open questions in cosmology if the no boundary proposal gave a complete and consistent picture of the early history of the universe with just semi-classical elements where extremely high curvature regimes are not reached at early times. The Hartle-Hawking saddle point and the saddle point approximation of the wavefunction $\Psi \approx e^{i \, S_{HH}}$ are indeed consistent in this sense. 
We thus took the point of view in this work of keeping this simple framework and looking for path integrals that do give the Hartle-Hawking wavefunction in the saddle point approximation. Our goal was to understand what is the meaning of the no boundary proposal if such path integrals exist and at the same time extend our understanding of gravitational path integrals which we believe is a matter of relevance on its own. We thus studied minisuperspace path integrals with Robin type of boundary conditions and showed how the Hartle-Hawking wavefunction can indeed be formulated in this fashion in chapter \ref{robincosmology}. In this case, rather than requiring the initial size of the universe to vanish, one fixes a linear combination of its initial size and momentum. The specification of the desired initial Euclidean momentum is essential to get rid of the unstable solutions with the related unsuppressed perturbations. First, we studied path integrals with what we called ``canonical'' Robin conditions: this Robin initial condition is specific to the minisuperspace approximation and has the nice feature that it can be interpreted as an initial coherent state with both initial size and momentum specified with a given uncertainty. In this case, the initial size of the universe, instead of being exactly zero, is associated with a gaussian peaked around zero, and one could say that, in a sense, there were already fluctuations of space and time from which our universe originated. This type of Robin initial condition could also be useful for a deeper semi-classical understanding of the regime of eternal inflation. We saw in section \ref{papersebastian} that if one wants to describe the inflaton jumping up the potential, in a eternal inflation-like homogeneous transition, one needs such type of Robin condition because the path integral 'needs to know' that the universe is already inflating (through the conjugate momentum) and where the scalar field is located on the potential (that is, the field value) when the transition happens. 
We discussed also, in section \ref{sec:hubblerate}, the possibility of a final covariant Robin condition (with an initial Neumann condition) for the Hartle-Hawking wavefunction, which corresponds to fixing the final Hubble rate of the universe. We consider this an appealing feature, being the expansion rate a measurable quantity. Both the canonical and the covariant implementations result in a well defined Hartle-Hawking wavefunction upon carefully specifying the initial Euclidean momentum. In the covariant case, however, the convergent contour does not correspond to the real lapse line but has a finite off-set with respect to it, for no boundary boundary conditions, a problem which does not arise in the canonical case. In the canonical implementation, sending the variance of the gaussian to infinity, the coherent state becomes a plane wave and the Robin condition a Neumann condition. The implementation of the no boundary path integral with Neumann boundary conditions can be given a covariant form (and correspond to no boundary term in four dimensions) and is convergent over the real $N$ line. For these reasons it represents, in our opinion, the best path integral representation of the Hartle-Hawking wavefunction. This is showed in section \ref{sec:noboundaryterm} where we illustrate how the Harte-Hawking wavefunction can be constructed using such Neumann conditions for both the closed FLRW model and the Bianchi IX model. The various well-defined implementations of the no boundary proposal presented in this work have important elements in common: they all require a departure from the idea of a sum over compact geometries and focus instead on the sign of the initial Euclidean momentum. The initial size of the geometries summed over is not fully specified and in fact geometries of all size are included in the sum. Note that, while Hartle and Hawking suggested to sum over compact regular geometries, the path integral with Dirichlet boundary conditions is full of singularities: many of the off-shell geometries shrink to zero size in an irregular way causing linear perturbations to blow up. This does not happen with Robin or Neumann conditions where only the Hartle-Hawking saddle point geometry starts at zero size and the initial momentum is picked in such a way to ensure a regular closing off. We can then think about the focus on the initial momentum as a regularity condition and thus the Hartle-Hawking wavefunction as the correct one not because it describes a universe which starts at zero size but because it makes it close off regularly. There is hence here a shift of focus in what is the crucial essence of the no-boundary construction from the 'nothing'-ness to the smoothness of the early universe. This intuition is reinforced by the study of black holes in anti-de Sitter space. There, a Neumann initial condition of the same type was needed to make sure that the geometries in the sum had no conical deficit, hence implementing regularity at the horizon. Path integrals for black holes were studied in section \ref{sec:blackholes} where we demonstrated that the Hawking-Page phase transition can be recovered in our framework using a Kantowsky-Sachs type of ansatz. In this case we imposed Dirichlet boundary conditions at the ``outer'' boundary (what was the final boundary in cosmology) and Neumann conditions in the interior (the initial condition in cosmology). The Dirichlet condition at the ``outer'' boundary has the meaning of fixing, via holography, the temperature of the corresponding canonical ensemble. We studied in section \ref{thermodynamics} the thermodynamic interpretation of the saddle points corresponding to AdS black holes finding that Neumann boundary conditions in the interior are in fact necessary in order to obtain the correct canonical partition function of the system. We interpret these results for black holes as a strong support in favour of the Neumann path integrals for the no boundary proposal. The no-boundary wavefunction and the black holes partition function with the Neumann condition are truly sums over regular geometries in the sense of that we explicitly require that there are no irregular geometries in the sum. We start seeing in fact some elements of universality of the Neumann condition in the interior as it is the only condition which gives physically meaningful results and can indeed be seen as a regularity condition in all studied cases. This intuition requires certainly further study which we leave for future work. \\

A deep understanding of the saddle point approximation of gravitational path integrals is of crucial importance in many aspects of theoretical physics. Indeed what we know about the connection between thermodynamics and event horizons is based on this construction \cite{Gibbons:1977mu, Hartle:1976tp}. We observe a renovated attention in recent times towards gravitational path integrals \cite{Almheiri:2019qdq, Penington:2019kki}. The latest breakthrough in this field is certainly the novel understanding of the role that quantum extremal surfaces and islands \cite{Engelhardt:2014gca, Almheiri:2019hni, Penington:2019kki, Almheiri:2019qdq} play in the black holes information paradox \cite{Almheiri:2019psf, Almheiri:2019hni, Hashimoto:2020cas,  Penington:2019npb} and there also similar questions come about \cite{Almheiri:2020cfm, Lewkowycz:2013nqa, Faulkner:2013ana, Dong:2017xht}. In general, the best understood realizations of the holographic principle focus on identification between the saddle point approximation of gravitational path integrals and QFT partition functions. In this sense we believe it will be illuminating to understand the role of the Neumann initial condition in other situations which have a well understood holographic dual extending our calculation for example to arbitrary dimensions or more general types of spacetimes. Another important concept in holography, which is of relevance for our work, is the understanding of specific types of CFT deformations as cutting the bulk spacetime at a finite radius where the imposed boundary conditions are deformed accordingly with respect to those at infinity \cite{McGough:2016lol, Guica:2019nzm, Balasubramanian:2012hb}. It would be nice to see this effect at work in specific examples using the techniques developed in our studies.
 See for example Ref. \cite{Caputa:2019pam} and \cite{Donnelly:2019pie} for a use of minisuperspace integrals in this context. There is also another reason why the study of path integrals for asymptotically anti-de Sitter space is of interest for us. Our work is heavily based on the use of the minisuperspace approximation. If, on the one hand, this allows us perform many calculation explicitly with a good handle on the technicalities of the path integral evaluation, on the other hand, the question remains of how much of the physical input is actually lost with such simplification. We have already mentioned how the ABJM superconformal field theory three-sphere partition function can be recovered exactly with a minisuperspace path integral \cite{Caputa:2018asc, Hirano:2019szi}. It is an interesting question to ask whether this is just a fortunate coincidence or the minisuperspace approximation does indeed already include all of the relevant information in specific physical situations.
We saw in section \ref{sec:blackholes} how when AdS spacetime is cut at a finite radius additional saddle points become relevant to the path integral. This is a genuine prediction of our minisuperspace calculation in the sense of that there is no way to make sense of black hole thermodynamics avoiding such contributions. Hence, through the study of the dual role of such saddle points, holography could in principle provide a playground for testing the soundness of the minisuperspace approximation itself. \\

Our understanding of the early universe cosmology through the mechanism of inflation relies on the assumption of Bunch-Davies initial conditions for the primordial fluctuations. From a theoretical point of view, this assumption is really natural only if inflation is infinite. %However, in order to match observations, inflation must last only around NUMBER of e-folds, making this assumption strictly speaking inconsistent. 
We know however that this cannot be the case, at the very least because %there is another relevant scale in cosmology?, namely the Planck scale, where 
at the Planck scale new physics is expected to kick in, making this assumption strictly speaking inconsistent.
In order to derive concrete predictions for any specific model of inflation to compare with astronomical data, one may in many cases relatively safely ignore the problem. With the opposite take on it, one could say that these types of inconsistencies put the entire framework of inflation in serious trouble~\cite{Martin:2000xs, Agrawal:2018own,  Bedroya:2019snp}.
%A question we ask in theoretical cosmology of the early universe is whether there is something of fundamental nature that is inconsistency is signaling and whether this  fact can give us useful hints for answering the question of what happened in the early universe/how we should really picture the early universe. Following this way of reasoning, if there exists a regime where quantum gravity is approximated by semi-classical general relativity, then the no boundary proposal tells us why perturbations should start out in the Bunch-Davies vacuum.
%(more to it/), (is signaling/giving)useful hints towards 
A question we ask in theoretical cosmology of the early universe is whether there is something of fundamental nature that this inconsistency is signaling and whether this fact can give us useful hints for uncovering how we should really picture the early universe. Following this way of reasoning, if there exists a regime where quantum gravity is approximated by semi-classical general relativity, then the no boundary proposal tells us why perturbations should start out in the Bunch-Davies vacuum.
If the no boundary proposal was really inconsistent has claimed in Refs.  \cite{Lehners:2018eeo, Feldbrugge:2017mbc}, then there would really be an issue with the Bunch-Davies assumption itself. Not in the sense that this would make it a wrong assumption but because, from a theoretical point of view, there would be really no justification or firm ground to support it. What we showed in this work is that the Hartle-Hawking wavefunction is in fact well-defined and consistent within the framework on semi-classical gravity. This is means that one can consistently derive the Bunch-Davies condition from quantum cosmology. At the same time, we showed that the Hartle-Hawking wavefunction represents possibly something very different from the nucleation of the universe out of nothing. The challenge is now to deeply understand what this quantum state means and whether the no boundary proposal could really explain the initial conditions of the universe, rather than simply describe them. Our intuition is that the Hartle-Hawking wavefunction might come out as the necessary answer to the gravitational path integral in the semi-classical framework because of the intrinsic properties of such integrals, in particular due to regularity requirements implemented through Neumann boundary conditions.

\clearpage

\clearpage 
%%%%%%%%%%%%%%%%%%%%%%%%%%%%%%%%%%%%%%%%%%%%%%%%%%%%%%%%%%%%%%%%%%%%%%%%%%%%%%%%%%%%

\addcontentsline{toc}{section}{Acknowledgments}
\section*{Acknowledgments}

I wish to express my gratitude to my supervisor Jean-Luc Lehners, who has been a constant source of inspiration and a guidance during these years. I am grateful for sharing with me his view on science and cosmology, for all the blackboard discussions and for helping me become an independent scientist and thinker.\\
It was a pleasure to collaborate with extraordinary scientists such as Laura Sberna, Michal Heller, Sebastian Bramberger, Neil Turok and Job Feldbrugge.\\
I would like to especially thank Paolo Benincasa and Michal Heller for being my mentors in this last year of important decisions.\\
I would also like to thank Ana Alonso-Serrano, Teresa Bautista Solans and all the Soapbox Science Berlin team for showing me that empowerment comes from sharing experiences and knowledge and for the effort we do together to increase the visibility and recognition of the work of women* in science.\\
Anika Rast, Darya Niakhaichyk, Matthias Blittersdorf, Christina Pappa and Brit Holland are the people that make the AEI a warm and welcoming place to be.\\
I am thankful to the Max Planck Institute for Gravitational Physics (Albert Einstein Institute) for their continued support, in particular to the director Hermann Nicolai and Axel Kleinschmidt and Hadi Godazgar, organizers of the IMPRS program.\\
This work has benefited from numerous discussions with the members of the theoretical cosmology group Caroline Jonas, Shane Farnsworth, Ana Alonso-Serrano, George Lavrelashvili, Jerome Quintin,  Enno Mallwitz and Marc Schneider.\\
I am especially grateful to my friend, office mate and collegue Sebastian Bramberger for the endless conversations about the universe, from the mathematics to the philosophy of it.\\
I would like to thank my friends, which all became experts in cosmology during this journey, especially Roukaya Dekhil which knew already all about the world and helped me understand so much.\\
I wish to end by thanking my family which supported me in every single step. This thesis is dedicated to Bianca e Francesco, because we do science to make the world a better place for the next generation.

\clearpage

\shipout\null
%%%%%%%%%%%%%%%%%%%%%%%%%%%%%%%%%%%%%%%%%%%%%%%%%%%%%%%%%%%%%%%%%%%%%%%%%%%%%%%%%%%%

\appendix

\section{ The path integral satisfies the WdW equation }\label{WdWproof}

We show in this section that the path integral with initial canonical Robin condition satisfies the homogeneous or inhomogeneous Wheeler--deWitt (WdW) equation depending on the defining integration contour. Let us consider the normalised initial state 
\begin{equation}
\Psi(q_0) = \sqrt{\frac{2}{ \pi i \beta}} \, e^{i \alpha q_0 + i   \frac{q_0^2}{ 2  \beta}}\,.
\end{equation}
Note that $\Psi(q_0)$ is a coherent state for $\beta = - i |\beta|$. 
The path integral with Robin boundary conditions can be written as follows 
\begin{equation}
\Psi(q_1) = \int dN \,  d q_0 \,  \delta q \, e^{ i S_D} \,  \Psi(q_0) \, ,
\end{equation} 
where $S_D$ is the appropriate action for the Dirichlet problem for gravity i.e. the Einstein-Hilbert bulk term plus the Gibbons-Hawking-York boundary term,
\begin{equation}
S_D= S_{EH}+S_{GHY} = 3 V_3 \int_0^1  \left[ - \frac{\dot{q}^2}{4 N} + N (1 - H^2 q)\right] \,.
\end{equation}
The integral can be written as follows
\begin{equation}
\Psi(q_1) = \sqrt{\frac{2}{ \pi i \beta}}  dN \,  d q_0 \, \int \delta q \, e^{i S_0}\,,
\end{equation}
where
\begin{equation}
S_0 =3 V_3  \int_0^1  [ - \frac{\dot{q}^2}{4 N} + N (1 - H^2 q)] + \alpha q_0 + \frac{q_0^2}{2 \beta}
\end{equation}
is the total action, including the initial canonical boundary term.
The functional integral over $q$ gives
\begin{equation}
\Psi(q_1)  =\sqrt{\frac{2}{ \pi i \beta_0}} \int d N \,  d q_0 \sqrt{\frac{i 3 V_3 }{4 N}} e^{i S}\,,
\end{equation}
with 
\begin{equation}
S = 3 V_3 \frac{1}{3} \left(\frac{H^4 N^3}{4} - \frac{3 (q_1 - q_0)^2}{4 N} + 3 N \Bigl( 1 - \frac{H^2}{2}(q_1 + q_0)\Bigr) \right) + \alpha q_0 + \frac{q_0^2}{2 \beta}\,.
\end{equation}
To implement the no-boundary proposal, we need to fix $\alpha = - 3 V_3  i  $. Integrating over $ q_0 $, we find
\begin{equation}
\Psi(q_1 ) =  - i \sqrt{\frac{3 V_3}{2 i }}\int \frac{d N}{\sqrt{2 N  - 3 V_3 \beta }} e^{i \overline{S}} \,.
\end{equation}
The action $ \overline{S} = S(q_1 , \overline{q}_0)$ is evaluated at the saddle point $\overline{q}_0 = \frac{3 V_3 \beta (q_1 - N(2 i + H^2 N ))}{3 V_3 \beta  - 2 N}$,
\begin{equation}
\begin{split}
\frac{\overline{S}}{3 V_3} = \frac{1}{6 \left(2 N - 3 V_3 \beta \right)} \Bigl(&H^4 N^4 - 6 V_3 \beta  H^4 N^3 + 6 N^2 (2 - H^2 ( i 3 V_3 \beta  +   q_1)) + \\
& + 18 V_3 \beta  H^2 q_1 N + 3 q_1 ( i 6  V_3 \beta  -  q_1) \Bigr) \, .
\end{split}
\end{equation}
To evaluate the WdW equation, we need to compute
\begin{align}
\frac{\partial^2 \Psi(q_1)}{\partial q_1^2} & =   - i \sqrt{\frac{3 V_3}{2 i }} \int \frac{dN }{\sqrt{2 N -  3 V_3 \beta }} \Bigl[i \frac{\partial^2 \overline{S}}{\partial q_1^2} - \Bigl( \frac{\partial \overline{S}}{\partial q_1}\Bigr)^2 \Bigr] e^{i \overline{S}} \, . %\quad \frac{\partial^2 \overline{S}}{\partial q_1^2}  = \frac{3 V_3}{\beta - 2 N }\,,
\end{align}
We find
\begin{equation}
\begin{split}
 \int \frac{dN }{\sqrt{2 N -3 V_3 \beta}} i   e^{i \overline{S}} \frac{\partial^2 \overline{S}}{\partial q_1^2}  = &- 3 i  V_3 \int \frac{d N}{[2 N -3 V_3 \beta]^{3/2}} \, e^{i \overline{S}} = \\
 =&- 3 i V_3  \Bigl[ \int \frac{d N}{\sqrt{2 N -3 V_3 \beta}} \, i \,  \frac{\partial \overline{S}}{\partial N} \, e^{i \overline{S}} - %\int dN \frac{\partial}{\partial N} 
 \Bigl( \frac{e^{i \overline{S}}}{\sqrt{2 N -3 V_3 \beta}}\Bigr)\Bigr|_{boundary} \Bigr]\,.
 \end{split}
\end{equation}
Therefore 
\begin{align}
\frac{\partial^2 \Psi(q_1)}{\partial q_1^2} =  &- i \sqrt{\frac{3 V_3}{2 i }}  \int \frac{dN}{\sqrt{2 N -3 V_3 \beta}} e^{i \overline{S}} \Bigl[ - \Bigl(\frac{\partial \overline{S}}{\partial q_1} \Bigr)^2 + 3 V_3  \frac{\partial \overline{S}}{\partial N}\Bigr] + \sqrt{\frac{(3 V_3)^3}{2 i }}  \Bigl( \frac{e^{i \overline{S}}}{\sqrt{2 N -3 V_3 \beta}}\Bigr)\Bigr|_{boundary} \nonumber \\
= & \, i \sqrt{\frac{(3 V_3)^5}{2 i }} \int \frac{dN}{\sqrt{2 N -3 V_3 \beta}}  \Bigl[(H^2 q_1 - 1) \, e^{i \overline{S}} \Bigl]+ \sqrt{\frac{(3 V_3)^3}{2 i }} \Bigl(\frac{e^{i \overline{S}}}{\sqrt{2 N -3 V_3 \beta}} \Bigr) \Bigr|_{boundary}
\end{align}
If we consider an integration contour which runs from $N \rightarrow - \infty$ to $N \rightarrow + \infty$ along the real line $\Psi (q_1 )$ satisfies the WdW equation. Indeed, in this case the boundary term vanishes,
\begin{equation}
\lim_{N \rightarrow \pm \infty} \frac{e^{i \overline{S}}}{\sqrt{2 N -  3 V_3 \beta }} = 0 \, . 
\end{equation}
We thus obtain that $\frac{\partial^2 \Psi(q_1)}{\partial q_1^2}  =   - 9 V_3^2 (H^2 q_1 - 1) \Psi(q_1)$.  

The other possibility is to take the contour to run from the singularity at $N^* = \frac{3 V_3 \beta}{2}$ to $N \rightarrow + \infty$.  In order to calculate the boundary term, notice that for $N \approx N^*$ the action diverges as
\begin{equation}
\frac{\overline{S}}{3 V_3} \approx - \frac{1}{32 \left(2 N -  3 V_3 \beta \right)} \left(4 q_1 -  \beta_0  (4 i + 3 V_3 \beta H^2)\right)^2 \,.
\end{equation}
It was shown in~\cite{DiTucci:2019dji} and in Section~\ref{sec:canonical} that the relevant case for the no-boundary proposal is when $\beta$ takes negative imaginary values, $\beta = - i |\beta|$. In this case the singularity $N^* $ lies on the negative imaginary axis and the thimble approaches it along the axis, so that $N \sim i \, n$ as $N \rightarrow N^*$. 
The boundary term at the singularity is then proportional to a Dirac delta function
\begin{equation}
\begin{split}
\lim_{N \rightarrow N^*}  \frac{e^{i \overline{S}}}{\sqrt{2 N -3 V_3 \beta }} & = \lim_{x \rightarrow 0 } \sqrt{2 \pi i } \frac{1}{\sqrt{2 \pi i  x}} \, e^{- \frac{i 3 V_3}{4 x} \left(q_1 -\frac{3 V_3 \beta}{4}(4 i +  3 V_3 \beta  H^2)\right)^2} \\ 
&= \sqrt{\frac{4 \pi}{3  i V_3 }} \, \delta ( q_1 - \frac{3 V_3 \beta}{4} (4 i + \beta_0 H^2)) \, ,
\end{split}
\end{equation} 
where $x = N - N^* $ is purely imaginary. Therefore
\begin{equation}
\frac{\partial^2 \Psi(q_1)}{\partial q_1^2}  =   - 9 V_3^2(H^2 q_1 - 1) \Psi(q_1) - 3 V_3 i  \sqrt{2 \pi}  \,  \delta ( q_1 - \frac{3 V_3 \beta}{4} (4 i + 3 V_3 \beta H^2))\,,
\end{equation}
i.e. $\Psi$ satisfies the inhomogeneous WdW equation.

\clearpage

\section{Fluctuation determinant for mixed Neumann-Dirichlet boundary conditions} \label{sec:determinant}

In evaluating our path integrals, we could make use of the fact that the actions were quadratic in the scale factors, thus allowing a decomposition of the path into a classical solution $\bar{q}$ and a fluctuation $Q$, i.e. $q(r) = \bar{q}(r) + Q(r),$ with the resulting path integral over $Q$ being of Gaussian form,
\begin{align}
F(N) = \int_{\dot{Q}(0)=0}^{Q(1)=0} D[Q] e^{i\int_0^1 dr \frac{\dot{Q}^2}{N}}\,,    
\end{align}
where we have neglected an unimportant numerical factor in the exponent. To ensure that the total scale factor $q$ satisfies the mixed Neumann-Dirichlet boundary conditions, the fluctuation must satisfy $\dot{Q}(0)=0$ and $Q(1)=0.$ Here we would like to determine the dependence of the above integral on the lapse $N.$ To do so, we will use a re-scaled coordinate $\tilde{r}=rN,$ with range $0 \leq \tilde{r} \leq N.$ The integral then becomes 
\begin{align}
F(N) &= \int_{Q_{,\tilde{r}}(0)=0}^{Q(1)=0} D[Q] e^{i\int_0^N d\tilde{r} {Q}_{,\tilde{r}}^2} \nonumber \\ &= \int_{Q_{,\tilde{r}}(0)=0}^{Q(1)=0} D[Q] e^{- i\int_0^N {Q}_{,\tilde{r}} \frac{d^2}{d\tilde{r}^2} {Q}_{,\tilde{r}}} = \sqrt{\frac{2}{\pi i}}\left[\text{det}\left(-\frac{d^2}{d\tilde{r}^2}\right)\right]^{-1/2}\,.
\end{align}
With the assumed boundary conditions, the operator $-\frac{d^2}{d\tilde{r}^2}$ satisfies the eigenvalue equation $-\frac{d^2}{d\tilde{r}^2} x_n = \lambda_n x_n$ with eigenfunctions $x_n$ and eigenvalues $\lambda_n,$
\begin{align}
    x_n = a_n \cos\left[\frac{(2n+1)\pi}{2N}\tilde{r} \right]\,, \quad \lambda_n = \left[\frac{(2n+1)\pi}{2N}\right]^2\,, \quad n \in \mathbb{N}\,.
\end{align}
The determinant is given by the product of all eigenvalues. We can evaluate it using zeta function regularisation (see e.g. Ref.~\cite{Grosche:1998yu}). Thus in analogy with the zeta function $\zeta(s)=\sum_{n \in \mathbb{N}}n^{-s}$ we define
\begin{align}
    \zeta_\lambda(s) \equiv \sum_{n \in \mathbb{N}} \lambda_n^{-s} = \left(\frac{2N}{\pi} \right)^{2s}\sum_{n\in \mathbb{N}} \frac{1}{(2n+1)^{2s}}\,.
\end{align}
The last term corresponds to the zeta function where one would sum only over odd terms. We can obtain this sum by subtracting the even terms,
\begin{align}
    1+\frac{1}{3^{2s}}+\frac{1}{5^{2s}}+\cdots & = 1+\frac{1}{2^{2s}}+\frac{1}{3^{2s}}+\cdots - [\frac{1}{2^{2s}}+\frac{1}{4^{2s}}+\cdots] \nonumber \\ & = 1+\frac{1}{2^{2s}}+\frac{1}{3^{2s}}+\cdots - \frac{1}{2^{2s}} [1+ \frac{1}{2^{2s}}+\frac{1}{3^{2s}}+\cdots]\,.
\end{align}
Hence we obtain 
\begin{align}
    \zeta_\lambda(s) = \left(\frac{2N}{\pi} \right)^{2s} \left( 1 - 2^{-2s}\right)\zeta(s)\,.
\end{align}
The zeta function can be analytically continued to $s=0,$ where the derivative $\zeta_\lambda^\prime(0)$ is related to the product of all $\lambda_n$ such that
\begin{align}
    \left[\text{det}\left(-\frac{d^2}{d\tilde{r}^2}\right)\right] = e^{-\zeta_\lambda^\prime(0)}=2\,,
\end{align}
where we have made use of $\zeta(0)=-\frac{1}{2}.$ In the end we find the remarkably simple result that 
\begin{align}
    F(N) = \frac{1}{\sqrt{\pi \, i}}\,.
\end{align}
In particular, note that the fluctuation determinant for the Neumann-Dirichlet problem does not contain any dependence on the lapse $N,$ unlike in the well known pure Dirichlet case where the determinant is proportional to $N^{-1/2}$ \cite{Grosche:1998yu}.

\clearpage

\shipout\null

%%%%%%%%%%%%%%%%%%%%%%%%%%%%%%%%%%%%%%%%%%%%%%%%%%%%%%%%%%%%%%%%%%%%%%%%%%%%%

\addcontentsline{toc}{section}{References}

\bibliography{Bibliotesi}

\end{document}